\documentclass[acmsmall,screen,nonacm]{acmart} 
\usepackage{listings}
\usepackage{pifont}
\usepackage{braket}
\usepackage{wrapfig}
\usepackage{subcaption} 
\usepackage{amsthm}
\usepackage[linesnumbered,ruled,vlined]{algorithm2e}
 \usepackage{tikz}
  \usetikzlibrary{arrows.meta,positioning}
  \definecolor{accent}{RGB}{128, 128, 128} 
  \definecolor{mark}{RGB}{0, 0, 139}

\newcommand{\cmark}{\ding{51}}
\newcommand{\xmark}{\ding{55}}

\newcommand{\RQC}{\textbf{RQC}}

\newcommand{\abort}{\mathbf{abort}}
\newcommand{\main}{\mathtt{main}}
\newcommand{\M}{\mathcal{M}}
\newcommand{\St}{\mathcal{S}}
\newcommand{\K}{\mathcal{K}}
\newcommand{\G}{\mathcal{G}}
\newcommand{\D}{\mathcal{D}}

\newcommand{\ctx}{\mathit{ctx}}
\newcommand{\Qpool}{\mathcal{Q}}
\newcommand{\Bbuf}{\mathcal{B}}
\newcommand{\args}{\mathrm{args}}
\newcommand{\alloc}{\mathrm{alloc}}
\newcommand{\activelabel}{\mathrm{active}}
\newcommand{\Vactive}{V_{\activelabel}}
\newcommand{\Eargs}{E_{\args}}
\newcommand{\Valloc}{V_{\alloc}}
\newcommand{\Outd}{\operatorname{Outd}}
\newcommand{\sys}{\mathit{sys}}

\newcommand{\Gar}{\mathit{Gar}}
\newcommand{\sz}{\mathit{sz}}
\newcommand{\mvq}{\mathsf{mvq}}
\newcommand{\rvq}{\mathsf{rvq}}
\DeclareMathOperator{\dom}{dom}
\DeclareMathOperator{\cod}{cod}
\DeclareMathOperator{\pos}{pos}
\DeclareMathOperator{\tr}{tr}
\DeclareMathOperator{\qinv}{qinv}
\DeclareMathOperator{\anc}{anc}

\DeclareMathOperator{\qv}{qv}
\DeclareMathOperator{\base}{base}
\DeclareMathOperator{\node}{node}
 
\newcommand{\outd}{\ensuremath{\operatorname{Outd}}}
\DeclareMathOperator{\Fun}{Fun}
\DeclareMathOperator{\Call}{Call}
\DeclareMathOperator{\SCC}{SCC}

\DeclareMathOperator{\Succ}{succ}
\newcommand{\contuses}{\ensuremath{\operatorname{contuses}}}
\newcommand{\idx}{idx}
\DeclareMathOperator{\fp}{fp}
\DeclareMathOperator{\req}{req}
\DeclareMathOperator{\re}{re}
\DeclareMathOperator{\push}{push}
\DeclareMathOperator{\pop}{pop}
\newcommand{\flag}{\ensuremath{\mathrm{flag}}}
\newcommand{\var}{\mathsf{var}} 
\newcommand{\getQ}{\ensuremath{\operatorname{get\_Q}}}
\newcommand{\getC}{\ensuremath{\operatorname{get\_C}}}

\newtheorem{notation}{Notation} 

\newtheorem{assumption}{Assumption}

\definecolor{keywordcolor}{RGB}{0, 0, 139}    % 深蓝紫色（关键字 def, if, return）
\definecolor{typecolor}{RGB}{122, 59, 46}    
\definecolor{commentcolor}{RGB}{128, 128, 128}  % 墨绿色（注释）
\definecolor{numbercolor}{RGB}{128, 128, 128} % 灰色（行号）
\definecolor{stringcolor}{RGB}{163, 80, 30}   % 橙色（字符串/数字）

\lstdefinelanguage{MyLang}{
    keywords={def, if, then, return, main, nat,sz, fi,qif,fiq,qinv,skip, and,else, Q,circ,bool},    % 蓝色粗体关键字
    keywordstyle=\color{keywordcolor}\bfseries,
    morekeywords=[2]{sum, aux, rz_suber_c_form, QFT, rz_suber_c,RQFT,X,beq,rz_adder_c_form,rz_adder,rz_adder',init_c,copy,rz_adder_full_form,RZ,Fsum,Faux,Fmain,F_find,find,DFS,F_DFS,F_aux1, F_aux2, locate, rz_power, rz_power_form,power_2,rz_mult_c},   % 第二类关键字（类型）
    keywordstyle=[2]\color{typecolor},
    sensitive=true,
    morecomment=[l]{//}, 
}

\AtBeginDocument{%
  }

\newcommand{\whl}[1]{{#1}}
\newcommand{\rev}[1]{{#1}}

\begin{document}

\title{ReOC: Compilation of Recursive Quantum Oracles with Recursion-Aware Uncomputation} 

\author{Huiling Wu}
\email{52275902009@stu.ecnu.edu.cn}
\affiliation{
  \institution{Shanghai Key Laboratory of Trustworthy Computing, East China Normal University}
  \city{Shanghai}
  \country{China}}

\author{Yuxin Deng}
\authornote{Corresponding author.}
\email{yxdeng@msg.sufe.edu.cn}
\affiliation{
  \institution{MoE Key Laboratory of Interdisciplinary Research of Computation and Economics, Shanghai University of Finance and Economics}
  \city{Shanghai}
  \country{China}}
\affiliation{
  \institution{Shanghai Key Laboratory of Trustworthy Computing, East China Normal University}
  \city{Shanghai}
  \country{China}}
% \affiliation{%

% \affiliation{%

% \affiliation{%

% \affiliation{%

% \affiliation{%

% \affiliation{%

% \affiliation{%

\renewcommand{\shortauthors}{Wu and Deng}

\begin{abstract}
 Quantum oracles are essential to many quantum algorithms, and their specifications may involve recursive control flow that depends on runtime quantum data. However, existing reversible compilation frameworks provide limited support for such quantum-controlled recursive structures. 
 
 We present ReOC, a compilation framework that transforms high-level recursive
  oracle specifications with quantum control flow into reversible quantum programs. The framework comprises RQIMP, a 
  high-level imperative source language for specifying recursive oracles, and  
  a method of compiling programs in that language into \(\RQC^{++}\), an existing high-level quantum recursive language with quantum control flow. 
  In this way, we avoid the tedious and error-prone process of directly writing quantum oracles in \(\RQC^{++}\). 
  
  To manage static storage under dynamic quantum control, ReOC uses an indexed static-register discipline to isolate live variables across recursion layers, enabling safe register reuse while controlling quantum storage usage. Furthermore, to address the exponential time blow-up caused by naive uncomputation in recursive settings, ReOC employs a recursion-aware uncomputation strategy: temporary variables from recursive calls are cleaned using deferred strategies to control time overhead, while those from non-recursive statements are cleaned eagerly to reduce space usage. For linear recursion, this strategy yields overhead linear in recursion depth, parameterized by the per-layer register footprint and primitive-operation costs.  Finally, we provide a mathematical proof of compilation correctness from RQIMP to \(\RQC^{++}\), establishing semantic preservation and correct uncomputation of temporary quantum variables. 
\end{abstract}

\maketitle

\section{Introduction}\label{sec_intro}

Some quantum algorithms can achieve substantial speedups over their best-known classical counterparts, as exemplified by Shor's factoring algorithm~\cite{shor1997} and Grover's search algorithm~\cite{Gro96}. A key source of such speedups is the ability to coherently evaluate classical functions in superposition, typically realized via quantum oracles. Although such oracles may correspond to familiar classical computations, they often account for a large fraction of the cost of the compiled quantum program. Therefore, efficient implementation of quantum oracles is crucial for
realizing practical quantum speedups.

Prior work on reversible and quantum-oracle construction ranges from low-level reversible
synthesis~\cite{10.1145/775832.775915,10.1145/1278349.1278355,
8a301c2571004c6bb12a10cf788b8b2b,shafaeiReversibleLogicSynthesis2013,
linRMDDSReedmullerDecision2014a,
saeediSynthesisOptimizationReversible2013}
  to high-level languages and compilers for generating reversible
circuits~\cite{oemer2000quantum,Abhari2012ScaffoldQP,
10.1145/3183895.3183901,thomsenFunctionalLanguageDescribing2012,
randReQWIREReasoningReversible2019,willeSyReCProgrammingLanguage2010,
10.1145/2491956.2462177,parent2015reversiblecircuitcompilationspace}. Notably, ReVerC~\cite{DBLP:conf/cav/AmyRS17} provides a verified compiler for
  space-efficient reversible circuits, while VQO~\cite{liVerifiedCompilationQuantum2022}
  provides high-assurance oracle
  compilation with efficient QFT-based arithmetic. However, in practical quantum algorithms, oracle specifications can involve
  complex structures, including loops and recursion whose execution
  path or recursion depth depends on the value of data in quantum superposition. Representative examples include oracles arising in quantum algorithms for
  graph problems~\cite{durrQuantumQueryComplexity2006,
  ambainisQuantumSpeedupsExponentialTime2018,
  klevickis_et_al:LIPIcs.TQC.2022.11}:
  \(
  \sum_k \ket{k}\ket{x} \mapsto \sum_k \ket{k}\ket{x \oplus g(k)},
  \)
  where the specification of \(g(k)\) may contain recursive subprocedures
  (e.g., recursive graph traversal such as depth-first search) whose control flow depends on 
  the superposed value \(k\). We refer to this setting as \emph{(dynamically) quantum-controlled recursion}; such specifications remain less directly supported by
  existing oracle-construction tools.  
  % motivating more direct support for implementing oracles with quantum-controlled recursion.
  At the same time, efficient execution of these algorithms
  typically requires such subprocedures to run in polynomial time, 
  which highlights the importance of efficient implementations of oracles with
  quantum-controlled recursion.
  
%The implementation of 
To implement
such quantum oracles, 
we need a quantum programming language capable of representing
    recursive structures with quantum control flow. In classical programming languages, recursion is a natural abstraction. It can
  be implemented through call stacks, program counters, and conditional jumps,
  and is thus natively supported by classical architectures.  In the quantum setting,  
  programs are typically represented as quantum circuits, i.e., fixed sequences of  
  logic gates whose structure cannot depend on data in quantum superposition, a limitation emphasized in prior work~\cite{yuanQuantumControlMachine2024}. Although emerging quantum programming languages and oracle compilers provide
  certain abstractions for quantum control flow, including quantum conditional constructs~\cite{bichselSilqHighlevelQuantum2020,
hirataQurtsAutomaticQuantum2025,voichickQunityUnifiedLanguage2023},
  quantum-controlled recursion remains less direct in these systems.
  Tower~\cite{yuanTowerDataStructures2022}, the first quantum programming language
  with random-access memory, supports quantum data structures and recursive
  programs over data in superposition, while requiring classical recursion bounds
  to ensure compilability to quantum circuits.  
  For our purposes, $\RQC^{++}$~\cite{yingVerificationRecursivelyDefined2024} is a
  natural target: it not only supports quantum conditional constructs and quantum-controlled
  recursive programs at the language level, but is also supported by the QRM
  framework~\cite{zhangQuantumRegisterMachine2025} through a compilation backend
  to a computational model with both quantum control flow and recursive procedure
  calls. 
  
  % but also provides %an existing
  % \yx{a} 
  % compilation %path 
  % \yx{method} to QRM~\cite{zhangQuantumRegisterMachine2025}, a
  % computational
  % model for realizing such dynamic recursive execution. 

%However, 
Although $\RQC^{++}$ serves as a useful target language for quantum oracles with dynamic
recursion, programming directly in $\RQC^{++}$ still poses two
main challenges: (1) the resulting programs are verbose and lack high-level
structured abstractions, making it difficult to preserve the structure and
intent of the original algorithm; and (2) the low-level programming process is
cumbersome, significantly increasing both the implementation burden and the risk of errors. 

To address these challenges, we introduce
a high-level source abstraction together with a
  compiler: programmers express recursive oracles in a
  classical-style structured language, while the compiler 
  transforms them into
  reversible $\RQC^{++}$ programs to realize dynamically controlled recursive
  execution. In this way, the high-level structure of the source program remains explicit
  to the programmer, while the complex and error-prone tasks are delegated to the compiler, thereby mitigating the practical difficulties of direct $\RQC^{++}$ programming.  However, this compilation process is nontrivial and faces two additional technical challenges. 
  % However, the compilation process is nontrivial and must overcome the following two issues.

\paragraph{Static storage under dynamic quantum control flow} In the $\RQC^{++}$/QRM execution model used here, quantum data are organized as
  statically allocated arrays, with no runtime allocation of fresh quantum data.
  Thus, the available quantum storage must be determined at compile time. However, when control flow depends on quantum data at runtime, the compiler cannot simply statically unroll all function calls. This raises a key problem: without static unrolling, how can the compiler
  statically manage quantum storage while preventing temporary quantum variables from
  different recursion layers from conflicting? 

A straightforward approach is to represent all temporaries as indexed quantum
  arrays and increment the index at each recursive call, so that each recursion
  layer uses independent storage. However, because quantum storage is expensive,
  indexing every temporary is overly conservative and may introduce unnecessary
  ancilla overhead. It is therefore important to identify which
  variables require per-layer isolation and which can be shared across
  recursion layers, so as to safely reuse the same static
  register pool across recursive layers, ensuring correct program
  execution while controlling ancilla overhead.

\paragraph{Uncomputation of temporary quantum variables} Unlike in classical programs, where temporary variables can be discarded once they are no longer needed, reversible quantum programs must explicitly restore such variables to their initial states through uncomputation. Starting from Silq~\cite{bichselSilqHighlevelQuantum2020}, the first quantum
  language to support safe automatic uncomputation, several languages and tools
  have developed automatic or optimization-oriented uncomputation  strategies~\cite{
  paradisUnqompSynthesizingUncomputation2021,
paradisReqompSpaceconstrainedUncomputation2024,
  hirataQurtsAutomaticQuantum2025,10.1007/978-3-031-38100-3_11,
sharmaOptimizingAncillaBasedQuantum2025,10.1145/3779212.3790134,
  khattarRiseConditionallyClean2025,venevModularSynthesisEfficient2024}. Since $\RQC^{++}$ does not automate this process, our compiler is responsible for this task. 

In general, the timing of uncomputing temporary quantum variables is not
  unique, and different choices directly affect resource overhead. Late uncomputation
   keeps variables alive and can incur significant space overhead, as in
  Bennett-style delayed cleanup~\cite{10.1147/rd.176.0525}. Early uncomputation  may introduce extra time
  overhead: in particular, in a linear recursive program, where each recursion level makes at most one recursive 
  subcall, naive early
  uncomputation can force each layer to recompute recursive subcalls, turning a
  source-level linear recursion into exponential recursive recomputation, as illustrated by prior work~\cite{yuanTowerDataStructures2022,venevModularSynthesisEfficient2024}. 
  % Although this blow-up can sometimes be avoided by manually rewriting programs
  % into a tail-recursive form~\cite{yuanTowerDataStructures2022}, such rewriting is
  % restrictive and is not a general solution for recursive programs. \whl{Venev et al.~\cite{venevModularSynthesisEfficient2024} propose the first modular
  % automatic method for correct and efficient uncomputation in expressive quantum
  % programs, avoiding redundant recomputation in recursive programs by deferring
  % cleanup.}
Together, these observations show that recursive uncomputation requires a
  careful choice of cleanup timing: we may need to delay cleanup in some
  recursive cases to avoid exponential recomputation, while still avoiding
  excessive delay to control space overhead.

% Together, these observations show that recursive uncomputation requires a careful choice
%   of cleanup timing:  avoid naive exponential recomputation in
%   linear-recursive cases while effectively managing space overhead.
  
% Moreover, although existing work studies various time-space trade-offs in
%   uncomputation~\cite{paradisUnqompSynthesizingUncomputation2021,paradisReqompSpaceconstrainedUncomputation2024,venevModularSynthesisEfficient2024,10.1007/978-3-031-38100-3_11,sharmaOptimizingAncillaBasedQuantum2025}, these techniques often do not systematically
%   distinguish cleanup timing for temporaries generated by different program structures. Therefore, it
%   is important to design an uncomputation strategy  so as to avoid naive exponential recomputation in
%   linear-recursive cases while effectively managing space overhead. 

Therefore, the core challenge is not merely to translate high-level recursive
oracle programs into $\RQC^{++}$ and generate uncomputation code, but to do so in
a way that preserves correctness while controlling both time and space overheads. 

% \subsection{ReOC: The Compilation of Recursive Quantum Oracles}
\vspace{0.3cm}
\noindent\textbf{This work.}
  We present ReOC, a structured compilation framework for
  high-level recursive quantum-oracle programs. ReOC compiles well-typed source programs involving dynamically controlled loops,
  recursion, and mutual recursion to $\RQC^{++}$ and automatically generates
  recursion-aware uncomputation code for temporary quantum variables. By targeting \(\RQC^{++}\), ReOC makes the generated programs compatible with
  the existing \(\RQC^{++}\)-to-QRM backend for dynamically controlled recursive
  execution. 
  
  % \whl{reusing the existing $\RQC^{++}$/QRM
  % dynamic-recursion backend}, while automatically generating recursion-aware
  % uncomputation code for temporary quantum variables. 

We first introduce RQIMP, a simple   
high-level  
language for specifying recursive quantum oracles. It supports dynamically controlled
  loops and (mutual) recursion, allowing programmers to write oracles in a style
  close to classical programming. The language is also equipped with a type
  system
  that restricts source programs so that the compiler can generate well-formed 
  $\RQC^{++}$ programs and establish uncomputation correctness for temporary variables.  

We then present a compiler from RQIMP to $\RQC^{++}$. 
 The two key ingredients are as follows.
 
\paragraph{Distinguishing shareable from non-shareable variables across recursion layers.}
We present an indexed static-register discipline for recursive 
  $\RQC^{++}$ procedures. To distinguish variables that require per-layer
  isolation from those that can be safely reused, the compiler records, for each
  function, the variables that remain live at each compilation point. When
  compiling a recursive call in a function \(f\), the variables live in \(f\) at
  that point are represented using indexed quantum arrays, and their
  indices are incremented when passed to the recursive invocation. 
  By contrast, variables that have already been cleaned require no such treatment,
  since they can be safely reused by subsequent recursion layers. 
  Furthermore, for variables
  introduced after the recursive call, the compiler assigns fresh index
  variables,
  so that index increments are confined to temporaries that actually require
  recursive-layer isolation. In this way, variables that must not be shared are
  isolated across recursion layers, while others may be safely reused. 

Through this strategy, the compiler ensures that the generated $\RQC^{++}$
  program can safely reuse the same static register pool across recursion layers, while keeping the ancilla usage of temporary
  variables
  under control. 

 \paragraph{Performing recursion-aware uncomputation for temporary quantum variables.} ReOC automatically generates recursion-aware uncomputation code. To avoid
  potential exponential recomputation while controlling space overhead, we
  propose
  a structure-guided, selective uncomputation strategy that applies different
  cleanup policies according to the origin of temporary variables.
  Specifically, we adopt the following selective cleanup policies:
  \begin{enumerate}
      \item For variables introduced by non-recursive statements, we adopt eager
      cleanup~\cite{DBLP:conf/cav/AmyRS17} to reduce space usage. Its core idea is to clean a variable immediately
  once its lifetime ends and it can be safely uncomputed. 
      \item For variables introduced by recursive-call statements, we adopt deferred 
      cleanup to avoid recomputation: for linear recursion, cleanup is deferred
      until after the entire recursion completes; for non-linear recursive
      structures, where a recursion layer may contain multiple recursive subcalls, cleanup is deferred until after each layer completes.
  \end{enumerate} 

For linear recursion, deferred cleanup incurs time and space overheads
  that are linear in the recursion depth, parameterized by the per-layer
  primitive-operation costs and register footprint. Eager cleanup of non-recursive temporaries further reduces the per-layer live
  footprint, helping control the constant factors in this linear bound. For non-linear recursion, the corresponding strategy avoids deferring all recursive-call
  garbage to the outermost call and provides a practical
  time-space trade-off.   

Finally, we provide a mathematical proof of correctness for the compilation from RQIMP to $\RQC^{++}$, establishing source-target semantic preservation and showing that all temporary quantum
  variables in the generated target program are properly uncomputed.

\vspace{0.3cm}
\noindent\textbf{Organization and Contributions.} For orientation, Section~\ref{sec:technical-overview} presents a running example
  that motivates the problem and illustrates the main compilation challenges. The subsequent sections
present our main contributions: 
\begin{itemize}
    \item Section~\ref{sec:source} introduces RQIMP, our high-level
    imperative language.

    \item Section~\ref{sec:key} presents two key compilation techniques:
    indexed static-register management and recursion-aware uncomputation.

    \item Section~\ref{sec:comp} presents the compilation from RQIMP programs
    to $\RQC^{++}$.

    \item Section~\ref{sec:veri} provides an outline of the mathematical
    proof of compiler correctness.
\end{itemize}
Finally, Section~\ref{sec:rela} reviews related work, and
Section~\ref{sec:con} concludes the paper. Relevant basic concepts of $\RQC^{++}$, together with further details and examples,
  are given in the appendices.

\section{A Running Example} 
\label{sec:technical-overview} 
In this section we introduce a running example to motivate the compilation of recursive quantum oracles and highlight the key challenges.

Consider a simple recursive oracle that computes a summation:
\(
\sum_k \ket{k}\ket{x} \mapsto \sum_k \ket{k}\ket{x \oplus \texttt{sum}(k)},
\)
where \texttt{sum} is defined recursively as:
\[
\texttt{sum}(k) =
\begin{cases}
0 & k = 0 \\
k + \texttt{sum}(k-1) & \text{otherwise}.
\end{cases}
\]
In this quantum oracle, the recursive control path is determined by the variable \(k\), which is in
superposition and thus not known at compile time as a classical value.
\begin{figure}[t] 
\centering
\begin{minipage}[t]{0.69\textwidth}
        \centering
\includegraphics[width=\linewidth]{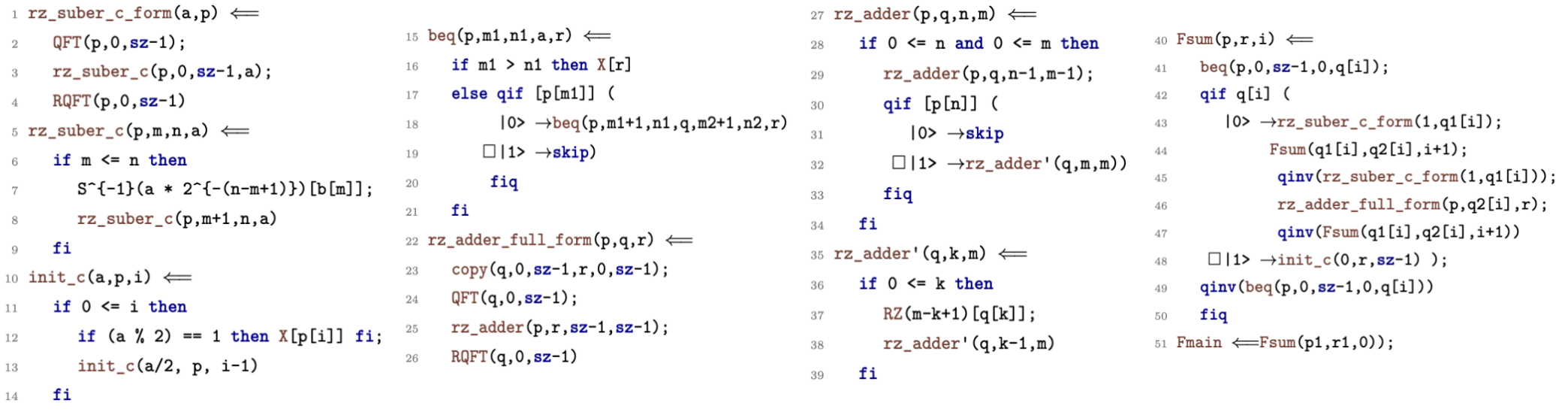}
        \caption{$\RQC^{++}$ program for $\mathtt{sum}$ (more
readable version in Appendix~\ref{app:compare-rqc}).   
}
        \label{fig:RQC_sum}
        \Description{...}
\end{minipage}
\hfill
\begin{minipage}[t]{0.30\textwidth}
\centering
\includegraphics[width=\linewidth]{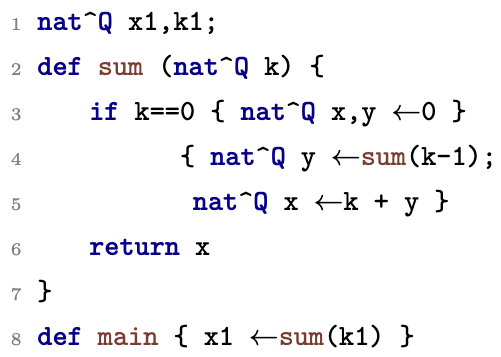}
\caption{RQIMP program for $\mathtt{sum}$.}
\label{fig:RQIMP_sum}
\end{minipage}
\end{figure} 
A direct implementation of this oracle in \(\RQC^{++}\), as shown in Figure~\ref{fig:RQC_sum}, 
produces verbose code that lacks high-level structured abstractions and obscures the original program semantics. By contrast, our high-level language allows the \texttt{sum} program to be written directly in the concise form shown in
  Figure~\ref{fig:RQIMP_sum}, which is intuitive and closely follows classical programming practice. 
Moreover, writing such code directly in \(\RQC^{++}\) is cumbersome and error-prone. 
Specifically, the main difficulty arises when recursive calls are performed while some quantum variables at the current recursion level remain active (i.e., used but have not yet been cleaned). If the target function is invoked without special handling, the next recursion layer will continue operating on these same variables. For instance, Figure~\ref{fig:RQC_sum_incor} shows a schematic $\RQC^{++}$ implementation of the \(\mathtt{sum}\) function, where quantum register \(p\) holds the input \(k\), \(r\) stores the return value \(x\), and \(q, q_1, q_2\) are temporaries that store the result of \((k==0)\), the value \((k-1)\), and the local variable \(y\), respectively. For readability, we abstract away the details of expression compilation and simply write \(\operatorname{circ}(e,q)\) to denote the quantum circuit that evaluates the expression \(e\) and stores the result in the quantum variable \(q\). In this formulation, the recursive call at line~5 causes the next layer to reuse variables such as \(q\) to store \((k==0)\), while they remain live---since they are not uncomputed until line~10 via \(\qinv(C)\) (the inverse of \(C\))---leading to data conflicts. To ensure that each recursive layer operates on a fresh, zero-initialized quantum state---thereby preserving correctness without overwriting existing values---we introduce an index variable (e.g., $i$) as an additional parameter. The correct formulation therefore takes the form shown in Figure~\ref{fig:RQC_sum_cor}, where variables are represented as indexed arrays and each recursive call increments the index $i$ to select a clean set of temporary variables. 
Here an indexed, or subscripted, quantum variable such as \(q[i]\) denotes an
  array element in \(\RQC^{++}\); the element may be a single qubit or a
  multi-qubit typed register, depending on its type.

However, a fundamental challenge remains: how to precisely determine which variables need to be indexed and which can be safely shared across recursion layers, so as to ensure correct execution while controlling quantum variable usage. While this issue can be handled manually for this simple example, it becomes significantly more complex for general recursive functions involving mutual recursion and multiple recursive paths. 

Furthermore, \(\RQC^{++}\) lacks an automatic uncomputation mechanism. As shown
in Figure~\ref{fig:RQC_sum_cor}, programmers are responsible for manually
  inserting uncomputation code (lines~6, 8, and 10). Beyond this programming burden, a plain uncomputation strategy can incur exponential overhead. In the implementation shown in Figure~\ref{fig:RQC_sum_cor}, \(F_{\mathtt{sum}}\) is invoked at line~5 and \(q_2[i]\) (the quantum variable for the non-return local variable \(y\)) is uncomputed at line~8, when its lifetime ends. Since each layer performs computation and uncomputation separately, the number of such invocations grows exponentially with \(k\).  As a result, the timing of
  uncomputation, its specific implementation, and the choice of when to apply it to
  control resource usage (e.g., to avoid exponential blow-up) are left to the
  programmer.  

In summary, $\RQC^{++}$ delegates the complex tasks of both recursive-state management and temporary-variable uncomputation entirely to the programmer. 
This not only significantly increases programming complexity, but also greatly increases the risk of errors. 

\begin{figure}
\begin{minipage}[t]{0.46\textwidth}
    \centering
   \begin{lstlisting}[language=MyLang, basicstyle=\ttfamily\scriptsize]
Fsum(p,r) <==
   circ(k==0, red(q)); 
   qif red(q) (
       |0> -> circ(k-1, red(q1)); 
            Fsum(red(q1),red(q2));  // (y <- sum(k-1))
            qinv(circ(k-1, red(q1))); 
            circ(k+y, r); // (x <- k+y)
            qinv(Fsum(red(q1),red(q2)))  
     square |1> -> circ(0, r) ); // (x <- 0)
   qinv(circ(k==0, red(q)));  
   fiq
   \end{lstlisting}
    \caption{$\RQC^{++}$ $\mathtt{sum}$: incorrect use of temporaries.}
    \label{fig:RQC_sum_incor}
    \end{minipage}
    \hfill
    \begin{minipage}[t]{0.53\textwidth}
    \centering
    \begin{lstlisting}[language=MyLang, basicstyle=\ttfamily\scriptsize]
Fsum(p,r,i) <==
   circ(k==0, red(q[i]));
   qif red(q[i]) (
      |0> -> circ(k-1, red(q1[i]));
          Fsum(red(q1[i]),red(q2[i]),red(i+1));
          qinv(circ(k-1, red(q1[i]))); // Uncompute q1[i] 
          circ(k+y, r); 
          qinv(Fsum(red(q1[i]),red(q2[i]),red(i+1))) // Uncompute q2[i]
    square |1> -> circ(0, r) );
   qinv(circ(k==0, red(q[i]))) // Uncompute q[i] 
   fiq
    \end{lstlisting}
    \caption{$\RQC^{++}$ $\mathtt{sum}$: correct use of temporaries.}
    \label{fig:RQC_sum_cor}
    \Description{...}
    \end{minipage}
\end{figure}  

  With our high-level language and compiler, programmers can express recursive logic in a familiar style, without managing low-level details such as quantum-variable usage, index management, or the timing and strategy for uncomputation. These tedious and error-prone tasks are fully handled by the compiler, which compiles the high-level program (Figure~\ref{fig:RQIMP_sum}) into the target form (Figure~\ref{fig:RQC_sum}) and systematically addresses the challenges arising from recursive transformation and uncomputation. In this way, the compiler preserves the high-level structure of the source program, while reducing the programmer's burden and supporting correctness and controlled resource usage in the generated quantum program. In Section~\ref{sec:key}, we show how the compiler addresses these challenges through two mechanisms: an indexed static-register discipline that distinguishes shareable from non-shareable variables across recursion layers, and a recursion-aware uncomputation strategy that avoids naive exponential recomputation while controlling space overhead.

\section{RQIMP: High-level Programming Language}
\label{sec:source}

In this section, we present an imperative high-level programming language (RQIMP) that enables 
programmers to write quantum oracles in a more concise and straightforward way, with particular support for recursive programs with dynamic control. We present its syntax, type system, and semantics. The
syntax and semantics of the target language $\RQC^{++}$ are provided in Appendix~\ref{app:tar}.

\subsection{Syntax} 
The RQIMP program syntax includes variable type declarations, values, lvalues, expressions, statements, and function declarations, which are described in Table~\ref{tab:syn}.  

\textit{Types.} The basic types comprise natural numbers and Booleans, each of which has two modes: the $\mathtt{C}$ mode indicates that the variable serves as a classical parameter of the oracle, while the $\mathtt{Q}$ mode represents quantum inputs to the oracle. Values of the two modes are implemented using classical variables (for the $\mathtt{C}$-mode) and quantum variables (for the $\mathtt{Q}$-mode) respectively. The types include unit, basic types, arrays, and products. We assume that natural numbers range over $[0, 2^{\sz}-1]$, where $\sz$ is the size of the machine word. This directly determines the number of qubits required for their representation in quantum compilation. 

\begin{table}[t]
    \centering
        \caption{RQIMP types and syntax}
    \begin{gather*}
    \begin{array}{llll}
      \mathrm{Mode}   &\mathtt{q}&:= & \mathtt{C} \ | \ \mathtt{Q}  \\
      \mathrm{Base type}   & \omega &:= & 
      \mathtt{nat} \ | \ \mathtt{bool} \ \\ 
      \mathrm{Type}   & \tau &:= & \mathtt{unit} \ | \ 
      \omega^{\mathtt{q}} \ | \ \mathtt{array} \ n \ \tau  \ | \ (\tau_1, \tau_2)  \ \\ 
 \\[0.1em]  
      \mathrm{Value}   &v &:= & () \ | \ l  \ | \ n \ | \ \mathtt{true} \ | \  \mathtt{false}  \    \\
       \mathrm{LValue}   & l &:= & 
      x\ | \ x[v] \ | \ \pi_1(x) \ | \  \pi_2(x) \ | \ (x_1, x_2)  \   \\ 
     \mathrm{Operator}   & \mathtt{aop} &:= & \mathtt{+} \ | \ \mathtt{-} \ | \ \mathtt{\times} \ | \ \  \mathtt{\hat{}} \  \ | \ \mathtt{/} \ | \ \mathtt{\%} \ | \ \oplus   \qquad  \mathtt{bop} := \  \mathtt{<} \ | \ \mathtt{==}   \\ 
      \mathrm{Exp}  & e &:= & \ v  \ |  \ e \ \mathtt{aop} \ e \ | \  e\ \mathtt{bop} \ e \ | \ \neg e \ | \  e \wedge e \\ 
      \mathrm{Statement}   & s &:=  &  \tau x\leftarrow e \ | \ l \leftarrow \mathtt{aop} \ e \ |  \  \tau x \leftarrow f (\overline{u})  \\ 
      & & & |  \   s; s \ | \  \mathtt{if} \ e \ \{s\} \ \{s\} \ | \ \mathtt{for} \ x \ v  \ \{s\} \ | \ \mathtt{while} \ e \ \{s\} \  \\
      \mathrm{FuncDef} & d &:= & \mathtt{def} \ f (\overline{\tau x}) \{ s; \mathtt{return} \ v\} \\ 
      \mathrm{Program} & P &:= & \overline{\tau z}; \overline{d} 
    \end{array}
    \end{gather*}
    \label{tab:syn}
    \vspace{-1.0em}
\end{table}

\textit{Values.} The language distinguishes syntactic values $v$, comprising unit, lvalues, and literals (natural numbers and Boolean constants). Lvalues consist of variables, array locations, pairs of variables, and projections of such pairs. 

\textit{Expressions.} Expressions include values, arithmetic expressions, and Boolean expressions. Arithmetic expressions are constructed from natural numbers using operators like addition, subtraction, multiplication, and exponentiation. Boolean expressions are formed by comparing arithmetic expressions and combining Boolean values with logical operators such as negation and conjunction. 

\textit{Statements.}
The statements include the assignment \(\tau x \leftarrow e\), which evaluates \(e\) and assigns the result to \(x\) (where \(\tau\) is the type annotation of \(x\)). The unary assignment \(l \leftarrow \mathtt{aop}\ e\) is equivalent to \(l \leftarrow l\ \mathtt{aop}\ e\). Function calls are written \(\tau x \leftarrow f(\overline{u})\), calling \(f\) with parameters \(\overline{u}\) and assigning the return value to \(x\). The assignment and function call require an explicit type annotation for \(x\), implying that \(x\) is assigned for the first time (i.e., \(x\) may not be assigned by these forms again). In contrast, the unary assignment allows mutation of an lvalue without a type annotation, meaning \(l\) has already been declared and may be modified multiple times by this form. These constraints will be detailed in the type system of the next subsection. Sequential statements, if statements, for loops, and while loops follow standard conventions. The if statement executes $s_1$ if $e$ evaluates to true, otherwise it executes $s_2$. The for loop statement $\mathtt{for} \ x \ v  \ \{s\}$ increments the value of $x$ from $0$ to $v$ with step $1$ and executes $s$ in each iteration. The while loop $\mathtt{while} \ e \ \{s\}$ repeatedly executes $s$ as long as $e$ evaluates to true. For any expression \(e\) and statement \(s\), we write \(\mathsf{var}(e)\) and \(\mathsf{var}(s)\) to denote the set of variables occurring in \(e\) and \(s\), respectively. Furthermore, \(\mathsf{mv}(s)\) denotes the set of variables modified by \(s\); \(\mathsf{rv}(s)\) denotes the set of variables read by \(s\). 

\textit{Functions.} The function definition \( d \) consists of formal parameter declarations \( \overline{\tau x} \) and an execution body, which includes statements \( s \) and returns an evaluation result \( v \).

Finally, a program \( P \) comprises a series of global variable declarations \( \overline{\tau z} \), followed by multiple function definitions \( \overline{d} \). The last function definition serves as the main function, which has no formal parameters and returns unit. Recursive and mutually recursive functions are supported by allowing the body of a function $f$ to contain calls to $f$ itself, which naturally gives rise to indirect cyclic dependencies among multiple functions through chains of calls.

\subsection{Type System}
The type system of RQIMP imposes restrictions on the use of the syntactic constructs introduced earlier, so that the compiler can generate well-formed $\RQC^{++}$ programs and ensure the correctness of uncomputation for temporary variables. Concretely, our type system defines the following three type judgments. 

\paragraph{Typing rules for expressions}
In Table~\ref{tab:exp_type}, we define type judgments for expressions in the form $\Gamma \vdash e : \tau$, 
which says that under some type environment \(\Gamma = \{x_1 : \tau_1, \dots, x_n : \tau_n\}\), expression $e$ has type $\tau$. Here, we use the following preorder $\sqsubseteq$ on modes:
\[ 
    \mathtt{C} \sqsubseteq \mathtt{q} \quad \text{for} \quad \mathtt{q} \in \{\mathtt{C}, \mathtt{Q}\}, \qquad \mathtt{Q} \sqsubseteq \mathtt{Q}.
\]
The operation $\mathtt{q}_1 \sqcup \mathtt{q}_2$ yields the greater element under the partial order $\sqsubseteq$. Then $M(\tau)$ takes the mode of type $\tau$, defined as follows: 
\[
    M(\omega^\mathtt{q})  =  \mathtt{q};  \quad  M(\mathtt{array}\ n\ \tau)  =  M(\tau); \quad  M(\tau_1, \tau_2)  = M(\tau_1) \sqcup M(\tau_2), 
\] 
and $T(\tau)$ extracts the type ignoring mode, defined as follows: 
\[
    T(\omega^\mathtt{q})  = \omega; \quad  T(\mathtt{array}\ n\ \tau)  =  T(\tau); \quad  T(\tau_1, \tau_2)  =  (T(\tau_1), T(\tau_2)).
\] 
For simplicity, we refer to such \textit{mode-ignored types} simply as \textit{types} when no ambiguity arises. 

\begin{table}[t]
    \centering
    \caption{Selected typing rules for expressions. The full definition is presented in Appendix~\ref{app:source} }
 \begin{gather*}
       \dfrac{\Gamma \vdash x: \mathtt{array} \ n \ \tau \quad \Gamma \vdash v: \mathtt{nat}^{\mathtt{q}} \quad \mathtt{q} \sqsubseteq M(\tau)}{\Gamma \vdash x[v] : \tau} \\ 
       \dfrac{\Gamma \vdash a_1: \mathtt{nat}^{\mathtt{q}_1} \quad \Gamma \vdash a_2: \mathtt{nat}^{\mathtt{q}_2} \quad \mathtt{q}= \mathtt{q}_1 \sqcup \mathtt{q}_2}{\Gamma \vdash a_1 \ \mathtt{aop} \ a_2: \mathtt{nat}^{\mathtt{q}}} \quad  \dfrac{\Gamma \vdash b_1: \mathtt{nat}^{\mathtt{q}_1} \quad \Gamma \vdash b_2: \mathtt{nat}^{\mathtt{q}_2} \quad \mathtt{q} = \mathtt{q}_1 \sqcup \mathtt{q}_2 }{\Gamma \vdash b_1 \ \mathtt{bop} \ b_2: \mathtt{bool}^{\mathtt{q}}} 
 \end{gather*}
  \label{tab:exp_type}
  \vspace{-1.0em}     
\end{table}

For any array $x$, we permit access 
to the data
$x[v]$ even when position $v$ is a quantum variable, provided the array itself is in $\mathtt{Q}$-mode. This allows the index to be a quantum variable, enabling data access under quantum control.  In arithmetic and Boolean expressions, if at least one operand is in $\mathtt{Q}$-mode, then the whole expression is also in $\mathtt{Q}$-mode.

\paragraph{Typing rules for statements} 
Following the context-transforming statement judgments used in
  Tower~\cite{yuanTowerDataStructures2022}, for a statement \(s\), we define the
  typing judgment \(\Xi;\Gamma \vdash s \dashv \Gamma'\) in
  Table~\ref{tab:stmt-type}. Here, \( \Xi \) represents a partial mapping from function names to tuples containing formal parameter lists, function body statements, return values, and a type environment recording the types of local and global variables that will be generated during type checking of the function body. This judgment indicates that under the function environment \(\Xi\) and type
  environment \(\Gamma\), statement \(s\) is well-typed and transforms the typing
  context to \(\Gamma'\). We define \(\text{dom}(\Gamma)\) as follows: 
\[
\text{dom}(\{x_1 : \tau_1, \dots, x_n : \tau_n\}) = \{x_1, \dots, x_n\}.
\] Further, we denote 
\[
\Gamma, x : \tau = \Gamma \cup \{x : \tau\},
\]
where it is assumed that \(x \notin \operatorname{dom}(\Gamma)\). Similarly, 
\(\Gamma, \Gamma' = \Gamma \cup \Gamma'\) under the assumption that 
\(\operatorname{dom}(\Gamma) \cap \operatorname{dom}(\Gamma') = \emptyset\).
Moreover, we write $\Gamma_\mathtt{Q}$ for the type context of quantum-mode variables in the context $\Gamma$, i.e.,
\[
\Gamma_\mathtt{Q}= \{\, (x : \tau) \in \Gamma \mid \ M(\tau)=\mathtt{Q} \,\}.
\]
For an expression or l-value \(u\), \(M(u)\) denotes the mode of its type under the current type environment.
\begin{table}[t]
    \centering
    \caption{Typing rules for statements}
    \label{tab:stmt-type}
    \begin{gather*} 
         \dfrac{ \begin{array}{c} 
         \Gamma \vdash e: \tau' \quad x \notin \text{dom}(\Gamma) \\  T(\tau)= T(\tau') \quad  M(\tau') \sqsubseteq M(\tau) 
         \end{array} }{ \Xi; \Gamma \vdash \tau x \leftarrow e \dashv \Gamma,x:\tau} \qquad  \qquad  
           \dfrac{\begin{array}{c} 
    \mathtt{aop} \in \{ +, -,\oplus \} \quad \Gamma \vdash l \ \mathtt{aop} \ e: \tau  \\  \mathsf{var}(l) \cap \mathsf{var}(e)=\emptyset \quad   M(e) \sqsubseteq M(l)
           \end{array} }{ \Xi; \Gamma \vdash l \leftarrow \mathtt{aop} \ e\dashv \Gamma } \\ 
              \dfrac{\Xi; \Gamma \vdash s_1 \dashv \Gamma' \quad \Xi; \Gamma' \vdash s_2 \dashv \Gamma'' }{ \Xi; \Gamma \vdash s_1 ; s_2 \dashv \Gamma''}  \quad \quad \quad   \dfrac{\Gamma \vdash e: \mathtt{bool}^{\mathtt{C}}   \quad \Xi; \Gamma \vdash s_1\dashv \Gamma' \quad \Xi; \Gamma \vdash s_2 \dashv \Gamma'}{ \Xi; \Gamma \vdash \mathtt{if} \ e \ \{s_1\} \ \{s_2\} \dashv \Gamma'}  \\ 
    \dfrac{
\begin{array}{c}
\Gamma \vdash e: \mathtt{bool}^{\mathtt{Q}} \quad \Xi; \Gamma \vdash s_1 \dashv \Gamma' \quad \Xi; \Gamma \vdash s_2 \dashv \Gamma' \\
\mathsf{var}(e) \cap (\mvq_{\Gamma'}(s_1) \cup \mvq_{\Gamma'}(s_2)) = \emptyset \quad \Gamma'' = \Gamma, (\Gamma' \setminus \Gamma)_{\mathtt{Q}}
\end{array}
}{
\Xi; \Gamma \vdash \mathtt{if}\; e \; \{s_1\} \; \{s_2\} \dashv \Gamma''
} \\  
     \dfrac{\begin{array}{c} 
     \Gamma \vdash e: \mathtt{bool}^\mathtt{Q} \\  \Xi; \Gamma \vdash s \dashv \Gamma'
          \quad \Gamma''=\Gamma, (\Gamma' \setminus \Gamma)_{\mathtt{Q}}
     \end{array}}{ \Xi; \Gamma \vdash \mathtt{while} \ e \  \{s\} \dashv \Gamma''}  \qquad  \qquad 
     \dfrac{ \begin{array}{c} 
     \Gamma \vdash x,v: \mathtt{nat}^\mathtt{Q} \\ \Xi; \Gamma \vdash s \dashv \Gamma'
          \quad \Gamma''=\Gamma, (\Gamma' \setminus \Gamma)_{\mathtt{Q}} 
     \end{array}}{ \Xi; \Gamma \vdash \mathtt{for} \ x \ v \  \{s\} \dashv \Gamma''} \\
\dfrac{ \Xi(f)=(\overline{\tau x}, s, \tau_r v, -) \quad M(\tau_r)=\mathtt{Q} \quad \Gamma \vdash \overline{u}: \overline{\tau} \quad  x_r \notin \dom(\Gamma) }{ \Xi; \Gamma \vdash  \tau_r x_r \leftarrow f(\overline{u}) \dashv \Gamma, x_r:\tau_r } 
    \end{gather*} 
\vspace{-0.8em}    
\end{table} 
For any statement \(s\), let \(\mathsf{mvq}_{\Gamma}(s)\) and
  \(\mathsf{mvc}_{\Gamma}(s)\) denote the quantum and classical subsets of
  \(\mathsf{mv}(s)\) under \(\Gamma\), respectively. When \(\Gamma\) is clear from the context, we omit the subscript. 

For the assignment \(\tau x \leftarrow e\), we require \(x \notin \mathrm{dom}(\Gamma)\) and \(\Gamma \vdash e : \tau'\). This ensures that \(x\) appears for the first time (assigned at most once by this form) and that all variables in \(e\) are already declared, which also implies that \(x\) does not occur in \(\mathsf{var}(e)\). Additionally, the type of \(x\) must match that of \(e\) (\(T(\tau) = T(\tau')\)), and the condition \(M(\tau') \sqsubseteq M(\tau)\) requires that if \(x\) is \(\mathtt{C}\)-mode then \(e\) must also be \(\mathtt{C}\)-mode. After the assignment, the binding \(x:\tau\) is added to \(\Gamma\), which prevents subsequent assignment via this form. However, we allow certain unary assignments \( l \leftarrow \mathtt{aop}\ e \) (with \(\mathtt{aop}\) restricted to reversible operations: $\mathtt{+}$, $\mathtt{-}$ and bitwise XOR $\mathtt{\oplus}$) under relaxed constraints. In such cases, the lvalue \(l\) may be modified multiple times by this unary assignment, provided all variables appearing in \(l\) are already declared and \(l\ \mathtt{aop}\ e\) is well-typed. The other requirements are similar to those of the aforementioned assignment statements. Sequential statements combine \( s_1 \) and \( s_2 \) by composing their contextual effects sequentially. 

In $\mathtt{if}$ statements, we require $s_1$ and $s_2$ to be well-typed and to introduce the same set of variables. For quantum-mode guards ($e : \mathtt{bool}^\mathtt{Q}$), reversibility demands that the branches do not modify any quantum variable occurring in $e$. Moreover, any modifications to classical variables within a branch are local to that branch and are not propagated beyond the conditional. Hence, the resulting context $\Gamma''$ contains only the quantum variables introduced in the branches. For $\mathtt{while}$ and $\mathtt{for}$ loops, the guard expression and loop variable are required to be in $\mathtt{Q}$-mode, capturing dynamic control flow dependent on quantum data. As in conditionals, any classical variables modified inside the loop body remain local and do not propagate beyond the loop.

 When typing a function call $\tau_r x_r \leftarrow f(\overline{u})$, we require the types of $\overline{u}$ to match the formal parameter types $\overline{\tau}$ and the type of $x_r$ to match the function's return type. Note that functions are only allowed to return $\mathtt{Q}$-mode variables; consequently, any classical variables within a function are treated as local and do not persist after the call. As with assignment, $x_r$ must be unassigned and all variables in $\overline{u}$ already declared. The placeholder “\(-\)” in $\Xi(f)$ indicates that the function body may not have been type‑checked yet.

For assignments, we impose the same restriction in classical mode as in quantum mode: a variable cannot be assigned more than once. While this may appear stricter, repeated assignments can always be simulated by introducing fresh variables via renaming, so this restriction does not reduce the language's expressive power. Loops that depend only on classical data correspond to static control flow and can be compiled by evaluating classical values at compile time and unrolling the loop structure, following standard practice. Since our focus is on compiling dynamic control flow, we consider only loops whose behavior depends on quantum data.

\paragraph{Typing rule for function declaration}
\begin{gather*}
  \Gamma'=\Gamma, \overline{x}:\overline{\tau} \quad \Xi; \Gamma' \vdash s \dashv \Gamma''  \quad \Xi(f)=(\overline{\tau x}, s, \tau_r v, -)\\ 
    \dfrac{ \mathsf{var}(v) \cap \dom(\Gamma') = \emptyset  \quad M(\tau_r)=\mathtt{Q} \quad \Gamma'' \vdash v : \tau_r  \quad  \dom(\Gamma) \cap \mathsf{mv}(s) =\emptyset }{\Xi; \Gamma \vdash  \mathtt{def} \ f(\overline{\tau x}) \{s ; \mathtt{return} \ v\} \triangleright \Xi[f \mapsto (\overline{\tau x}, s, \tau_r v, \Gamma'')]} 
\end{gather*}
The type checking rule for functions is \(\Xi; \Gamma \vdash d \triangleright \Xi'\), where \(\Xi\) and \(\Xi'\) are function environments. Here we only give
the rule for non-main functions; the rule for the main function is deferred to 
Appendix~\ref{app:source}.
 The rule first extends the type environment \(\Gamma\) for global variables with the formal parameters and their types to obtain \(\Gamma'\). We require the function body \(s\) to be well-typed, and the return value \(v\) (which must be in \(\mathtt{Q}\)-mode) to be distinct from formal parameters and global variables; formally, \(\mathsf{var}(v) \cap \dom(\Gamma') = \emptyset\).  
Moreover, to avoid global-state interference that would obstruct the
  uncomputation of local quantum variables, we impose the following 
constraint on 
  global variables. For every non-main function body \(s\), we require
  \(\dom(\Gamma) \cap \mathsf{mv}(s)=\emptyset\), ensuring that no global variable
  is modified by a non-main function. For the main function body, we
  require each global variable to be either read-only or write-only within the main function. This avoids mutual dependencies between global  and local
  variables. A more detailed explanation is provided in Appendix~\ref{app:source}. Ultimately, the resulting \(\Xi'\) incorporates the function's information for \(d\).   

Finally, the type checking of a program \(\overline{\tau z};\overline{d}\) proceeds as follows. First, we build a typing context \(\Gamma = \{ \overline{z:\tau}\}\) from the global declarations \(\overline{\tau z}\), where \(M(\tau)=\mathtt{Q}\) for every global declaration \(\tau z\). Starting from the empty function context \(\Xi_0 = \emptyset\), we iterate over each declaration \(d\) in \(\overline{d}\). For each function \(f\), we add its signature \((\overline{\tau x}, s, v)\),
  with the return type \(\tau_r\) obtained by a preliminary scan of the declaration of the returned
  variable \(v\), to the current context, so that \(\Xi(f) = (\overline{\tau x}, s, \tau_r v, -)\). After constructing the full \(\Xi\), we check each function body \(s\) under \(\Xi\) and \(\Gamma\) using the function declaration typing rule. If all checks pass, the program is well‑typed. These well-typedness conditions, covering
  conditionals, loops, function calls, and global variables, are imposed to support
  the generation of well-formed \(\RQC^{++}\) programs; see Appendix~\ref{app:proof-main} for the corresponding argument. 

\subsection{Semantics}
\label{subsec:source-sem}
We define small-step operational semantics for RQIMP. A program state \(\sigma\) is a mapping from variable identifiers to literal values. For a given state \(\sigma\) and expression \(e\), we write \(\sigma(e)\) for the result of evaluating \(e\) under \(\sigma\), and assume that expression evaluation is an atomic operation. A configuration \(\langle s, \sigma \rangle\) represents the current program fragment \(s\) to be executed and the current state \(\sigma\). Given a function environment \(\Xi\), program execution is characterized by a transition relation between configurations, written as \(\langle s, \sigma \rangle \to_{\Xi} \langle s', \sigma' \rangle\),
meaning that under the function environment \(\Xi\) and the current state \(\sigma\), the program \(s\) takes one step of execution, yielding the remaining program \(s'\) and updating the state to \(\sigma'\). We write \(\to_{\Xi}^{*}\) for the reflexive transitive closure of \(\to_{\Xi}\) and \(\downarrow\) for the empty program (terminal state). For brevity, we use the symbol \(\mathbf{Conf}\) to denote the set of all configurations. Thus, the operational semantics of the RQIMP language for a given \(\Xi\) is the transition relation \(\to_{\Xi} \subseteq \mathbf{Conf} \times \mathbf{Conf}\) defined by the transition rules presented in Table~\ref{tab:opsem}. In the rules, \(L \leftarrow \sigma(L)\) abbreviates the simultaneous
  restoration \(x \leftarrow \sigma(x)\) for all \(x \in L\).

  The conditional statement selects the appropriate branch based on the value of the guard \(e\). For quantum-mode guards, RQIMP treats classical updates inside each branch as branch-local: after the conditional, only quantum-variable updates are retained, while classical variables modified in the branch are restored to their pre-branch values. 
  In the function-call rule, the formal parameters, local variables, and return variable of the callee are assumed to be freshly alpha-renamed before the body is expanded,
  so that they do not conflict with variables in the caller. A function call \(\tau x \leftarrow f(\overline{u})\) first binds the formal parameters \(\overline{y}\) to the
  values of the actual arguments \(\overline{u}\) in the current state, then executes the function body \(s\), and finally assigns the value of the return variable \(v\) to
  \(x\). After the return value is transferred, the callee's local state is restored. 
  
\begin{table}[t]
\centering
\caption{Selected rules for the operational semantics of RQIMP. The full definition is presented in Appendix~\ref{app:source}
}
\label{tab:opsem} 
\begin{gather*}
    \frac{\begin{array}{c}
       M(e)=\mathtt{Q} \quad \sigma(e) = \mathtt{true}  \quad L = \mathsf{mvc}(s_1)
    \end{array}}
        {\langle \mathtt{if} \  e \; \{s_1\} \; \{s_2\}, \sigma \rangle \to_{\Xi} \langle s_1; L \gets \sigma(L), \sigma \rangle} \quad 
    \frac{\begin{array}{c}
       \Xi(f) = (\overline{\tau y}, s, \tau_r v, \Gamma) \quad  L =\mathsf{mv}(s) 
    \end{array}} 
        {\langle \tau x \leftarrow f(\overline{u}), \sigma \rangle \to_{\Xi}\langle \overline{y} \gets \overline{u}; s; x \gets v; L \gets \sigma(L), \sigma \rangle} \end{gather*} 
\vspace{-1.0 em}    
\end{table}

\subsection{Soundness}

\begin{definition}[Store Consistency]
   Let $\Gamma$ be a type environment and $\sigma$ be a store. We say $\sigma$ is consistent with $\Gamma$, denoted as $\sigma \models \Gamma$, if and only if
   \(
   \text{dom}(\Gamma) = \text{dom}(\sigma)
   \) and \(
   \forall x \in \text{dom}(\Gamma).\ \sigma(x): \Gamma(x),
   \) where $v: \tau$ means that $v$ has type $\tau$.
\end{definition}

\begin{definition}[Valid Configuration]
Given a function environment \(\Xi\), a configuration \(\langle s, \sigma \rangle\) is valid under \(\Xi\), initial type context \(\Gamma\) and final type context \(\Gamma'\), denoted \(\Xi; \Gamma \vdash \langle s, \sigma \rangle \dashv \Gamma'\), if and only if \(\sigma \models \Gamma\) and \(\Xi; \Gamma \vdash s \dashv \Gamma'\). 
\end{definition} 

Type soundness is divided into two parts: progress and preservation.

\begin{theorem}[Preservation]
If \(\Xi; \Gamma \vdash \langle s, \sigma \rangle \dashv \Gamma'\) and \(\langle s, \sigma \rangle \rightarrow_{\Xi}  \langle s', \sigma' \rangle\), then there exists a context \(\Gamma''\) such that \(\Xi; \Gamma' \vdash \langle s', \sigma' \rangle \dashv \Gamma''\). 
\end{theorem}

\begin{proof}
By induction on the derivation of the operational semantics step.
\end{proof}

\begin{theorem}[Progress]
If \(\Xi; \Gamma \vdash \langle s, \sigma \rangle \dashv \Gamma'\), then either \(s= \ \downarrow\), or there exists a configuration \(c\) such that \(\langle s, \sigma \rangle \rightarrow_{\Xi} \ c\). 
\end{theorem}

\begin{proof}
By induction on the definition of valid configurations \(\Xi; \Gamma \vdash \langle s, \sigma \rangle \dashv \Gamma'\). 
\end{proof}

\section{Key Techniques} 

\label{sec:key}
This section presents the two key techniques used by the compiler to address the
challenges illustrated in Section~\ref{sec:technical-overview}: indexed
static-register management for recursion-layer isolation, and recursion-aware
uncomputation for controlling recomputation and storage overhead. 

\subsubsection*{Indexed Static-Register Management} 
 To precisely distinguish variables requiring per-layer isolation from those that can be safely shared across recursive layers, we maintain a call stack ($\St$) together with a data structure (\(\Vactive\)) that tracks, for each function, the set of quantum variables that are currently live. During compilation, all variables are initially represented as non-indexed quantum variables.  When a recursive invocation of a function $f$ occurs, we ensure that all live quantum variables of $f$ in the current layer are represented as indexed variables of the form $q[i], p[j], r[k]$ through systematic variable renaming. The recursive call then increments these indices (e.g., passing $i+1, j+1, k+1$) to guarantee that different recursion layers operate over disjoint quantum variables. In this way, variables that have already been cleaned are left unrenamed and can be reused
  in subsequent layers.

\begin{figure}
    \centering
\includegraphics[width=0.8\linewidth]{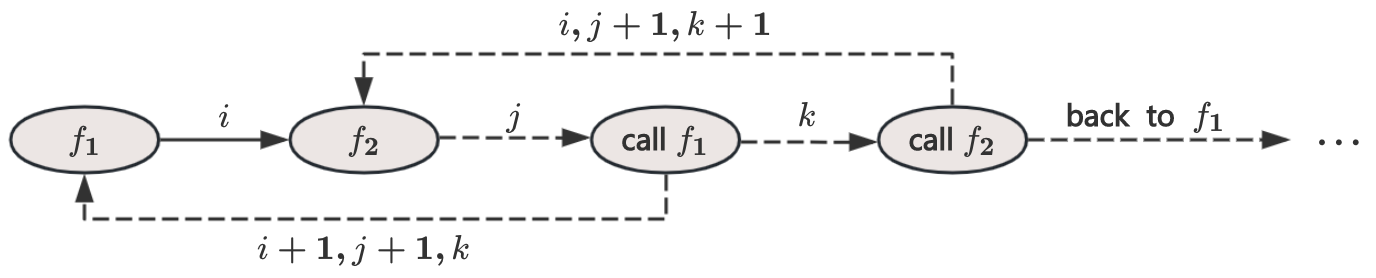}
   \caption{A diagram depicting a mutual recursion example: nodes represent function names or call sites, solid arrows (\(a \rightarrow b\)) indicate function dependency, and dashed arrows (\(a \dashrightarrow b\)) show the execution order inside \(f_2\).} 
    \label{fig:mul_re_exam}
    \Description{...}
     \vspace{-0.6em}
\end{figure}

Moreover, complex recursive structures can contain multiple paths. In Figure~\ref{fig:mul_re_exam}, \(f_1\) depends on \(f_2\), and \(f_2\) sequentially calls both \(f_1\) and itself in its body, forming two distinct recursion patterns: \(f_1 \to f_2 \to f_1\) and \(f_2 \to f_2\). 
To reduce unnecessary quantum-register usage, we assign independent index variables to different recursive paths. Specifically, quantum variables that are used between the calls to \(f_1\) and \(f_2\) and that remain live are accessed via index \(k\), while other live variables are marked with indices \(i\) and \(j\). When \(f_1\) is called recursively, only \(i\) and \(j\) need to be incremented; \(k\) remains unchanged. Since the variables used between the calls to \(f_1\) and \(f_2\) are still clean at the time of \(f_1\)'s recursive call, they can be safely reused by subsequent recursive layers. Similarly, calling \(f_2\) only requires incrementing \(j\) and \(k\); the variables marked by \(i\) can also be safely reused at the next recursive layer of the call to \(f_2\).  To implement this mechanism, the compiler maintains an index supplier \(\iota\) that records an index variable for renaming quantum variables. After each compiled recursive call of a function \(f\), \(\iota\) is updated to a fresh index variable associated with the corresponding mutually recursive group  of functions containing \(f\), and this fresh index variable is then used for subsequent invocations.  

In summary, our mechanism guarantees that distinct recursive layers operate on disjoint quantum variables and that different paths correspond to independent index variables, which ensures the correct 
recursive-layer isolation while controlling register reuse.  This mechanism is implemented in the \(\textsc{Compile\_Recur}\) algorithm 
(Section~\ref{sec_com_stmt}) to handle recursive call statements. 

Moreover, note that $\mathtt{sum}$ in Figure~\ref{fig:RQC_sum_cor} is compiled into a target function that operates on shared quantum array variables (e.g., \(q, q_1,q_2\)). Consequently, whenever this function is invoked, its execution always acts upon this same set of quantum variables. However, these auxiliary variables are not necessarily clean and available at every call site. Therefore, to enable safe reuse of procedures, the quantum variables used internally in a function are added to the procedure's formal parameters. When compiling a function call, suitable quantum variables can be allocated to these parameters according to the current availability of qubits. Moreover, for mutually recursive functions, resource requirements may be known only after the whole group is compiled. We therefore treat the mutually recursive group as a single compilation unit: only after all functions in the group have been compiled do we uniformly add all auxiliary quantum variables used in the unit to the parameter lists of all involved functions. This mechanism is  
implemented
in the function compilation algorithm \(\textsc{Compile\_Fun}\) (Section~\ref{sec_com_fun}). We note that the original $\RQC^{++}$ only permits classical expressions as parameters. Lifting this restriction to also allow quantum variables as parameters is sound, since quantum arguments are interpreted with call-by-reference semantics and can be compiled to classical address/index information for statically allocated quantum storage, rather than to copies of quantum data; see Appendix~\ref{app:tar}. 

We now illustrate how the above mechanism is applied for the running example. The source program is first simplified to the form in Figure~\ref{fig:sum_source}, and the final compiled functions are given in Figure~\ref{fig:sum_target}. Here, \texttt{Fsum} and \texttt{Faux} correspond to \(F_{\mathtt{sum}}\) and \(F_{\mathtt{aux}}\) (which correspond to source programs $\mathtt{sum}$ and $\mathtt{aux}$, respectively). The temporaries $q,q_1,q_2$ used in the mutually recursive procedures $F_{\mathtt{sum}}$ and $F_{\mathtt{aux}}$ are added to the parameter lists of both functions only after both have been compiled. When $F_{\mathtt{sum}}$ is invoked by an external function (line~13), fresh quantum variables are allocated from the register pool and bound to these parameters. The recursion indices $i,j$ select distinct elements of these arrays for each layer: the initial call passes $i, j= 0$ (line~13), and each recursive invocation of $F_{\mathtt{sum}}$ (line~3) supplies $i+1$, $j+1$, guaranteeing that successive layers operate on disjoint slots of $q, q_1,q_2$. Note that the mutually recursive functions \(F_{\mathtt{aux}}\) and \(F_{\mathtt{sum}}\) form a linear-recursive structure, so the two index variables \(i\) and \(j\) do not provide an immediate advantage in this example. However, if \(F_{\mathtt{aux}}\) recursively calls itself within its own body, only the index \(j\) needs to be incremented, i.e., passing \((i,j+1)\), thereby avoiding unnecessary increments of variables associated only with \(F_{\mathtt{sum}}\). 

\begin{figure}[t] 
\begin{minipage}[b]{0.4\linewidth}
    \centering 
    \begin{lstlisting}[language=MyLang, basicstyle=\ttfamily\scriptsize]
nat^Q x1, k1;
def aux(nat^Q k) {
    nat^Q y1 <- k-1;
    nat^Q y <- sum(y1);
    nat^Q x <- k + y;
    return x }
def sum(nat^Q k) {
    bool^Q y1 <- (k==0);
    if y1 { nat^Q x <- 0 }
           { nat^Q x <- aux(k) }
    return x }
def main { 
    x1 <- sum(k1);
    return () }
\end{lstlisting}
    \caption{The source program of $\mathtt{sum}$.}
    \label{fig:sum_source}
\end{minipage}
\hfill 
\begin{minipage}[b]{0.52\linewidth}
    \centering
    \begin{lstlisting}[language=MyLang, basicstyle=\ttfamily\scriptsize]
Faux(p,r,red(i,j,q,q1,q2)) <==
   circ(k-1, q1[j]); 
   Fsum(q1[j],q2[j],red(i+1,j+1,q,q1,q2));
   qinv(circ(k-1, q1[j])); // Uncompute q1[j]
   circ(k+y, r)
Fsum(p,r,red(i,j,q,q1,q2)) <==
   circ(k==0, q[i]);
   qif q[i] |0> -> Faux(p,r,red(i,j,q,q1,q2))
           square |1> -> circ(0,r) 
   fiq
   qinv(circ(k==0, q[i])); // Uncompute q[i]
Fmain <==
   Fsum(p1,r',red(0,0,q,q1,q2));
   copy(r',r1); // copy r' to r1
   qinv(Fsum(p1,r',red(0,0,q,q1,q2))); // Uncompute q2,r'
\end{lstlisting}
    \caption{The target program of $\mathtt{sum}$.}
    \label{fig:sum_target}
    \Description{...}
\end{minipage}
\vspace{-1.0em}
\end{figure}

% \begin{figure}[t]
%     \centering
% \[
% \begin{array}{cc}
% \begin{aligned}
% 1: &\  F_{\mathtt{f}}(p, r, \mathcolor{red}{i, j, q, q_1, q_2})\Leftarrow \\
% 2: &\ \quad (k-1,\ p); \\
% 3: &\ \quad F_{\mathtt{sum}}(p, q_1[j], \mathcolor{red}{i+1, j+1, q, q_1}); \\
% 4: & \ \quad \mathcolor{blue}{\qinv((k-1,\ p));} \text{// Uncompute $p$}; \\  
% 5: &\  \quad (y+k, r) \\ 
% \end{aligned} 
% & 
% \begin{aligned}
% 6: & \ F_{\mathtt{sum}}(p, r, \mathcolor{red}{i, j, q, q_1, q_2}) \Leftarrow \\
% 7: & \ \quad (k=0,\ q[i]); \\
% 8: & \ \quad \textbf{qif}\  q[i] \ |0\rangle \rightarrow F_{\mathtt{f}}(p, r, \mathcolor{red}{i, j, q, q_1, q_2}) \\
% 9: & \ \quad \qquad \quad |1\rangle \rightarrow \mathit{Init\_c}(0, r,\sz-1) \ \textbf{fiq} \\ 
% 10: & \ \quad \mathcolor{blue}{\qinv((k=0,\ q[i]));} \text{\quad // Uncompute $q[i]$}  \\ 
% \end{aligned}
% \end{array} 
% \]
% \[
% \begin{aligned}
% 11: &\  F_{\mathtt{main}} \Leftarrow \\
% 12: &\ \quad F_{\mathtt{sum}}(p, r', \mathcolor{red}{0, 0,q, q_1}); \\
% 13: &\  \quad \mathit{copy}(p', r); \\ 
% 14: & \ \quad \mathcolor{blue}{\qinv(F_{\mathtt{sum}}(p, r' ,\mathcolor{red}{0, 0, q, q_1}));} \text{ \quad // Uncompute $q_1, \ r'$}  
% \end{aligned} \]
% \caption{Target procedure of running example}
% \Description{A quantum program illustrating the compilation of recursive procedures with uncomputation.}
%     \label{fig:run_ex} 
% \end{figure} 

\subsubsection*{Recursion-Aware Uncomputation Strategy}

To reduce the time overhead caused by recomputation of recursive calls under a
plain uncomputation strategy, we defer cleanup for local variables introduced by
  recursive-call statements, while using eager cleanup~\cite{DBLP:conf/cav/AmyRS17}
  for variables introduced by non-recursive statements to reduce qubit usage. To better illustrate the advantages of this approach, we first present a brief analysis. 

As described in Section~\ref{sec:technical-overview}, \texttt{sum} is a linear
  recursive program. However, implementing it with per-layer uncomputation, as shown in
  Figure~\ref{fig:RQC_sum_cor}, leads to exponential growth in the number of
  recursive invocations. To avoid this exponential blow-up, we adopt a deferred cleanup strategy for
  variables introduced by recursive-call statements: we treat each linear
  (mutual-)recursive component as an integrated unit, keep those variables live
  across recursion layers until the top-level call finishes, and then uncompute
  them uniformly. For example, in Figure~\ref{fig:sum_target}, when \(F_{\mathtt{sum}}\) is called recursively (line~3), the non-return local variable \(y\) in \(F_{\mathtt{aux}}\) retains its quantum variable \(q_2[j]\) uncleaned across levels; cleanup is deferred until after \(F_{\mathtt{sum}}\) returns and is performed by the main function (line~15). Other ancillae for non-recursive variables (e.g., \(q[i]\), \(q_1[j]\)) are uncomputed eagerly when no longer needed in each layer.  

This eliminates per-layer self-invocation for uncomputation, avoiding the exponential recomputation pattern and yielding linear recursive-invocation overhead, with the resulting time overhead parameterized by the primitive-operation costs of each layer. Moreover, the additional
  storage remains proportional to the recursion depth. For general linear recursion, let \(n\) denote the recursion depth. We measure space by the additional typed quantum-register footprint of the generated target program, with register widths determined by the source types. In each recursion layer, suppose that the pre-recursive computation uses \(N\) temporary registers and leaves \(K\) registers live when the recursive call is made, while the post-recursive computation uses \(M\) temporary registers that are cleaned after use and can therefore be reused across layers. Under this per-layer accounting, the additional space required by the generated program is bounded by 
\(
(N+(n-1)K)+n+M,
\)
where the \(+n\) term accounts for one per-layer temporary associated with the recursive call. Thus, for fixed per-layer quantities \(K\), \(N\), and \(M\), the additional quantum-register footprint is \(O(n(K+1)+N+M)\), i.e., linear in the recursion depth and parameterized by the per-layer register footprint. Moreover, since non-recursive temporaries in the pre- and post-recursive computations are cleaned eagerly, ReOC reduces the per-layer register footprint reflected in \(K\), \(N\), and \(M\), especially the live-across-call component \(K\), while controlling this linear-depth storage bound. 

\begin{wrapfigure}{r}{0.5\textwidth}
      \centering
         \includegraphics[width=\linewidth]{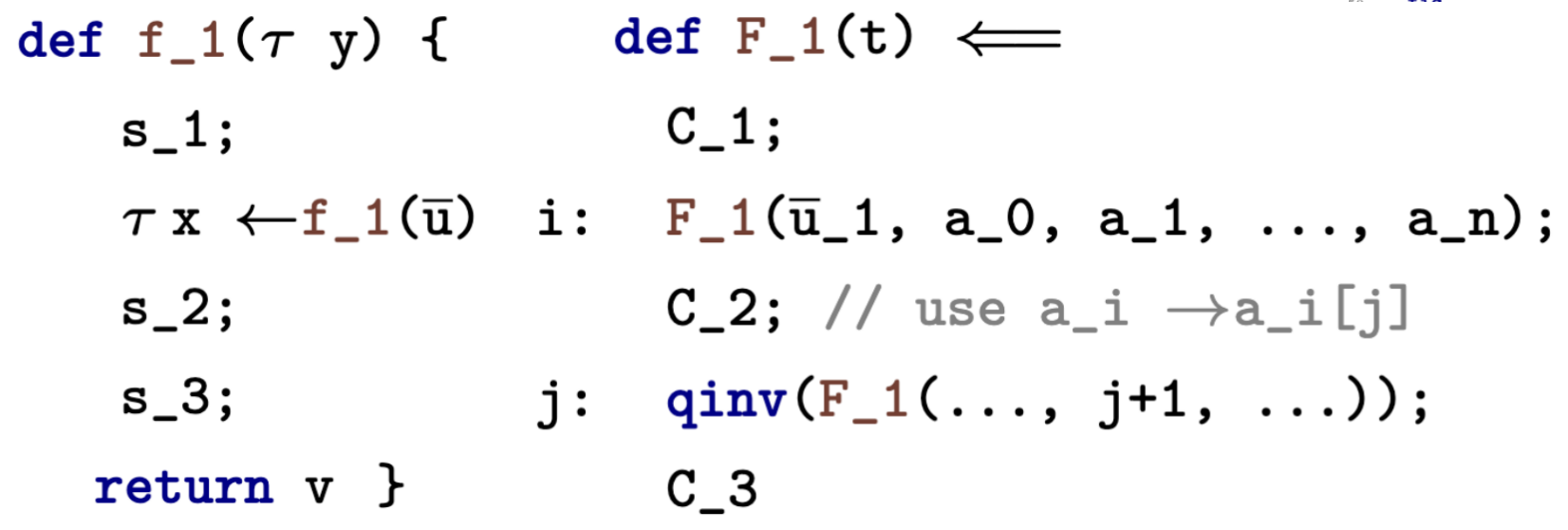}
        \caption{Source and target programs of $\mathtt{f_1}$.}
        \label{fig:f1-source-target}
        \Description{...} 
\end{wrapfigure}

 For more complex non-linear recursive structures, deferring all ancillae for recursive calls until the
  top level may lead to space explosion. For these cases, we perform cleanup in each layer to mitigate this issue. In particular, we place the cleanup at the end of each layer's function body to prevent redundant recomputation within the same layer that would arise from earlier cleanup. For example, in Figure~\ref{fig:f1-source-target}, suppose \(\mathtt{f_1}\) is a non‑linear recursive function. 
  At line~\(i\), a recursive call \(x \gets \mathtt{f_1}(\overline{u})\) in the target program uses quantum variable \(a_0\) and temporaries \(a_1,\dots,a_n\); it is uncomputed at line~\(j\) when no longer needed, leaving a block \(C_3\) still to execute. If some variables \(y\) and \(z\) depend on \(x\) and have not been uncomputed at line~$j$, uncomputing either \(y\) or \(z\) inside \(C_3\) would force recomputation of \(x\), incurring extra time overhead.

Furthermore, from a spatial perspective, delaying cleanup of recursive calls until the end of each layer is often more efficient than earlier cleanup. Admittedly, earlier cleanup may allow \(a_0\) to be reused in \(C_3\); however, if an ancilla \(a_i\) is reused after line \(i\) and not yet cleaned at line \(j\), it may be clean during forward computation (allowing sharing across layers) but becomes dirty during uncomputation. Thus, uncomputing \(x\) requires a fresh \(a_i\) per layer, turning it into a quantum array, which increases space consumption. Conversely, cleaning up uniformly at the end of the function body ensures that all required ancillae for recursive calls are clean, which also simplifies the implementation of uncomputation.

To this end, we introduce a cleanup mechanism based on variable roles, encoded by a flag associated with each variable node in the variable dependency graph (see Section~\ref{sec:dg-ana}): 
\begin{itemize}
   \item \(x.\flag = 2\): non‑return local variables arising from linear recursive-call statements. Cleanup is deferred to an external caller after the entire recursion finishes (i.e., not cleaned within any recursive layer), to control time complexity. 
    \item \(x.\flag=1\): non‑return local variables arising from recursive-call statements in non-linear recursive structures. Cleanup is performed at the end of each recursive layer to avoid deferring all recursive-call temporaries to the outermost call, while reducing redundant recomputation and limiting temporary-register overhead. 
    \item \(x.\flag = 0\): other variables generated by non‑recursive statements. Eager cleanup is applied to minimize space usage. 
\end{itemize}  

\section{Compilation from RQIMP to \texorpdfstring{\(\RQC^{++}\)}{RQC++}}
\label{sec:comp}

\begin{wrapfigure}{r}{0.42\linewidth}
    \centering
    \includegraphics[width=\linewidth]{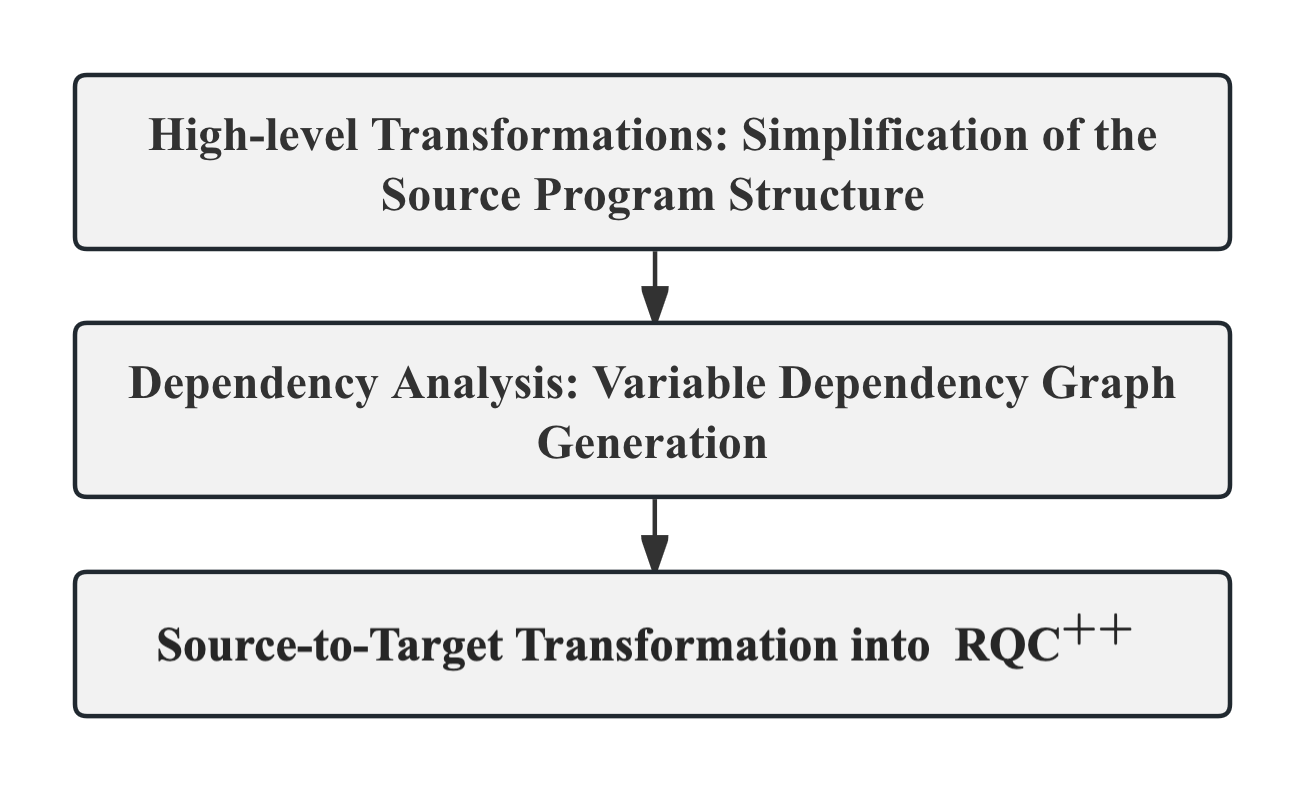} 
    \caption{The compilation process.}
    \label{fig:com_pro}
\end{wrapfigure}   

In this section, we show how to compile an RQIMP program \(P\) into \(\RQC^{++}\) using the key techniques described above. Figure~\ref{fig:com_pro} summarizes the compilation process.  

(1) First, a series of high‑level transformations are applied to the original program \(P\) to obtain \(P_h\), simplifying the program structure and facilitating further compilation; see Section~\ref{sec:hl-trans}.

(2) After simplification, we construct dependency graphs that capture data dependencies and variable lifetimes, enabling the compiler to determine when temporary variables become dead and can be safely uncomputed; see Section~\ref{sec:dg-ana}.

(3) The transformed program is then compiled into \(\RQC^{++}\), where translation of program constructs and uncomputation of temporaries take place. 

The remainder of this section is devoted to describing these passes carefully.    

\subsection{High-Level Transformations}
\label{sec:hl-trans} 
We apply a series of high-level transformations to the original program $P$. These transformations simplify the structure of \(P\), making it more amenable to subsequent compilation steps. The resulting program remains within a syntactic subset of RQIMP. Concretely, we replace each branch of a conditional with a call to a fresh function,  transform loops into recursive functions, and replace complex expressions in guards, function call arguments, and return values with variables via temporary assignments. The detailed transformations are provided in Appendix~\ref{app:high-tr}.  After applying these transformations, the transformed program uses the following
  core statement grammar: 
\[
\begin{array}{rlll}
 \mathrm{Statement}   & s &:=  &  \tau x\leftarrow e \ | \ l \leftarrow \mathtt{aop} \ e \ |  \  \tau x \leftarrow f (\overline{x}) \ |  \   s; s \ | \  \mathtt{if} \ x \ \{s\} \ \{s\}. 
\end{array}
\]
For example, after applying the high‑level transformations, $\mathtt{sum}$ becomes the mutually recursive program with $\mathtt{aux}$ shown in Figure~\ref{fig:sum_source}.

\subsection{Dependency Analysis}
\label{sec:dg-ana}
The eager cleanup strategy determines when to clean variables based on their dependency relationships. To capture these dependencies explicitly, we construct a function call graph \(G_{\mathrm{call}}\) and, for each function \(f\), a variable dependency graph \(G_{\mathrm{fun}}(f)\).  In \(G_{\mathrm{call}}\), every function is a node; a directed edge from \(f_1\) to \(f_2\) is added for each call from \(f_1\) to \(f_2\) (thus multiple calls produce parallel edges).   In \(G_{\mathrm{fun}}(f)\), nodes are of two kinds: each \(\mathtt{Q}\)-mode variable and each statement. Predefined dependency subgraphs for specific statement types are shown in Appendix~\ref{secA_dg}. For example, the statement \(\tau x \gets f(y)\) yields the following dependency graph.   
\begin{figure}[h] 
    \centering
    \vspace{-0.6em}
    \includegraphics[width=0.25\linewidth, height=0.7cm, keepaspectratio]{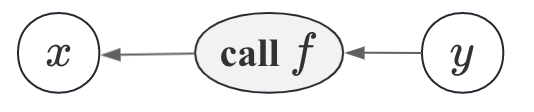} 
    \vspace{-0.8em}
    \Description{...}
\end{figure}

Next, to capture recursive characteristics, each variable node \(x\) is assigned a flag, with \(x.\flag=0,1,2\) indicating the different uncomputation timings described in Section~\ref{sec:key}. 

Furthermore, some local variables must be cleaned together when they are mutually dependent; cleaning only one of them may be ineffective. To address such cases, we merge all variables belonging to the same strongly connected component (i.e., mutually dependent variables) into a composite node. Such a node indicates that its variables form an atomic unit: collective uncomputation is performed only when all variables inside the unit are no longer in use. A single variable that is not merged is treated as a composite node containing exactly one variable (the degenerate case).  This node partitioning ensures that no mutual dependencies remain among nodes. Moreover, we guarantee that the composite nodes containing global variables or the return value are disjoint from those containing other local variables (see Appendix~\ref{secA_dg} for details). Consequently, we can correctly clean up the quantum variables associated with source local variables, while also ensuring that globals and the return value are not affected.

 For each composite node $w$, we define its flag $w.\flag$ as follows: if there exists at least one variable $x$ in $w$ with $x.\flag = 1$, then $w.\flag = 1$; otherwise $w.\flag = 0$. Finally, all composite nodes with $\flag=1$ are merged into one composite node, ensuring that at most one composite node with $\flag=1$ remains after the merge and allowing cleanup of the entire group to be deferred until the end of the function body. The uniqueness of this node ensures that, at the time of its uncomputation, every other node has already been cleaned, thereby facilitating the uncomputation process; see Subsection~\ref{sec:clean_temp}. 
 
 It is important to note that a node with \(\flag=1\) cannot contain any variable with \(\flag=2\) (the two flags are mutually exclusive). Conversely, during the cleanup phase, we will ensure that a node with \(\flag=0\) only cleans the variables with \(\flag\neq 2\) inside it, ignoring those with \(\flag=2\). Consequently, variables marked \(\flag=2\) are never cleaned within the current function; their uncomputation is deferred until the outermost recursive call finishes.

\subsection{Source-to-Target Program Transformation}
\label{sec:source-tar-trans}
In this subsection, we describe how to translate the simplified program into the corresponding $\RQC^{++}$ target program. We first present a high-level sketch of the compilation; further details are given in the following subsections. 

 The compiler operates over \textit{static input parameters} \(\K\) and a \textit{dynamic compilation context} \(\ctx\): the former is read‑only during compilation, while the latter carries the mutable state of the algorithm. 

\paragraph{Static input parameters.}
Let
\[
\K = (\Xi,\ \Gamma,\ \G = (G_{\mathrm{call}}, G_{\mathrm{fun}}),\;  \Theta, \; \rho_{\mathrm{glob}})
\]
be the static input parameters. The function environment \(\Xi\) and the global type 
environment \(\Gamma\) are obtained during type checking 
(Section~\ref{sec:source}).  \(\G = (G_{\mathrm{call}}, G_{\mathrm{fun}})\) is the
program dependency graph (Section~\ref{sec:dg-ana}), constructed after the high‑level transformations.  
\(\Theta\) is the function name mapping that maps each source function
\(f \in \Xi\) to its target name \(F_f\) in $\RQC^{++}$.  \(\rho_{\mathrm{glob}}\)
is the global variable mapping that links source global variables to
their allocated quantum variables. \(\Xi\), \(\Gamma\) and \(\G\) are extracted once from the source program
by static analysis and serve as inputs to the entire compilation
algorithm; \(\Theta\) and \(\rho_{\mathrm{glob}}\) are initialized when
compilation begins. 

\paragraph{Dynamic compilation context.}
The dynamic compilation context is a tuple 
\[
\ctx = (\D, \St, \M, \Qpool, \iota, f_{\mathrm{cur}}, \Bbuf) 
\] 
that captures all information from the global level down to the current point of focus, where: 
\begin{itemize}
    \item \(\D\) (Target declaration set) collects the $\RQC^{++}$ declarations of functions that have been compiled.
    \item \(\St\) (Compilation stack) records contexts suspended due to function calls. Elements are either a suspended function name \(f\) or a pair \((g, \Vactive)\), where \(\Vactive\) is the active-variable fragment of function \(g\) at the suspension point (defined in Section~\ref{sec_com_fun}). 
 \item \(\M\) (Function information map) stores the local compilation state \(m = \M(f)\) for each suspended \(f\), excluding the active-variable fragment \(\Vactive\), which is kept in \(\St\) when suspended. 
 \item \(\Qpool\) (Quantum variable pool) is the set of currently available
      idle quantum variables---a sufficiently large pool of fresh quantum variables, initially excluding \(\cod(\rho_{\mathrm{glob}})\). 
  \item \(\iota\) (Index supplier) records the current index variable names and is initialized to a fresh variable \(i\), which is used when handling recursive calls.  
    \item \(f_{\mathrm{cur}}\) (Current function) is the name of the source function being compiled.
        \item $\Bbuf$ (current buffer) denotes the transient compilation state associated with $f_{\mathrm{cur}}$. It contains the pair $(\Vactive, m)$, where $m$ stores the remaining compilation record for the current function, such as the compiled $\RQC^{++}$ code of its body. Further details are provided in Section~\ref{sec_com_fun}. 
\end{itemize}
The compilation context manages the compilation state as follows. During statement compilation, quantum variables are allocated from \(\Qpool\) for
source-level variables and ancillae; once a quantum variable is no longer needed, it is uncomputed and returned to \(\Qpool\) for reuse. Upon encountering a call to an as‑yet‑uncompiled function \(f\), the current function is suspended. Its active-variable fragment \(\Vactive\) is pushed onto the stack \(\St\), and its remaining local record \(m\) is stored in \(\M\). The function name \(f\) is then pushed onto \(\St\) to mark the start of its compilation, and the focus shifts to \(f\) by setting \(f_{\mathrm{cur}} = f\) and beginning its compilation. The current buffer \(\Bbuf\) is updated during the compilation of the current function. After \(f\) is fully compiled, its resulting declaration is recorded in \(\D\); subsequent calls to \(f\) can be resolved by looking up \(\Theta\) and retrieving the pre‑compiled declaration from \(\D\). Further notation will be introduced as needed in the following subsections.

The overall compilation process is formalized in Algorithm~\ref{alg:compile_program} (\textsc{Compile\_Prog}), which starts from the main function. In the initialization step, the compilation context \(\ctx\) is set up as
follows: \(\D\), \(\M\) and $\St$ are empty, the remaining global components are
initialized as described above, and the current buffer \(\Bbuf\) is prepared as in function compilation 
(the \textsc{Compile\_Fun} algorithm in Section~\ref{sec_com_fun}).  Then \(\mathtt{main}\) is pushed onto the stack \(\St\) and \(f_{\mathrm{cur}} \gets \mathtt{main}\) to start compilation. The algorithm then invokes \(\textsc{Compile\_Body}\) (Section~\ref{sec:com_body}) to
obtain the compiled circuit for the body %\(s\) 
and finally adds the resulting declaration to \(\D\). 

\begin{algorithm}[!tbp]
\caption{\textsc{Compile\_Prog}: Compiling the Whole Source Program}
\label{alg:compile_program}
\KwIn{Function environment $\Xi$, global type environment $\Gamma$, program dependency graph $\G$} 
\KwOut{Compiled $\RQC^{++}$ program declaration set $\D$}
\SetKwProg{Fn}{Function}{}{}
\Fn{\textsc{Compile\_Prog}($\Xi$, $\Gamma$, $\G$)}{ 
    Initialize \(\Theta\) and \(\rho_{\mathrm{glob}}\); \(\K \gets (\Xi,\ \Gamma,\ \G,\ \Theta,\ \rho_{\mathrm{glob}})\) \tcp*{initialize static parameters}
    Initialize \(\ctx\); \(\push(\St, \mathtt{main})\); \(f_{\mathrm{cur}} \gets \mathtt{main}\) \tcp*{initialize compilation context}  
    \(\ctx \gets \textsc{Compile\_Body}(\K, \ctx, s_{\mathtt{main}})\) \tcp*{compile main body: $s_{\mathtt{main}}$}
    \(\D \gets \D \cup \{\Theta({\mathtt{main}}) \Leftarrow \ctx.\Bbuf.m.C\}\)  \tcp*{register declaration}
    \Return \(\D\)\;
}

\end{algorithm} 

\begin{wrapfigure}{r}{0.38\linewidth}
\centering
\definecolor{algblue}{RGB}{127,144,181}
\definecolor{algorange}{RGB}{244,199,151}
\definecolor{algred}{RGB}{186,111,101}
\definecolor{algpink}{RGB}{230,191,203}
\definecolor{alglightblue}{RGB}{198,215,232}
\vspace{-1.5ex}
\resizebox{\linewidth}{!}{%
\begin{tikzpicture}[
  font=\scriptsize\bfseries,
  algnode/.style={
    draw=black!70,
    rounded corners=2pt,
    minimum height=5.5mm,
    minimum width=24mm,
    align=center,
    inner xsep=3pt
  },
  dep/.style={-Stealth, semithick, draw=black!80}
]
\node[algnode, fill=algblue] (prog) at (0,0)
  {\textsc{Compile\_Prog}};
\node[algnode, fill=algorange] (body) at (0,-0.90)
  {\textsc{Compile\_Body}};
\node[algnode, fill=algred] (com) at (-1.70,-1.85)
  {\textsc{Compile\_Com}};
\node[algnode, fill=algpink] (fun) at (-1.70,-2.75)
  {\textsc{Compile\_Fun}};
\node[algnode, fill=algred] (clean) at (1.70,-1.85)
  {\textsc{Clean\_Temp}};
\node[algnode, fill=alglightblue] (recur) at (1.70,-2.75)
  {\textsc{Compile\_Recur}};

\draw[dep] (prog) -- (body);
\draw[dep] (body) -- (com);
\draw[dep] (body) -- (clean);
\draw[dep] (com) -- (fun);
\draw[dep] (com.east) -- ++(.45,0) |- (recur.west);
\draw[dep] (fun.west) -- ++(-.70,0) |- (body.west);
\end{tikzpicture}
}
\caption{Structure of the compilation algorithms.}
\label{fig:compile-alg-structure}
\Description{A flow diagram showing Compile_Prog calling Compile_Body, which invokes Compile_Com and Clean_Temp; Compile_Com may invoke Compile_Fun and Compile_Recur, and Compile_Fun returns to Compile_Body.}
\vspace{-2ex}
\end{wrapfigure}
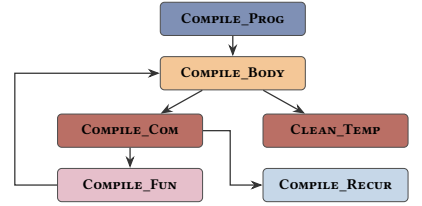

Internally, \(\textsc{Compile\_Body}\)  compiles the statements of the function body inductively into the corresponding \(\RQC^{++}\) program. As shown in Figure~\ref{fig:compile-alg-structure}, it relies on two main sub‑processes, \textsc{Compile\_Com} and \textsc{Clean\_Temp}. 

\(\textsc{Compile\_Com}\) (Section~\ref{sec_com_stmt}) is called to translate individual statements. Recursive call statements are delegated to \(\textsc{Compile\_Recur}\) (Section~\ref{sec_com_stmt}), which implements the aforementioned variable indexing mechanism to ensure correct usage of temporaries in recursive calls. When a call to a yet‑uncompiled function is encountered, \(\textsc{Compile\_Fun}\) (Section~\ref{sec_com_fun}) is invoked recursively to compile that function on the fly, after which compilation resumes at the original call site.

 When a temporary variable reaches the end of its lifetime, we decide whether to clean it based on the flags in the dependency graph. Uncomputation is performed by calling \textsc{Clean\_Temp} (Section~\ref{sec:clean_temp}), which inserts a circuit that restores the variable to a clean state (\(\ket{0}\)) for reuse.

Recall the running example: compilation starts from \texttt{main}. When a call to \texttt{sum} is encountered in the body of \texttt{main}, the compilation of \texttt{main} is suspended and the compiler proceeds to compile \texttt{sum}. Similarly, when a call to \texttt{aux} is encountered during the compilation of \texttt{sum}, the compilation of \texttt{sum} is suspended while the compiler first compiles \texttt{aux}.

\subsection{Compilation of Functions}
\label{sec_com_fun}
In this subsection, we present the function-compilation algorithm, which transforms source function declarations into target declarations. We begin by detailing the structure of the compilation buffer for the current function. The current buffer, denoted \(\Bbuf = (\Vactive, m)\), encapsulates the transient state of the function currently being compiled, namely \(f_{\mathrm{cur}}\). 

\begin{enumerate}
    \item \(\Vactive\) (Active-variable set): the active-variable fragment newly generated since the current function was last
    resumed, consisting of temporary quantum variables that have been used but not
    yet cleaned. 
    \item \(m = (C, \rho_{\mathrm{loc}}, \kappa, \delta, \Eargs, \Valloc)\):  
    \begin{itemize} 
        \item \(C\) (Code segment): the $\RQC^{++}$ program code already generated for compiling the current function body.
        \item \(\rho_{\mathrm{loc}}\) (Local variable mapping): a mapping from the local variables (including input parameters) of the current function to their corresponding quantum variables. 
        \item \(\kappa\) (Cleanup circuit mapping): a mapping from nodes of the variable dependency graph to their cleanup circuits. A node may group multiple variables that must be cleaned together; applying the inverse of this circuit uncomputes the quantum variables associated with the variables in that node (see Section~\ref{sec:com_body} for details). 
        \item \(\delta\) (Residual out-degree mapping): a mapping from variable dependency graph nodes to their current residual out-degrees, initialized from the dependency graph and decremented during compilation. 
        \item \(\Eargs\) (Extra-parameter set): the set of additionally introduced parameters, such as indices and temporary variables, for the target version of the current function. 
        \item \(\Valloc\) (Allocated-variable set): the set of all quantum-variable base names that have been allocated from \(\Qpool\) for compiling the current function. This set is later merged into \(\Eargs\) after function compilation completes.
        % \item \(R_{\mathrm{stmt}}\) (Remaining statements): the sequence of source statements of \(f_{\mathrm{cur}}\) that have not yet been compiled. 
    \end{itemize}
\end{enumerate} 

\textbf{Remark.}  The \(\Vactive\) component in the current buffer represents the latest active-variable fragment of \(f_{\mathrm{cur}}\), while the fragments pushed onto the stack \(\St\) when \(f_{\mathrm{cur}}\) was previously suspended remain in \(\St\). Thus the complete active-variable set for \(f_{\mathrm{cur}}\) is the stack-ordered concatenation of all its fragments associated with \(f_{\mathrm{cur}}\) on \(\St\) followed by the current buffer's \(\Vactive\). Note that the stack \(\St\) stores, in order, either a function name (marking the start of its compilation) or a suspended pair \((g, \Vactive)\). This design preserves the actual usage order of active variables relative to suspended functions in \(\St\), which is essential for identifying the used but uncleaned variables of the current layer for each function, especially for mutually recursive functions.

\begin{wrapfigure}{r}{0.42 \linewidth} 
    \centering
    \includegraphics[width=\linewidth]{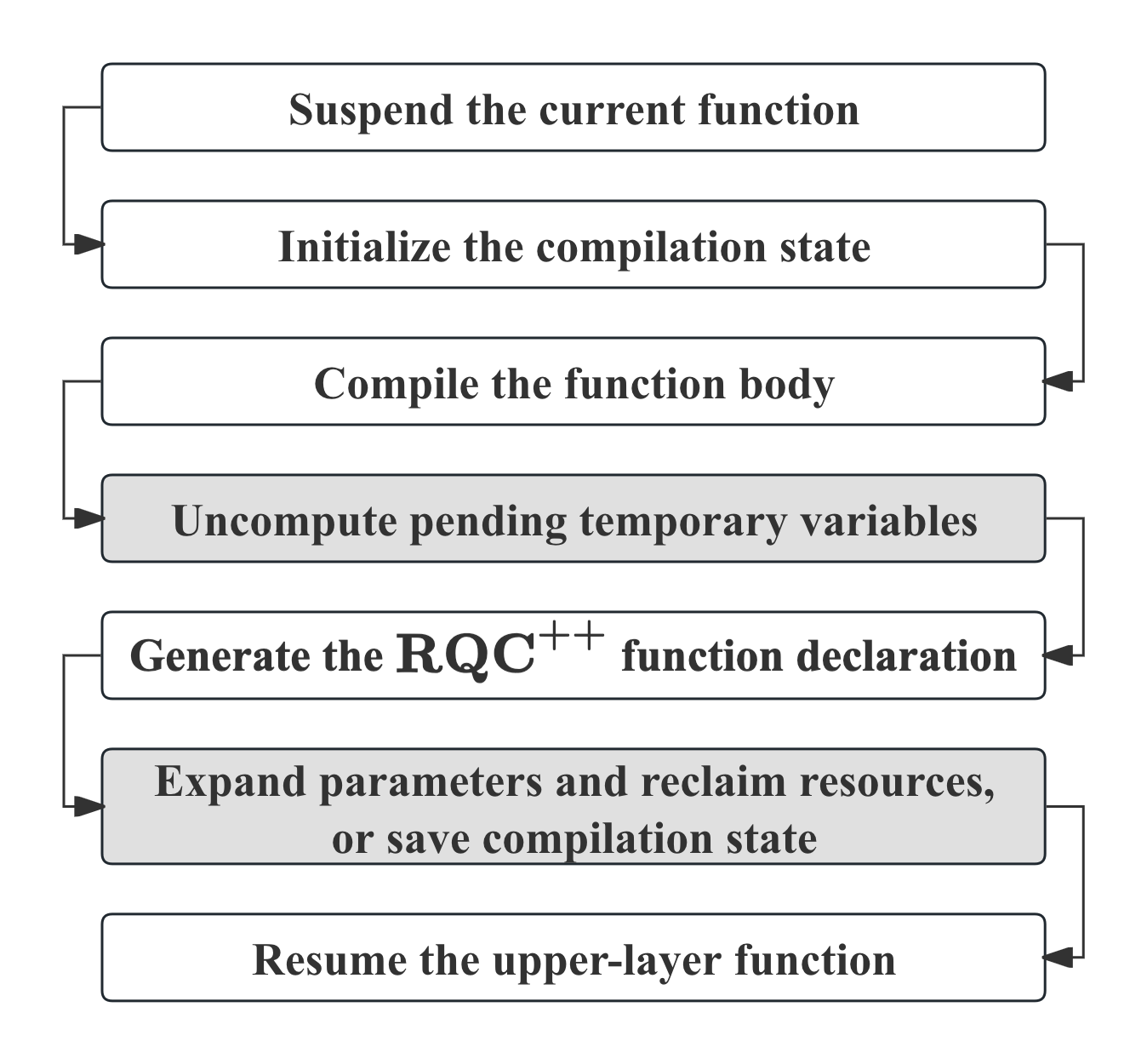}
    \caption{The Workflow for Compilation of Functions.}  
    \label{fig:com_fun}
\end{wrapfigure}   

The function-compilation algorithm \(\textsc{Compile\_Fun}\) compiles a function \(f\) by operating on the compilation context \(\ctx\). It proceeds in seven steps, as shown in Figure~\ref{fig:com_fun}; the complete formal algorithm definitions are given in Appendix~\ref{app:com_fun}.  At a high level, it follows the standard pattern of compiling a function:
  saving the current context, initializing the new function context, compiling
  the body, generating and recording the target declaration, and then resuming the
  suspended context.  
The additional steps are explained below.

Upon completion of the function-body compilation by \(\textsc{Compile\_Body}\), some temporary variables may still require cleanup but remain uncomputed, namely temporary quantum variables that correspond to local variables generated by non-linear recursive structures (whose nodes have \(\flag=1\)). According to the uncomputation strategy described above, these variables are uncomputed after the function body finishes. The \(\textsc{Clean\_Temp}\) algorithm (Section~\ref{sec:clean_temp}) is then invoked to uncompute them.

In Step~6, after generating the target function, the algorithm distinguishes two cases. Let \(g\) be the function suspended in the first step, and let $\SCC(g)$ denote the strongly connected component containing $g$ in the call-dependency graph, i.e., $g$'s mutual-recursion group. If \(f \notin \SCC(g)\), then \(f\) and the functions following it on \(\St\) constitute a distinct mutual-recursion group \(\SCC(f)\), indicating that the compilation of this group has completed. The internal quantum variables recorded in the \(\Valloc\) sets of these functions are collected as shared extra parameters and added consistently to their declarations and all call sites in \(\D\). The resources consumed by these functions are then returned to \(\Qpool\), and their compilation information is cleared from the stack \(\St\) and the mapping \(\M\). If \(f \in \SCC(g)\), then \(f\) remains on the stack and its compilation information is preserved for later recursive calls and parameter expansion. 
 
We use part of the compilation flow of function \(\mathtt{aux}\) in Figure~\ref{fig:sum_source} to illustrate some of these steps. The suspension
and initialization steps are shown below, where ``-'' denotes an ignored component.

\begin{figure}[h] 
    \centering

  \resizebox{0.99\linewidth}{!}{
\begin{tikzpicture}[
  font=\sffamily,
  cell/.style={draw=black!45, line width=.12mm, minimum height=9.3mm, align=center, inner xsep=5pt, inner ysep=2pt, fill=white},
  changed/.style={draw=accent!65, line width=.12mm, minimum height=9.3mm, align=center, inner xsep=5pt, inner ysep=2pt, fill=accent!8},
  head/.style={font=\sffamily\bfseries\normalsize},
  val/.style={font=\sffamily\normalsize},
  title/.style={font=\sffamily\bfseries\large},
  lab/.style={font=\sffamily\small},
  arr/.style={-{Latex[length=2.4mm]}, line width=.22mm, draw=accent!70}
]
\def\xzero{0}
\def\ybefore{-0.78}
\def\yafter{-2.63}
\def\h{0.78}
\def\wD{0.72}
\def\wS{3.05}
\def\wM{2.55}
\def\wQ{0.72}
\def\wI{0.72}
\def\wF{1.05}
\def\wB{8.95}

\node[cell, minimum width=\wD cm] (D1) at (1.36,\ybefore)
  {\large \(\mathcal{D}\)\\[-.5mm]\normalsize -};
\node[changed, minimum width=\wS cm, right=0cm of D1] (S1)
  {\large \(\mathcal{S}\)\\[-.5mm]\normalsize \([\mathtt{main};\mathtt{sum}]\)};
\node[changed, minimum width=\wM cm, right=0cm of S1] (M1)
  {\large \(\mathcal{M}\)\\[-.5mm]\normalsize \([\mathtt{main}\mapsto -]\)};
\node[cell, minimum width=\wQ cm, right=0cm of M1] (Q1)
  {\large \(\mathcal{Q}\)\\[-.5mm]\normalsize -};
\node[cell, minimum width=\wI cm, right=0cm of Q1] (I1)
  {\large \(\iota\)\\[-.5mm]\normalsize \(j\)};
\node[changed, minimum width=\wF cm, right=0cm of I1] (F1)
  {\large \(f_{\mathrm{cur}}\)\\[-.5mm]\normalsize \(\mathtt{sum}\)};
\node[changed, minimum width=\wB cm, right=0cm of F1] (B1)
  {\large \(\mathcal{B} \)\\[-.5mm]
   \normalsize \(V_{\mathrm{active}}=\{q[i]\},\,
   m=(C,\rho_{\mathrm{loc}}:\{k \mapsto p, x \mapsto r, y_1 \mapsto q[i]\}, \kappa,\delta,E_{\mathrm{args}}:\{i\}, V_{\mathrm{alloc}}:\{q\})\)};

\node[lab, text=red, anchor=west] at (5,-1.67)
  {Suspend current function and initialize the callee: };
\draw[arr] (12,-1.3) -- (12,-2.08);
\node[lab, text=red, anchor=west] at (12.3,-1.67)
  {push \([(\mathtt{sum},V_{\mathrm{active}}),\mathtt{aux}]\), store \(m\), and initialize \(\mathcal{B}\)};

% After row cells
\node[cell, minimum width=\wD cm] (D2) at (1.35,\yafter)
  {\large \(\mathcal{D}\)\\[-.5mm]\normalsize -};
\node[changed, minimum width=\wS cm, right=0cm of D2] (S2)
  {\large \(\mathcal{S}\)\\[-.5mm]\normalsize \([\mathtt{main};\mathtt{sum};\textcolor{mark}{(\mathtt{sum},\{q[i]\});\mathtt{aux}}]\)};
\node[changed, minimum width=\wM cm, right=0cm of S2] (M2)
  {\large \(\mathcal{M}\)\\[-.5mm]\normalsize \([\mathtt{main}\mapsto -,\,\textcolor{mark}{\mathtt{sum}\mapsto m}]\)};
\node[cell, minimum width=\wQ cm, right=0cm of M2] (Q2)
  {\large \(\mathcal{Q}\)\\[-.5mm]\normalsize -};
\node[cell, minimum width=\wI cm, right=0cm of Q2] (I2)
  {\large \(\iota\)\\[-.5mm]\normalsize \(j\)};
\node[changed, minimum width=\wF cm, right=0cm of I2] (F2)
  {\large \(f_{\mathrm{cur}}\)\\[-.5mm]\normalsize \(\textcolor{mark}{\mathtt{aux}}\)};
\node[changed, minimum width=\wB cm, right=0cm of F2] (B2) 
  {\large \(\mathcal{B}\)\\[-.5mm]
   \normalsize \(\textcolor{mark}{V_{\mathrm{active}}=\{\},\,
   m=(I,\{k\mapsto p,x\mapsto r\},\{\},\{y_1\mapsto 1, k \mapsto 2\},\{i\},\{\})}\)};
\end{tikzpicture}}
\Description{Compilation-context transition for suspending the current function and initializing the callee.}
\end{figure} 

The compiler pushes the current \(\Vactive\) onto the stack \(\St\), stores the remaining compilation state \(m\) in the mapping \(\M\), and pushes the name of the function to be compiled, \(\mathtt{aux}\), onto the top of \(\St\). It then initializes the local buffer by setting \(C=I\), initializing \(\Vactive\) and \(\Valloc\) to empty sets, setting \(\kappa\) to an empty mapping, initializing the residual out-degree map \(\delta\) from the variable dependency graph, and assigning quantum variables to the formal parameter \(k\) and return variable \(x\), with the corresponding mappings recorded in \(\rho_{\mathrm{loc}}\). Since \(\mathtt{aux}\) and the suspended function \(\mathtt{sum}\) belong to the same mutual-recursion group, \(\mathtt{aux}\) initializes its extra-parameter set \(\Eargs\) to that of \(\mathtt{sum}\), namely \(\{i\}\), to ensure consistency within the group. In contrast, when compiling a function outside the current mutual-recursion group, \(\Eargs\) is initialized to \(\emptyset\). Finally, the current function marker is updated to \(f_{\mathrm{cur}}=\mathtt{aux}\), and then compilation of the function body begins. 

After compiling the function body and cleaning up the remaining temporary variables, the compiler generates
  the target declaration of \(\mathtt{aux}\) and records it in \(\D\). Its parameters include the input, the
  output, and the current extra-parameter set \(\{i,j\}\), and its body is the code \(C_{\mathtt{aux}}\) recorded in the current buffer. Then the  context updates in Steps~6 and~7 are shown as follows. 

\begin{figure}[h]
   \centering
   \resizebox{0.99\linewidth}{!}{
\begin{tikzpicture}[
  font=\sffamily,
  cell/.style={draw=black!45, line width=.12mm, minimum height=9.3mm, align=center, inner xsep=5pt, inner ysep=2pt, fill=white},
  changed/.style={draw=accent!65, line width=.12mm, minimum height=9.3mm, align=center, inner xsep=5pt, inner ysep=2pt, fill=accent!8},
  title/.style={font=\sffamily\bfseries\large},
  lab/.style={font=\sffamily\normalsize},
  arr/.style={-{Latex[length=2.4mm]}, line width=.22mm, draw=accent!70}
]

\def\ybefore{-0.78}
\def\yafter{-2.8}
\def\ylast{-4.48}
\def\wD{0.72}
\def\wS{3.05}
\def\wM{2.55}
\def\wQ{0.72}
\def\wI{0.72}
\def\wF{1.05}
\def\wB{8.95}

\node[changed, minimum width=\wD cm, anchor=west] (bD1) at (0.7,\ybefore)
  {\large \(\mathcal{D}\)\\[-.5mm]\normalsize \(\textcolor{mark}{\{F_{\mathtt{aux}}(p,r,i,j)\Leftarrow C_{\mathtt{aux}}\}}\)};
\node[changed, minimum width=\wS cm, right=0cm of bD1] (bS1)
  {\large \(\mathcal{S}\)\\[-.5mm]\normalsize \([\mathtt{main};\mathtt{sum};(\mathtt{sum},\{q[i]\});\mathtt{aux}]\)};
\node[changed, minimum width=\wM cm, right=0cm of bS1] (bM1)
  {\large \(\mathcal{M}\)\\[-.5mm]\normalsize \([\mathtt{main}\mapsto -,\,\mathtt{sum}\mapsto -]\)};
\node[cell, minimum width=\wQ cm, right=0cm of bM1] (bQ1)
  {\large \(\mathcal{Q}\)\\[-.5mm]\normalsize -};
\node[cell, minimum width=\wI cm, right=0cm of bQ1] (bI1)
  {\large \(\iota\)\\[-.5mm]\normalsize \(k\)};
\node[changed, minimum width=\wF cm, right=0cm of bI1] (bF1)
  {\large \(f_{\mathrm{cur}}\)\\[-.5mm]\normalsize \(\mathtt{aux}\)};
\node[changed, minimum width=\wB cm, right=0cm of bF1] (bB1)
  {\large \(\mathcal{B}\)\\[-.5mm]\normalsize
   \((V_{\mathrm{active}}=\{q_2[j],q_2[j+1{:})\},\,m=(C_{\mathtt{aux}},-,-,-,\{i,j\},\{q_1,q_2\}))\)};

\node[lab, text=red, anchor=center] at (8.25,-1.72)
  {\large Save compilation state and resume the upper-layer function:};  
\draw[arr] (13.5,-1.3) -- (13.5,-2.0);
\node[lab, text=red, anchor=center] at (17.5,-1.72)
  {\large save \(m\) and \((\mathtt{aux},V_{\mathrm{active}})\), and restore \(\mathcal{B}\) for $\mathtt{sum}$};

% After row cells
\node[cell, minimum width=.72cm, anchor=west] (bD2) at (0.7,\yafter)
  {\large \(\mathcal{D}\)\\[-.5mm]\normalsize \(-\)};
\node[changed, minimum width=9 cm, right=0cm of bD2] (bS2) 
  {\large \(\mathcal{S}\)\\[-.5mm]\normalsize \([\mathtt{main};\mathtt{sum};(\mathtt{sum},\{q[i]\});\mathtt{aux};\textcolor{mark}{(\mathtt{aux}, \{q_2[j],q_2[j+1{:})\})}]\)};
\node[changed, minimum width=6 cm, right=0cm of bS2] (bM2)
  {\large \(\mathcal{M}\)\\[-.5mm]\normalsize \([\mathtt{main}\mapsto -,\,\mathtt{sum}\mapsto -,\,\textcolor{mark}{\mathtt{aux}\mapsto m}]\)};
\node[cell, minimum width=\wQ cm, right=0cm of bM2] (bQ2)
  {\large \(\mathcal{Q}\)\\[-.5mm]\normalsize -};
\node[cell, minimum width=\wI cm, right=0cm of bQ2] (bI2)
  {\large \(\iota\)\\[-.5mm]\normalsize \(k\)}; 
\node[changed, minimum width=\wF cm, right=0cm of bI2] (bF2)
  {\large \(f_{\mathrm{cur}}\)\\[-.5mm]\normalsize \(\textcolor{mark}{\mathtt{sum}}\)};
\node[changed, minimum width=5.6cm, right=0cm of bF2] (bB2)
  {\large \(\mathcal{B}\)\\[-.5mm]\normalsize \(\textcolor{mark}{(V_{\mathrm{active}}=\{\},\,m=\mathcal{M}(\mathtt{sum}))}\)};
\end{tikzpicture}}
\Description{Compilation-context transition for saving a compiled function state and resuming the suspended caller.}
\end{figure} 
Since \(\mathtt{aux} \in \SCC(\mathtt{sum})\), its compilation information is preserved in \(\St\) and
  \(\M\), and the compiler restores the compilation state of \(\mathtt{sum}\), where only \(m\) is restored to the
  current buffer, while \(\Vactive\) remains on the stack. In contrast, after \(\mathtt{sum}\) is compiled, since \(\mathtt{sum} \notin \SCC(\mathtt{main})\), \(\mathtt{sum}\) together with the already compiled function \(\mathtt{aux}\) forms a completed mutual-recursion group. The temporary variables \(q\), \(q_1\), and \(q_2\) are then added consistently to the parameter lists of both \(\mathtt{sum}\) and \(\mathtt{aux}\) to form the final function declaration as shown in Figure~\ref{fig:sum_target}.

\subsection{Compilation of Statements}
\label{sec_com_stmt}
This subsection presents the \(\textsc{Compile\_Com}\) algorithm, which translates individual statements within a function body, including assignments, unary operations, conditionals, and function calls, into target code. The compilation of assignment statements, including unary assignments, mainly relies on expression translation, in which all operations are predefined as corresponding
 \(\RQC^{++}\) programs, with arithmetic operations implemented using QFT-based methods~\cite{liVerifiedCompilationQuantum2022}. The complete compilation algorithm is given in Appendix~\ref{app:com-stmt}; here, we focus only on representative function calls. We distinguish three cases: recursive calls, calls to already compiled functions, and calls to functions not yet compiled.

For a call statement \(\tau x \gets f(\overline{u})\), 
if \(f\) already appears on the stack \(\St\), a recursive call is in progress. Such calls are handled by the \textsc{Compile\_Recur} algorithm, as defined in Appendix~\ref{app:com-stmt}. The core idea is to first ensure that all used but uncleaned variables of the current recursive layer of the callee $f$ are already in subscripted form, thereby enabling the compiler to isolate the quantum variables of different layers via distinct indices. 

For readability, we still use the compilation of function \(\mathtt{aux}\) in Figure~\ref{fig:sum_source} to illustrate the process. As discussed earlier, during the initialization phase of compiling \(\mathtt{aux}\), its parameter list is initialized to \(\{i\}\), inherited from that of \(\mathtt{sum}\). When compiling the body of \(\mathtt{aux}\), the compiler encounters a recursive call to \(\mathtt{sum}\): \lstinline[language=MyLang]{y <- sum(y1)}, while \(\mathtt{sum}\) has already been entered and remains suspended in \(\St\). The subsequent transformation flow is shown in the following diagram. 

\begin{figure}[h] 
    \centering
   
  \resizebox{0.99\linewidth}{!}{
\begin{tikzpicture}[
  font=\sffamily,
  cell/.style={draw=black!45, line width=.12mm, minimum height=9.3mm, align=center, inner xsep=5pt, inner ysep=2pt, fill=white},
  changed/.style={draw=accent!65, line width=.12mm, minimum height=9.3 mm, align=center, inner xsep=5pt, inner ysep=2pt, fill=accent!8},
  title/.style={font=\sffamily\bfseries\large},
  lab/.style={font=\sffamily\normalsize},
  arr/.style={-{Latex[length=2.4mm]}, line width=.22mm, draw=accent!70}
]

\def\ybefore{-0.78}
\def\yafter{-2.55}
\def\ylast{-4.32}
\def\wD{.72}
\def\wS{4.35}
\def\wM{5.25}
\def\wQ{.72}
\def\wI{.72}
\def\wF{1.05}
\def\wB{8.15}

% Before row cells
\node[cell, minimum width=\wD cm] (cD1) at (2.00,\ybefore)
  {\large \(\mathcal{D}\)\\[0mm] \normalsize \(-\)};
\node[changed, minimum width=\wS cm, right=0cm of cD1] (cS1)
  {\large \(\mathcal{S}\)\\[0mm] \normalsize \([\mathtt{main};\mathtt{sum};(\mathtt{sum},\{q[i]\});\mathtt{aux}]\)};
\node[changed, minimum width=\wM cm, right=0cm of cS1] (cM1)
  {\large \(\mathcal{M}\)\\[0mm]\normalsize \([-,\,\mathtt{sum}\mapsto (-,-,-,\{i\},-)]\)};
\node[cell, minimum width=\wQ cm, right=0cm of cM1] (cQ1)
  {\large \(\mathcal{Q}\)\\[0mm]\normalsize  \(-\) };
\node[changed, minimum width=\wI cm, right=0cm of cQ1] (cI1)
  {\large \(\iota\)\\[0mm] \normalsize \(j\)};
\node[cell, minimum width=\wF cm, right=0cm of cI1] (cF1)
  {\large \(f_{\mathrm{cur}}\)\\[0mm] \normalsize\(\mathtt{aux}\)};
\node[changed, minimum width=\wB cm, right=0cm of cF1] (cB1)
  {\large \(\mathcal{B}\)\\[0mm]\normalsize 
   \((V_{\mathrm{active}}=\{q_1,q_2\},\,m=(C,-,-,\{y_1\mapsto 1, k \mapsto 1\},\{i\},-))\)};

\node[lab, text=red, anchor=west, align=left] at (5.75,-1.60)
  {\normalsize add the index variable from $\iota$ to parameters,}; 
\draw[arr] (12.45,-1.22) -- (12.45,-1.92);
\node[lab, text=red, anchor=west, align=left] at (12.65,-1.60)
  {\normalsize subscript live variables and update $\iota$ to a fresh index};

% After row cells
\node[cell, minimum width=\wD cm] (cD2) at (1.70,\yafter)
  {\large \(\mathcal{D}\)\\[0mm]\normalsize \(-\)};
\node[changed, minimum width=3.95cm, right=0cm of cD2] (cS2)
  {\large \(\mathcal{S}\)\\[0mm]\normalsize \([\mathtt{main};\mathtt{sum};(\mathtt{sum},\{q[i]\});\mathtt{aux}]\)};
\node[changed, minimum width=4.65cm, right=0cm of cS2] (cM2)
  {\large \(\mathcal{M}\)\\[0mm]\normalsize \([-,\,\mathtt{sum}\mapsto (-,-,-,\textcolor{mark}{\{i,j\}},-)]\)};
\node[cell, minimum width=\wQ cm, right=0cm of cM2] (cQ2)
  {\large \(\mathcal{Q}\)\\[0mm]\normalsize \(-\)};
\node[changed, minimum width=\wI cm, right=0cm of cQ2] (cI2)
  {\large \(\iota\)\\[0mm] \normalsize \(\textcolor{mark}{k}\)};
\node[cell, minimum width=\wF cm, right=0cm of cI2] (cF2)
  {\large\(f_{\mathrm{cur}}\)\\[0mm]\normalsize \(\mathtt{aux}\)};
\node[changed, minimum width=7.75cm, right=0cm of cF2] (cB2)
  {\large \(\mathcal{B}\)\\[0mm]\normalsize
   \((\textcolor{mark}{V_{\mathrm{active}}=\{q_1[j],q_2[j]\}},\,m=(C,-,-,\{y_1\mapsto 1, k \mapsto 1\},\textcolor{mark}{\{i,j\}},-))\)};

% Recursive-call step
\node[lab, text=red, anchor=west] at (3.75,-3.40) 
  {\normalsize generate recursive call to $\Theta(f)$ with incremented indices,};
\draw[arr] (12.45,-3.00) -- (12.45,-3.70);
\node[lab, text=red, anchor=west] at (12.6,-3.40) 
  {\normalsize and add uncleaned  variables to $V_{\activelabel}$};

   \node[cell, minimum width=\wD cm] (cD3) at (2.8,\ylast)
  {\large \(\mathcal{D}\)\\[0mm]\normalsize \(-\)};
\node[cell, minimum width=.72cm, right=0cm of cD3] (cS3)
  { \large \(\mathcal{S}\)\\[0mm]\normalsize \(-\) };
\node[cell, minimum width=.72cm, right=0cm of cS3] (cM3)
  {\large \(\mathcal{M}\)\\[0mm]\normalsize \(-\) };
\node[cell, minimum width=\wQ cm, right=0cm of cM3] (cQ3)
  {\large \(\mathcal{Q}\)\\[0mm]\normalsize \(-\)};
\node[cell, minimum width=\wI cm, right=0cm of cQ3] (cI3)
  { \large \(\iota\)\\[0mm]\normalsize \(k\)};
\node[cell, minimum width=\wF cm, right=0cm of cI3] (cF3)
  {\large \(f_{\mathrm{cur}}\)\\[0mm] \normalsize \(\mathtt{aux}\)};
\node[changed, minimum width=\wB cm, right=0cm of cF3] (cB3)
  {\large \(\mathcal{B}\)\\[0mm] \normalsize
   \((\textcolor{mark}{V_{\mathrm{active}}=\{q_1[j],q_2[j], q_2[j+1{:})\}},\,m=(\textcolor{mark}{C;F_{\mathtt{sum}}(q_1[j],q_2[j],i+1,j+1)},-,-,\{y_1\mapsto 1, k \mapsto 1\},\{i,j\},-))\)};  
\end{tikzpicture}} 
    % \caption{Workflow of compiling a recursive call statement} 
    % \label{fig:com_recur}
    % \vspace{-1.0 em}  
    \Description{...} 
\end{figure} 

First, the current index \(j\) from \(\iota\) is added to the extra parameter lists \(\Eargs\) of all functions in the mutual-recursion group \(\SCC(\mathtt{aux})\), namely \(\mathtt{aux}\) and \(\mathtt{sum}\). At this point, both functions have the extra parameter list \(\{i,j\}\). Then, this index variable is used to transform the relevant variables into subscripted quantum variables. Concretely, the used but uncleaned variables in the current layer of the callee $\mathtt{sum}$ are the union of the \(\Vactive\) fragments after \(\mathtt{sum}\) on \(\St\) and the current buffer's \(\Vactive\), namely \(q[i], q_1, q_2\), where \(q_1\) and \(q_2\) correspond to the non-return local variables \(y_1\) and \(y\) of the current function \(\mathtt{aux}\). Since \(q[i]\) is already subscripted, the compiler only needs to replace \(q_1\) and \(q_2\) with \(q_1[j]\) and \(q_2[j]\). This renaming also needs to propagate through the compilation state of every function in \(\SCC(\mathtt{aux})\), so that all affected occurrences are replaced consistently. The compiler then updates \(\iota\) from \(j\) to a fresh index variable \(k\) for the next recursive call. 

Finally, in the generated call to the target function, to ensure that all ancillae used by the next layer are clean, every index associated with the live variables of the callee, namely \(i\) in
  \(q[i]\) and \(j\) in \(q_1[j]\) and \(q_2[j]\), is incremented by one:
\lstinline[language=MyLang]{Fsum(q1[j],q2[j],i+1,j+1) // y <- sum(y1)}.
Moreover, since this is a linear recursion, the quantum variables corresponding to \(y\) in the next and deeper layers remain uncleaned after the call returns; therefore, \(q_2[j+1{:})\) is added to \(\Vactive\).

% If function \(f\) has not yet been compiled (i.e., \(f \notin (\Theta^{-1}(\dom(\D)) \cup \St)\)), we invoke Algorithm \(\textsc{Compile\_Fun}\) (Section~\ref{sec_com_fun}) to compile it. Note that this invocation is recursive: \(\textsc{Compile\_Fun}\) depends on \(\textsc{Compile\_Body}\), which in turn depends on the current algorithm. However, this recursion is well-founded because we can define the following measure:
% \[
% \mu(\ctx) \triangleq \bigl| \dom(\Xi) \setminus \bigl( \Theta^{-1}(\dom(\ctx.\D)) \cup \ctx.\St \bigr) \bigr|,
% \]
% i.e., the number of uncompiled source functions. Before calling \(\textsc{Compile\_Body}\) to compile \(f\)'s body, \(\textsc{Compile\_Fun}\) pushes \(f\) onto \(\St\). When a recursive call occurs, the measure \(\mu\) strictly decreases, thereby ensuring termination. 

If \(f\) has already been compiled and the call is non-recursive, the compiled procedure \(\Theta(f)\) is retrieved directly from the declaration set \(\D\). In this case, we need to initialize its classical parameters (if any) and allocate fresh quantum variables from the variable pool \(\Qpool\) as ancillae. For the running example, after compiling \(\mathtt{sum}\), the compiler returns to its call site and compiles \lstinline[language=MyLang]{x <- sum(k)} as follows, where \(p_1, r_1\) record the values of global variables \(k_1, x_1\).
 \begin{center}
  \lstinline[language=MyLang]
  ! Fsum(p1,r',0,0,q,q1,q2); copy(r',r1); qinv(Fsum(p1,r',0,0,q,q1,q2)) !
  \end{center} 
It allocates \(q, q_1, q_2\) as ancillae and sets the recursion indices \(i,j\) to \(0\).  
Moreover, since \(F_{\mathtt{sum}}\) leaves \(q_2\) uncleaned, its return value is first stored in \(r'\), then copied to \(r_1\), 
and finally the inverse of \(F_{\mathtt{sum}}\) is executed to clean all ancillae. 
Otherwise, if the function \(f\) has no uncleaned ancillae, the return value can be assigned directly to the quantum variable corresponding to $x$, as in \lstinline[language=MyLang]{Fsum(p1,r1,0,0,q,q1,q2)}.

If function \(f\) has not yet been compiled, we invoke \(\textsc{Compile\_Fun}\) (Section~\ref{sec_com_fun}) to compile it. Although this creates mutual recursion between \(\textsc{Compile\_Fun}\) and 
the algorithm \(\textsc{Compile\_Com}\), it is well-founded: the number of uncompiled source functions provides a strictly decreasing measure. Moreover, if the callee \(f\) is in the same mutual-recursion group as the current function, the compiler performs a renaming step analogous to the recursive-call case before invoking \(\textsc{Compile\_Fun}\) on \(f\). For example, when compiling \(\mathtt{sum}\), the compiler encounters a call \lstinline[language=MyLang]{x <- aux(k)} to \(\mathtt{aux}\), at a point where \(\mathtt{aux}\) has not yet been compiled. Since \(\mathtt{aux}\) belongs to the same mutual-recursion group as \(\mathtt{sum}\), the extra parameter list of \(\mathtt{sum}\) is first updated to \(\{i\}\), and the currently uncleaned ancillary variable \(q\) in \(\mathtt{sum}\) is renamed to \(q[i]\). The index supplier \(\iota\) is then updated to a fresh index variable \(j\). After these steps, the algorithm invokes \textsc{Compile\_Fun} to compile $\mathtt{aux}$, as described in Section~\ref{sec_com_fun}.  

% After \(\mathtt{aux}\) has been compiled and returned to \(\mathtt{sum}\), the call to \(F_\mathtt{aux}\) can be inserted directly, i.e., \lstinline[language=MyLang]{Faux(p,r,i,j) // x <- aux(k)}, where the first and second arguments are input and output, respectively, and the remaining arguments are the \(\Eargs\) of \(\mathtt{aux}\) at the current compilation time. Neither temporary reallocation, as in Case~2, nor index increment, as in Case~3, is required, because during the compilation of \(\mathtt{aux}\), its internal temporaries are freshly allocated from \(\Qpool\) and are guaranteed to be initialized in the state \(\ket{0}\).

%  \section{Uncomputation} 
%  \label{sec:uncom}
% In the preceding discussion in Section~\ref{sec:technical-overview}, we have detailed the cleanup strategies for different scenarios (eager vs. deferred). These strategies rely on the variable-dependency structure introduced in Section~\ref{sec:dg-ana}. We now present how they are concretely implemented in our compilation framework. 

% The eager cleanup method builds upon prior work~\cite{DBLP:conf/cav/AmyRS17}, with some improvements that will be provided in the following sections. 

\subsection{Compilation of Function Bodies}
\label{sec:com_body}

\begin{wrapfigure}{r}{0.42\linewidth}
    \centering
    \includegraphics[width=\linewidth]{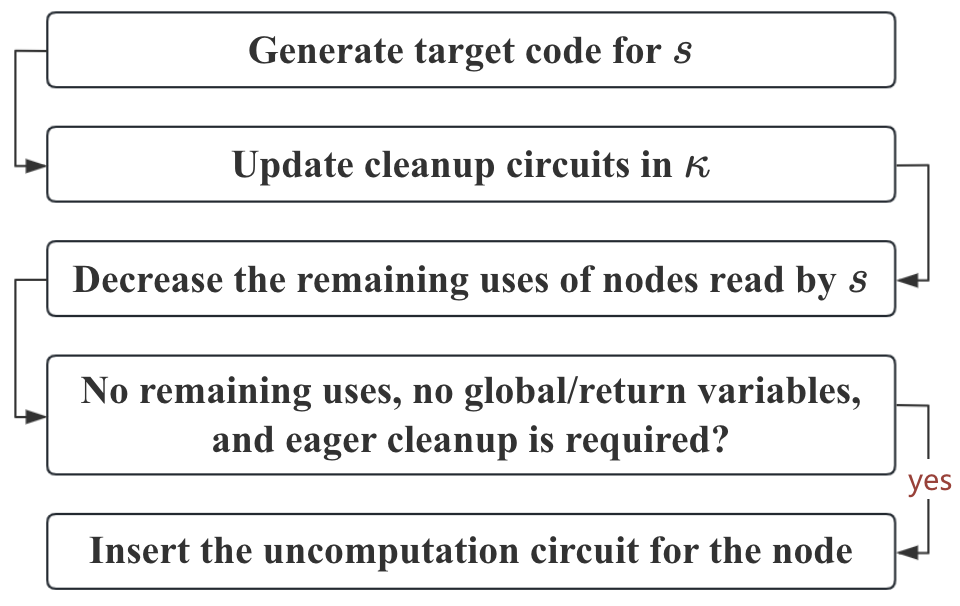}
    \caption{The workflow for compilation of function bodies} 
    \label{fig:com_body} 
\end{wrapfigure}

This subsection presents \textsc{Compile\_Body}, the algorithm for compiling function bodies. Figure~\ref{fig:com_body} shows the compilation process for the single-statement case; sequential statements are handled inductively. A formal definition is given in Appendix~\ref{app:body}, and we explain the main steps below.

Each node in the variable dependency graph is associated with a cleanup circuit, stored in a data structure \(\kappa\). Concretely, \(\kappa(w)\) denotes a circuit whose inverse resets all variables in node \(w\) with \(\flag \neq 2\) to their initial values without affecting variables outside \(w\).

The algorithm first invokes \textsc{Compile\_Com} to generate the target circuit $C_s$ for statement \(s\) and then updates the cleanup circuits of the relevant nodes in $\kappa$. For instance, if the statement \(s\) is not a linear recursive call, then the
  generated circuit $C_s$ is appended to the cleanup circuit of the node containing
  \(\mvq(s)\), i.e.,
  \(\kappa \gets \kappa[\node(\mvq(s)) \mapsto \kappa(\node(\mvq(s))); C_s]\), where $\node(x)$ denotes the node containing variable $x$.  Moreover, any
  other node \(u\) that depends on this variable must be updated accordingly to preserve the correctness of later uncomputation, namely
  \(\kappa \gets \kappa[u \mapsto \qinv(C_s); \kappa(u); C_s]\). The detailed update rules are 
  given in Appendix~\ref{app:body}. We ensure that linear recursive calls never appear in the cleanup circuit of any node, thus avoiding the time overhead of redundant uncomputation. At the same time, during uncomputation, executing the inverse of the
  cleanup circuit with all required ancillae clean correctly restores all variables in the node with \(\flag \neq 2\).  
% If the statement $s$ is not a linear recursive call, then the generated circuit is appended to the cleanup circuit of the node containing \(\mvq(s)\); that is, \(\kappa(\node(\mvq(s))) \gets (\kappa(\node(\mvq(s))); C_1)\). Furthermore, if a quantum variable within a cleanup circuit is later modified or cleaned, directly performing uncomputation may fail to restore the initial state. Therefore, we require that whenever a quantum variable is modified or cleaned by a non‑linear recursive statement, the circuit implementing this operation must be updated in the cleanup circuits of all other nodes \(w\) that depend on this variable; namely, \(\kappa(w) \gets (\qinv(C_1); \kappa(w); C_1)\). 

% In summary, we ensure that linear recursive calls never appear in the cleanup circuit of any node, thus avoiding the time overhead of redundant uncomputation. At the same time, executing the inverse of the cleanup circuit (i.e., in reverse computation order) during uncomputation correctly restores all variables in the node with \(\flag \neq 2\).  

After updating the cleanup circuits, we decrement the out‑degree $\delta(w)$ of every node \(w\) associated with a \(\mathtt{Q}\)-mode variable read by \(s\). If $\delta(w)$ reaches zero for such a node (excluding global variables and the return value), then all computations involving variables within that node are complete, and the node becomes eligible for cleanup. Subsequent handling depends on the node's flag. 
% \vspace{-1.6em}   
\begin{itemize}
    \item If \(w.\flag = 0\), the node may contain local variables $x$ that require eager cleanup ($x.\flag=0$); cleanup is triggered immediately by invoking \textsc{Clean\_Temp} (described in the next subsection). 
    \item If \(w.\flag = 1\), cleanup is deferred and performed uniformly after the function body completes in each recursive layer (in the \textsc{Compile\_Fun} algorithm). Consequently, such nodes are skipped during the eager cleanup phase to implement the deferred uncomputation mechanism. 
\end{itemize}
Recall the running example. After the recursive call \lstinline[language=MyLang]{y <- sum(y1)}, the out-degree map \(\delta\) changes from \(\{y_1\mapsto 1, k\mapsto 1\}\) to \(\{y_1\mapsto 0, k\mapsto 1\}\). Since \(y_1\) is a non-return local variable generated by a non-recursive statement, its associated quantum variable \(q_1[j]\) is eagerly uncomputed via \lstinline[language=MyLang]{qinv(circ(k-1, q1[j]))} (at line~4 of Figure~\ref{fig:sum_target}), performing the inverse of its computation.

 \subsection{Uncomputation} 

This subsection details the uncomputation process for a given node \(w\). The process is implemented by the $\textsc{Clean\_Temp}$ algorithm; its workflow is shown in Figure~\ref{fig:clean_temp}. A more formal definition is given in Appendix~\ref{app:body}. We outline the essential aspects of each step below.

\textbf{Step 1.} As presented in the last subsection, \(\kappa(w)\) records the computation sequence for node \(w\). The core idea is therefore to append the inverse of \(\kappa(w)\) to the compiled code \(\ctx.\Bbuf.m.C\). When \(w.\flag=0\), however, additional handling is required.
  Some ancillae appearing in \(\kappa(w)\) may have been reclaimed immediately after computation and reallocated before
  uncomputation, so they may no longer be clean when the uncomputation is performed. Consequently, the ancillae in the circuit $\kappa(w)$ must be replaced by clean variables allocated from the quantum variable pool $\Qpool$. Since we ensure that no recursive call appears in a cleanup circuit with \(w.\flag=0\), all ancillae used in \(\kappa(w)\) are contained in the set of variables that appear 
  explicitly in that circuit, which makes this substitution syntactically feasible. 
  When \(w.\flag=1\), the uncomputation occurs in the unified phase at the end of the function body. By
  construction, all quantum variables for other nodes have already been cleaned at this point, so the ancillae in the cleanup
  circuit \(\kappa(w)\) of \(w\) are already clean; hence, the  circuit $\kappa(w)$ can be inserted directly without variable replacement. 

  \label{sec:clean_temp}
\begin{wrapfigure}{r}{0.4\linewidth}
    \centering
    \includegraphics[width=\linewidth]{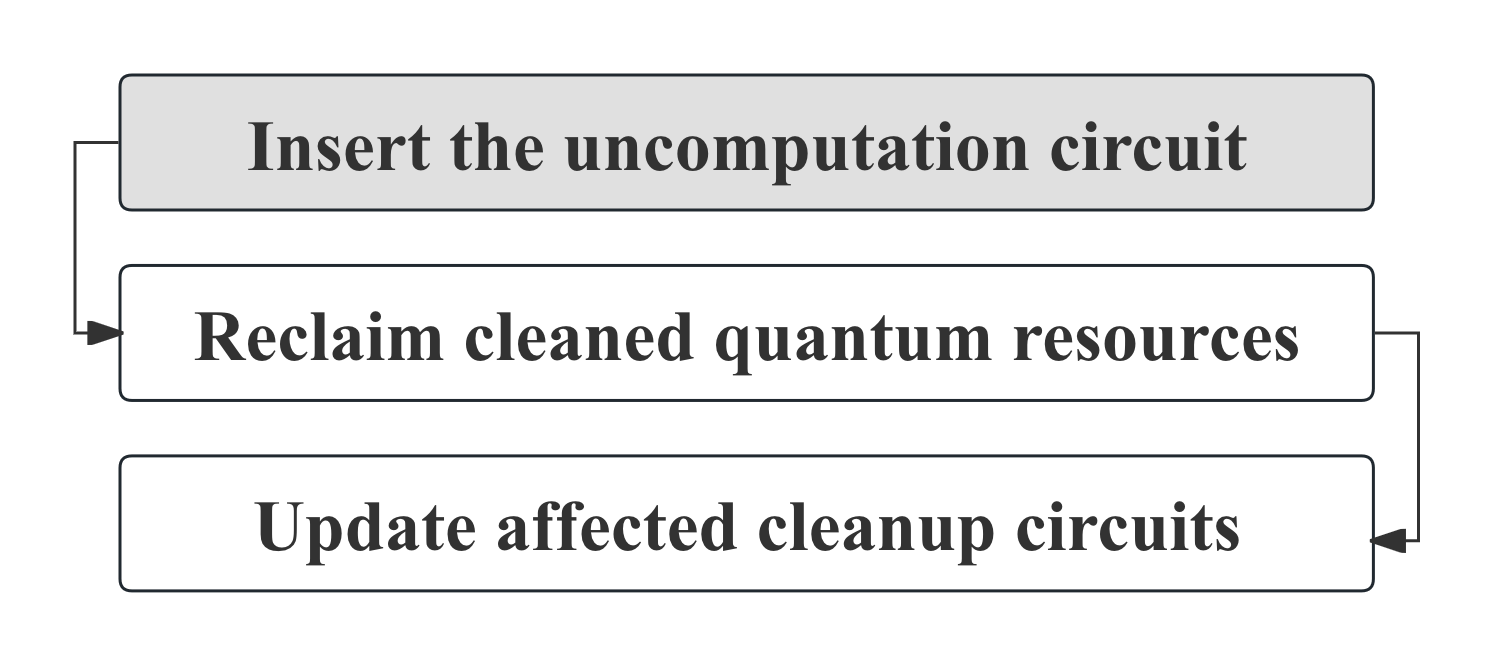}
    \caption{The workflow for the uncomputation of temporary quantum variables.}
    \label{fig:clean_temp} 
    \Description{...} 
\end{wrapfigure}  
  
  \textbf{Step 2.} The inserted uncomputation circuit restores each variable \(y\) in \(w\) with \(y.\flag \neq 2\) to its initial value. Therefore, for each such source variable that is not an input parameter, meaning that the state of its quantum variable \(\rho_{\mathrm{loc}}(y)\)
  has been restored to \(\ket{0}\), \(\rho_{\mathrm{loc}}(y)\) is removed from all relevant structures, including \(\rho_{\mathrm{loc}}\) and \(\Vactive\), and
  then returned to \(\Qpool\) for reuse.

  \textbf{Step 3.} Since the values of the variables associated with node \(w\) have been modified, all cleanup circuits \(\kappa(u)\) of other nodes \(u\) that depend on the value of these variables must be updated accordingly. This applies when \(w.\flag=0\). When \(w.\flag=1\), all other nodes have already been cleaned, and no further updates to the cleanup circuits of other nodes are required.

\section{Compilation Correctness} \label{sec:veri}
In this section we prove the correctness of our compiler. We present here only the principal theorems and proof outlines; the auxiliary lemmas and detailed proofs are
  given in Appendix~\ref{app:veri}.

  Let \(\Xi\) be the function environment and \(\Gamma\) the global type
  environment obtained from type checking the source program, let \(\G\) be its
  dependency graph, and let \(\D\) be the target program declarations generated by
  \[
    \D = \textsc{Compile\_Prog}(\Xi,\Gamma,\G).
  \] 
  Let \(s_{\main}\) be the body of the source main procedure defined in \(\Xi\), and let \(C_{\main}\) be the body of the target main procedure declared in \(\D\). We now
  prove that \(C_{\main}\) is semantically equivalent to \(s_{\main}\), where this equivalence includes both preservation of the program output and correct uncomputation of
  temporary variables.

  To formalize the correctness we first introduce a few key concepts; full details are given in Appendix~\ref{app:nota-def}.
  For a state \(\sigma\) of the source language, its encoding as a state of the target language under a variable mapping \(\rho\) is written \(\mathcal{E}(\sigma, \rho)\)
  and defined by
  \[
  \mathcal{E}(\sigma, \rho) \triangleq
  \bigl( \sigma_{c},\;
         \otimes_{x \in \dom(\rho)} \ket{\sigma(x)}_{\rho(x)}
  \bigr),
  \]
  where \(\sigma_{c}\) is the restriction of \(\sigma\) to its classical-mode variables. In the following, we omit $\sigma_c$ 
  and focus solely on the quantum part. For a target program \(C\) and a source program \(s\), we write \(C^{k}\) and
  \(s^{k}\) for their \(k\)-th syntactic approximations, respectively; the formal definition is given in Appendix~\ref{app:nota-def}.

 \whl{Let \(\Sigma\) denote the state space of source states whose domains
  consist of the global variables, and let \(\mathcal{H}\) denote the
  Hilbert space of the target quantum system.}  We say that the generated target program \(C_{\mathtt{main}}\) is semantically equivalent to the source program \(s_{\mathtt{main}}\), written \(C_{\mathtt{main}} \approx s_{\mathtt{main}}\), \rev{if, for every initial source state \(\sigma_0 \in \Sigma\), either of the
  following holds:}
  \begin{enumerate}
    \item \rev{both} configurations $\langle s_{\mathtt{main}},\sigma_0\rangle$ and $\langle C_{\mathtt{main}}, \, \mathcal{E}(\sigma_0,\rho_{\mathrm{glob}})
    \otimes \ket{0}_{\mathrm{rem}} \rangle$ diverge;
    \item
    \( \langle s_{\mathtt{main}},\sigma_0\rangle \to_{\Xi}^{*}
      \langle \downarrow,\sigma\rangle
    \) for some $\sigma$ and 
    \(
      \langle C_{\mathtt{main}}, \, \mathcal{E}(\sigma_0,\rho_{\mathrm{glob}})
    \otimes \ket{0}_{\mathrm{rem}} \rangle \to_{\D}^{*}
      \langle \downarrow,
      \mathcal{E}(\sigma,\rho_{\mathrm{glob}})\otimes\ket{0}_{\mathrm{rem}}
      \rangle,
    \) where \(\mathrm{rem}\) denotes all auxiliary target registers.

%    \item More generally, the result sequences obtained by executing the approximants
%  \(\{s_{\main}^{\,k}\}_{k\ge 0}\) and \(\{C_{\main}^{\,k}\}_{k\ge 0}\) either both
%  diverge, or both converge to equivalent limits: if the source result sequence
%  converges to \(\sigma\), then the target result sequence converges to
%  \( \mathcal{E}(\sigma,\rho_{\mathrm{glob}})\otimes\ket{0}_{\mathrm{rem}} .
%  \) 
  \end{enumerate}

   % if the approximation-result sequence of \(s_{\main}\)
    % from \(\sigma_0\) converges to \(\sigma\), then the approximation-result
    % sequence of \(C_{\main}\) from \(\ket{\psi_0}\) converges to
    % \[
    %   \mathcal{E}(\sigma,\rho_{\mathrm{glob}})
    %   \otimes\ket{0}_{\mathrm{rem}} .
    % \]

    % \item If the source approximation-result sequence diverges, then the target
    % approximation-result sequence also diverges, and conversely.

  % Let \(\Sigma\) denote the state space of source states whose domains
  % consist of the global variables, equipped with the discrete topology, and let
  % \(\mathcal{H}\) denote the Hilbert space of the target quantum system, equipped
  % with the norm topology. 
  % We say that the generated target program \(C_{\main}\) is semantically equivalent to the source program \(s_{\main}\), written 
  % \(C_{\main} \approx s_{\main}\), if the following condition holds. On every common input state, executing \(C_{\main}\) with all auxiliary registers
  % initialized to \(\ket{0}\) produces the same output on the program variables as \(s_{\main}\), while restoring all auxiliary registers to \(\ket{0}\).

  % \(
  % \ket{\psi_0} = \mathcal{E}(\sigma_0, \rho_{\mathrm{glob}}) \otimes \ket{0}_{\mathrm{rem}},\)
  % then
  % , where \(\mathrm{rem}\) denotes all remaining auxiliary registers. 

 \rev{
 \begin{lemma}[Determinism of the operational semantics]
 \label{lem:semantic-determinism}
 The operational semantics of both RQIMP and \(\RQC^{++}\) are deterministic.
\begin{itemize}
\item For RQIMP, let \(s\) be an RQIMP statement and let
 \(\sigma,\sigma_1,\sigma_2\) be source states. If
 \(
 \langle s,\sigma\rangle \to_{\Xi}^{*}
 \langle\downarrow,\sigma_1\rangle\)
 and 
 \(\langle s,\sigma\rangle \to_{\Xi}^{*}
 \langle\downarrow,\sigma_2\rangle,
 \)
 then \(\sigma_1=\sigma_2\).

  \item For \(\RQC^{++}\), let \(C\) be an \(\RQC^{++}\) program and let
  \(\ket{\psi},\ket{\phi_1},\ket{\phi_2}\) be target quantum states. If
 \(
 \langle C,\ket{\psi}\rangle \rightarrow_{\D}^{*}
 \langle\downarrow,\ket{\phi_1}\rangle\)
 and 
 \(\langle C,\ket{\psi}\rangle \rightarrow_{\D}^{*}
 \langle\downarrow,\ket{\phi_2}\rangle,
 \)
 then \(\ket{\phi_1}=\ket{\phi_2}\).

\end{itemize}
 \end{lemma}
 }

 We also use the standard finite-unfolding property for both languages:
  a recursive program terminates with a given result if and only if some finite
  syntactic approximant terminates with the same result.

  \begin{theorem}[Program Compilation Correctness]
    \label{theo:correctness-main}
    The compiled target program is equivalent to the source program:
    \(
      C_{\main} \approx s_{\main}.
    \)
  \end{theorem}

  \begin{proof}[Sketch of proof]
    Fix an arbitrary initial global source state \(\sigma_0 \in \Sigma\), and let
    \(
      \ket{\psi_0}
      =
      \mathcal{E}(\sigma_0,\rho_{\mathrm{glob}})
      \otimes \ket{0}_{\mathrm{rem}} \in \mathcal{H}.
    \)
    We consider two cases.

    First, suppose that the source program terminates:
    \(
      \langle s_{\main},\sigma_0\rangle
      \to_{\Xi}^{*}
      \langle\downarrow,\sigma\rangle
    \)
    for some \(\sigma\). By finite-unfolding adequacy, there exists
    \(k \geq 1\) such that
    \(
      \langle s_{\main}^{\,k},\sigma_0\rangle
      \to_{\Xi}^{*}
      \langle\downarrow,\sigma\rangle.
    \)
    Applying the correctness theorem for function bodies
    (Theorem~\ref{theo:body}) yields
    \[
      \langle C_{\main}^{\,k},\ket{\psi_0}\rangle
      \rightarrow_{\D}^{*}
      \left\langle
        \downarrow,\,
        \mathcal{E}(\sigma,\rho_{\mathrm{glob}})
        \otimes\ket{0}_{\mathrm{rem}}
      \right\rangle.
    \]
    By finite-unfolding adequacy for the target language, it follows that
    \[
      \langle C_{\main},\ket{\psi_0}\rangle
      \rightarrow_{\D}^{*}
      \left\langle
        \downarrow,\,
        \mathcal{E}(\sigma,\rho_{\mathrm{glob}})
        \otimes\ket{0}_{\mathrm{rem}}
      \right\rangle.
    \]

    Second, suppose that
    \(\langle s_{\main},\sigma_0\rangle\) diverges. Assume, for contradiction,
    that \(\langle C_{\main},\ket{\psi_0}\rangle\) terminates. By
    finite-unfolding adequacy for the target language, there exists some finite
    \(k \geq 1\) such that \(C_{\main}^{\,k}\) terminates from
    \(\ket{\psi_0}\). By Lemma~\ref{lem:finite-reflection}, there exists a source
    state \(\sigma'\) such that
    \[
      \langle s_{\main}^{\,k},\sigma_0\rangle
      \to_{\Xi}^{*}
      \langle\downarrow,\sigma'\rangle.
    \]
    Finite-unfolding adequacy for the source language then implies that
    \(s_{\main}\) terminates from \(\sigma_0\), contradicting the assumption.
    Therefore \(C_{\main}\) diverges from \(\ket{\psi_0}\) as well.

    Thus, from every initial source state, the source and target programs either
    diverge together or terminate with corresponding results. Hence
    \(C_{\main} \approx s_{\main}\).
  \end{proof}

  The extension from basis-state encodings to coherent superpositions follows
  from the unitarity of well-formed \(\RQC^{++}\) programs; see
  Appendix~\ref{app:proof-main}.

\section{Related Work}
\label{sec:rela}

Compiling classical specifications into reversible quantum implementations is a central task in quantum algorithm design and quantum compilation. Early studies primarily focused on reversible circuit synthesis~\cite{10.1145/775832.775915,10.1145/1278349.1278355,8a301c2571004c6bb12a10cf788b8b2b,shafaeiReversibleLogicSynthesis2013,linRMDDSReedmullerDecision2014a}, i.e., automatically generating reversible circuits from given Boolean functions or truth tables; see~\cite{saeediSynthesisOptimizationReversible2013} for a survey. Such methods typically operate on function- or circuit-level descriptions, making it difficult to handle programs with complex control flow directly. We next review representative high-level quantum and
reversible languages, compilers, and uncomputation techniques most relevant to our setting.

\subsection{Quantum Programming Languages and Compilers}

Quantum programming languages often provide high-level abstractions that can be used to construct quantum oracles. Among them, languages such as QCL~\cite{oemer2000quantum}, Q\#~\cite{10.1145/3183895.3183901}, Scaffold~\cite{Abhari2012ScaffoldQP}, and Quipper~\cite{10.1145/2491956.2462177} mainly raise the abstraction level for quantum circuit construction. On the reversible-circuit side, Thomsen's language~\cite{thomsenFunctionalLanguageDescribing2012} and 
  SyReC~\cite{willeSyReCProgrammingLanguage2010} provide high-level descriptions for reversible circuit design based on synthesis techniques. When control depends on data in superposition, these circuit-oriented settings typically encode it using lower-level circuit constructs, such as
  controlled gates. Such circuit-level structure does not naturally capture dynamic control flow whose execution path depends on runtime quantum data. Moreover, possible execution paths are typically represented by static unrolling at compile time, leading to longer compilation times, worst-case circuit sizes, and more difficult uncomputation optimization, particularly for
  complex recursive structures where the number of possible execution paths can grow rapidly.

 More recent quantum languages and models provide higher-level abstractions for quantum control flow, such as quantum if-statements whose branches may be controlled by data in superposition. For example, Qunity~\cite{voichickQunityUnifiedLanguage2023} proposes a classical-quantum unified syntax and type framework, allowing programmers to express 
 quantum algorithms in a manner similar to writing classical programs. Silq~\cite{bichselSilqHighlevelQuantum2020} allows programmers to construct oracles directly using quantum primitives and was among the first quantum programming languages to provide safe automatic uncomputation, enforced by its type system.
  Qurts~\cite{hirataQurtsAutomaticQuantum2025} introduces Rust-style ownership and lifetime annotations for quantum variables, enabling structured and safe uncomputation at
  the program level, and employs a reversible pebble-game strategy to support flexible uncomputation strategies. Nevertheless, such high-level languages generally provide limited native support for recursive quantum control flow. Tower~\cite{yuanTowerDataStructures2022}, the first quantum programming language
  with random-access memory, is a notable exception, supporting quantum data
  structures and certain forms of quantum-controlled recursion. However, it requires recursion to be classically bounded so that recursive calls can be unrolled into static quantum circuits at compile time, thereby inheriting the circuit-level limitations discussed above. Moreover, temporary-variable cleanup remains a programmer responsibility,
  increasing the programming burden and the risk of errors.

In comparison, \(\RQC^{++}\)~\cite{yingVerificationRecursivelyDefined2024} supports quantum-controlled recursive quantum programs at the language level and is equipped with an implemented QRM backend that compiles \(\RQC^{++}\) programs to a quantum computational model with both quantum control flow and recursive procedure calls~\cite{zhangQuantumRegisterMachine2025}.
Compared with QCM~\cite{yuanQuantumControlMachine2024}, another computational model that supports quantum control flow, QRM addresses the synchronization problem by combining partial evaluation of quantum control flow with runtime mechanisms, rather than manually inserting \texttt{nop} instructions into low-level programs. This
avoids changing the static program text and makes QRM particularly suitable for recursive programs, whose dynamic computation length cannot in general be predetermined from the static program text. 

Although \(\RQC^{++}\) supports dynamically controlled recursive quantum programs at both the language and semantic levels, writing such programs directly in \(\RQC^{++}\) remains verbose, cumbersome, and error-prone. In particular, \(\RQC^{++}\) does not automatically manage the cleanup of temporary variables. Taken together, the above systems motivate the combination needed for our setting: a higher-level source language for recursive oracle programming, a compiler to \(\RQC^{++}\), and automatic recursion-aware uncomputation for temporary variables. ReOC provides this combination while retaining the dynamic-recursion semantics of \(\RQC^{++}\) and using recursion-aware strategies to control time and space overheads. 

We next turn to some related reversible and quantum-oracle compilers.  Revs~\cite{parent2015reversiblecircuitcompilationspace} implements an eager-cleaning strategy based on dependency-graph analysis to reduce the number of ancillary qubits. ReVerC~\cite{DBLP:conf/cav/AmyRS17} builds on this eager-cleaning strategy and provides a verified reversible circuit compiler for the Revs language.  Our eager-cleanup mechanism builds on these ideas, but both Revs and ReVerC operate under static control flow. In addition, ReQWire~\cite{randReQWIREReasoningReversible2019}  provides a verification framework for temporary-variable uncomputation, enabling systematic conversion from classical Boolean formulas to reversible quantum circuits. VQO~\cite{liVerifiedCompilationQuantum2022} presents a verified framework for compiling quantum oracles. Its QFT-based arithmetic oracles require fewer qubits than classical-gate constructions, and our arithmetic operations adopt a similar QFT-based implementation style. Overall, these verification-oriented frameworks are not primarily aimed at dynamically controlled recursive oracles. ReOC is complementary: it targets this setting with a mathematical correctness proof, while machine-checked verification is left for future work. 
  
A comparison of the above quantum languages and compilers is given in Table~\ref{tab:tab_compare}.

\begin{table}[t]
\centering
\caption{Comparison with representative quantum languages and compilers. Parenthesized features are partially (but not fully) supported.}
\footnotesize
\setlength{\tabcolsep}{2pt} 
\begin{tabular}{lcccc}
\toprule
\textbf{Language / Compiler} 
& \begin{tabular}{c}
\textbf{Quan. Ctrl.} \\
\textbf{Branch}
\end{tabular}
& \begin{tabular}{c}
\textbf{Quan. Ctrl.} \\
\textbf{Recur.}
\end{tabular}
& \begin{tabular}{c}
\textbf{Auto} \\
\textbf{Uncomp.}
\end{tabular}
& \begin{tabular}{c}
\textbf{Recur.-Aware} \\
\textbf{Uncomp.}
\end{tabular} \\
\midrule

ReVerC~\cite{DBLP:conf/cav/AmyRS17}     & \xmark & \xmark & (\cmark) & \xmark \\
ReQWire~\cite{randQWIREPracticeFormal2018}    & \xmark & \xmark & (\xmark) & \xmark \\
Silq~\cite{bichselSilqHighlevelQuantum2020}      & \cmark & \xmark & \cmark & \xmark \\
VQO~\cite{liVerifiedCompilationQuantum2022}       & \cmark & \xmark & \cmark & \xmark \\
Qunity~\cite{voichickQunityUnifiedLanguage2023}     & \cmark & \xmark & \xmark & \xmark \\
Qurts~\cite{hirataQurtsAutomaticQuantum2025}      & \cmark & \xmark & \cmark & \xmark \\
Tower~\cite{yuanTowerDataStructures2022}     & \cmark & (\cmark) & \xmark & \xmark \\
$\RQC^{++}$~\cite{yingVerificationRecursivelyDefined2024} & \cmark & \cmark & \xmark & \xmark \\
\textbf{ReOC (this work)}
          & \cmark
          & \cmark
          & \cmark
          & \cmark \\ 
\bottomrule 
\end{tabular}
\label{tab:tab_compare}
\vspace{-0.6em}  
\end{table}

\subsection{Uncomputation Strategies}
Several works have proposed dedicated strategies for ancilla management, complementing type-driven automatic uncomputation, as in Silq, and compiler-level cleanup optimizations. 

SQUARE~\cite{dingSQUAREStrategicQuantum2020} presents a heuristic compilation framework for strategic ancilla reuse in modular quantum programs, selectively applying
uncomputation to balance qubit usage, execution time, and communication costs.   Unqomp~\cite{paradisUnqompSynthesizingUncomputation2021} is the first procedure to automatically synthesize uncomputation for a given quantum circuit. It represents circuits as graphs and identifies appropriate positions to insert uncomputation gates. Reqomp~\cite{paradisReqompSpaceconstrainedUncomputation2024} extends this line of work to space-constrained settings by tracking clean-ancilla availability, allowing recomputation and ancilla recycling to trade gate count for reduced ancilla usage.  
% In addition, the work~\cite{11000238} proposes a partial uncomputation approach based on Unqomp that selectively reverses only a suffix of the gate sequence, enabling the uncomputation of a maximal subset of ancilla qubits even in the presence of cyclic dependencies. 
Qrisp~\cite{10.1007/978-3-031-38100-3_11} builds upon
Unqomp to provide a high-level programming framework that supports automatic uncomputation. 
SPARE~\cite{sharmaOptimizingAncillaBasedQuantum2025} is a rewrite-based optimizer targeting compute--uncompute patterns in circuits compiled from
high-level languages.  Su et al.~\cite{10.1145/3779212.3790134} extend automatic uncomputation to dirty qubits, and Khattar et al.~\cite{khattarRiseConditionallyClean2025} introduce conditionally clean ancillae, broadening the design space between clean and dirty ancillae. 

Among these works, the most closely related is Venev et al.~\cite{venevModularSynthesisEfficient2024}, which introduces the first modular automatic approach for synthesizing correct and efficient uncomputation for expressive quantum programs. Their garbage-mode transformation avoids redundant recomputation in recursive programs by propagating intermediate values, including those from non-recursive computations, as garbage and disposing them into a garbage bin, thereby delaying their cleanup and preserving the original asymptotic running time.  ReOC instead uses an origin- and structure-sensitive policy: recursive-call temporaries are deferred, while non-recursive temporaries are eagerly uncomputed to limit unnecessary space accumulation; for non-linear recursive structures, cleanup is performed at the end of each layer rather than only at the outermost call. Moreover, 
the aforementioned work focuses on 
classically controlled recursion, while ReOC addresses recursive control flow driven by runtime quantum data. 

\section{Conclusion and Future Work}
\label{sec:con}

We have presented ReOC, a compilation framework for high-level recursive quantum-oracle specifications. ReOC introduces RQIMP, a %restricted 
simple
imperative source language
  for writing oracles with dynamically controlled loops, recursion, and mutual recursion, and compiles well-typed RQIMP programs to \(\RQC^{++}\). By targeting \(\RQC^{++}\), ReOC reuses the existing QRM backend for dynamic recursive execution, while providing a source-level abstraction and reducing the programming burden and risk of errors.  The compiler addresses static-register management under dynamic control using an indexed discipline, which enables safe reuse of the static register pool across recursion layers while controlling quantum storage usage. It also combines eager cleanup for non-recursive temporaries to control space overhead with deferred cleanup for recursive-call temporaries to control time overhead, in particular avoiding exponential recomputation for linear recursion. ReOC also comes with a mathematical correctness proof for the compiler, covering semantic preservation and uncomputation of temporary quantum variables. 
  % Overall, ReOC compiles high-level recursive oracle programs to a dynamic-control quantum target, preserving correctness while controlling overheads through static-storage
  % management and uncomputation strategies. 

  As for future work, we plan to refine the uncomputation strategy for non-linear recursive structures, extend the framework to support richer abstract data
  structures, and develop an implementation integrated with the downstream \(\RQC^{++}\)/QRM backend, with the longer-term goal of machine-checked verification.  

% We have presented a compilation framework that, for the first time, enables the systematic translation of classical recursive programs—including loops, mutual recursion, and complex control flows—into reversible quantum circuits that genuinely support dynamic quantum control. The framework centres on RQIMP, a high-level imperative language for quantum oracles, and compiles it to $\RQC^{++}$, thereby reusing an existing backend to Quantum Register Machine and decoupling reversibilization from low-level implementation details. To address the exponential blowup caused by naive uncomputation in recursive settings, we proposed a differentiated uncomputation strategy: temporary variables generated by recursive calls are deferred, while those from non‑recursive statements are cleaned eagerly. For linear recursion, this strategy achieves linear time and space complexity, preserving the asymptotic efficiency of the original algorithm. The correctness of the compiler—including semantic preservation and proper cleanup of all temporaries—has been established through rigorous mathematical proofs.  

% In future work, we aim to refine the uncomputation strategy for more complex recursive structures (e.g., tree recursion), extend the framework to support dynamic data structures, and integrate with verified quantum hardware simulators for end‑to‑end validation.

\bibliographystyle{ACM-Reference-Format}
\bibliography{ref}

\appendix
\section{Comparing RQIMP to $\RQC^{++}$}  
\label{app:compare-rqc}
 Figure~\ref{fig:full-rqc} provides the full versions of the programs shown in Figure~\ref{fig:RQC_sum}, illustrating the burden of writing the oracle directly in \(\RQC^{++}\). Here, \texttt{Fsum} is 
  intentionally implemented in an unoptimized form, using naive uncomputation; the ReOC-generated version with recursion-aware uncomputation is discussed in the main text. The implementation of
  expression-level subroutines, such as \texttt{rz\_suber\_c\_form}, is further explained in Appendix~\ref{sec_com_exp_assgn}. 

\begin{figure}[t]
    \centering
    \begin{minipage}[t]{0.49\linewidth}
      \begin{lstlisting}[language=MyLang, basicstyle=\ttfamily\footnotesize]
rz_suber_c_form(a,p) <==  
// Decrement quantum p by classical a
   QFT(p,0,sz-1);
   rz_suber_c(p,0,sz-1,a); 
   RQFT(p,0,sz-1)
rz_suber_c(p,m,n,a) <== 
   if m <= n then
      S^{-1}(a * 2^{-(n-m+1)})[b[m]];
      rz_suber_c(p,m+1,n,a)
   fi
init_c(a,p,i) <== 
// Initialize quantum p to classical a
   if 0 <= i then
      if (a % 2) == 1 then X[p[i]] fi;
      init_c(a/2, p, i-1)
   fi
beq(p,m1,n1,a,r) <== 
// Compare quantum p with classical a
   if m1 > n1 then  X[r]
   else qif [p[m1]] (
         |0> -> beq(p,m1+1,n1,q,m2+1,n2,r)
       square |1> -> skip) 
        fiq
   fi
rz_adder_full_form(p,q,r) <== 
// Add quantum p and q, store result in r
   copy(q,0,sz-1,r,0,sz-1);
   QFT(q,0,sz-1);
   rz_adder(p,r,sz-1,sz-1);
   RQFT(q,0,sz-1)
\end{lstlisting}
    \end{minipage} 
    \begin{minipage}[t]{0.49\linewidth}
        \begin{lstlisting}[language=MyLang, basicstyle=\ttfamily\footnotesize, firstnumber=31, ] 
rz_adder(p,q,n,m) <==  
// Add quantum p into q
   if 0 <= n and 0 <= m then
      rz_adder(p,q,n-1,m-1);
      qif [p[n]] (
          |0> -> skip
        square |1> -> rz_adder'(q,m,m))
      fiq
   fi
rz_adder'(q,k,m) <== 
   if 0 <= k then
      RZ(m-k+1)[q[k]];
      rz_adder'(q,k-1,m)
   fi
Fsum(p,r,i) <== // x <- sum(k) 
   beq(p,0,sz-1,0,q[i]); 
   // Test k==0 and store the result in q[i]
   qif q[i] (
      |0> -> rz_suber_c_form(1,q1[i]); 
      // Compute k-1 and store it in q1[i] 
             Fsum(q1[i],q2[i],i+1); 
             //  y <- sum(k-1)
             qinv(rz_suber_c_form(1,q1[i])); 
             // Uncompute q1[i]
             rz_adder_full_form(p,q2[i],r); 
             // x<-y+k
             qinv(Fsum(q1[i],q2[i],i+1)) 
             // Uncompute q2[i]
     square |1> -> init_c(0,r,sz-1) ); // x <-0
   qinv(beq(p,0,sz-1,0,q[i]))  
   // Uncompute q[i] 
   fiq
Fmain <== 
   Fsum(p1,r1,0)); // x <-sum(k) 
        \end{lstlisting}
    \end{minipage}
    \caption{The full $\RQC^{++}$ program for $\mathtt{sum}$ (enlarged version of Figure~\ref{fig:RQC_sum})}
    \label{fig:full-rqc}
    \Description{...} 
\end{figure}

\section{Target Languages} 
\label{app:tar}
In this section, we briefly recall the syntax, operational semantics and Hoare proof system of $\RQC^{++}$ programs; for further details we refer the reader to~\cite{yingVerificationRecursivelyDefined2024}. 

\subsection{Quantum Variables and Alphabet}
Similar to classical programming languages where arrays can be declared, $\mathbf{RQC}^{++}$ supports quantum array types. It distinguishes two kinds of quantum variables:

\begin{itemize}
\item \textbf{Simple quantum variables}: each has a basic quantum type $\mathcal{H}$ (e.g., a qubit).
\item \textbf{Array quantum variables}: each has a higher type $T_1\times\cdots\times T_n\to\mathcal{H}$, where $T_1,\ldots,T_n$ are classical types (e.g., \texttt{integer}). Such a variable denotes a tensor product $\bigotimes_{v_1\in T_1,\ldots,v_n\in T_n}\mathcal{H}_{v_1,\ldots,v_n}$ with $\mathcal{H}_{v_1,\ldots,v_n}=\mathcal{H}$.
\end{itemize}

For an array quantum variable $q$, an element $q[s_1,\ldots,s_n]$ — where each $s_i$ is a classical expression of type $T_i$ — is called a \emph{subscripted quantum variable} and has type $\mathcal{H}$. For example, $q[2x-y,\,3x+4y-75]$ is a valid subscripted qubit when $q$ is a two‑dimensional qubit array.

To see how quantum arrays and classical parameters are used in specifying quantum states, consider the quantum Fourier transform. The output state of the quantum Fourier transform on a quantum array section \(q[m:n]\) can be expressed by the following parameterised quantum state, where \(j\) is a classical bit array:
\[
|\Psi\rangle(j,m,n) \;=\; \frac{1}{\sqrt{2^{n-m+1}}}\bigotimes_{l=1}^{n-m+1}\Bigl(|0\rangle + e^{2\pi i\,0.j[n-l+1:n]}|1\rangle\Bigr),
\]
with \(0.j[n-l+1:n]\) denoting the binary fraction formed by the bits \(j[n-l+1],\dots,j[n]\). Here the classical parameters \(j,m,n\) serve both to index into the quantum array \(q\). 

With these quantum data types in place, the full alphabet of $\mathbf{RQC}^{++}$ consists of classical variables $x, y, \ldots$, quantum variables $q, r, \ldots$, procedure identifiers $P, Q, \ldots$, elementary unitary gates (e.g., $\{H,T,CNOT\}$), and elementary classical arithmetic and Boolean operators. Variables can be of array types (both classical and quantum), e.g., $x[t]$ or $q[t]$, where the subscript $t$ is a classical expression; for simplicity, we restrict to one‑dimensional arrays and require that subscripts contain no nested subscripts. A sequence $\overline{q} = q_1, \ldots, q_n$ of distinct quantum variables is called a \emph{quantum register} and denotes a composite quantum variable consisting of subsystems $q_1, \ldots, q_n$; its type is defined as the tensor product $T(\overline{q}) = T(q_1) \otimes \cdots \otimes T(q_n)$. Similarly, $\overline{x} = x_1, \ldots, x_n$ denotes a list of classical variables.

\subsection{Syntax} 
A program is a set $\mathcal{D}$ of \emph{procedure declarations} of the form
\[
P(\overline{u}) \Leftarrow C,
\]
where $\overline{u}=u_1,\dots,u_n$ is a list of \emph{formal parameters} (classical simple variables) and $C$ is a \emph{statement} (the procedure body). One procedure, say $P_{\text{main}}$, is designated as the entry point. Recursion is supported by allowing \(C\) to contain calls to \(P\) itself. Statements $C$ are generated by the following grammar:
\[
\begin{array}{rcl}
C &::=& \mathbf{skip}
      \mid \overline{x} := \overline{t}
      \mid U[\overline{q}]
      \mid C_1;C_2
      \mid P(\overline{t})
      \mid \mathbf{if}\;b\;\mathbf{then}\;C_1\;\mathbf{else}\;C_2\;\mathbf{fi}\\
   & & \mid \mathbf{while}\;b\;\mathbf{do}\;C\;\mathbf{od}
      \mid \mathbf{begin\;local}\; \overline{x}:= \overline{t};C\;\mathbf{end}
      \mid \mathbf{qif}[q]\bigl(|0\rangle\to C_0\bigr)\square\bigl(|1\rangle\to C_1\bigr)\mathbf{fiq},
\end{array}
\]
where $\overline{x}$ is a list of classical variables, $\overline{t}$ a list of classical expressions, $b$ a Boolean expression, $U$ an elementary gate, and $q$ and $\overline{q}$ can be simple
or subscripted quantum variables.
The unitary gate $U[\overline{q}]$ applies the elementary quantum gate $U$ on quantum variables $\overline{q}$.  The quantum if‑statement $\mathbf{qif}\ldots\mathbf{fiq}$ creates quantum control flow.  
Intuitively, if quantum coin $q$ is in the state $\ket{0}$, then program $C$ behaves like $C_0$, and if $q$ is in
the state $\ket{1}$, then $C$ behaves like $C_1$. Two branches can run in superposition.  The statement $\mathbf{begin\;local}\;\overline{x}:= \overline{t};C\;\mathbf{end}$ declares a list of classical variables $\overline{x}$ as \emph{local variables} within the block. They are initialised to the values of $\overline{t}$ at the start, and restored to their original values after $C$ finishes.

For each quantum circuit $C$ in $\mathbf{RQC}^{++}$, the set $\qv(C)$ of quantum variables appearing in $C$ is defined inductively as Figure~\ref{fig:quan_var}. 

\begin{figure}[t]

\begin{itemize}
\item If $C \equiv \mathbf{skip}$ or $C \equiv \overline{x} := \overline{t}$, then $\qv(C) = \emptyset$.

\item If $C \equiv U[\overline{q}]$, then $\qv(C) = \overline{q}$ (the list of quantum variables that $U$ acts on).

\item If $C \equiv C_1;C_2$, then $\qv(C) = \qv(C_1) \cup \qv(C_2)$.

\item If $C \equiv \mathbf{if}\;b\;\mathbf{then}\;C_1\;\mathbf{else}\;C_2\;\mathbf{fi}$, then $\qv(C) = \qv(C_1) \cup \qv(C_2)$.

\item If $C \equiv \mathbf{qif}[q](|0\rangle \to C_0)\;\square\;(|1\rangle \to C_1)\;\mathbf{fiq}$, then $\qv(C) = \{q\} \cup \qv(C_0) \cup \qv(C_1)$.

\item If $C \equiv \mathbf{begin\;local}\;\overline{x} := \overline{t};C'\;\mathbf{end}$, then $\qv(C) = \qv(C')$.

\item If $C \equiv P(\overline{t})$ is a procedure call, and $P$ is declared by $P(\overline{u}) \Leftarrow C_P$, then $\qv(C) = \qv(C_P)$.
\end{itemize} 
\caption{Quantum variables of \(\RQC^{++}\) commands}
\label{fig:quan_var}
\Description{...}
\end{figure}

To guarantee well‑defined semantics, the language imposes three \emph{well‑formedness conditions}
\begin{enumerate}
\item \label{wf-rqc-coin} \textbf{External quantum coin}: In every $\mathbf{qif}[q](|0\rangle \to C_0)\;\square\;(|1\rangle \to C_1)\;\mathbf{fiq}$, the coin $q$ does not appear in $C_0$ or $C_1$ (i.e., $q \notin \qv(C_0) \cup \qv(C_1)$). 
\item \label{wf-rqc-branch} \textbf{No free changed variables in qif branches}: Classical variables that are assigned inside $C_0$ or $C_1$ must be declared local; otherwise their changes could conflict between branches.
\item \label{wf-rqc-proc} \textbf{No free changed variables in procedure bodies}: Procedure bodies may only modify local classical variables, ensuring that recursive calls are safe when used inside quantum branches.
\end{enumerate}

In the original syntax of $\mathbf{RQC}^{++}$, the parameters $\overline{t}$ of a procedure call are restricted to classical expressions; quantum variables are not allowed. However, in the compilation process, to enable the same procedure to be flexibly applied to different quantum variables (without hardwiring the procedure to fixed quantum variables), we extend the language as follows: quantum variables are allowed as parameters in procedure calls (e.g., one can pass $q$ as an argument to a recursive procedure). The semantics of this extension is call‑by‑reference, meaning that the caller and the callee share the same quantum data object rather than making a physical copy of the quantum state. 

The justification of this extension is that once the program is compiled, all quantum variables (including array elements) can be statically allocated into a large multi‑dimensional quantum storage space, where each variable name corresponds to the base address and length index of a contiguous fragment (represented by classical expressions). Therefore, passing a quantum variable is equivalent to passing its index information (i.e., classical expressions), which can be fully simulated by the standard classical call‑by‑value mechanism where only classical expressions are permitted as parameters. Therefore this extension is merely a syntactic convenience; it does not affect the correctness of the program nor hinder compilation to low‑level quantum instruction sets.  

With this extension, the set of quantum variables for a procedure call is redefined as \(\qv(P(\overline{t})) = \qv(C_P) \cup \qv(\overline{t})\), where \(\qv(\overline{t})\) collects the quantum variables occurring in the actual parameters. In addition to these standard notions, we also introduce a new auxiliary function $\qinv(C)$ that denotes the inverse of program $C$,  defined by structural induction as Figure~\ref{fig:quan_inv}.
\begin{figure}[t]
\[
\begin{array}{r@{\;}l@{\;}l}
\qinv(\mathbf{Skip})
    &:=& \mathbf{Skip} \\[0.4em]
\qinv(\overline{x} := \overline{x} \pm \overline{t}) &:=& \overline{x} := \overline{x} \mp \overline{t} \quad 
\qinv(\overline{x}:=\overline{t})
    := \overline{x}:=\overline{x}\oplus \overline{t} \ (\text{for all other assignments})   \\[0.4em]
\qinv(U[q])
    &:=& U^{-1}[q] \\[0.4em]
\qinv(c_1;c_2)
    &:=& \qinv(c_2);\,\qinv(c_1) \\[0.4em]
\qinv(\mathbf{if}\ b\ \mathbf{then}\ c_1\ \mathbf{else}\ c_2\ \mathbf{fi})
    &:=& \mathbf{if}\ b\ \mathbf{then}\ \qinv(c_1)\ \mathbf{else}\ \qinv(c_2)\ \mathbf{fi} \\[0.4em]
\qinv(\mathbf{qif}\ q\ \mathbf{then}\ c_1\ \mathbf{else}\ c_2\ \mathbf{fiq})
    &:=& \mathbf{qif}\ q\ \mathbf{then}\ \qinv(c_1)\ \mathbf{else}\ \qinv(c_2)\ \mathbf{fiq} \\[0.4em]
\qinv(\mathbf{begin}\ \mathbf{local}\ \overline{x}:=\overline{t};\,c\ \mathbf{end})
    &:=& \mathbf{begin}\ \mathbf{local}\ \overline{x}:=\overline{t};\,\qinv(c)\ \mathbf{end} \\[0.4em]
\qinv(P(t))
    &:=& P'(t),
    \quad\text{where } P(\overline{u})\Leftarrow C_p \text{ and } P'(\overline{u})\Leftarrow \qinv(C_p).
\end{array}
\]
\caption{Inverse transformation of \(\RQC^{++}\) commands} 
\label{fig:quan_inv}
\Description{...}
\end{figure}
A brief remark is in order. In our language, classical variables are allowed to be assigned only once. Consequently, the assignment \(\overline{x}:=\overline{t}\) is equivalent to its reversible form \(\overline{x}:=\overline{x}\oplus\overline{t}\) (since initially \(\overline{x}\) can be regarded as zero). Based on this equivalence, we directly define the inverse of the assignment as \(\overline{x}:=\overline{x}\oplus\overline{t}\). 
Moreover, note that \(\qv(\qinv(C)) = \qv(C)\) holds for every \(C\) by definition. 

\subsection{Operational Semantics}

A \emph{configuration} is a triple $(C,\sigma,|\psi\rangle)$, where:
\begin{itemize}
\item $C$ is the remaining statement to be executed (or $C = \downarrow$ for termination; we identify $\downarrow;C' \equiv C'$);
\item $\sigma$ is the current classical state (a mapping from classical variables to values);
\item $|\psi\rangle$ is the current quantum state over all quantum variables.
\end{itemize}
Let $\mathcal{C}$ be the set of configurations. The operational semantics is defined as a transition relation $\rightarrow\subseteq\mathcal{C}\times\mathcal{C}$ by the rules in Figure~\ref{fig:opsem}. In each rule, the premise (above the line) is assumed to hold, and the conclusion (below the line) is a transition.

\begin{figure}[t]
\centering
\[
\begin{array}{l}
\text{(SK)}\quad (\mathbf{skip},\;\sigma,\;|\psi\rangle)\;\rightarrow\;(\downarrow,\;\sigma,\;|\psi\rangle)\\
\text{(AS)}\quad (\overline{x} := \overline{t},\;\sigma,\;|\psi\rangle)\;\rightarrow\;(\downarrow,\;\sigma[\overline{x}:=\sigma(\overline{t})],\;|\psi\rangle)\\
\text{(GA)}\quad \dfrac{\sigma\models Dist(\overline{q})}{(U[\overline{q}],\;\sigma,\;|\psi\rangle)\;\rightarrow\;(\downarrow,\;\sigma,\;(U_{\overline{q}}\otimes\mathbb{I})|\psi\rangle)}\\
\text{(SC)}\quad \dfrac{(C_1,\sigma,|\psi\rangle)\rightarrow(C_1',\sigma',|\psi'\rangle)}
{(C_1;C_2,\;\sigma,\;|\psi\rangle)\;\rightarrow\;(C_1';C_2,\;\sigma',\;|\psi'\rangle)}\\[12pt]
\text{(IF-T)}\quad \dfrac{\sigma\models b}{(\mathbf{if}\;b\;\mathbf{then}\;C_1\;\mathbf{else}\;C_2\;\mathbf{fi},\;\sigma,\;|\psi\rangle)\;\rightarrow\;(C_1,\;\sigma,\;|\psi\rangle)}\\[12pt]
\text{(IF-F)}\quad \dfrac{\sigma\models\neg b}{(\mathbf{if}\;b\;\mathbf{then}\;C_1\;\mathbf{else}\;C_2\;\mathbf{fi},\;\sigma,\;|\psi\rangle)\;\rightarrow\;(C_2,\;\sigma,\;|\psi\rangle)}\\[12pt]
\text{(WHILE-T)}\quad \dfrac{\sigma\models b}{(\mathbf{while}\;b\;\mathbf{do}\;C\;\mathbf{od},\;\sigma,\;|\psi\rangle)\;\rightarrow\;(C;\;\mathbf{while}\;b\;\mathbf{do}\;C\;\mathbf{od},\;\sigma,\;|\psi\rangle)}\\[12pt] 
\text{(WHILE-F)}\quad \dfrac{\sigma\models\neg b}{(\mathbf{while}\;b\;\mathbf{do}\;C\;\mathbf{od},\;\sigma,\;|\psi\rangle)\;\rightarrow\;(\downarrow,\;\sigma,\;|\psi\rangle)}\\[12pt]
\text{(QIF)}\quad \dfrac{
|\psi\rangle=\alpha_0|0\rangle_q|\theta_0\rangle+\alpha_1|1\rangle_q|\theta_1\rangle, \quad 
(C_i,\sigma,|\theta_i\rangle)\rightarrow^{*}(\downarrow,\sigma',|\theta'_i\rangle)\quad(i=0,1)
}{
(\mathbf{qif}[q](|0\rangle\to C_0)\square(|1\rangle\to C_1)\mathbf{fiq},\;\sigma,\;|\psi\rangle)
\;\rightarrow\;
(\downarrow,\;\sigma',\;\alpha_0|0\rangle_q|\theta'_0\rangle+\alpha_1|1\rangle_q|\theta'_1\rangle)
}\\[12pt]
\text{(BS)}\quad 
(\mathbf{begin\;local}\;\overline{x} := \overline{t};C\;\mathbf{end},\;\sigma,\;|\psi\rangle)
\;\rightarrow\;
(\overline{x}:= \overline{t};C;\overline{x} := \sigma(\overline{x}),\;\sigma,\;|\psi\rangle)\\[12pt]
\text{(RC)}\quad \dfrac{P(\overline{u})\Leftarrow C\in\mathcal{D}}
{(P(\overline{t}),\;\sigma,\;|\psi\rangle)\;\rightarrow\;
(\mathbf{begin\;local}\;\overline{u} := \overline{t};C\;\mathbf{end},\;\sigma,\;|\psi\rangle)}
\end{array}
\]
\caption{Transition rules for the operational semantics of $\mathbf{RQC}^{++}$.}
\Description{...} 
\label{fig:opsem}
\end{figure}

\noindent
A few explanations of the rules are in order.

\begin{itemize}
\item \textbf{(GA)}: The condition $\sigma\models Dist(\overline{q})$ ensures that the list $\overline{q}$ denotes distinct quantum subsystems in the classical state $\sigma$ (e.g., no two subscripted names collide). If $\overline{q}=q_1$, $Dist(\overline{q})$ is always true. For $\overline{q}=q_1,q_2$, it requires $q_1\neq q_2$ and, for subscripted variables, that their indices differ. The gate $U$ acts on the quantum register $\overline{q}$ while the identity acts on all other quantum variables.

\item \textbf{(QIF)}: This is the quantum conditional. The quantum coin $q$ must be external to $C_0$ and $C_1$, which guarantees that the current state can be decomposed as shown. The two branches must both terminate in the same classical state $\sigma'$ (so that classical variables remain classical). The \( \rightarrow^{*} \) denotes the reflexive and transitive closure of the relation \( \rightarrow \).
\item \textbf{(BS)}: The block statement first initialises the local variables $\overline{x}$ to the values of $\overline{t}$, then executes $C$, and finally restores the original values of  $\overline{x}$ (which are $\sigma(\overline{x})$). This makes the block reversible with respect to classical state changes.

\item \textbf{(RC)}: A procedure call $P(\overline{t})$ is replaced by the body $C$ of $P$, with the formal parameters $\overline{u}$ replaced by the actual parameters $\overline{t}$ inside a block that localises $\overline{u}$. 

\end{itemize}

The semantics is deterministic: for any well‑formed program and initial classical state, the classical part of the computation follows a unique path, while the quantum part evolves unitarily. The three well‑formedness conditions (external quantum coin, absence of free changed variables in qif branches, and absence of free changed variables in procedure bodies) ensure that the (QIF) rule can be applied consistently and that classical variables never become superposed.  

Furthermore, Lemma~2.1 in~\cite{yingVerificationRecursivelyDefined2024} shows that for any fixed classical state \(\sigma\), if the quantum circuit \(C\) terminates on some input state, it terminates on all input states. Hence the denotational semantics of \(C\) can be defined as the mapping  
 \[
[\![C]\!]_\sigma : \mathcal{H}_{\qv(C)} \to \mathcal{H}_{\qv(C)},\qquad  [\![C]\!]_\sigma(|\phi\rangle) = |\psi\rangle\ \text{ whenever }\ (C,\sigma,|\phi\rangle)\to^*(\downarrow,\sigma',|\psi\rangle),\]  
where \([\![C]\!]_\sigma\) is well-defined and a unitary operator on the Hilbert space \(\mathcal{H}_{\qv(C)}\). Moreover, when the behaviour of \(C\) does not depend on the classical state (e.g., when \(C\) contains no classical variables or the classical state is irrelevant), we write simply 
\[
[\![C]\!]: \mathcal{H}_{\qv(C)} \to \mathcal{H}_{\qv(C)},\qquad  [\![C]\!] (|\phi\rangle) = |\psi\rangle\ \text{ whenever }\ (C, |\phi\rangle)\to^*(\downarrow,|\psi\rangle).\] 

\subsection{Hoare Logic} 
\label{sec:hoare-logic-summary}

\subsubsection{Parameterised Quantum States}
 A parameterised quantum state $|\phi\rangle$ is a (partial) function that maps a classical state $\sigma$ to a pure quantum state $|\phi\rangle(\sigma)$ over the quantum variables determined by $\sigma$. The equality $|\phi\rangle = |\psi\rangle$ under a classical condition $A$ is written $A \models |\phi\rangle = |\psi\rangle$ and means that for every $\sigma\models A$, $|\phi\rangle(\sigma)=|\psi\rangle(\sigma)$. Operations on ordinary quantum states extend pointwise.

\subsubsection{Quantum Hoare Triples}
A Hoare triple has the form
\[
\{A,\;|\phi\rangle\}\;C\;\{B,\;|\psi\rangle\},
\]
where $C\in\mathbf{RQC}^{++}$, $A,B$ are first‑order logical formulas over classical variables, and $|\phi\rangle,|\psi\rangle$ are parameterised quantum states.  

Intuitively, if the execution starts in a classical state $\sigma$ satisfying $A$ and the quantum state is $|\phi\rangle(\sigma)$, then whenever $C$ terminates (for total correctness termination is required; for partial correctness it is not), the final classical state $\sigma'$ satisfies $B$ and the final quantum state equals $|\psi\rangle(\sigma')$.  
Two flavours are considered: \emph{partial correctness} ($\models_{par}$) and \emph{total correctness} ($\models_{tot}$); the latter additionally guarantees termination. Formally, the semantics are defined as follows.

Given an interpretation \(\mathbb{I}\) of the assertion language, the truth of a Hoare triple \(\{A,|\phi\rangle\} C \{B,|\psi\rangle\}\) is defined as follows.
\begin{itemize}
    \item Partial correctness \(\mathbb{I} \models_{\mathit{par}} \{A,|\phi\rangle\} C \{B,|\psi\rangle\}\):  
  For every classical state \(\sigma\) with \(\sigma \models_{\mathbb{I}} A\), if \((C,\sigma,|\phi\rangle(\sigma)) \to^* (\downarrow,\sigma',|\psi'\rangle)\) for some \(\sigma',|\psi'\rangle\), then  
  \(\sigma' \models_{\mathbb{I}} B\) and \(|\psi'\rangle = |\psi\rangle(\sigma')\).
  \item Total correctness \(\mathbb{I} \models_{\mathit{tot}} \{A,|\phi\rangle\} C \{B,|\psi\rangle\}\):  
  For every classical state \(\sigma\) with \(\sigma \models_{\mathbb{I}} A\), there exist \(\sigma',|\psi'\rangle\) such that  
  \((C,\sigma,|\phi\rangle(\sigma)) \to^* (\downarrow,\sigma',|\psi'\rangle)\), and moreover  
  \(\sigma' \models_{\mathbb{I}} B\) and \(|\psi'\rangle = |\psi\rangle(\sigma')\).
\end{itemize}
Partial correctness $\models_{\mathit{par}} \{A,|\phi\rangle\} C \{B,|\psi\rangle\}$ means that the Hoare triple is true under all interpretations $\mathbb{I}$ of the assertion language. The notation $\models_{\mathit{tot}}$ is defined analogously for total correctness.

\subsubsection{Proof System (Selected Rules)}
The proof system consists of structural rules, rules for non‑recursive constructs, rules for parameterisation, rules for quantum recursion, and auxiliary rules. We list representative rules in Figure~\ref{fig:proof-rules}.

\begin{figure}[t]
\centering
\[
\begin{array}{l}
\text{(SK-P)}\quad \dfrac{}{\{A,|\phi\rangle\}\;\mathbf{skip}\;\{A,|\phi\rangle\}} \\[12pt]
\text{(AS-P)}\quad \dfrac{}{\{A[\overline{x}:=\overline{t}],|\phi\rangle\}\;\overline{x}:=\overline{t}\;\{A,|\phi\rangle\}} \\[12pt]
\text{(GA-P)}\quad \dfrac{}{\{A,U^\dagger|\phi\rangle\}\;U[\overline{q}]\;\{A,|\phi\rangle\}} \\[12pt]
\text{(SC-P)}\quad \dfrac{\{A,|\phi\rangle\}C_1\{B,|\psi\rangle\}\quad\{B,|\psi\rangle\}C_2\{D,|\theta\rangle\}}
{\{A,|\phi\rangle\}C_1;C_2\{D,|\theta\rangle\}} \\[12pt]
\text{(IF-P)}\quad \dfrac{\{A\land b,|\phi\rangle\}C_1\{B,|\psi\rangle\}\quad\{A\land\neg b,|\phi\rangle\}C_2\{B,|\psi\rangle\}}
{\{A,|\phi\rangle\}\;\mathbf{if}\;b\;\mathbf{then}\;C_1\;\mathbf{else}\;C_2\;\mathbf{fi}\;\{B,|\psi\rangle\}} \\[12pt]
\text{(QIF-P)}\quad \dfrac{
\{A,|\phi_i\rangle\}C_i\{B,|\psi_i\rangle\}\;(i=0,1) \quad cv(q)\cap change(C_i)=\emptyset 
}{
\{A,\;\alpha_0|0\rangle_q|\phi_0\rangle+\alpha_1|1\rangle_q|\phi_1\rangle\}\;
\mathbf{qif}[q](|0\rangle\to C_0)\square(|1\rangle\to C_1)\mathbf{fiq}\;
\{B,\;\alpha_0|0\rangle_q|\psi_0\rangle+\alpha_1|1\rangle_q|\psi_1\rangle\}
} \\[12pt]
\text{(Recursion-Par)} \quad 
\dfrac{
\begin{array}{l}
\{A_i,|\phi_i\rangle\}P_i(\overline{t_i})\{B_i,|\psi_i\rangle\}\;(i=1,\dots,n)\;\vdash\;\{A,|\phi\rangle\}C\{B,|\psi\rangle\}\\[2pt]
\{A_i,|\phi_i\rangle\}P_i(\overline{t_i})\{B_i,|\psi_i\rangle\}\;(i=1,\dots,n)\;\vdash\\
\qquad \{\!A_j,|\phi_j\rangle\!\}\;\mathbf{begin\;local}\;\overline{u_j}:=\overline{t_j};C_j\;\mathbf{end}\;\{\!B_j,|\psi_j\rangle\!\}\quad(j=1,\dots,n)
\end{array}
}{
\{A,|\phi\rangle\}C\{B,|\psi\rangle\}
}\\[20pt]
\text{(Recursion-Tot)} \quad 
\dfrac{
\begin{array}{l}
\{A_i\land r<z,|\phi_i\rangle\}P_i(\overline{t_i})\{B_i,|\psi_i\rangle\}\;(i=1,\dots,n)\;\vdash\;\{A,|\phi\rangle\}C\{B,|\psi\rangle\}\\[2pt]
\{A_i\land r<z,|\phi_i\rangle\}P_i(\overline{t_i})\{B_i,|\psi_i\rangle\}\;(i=1,\dots,n)\;\vdash\\
\qquad \{\!A_j\land r=z,|\phi_j\rangle\!\}\;\mathbf{begin\;local}\;\overline{u_j}:=\overline{t_j};C_j\;\mathbf{end}\;\{\!B_j,|\psi_j\rangle\!\}\quad(j=1,\dots,n)\\[2pt]
A_i\to r\ge0\quad(i=1,\dots,n)
\end{array}
}{
\{A,|\phi\rangle\}C\{B,|\psi\rangle\}
}
\end{array}
\]
\caption{Selected Proof Rules for Program Constructs}
\label{fig:proof-rules}
\Description{...} 
\end{figure} 

In (QIF-P), \(cv(q)\) denotes the set of classical variables appearing in the quantum coin \(q\) (empty if \(q\) is a simple variable); \(change(C_i)\) is the set of classical variables modified by \(C_i\). The side condition ensures that the two quantum branches do not change classical variables in conflicting ways, so that the classical state remains well‑defined after merging.

Let \(\mathcal{D}=\{P_i(\overline{u_i})\Leftarrow C_i\mid i=1,\dots,n\}\) be a set of procedure declarations. For a set \(\mathcal{A}\) of Hoare triples, the judgement
\[
\mathcal{A} \;\vdash\;\{A,|\phi\rangle\}C\{B,|\psi\rangle\}
\]
means that the triple can be derived from the assumptions \(\mathcal{A}\) using the rules for non‑recursive constructs. In (Recursion-Tot), \(r\) is an integer ranking function and \(z\) is a fresh integer variable not occurring in \(r\), \(C_i\) or any free variables in \(A_i,B_i\). The condition \(A_i\to r\ge0\) guarantees non‑negativity, enabling induction on \(r\) to prove termination.

\subsubsection{Soundness and (Relative) Completeness}
\begin{theorem}[Soundness]
For any $C\in\mathbf{RQC}^{++}$ and any interpretation $\mathbb{I}$ of the assertion language:
\begin{enumerate}
\item If $\vdash_{\mathit{par}}\{A,|\phi\rangle\}C\{B,|\psi\rangle\}$ then $\mathbb{I}\models_{\mathit{par}}\{A,|\phi\rangle\}C\{B,|\psi\rangle\}$.
\item If $\vdash_{\mathit{tot}}\{A,|\phi\rangle\}C\{B,|\psi\rangle\}$ then $\mathbb{I}\models_{\mathit{tot}}\{A,|\phi\rangle\}C\{B,|\psi\rangle\}$.
\end{enumerate}
\end{theorem}
Here, \(\vdash_{\mathit{par}}\{A,|\phi\rangle\} C \{B,|\psi\rangle\}\) and \(\vdash_{\mathit{tot}}\{A,|\phi\rangle\} C \{B,|\psi\rangle\}\) mean that the Hoare triple can be derived in the proof system for partial and total correctness, respectively. 

\begin{theorem}[Relative completeness]
Let $\mathbb{I}$ be any interpretation of the assertion language and let $\mathit{Th}(\mathbb{I})$ be the set of all true assertions in $\mathbb{I}$. Then:
\begin{enumerate}
\item If $\mathbb{I}\models_{\mathit{par}}\{A,|\phi\rangle\}C\{B,|\psi\rangle\}$ then $\mathit{Th}(\mathbb{I})\vdash_{\mathit{par}}\{A,|\phi\rangle\}C\{B,|\psi\rangle\}$.
\item If $\mathbb{I}\models_{\mathit{tot}}\{A,|\phi\rangle\}C\{B,|\psi\rangle\}$ then $\mathit{Th}(\mathbb{I})\vdash_{\mathit{tot}}\{A,|\phi\rangle\}C\{B,|\psi\rangle\}$. 
\end{enumerate}
\end{theorem}

\section{Detailed Description of RQIMP Statements}
\label{app:source}
For brevity, the main text covers only core types and semantics; this section gives the full set of rules and formal definitions for all informally used symbols.

For a statement \(s\), we define \(\mathsf{mv}(s)\) (modified variables) and \(\mathsf{rv}(s)\) (read variables) as follows.
\textbf{Modified variables \(\mathsf{mv}(s)\).} The set of variables (left‑values) that are assigned to by \(s\). Here \(\mathrm{base}(l)\) returns the set of base names of \(l\) (i.e., the variable names obtained by stripping all index subscripts). For example, for \(l = x[v]\), \(\mathrm{base}(l) = \{x\}\).
\[
\begin{array}{l}
\mathsf{mv}(\tau x \leftarrow e) = \{x\},\ \ 
\mathsf{mv}(l \leftarrow \mathtt{aop} \ e) = \base(l),\ \ 
\mathsf{mv}(\tau x \leftarrow f(\overline{u})) = \{x\},\\[4pt]
\mathsf{mv}(s_1;s_2) = \mathsf{mv}(s_1) \cup \mathsf{mv}(s_2),\ \ 
\mathsf{mv}(\mathtt{if}\;e\;\{s_1\}\;\{s_2\}) = \mathsf{mv}(s_1) \cup \mathsf{mv}(s_2),\\[4pt]
\mathsf{mv}(\mathtt{for}\;x\; v\; \{s\}) = \mathsf{mv}(s) \cup \{x\},\ \ 
\mathsf{mv}(\mathtt{while}\;e\;\{s\}) = \mathsf{mv}(s).
\end{array}
\]
\textbf{Read variables \(\mathsf{rv}(s)\).} The set of variables (appearing in expressions or as sources) that are read by \(s\).
\[
\begin{array}{l}
\mathsf{rv}(\tau x \leftarrow e) = \var(e),\quad 
\mathsf{rv}(l \leftarrow \mathtt{aop} \  e) = \var(l) \cup \var(e),\quad 
\mathsf{rv}(\tau x \leftarrow f(\overline{u})) = \bigcup_{u\in\overline{u}} \var(u),\\[4pt]
\mathsf{rv}(s_1;s_2) = \mathsf{rv}(s_1) \cup \mathsf{rv}(s_2),\quad 
\mathsf{rv}(\mathtt{if}\;e\;\{s_1\}\;\{s_2\}) = \var(e) \cup \mathsf{rv}(s_1) \cup \mathsf{rv}(s_2),\\[4pt]
\mathsf{rv}(\mathtt{for}\;x\; v\;\{s\}) = \mathsf{rv}(s) \cup \var(v) \cup \{x\}, \quad 
\mathsf{rv}(\mathtt{while}\;e\;\{s\}) = \var(e) \cup \mathsf{rv}(s).
\end{array}
\]
\textbf{Read-only variables \(\mathsf{ro}(s)\).} For every statement \(s\), we define \(\mathsf{ro}(s)\) as follows:
\begin{itemize}
    \item If \(s\) is of the form \(l \leftarrow \mathtt{aop}\ e\), then \(\mathsf{ro}(s) =  \mathsf{rv}(s) \setminus  \mathsf{mv}(s) \).
    \item For all other statements, \(\mathsf{ro}(s)\) is defined analogously to \(\mathsf{rv}(s)\) (i.e., using the same rules, replacing \(\mathsf{rv}\) by \(\mathsf{ro}\) for compound statements).
\end{itemize} 

\paragraph{Type system.}
Let \(\Gamma\) be a type environment. We define the set of modified quantum variables of a statement \(s\) under \(\Gamma\), denoted \(\mvq_{\Gamma}(s)\), as the subset of \(\mathsf{mv}(s)\) consisting of those variables whose type has mode \(\mathtt{Q}\). The definition relies on \(\Gamma\) to determine the mode of each variable. When \(\Gamma\) is clear from the context, we omit the subscript and simply write \(\mvq(s)\).  We write \(\mathsf{mvc}(s) = \mathsf{mv}(s) \setminus \mvq(s)\) for the modified classical variables. 
\[
\begin{array}{l}
\mvq(\tau x \leftarrow e) = \{x\}\ \text{if}\ M(x)=\mathtt{Q}\ \text{else}\ \emptyset,\qquad
\mvq(l \leftarrow \mathtt{aop}\ e) =
\begin{cases}
\base(l), & \text{if } M(l)=\mathtt{Q},\\
\emptyset, & \text{otherwise},
\end{cases}\\[4pt] 
\mvq(\tau x \leftarrow f(\overline{u})) = \{x\}\ \text{if}\ M(x)=\mathtt{Q}\ \text{else}\ \emptyset,\qquad
\mvq(s_1;s_2) = \mvq(s_1) \cup \mvq(s_2),\\[4pt]
\mvq(\mathtt{if}\;e\;\{s_1\}\;\{s_2\}) = \mvq(s_1) \cup \mvq(s_2),\qquad
\mvq(\mathtt{for}\;x\; v\; \{s\}) = \mvq(s) \cup \{x\},\\[4pt]
\mvq(\mathtt{while}\;e\;\{s\}) = \mvq(s).
\end{array}
\] 
The definition of \(\mathsf{rvq}_{\Gamma}(s)\) — the quantum variables read by statement \(s\) under \(\Gamma\) — is analogous.   

The complete typing rules for expressions are given in Table~\ref{tab:exp_type_app}; they are mostly standard and are not elaborated here. 
\begin{table}[t]
    \centering
    \caption{Typing rules for expressions.}
 \begin{gather*}
  \dfrac{}{\Gamma \vdash () : \mathtt{unit}} \qquad
  \dfrac{ }{\Gamma, x: \tau  \vdash x : \tau} \qquad 
     \dfrac{}{\Gamma \vdash n : \mathtt{nat}^\mathtt{C}} \qquad  \dfrac{b \in \{\mathtt{true}, \mathtt{false} \} }{\Gamma \vdash b : \mathtt{bool}^\mathtt{C}} \\ 
       \dfrac{\Gamma \vdash x: \mathtt{array} \ n \ \tau \quad \Gamma \vdash v: \mathtt{nat}^{\mathtt{q}} \quad \mathtt{q} \sqsubseteq M(\tau)}{\Gamma \vdash x[v] : \tau} \\ 
       \quad \dfrac{ \Gamma \vdash x_1: \tau_1 \quad \Gamma \vdash x_2: \tau_2 }{\Gamma \vdash (x_1,x_2) : (\tau_1,\tau_2)} \qquad \dfrac{ \Gamma \vdash x: (\tau_1,\tau_2)  }{\Gamma \vdash \pi_{i}(x) : \tau_i} \\ 
       \dfrac{\Gamma \vdash a_1: \mathtt{nat}^{\mathtt{q}_1} \quad \Gamma \vdash a_2: \mathtt{nat}^{\mathtt{q}_2} \quad \mathtt{q}= \mathtt{q}_1 \sqcup \mathtt{q}_2 }{\Gamma \vdash a_1 \ \mathtt{aop} \ a_2: \mathtt{nat}^{\mathtt{q}}} \quad  \dfrac{\Gamma \vdash b_1: \mathtt{nat}^{\mathtt{q}_1} \quad \Gamma \vdash b_2: \mathtt{nat}^{\mathtt{q}_2} \quad \mathtt{q} = \mathtt{q}_1 \sqcup \mathtt{q}_2 }{\Gamma \vdash b_1 \ \mathtt{bop} \ b_2: \mathtt{bool}^{\mathtt{q}}}   \\ 
        \dfrac{\Gamma \vdash b: \mathtt{bool}^{\mathtt{q}}}{ \Gamma \vdash \neg b : \mathtt{bool}^{\mathtt{q}}} \qquad \dfrac{\Gamma \vdash b_1: \mathtt{bool}^{\mathtt{q}_1} \quad \Gamma \vdash b_2: \mathtt{bool}^{\mathtt{q}_2} \quad \mathtt{q} = \mathtt{q}_1 \sqcup \mathtt{q}_2 }{ \Gamma \vdash b_1 \wedge b_2 : \mathtt{bool}^{\mathtt{q}}} 
 \end{gather*}
  \label{tab:exp_type_app}
\end{table}
The full typing rules are given in Table~\ref{tab:type_fun}; we explain only the key points below. To avoid global-state interference that would obstruct the
uncomputation of local quantum variables, we impose the following discipline on
global variables. For every non-main function body \(s\), we require
\(\dom(\Gamma) \cap \mathsf{mv}(s)=\emptyset\), ensuring that no global variable is modified by a non-main function.  To explain the necessity of this global-variable discipline, consider a function
  call \(x \gets f(y)\). If the callee \(f\) modifies a global variable during its
  execution, then this modification remains after the function call returns. In
  this case, if we later need to uncompute the quantum system associated with the
  local variable \(x\), we can no longer do so simply by executing the inverse of
  the target function \(F_f\). There are two reasons. First, the global state on
  which the inverse execution of \(F_f\) relies may already differ from the global
  state at the time of the forward call. Second, even if one forcibly applies
  \(\qinv(F_f)\), this inverse execution would undo the side effect of \(f\) on
  the global variable, thereby changing the global state that should have been
  preserved in the calling context. Therefore, we require non-main functions not
  to modify global variables, ensuring that the uncomputation of local variables
  does not carry externally visible global side effects.

  The situation for the main function is slightly different. Since the main
  function is not called as an ordinary function from another call site, the above
  problem of cleaning a variable by inverse execution after a function call
  does not arise. Nevertheless, global variables in the main function still need
  to satisfy a consistent access discipline. Consider the following program
  fragment, where \(x\) is a quantum-mode local variable and \(y\) is a
  quantum-mode global variable:
  \[
  x \gets y; \qquad y \gets y + x .
  \]
  Here, the local variable \(x\) first depends on the global variable \(y\), and
  then the global variable \(y\) depends on the updated local variable \(x\). This
  creates a cyclic dependency between the local variable and the global variable.
  If we later need to uncompute the quantum variable associated with \(x\), then, because the value of \(y\) has already changed, the original inverse operation can no longer correctly restore \(x\). If we instead temporarily revert \(y\) to enable the uncomputation of \(x\), the updated value of \(x\) prevents us from correctly recomputing the result of \(y\). Therefore, neither ordering can avoid affecting the state of the  global variable \(y\). 

  Therefore, in the main function, we require each global variable to maintain a
  single access role: it is used either only as an input to be read or only as an
  output to be written, but not both. Formally, letting \(s\) be the main
  function body, we require
  \[
  \dom(\Gamma) \cap (\mathsf{mv}(s) \cap \mathsf{ro}(s)) = \emptyset.
  \]
  This condition prevents cyclic dependencies from forming between global
  variables and local variables, ensuring that the uncomputation of local quantum
  variables does not disturb the global quantum state.  For a function $f \in \Xi$, let $\fp(f)$, $\re(f)$, $s_f$, and $\Gamma_f$ denote its formal parameters, return value, body, and type environment; the subscript may be omitted when \(\Xi\) is clear. When no confusion arises, we identify a list or tuple with the set of its elements. 
\begin{table}[t]
    \centering
    \begin{gather*}
  \Gamma'=\Gamma, \{\overline{x}:\overline{\tau}\} \quad \Xi; \Gamma' \vdash s \dashv \Gamma''  \quad \Xi(f)=(\overline{\tau x}, s, \tau_r v, -) \\ 
    \dfrac{ \mathsf{var}(v) \cap \dom(\Gamma') = \emptyset  \quad M(\tau_r)=\mathtt{Q} \quad \Gamma'' \vdash v : \tau_r  \quad  \dom(\Gamma) \cap \mathsf{mv}(s) =\emptyset}{\Xi; \Gamma \vdash  \mathtt{def} \ f(\overline{\tau x}) \{s ; \mathtt{return} \ v\} \triangleright \Xi[f \mapsto (\overline{\tau x}, s, \tau_r v, \Gamma'')]} \qquad  
      \\ 
      \vspace{3.0pt} 
    \dfrac{ \quad \Xi; \Gamma \vdash s \dashv \Gamma'  \quad  \dom(\Gamma) \cap (\mathsf{mv}(s) \cap \mathsf{ro}(s)) = \emptyset }{\Xi; \Gamma \vdash  \mathtt{def} \ \mathtt{main} \{s ; \mathtt{return} \ ()\}  \triangleright \Xi[\mathtt{main} \mapsto ((), s, (), \Gamma')]} \\
\end{gather*} 
\caption{Typing rules for function declarations}
\label{tab:type_fun}
\end{table}

\paragraph{Semantics.} As mentioned in Section~\ref{subsec:source-sem}, the operational semantics of RQIMP is defined via transitions between configurations; the complete semantic rules are given in Table~\ref{tab:opsem-app}. 
Most of the above rules are standard. The assignment statement \(\tau x \leftarrow e\) updates the value of variable \(x\) to \(\sigma(e)\); the unary assignment \(l \leftarrow \mathtt{aop} \; e\) updates the value of the lvalue \(l\) to \(\sigma(l) \; \mathtt{aop} \; \sigma(e)\). The conditional statement selects the appropriate branch based on the value of the guard \(e\). When the guard \(e\) is evaluated in quantum mode, after execution only the modifications to quantum variables are retained, while classical variables modified in the branch are restored to their pre-branch values. A function call \(\tau x \leftarrow f(\overline{u})\) first binds the formal parameters \(\overline{y}\) to the values of the actual arguments \(\overline{u}\) in the current state, then executes the function body \(s\) in the updated state, and finally assigns the value of the return variable \(v\) to \(x\). After the return value is transferred, the callee's local state is restored.  The loop statement \(\mathtt{for } \ x \; v \; \{s\}\) starts with \(x = 0\) and repeatedly executes the loop body \(s\) while incrementing \(x\) until \(x > v\). Similarly, only modifications to quantum variables are preserved across iterations; classical variable updates are reverted after the entire statement finishes. The semantics for while-loops are defined analogously.

\begin{table}[t]
\centering
\caption{Operational semantics of RQIMP.}
\label{tab:opsem-app} 
\begin{gather*}
    \frac{}{\langle \tau x \leftarrow e, \sigma \rangle \to_{\Xi} \langle \downarrow, \sigma[x \mapsto \sigma(e)] \rangle} \quad 
    \frac{}{\langle l \leftarrow \mathtt{aop} \; e, \sigma \rangle \to_{\Xi} \langle \downarrow, \sigma[l \mapsto \sigma(l) \; \mathtt{aop} \; \sigma(e)] \rangle} \\
    \frac{\langle s_1, \sigma \rangle \to_{\Xi} \langle s_1', \sigma' \rangle}{\langle s_1; s_2, \sigma \rangle \to_{\Xi} \langle s_1'; s_2, \sigma' \rangle} \quad
    \frac{}{\langle \downarrow; s_2, \sigma \rangle \to_{\Xi} \langle s_2, \sigma \rangle} \\
    \frac{M(e)=\mathtt{C} \quad \sigma(e) = \mathtt{true}}{\langle \mathtt{if} \ e \; \{s_1\} \; \{s_2\}, \sigma \rangle \to_{\Xi} \langle s_1, \sigma \rangle} \qquad
    \frac{M(e)=\mathtt{C} \quad \sigma(e) = \mathtt{false}}{\langle \mathtt{if} \ e \; \{s_1\} \; \{s_2\}, \sigma \rangle \to_{\Xi} \langle s_2, \sigma \rangle} \\
   \frac{M(e)=\mathtt{Q} \quad \sigma(e) = \mathtt{true} \quad L = \mathsf{mvc}(s_1)}
  {\langle \mathtt{if} \ e \; \{s_1\} \; \{s_2\}, \sigma \rangle \to_{\Xi}
   \langle s_1; L \gets \sigma(L), \sigma \rangle}
  \qquad
  \frac{M(e)=\mathtt{Q} \quad \sigma(e) = \mathtt{false} \quad L = \mathsf{mvc}(s_2)} {\langle \mathtt{if} \ e \; \{s_1\} \; \{s_2\}, \sigma \rangle \to_{\Xi} \langle s_2; L \gets \sigma(L), \sigma \rangle} \\ 
\frac{
    \Xi(f) = (\overline{\tau y}, s, \tau_r v, \Gamma)
    \quad 
    L = \mathsf{mv}(s)
  }
  {\langle \tau x \leftarrow f(\overline{u}), \sigma \rangle \to_{\Xi}
   \langle \overline{y} \gets \overline{u}; s; x \gets v; L \gets \sigma(L), \sigma \rangle} \\ 
\frac{\sigma(e)=\mathtt{true} \quad L = \mathsf{mvc}(s)}
    { \langle \mathtt{while} \ e \; \{s\}, \sigma \rangle \to_{\Xi}
      \langle s; \mathtt{while} \ e \; \{s\}; L \gets \sigma(L), \sigma \rangle }
\qquad
\frac{\sigma(e)=\mathtt{false}}
{ \langle \mathtt{while} \ e \; \{s\}, \sigma \rangle \to_{\Xi} \langle \downarrow, \sigma \rangle }\\ 
\frac{L = \mathsf{mvc}(s)}
  {\langle \mathtt{for} \ x \; v \; \{s\}, \sigma \rangle \to_{\Xi}
   \langle \mathtt{forIter} \ x \; v \; \{s\}; L \gets \sigma(L), \sigma[x \mapsto 0] \rangle} \quad 
\frac{\sigma(x) > v}
  {\langle \mathtt{forIter} \ x \; v \; \{s\}, \sigma \rangle \to_{\Xi}
   \langle \downarrow, \sigma \rangle} \\ 
\frac{\sigma(x) \leq v}
  {\langle \mathtt{forIter} \ x \; v \; \{s\}, \sigma \rangle \to_{\Xi} \langle s; x \gets x+1; \mathtt{forIter} \ x \; v \; \{s\}, \sigma \rangle} 
\end{gather*}
\end{table}

\section{Compilation Details}

This section provides additional details of the compilation pipeline that are omitted from the main text. Our compiler translates the high-level language RQIMP into reversible quantum circuits in $\RQC^{++}$. The downstream compilation chain from $\RQC^{++}$ to the low-level instruction set QINS of the Quantum Register Machine (QRM) has been fully implemented in~\cite{zhangQuantumRegisterMachine2025}. Thus, by compiling RQIMP source programs to $\RQC^{++}$, our compiler can directly reuse this mature backend. We first clarify how the indexed variables used by
the compiler should be interpreted with respect to finite static storage.

\paragraph{Finite storage interpretation of indexed variables.}
The compiler uses indexed variables such as \(q[i]\), \(q[i+1]\), and \(q[i:)\)
as a target-level abstraction for the storage locations required by different
recursion layers. These indices do not represent dynamic allocation of new
quantum registers during execution, nor do they mean that the target execution
relies on infinite quantum storage. Instead, they make explicit which storage
locations must be separated across recursion layers. In the full
\(\RQC^{++}\)-to-QRM compilation chain, the existing QRM backend realizes this
abstract indexed-storage discipline as a finite QRAM layout. Concretely, the
\(\RQC^{++}\) program is first compiled into the low-level QINS program,
together with the metadata needed for the QRM memory layout. Then, for a given
input, QRM performs partial evaluation of the quantum control flow to construct
the corresponding qif table and determine a finite execution length
\(T_{\mathrm{exe}}\); if this length exceeds the practical execution bound, it
reports timeout. During execution, QRM loads the compiled program, symbol table,
qif table, quantum data, and stack into QRAM; the backend then allocates finite
variable and stack sections according to the above finite execution bound, so
that they are sufficient for the indexed variables accessed during the bounded
execution. Thus, ReOC generates \(\RQC^{++}\) code with an explicit indexed
static-storage discipline, while the concrete finite low-level memory layout is
supplied by the downstream QRM backend. Further details of the QRM compilation
and execution model can be found in~\cite{zhangQuantumRegisterMachine2025}.

\subsection{Detailed Rules for High-Level Transformations}
\label{app:high-tr}
This subsection provides a detailed account of the high‑level transformations outlined in Section~\ref{sec:comp}. After these transformations, the program conforms to a strict syntactic subset of RQIMP, while the type system is temporarily relaxed to enable reversibilization and the subsequent uncomputation of temporary variables.

\paragraph{Replacing quantum branches by procedure calls}
For a conditional statement \(\mathtt{if}\ e\ \{s_1\}\ \{s_2\}\) with $M(e)=\mathtt{Q}$, we extract each branch into a separate procedure. Specifically, for branch \(s_1\) we generate a procedure declaration: 
\[
\mathtt{def} \ f_1(\overline{\tau y}) \ \{ s_1;\ \mathtt{return}\ v \}
\]
where:
\begin{itemize}
  \item \(\overline{\tau y}\) lists the types and names of external variables used in \(s_1\) (i.e., variables declared before the if-statement) and \(\overline{\tau}\) are their corresponding types;
  \item \(v\) denotes a tuple of variables whose values are updated in $s_1$ and needed after the conditional (e.g., those variables $x$ such that \(x \in \mvq(s_1)\) and are used later).
\end{itemize}
The original branch is then replaced by a call \(\tau x \leftarrow f_1(\overline{y})\), where \(\overline{y}\) is the list of actual arguments corresponding to \(\overline{\tau y}\), and \(x\) is a fresh variable that receives the returned tuple (its type \(\tau\) is derived from the types of \(v\)). If an argument in \(\overline{y}\) occurs in \(v\), it is first copied to a fresh temporary variable and replaced in \(v\) accordingly. An analogous procedure \(f_2\) is created for the other branch \(s_2\). This transformation ensures that the quantum control flow is encapsulated in procedures, isolating the quantum effects. We also apply this transformation to $\mathtt{C}$‑mode conditionals, turning each branch into a non‑recursive function call. Although such calls may return classical variables (disallowed by our type system in RQIMP), they are introduced solely to enable later uncomputation of temporary variables and can be inlined afterwards.  If a branch consists of a single statement, no transformation is applied.

\paragraph{Converting loops into recursive procedure calls}
For a loop \(\mathtt{for}\ i\ w\ \{s\}\), we encode the iteration as a recursive procedure:
\[
\mathtt{def} \ f(\tau_1 i, \tau_2 \mathtt{var}(w), \overline{\tau y}) \ \{ s;\ \mathtt{if}\ i < w\ \{\tau x \leftarrow f(i+1, \mathtt{var}(w), \overline{y})\}\ \{ \tau x \leftarrow v \};\ \mathtt{return}\ x \}
\]
Here:
\begin{itemize}
  \item \(\tau_1\) and \(\tau_2\) are the types of \(i\) and \(\mathtt{var}(w)\), respectively;
  \item \(\overline{\tau y}\) again lists the external variables of the loop body \(s\);
  \item \(v\) is a tuple of quantum variables that accumulate the results (i.e., those variables $x$ such that \(x \in \mvq(s)\) and are used after the loop);
  \item \(x\) is a fresh variable (with type \(\tau\) derived from $v$) that will hold the final tuple after all iterations.
\end{itemize}
The original loop is then replaced by a call \(\tau x \leftarrow f(0, \mathtt{var}(w), \overline{y})\). The translation of while-loops proceeds analogously, except that the generated call is guarded by the loop condition so that zero-iteration loops are handled by the else branch. 

After the above transformations for branches and loops, any subsequent use of a variable that originally appears in \(v\) must be replaced by the corresponding component of the returned tuple \(x\).

\paragraph{Removing expressions in conditions and parameters}
The third step eliminates complex expressions that appear as guards in conditional statements, as arguments in procedure calls, and as components of return values. After this step, all conditions and call arguments become simple variables.

\begin{itemize}
  \item \textbf{Conditional statements:} For every \(\mathtt{if}\ e\ \{s_1\}\ \{s_2\}\), we introduce a fresh variable \(x\) of the same type as \(e\) (which is \(\mathtt{bool}^\mathtt{C}\) or \(\mathtt{bool}^\mathtt{Q}\)) and rewrite: 
  \[
  \mathtt{if}\ e\ \{s_1\}\ \{s_2\} \quad\Longrightarrow\quad \tau x \leftarrow e;\ \mathtt{if}\ x\ \{s_1\}\ \{s_2\}
  \]
  where \(\tau\) is the type of \(e\). This replaces the original guard expression with a simple variable.

  \item \textbf{Procedure calls:} For every call \(\tau x \leftarrow f(\overline{t})\) where \(\overline{t}\) is a list of expressions (possibly containing operators), we introduce a fresh list of variables \(\overline{y}\) with the same types \(\overline{\tau'}\) as \(\overline{t}\) and rewrite:
  \[
  \tau x \leftarrow f(\overline{t}) \quad\Longrightarrow\quad \overline{\tau' y} \gets \overline{t};\ \tau x \leftarrow f(\overline{y})
  \]
  Here \(\overline{\tau' y} \gets \overline{t}\) denotes a sequence of assignments (one per element) that evaluate the arguments and store them in the fresh variables. 
  \item \textbf{Return value:} For each component \(v_i\) of the return value \(v\) that is an array location or a projection, we introduce a fresh binding \(\tau x \leftarrow v_i\) and substitute \(x\) for \(v_i\) in \(v\). 
\end{itemize}

After applying these transformations, the syntax of a program in RQIMP is reduced to the following core grammar:
\[
\begin{array}{rlll}
 \mathrm{Statement}   & s &:=  &  \tau x\leftarrow e \ | \ l \leftarrow \mathtt{aop} \ e \ |  \  \tau x \leftarrow f (\overline{x}) \ |  \   s; s \ | \  \mathtt{if} \ x \ \{s\} \  \{s\}. \\
\end{array}
\]

\subsection{Details for Dependency Graph}
\label{secA_dg} 
\begin{figure}
    \centering
    \includegraphics[width=1\linewidth]{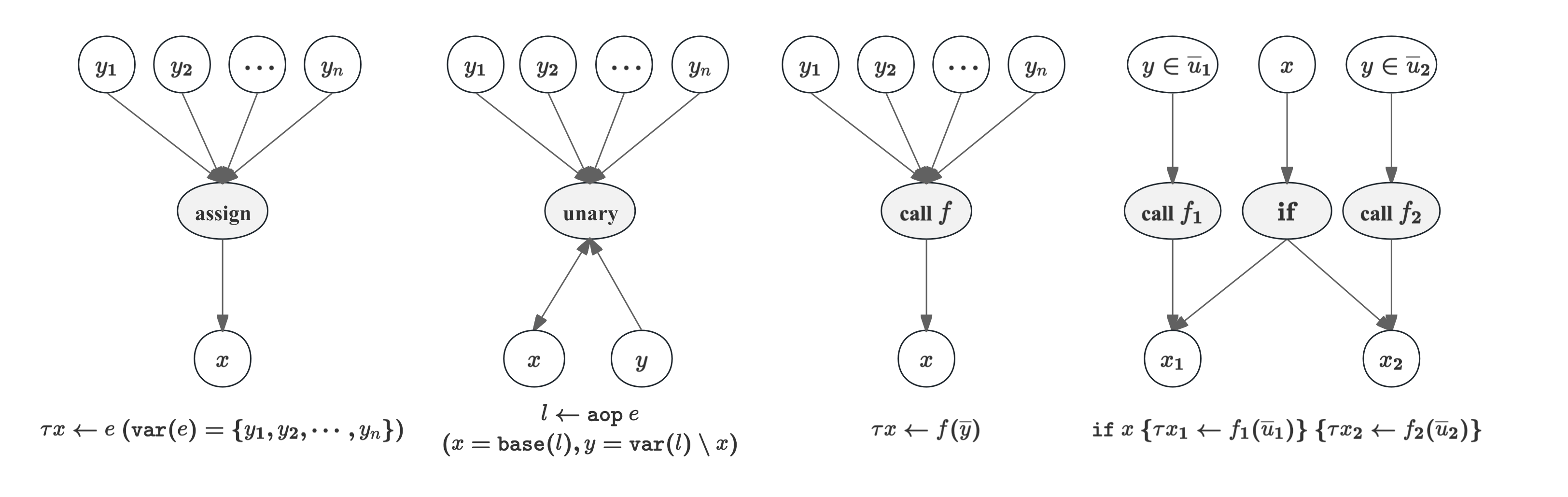}
    \caption{Dependency subgraphs for statement types.}
    \Description{...}
    \label{fig:var_dg}
\end{figure}

In this subsection, we provide further details on the construction of the dependency graph. Predefined dependency subgraphs for specific statement types are illustrated in Figure~\ref{fig:var_dg}, where grey nodes represent statement nodes and white nodes represent variable nodes with $\mathtt{Q}$-mode. For a sequential composition \(s_1; s_2\), the dependency subgraph is constructed by concatenating the subgraphs of \(s_1\) and \(s_2\) in order. 
When constructing the dependency graph, if a variable node involved in a statement does not yet exist, we create the node along with its dependency edges and the statement node; if the node already exists, we extend the graph by adding only the corresponding edges and the statement node. These construction steps are straightforward and are therefore omitted here.  

Next, to capture recursive characteristics, each variable node is assigned a flag. Given a variable dependency graph \(G_{\mathrm{fun}}(f)\), for each non-return local variable \(x\) in it, we locate its direct predecessor statement nodes. If any of these predecessor nodes contains a call to some function \(f_1\), we examine \(\SCC(f_1)\). If \(\SCC(f_1)\) contains \(f\) and has exactly one cycle, we set \(x.\flag = 2\); if it contains \(f\) and has multiple cycles, we set \(x.\flag = 1\). If neither of the above cases applies, we set \(x.\flag = 0\). 

Furthermore, some local variables must be cleaned together when they are mutually dependent; cleaning only one of them may be ineffective. To address such cases, we merge all variables belonging to the same strongly connected component (i.e., mutually dependent variables) into a composite node. Such a node indicates that its variables form an atomic unit: collective uncomputation is performed only when all variables inside the unit are no longer in use. A single variable that is not merged is treated as a composite node containing exactly one variable (the degenerate case). 

However, since a return value's quantum variable is never cleaned, if a component \(x\) of the return value is mutually dependent with another local variable that needs to be cleaned, this dependency would prevent the latter from being cleaned. To resolve this, before merging nodes, we preprocess each such component \(x\): insert \(\tau y \leftarrow x\) just before the \(\mathtt{return}\ v\) statement, and replace \(x\) in the return tuple \(v\) with \(y\). The newly introduced variable \(y\) has no mutual dependency with other variables (since \(y\) is fresh), thus it does not block the cleanup of other variables. Moreover, the source type system ensures no mutual dependency between global and local variables. Consequently, we guarantee that nodes containing globals or the return value are disjoint and mutually independent from nodes containing other local variables. Furthermore, the source-level type system guarantees that for any function call \(\tau x \leftarrow f(\overline{u})\), the body of \(f\) never modifies the values of global variables. Together, these two facts ensure the correct cleanup of quantum variables for local variables without affecting globals or the return value.

For each composite node $w$, we define its flag $w.\flag$ as follows: if there exists at least one variable $x$ in $w$ with $x.\flag = 1$, then $w.\flag = 1$; otherwise $w.\flag = 0$. Finally, all composite nodes with $\flag=1$ are merged into a single composite node, ensuring that there is only one node satisfying $\flag=1$ after the merge and allowing cleanup of the entire group to be deferred until the end of the function body.
 
For a variable dependency graph, we write \(\node(x)\) for the composite node containing \(x\), \(\var(w)\) for the set of variables in node \(w\), and \(\mathrm{pred}^+(w)\) for the set of strict predecessors of \(w\) in the transitive closure (i.e., not including \(w\)). 

% Furthermore, we define a mapping \(\mathrm{Outd}\) that assigns to each node \(v\) its out‑degree, denoted \(\mathrm{Outd}(v)\), which is the sum of the out‑degrees of the variables inside \(v\).  

\subsection{Details of Compilation Functions}
\label{app:com_fun}

\begin{algorithm}[!tbp]
\caption{\textsc{Compile\_Fun}: Compiling a Function Declaration}
\label{com_fun}
\KwIn{
    Static input parameters $\K$,
    compilation context $\ctx$,
    function name $f$, fresh quantum registers $\{\overline{p}, \overline{r}\}$
}
\KwOut{Updated context $\ctx$}
\SetKwProg{Fn}{Function}{}{}
\Fn{\textsc{Compile\_Fun}($\K, \ctx, f, \{\overline{p}, \overline{r}\}$)}{
    $g \gets f_{\mathrm{cur}}$; \ $\push((g, \Vactive), \St)$; \ $\M \gets \M[g \mapsto (C, \rho_{\mathrm{loc}}, \kappa, \delta, \Eargs, \Valloc)]$; \ $\push(f, \St)$\; 
    % $\{\overline{\tau x}, s, v, \Gamma'\} \gets \Xi(f)$; \quad $\K \gets \K[\K \mapsto \Gamma']$\; 
    \If{$f \notin \SCC(f_{\mathrm{cur}})$}{
        $\Qpool_{\mathrm{temp}} \gets \{ q\in \Qpool \mid \base[q] \neq q\}$;\ $\Qpool' \gets \Qpool \setminus \Qpool_{\mathrm{temp}}$\;
$n_f \gets |\fp(f)|+|\re(f)|$; \ $\{\overline{p},\overline{r}\} \gets \pop(\Qpool', n_f)$\;
        $\Eargs' \gets \emptyset$\;
    }
    \Else {
        $\Eargs' \gets \Eargs$;\  $\Qpool' \gets \Qpool$\;
    } 
    $\Qpool \gets \Qpool'$; \ $f_{\mathrm{cur}} \gets f$; \ $\Vactive \gets \emptyset$; \ $C \gets I$; \ $\rho_{\mathrm{loc}} \gets \{(\fp(f),\re(f))\mapsto (\overline{p},\overline{r})\}$\;
    $\kappa \gets \emptyset$; \ $\delta \gets \outd_{G_{\mathrm{fun}}(f)}$;\ $\Eargs \gets \Eargs'$; \ $\Valloc \gets \emptyset$\; 
    $\ctx \gets \textsc{Compile\_Body}(\K, \ctx, s_f)$\;
    $W \gets \{\node(y)\mid y \in \dom(\rho_{\mathrm{loc}}) \setminus \re(f) \wedge \node(y).\flag =1\}$\;
    $\ctx \gets \textsc{Clean\_Temp}(\K, \ctx, w)$ for each $w \in W$\; 
    $\D \gets \D \cup \{\Theta(f)(\getC(\fp(f)), \rho_{\mathrm{loc}}(\getQ(\fp(f))), \rho_{\mathrm{loc}}(\re(f)), \Eargs) \Leftarrow C\}$\;
    \If{$f \notin \SCC(g)$}{
        \For{each $h \in \St$ with $\pos(h) \geq \pos(f)$}{
            $E_{\args,h} \gets E_{\args,h} \cup V_{\alloc,\succeq f}$\; 
            Add new parameters in \(E_{\args,h}\) to \(\Theta(h)\)'s decl \& calls in \(\D\)\; 
            $\push (\Qpool, (\cod(\rho_{\mathrm{loc},h})\cup V_{\activelabel,h}))$\;
             Remove $h$ from $\St$ and $\M$\;
        }
        $\Qpool \gets \{\base[q] \mid q \in \Qpool\} \cup \Qpool_{\mathrm{temp}}$ \tcp*{remove all subscripts in $\Qpool$}
    }
    \Else{
      $\St \gets \push(\St,(f,\Vactive))$; \ $\M \gets \M[f \mapsto m]$\; 
    }
    $f_{\mathrm{cur}} \gets g; \  \Vactive \gets \emptyset; \ m=(C, \rho_{\mathrm{loc}}, \kappa, \delta, \Eargs, \Valloc) \gets \M(g)$\; 
    \Return $\ctx$ }
\end{algorithm}

Here we present the details of compiling a function declaration, as given in Algorithm~\ref{com_fun}. For brevity, we introduce the following notations. Complete notation and formal definitions are in Appendix~\ref{app:nota-def}.  For any structure $S$, we write $S[x \mapsto y]$ for the structure obtained by replacing $x$ with $y$ in $S$, leaving everything else unchanged. For instance, in a mapping this updates the binding of $x$ to $y$; in a tuple or set, it replaces the element $x$ by $y$. We write $|\cdot|$ for the size of a list or set, and $\pop(\Qpool, n)$ for popping $n$ fresh registers from pool $\Qpool$, yielding $\overline{p} = (p_1, \dots, p_n)$. We use \(\base[q]\) for the base name of a single indexed quantum register \(q\), while \(\base(S)\) denotes the set of base names of all registers in a set \(S\). We also use \(\getC(\overline{t})\) and \(\getQ(\overline{t})\) for the $\mathtt{C}$‑mode and $\mathtt{Q}$‑mode variables in parameter list $\overline{t}$, respectively. Let $\SCC(f)$ denote the strongly connected component containing $f$ in the call-dependency graph, i.e., $f$'s mutual-recursion closure. For a variable dependency graph \(G\), \(\outd_G\) denotes the initial out-degree map induced by \(G\). 

Unless otherwise specified, compilation variables in each algorithm refer to the corresponding fields in \(\ctx\) (e.g., \(\Qpool\) stands for \(\ctx.\Qpool\)). We also index compilation variables by function names (e.g., \(E_{\args,h} \triangleq \M(h).\Eargs\) for \(h \in \St\) with \(h \neq f_{\mathrm{cur}}\)), and write \(\star_{\succeq f}\) for the aggregation of any field \(\star \in m\) over \(f\) and all functions succeeding \(f\) on \(\St\). Other primed or subscripted names (e.g., \(\Qpool'\), \(\Qpool_{\mathrm{temp}}\), \(\Eargs'\)) are local variables or values used by the algorithm, not \(\ctx\) fields. Updates to compilation variables implicitly modify \(\ctx\); for example, \(f_{\mathrm{cur}} \gets g\) means \(\ctx \gets \ctx[f_{\mathrm{cur}} \mapsto g]\). The notation \(\pos(f)\) denotes the position of function \(f\) on the stack \(\St\).  

The algorithm operates on the compilation context \(\ctx\) and proceeds in seven steps:

\begin{enumerate}
    \item  \textbf{Suspend the current function (line~2).} The name of the previously active function is saved as \(g\). The compiler pushes the \(\Vactive\) component in the current buffer onto the stack \(\St\), and stores the remaining compilation state \(m\) in the mapping \(\M\). It then pushes the name of the function to be compiled, \(f\), onto the top of \(\St\), marking the point at which the compilation of \(f\) begins.  

\item \textbf{Initialize the compilation state for \(f\) (lines~3--10).}  
The compiler initializes the local buffer by setting \(C=I\), initializing \(\Vactive\) and \(\Valloc\) to empty sets, initializing \(\kappa\) to an empty mapping, and setting \(\delta\) to the initial out-degree map \(\outd_{G_{\mathrm{fun}}(f)}\). Formal parameters \(\fp(f)\) and return values \(\re(f)\) are mapped to the input registers \(\overline{p},\overline{r}\), with the corresponding mappings recorded in \(\rho_{\mathrm{loc}}\). 

The treatment of the quantum register pool \(\Qpool\) and the extra-parameter set \(\Eargs\) depends on the recursion group. If \(f \notin \SCC(f_{\mathrm{cur}})\) (i.e., not mutually recursive with \(f_{\mathrm{cur}}\)), all subscripted quantum registers are temporarily removed from \(\Qpool\) and stored in \(\Qpool_{\mathrm{temp}}\), isolating namespaces across distinct recursion groups (they are restored after the group finishes). Moreover, the input registers \(\overline{p}\) and \(\overline{r}\) are re-evaluated using the updated \(\Qpool\), not taken directly from the caller. If \(f \in \SCC(f_{\mathrm{cur}})\), then \(E_{\args,f}\) inherits the full contents of the current function's \(\Eargs\), ensuring consistent parameters inside the mutual-recursion group; otherwise, \(\Eargs\) is initialized to \(\emptyset\).

Moreover, the current function marker is updated to \(f_{\mathrm{cur}}=f\), so that subsequent compilation proceeds in the context of \(f\).  

\item \textbf{Compile the function body (line~11).} Invoke the function body compilation algorithm \(\textsc{Compile\_Body}(\K, \ctx, s_f)\) (Algorithm~\ref{com_body} in Section~\ref{sec:com_body}) to obtain the updated compilation context \(\ctx\), where \(C\) in \(\ctx\) denotes the quantum circuit corresponding to the function body \(s_f\) of function \(f\). 
  
 \item \textbf{Clean up temporary variables (line~12--13).} Upon completion of the function body compilation, any temporary quantum variables in \(f\) pending cleanup but not yet uncomputed (i.e., non-return values in \(\dom(\rho_{\mathrm{loc}})\) whose nodes have \(\flag=1\)) are uncomputed by the \(\textsc{Clean\_Temp}\) algorithm (Algorithm~\ref{clean_temp} in Section~\ref{sec:clean_temp}). 

\item \textbf{Generate the function declaration (line~14).}  The compilation result takes the form \[
  \Theta(f)\bigl(\getC(\fp(f)), \rho_{\mathrm{loc}}(\getQ(\fp(f))), \rho_{\mathrm{loc}}(\re(f)), \Eargs\bigr) \Leftarrow C, 
  \] and is added to the set \(\D\) of \(\RQC^{++}\) function declarations. In addition, the function body \(C\) is wrapped as
  \(\mathbf{begin}\ \overline{y}:=0 \ C\ \mathbf{end}\), where \(\overline{y}\)
  denotes the local classical variables occurring in \(C\). This confines
  classical updates to the generated procedure body and ensures the
  procedure-locality requirement of \(\RQC^{++}\). Upon return, these local classical variables are automatically uncomputed in the sense of the local-block
  semantics of \(\RQC^{++}\). 

\item \textbf{Parameter expansion and resource reclamation (lines~15--23).} 
The algorithm distinguishes two cases. 
\begin{itemize}
\item Case A: \(f \notin \SCC(g)\). If \(f \notin \SCC(g)\), then \(f\) and the functions following it on \(\St\) constitute a distinct mutual-recursion group \(\SCC(f)\), indicating that the compilation of this group has completed. The internal quantum variables recorded in the \(\Valloc\) sets of all functions in this group are combined into a shared extra-parameter set \(V_{\alloc,\succeq f}\) and synchronously appended to the parameter lists of all involved functions, both in their declarations and at all call sites in \(\D\). Afterwards, the resources consumed by these functions are returned to \(\Qpool\), and their compilation information is cleared from the stack \(\St\) and the mapping \(\M\). 

Finally, since the compilation of the mutual recursion group \(\SCC(f)\) has completed, the isolated indexed registers are reintegrated: subscripts are removed from all quantum variables in \(\Qpool\) (e.g., \(q[i] \mapsto q\)), turning them into base names that can be reused by other functions, and the temporarily removed array variables \(\Qpool_{\mathrm{temp}}\) are returned to \(\Qpool\), thereby restoring the unified register pool for subsequent compilation. 

\item Case B: \(f \in \SCC(g)\). This indicates that \(f\) depends on a function that has been compiled but not yet completed; consequently, its final parameter list cannot be determined until the entire group is compiled. At this point, \(f\) remains on the stack and its compilation information is preserved for later recursive calls and parameter expansion.
\end{itemize}

\item \textbf{Resume the compilation of the upper-layer function (line~24).} Finally, compilation of \(g\) is resumed by loading its state \(m\) into the current buffer, with its active-variable fragment \(\Vactive\) remaining on the stack. Thus, the \(\Vactive\) in the buffer corresponds to the content compiled since the most recent resumption, while the previously suspended active-variable fragment stays on the stack, so as to accurately record the actual usage order of unclean variables relative to the compilation time of each unfinished function in the stack \(\St\).
\end{enumerate}

\subsection{Details of Compilation Statements}  
\label{app:com-stmt}
In this section, we provide the details of the translation for statements, including assignments, unary operations, and conditionals within a function body, as described in Algorithm~\ref{Com_state}. 

We first set \(\Gamma' = \Gamma_{f_{\mathrm{cur}}}\) and write \(\rho \triangleq \rho_{\mathrm{glob}} \cup \rho_{\mathrm{loc}}\) for the combined variable mapping. For any statement \(s\), if its written \(\mathtt{Q}\)-mode variables \(\mvq_{\Gamma'}(s)\) (the subscript \(\Gamma'\) is omitted in the algorithm for brevity) are not yet contained in \(\dom(\rho)\), we allocate fresh quantum variables from \(\Qpool\) for the missing variables, with the number of variables determined by their types in \(\Gamma'\), and record the new mappings in \(\rho_{\mathrm{loc}}\). For each variable in \(\mvq(s)\), allocation follows its type in \(\Gamma'\): if a variable \(y\) is of pair type, it is assigned two variables \(p_1, p_2\); otherwise, a single variable is assigned. When accessing a variable via \(\rho_{\mathrm{loc}}(y)\), if \(y\) is a projection \(\pi_1(x)\), we take the first variable of \(\rho_{\mathrm{loc}}(x)\). Then \(\rho_{\mathrm{loc}}(\mvq(s))\) is added to the active-variable set \(\Vactive\) unless it is already present or is an input parameter. Subsequent compilation depends on the type of the statement.   

\begin{algorithm}[!tbp]
\caption{\textsc{Compile\_Com}: Compiling Basic Statements}
\label{Com_state}
\KwIn{ Static input parameters $\K$, 
    compilation context $\ctx$, statement $s$}
\KwOut{Updated compilation context $\ctx$ and the generated circuit $C'$ for $s$}
\SetKwProg{Fn}{Function}{}{}
\Fn{\textsc{Compile\_Com}($\K,\ctx,s$)}{
    \If{$\mvq(s) \nsubseteq \dom(\rho)$}{
        $n_s \gets$ the number of variables required by \(\mvq(s)\) under \(\Gamma'\)\;
        $\overline{p} \gets \pop(\Qpool,n_s)$\;
        $\rho_{\mathrm{loc}} \gets \rho_{\mathrm{loc}}[\mvq(s) \mapsto \overline{p}]$\; 
        $V_{\mathrm{alloc}} \gets V_{\mathrm{alloc}} \cup \overline{p}$
    }
    \If{$\rho_{\mathrm{loc}}(\mvq(s)) \setminus \rho_{\mathrm{loc}}(\fp(f_{\mathrm{cur}})) \not\subseteq \Vactive$} {
        $\Vactive \gets \Vactive \cup \rho_{\mathrm{loc}}(\mvq(s)) \setminus \rho_{\mathrm{loc}}(\fp(f_{\mathrm{cur}}))$\;  
    } 
    \uIf{$s = \tau x \leftarrow e$ or $l \leftarrow \mathtt{aop} \ e$}{
        $\Gamma';\rho;\Qpool \vdash s \leadsto C' \dashv \Qpool'$; \quad  $\ctx \gets \ctx[\Qpool \mapsto \Qpool',\; C \mapsto C; C']$\; $V_{\mathrm{alloc}} \gets V_{\mathrm{alloc}} \cup (\base(\qv(C)) \setminus \cod(\rho))$ 
    }
    \ElseIf{$s = \mathtt{if}\ x\ \{s_1\} \wedge M(x)=\mathtt{Q}$}   
    {
        $(\ctx, C_1) \gets \textsc{Compile\_Com}(\K,\ctx,s_1)$\;
        $C' \gets \mathbf{qif}\ \rho(x) \ \ket{1} \mapsto C_1 \ \square \ \ket{0} \mapsto \mathbf{Skip}$\;
        $\ctx \gets \ctx[C \mapsto C[C_1 \mapsto C']]$\;
    }
    \ElseIf{$s = \tau x \leftarrow f(\overline{u})$}{
        \uIf{$f \notin \Fun(\St)$ and $f \in \Theta^{-1}(\dom(\D))$}  
        % \tcp*{already compiled function}
        {
	            $\{\Theta(f),\overline{t},B\} \gets$ retrieve the declaration of \(\Theta(f)\) from \(\D\)\;
	            $n_0 \gets \vert\getQ(\overline{t})\vert-\vert\getQ(\overline{u})\vert$\;
	            $n_1 \gets \vert\getC(\overline{t})\vert-\vert\getC(\overline{u})\vert$\;
	            $\overline{q} \gets \pop(\Qpool,n_0)$ \;
	            \uIf{there exists a variable \(y\) in \(\SCC(f)\) with \(y.\flag=2\)}{
                $r \gets \pop(\Qpool)$\; 
	                $C' \gets \Theta(f)\bigl(\getC(\overline{u}),\ \rho(\getQ(\overline{u})),\ r,\ 0^{n_1},\ \overline{q}\bigr); \ \mathit{copy}(r,\rho(x));\ \qinv(\Theta(f))$\;
	                $\push(r,\Qpool)$\;
	            }
	            \Else{
	                $C' \gets \Theta(f)\bigl(\getC(\overline{u}),\ \rho(\getQ(\overline{u})),\ \rho(x),\ 0^{n_1},\ \overline{q}\bigr)$\;
	            }
	            $\ctx \gets \ctx[C \mapsto C;C']$; \  $\push(\overline{q},\Qpool)$\;
	        }
	        \ElseIf{$f \in \Fun(\St)$} 
        {
	            $(\ctx, C') \gets \textsc{Compile\_Recur}(\K, \ctx, x\leftarrow f(\overline{u}))$\;
	        }
        \ElseIf{$f \notin \Fun(\St)$ and $f \notin \Theta^{-1}(\dom(\D))$} 
        {
	            \uIf{$f \in \SCC(f_{\mathrm{cur}})$} 
	            {
	                \text{update variable names as in Steps~1 and~2 of Algorithm~\ref{com_re}}\; 
                $\ctx \gets \textsc{Compile\_Fun}(\K, \ctx, f,\{\rho(\getQ(\overline{u})), \rho(x)\})$\;
	                $C' \gets \Theta(f)\bigl(\getC(\overline{u}),\ \rho(\getQ(\overline{u})),\ \rho(x),\ E_{\args,f} \bigr)$\;
	                $\ctx \gets \ctx[C \mapsto C;C']$\; 
               \uIf{$x.\flag=2$}{  $\ctx \gets \ctx[\Vactive \mapsto  \Vactive \cup \Gar_f]$\; } 
	            }
	            \Else{
                $\ctx \gets \textsc{Compile\_Fun}(\K,\ctx, f,\{\rho(\getQ(\overline{u})), \rho(x)\})$\;
	                $(\ctx, C') \gets \textsc{Compile\_Com}(\K,\ctx,s)$\; 
	            }
        }
    }
    \Return $(\ctx, C')$\;
}
\end{algorithm}

\begin{itemize}
    \item \textbf{Assignment and unary operation} \(\tau x \leftarrow e\) or \(l \leftarrow \mathtt{aop}\ e\): These statements are compiled following the rules for assignment statements in Appendix~\ref{sec_com_exp_assgn}.  The judgment $\Gamma'; \rho; \Qpool \vdash s \leadsto C \dashv \Qpool' $ states that under the given type environment \(\Gamma'\), variable mapping \(\rho\), and register pool \(\Qpool\), the statement \(s\) compiles to circuit \(C\) and updates the register pool to \(\Qpool'\).
    \item \textbf{Single-branch quantum conditional} \(\mathtt{if}\ x\ \{s_1\}\) with $M(x)=\mathtt{Q}$: Recursively compile the branch body \(s_1\) to obtain \(C_1\) and the updated compilation context \(\ctx\); then form the quantum conditional \(\mathbf{qif}\ \rho(x)\ \ket{1} \mapsto C_1 \ \square \ \ket{0} \mapsto \mathbf{Skip}\) and replace \(C_1\) in the updated \(\ctx\) with it.  We only show the case with a single branch whose guard is in \(\mathtt{Q}\)-mode here; multiple branches can be decomposed into separate conditional statements (e.g., \(\mathtt{if}\ x\ \{s_1\}\ \{s_2\}\) is handled as \(\mathtt{if}\ x\ \{s_1\}\) followed by \(\mathtt{if}\ \neg x\ \{s_2\}\)). For a \(\mathtt{C}\)-mode guard, the compilation is similar, replacing the quantum conditional \(\mathbf{qif}\) with a classical conditional \(\mathbf{if}\ x\ \mathbf{then}\ C_1\ \mathbf{fi}\). 
\end{itemize}

The function call statement \(\tau x \leftarrow f(\overline{u})\) is more involved; its handling depends on the status of the called function \(f\). Here \(\Fun(\St)\) denotes the set of function names currently recorded on the stack \(\St\):

 \textbf{Case 1: \(f\) is already compiled (\(f \in \Theta^{-1}(\dom(\D))\) and \(f \notin \Fun(\St)\))}.  
    Retrieve the compiled function \(\Theta(f)\) from the declaration set \(\D\). The formal parameters $\overline{t}$ of \(\Theta(f)\) may include recursion markers and additional quantum variables for internal temporaries. For these extra parameters, we initialize their classical parts (if any) and allocate clean 
	    quantum variables \(\overline{q}\) from \(\Qpool\) to serve as arguments. The return value is received by \(\rho(x)\), yielding the call: 
    \[
	    \Theta(f)\bigl(\getC(\overline{u}),\; \rho(\getQ(\overline{u})),\; \rho(x),\; 0^{n_1},\; \overline{q}\bigr).
	    \]
	    However, as mentioned in the previous subsection, \(y.\flag=2\) for some variable \(y\) implies that \(y\) is not cleaned within the execution of the recursive function to which it belongs. Consequently, if there exists a variable in \(\SCC(f)\) with \(\flag=2\), meaning the function contains uncleaned temporary variables that persist after the execution of \(\Theta(f)\) (see Section~\ref{sec:clean_temp} for details), we allocate an additional quantum variable \(r\) to receive the return value of \(\Theta(f)\). We then copy \(r\) to \(\rho(x)\) (see Appendix~\ref{sec_com_exp_assgn} for details of $\mathit{copy}$) and apply the inverse of \(\Theta(f)\), denoted \(\qinv(\Theta(f))\), to uncompute all uncleaned temporaries and \(r\) itself. After the call returns, the internal auxiliary quantum variables \(\overline{q}\) have been restored to the \(\ket{0}\) state and can be released back to \(\Qpool\) for reuse. 

  % For example, in Figure~\ref{fig:sum_target}, the main function calls \(F_{\mathtt{sum}}\) at line~12, allocating temporary variables \(q\), \(q_1\), \(q_2\) and initializing the recursion index \(i\) to \(0\). Since \(F_{\mathtt{sum}}\) contains an uncleaned temporary variable \(q_2\), the return value is temporarily stored in a temporary register \(p'\), then copied to the final result register \(r\), and finally the inverse of \(F_{\mathtt{sum}}\) is executed to clean all temporary variables.

  \textbf{Case 2: Recursive call (\(f \in \Fun(\St)\))}.  
    In this case, \(f\) is still under compilation but unfinished, indicating a recursive call. The compilation of recursive calls is illustrated by Algorithm~\ref{com_re} (explained below).  

 \textbf{Case 3: \(f\) is not yet compiled (\(f \notin \Fun(\St)\) and \(f \notin \Theta^{-1}(\dom(\D))\))}.  
	    We first need to compile function \(f\) by invoking \(\textsc{Compile\_Fun}\) (Algorithm~\ref{com_fun}) with parameters \(\rho(\getQ(\overline{u}))\) and \(\rho(x)\), which indicates that \(f\)'s \(\mathtt{Q}\)-mode formal parameters and its return variables may be mapped to \(\rho(\getQ(\overline{u}))\) and \(\rho(x)\). Note that this invocation is recursive: \(\textsc{Compile\_Fun}\) depends on \(\textsc{Compile\_Body}\), which in turn depends on the current algorithm. However, this recursion is well-founded because we can define the following measure:
\[
\mu(\ctx) \triangleq \bigl| \dom(\Xi) \setminus \bigl( \Theta^{-1}(\dom(\ctx.\D)) \cup \ctx.\St \bigr) \bigr|,
\]
i.e., the number of uncompiled source functions. Before calling \(\textsc{Compile\_Body}\) to compile \(f\)'s body, \(\textsc{Compile\_Fun}\) pushes \(f\) onto \(\St\). When a recursive call occurs, the measure \(\mu\) strictly decreases, thereby ensuring termination. 

The compilation proceeds in two cases: 
    
\begin{itemize}
\item 
If \(f \in \SCC(f_{\mathrm{cur}})\), then any call to \(f\) may eventually recurse back to \(f_{\mathrm{cur}}\) or earlier functions on the stack. To handle this correctly, we prepare the compilation state in advance, similarly to Case~2: before compiling \(f\), we apply the variable-name update from Step~1 and Step~2 of Algorithm~\ref{com_re} (i.e., add the index parameter, convert uncleaned variables to subscripted form and update $\iota$). This ensures that variables of suspended functions in the group are already subscripted when a recursive call occurs. 

After \(\textsc{Compile\_Fun}\) completes, we directly insert the call to \(\Theta(f)\). Here no allocation of temporaries (as in Case~1) or index increment (as in Case~2) is needed, since during \(f\)'s compilation its internal temporaries are allocated fresh from \(\Qpool\) and are guaranteed to be initially \(\ket{0}\); thus they are ready for immediate use in the call. Finally, as in Algorithm~\ref{com_re}, if \((x.\flag=2)\), indicating that uncleaned ancillae exist when returning from the next layer of \(\Theta(f)\), we push them, denoted \(\Gar_f\), into \(\Vactive\). 

For example, when compiling \(\mathtt{sum}\), the compiler encounters a call to \(\mathtt{aux}\) at a point where \(\mathtt{aux}\) has not yet been compiled. Since \(\mathtt{aux}\) belongs to the same mutual-recursion group as \(\mathtt{sum}\), the parameter list of \(\mathtt{sum}\) is first updated to \(\{i\}\), and the currently uncleaned ancillary variable \(q\) in \(\mathtt{sum}\) is renamed to \(q[i]\). The index variable \(\iota\) is then updated to a fresh index variable \(j\). After \(\mathtt{aux}\) has been compiled and returned to \(\mathtt{sum}\), the call to \(F_\mathtt{aux}\) can be inserted directly, i.e., \lstinline[language=MyLang]{Faux(p,r,i,j) // x <- aux(k)}, where the first and second arguments are the input and output, respectively, and the remaining arguments are the \(\Eargs\) of \(\mathtt{aux}\) at the current compilation time; the parameters \(q, q_1, q_2\) will be added later during parameter expansion.  

    \item Otherwise, we first call \(\textsc{Compile\_Fun}\) to compile \(f\); once \(f\) is fully compiled, we re-invoke \(\textsc{Compile\_Com}\) (Algorithm~\ref{Com_state}) to process the current call statement \(s\).

    In Figure~\ref{fig:sum_target}, when compiling the $\mathtt{main}$ function, a call to $\mathtt{sum}$ is encountered (line~13) before $\mathtt{sum}$ has been compiled. Since $\mathtt{main}$ and $\mathtt{sum}$ are not mutually recursive, no special handling is required. Once $\mathtt{sum}$ is fully compiled—meaning all functions it depends on have also been compiled—its parameter list is already expanded, and the compilation proceeds as in Case~1.
\end{itemize}

\begin{algorithm}[!tbp]
\caption{\textsc{Compile\_Recur}: Compiling Recursive Call Statements}
\label{com_re} 
\KwIn{
    Static input parameters $\K$,
    compilation context $\ctx$,
    recursive call statement $\tau x \leftarrow f(\overline{u})$
}
\KwOut{Updated compilation context $\ctx$ and generated circuit $C_s$ for the recursive call}
\SetKwProg{Fn}{Function}{}{}
\Fn{\textsc{Compile\_Recur}($\K, \ctx, \tau x \leftarrow f(\overline{u})$)}{
    \ForAll{$h \in \Fun(\St)$ with $h \in \SCC(f_{\mathrm{cur}})$}{
        $\ctx \gets \ctx[E_{\args,h}  \mapsto E_{\args,h} \mathbin{+\!+} [\iota]]$ \tcp*{add $\iota$ to $E_{\args,h}$}
    }
    $S \gets \{ q\in V_{\activelabel,f_{\mathrm{cur}}} \setminus \rho_{\mathrm{loc}}(\{x\} \cap \re(f_{\mathrm{cur}})) \mid \base[q] = q\}$\;   
    \ForAll{$h \in \Fun(\St)$ with $h \in \SCC(f_{\mathrm{cur}})$}{
            $\ctx \gets \ctx[(C_h,V_{\activelabel,h},\rho_{\mathrm{loc},h},\kappa_h) \mapsto (C_h,V_{\activelabel,h},\rho_{\mathrm{loc},h},\kappa_h)[r \mapsto r[\iota]]]$ for all $r \in S$\;}  
    Update \(\iota\) to the next index in alphabetical order\;
    $c \gets \{\var(\idx[q]) \mid q \in V_{\activelabel,\succeq f}\}$ \tcp*{$\idx[q]$: the index of $q$ ($0$ if no subscript)}
    $C_s \gets \Theta(f)(\getC(\overline{u}), \rho(\getQ(\overline{u})),\rho(x), E_{\args,f}[c \mapsto c+1])$\;  
    $\ctx \gets \ctx[C \mapsto C;C_s]$ \tcp*{append generated circuit $C_s$ to code $C$}
    \uIf{$x.\flag=2$}{  $\ctx \gets \ctx[\Vactive \mapsto  \Vactive \cup \Gar_f]$ \tcp*{add uncleaned variables to $\Vactive$} }
    \Return $(\ctx, C_s)$
}
\end{algorithm} 
Algorithm~\ref{com_re} (\textsc{Compile\_Recur}) presents the compilation process for a recursive statement. The core idea is to first ensure that all used but uncleaned variables of the current layer of \(f\) are already in subscripted form, thereby enabling us to isolate the quantum variables of different layers via distinct indices. We explain the steps in more detail below. 

\begin{enumerate}
    \item First, we add \(\iota\) as an extra formal parameter to the parameter lists of all functions in the mutual-recursion group \(\SCC(f_{\mathrm{cur}})\) and use this index variable to
  transform the relevant variables into subscripted quantum variables. Concretely, the uncleaned variables in the current layer of \(f\) at this point are the union of the
  \(\Vactive\) fragments after \(f\) on \(\St\) and the current buffer's \(\Vactive\), denoted \(V_{\activelabel,\succeq f}\). Since the parts that do not belong to the current function are
  already in subscripted form (by case~3 of Algorithm~\ref{Com_state}, which ensures that the suspended parts of the same mutual-recursion group have been renamed), we only need to replace
  each non-subscripted quantum variable \(q\) in the current function's \(\Vactive\) with the subscripted element \(q[\iota]\), with the sole exception that the variable \(x\)
  is left unchanged if and only if \(x\) is the return value of the current function. This renaming is propagated through the compilation state of every function in
  \(\SCC(f_{\mathrm{cur}})\), so that all occurrences of the affected variables in \(C\), \(\Vactive\), \(\rho_{\mathrm{loc}}\), and \(\kappa\) are replaced consistently.  
  % We assume that initially every variable in \(\xi\) is a non‑subscripted quantum variable, regarded as its subscripted form \(q[0]\).  
\item Additionally, update \(\iota\) to a fresh index variable for the next recursive call. 
\item Finally, generate a call to \(\Theta(f)\). To ensure that all ancillae used by the next layer are clean, every index variable appearing in \(V_{\activelabel,\succeq f}\) is incremented in the call parameters (e.g., \(j\) becomes \(j+1\)). Moreover, if \((x.\flag=2)\), indicating that uncleaned ancillae exist when returning from the next layer of \(\Theta(f)\), we push them, denoted \(\Gar_f\), into \(\Vactive\). 
  \end{enumerate}    

% Recall the running example in Figure~\ref{fig:sum_source} and Figure~\ref{fig:sum_target}. As discussed earlier, during the initialization phase of compiling \(\mathtt{f}\), its parameter list is initialized to \(\{i\}\), inherited from that of \(\mathtt{sum}\). When compiling the body of \(\mathtt{f}\), the compiler encounters a recursive call to \(\mathtt{sum}\), while \(\mathtt{sum}\) has already been compiled but remains suspended. Hence, the current index \(j\) from \(\IdxS\) is added to the parameter lists of both functions. At this point, both functions have the parameter list \(\{i,j\}\). The uncleaned ancillae of the current function \(q_1\) and \(q_2\), where \(q_2\) corresponds to the non-return variable \(y\), are then replaced by their indexed forms \(q_1[j]\) and \(q_2[j]\). In the generated call to the target function, all indices of quantum variables used in the current layer of \(\mathtt{sum}\), namely \((i,j)\) in \(q[i], q_1[j],q_2[j]\), are incremented by one: \lstinline[language=MyLang]{F_sum(q1[j],q2[j],red(i+1),red(j+1))  // y <- sum(y_1)}. 
% Since this is linear recursion, the quantum variables corresponding to \(y\) in the next and deeper layers remain uncleaned after the call returns; therefore, \(Gar_{\mathtt{sum}} = q_2[j+1:]\) is added to \(ucv\). 

In summary, using the stack order of \(\Vactive\) and function positions, we determine exactly which active variables of the current layer need to be converted to subscripted form with incremented indices. This avoids both inter-layer conflicts and the overhead of subscripting or incrementing redundant variables.

\subsection{Details for the Compilation of Bodies and Cleanup of Temporary Variables} 
\label{app:body} 

\begin{algorithm}[!tbp]
\caption{\textsc{Compile\_Body}: Compiling Function Bodies}
\label{com_body}
\KwIn{Static input parameter $\K$, compilation context $\ctx$, statement $s$}
\KwOut{Updated compilation context $\ctx$}
\SetKwProg{Fn}{Function}{}{}
\Fn{\textsc{Compile\_Body}($\K, \ctx, s$)}{
    \uIf{$s = \tau x \leftarrow e$ \textbf{or} $s = l \leftarrow \mathtt{aop}\ e$ \textbf{or} $s = \tau x \leftarrow f(\overline{u})$ \textbf{or} $s = \mathtt{if}\ x\ \{s_1\}$}{
        $(\ctx, C_s) \gets \textsc{Compile\_Com}(\K, \ctx, s)$\; 
        \If{$\neg (\mvq(s).\flag = 2 \wedge \Call(s) \neq \emptyset)$}{
      \If{$\mvq(s) \not\subseteq \dom(\rho_{\mathrm{glob}})$}{
    $\kappa \gets \kappa[\node(\mvq(s)) \mapsto (\kappa(\node(\mvq(s))); C_s)]$}
    \ForAll{$u  \in \dom(\kappa) \mid \mvq(s) \in \mathrm{pred}^+(u)$}{$\kappa  \gets \kappa[u \mapsto (\qinv(C_s); \kappa(u); C_s)]$;}
} 
        \ForAll{$w \mid w = \node(x),\ x \in \rvq(s)$}{
            $\delta \gets \delta[w \mapsto \delta(w) - 1]$\;
            \If{$\delta(w) = 0 \land w.\flag = 0 \land \var(w) \cap (\dom(\rho_{\mathrm{glob}}) \cup \re(f_{\mathrm{cur}})) =\emptyset$}{
                $\ctx \gets \textsc{Clean\_Temp}(\K, \ctx, w)$\;
            } 
        } 
    }
    \ElseIf{$s = s_1; s_2$}{
        $\ctx \gets \textsc{Compile\_Body}(\K, \ctx, s_1)$\;
        $\ctx \gets \textsc{Compile\_Body}(\K, \ctx, s_2)$\;
    }
  \Return $\ctx$ 
}
\end{algorithm} 

% \begin{algorithm}[!tbp]
% \caption{Compile\_body}
% \label{com_body_app}
% \KwIn{Static input parameter $\K$, compilation context $\ctx$, statement $s$}
% \KwOut{Updated compilation context $\ctx$}
% \SetKwProg{Fn}{Function}{}{}
% \Fn{\textsc{Compile\_body}($\K, \ctx, s$)}{
%       $\cdots$\; 
%         updtate $\kappa$ for related node: \\ 
%         \If{$\neg (\mvq(s).\flag = 2 \wedge \Call(s) \neq \emptyset)$}{
%       \If{$\mvq(s) \not\subseteq \dom(\rho_{\mathrm{glob}})$}{
%     $\kappa \gets \kappa[\node(\mvq(s)) \mapsto (\kappa(\node(\mvq(s))); C_s)]$}
%     \ForAll{$w  \in \dom(\kappa) \mid \mvq(s) \in \mathrm{pred}^+(w)$}{$\kappa  \gets \kappa[w \mapsto (\qinv(C_s); \kappa(w); C_s)]$;}
% } 
% $\cdots$
% }
% \end{algorithm}

The algorithm for compiling function bodies is presented in Algorithm~\ref{com_body}. In Section~\ref{sec:com_body}, we deferred a detailed explanation of how $\kappa$ is updated during the compilation of function bodies. We now fill in these missing details. 

If the statement is not a linear recursive call — i.e., it is not the case that \(\mvq(s).\flag = 2\) and \(s\) contains a function call — we update the corresponding cleanup circuits as follows. First, note that \(\mvq(s)\) contains exactly one element by definition. If that element is not a global variable (global variables do not need cleanup, and therefore no update is required), then the generated circuit is appended to the cleanup circuit of the node containing \(\mvq(s)\) (line~6). This ensures that cleanup circuits are constructed strictly in the order of computation. Since the quantum values associated with nodes are generated sequentially by their computation circuits, executing the inverse of this cleanup circuit (i.e., in reverse computation order) during uncomputation correctly cleans all variables with \(\flag \neq 2\).  

Moreover, if quantum variables within a cleanup circuit are later modified or cleaned, directly performing uncomputation may fail to restore the initial state. Therefore, we require that whenever a quantum variable corresponding to a variable (including global variables, since other systems may depend on the value of that global variable) is modified or cleaned, the circuit implementing this operation must be updated in the cleanup circuits of all other nodes \(u\) satisfying \(\mvq(s) \in \mathrm{pred}^+(u)\) (line~8).

Note that if the statement is a linear recursive call, the source type system guarantees that \(\mvq(s)\) is written for the first time, meaning that no other variable depends on the value of \(\mvq(s)\) before that write. Consequently, for any variable \(u\) that depends on \(\mvq(s)\) (i.e., \(\mvq(s) \in \mathrm{pred}^*(u)\), where \(\mathrm{pred}^*(u)\) denotes the set of predecessors of \(u\) in the reflexive transitive closure, \(\mathrm{pred}^*(u) = \mathrm{pred}^+(u) \cup \{u\}\)), and that uses \(\rho_{\mathrm{loc}}(\mvq(s))\), it merely employs this value as an auxiliary variable. In this case, we can always replace this auxiliary variable with another quantum variable from \(\Qpool\), or alternatively allocate dedicated quantum variables for variables with \(\flag = 2\) at the beginning of compilation to prevent them from being used as auxiliary variables in other computations. Therefore, even if \(\kappa(u)\) and \(\rho_{\mathrm{loc}}(\mvq(s))\) intersect, we can always eliminate the intersection by the methods described above without altering the semantics; hence, without loss of generality, we may assume they are disjoint.

Through this mechanism, we ensure both the correctness of the cleanup circuit for each node and that no linear recursive statement appears in any node's cleanup circuit. Furthermore, variables with \(\flag = 2\) employ a deferred cleanup strategy — they are never cleaned within the current function — so their cleanup does not rely on the cleanup circuit of the node. In summary, the above update operations are sufficient. Consequently, given that all temporary variables are clean, the cleanup circuit of every node correctly cleans all quantum variables corresponding to variables with \(\flag \neq 2\), without involving the computation of linear recursive statements.
 
% \begin{remark}
%   The above algorithm presents a basic update framework; in practice, further optimisation can reduce redundant computations. 
  
%   For any node \(v\) that depends on \(\node(\mvq(s))\), it suffices to ensure that the variables in \(\kappa(v)\) are disjoint from \(\rho_{\mathrm{loc}}(\mvq(s))\). In line~7 of the algorithm, the condition \(\mvq(s) \in \mathrm{pred}^*(v)\) may be too strong: if some node on the path between \(\node(\mvq(s))\) and \(w\) has not yet been cleaned, then \(w\) does not directly depend on \(\node(\mvq(s))\) at the current moment, which already guarantees \(\qv(\kappa(v)) \cap \rho_{\mathrm{loc}}(\mvq(s)) = \emptyset\). Therefore, we only need to update those nodes \(v\) satisfying \(\mvq(s) \notin \mathrm{pred}^*(v)\) and for which all nodes on the path between \(\node(\mvq(s))\) and \(w\) have already been cleaned.  
  
%   On the other hand, if \(\mvq(s)\) is not in \(\dom(\rho)\) of the input compilation context (e.g., in the linear recursive call case), we can simply replace every occurrence of \(\rho_{\mathrm{loc}}(\mvq(s))\) and may even avoid updating \(C_s\) into the corresponding cleanup circuits.
% \end{remark}
% \subsection{}  
% \label{app:clean_temp}

Algorithm~\ref{clean_temp} details the cleanup process for a given node $w$.  Here $\req(\kappa(w)) \triangleq \qv(\kappa(w)) \setminus \rho(\operatorname{pred}^*(w))$, with $\qv(\kappa(w))$ the set of variables in $\kappa(w)$ and $\rho = \rho_{\mathrm{glob}} \cup \rho_{\mathrm{loc}}$. Since most of the explanation has already been provided in the main text, we do not repeat it here. 

\begin{algorithm}[!tbp]
\caption{\textsc{Clean\_Temp}: Cleaning Up Temporary Variables}
\label{clean_temp}
\KwIn{Static input parameter $\K$, compilation context $\ctx$, node $w$ to be cleaned}
\KwOut{Updated compilation context $\ctx$}
\SetKwProg{Fn}{Function}{}{}
\Fn{\textsc{Clean\_Temp}($\K, \ctx, w$)}{
    \If{$w.\flag=0 \wedge \exists x\in \var(w), x.\flag \neq 2$}{
        $\overline{p} \gets \pop(\Qpool, |\req(\kappa(w))|)$; \ $\Valloc \gets \Valloc \cup \base(\overline{p})$\; 
        $C_{\mathrm{cl}} \gets \qinv(\kappa(w)[\req(\kappa(w)) \mapsto \overline{p}])$; $\ctx \gets \ctx[C\mapsto C; C_{\mathrm{cl}}]$ \tcp*{clean circuit}
        $\push(\overline{p}, \Qpool)$\;
        \ForAll{$y \in \var(w) \mid \rho_{\mathrm{loc}}(y) \in V_{\activelabel,f_{\mathrm{cur}}} \land y.\flag \neq 2$}{
               \(\rho_{\mathrm{loc},\succeq f_{\mathrm{cur}}} \gets \rho_{\mathrm{loc},\succeq f_{\mathrm{cur}}} \setminus \{(-,\rho_{\mathrm{loc}}(y))\}\); \ $V_{\activelabel,\succeq f_{\mathrm{cur}}}\gets (V_{\activelabel,\succeq f_{\mathrm{cur}}} \setminus \rho_{\mathrm{loc}}(y))$; \ $\push(\rho_{\mathrm{loc}}(y), \Qpool)$\; 
                }
        \ForAll{$u  \in \dom(\kappa) \mid w \in \mathrm{pred}^+(u)$}{
            $\kappa \gets \kappa[u \mapsto (\qinv(C_{\mathrm{cl}});\kappa(u); C_{\mathrm{cl}})]$\;
        }
    }
    \ElseIf{$w.\flag=1$}{
        $C_{\mathrm{cl}} \gets \qinv(\kappa(w))$; $\ctx \gets \ctx[C \mapsto C; C_{\mathrm{cl}}]$\;
     \ForAll{$y \in \var(w) \mid \rho_{\mathrm{loc}}(y) \in V_{\activelabel,f_{\mathrm{cur}}}$}{ 
     \(\rho_{\mathrm{loc},\succeq f_{\mathrm{cur}}} \gets \rho_{\mathrm{loc},\succeq f_{\mathrm{cur}}} \setminus \{(-,\rho_{\mathrm{loc}}(y))\}\); \ $V_{\activelabel,\succeq f_{\mathrm{cur}}}\gets (V_{\activelabel,\succeq f_{\mathrm{cur}}} \setminus \rho_{\mathrm{loc}}(y))$; \ $\push(\rho_{\mathrm{loc}}(y), \Qpool)$\; } 
    }
    \Return $\ctx$
}
\end{algorithm}  
 
\subsection{Expression and Assignment Statement Compilation}
% \label{secA_Compe}
\label{sec_com_exp_assgn}
We now introduce the compilation judgment for expressions. For an expression \(e\) that is to be evaluated and whose result is to be stored in a target quantum variable \(p\), we write:
\begin{equation}
\Gamma; \rho; \Qpool \vdash (e, p) \leadsto u \dashv \Qpool',
\end{equation}
where \(\Gamma\) is the type environment, \(\rho\) maps source variables to their corresponding quantum variables, \(\Qpool\) is the pool of available quantum registers, and \(u\) is the generated RQC\(^{++}\) circuit that implements the evaluation of \(e\) on the system \(p\). The judgment indicates that under the given context, the expression \(e\) compiles to circuit \(u\), and the register pool is updated to \(\Qpool'\). The detailed rules for expression compilation are provided in Tables~\ref{tab:expr-compile} and~\ref{tab:expr-compile-bin}.  

The operators $\mathit{init\_c}$, $\mathit{copy}$, $\mathit{locate}$, and others appearing in Table~\ref{tab:expr-compile} are realized as concrete $\RQC^{++}$ programs, deferred to the end of this section.  For array access $x[v]$ with a quantum index $v$, we first compute $v \times |\tau|$ to obtain the starting bit position of the element and store the result in $q$, where $|\tau|$ denotes the bit width of type $\tau$ (e.g., $|\mathtt{nat}| = \sz$, $|\mathtt{bool}| = 1$). The $\mathit{locate}$ operation then uses the value in quantum register $q[0:\mathit{sz}-1]$ to address into the target array $\rho(x)[0:2^{\sz}-1]$. If the resolved position is at most $n \times |\tau| - 1$ (the maximum valid index of array $x$), it extracts the $|\tau|$ bits beginning at that position and assigns them to register $p$; otherwise, $p$ retains its initial value. 

The construct $\mathrm{get\_uaop}(\mathtt{aop}, \ast)$ yields the $\RQC^{++}$ program implementing the unary operation $\mathtt{aop}$, also deferred to the end of this section; the annotation $\ast \in \{\mathtt{C}, \mathtt{Q}\}$ distinguishes classical from quantum operands, as the implementation differs by mode. For instance, $\mathrm{get\_uaop}(+, \mathtt{Q})(q, p) = \mathit{rz\_adder\_form}(q, p)$ adds the value in system $q$ into system $p$. Similarly, $\mathrm{get\_binaop}(\mathtt{aop}, \ast, \ast)$ produces the binary variant; e.g., $\mathrm{get\_binaop}(+, \mathtt{Q}, \mathtt{Q})(p, q, r) = \mathit{rz\_adder\_full\_form}(p, q, r)$ adds the values in systems $p$ and $q$, placing the result in system $r$. The construct $\mathrm{get\_bop}(\mathtt{bop}, \ast ,\ast)$ is defined analogously for Boolean operations. 

For expressions of the form $\mathtt{aop}\ e$, we give three cases: $e$ a classical variable, $e$ a quantum variable, and $e$ a compound expression. For $e_1\ \mathtt{aop}\ e_2$ and $e_1\ \mathtt{bop}\ e_2$, we present only the case where $e$ is compound; the cases for classical and quantum variables follow the same pattern and are elided.

\begin{table}[!tbp]
\centering
\caption{Expression compilation rules}
\label{tab:expr-compile}
\[
\begin{array}{c|c}
\hline
\mathrm{Exp} & \mathrm{Compilation} \\[2ex]
\hline
\\
& \displaystyle \dfrac{}{\Gamma; \rho; \Qpool \vdash ((), p) \leadsto \mathbf{skip} \dashv \Qpool} 
\qquad
\displaystyle \dfrac{\Gamma \vdash n:\mathtt{nat}^{\mathtt{C}}}{\Gamma; \rho; \Qpool \vdash (n, p) \leadsto \mathrm{init\_c}(n,p,\sz-1) \dashv \Qpool} \\[2ex]
v & 
\displaystyle \dfrac{\Gamma \vdash x:\tau \quad M(\tau)=\mathtt{Q}}{\Gamma; \rho; \Qpool \vdash (x, p) \leadsto \mathit{copy}(\rho(x), 0, \sz-1, p, 0, \sz-1) \dashv \Qpool} 
\\ 
& 
\displaystyle \dfrac{\Gamma \vdash x:\tau \quad M(\tau)=\mathtt{C}}{\Gamma; \rho; \Qpool \vdash (x, p) \leadsto \mathit{init\_c}(x,p,\sz-1) \dashv \Qpool} \\[2ex]
& 
\displaystyle \Gamma \vdash x:\mathtt{array} \ n \ \tau \quad M(\tau)=\mathtt{Q} \quad \Gamma \vdash v:\mathtt{nat}^\mathtt{Q} \quad (q, \Qpool_0) = \pop(\Qpool)\\  
& 
\displaystyle \dfrac{
    \Gamma; \rho; \Qpool_0 \vdash (v \times \vert\tau\vert, q) \leadsto u_1 \dashv \Qpool_1 \quad 
   u= \mathit{locate}(q,0,\sz-1,\rho(x),0,2^{\sz}-1,p, \vert\tau\vert,n \cdot \vert\tau\vert-1)
   }{
   \Gamma; \rho; \Qpool \vdash (x[v], p) \leadsto u_1; u; \qinv(u_1) \dashv (\push(q, \Qpool_1))
   } \\[2ex]
& 
\displaystyle \dfrac{
   \Gamma \vdash x:\mathtt{array} \ n \ \tau \quad M(\tau)=\mathtt{Q} \quad \Gamma \vdash v:\mathtt{nat}^\mathtt{C} \quad 
   }{
   \Gamma; \rho; \Qpool \vdash (x[v], p) \leadsto \mathit{copy}(\rho(x), v \cdot \vert\tau\vert, (v+1)\cdot \vert\tau\vert-1, p, 0, \vert\tau\vert-1) \dashv \Qpool
   } \\[2ex]
& 
\displaystyle \dfrac{ x=(x_1,x_2) \quad
   \Gamma; \rho; \Qpool \vdash (x_1, p) \leadsto u \dashv \Qpool'
   }{
   \Gamma; \rho; \Qpool \vdash (\pi_1(x), p) \leadsto u \dashv \Qpool'
   }
\qquad
\displaystyle \dfrac{ x=(x_1,x_2) \quad
   \Gamma; \rho; \Qpool \vdash (x_2, p) \leadsto u \dashv \Qpool'
   }{
   \Gamma; \rho; \Qpool \vdash (\pi_2(x), p) \leadsto u \dashv \Qpool'
   } \\[2ex]
& 
\displaystyle \dfrac{
   \Gamma; \rho; \Qpool \vdash (v_1, p_1) \leadsto u_1 \dashv \Qpool_1
   \quad
   \Gamma; \rho; \Qpool_1 \vdash (v_2, p_2) \leadsto u_2 \dashv \Qpool_2
   }{
   \Gamma; \rho; \Qpool \vdash ((v_1, v_2), p_1 p_2) \leadsto u_1; u_2 \dashv \Qpool_2
   } \\[2ex]
& 
\displaystyle \dfrac{}{\Gamma; \rho; \Qpool \vdash (\mathbf{true}, p) \leadsto X[p] \dashv \Qpool}
\qquad
\displaystyle \dfrac{}{\Gamma; \rho; \Qpool \vdash (\mathbf{false}, p) \leadsto \mathbf{skip} \dashv \Qpool} \\[2ex]
\\
\hline
\\
 & 
\displaystyle \dfrac{\Gamma \vdash e:\mathtt{nat}^{\mathtt{C}}}{
   \Gamma; \rho; \Qpool \vdash (\mathtt{aop} \ e, p) \leadsto \mathrm{get\_uaop}(\mathtt{aop},\mathtt{C})(a, p) \dashv \Qpool
   } \\[2ex]
\mathtt{aop} \ e & 
\displaystyle \dfrac{ \Gamma \vdash x:\mathtt{nat}^{\mathtt{Q}}  }{
   \Gamma; \rho; \Qpool \vdash (\mathtt{aop} \ x, p) \leadsto \mathrm{get\_uaop}(\mathtt{aop},\mathtt{Q})(\rho(x), p) \dashv \Qpool 
   } \\[2ex]
& 
\displaystyle \Gamma \vdash e:\mathtt{nat}^{\mathtt{Q}} \\
& 
\displaystyle \dfrac{
    (r,\Qpool_0) = \pop(\Qpool) 
   \quad \Gamma; \rho; \Qpool_0 \vdash (e, r) \leadsto u_1 \dashv \Qpool_1
   \quad u_2 = \mathrm{get\_uaop}(\mathtt{aop},\mathtt{Q})(r, p)
   }{
   \Gamma; \rho; \Qpool \vdash (\mathtt{aop} \ e, p) \leadsto u_1; u_2; \qinv(u_1) \dashv \push(r, \Qpool_1)
   } \\
   \\ 
\hline
\end{array}
\]
\end{table}

\begin{table}[!tbp]
\centering
\caption{Expression compilation rules (binary operations)}
\label{tab:expr-compile-bin}
\[
\begin{array}{c|c}
\hline
\mathrm{Exp} & \mathrm{Compilation} \\[2ex]
\hline
\\ 
&  \Gamma \vdash e_1:\mathtt{nat}^{\mathtt{Q}} \quad \Gamma \vdash e_2:\mathtt{nat}^{\mathtt{Q}} \quad (p_1, \Qpool_1) = \pop(\Qpool)  \quad (p_2,\Qpool_2) = \pop(\Qpool_1) \\ 
e_1 \ \mathtt{aop} \ e_2 
& \Gamma; \rho; \Qpool_2 \vdash (e_1, p_1) \leadsto u_1 \dashv \Qpool_3 \quad 
  \Gamma; \rho; \Qpool_3 \vdash (e_2, p_2) \leadsto u_2 \dashv \Qpool_4 \\
& \dfrac{ u = \mathrm{get\_binaop}(\mathtt{aop},\mathtt{Q},\mathtt{Q})(p_1, p_2, p)}
      { \Gamma; \rho; \Qpool \vdash (e_1 \ \mathtt{aop} \ e_2, p) 
        \leadsto (u_1; u_2; u; \qinv(u_1; u_2)) 
        \dashv (\push(p_1,\push(p_2,\Qpool_4)))} \\
        \\ 
\hline
\\ 
&  \Gamma \vdash e_1:\mathtt{nat}^{\mathtt{Q}} \quad \Gamma \vdash e_2:\mathtt{nat}^{\mathtt{Q}}  \quad  (p_1, \Qpool_1) = \pop(\Qpool)  \quad (p_2,\Qpool_2) = \pop(\Qpool_1)  \\ 
e_1 \ \mathtt{bop} \ e_2  
&  \Gamma; \rho; \Qpool_2 \vdash (e_1, p_1) \leadsto u_1 \dashv \Qpool_3 \quad 
  \Gamma; \rho; \Qpool_3 \vdash (e_2, p_2) \leadsto u_2 \dashv \Qpool_4 \\ 
& \dfrac{ u = \mathrm{get\_bop}(\mathtt{bop},\mathtt{Q},\mathtt{Q})(p_1, p_2, p) }
      { \Gamma; \rho; \Qpool \vdash (e_1 \ \mathtt{bop} \ e_2, p) 
        \leadsto (u_1; u_2; u; \qinv(u_1; u_2)) 
        \dashv (\push(p_1,\push(p_2,\Qpool_4))) } \\
        \\ 
\hline
\\ 
\neg e & \displaystyle \dfrac{ \Gamma \vdash e:\mathtt{bool}^{\mathtt{Q}} \quad 
                  \Gamma; \rho; \Qpool \vdash (e, r) \leadsto u \dashv \Qpool' }
                { \Gamma; \rho; \Qpool \vdash (\neg e, r) \leadsto (u; X[r]) \dashv \Qpool' } \\[2ex]
                \\ 
\hline
\\ 
& \displaystyle \Gamma \vdash e_1:\mathtt{bool}^{\mathtt{Q}} \quad \Gamma \vdash e_2:\mathtt{bool}^{\mathtt{Q}} \quad (p_1, \Qpool_1) = \pop(\Qpool)  \quad (p_2,\Qpool_2) = \pop(\Qpool_1)  \\
e_1 \wedge e_2  
& \displaystyle \Gamma; \rho; \Qpool_2 \vdash (e_1, p_1) \leadsto u_1 \dashv \Qpool_3 \quad 
  \Gamma; \rho; \Qpool_3 \vdash (e_2, p_2) \leadsto u_2 \dashv \Qpool_4 \\
& \displaystyle \dfrac{ u = \mathbf{qif}[p_1](\ket{0} \rightarrow \mathbf{skip} \;\Box\; \ket{1} \rightarrow 
            \mathbf{qif}[p_2](\ket{0} \rightarrow \mathbf{skip} \;\Box\; \ket{1} \rightarrow X[p]) \mathbf{fiq}) \mathbf{fiq} }
      { \Gamma; \rho; \Qpool \vdash (e_1 \wedge e_2, p) 
        \leadsto (u_1; u_2; u; \qinv(u_1; u_2)) 
        \dashv (\push(p_1,\push(p_2,\Qpool_4)))} \\
        \\ 
\hline
\end{array}
\]
\end{table}

For assignment and unary-assignment statements, we adopt a similar formalism, captured by the judgment:
\begin{equation}
\Gamma; \rho; \Qpool \vdash s \leadsto u \dashv \Qpool',
\end{equation}
where \(s\) is the assignment statement being compiled, and \(u\) is the resulting RQC\(^{++}\) circuit. This judgment states that under the given type environment \(\Gamma\), variable mapping \(\rho\), and register pool \(\Qpool\), the statement \(s\) compiles to circuit \(u\) and updates the register pool to \(\Qpool'\).

This formalization underpins our compilation algorithms. For assignment operations, when both the declared variable \(\tau x\) and the expression \(e\) are in classical mode ($\mathtt{C}$‑mode), the rule directly yields a classical assignment \(x := e\) in the target language. If \(x\) is in quantum mode ($\mathtt{Q}$‑mode), the expression \(e\) is compiled into a quantum circuit \(u\) that operates on the quantum variable associated with \(x\); this is precisely captured by the auxiliary judgment introduced above: \(\Gamma; \rho; \Qpool \vdash (e, \rho(x)) \leadsto u \dashv \Qpool'\). Here \(\rho(x)\) gives the quantum variable assigned to \(x\). Unary assignment statements are handled similarly.

 \begin{gather*}
        \dfrac{M(\tau)=\mathtt{C} }{ \Gamma;\rho;\Qpool \vdash (\tau x\leftarrow e) \leadsto (x:=e) \dashv \Qpool} \quad \dfrac{M(\tau)=\mathtt{Q} \quad \Gamma; \rho; \Qpool \vdash (e, \rho(x)) \leadsto u \dashv \Qpool' }{\Gamma;\rho;\Qpool \vdash (\tau x \leftarrow e) \leadsto u \dashv \Qpool'} \\  
        \dfrac{\Gamma \vdash l :\tau \quad M(\tau)=\mathtt{C}}{\Gamma;\rho;\Qpool \vdash (l \leftarrow \mathtt{aop} \ e) \leadsto (l:=l \;\mathtt{aop}\; e) \dashv \Qpool} \quad \dfrac{\Gamma \vdash l :\tau \quad M(\tau)=\mathtt{Q} \quad \Gamma; \rho; \Qpool \vdash (\mathtt{aop} \ e, \rho(l)) \leadsto u \dashv \Qpool'}{ \Gamma;\rho;\Qpool \vdash (l \leftarrow \mathtt{aop} \ e) \leadsto u \dashv \Qpool'}  \\ 
    \end{gather*}

We now present the quantum circuits implementing arithmetic and Boolean operations, beginning with arithmetic operations.

\paragraph{Arithmetic operators.}
We use $\textbf{RQC}^{++}$ to implement a QFT-based adder. 
\[
    \begin{aligned}
        &\ket{a}_p \rightarrow \ket{a}_p \\
        &\ket{b}_q \rightarrow \ket{(a+b) \mathbin{\%} 2^{\sz}}_q
    \end{aligned}
\]
The transformation is shown as follows.
 The circuit employs this approach to perform the unary operation $b \leftarrow \mathtt{+} \ a$, where $a$ and $b$ are encoded into the quantum variables $p$ and $q$, respectively. The rotation gate $RZ(\theta)$ is 
 \[ \left[\begin{array}{cc}
      1 & 0 \\ 
      0 & e^{\frac{2\pi i}{2^{\theta}}}
 \end{array} \right].\]
We assume that the bits are indexed from the most significant bit to the least significant one as \(0\) through \(\sz-1\). \(\mathit{QFT}(q, 0, \mathit{sz}-1)\) performs the quantum Fourier transform on the qubits \(q[0:\mathit{sz}-1]\). 
\[
\begin{array}{ccc}
    \begin{aligned}
    &\mathit{rz\_adder\_form}(p,q) \Longleftarrow \\
    & \quad \mathit{QFT}(q,0, \mathit{sz}-1);\\
    & \quad \mathit{rz\_adder}(p,q,\mathit{sz}-1,\mathit{sz}-1); \\ 
    & \quad \mathit{RQFT}(q,0,\mathit{sz}-1);\\
\end{aligned}
&  \begin{aligned}
        &\mathit{rz\_adder}(p,q,n,m) \Longleftarrow \\ 
        & \quad  \mathbf{if} \ 0 \leq n \ \mathbf{and} \ 0 \leq m \ \mathbf{then} \\ 
        &  \qquad \mathit{rz\_adder}(p,q,n-1,m-1); \\
       &  \qquad \mathbf{qif}[p[n]] (\ket{0} \rightarrow \mathbf{skip}) \\ 
         &  \qquad  \qquad \qquad \square \ket{1} \rightarrow \mathit{rz\_adder'}(q,m,m); \\ 
         &   \qquad \mathbf{fiq} \\ 
        & \quad \mathbf{fi}
    \end{aligned}
    &
     \begin{aligned}
        &\mathit{rz\_adder'}(q,k,m)  \Longleftarrow \\ 
        & \quad \mathbf{if} \ 0 \leq k \ \mathbf{then} \\ 
         & \quad \quad RZ(m-k+1)[q[k]]; \\ 
         & \quad \quad \mathit{rz\_adder'}(q,k-1,m); \\ 
         & \quad \mathbf{fi}
    \end{aligned}
\end{array}
\] 

\[
\begin{array}{cc}
\begin{aligned}
    \mathit{QFT}(q,m,n) &\Leftarrow \\ 
     & \mathbf{if} \ m \leq n \ \mathbf{then} \\ 
     & \qquad  H[q[m]] \\ 
     & \qquad \mathit{rotate}(q,m,m+1,n); \\ 
     & \qquad \mathit{QFT}(q,m+1,n);\\ 
     & \mathbf{fi} 
\end{aligned} & 
\begin{aligned}
    \mathit{rotate}&(q,m,l,r) \Leftarrow \\ 
     & \mathbf{if} \ l \leq r \ \mathbf{then}  \\
     & \qquad \mathbf{qif}[q[l]](\ket{0}\rightarrow \mathbf{skip}) \\ 
     & \qquad \qquad \quad \square \ket{1} \rightarrow RZ(l-m+1)[q[m]] \\
     & \qquad \mathbf{fiq} \\
     & \qquad \mathit{rotate}(q,m,l+1,r); \\ 
     & \mathbf{fi} 
\end{aligned}
\end{array}
\]

If $a$ is classically known, it can be implemented in a simpler manner as follows, where the rotation gate $S(\theta)$ is 
 \[ \left[\begin{array}{cc}
      1 & 0 \\ 
      0 & e^{2\pi i \theta}
 \end{array} \right].\]
\[
    \begin{aligned}
        &\ket{b}_p \rightarrow \ket{(b+a) \mathbin{\%} 2^{\sz}}_p
        (\mbox{$a$ is classically known})
    \end{aligned}
\]
\[
\begin{array}{cc}
\begin{aligned}
    &\mathit{rz\_adder\_c\_form}(a,p) \Longleftarrow \\
    & \quad \mathit{QFT}(p,0,\sz-1);\\
    &\quad \mathit{rz\_adder\_c}(p,0,\sz-1,a); \\ 
    & \quad \mathit{RQFT}(p,0,\sz-1);\\
\end{aligned} \qquad &
    \begin{aligned}
        & \mathit{rz\_adder\_c}(p,m,n,a) \Longleftarrow \\ 
        & \quad \mathbf{if} \ m \leq n \ \mathbf{then} \\ 
         & \qquad S(a \cdot 2^{-(n-m+1)})[p[m]]; \\ 
         &\qquad \mathit{rz\_adder\_c}(p,m+1,n,a); \\ 
         & \quad \mathbf{fi}
    \end{aligned}
\end{array}
\]

The subtractors can be implemented analogously. In fact, $\mathit{rz\_suber}$ and $\mathit{rz\_suber\_c}$ are the inverses of $\mathit{rz\_adder}$ and $\mathit{rz\_adder\_c}$, respectively. 

\[
    \begin{aligned}
        &\ket{a}_p \rightarrow \ket{a}_p \\
        &\ket{b}_q \rightarrow \ket{(b-a) \mathbin{\%} 2^\sz}_q 
    \end{aligned}
\]

\[
\begin{array}{ccc}
\begin{aligned}
    & \mathit{rz\_suber\_form}(p,q) \Longleftarrow \\
    & \quad \mathit{QFT}(q,0,\sz-1);\\
    & \quad \mathit{rz\_suber}(p,q,\sz-1,
    \sz-1); \\ 
    & \quad  \mathit{RQFT}(q,0,\sz-1);\\
\end{aligned}
&
    \begin{aligned}
        & \mathit{rz\_suber}(p,q,n,m) \Longleftarrow \\ 
        & \quad \mathbf{if} \ 0 \leq n \ \mathbf{and} \ 0 \leq m \ \mathbf{then} \\ 
       & \qquad \mathbf{qif}[p[n]] (\ket{0} \rightarrow \mathbf{skip}) \\ 
         &\qquad  \qquad \quad \square \ket{1} \rightarrow \mathit{rz\_suber'}(q,m,m); \\ 
         & \qquad \mathbf{fiq} \\ 
        & \qquad \mathit{rz\_suber}(p,q,n-1,m-1); \\
        & \quad \mathbf{fi}
    \end{aligned}
&
    \begin{aligned}
        & \mathit{rz\_suber'}(q,k,m) \Longleftarrow \\ 
        & \quad \mathbf{if} \ 0 \leq k \ \mathbf{then} \\ 
         & \qquad \mathit{RZ}^{-1}(m-k+1)[q[k]]; \\ 
         &\qquad \mathit{rz\_suber'}(q,k-1,m); \\ 
         & \quad \mathbf{fi}
    \end{aligned}
\end{array}
\]

\[
    \begin{aligned}
        &\ket{b}_p \rightarrow \ket{(b-a) \mathbin{\%}2^{\sz}}_p
        (\mbox{$a$ is classically known})
    \end{aligned}
\]
\[
\begin{array}{cc}
\begin{aligned}
    & \mathit{rz\_suber\_c\_form}(a,p) \Longleftarrow \\
    & \quad \mathit{QFT}(p,0,\sz-1);\\
    & \quad \mathit{rz\_suber\_c}(p,0,\sz-1,a); \\ 
    &\quad \mathit{RQFT}(p,0,\sz-1);\\
\end{aligned}
&
    \begin{aligned}
        & \mathit{rz\_suber\_c}(p,m,n,a)  \Longleftarrow \\ 
        & \quad \mathbf{if} \ m \leq n \ \mathbf{then} \\ 
         & \qquad S^{-1}(a \cdot 2^{-(n-m+1)})[p[m]]; \\ 
         &\qquad \mathit{rz\_suber\_c}(p,m+1,n,a); \\ 
         &\quad \mathbf{fi}
    \end{aligned}
\end{array}
\]
Adders and subtractors for binary operations can be implemented in a straightforward manner as follows. Here Figures (a) and (b) implement quantum addition, where (a) operates on two quantum registers and (b) takes a classically known addend $a$. Figures (c) and (d) implement quantum subtraction, with (c) operating on two quantum registers and (d) taking a classically known subtrahend $a$.

\begin{minipage}[b]{0.48\textwidth}
\centering
\[
\begin{aligned}
    &\ket{a}_p \rightarrow \ket{a}_p\\
    &\ket{b}_q \rightarrow \ket{b}_q \\ 
    &\ket{0}_r \rightarrow \ket{(a+b) \mathbin{\%} 2^{\sz}}_r
\end{aligned}
\]
\[
\begin{aligned}
    &\mathit{rz\_adder\_full\_form}(p,q,r) \Longleftarrow \\
    & \quad \mathit{copy}(q,0,\sz-1,r,0,\sz-1);\\
    & \quad \mathit{rz\_adder\_form}(p,r)
\end{aligned}
\]
\centerline{(a) Both quantum}
\end{minipage}
\hfill
\begin{minipage}[b]{0.48\textwidth}
\centering
\[
\begin{aligned}
    &\ket{b}_q \rightarrow \ket{b}_q \\ 
    &\ket{0}_r \rightarrow \ket{(b+a) \mathbin{\%} 2^{\sz}}_r
\end{aligned}
\]
\[
\begin{aligned}
    &\mathit{rz\_adder\_c\_full\_form}(a,q,r) \Longleftarrow \\
   & \quad \mathit{copy}(q,0,\sz-1,r,0,\sz-1);\\
    & \quad \mathit{rz\_adder\_c\_form}(a,r)
\end{aligned}
\]
\centerline{(b) $a$ classical}
\end{minipage}

\noindent
\begin{minipage}[b]{0.48\textwidth}
\centering
\[
\begin{aligned}
    &\ket{a}_p \rightarrow \ket{a}_p\\
    &\ket{b}_q \rightarrow \ket{b}_q \\ 
    &\ket{0}_r \rightarrow \ket{(b-a) \mathbin{\%} 2^{\sz} }_r
\end{aligned}
\]
\[
\begin{aligned}
    &\mathit{rz\_suber\_full\_form}(p,q,r) \Longleftarrow \\
   & \quad \mathit{copy}(q,0,\sz-1,r,0,\sz-1);\\
    & \quad \mathit{rz\_suber\_form}(p,r)
\end{aligned}
\]
\centerline{(c) Both quantum}
\end{minipage}
\hfill
\begin{minipage}[b]{0.48\textwidth}
\centering
\[
\begin{aligned}
    &\ket{b}_q \rightarrow \ket{b}_q \\ 
    &\ket{0}_r \rightarrow \ket{(b-a) \mathbin{\%} 2^{\sz} }_r
\end{aligned}
\]
\[
\begin{aligned}
    &\mathit{rz\_suber\_c\_full\_form}(a,q,r) \Longleftarrow \\
    & \quad \mathit{copy}(q,0,\sz-1,r,0,\sz-1);\\
    & \quad \mathit{rz\_suber\_c\_form}(a,r)
\end{aligned}
\]
\centerline{(d) $a$ classical}
\end{minipage}

 We then proceed to the multiplier and power operator. Since their unary operations are not permitted, we focus solely on their binary operations. 

\textbf{Multiplier.} We present two implementations of the operator $\times$, depending on whether $a$ is classically known.  The definition of $\mathit{rz\_mult\_c\_form}$ is analogous to that of $\mathit{rz\_mult\_form}$, and is therefore omitted here. 
\[
\begin{array}{ccc}
 \begin{aligned}
        & \ket{a}_p \rightarrow \ket{a}_p\\
        & \ket{b}_q \rightarrow \ket{b}_q \\ 
        & \ket{0}_r \rightarrow \ket{(b \cdot a) \mathbin{\%} 2^{\sz}}_r \\
    \end{aligned} 
&
\begin{aligned}
    &\mathit{rz\_mult\_form}(p,q,r) \Leftarrow \\
    & \quad \mathit{QFT}(r,0,\sz-1);\\
    & \quad \mathit{rz\_mult}(p,\sz-1,q,r); \\ 
    & \quad \mathit{RQFT}(r,0,\sz-1);\\
\end{aligned}
&
\begin{aligned}
    &\mathit{rz\_mult}(p,n,q,r) \Leftarrow \\
    & \quad \mathbf{if} \ 0 \leq n \ \mathbf{then} \\
    & \qquad \mathbf{qif} [p[n]] (\ket{0}\rightarrow \mathbf{skip})\\ 
    & \qquad \qquad \quad \square \ket{1}
    \rightarrow \mathit{rz\_adder}(q,r,\sz-1,n)\\ 
    & \qquad \mathbf{fiq} \\
    &\qquad \mathit{rz\_mult}(p,n-1,q,r);\\
    & \quad \mathbf{fi}
\end{aligned}
\end{array}
\] 

\[
\begin{array}{cc}
  \begin{aligned}
        & \ket{b}_p \rightarrow \ket{b}_p\\
        & \ket{0}_q \rightarrow \ket{(b \cdot a) \mathbin{\%} 2^{\sz} }_q \ \text{(a is classically known)} \\ 
    \end{aligned} &
\begin{aligned}
    &\mathit{rz\_mult\_c}(p,n,a,q) \Leftarrow \\
    & \quad \mathbf{if} \ 0 \leq n \ \mathbf{then} \\
    & \qquad \mathbf{qif} [p[n]] (\ket{0}\rightarrow \mathbf{skip})\\ 
    & \qquad \qquad \quad \square \ket{1}
    \rightarrow \mathit{rz\_adder\_c}(q,0,\sz-1,a)\\ 
    & \qquad \mathbf{fiq} \\
    &\qquad \mathit{rz\_mult\_c}(p,n-1,2\cdot a,q);\\
    & \quad \mathbf{fi}
\end{aligned}
\end{array} 
\] 

\textbf{Power Operator.}
Similarly, we present three implementations of the power operator, depending on which operand is classically known. Hereafter, unless stated otherwise, $\mathit{copy}(s, t)$ stands for $\mathit{copy}(s, 0, \sz-1, t, 0, \sz-1)$ for any $s, t$. 

\subparagraph{Case 1: Both $a$ and $x$ quantum.} We provide a detailed introduction and resource analysis for this program in Section~\ref{app:case}.

\[
\begin{aligned}
    \ket{a}_p\ket{x}_q  \ket{0}_r\rightarrow \ket{a}_p \ket{x}_q\ket{(a^x) \mathbin{\%} 2^{\sz}}_r
\end{aligned}
\]

\[
\begin{array}{ll}
\begin{aligned}
    &\mathit{rz\_power\_form}(p,q,r,r_1,r_2,r_3,r_4) \Longleftarrow \\
    & \quad \mathit{rz\_power}(p,q,\sz-1,r_1,r_2,r_3,r_4); \\ 
    & \quad \mathit{copy}(r_1,r);\\ 
    & \quad \qinv(\mathit{rz\_power}(p,q,\sz-1,r_1,r_2,r_3,r_4)); \quad (\mathrm{uncompute} \ r_1)
\end{aligned}
&
\begin{aligned}
    &\mathit{power_2}(p,q,r) \Longleftarrow \\ 
    & \quad \mathit{copy}(p,r); \\ 
    & \quad \mathit{rz\_mult\_form}(p,r,q); \\ 
    & \quad \qinv(\mathit{copy}(p,r))
\end{aligned}
\end{array}
\]

\[
\begin{aligned}
    &\mathit{rz\_power}(p,q,n,r,r_1,r_2,r_3) \Longleftarrow \\ 
    & \quad \mathbf{if} \ 0 \leq n \ \mathbf{then} \\ 
    & \qquad \mathit{rz\_power}(p,q,n-1,r_1[n-1],r_1,r_2,r_3);\\ 
    & \qquad \mathit{power_2}(r_1[n-1],r_2,r_3); \\ 
    & \qquad \mathbf{qif} [q[n]] (\ket{0} \rightarrow \mathit{copy}(r_2,r)) \\ 
    & \qquad \qquad \quad \square \ket{1} \rightarrow \mathit{rz\_mult\_form}(p,r_2,r); \\ 
    & \qquad \mathbf{fiq} \\
    & \qquad \qinv(\mathit{power_2}(r_1[n-1],r_2,r_3)); \quad (\mathrm{uncompute} \ r_2) \\
    & \quad \mathbf{else} \ \mathit{init\_c}(1,r,\sz-1); \\
    & \quad \mathbf{fi}
\end{aligned}
\] 

\subparagraph{Case 2: $a$ quantum, $x$ classical.}

\[
\begin{aligned}
    \ket{a}_p\ket{0}_r \rightarrow \ket{a}_p \ket{(a^x) \mathbin{\%} 2^{\sz}}_r
\end{aligned} 
\] 
\[
\begin{array}{cc}
\begin{aligned}
    &\mathit{rz\_power\_c\_form}(p,x,r,r_1,r_2,r_3,r_4) \Longleftarrow \\
    & \quad \mathit{rz\_power\_c}(p,x,r_1,0,r_2,r_3,r_4); \\ 
    & \quad \mathit{copy}(r_1,r);\\ 
    & \quad \qinv(\mathit{rz\_power\_c}(p,x,r_1,0,r_2,r_3,r_4));
\end{aligned} &
\begin{aligned}
    &\mathit{rz\_power\_c}(p,x,r,i,r_1,r_2,r_3) \Longleftarrow \\ 
    & \quad \mathbf{if} \ 0 < x \ \mathbf{then} \\ 
    & \qquad \mathit{rz\_power\_c}(p, x/2,r_1[i],r_1,r_2,r_3);\\ 
    & \qquad \mathit{power_2}(r_1[i],r_2,r_3); \\ 
    & \qquad \mathbf{if} \ \mathit{odd}(x) \ \mathbf{then} \\
    & \qquad \qquad \mathit{rz\_mult\_form}(p,r_2,r); \\ 
    & \qquad \mathbf{else} \ \mathit{copy}(r_2,r); \\
    & \qquad \mathbf{fi} \\ 
    & \qquad \qinv(\mathit{power_2}(r_1[i],r_2,r_3)); \\ 
    & \quad \mathbf{else} \ \mathit{init\_c}(1,r,\sz-1);  \\
    & \quad \mathbf{fi}
\end{aligned}
\end{array}
\]

\subparagraph{Case 3: $a$ classical, $x$ quantum.} The definition of $\mathit{rz\_power\_base\_c\_form}$ is analogous to that of $\mathit{rz\_power\_form}$, and is therefore omitted here. 

\[
\begin{aligned}
    \ket{x}_q \ket{0}_r \rightarrow \ket{x}_q \ket{(a^x) \mathbin{\%} 2^{\sz}}_r 
\end{aligned}
\]
\[
\begin{aligned}
    &\mathit{rz\_power\_base\_c}(a,q,n,r,r_1,r_2,r_3) \Longleftarrow \\ 
    & \quad \mathbf{if} \ 0 \leq n \ \mathbf{then} \\ 
    & \qquad \mathit{rz\_power\_base\_c}(a,q,n-1,r_1[n-1],r_1,r_2,r_3);\\ 
    & \qquad \mathit{power_2}(r_1[n-1],r_2,r_3); \\ 
    & \qquad \mathbf{qif} [q[n]] (\ket{0} \rightarrow \mathit{copy}(r_2,r)) \\ 
    & \qquad \qquad \quad \square \ket{1} \rightarrow \mathit{rz\_mult\_c\_form}(r_2,a,r); \\ 
    & \qquad \mathbf{fiq} \\
    & \qquad \qinv(\mathit{power_2}(r_1[n-1],r_2,r_3)); \quad (\mathrm{uncompute} \ r_2) \\
    & \quad \mathbf{else} \ \mathit{init\_c}(1,r,\sz-1);  \\ 
    & \quad \mathbf{fi}
\end{aligned}
\]

\textbf{Modulo Operator and Divider.} 
Below we describe the implementations of the modulo and integer-division
  operators. We assume throughout that the divisor \(y\) is nonzero; the
  semantics and correctness claims for both operators are restricted to this
  domain. Writing
  \[
  x = a \cdot y + b,
  \qquad
  a = x/y,
  \qquad
  0 \leq b < y,
  \]
  the desired transformations are
  \[
  \begin{aligned}
  \ket{x}_p\ket{y}_q\ket{0}_{\mathit{re}}
  &\rightarrow
  \ket{x}_p\ket{y}_q\ket{x \mathbin{\%} y}_{\mathit{re}},\\
  \ket{x}_p\ket{y}_q\ket{0}_{\mathit{qt}}
  &\rightarrow
  \ket{x}_p\ket{y}_q
  \ket{x/y}_{\mathit{qt}}.
  \end{aligned}
  \]

  Here we consider the case in which both the dividend and divisor are quantum,
  and present the modulo implementation in detail. Division and modulo are
  implemented by successive trial subtraction. Starting from the most
  significant nonzero bit of the divisor \(y\), we construct the largest shifted
  multiple \(M = y \cdot 2^m\) that does not exceed \(2^{\mathit{sz}}-1\).
  We use \(M\) as the initial trial subtrahend and repeatedly halve it, comparing
  it with the current remainder at each step. If a subtraction succeeds, we
  retain the difference and record a quotient bit \(1\); otherwise, we restore
  the minuend and record \(0\). The first trial uses an unsigned comparison because the normalized initial
  subtrahend may exceed \(2^{\sz-1}\). After this trial, the invariant
  \(0\le R<2T\) with \(T\le2^{\sz-1}\) makes the most-significant-bit test sound for all subsequent trials.
  The quotient bits accumulate in an ancilla
  register \(p'\). For modulo, after \(p\) is returned to the computational
  basis, the final remainder is copied from \(p\) to \(\mathit{re}\) before
  uncomputation. Division is implemented analogously, except that \(p'\) is
  copied to the output register \(\mathit{qt}\), rather than copying the
  remainder from \(p\) to \(\mathit{re}\), before uncomputation. 
  
\[
\begin{array}{cc} 
    \begin{aligned}
    &\mathit{rz\_mod\_form}(p,q,p',\mathit{re},r)\Longleftarrow \\
    &\quad \mathit{QFT}(p,0,\sz-1);\\ 
    &\quad \mathit{extend}(p,q,0,p',\mathit{re},r); \\ 
    &\quad \mathit{RQFT}(p,0,\sz-1)\\ 
    & \\ 
    &\mathit{extend}(p,q,m,p',\mathit{re},r) \Longleftarrow \\
    &\quad \mathbf{if} \ m < \sz \ \mathbf{then} \\
    &\qquad \mathbf{qif}[q[m]]\\ 
    & \qquad \quad (\ket{1}\rightarrow \mathit{copy}(q,m,\sz-1,r,0,(\sz-1-m));\\ 
    &\qquad\qquad   \mathit{rz\_mod}(p,r,0,m,p'); \\
    & \qquad \qquad  \mathit{RQFT}(p,0,\sz-1); \\ 
    &\qquad\qquad  \mathit{copy}(p,\mathit{re});\\
     & \qquad \qquad  \mathit{QFT}(p,0,\sz-1); \\ 
    &\qquad\qquad \qinv(\mathit{rz\_mod}(p,r,0,m,p')); \\
    &\qquad\qquad   \qinv(\mathit{copy}(q,m,\sz-1,r,0,(\sz-1-m))); \\
    &\qquad \quad  \square \ket{0} \rightarrow \mathit{extend}(p,q,m+1,p',\mathit{re},r);\\ 
    &\qquad \mathbf{fiq}\\
    &\quad \mathbf{fi}
\end{aligned}
&
\begin{aligned}
    &\mathit{rz\_mod}(p,r,i,m,p') \Longleftarrow \\
    &\quad \mathbf{if}\ i=0\ \mathbf{then}\\
  &\qquad \mathit{RQFT}(p,0,\sz-1);\\
  &\qquad \mathit{blt}
     (p,0,\sz-1,r,0,\sz-1,p'[\sz-1-m]);\\
  &\qquad X[p'[\sz-1-m]];\\ 
  &\qquad \mathit{QFT}(p,0,\sz-1);\\
  &\qquad 
    \mathbf{qif}[p'[\sz-1-m]]
    (\ket0\rightarrow\mathbf{skip}\\
  &\qquad\qquad\qquad 
     \square\ket1\rightarrow
     \mathit{rz\_suber}(r,p,\sz-1,\sz-1));\\
  &\qquad \mathbf{fiq};\\
  & \qquad \mathit{rz\_mod}(p,r,i+1,m,p')\\ 
    &\quad \mathbf{else \ if} \ 0< i \leq m \ \mathbf{then} \\
    &\qquad \mathit{rz\_suber}(r,p,\sz-1-i,\sz-1);\\ 
    &\qquad \mathit{RQFT}(p,0,\sz-1); \\
    &\qquad \mathbf{qif}[p[0]](\ket{0}\rightarrow \mathbf{skip}\\
    &\qquad\quad\qquad \square \ket{1} \rightarrow X[p'[\sz-1-(m-i)]])\\
    &\qquad \mathbf{fiq}\\
    &\qquad \mathit{QFT}(p,0,\sz-1);\\
    &\qquad \mathbf{qif}[p'[\sz-1-(m-i)]](\ket{0}\rightarrow \mathbf{skip} \\
    &\qquad\quad\qquad \square \ket{1} \rightarrow \mathit{rz\_adder}(r,p,\sz-1-i,\sz-1))\\ 
    &\qquad \mathbf{fiq}\\
    &\qquad X[p'[\sz-1-(m-i)]];\\
    &\qquad \mathit{rz\_mod}(p,r,i+1,m,p')\\ 
    &\quad \mathbf{fi}
\end{aligned}
\end{array}
\]

% \[
%      \begin{aligned}
%         rz\_moder'(p,q,n,ex)& \Longleftarrow \\
%         & rz\_sub(q,p);\\ 
%         &  RQFT(p,0,N); \\
%         &  \mathbf{qif}[p[0]](\ket{0}\rightarrow \mathbf{skip}\\
%         &  \quad \qquad \square \ket{1} \rightarrow X[ex[N-m]])\\
%         &  \mathbf{fiq}\\
%         &  QFT(p,0,N);\\
%         &  \mathbf{qif}[ex[N-m]](\ket{0}\rightarrow \mathbf{skip} \\
%         &  \quad \qquad \square \ket{1} \rightarrow rz\_add(q,p));\\ 
%         &  \mathbf{fiq}\\
%         &  X[ex[N-m]];\\
%     \end{aligned}
% \]

\textbf{Other operators.} In what follows, we assume all registers are initialised to $\ket{0}$ unless stated otherwise. The operation \(\mathit{copy}(p,m_1,n_1,q,m_2,n_2)\) applies bitwise CNOTs from the slice \(p[m_1:n_1]\) to a clean target slice \(q[m_2:n_2]\). It realizes the basis-copy map \(\sum_x \alpha_x |
  x\rangle|0\rangle \mapsto \sum_x \alpha_x |x\rangle|x\rangle\), and therefore does not clone an arbitrary unknown quantum state.  The operation $\mathit{init\_c}(a, p, i)$ encodes the classical value or variable $a$ into qubits $p[0:i]$. The operation \(\mathit{xor}(p,m_1,n_1,q,m_2,n_2,r,m_3,n_3)\) performs bitwise XOR on slices \(p[m_1:n_1]\) and \(q[m_2:n_2]\), storing the result in \(r[m_3:n_3]\). 
\[
\begin{array}{cc}
    \begin{aligned}
        \mathit{copy}&(p,m_1,n_1,q,m_2,n_2) \Longleftarrow \\ 
         & \mathbf{if} \  m_1 \leq n_1 \  \mathbf{then} \\
        & \quad \mathbf{qif}[p[m_1]] (\ket{0}\rightarrow \mathbf{skip} \ \square \  
        \ket{1} \rightarrow 
        X[q[m_2]]) \ \mathbf{fiq}; \\ 
        &\quad \mathit{copy}(p,m_1+1,n_1,q,m_2+1,n_2) \\ 
        & \mathbf{fi} 
    \end{aligned} &
    \begin{aligned}
    \mathit{init\_c}&(a, p, i) \Longleftarrow \\
    & \quad \mathbf{if} \ 0 \leq i \ \mathbf{then} \\
    & \qquad \mathbf{if} \ (a \mathbin{\%} 2) = 1 \ \mathbf{then} \\
    & \qquad \quad X[p[i]]; \\
    & \qquad \mathbf{fi}; \\
    & \qquad \mathit{init\_c}(a/2, p, i-1) \\
    & \quad \mathbf{fi}
\end{aligned} 
\end{array}
\]

\[
    \begin{aligned}
        \mathit{xor}&(p,m_1,n_1,q,m_2,n_2,r,m_3,n_3)
         \Longleftarrow  \\
        & \mathbf{if} \  m_1 \leq n_1 \  \mathbf{then} \\
         &\quad \mathbf{qif}[p[m_1]](
         \ket{0}\rightarrow \mathbf{qif}[q[m_2]](\ket{0}\rightarrow \mathbf{skip} \ \square \  \ket{1} \rightarrow X[r[m_3]] )\ \mathbf{fiq} \\
        & \quad \qquad \ \ \quad  \square  \ket{1} \rightarrow
        \mathbf{qif}[q[m_2]](\ket{0}\rightarrow X[r[m_3]] \ \square \  \ket{1} \rightarrow \mathbf{skip})  \ \mathbf{fiq}) \\ 
         & \quad \mathbf{fiq};\\
         & \quad \mathit{xor} (p,m_1+1,n_1,q,m_2+1,n_2,r,m_3+1,n_3) \\ 
         & \mathbf{fi} 
    \end{aligned}
\]

 In $\mathit{locate}$, we assume the input parameters satisfy $(r-l+1) = 2^{n-m+1}$, and that the $\mathit{tar}$ system spans $[0: N]$. The function performs a binary search over $\mathit{tar}[l:r]$ to locate the position specified by the value in $q[m:n]$. If the resolved position is at most $N$, it extracts the \(b\) bits beginning at that position and assigns them to register $\mathit{re}$; otherwise, $\mathit{re}$ retains its initial value.  

\[
\begin{array}{cc}
\begin{aligned}
    & \mathit{locate}(q,m,n,\mathit{tar},l,r,\mathit{re},b,N) \Leftarrow \\ 
     & \quad \mathbf{if} \ m \le n \  \mathbf{then} \\
      &\quad \quad \mathbf{begin} \ \mathbf{local} \  k:= \lfloor (l+r)/2 \rfloor; \\ 
      & \quad \qquad \mathbf{qif}[q[m]](\ket{0} \rightarrow \mathit{locate}(q,m+1,n,\mathit{tar},l,k,\mathit{re},b,N) \\ 
      & \quad \qquad \qquad \qquad   \square
      \ket{1} \rightarrow \mathit{locate}(q,m+1,n,\mathit{tar},k+1,r,\mathit{re},b,N)) \\
      & \quad \qquad \mathbf{fiq} \\
      & \quad \quad \mathbf{end} \\ 
      & \quad \mathbf{else} \
         \mathbf{if} \ l \leq N \ \mathbf{then} \ 
        \mathit{copy}(\mathit{tar}, l, l+b-1, \mathit{re}, 0, b-1) \  
         \mathbf{else} \ \mathbf{skip} \\ 
      & \quad \mathbf{fi} 
\end{aligned} 
% &
% \begin{aligned}
%     loc&ate(q,m,n,l,r,re,k_0)) \Leftarrow \\ 
%      & \mathbf{if} \ m \le n \  \mathbf{then} \\
%       & \quad \mathbf{begin} \ \mathbf{local} \  k:= \lfloor (l+r)/2 \rfloor; \\ 
%       & \qquad \mathbf{qif}[q[m]](\ket{0} \rightarrow locate(q,m+1,n,l,k) \\ 
%       & \qquad \qquad \quad  \square
%       \ket{1} \rightarrow locate(q,m+1,n,k+1,r)) \\
%       & \qquad \mathbf{qif} \\
%       & \quad \mathbf{end} \\ 
%       & \mathbf{else} \ Swap(M, re, l, l+k_0, 0, k_0);\\
%       & \mathbf{end}
% \end{aligned} 
\end{array}
\]

\paragraph{Boolean operations.}
 We now present the implementation of Boolean operations in $\RQC^{++}$. We develop the comparators based on bitwise comparison. Here, we solely consider the scenario where $p$ and $q$ have the same bit length and $r$ has one qubit. 

\[
    \begin{aligned}
        &\ket{x}_p \ket{y}_q \ket{0}_r \rightarrow \ket{x}_p \ket{y}_q \ket{x = y}_r\\
    \end{aligned}
\]
\[
    \begin{aligned}
        \mathit{beq}&(p,m_1,n_1,q,m_2,n_2,r) \Longleftarrow   \\
         & \mathbf{if} \  m_1 > n_1 \  \mathbf{then} \ X[r] \ \mathbf{else} \\ 
         & \quad \mathbf{qif}[p[m_1]](
         \ket{0}\rightarrow \mathbf{qif}[q[m_2]](\ket{0}\rightarrow \mathit{beq} (p,m_1+1,n_1,q,m_2+1,n_2,r) \ \square \ket{1} \rightarrow \mathbf{skip} )\ \mathbf{fiq} \\
         & \quad \qquad \ \ \quad \square  \ket{1} \rightarrow
        \mathbf{qif}[q[m_2]](\ket{0}\rightarrow \mathbf{skip}\  \square \ket{1} \rightarrow 
        \mathit{beq} (p,m_1+1,n_1,q,m_2+1,n_2,r))  \ \mathbf{fiq}) \\ 
        & \quad  \mathbf{fiq}\\
         & \mathbf{fi} 
    \end{aligned}
\]
\[ \ket{x}_p \ket{y}_q \ket{0}_r \rightarrow \ket{x}_p \ket{y}_q \ket{x < y}_r\\\]
\[
    \begin{aligned}
        \mathit{blt}&(p,m_1,n_1,q,m_2,n_2,r) \Longleftarrow   \\
          & \mathbf{if} \  m_1 \leq n_1 \  \mathbf{then} \\ 
        &\quad \mathbf{qif}[p[m_1]](
         \ket{0}\rightarrow \mathbf{qif}[q[m_2]](\ket{0}\rightarrow \mathit{blt}(p,m_1+1,n_1,q,m_2+1,n_2,r) \square \ket{1} \rightarrow  X[r] )\ \mathbf{fiq} \\
         &\quad \qquad \ \ \quad \square  \ket{1} \rightarrow
        \mathbf{qif}[q[m_2]](\ket{0}\rightarrow \mathbf{skip} \ \square \ket{1} \rightarrow \mathit{blt}(p,m_1+1,n_1,q,m_2+1,n_2,r) )  \ \mathbf{fiq}) \\ 
        & \quad \mathbf{fiq} \\ 
         & \mathbf{fi}
    \end{aligned}
\]

 \section{Case Studies} 
 \label{app:case} 
 In addition to the running example, this section presents several typical recursive program examples to demonstrate how our compilation framework translates RQIMP source programs into $\RQC^{++}$ target code, and analyzes their resource overhead and theoretical complexity.

\subsection{Recursive Exponentiation — Power Function}

The first case study is a recursively implemented power function $a^x$, shown in Figure~\ref{fig:power}. It employs the repeated squaring strategy: $a^x = (a^{x/2})^2 \cdot a^{(x \bmod 2)}$.

The target program consists of two procedures, $\mathit{rz\_power}$ and $\mathit{rz\_power\_form}$, which together realize the reversible compilation of the recursive exponentiation. We stipulate that the base $a$ is stored in a quantum register $p$, the exponent $x$ in register $q$, and the final result is placed in register $r$.

\begin{figure}[t]
\centering
\begin{minipage}{0.8\linewidth}
\begin{lstlisting}[language=MyLang,basicstyle=\ttfamily\small,mathescape=true]
rz_power(p,q,n,r,r1,r2,r3) <==
  if 0 <= n then
    rz_power(p,q,n-1,r1[n-1],r1,r2,r3); // Compute a^(x/2)
    power_2(r1[n-1],r2,r3); // Compute  (a^(x/2))^2
    qif q[n]
      |0> -> copy(r2,r)
    $\square$ |1> -> rz_mult_form(p,r2,r) // Compute ((a^(x/2))^2) * a 
    fiq;
    qinv(power_2(r1[n-1],r2,r3)); // Uncompute r2
  else init_c(1,r,sz-1);
  fi

rz_power_form(p,q,r,r1,r2,r3,r4) <==
  rz_power(p,q,sz-1,r1,r2,r3,r4);
  copy(r1,r);
  qinv(rz_power(p,q,sz-1,r1,r2,r3,r4)). // Uncompute r2, r1
\end{lstlisting}
\end{minipage}
    \caption{Quantum program implementing recursive exponentiation}
    \label{fig:power}
    \Description{...}
\end{figure} 
% \begin{figure}[t]
% \centering
%  \[
% \begin{aligned}
% 1:  &\ \mathit{rz\_power}(p,q,n,r,r_1,r_2,r_3)\Longleftarrow \\ 
%  2:  &  \qquad \mathbf{if} \ 0 \leq n \ \mathbf{then} \\ 
%   3:      & \qquad \qquad \mathit{rz\_power}(p,q,n-1,r_1[(n-1)\cdot\sz:n\cdot\sz-1],r_1,r_2,r_3);\\ 
%    4:     & \qquad \qquad \mathit{power_2}(r_1[(n-1)\cdot\sz:n\cdot\sz-1],r_2,r_3); \\ 
%     5:    & \qquad \qquad \mathbf{qif} [q[n]] (\ket{0} \rightarrow \mathit{copy}(r_2,r)) \\ 
%      6:   & \qquad \qquad \qquad \quad \square \ket{1} \rightarrow \mathit{rz\_mult\_form}(p,r_2,r); \\ 
%      7:   & \qquad \qquad \mathbf{fiq} \\
%       8:  & \qquad \qquad \qinv(\mathit{power_2}(r_1[(n-1)\cdot\sz:n\cdot\sz-1],r_2,r_3)); \quad (\mathrm{Uncompute} \ r_2) \\ 
%         9:  & \qquad \mathbf{else} \ \mathit{init\_c}(1,r,\sz-1);  \\
%        10: & \qquad \mathbf{fi} \\ 
% 11: &\ \mathit{rz\_power\_form}(p,q,r,r_1,r_2,r_3,r_4) \Longleftarrow \\
%    12: &\qquad \mathit{rz\_power}(p,q,\sz-1,r_1,r_2,r_3,r_4); \\ 
%   13:  &\qquad \mathit{copy}(r_1,r);\\ 
%  14:   &\qquad \qinv(\mathit{rz\_power}(p,q,\sz-1,r_1,r_2,r_3,r_4)). \quad (\mathrm{Uncompute} \ r_2, r_1)\\ 
%     \end{aligned}
% \]
%     \caption{Quantum program implementing recursive exponentiation}
%     \label{fig:power}
%     \Description{...}
% \end{figure} 

In $rz\_power$, the procedure sequentially accesses each bit of the quantum register $q$ that stores the exponent $x$:
\begin{itemize}
    \item Base case (Line~10): $\mathit{init\_c}(1,r,
    \sz-1)$ initializes the result register $r$ to $1$. 
    \item  Recursive case:
    \begin{itemize}
        \item First, in Line~3, \(\mathit{rz\_power}\) is invoked to compute the power of the lower bits, e.g., \(a^{x/2}\); the result is stored in \(r_1[n-1]\). Since this is a linear recursion, our uncomputation strategy dictates that the result value of each recursive level be retained until the entire recursion completes. 
        \item Next, in Line~4, \(\mathit{power}_2\) computes the square, i.e., \((a^{x/2})^2\), and temporarily stores the result in \(r_2\); this operation requires an auxiliary quantum variable $r_3$, which is cleaned up after the \(\mathit{power}_2\) computation completes.
        The concrete implementation of $\mathit{power}_2$ is straightforward and provided in Appendix~\ref{sec_com_exp_assgn}. 
\item  In Lines~5--8, the quantum conditional statement \(\mathbf{qif}[q[n]]\) decides, depending on the current bit of the exponent \(x\) (i.e., \(x \bmod 2\)), whether to multiply the base \(a\) (stored in \(p\)) into the final result \(r\). 
\item  Finally, the temporary register \(r_2\) is cleaned in Line~9, whereas the recursive result \(r_1[n-1]\) is kept until the entire exponentiation completes.  
    \end{itemize}
\end{itemize}

$rz\_power\_form$ presents the complete reversible implementation of the power function, utilizing temporary variables $r_1$ through $r_4$, where $r_2$ is a quantum array. It first calls $rz\_power$ (Line~14) to compute $a^x$, storing the result temporarily in $r_1$, and supplies the auxiliary variables $r_2$–$r_4$ to $rz\_power$ (corresponding to $rz\_power$'s $r_1$–$r_3$). After the computation, since $rz\_power$ still contains uncleaned temporary variables (notably $r_2$), we copy the result from $r_1$ to the final result register $r$ (Line~15) and then execute the inverse of $rz\_power$ (Line~16) to uniformly uncompute all remaining temporaries including $r_1$. 

\textbf{Resource consumption analysis.} The recursion depth of this implementation is $\sz$ (the machine word length). The temporary variables used are $r_1$, $r_3$, $r_4$ (each a single register of $\sz$ qubits) and a quantum array $r_2$, where each element consists of $\sz$ qubits; with $\sz$ levels of recursion, $r_2$ occupies $\sz^2$ qubits. Thus the additional storage is $O(\sz^2)$, approximately $(\sz+3)\cdot\sz$ qubits in practice. The recursive-invocation overhead is linear in the recursion depth, while the total time is parameterized by the primitive-operation costs at each level, including the calls to \(\mathit{power_2}\), multiplication, copying, and their inverse computations. This implementation therefore realizes quantum recursive exponentiation with recursion overhead comparable to VQO, rather than incurring exponential recomputation.

\subsection{Find Function (Union-Find)}

The second case study is the find operation in the union-find data structure, depicted in Figure~\ref{fig:find}. This procedure recursively locates the root of the set containing element $x$; here we consider only the implementation without path compression.

\begin{figure}[t]
    \centering
\begin{minipage}[t]{0.43\linewidth}
\centering
\textbf{Source Procedure}
\begin{minipage}{0.82\linewidth}
\begin{lstlisting}[language=MyLang,basicstyle=\ttfamily\scriptsize,mathescape=true]
def find(nat^Q x, array n nat^Q root) {
  if (root[x] != x)
    { nat^Q y <- find(root[x],root) }
    { nat^Q y <- x }
  return y
}
\end{lstlisting}
\end{minipage}
\end{minipage}
\hfill
\begin{minipage}[t]{0.54\linewidth}
\centering
\textbf{Target Procedure}
\begin{minipage}{\linewidth}
\begin{lstlisting}[language=MyLang,basicstyle=\ttfamily\scriptsize]
F_find(p1,p2,r,i,q1,q2) <==
   rz_mult_c(p1,sz-1,sz,q1[i]);
   locate(q1[i],0,sz-1,p2,0,2^(sz-1),q2[i],sz,n*sz-1);
   qinv(rz_mult_c(p1,sz-1,sz,q1[i]));
   // Compute root[x] and store it in q2[i]
   beq(q2[i],0,sz-1,p1,0,sz-1,q1[i]); 
   // Test root[x] = x and store the result in q1[i]
   X[q1[i]]; // root[x] != x
   qif q1[i] 
      |1> -> F_find(q2[i],p2,r,i+1,q1,q2);
    square |0> -> copy(p1,0,sz,r,0,sz) 
   fiq; 
   qinv(beq(q2[i],0,sz-1,p1,0,sz-1,q1[i])); 
   // Uncompute q1[i] 
   rz_mult_c(p1,sz-1,sz,q1[i]);
   qinv(locate(q1[i],0,sz-1,p2,0,2^(sz-1),q2[i],sz,n*sz-1)); 
   qinv(rz_mult_c(p1,sz-1,sz,q1[i])); 
   // Uncompute q2[i]
\end{lstlisting}
\end{minipage}
\end{minipage}
    \caption{The source and target procedures implementing the find operation}
    \label{fig:find}
    \Description{...}
\end{figure} 

The target program uses \(p_1, p_2\) to store the input \(x\) and \(root\) value, respectively, and stores the return value \(y\) in \(r\). It also employs several temporary variables, namely \(q_1\) and \(q_2\). At each recursive level, the procedure proceeds as follows.

\begin{itemize}
    \item 
First (Lines~2--4), the value of \(\mathit{root}[x]\) is retrieved and stored in the temporary variable \(q_2[i]\). The computation requires a temporary variable \(q_1[i]\), which is immediately uncomputed and cleaned after the computation completes, allowing it to be reused later.

\item Next (Line~6), the program tests whether \(x\) equals its parent and stores the result in \(q_1[i]\).

\item Lines~9--12 then decide, based on the value of \(q_1[i]\), whether to proceed with the recursive call to \(F_{\mathtt{find}}\).

\item Since \(q_1[i]\) and \(q_2[i]\) at the current level are still needed later in that level when the recursive call occurs (Line~7), we must ensure that the temporaries used by the next recursive level are clean. To this end, we adopt quantum arrays to allocate separate registers for each level: level \(i\) uses \(q_1[i]\) and \(q_2[i]\), and the next level (\(i+1\)) uses \(q_1[i+1]\) and \(q_2[i+1]\). All these variables originate from non-recursive statements, so they are cleaned early after their lifetimes conclude via inverse computation.
At Line~13, \(q_1[i]\) is cleaned, and Lines~15--17 clean \(q_2[i]\). Since \(\textsc{find}\) is tail-recursive, the result \(y\) of the recursive call serves directly as the return value of the current level, requiring no extra state to be preserved; consequently, no special handling of temporaries for the recursive call is needed.
\end{itemize}

As a result, the recursive-invocation overhead is linear in the recursion depth, which is at most \(n\) for a union-find structure with \(n\) nodes, while the total time is parameterized by the primitive-operation costs at each level, including array access, equality testing, copying, and their inverse computations. The quantum variables $p_3$ and $r$ are both quantum arrays; each level requires \(\sz\) qubits for each array element, leading to a total space cost of \(2\cdot \sz \cdot n\) qubits.

\subsection{Graph Traversal — Depth-First Search (DFS)}

The third case study is depth-first search (DFS) on graphs. Given a graph \(G\) represented as an adjacency list array, where each element is a pair consisting of a node identifier and an array of its neighboring nodes, and an array \(d\) recording the out-degree of each node, the algorithm determines whether there exists a path from a given node \(v\) to another node \(w\), with \(\mathit{visited}\) serving as the visitation marker array. The program traverses the adjacency list recursively, exploring each unvisited neighbor of $v$ until the target node $w$ is found or all reachable nodes have been explored, as illustrated in Figure~\ref{fig:dfs}.

\textbf{Source Program Structure:}
\begin{itemize}
    \item Mark the current node as visited (Line~5): \(\mathit{visited}[v] \leftarrow \neg \mathit{visited}[v]\);
    \item Base case (Line~7): if \(v = w\), a path is found, return \(\mathbf{true}\); 
    \item Recursive case (Line~8--17): for each unvisited neighbor \(nv\) of the current node (iterating \(i\) from \(0\) to \(d[v]-1\)), recursively invoke \(DFS(G, d, nv, w, \mathit{visited})\) to determine reachability from \(nv\) to \(w\), storing the result temporarily in \(x\):
    \begin{itemize}
        \item If \(x\) is \(\mathbf{true}\), exit the loop early (set \(i = d[v]\));
        \item Otherwise, continue to the next neighbor;
    \end{itemize}
    \item If no neighbor yields a path, return \(\mathbf{false}\).
\end{itemize} 
% For a detailed algorithmic description of DFS, see [xx].

\begin{figure}[t]
    \centering
\begin{minipage}[t]{0.43\linewidth}
\centering
\textbf{Source Procedure}
\begin{minipage}{0.92\linewidth}
\begin{lstlisting}[language=MyLang,basicstyle=\ttfamily\scriptsize,mathescape=true]
def DFS(array n (nat^Q,array n nat^Q) G,
        array n nat^Q d,
        nat^Q v, nat^Q w,
        array n bool^Q visited) {
  visited[v] <- not visited[v];
  if v==w
    { bool^Q x <- true; }
    { nat^Q i <- 0;
      for i d[v]-1 {
        nat^Q nv <- pi2(G[v])[i];
        if visited[nv]==0
          { bool^Q x <- DFS(G,d,nv,w,visited);
            if x { i <- d[v] } 
                  {bool^Q x <- false}
          }
      }
    }
  return x
}
\end{lstlisting}
\end{minipage}
\end{minipage}
\hfill
\begin{minipage}[t]{0.54\linewidth}
\centering
\textbf{Target Procedure}
\begin{minipage}{\linewidth}
\begin{lstlisting}[language=MyLang,basicstyle=\ttfamily\scriptsize]
F_DFS(p1,p2,p3,p4,p5,r,j,k,q1,q2,q3,q4,q5) <==
  circ(not visited[v],p5); // Evaluate not visited[v]
  circ(v==w,q1[j]); // Test v==w and store result in q1[j]
  qif q1[j]
     |0> -> circ(0,p0);   
           F_aux1(p0,p1,p2,p3,p4,p5,r,j,k,q1,q2,q3,q4,q5); 
   square |1> -> X[r]
  fiq;
  qinv(circ(v==w,q1[j])); // Uncompute q1[j]
  qinv(circ(not visited[v],p5)); // Uncompute p5 

F_aux1(p0,p1,p2,p3,p4,p5,r,j,q1,q2,q3,q4,q5) <==
  circ(pi2(G[v])[i],q2[k]);  // Store nv in q2[k] 
  circ(visited[nv]==0,q3[k]); 
  // Test visited[nv]==0 and store result in q3[k] 
  qif q3[k]
     |1> -> F_aux2(p0,p1,p2,q2[k],p4,p5,q4[k],q5[k],j,k,q1,q2,q3,q4,q5)
   square |0> -> skip
  fiq;
  qinv(circ(visited[nv]==0,q3[k])); // Uncompute q3[k]
  qinv(circ(pi2(G[v])[i],q2[k]));   // Uncompute q2[k]
  circ(i <= (d[v]-1),q3[k]); 
  // Compute (i <= (d[v]-1)) and store in q3[k] 
  qif q3[k]
     |0> -> copy(q4[k],r)
   square |1> -> circ(i+1,q5[k]);
           F_aux1(q5[k],p1,p2,p3,p4,p5,r,j,k+1,q1,q2,q3,q4,q5)
           qinv(circ(i+1,q5[k])); 
  fiq;
  qinv(circ(i <= d[v]-1,q3[k])); // Uncompute q3[k]
  qinv(Lines 13--21);     // Uncompute q4[k], q5[k]

F_aux2(p0,p1,p2,q2[k],p4,p5,q4[k],q5[k],j,k,q1,q2,q3,q4,q5) <== 
FDFS(q1,p2,q2[k],p4,p5,q4[k],j+1,k+1,q1,q2,q3,q4,q5);
  qif q4[k]
     |1> -> circ(d[v],q5[k]) 
   square |0> -> circ(i, q5[k])
  fiq;
\end{lstlisting}
\end{minipage}
\end{minipage}
    \caption{The source and target procedures implementing DFS}
    \label{fig:dfs}
    \Description{...}
\end{figure}
\textbf{Target Program Structure:}

The target program comprises three mutually recursive procedures: \(\mathit{F\_DFS}\) (corresponding to the recursive function of $\mathit{DFS}$ in the source program) and \(F\_aux1\) (corresponding to the recursive implementation of the loop in lines 9–16 of the source program), along with \(\mathit{F\_aux2}\) (corresponding to the branch in lines 12–15). The quantum variable mapping is as follows:
\begin{itemize}
    \item \(\overline{p}\) (i.e., \(p_1\) through \(p_5\)) correspond to source variables \(G, d, v, w, \mathit{visited}\), respectively;
    \item The final result \(x\) is stored in register \(r\);
    \item The initial value of loop variable \(i\) is recorded in \(p_0\), and after line 17 of the target program, it is stored in \(q_5[k]\);
    \item \(q_1, q_2, q_3, q_4, q_5\) are temporary variables required by the functions, implemented as quantum arrays where each recursion level uses independent array elements (indices \(j\) and \(k\)).
\end{itemize}

\textbf{Implementation of \(F\_DFS\):}
\begin{itemize}
    \item Temporary variable \(q_1[j]\) stores the result of the comparison \(v == w\) (Line~3);
    \item A quantum branch is performed based on the value of \(q_1[j]\):
    \begin{itemize}
        \item If unequal (\(|0\rangle\) branch, false), proceed to \(F\_aux1\) for recursive handling (Line~5--6);
        \item If equal (\(|1\rangle\) branch), directly set result \(r\) to \(\mathbf{true}\) (Line~7);
    \end{itemize}
    \item After the branch, \(q_1[j]\) is immediately cleaned via inverse computation, and \(p_5\) (corresponding to \(\mathit{visited}\)) is restored (Line~9--10).
\end{itemize}

\textbf{Implementation of \(F\_aux1\) (recursive process for the loop):}
\begin{itemize}
    \item Use \(q_2[k]\) to store the current neighbor \(nv = \pi_2(G[v])[i]\) (Line~13);
    \item Use \(q_3[k]\) to record whether this neighbor has been visited (i.e., \(visited[nv] == 0\)) (Line~14);
    \item Based on the value of \(q_3[k]\), decide whether to explore this neighbor (Line~16--19):
    \begin{itemize}
        \item If unvisited (\(|1\rangle\) branch), enter \(F\_aux2\) to process the recursive call for this neighbor; this recursive procedure returns two values, recorded in \(q_4[k]\) and \(q_5[k]\), corresponding to the updated \(x\) and \(i\) in the original function;
        \item If already visited, skip;
    \end{itemize}
    \item After the branch, immediately clean \(q_3[k]\) and \(q_2[k]\) (Line~20--21);
    \item Subsequently, determine whether to continue the loop (Line~22--29): reuse \(q_3[k]\) to store the result of \(i \le d[v]-1\), and decide whether to continue invoking \(F\_aux1\) for the next neighbor (with index \(k\) incremented to ensure clean temporaries for the next level) or to exit the loop;
    \item After the loop, clean \(q_3[k]\) and the recursive return values \(q_4[k]\) and \(q_5[k]\) again (Line~30--31). 
\end{itemize}

\textbf{Implementation of \(F\_aux2\) (branch for neighbor exploration):}
\begin{itemize}
    \item Invoke \(F\_DFS\) to explore neighbor \(nv\), with indices \(k\) and \(j\) incremented; the return value \(x\) of the original function is stored in \(q_4[k]\) (Line~34);
    \item Based on the value of \(q_4\) (indicating whether a path was found), update the loop variable \(i\) (stored in a new quantum variable \(q_5[k]\)) (Line~35--38); if found, set \(i = d[v]+1\) to exit the loop early;
    \item This function has no additional temporary variables, so no cleanup is required.
\end{itemize}

\textbf{Temporary Variable Management Strategy:}
\begin{itemize}
    \item In \(F\_DFS\), \(q_1[j]\) is generated by non-recursive statements and is cleaned immediately after use (line~9);
    \item In \(F\_aux1\), \(q_2[k]\) and \(q_3[k]\) are also generated by non-recursive statements and are cleaned immediately after each basic block, enabling reuse (e.g., \(q_3[k]\) is used successively for visitation flags and loop conditions);
    \item The recursive return values \(q_4[k]\) and \(q_5[k]\) are uniformly cleaned after each level of \(F\_aux1\) (line~31). This "per-level immediate cleanup" strategy—which requires both forward computation and inverse computation at each level—may lead to exponential time overhead in worst-case scenarios (e.g., when the graph is a linear chain). Conversely, deferring the cleanup of \(q_4[k]\) and \(q_5[k]\) until the entire DFS recursion completes may cause exponential space growth with depth in tree-like recursive scenarios. Designing more refined optimization strategies to balance time and space efficiency in complex recursive structures remains an important direction for future work. This work focuses primarily on optimizing linear recursion; nevertheless, this case study successfully demonstrates the compiler's ability to handle complex recursive programs—different recursive paths use independent index variables, and temporaries from non-recursive statements are promptly cleaned after their lifetimes, effectively limiting space overhead.
\end{itemize}

\section{Compilation Correctness}
\label{app:veri}

In this subsection, we present the proof of correctness theorems stated in Section~\ref{sec:veri}.  We first define the necessary notations and basic concepts (Subsection~\ref{app:nota-def}), then specify valid static input parameters and compilation contexts (Subsection~\ref{app:valid}), and finally present the proof step by step, covering the correctness of:
\begin{itemize}
  \item Algorithm~\ref{clean_temp} for uncomputation of temporary variables (Theorem~\ref{theo:correctness-cleanup}, Subsection~\ref{app:proof-cleanup});
  \item Algorithm~\ref{com_body} for compilation of function bodies (Theorem~\ref{theo:correctness-body}, Subsection~\ref{app:proof-body});
  \item Algorithm~\ref{com_fun} for compilation of functions (Theorem~\ref{theo:correctness-fun}, Subsection~\ref{app:proof-fun});
  \item Algorithm~\ref{Com_state} for compilation of statements (Theorem~\ref{theo:correctness-com}, Subsection~\ref{app:proof-com});
  \item Algorithm~\ref{alg:compile_program} for the overall compilation framework (Main Theorem~\ref{theo:main}, Subsection~\ref{app:proof-main})
\end{itemize}  Since the correctness of the classical part is trivial, we focus on the
quantum part in the proof.  

\subsection{Notations and Definitions}
\label{app:nota-def}
To simplify the subsequent formal presentation, this subsection provides formal definitions of several notations. 

\paragraph{Compilation state.}
We first introduce some notational conventions for the compilation state.
We use shorthand notations for the components of the current buffer \(\ctx.\Bbuf=(\Vactive,m)\). Let
\[
\ctx.\Vactive \triangleq \ctx.\Bbuf.\Vactive,\qquad
\ctx.C\triangleq \ctx.\Bbuf.m.C, 
\]
and similarly for the remaining components of \(\ctx.\Bbuf.m\).
Furthermore, for any compilation variable \(\star\) in \(\Bbuf\), we introduce the auxiliary notations
\(\star_{\prec f}\), \(\star_{\succeq f}\), and \(\star_f\), based on the compilation stack \(\ctx.\St\) and the function information map \(\ctx.\M\). Recall that elements of \(\ctx.\St\) are either a function name \(f\) or a pair \((h,\Vactive)\) with \(\Vactive\). We assume that \(\ctx.\St\) is linearly ordered from bottom (index \(1\)) to top (index \(|\ctx.\St|\)).
Let
\[
\Fun(\ctx.\St)\triangleq
\{\, f \mid \exists i,\ 1\le i\le |\ctx.\St|,\ \ctx.\St[i]=f \,\}.
\]
That is, \(\Fun(\ctx.\St)\) consists exactly of those functions that occur in the stack as standalone markers. 
For each \(f\in \Fun(\ctx.\St)\), let \(\pos_{\St}(f)\) denote the unique index \(i\) such that \(\ctx.\St[i]=f\) under stack $\St$.
When the stack is clear from the context, we simply write \(\pos(f)\) for \(\pos_{\St}(f)\).
For any \(1\le i\le j\le |\ctx.\St|\), let \(\ctx.\St[i:j]\) denote the subsequence of \(\ctx.\St\) from index \(i\) to \(j\); if \(i>j\), then \(\ctx.\St[i:j]\) is the empty sequence. For any \(f \in \Fun(\ctx.\St)\), let
\[
\ctx.\St_{\prec f} \triangleq \ctx.\St[1 : \pos(f)-1],
\qquad
\ctx.\St_{\succeq f} \triangleq \ctx.\St[\pos(f) : |\ctx.\St|].
\]

For any function \(f\in \Fun(\ctx.\St)\), we define
\[
\begin{aligned}
\ctx.V_{\activelabel,\prec f}
&\triangleq
\bigcup_{\substack{1\le j<\pos(f)\\ \ctx.\St[j]=(h,\Vactive)}} \Vactive, \qquad 
\ctx.V_{\activelabel,\succeq f} \triangleq
\Bigl(\bigcup_{\substack{\pos(f) \leq j\le |\ctx.\St|\\ \ctx.\St[j]=(h,\Vactive)}} \Vactive\Bigr)\cup \ctx.\Vactive,\\[1mm]
\ctx.V_{\activelabel,f}
&\triangleq
\Bigl(\bigcup_{\substack{\pos(f)< j\le |\ctx.\St|\\ \ctx.\St[j]=(f,\Vactive)}} \Vactive\Bigr)\cup 
\begin{cases}
\ctx.\Vactive & \text{if } f=f_{\mathrm{cur}},\\
\emptyset & \text{otherwise}.
\end{cases}
\end{aligned}
\]
More generally, for any \(f,g\in \Fun(\ctx.\St)\) with \(\pos(f)\le \pos(g)\), we define
\[
\ctx.V_{\activelabel,[f,g)}
\triangleq
\bigcup_{\substack{\pos(f) \leq j<\pos(g)\\ \ctx.\St[j]=(h,\Vactive)}} \Vactive.
\]
Thus, \(\ctx.V_{\activelabel,[f,g)}\) denotes the sequential composition of all suspended code fragments stored strictly between the markers \(f\) and \(g\) in the stack. As a special case, we write
\[
\ctx.V_{\activelabel,[f,)} \triangleq \ctx.V_{\activelabel,\succeq f}.
\]
% Analogous notations are defined for the other components of \(d\), for instance \(ctx.ucv_{\prec f}\), \(ctx.ucv_{\succeq f}\), and \(ctx.V_{\activelabel,f}\).

For any compilation variable \(\star\) in \(m\), we write
\[
\ctx.\star_f \triangleq
\begin{cases}
\ctx.\M(f).\star & \text{if } f\neq f_{\mathrm{cur}},\\
\ctx.\star & \text{if } f=f_{\mathrm{cur}}.
\end{cases}
\]

Moreover, for any function \(f\in \Fun(\ctx.\St)\), we define
\[
\begin{aligned}
\ctx.\star_{\prec f}
&\triangleq
\langle\, \ctx.\star_h \mid h\in \Fun(\ctx.\St),\ \pos(h)<\pos(f)\,\rangle,\\[1mm]
\ctx.\star_{\succeq f}
&\triangleq
\langle\, \ctx.\star_h \mid h\in \Fun(\ctx.\St),\ \pos(h)\ge \pos(f)\,\rangle.
\end{aligned}
\]
Both sequences are ordered by increasing stack position. When \(\star\) is set-valued, we also use the same notation to denote the union of the corresponding sequence when this causes no ambiguity. When \(\star\) is map-valued, the notation refers to the corresponding sequence of maps; operations such as \(\dom(\cdot)\) and \(\cod(\cdot)\) are lifted pointwise to such sequences. In particular,
\[
\cod(\ctx.\rho_{\mathrm{loc},\succeq f})
\triangleq
\bigcup_{\substack {h\in \Fun(\ctx.\St),\\ \pos(h) \geq \pos(f)}} \, \cod(\ctx.\rho_{\mathrm{loc},h}) . \] 

Recall some notation that may otherwise be confusing. Let \(\rho \triangleq \rho_{\mathrm{glob}} \cup \rho_{\mathrm{loc}}\) denote the combined variable mapping. For a given function \(f\), we write \(\rho_{\mathrm{loc},f}\) for its local mapping, and define \(\rho_f \triangleq \rho_{\mathrm{glob}} \cup \rho_{\mathrm{loc},f}\). When \(f\) is the current function \(f_{\mathrm{cur}}\), the subscript may be omitted.

% To simplify the description of global properties, we define
% \[
% ctx.C_{\all}
% \triangleq
% \bigl\langle\, C \mid \exists j,\ ctx.\St[j]=(h,(C,ucv)) \,\bigr\rangle, \]
% \[
% ctx.\rho_{\all}
% \triangleq
% ctx.\rho_g
% \cup
% \Bigl(\bigcup_{h\in \Fun(ctx.\St)\setminus\{ctx.curf\}} ctx.\M(h).\rho_l\Bigr)
% \cup
% ctx.\rho_l.
% \]
% Here, \(ctx.C_{\all}\) denotes the sequence of all code fragments stored in the stack in stack order, and \(ctx.\rho_{\all}\) denotes the union of all active variable mappings, including both the global mapping and the local mappings of all active functions.
% Other components of \(d\) and \(m\) are defined similarly when needed.

\paragraph{Higher‑index extension.}  
 For any quantum variable \(q\), we define:
\begin{itemize}
    \item \(\base[q]\): the base name of \(q\).  If \(q = b[i]\), then \(\base[q] = b\); otherwise \(\base[q] = q\).
    \item \(\idx[q]\): the index term of \(q\).  If \(q = b[i]\), then \(\idx[q] = i\); otherwise \(\idx[q] = 0\).
\end{itemize}
For an index term \(i\), we write \(\var(i)\) for the set of (source‑level) variables that appear in \(i\);
if \(i\) is a constant, then \(\var(i) = \emptyset\).

In set‑theoretic expressions we overload \(\base(q)\) and \(\idx(q)\) 
to stand for the singletons \(\{\base[q]\}\) and \(\{\idx[q]\}\). 
These liftings extend pointwise to sets of quantum variables:
for any \(S\) of quantum variables, 
\[
\base(S) \triangleq \bigcup_{q \in S} \base(q),\qquad
\idx(S)  \triangleq \bigcup_{q \in S} \idx(q).
\]

For any set \(S\) of quantum variables, its \emph{higher‑index extension}, written \(S[+]\), is defined as
\[
S[+] \triangleq \{\, b[t] \mid \exists\, b[k] \in S \text{ with } t > k \,\}.
\]
Equivalently, \(S[+]\) contains exactly those indexed quantum variables whose base names
coincide with those of indexed variables in \(S\) and whose indices are strictly larger.
A non‑indexed variable \(r \in S\) is treated as \(r[0]\) for the purpose of this definition.
For example, if \(S = \{p[j], q[i]\}\), then
\[
S[+] = \{p[j+1], p[j+2], \dots\} \cup \{q[i+1], q[i+2], \dots\},
\]
which may be abbreviated as \(\{p[j+1{:}),\, q[i+1{:})\}\).

\paragraph{Garbage variables.} 
Recall that the \(G_{\mathrm{call}}\) is the function call graph and that  variable-dependency information in \(\ctx\) is given by
\( G_{\mathrm{fun}},\) which maps each function to its local variable dependency graph; write \(G_f \triangleq G_{\mathrm{fun}}(f)\). We write \(\mathsf{Node}(G_f)\) for the set of composite nodes in \(G_f\), and \(\mathsf{Var}(G_f)\) for the set of variable nodes in \(G_f\). For any \(f\), let
\[
\mathsf{Node}_{\SCC(f)}
\triangleq
\biguplus_{h \in \SCC(f)} \mathsf{Node}(G_h),
\qquad
 \mathsf{Var}_{\SCC(f)}
\triangleq
\biguplus_{h \in \SCC(f)} \mathsf{Var}(G_h),
\]
where \(\biguplus\) denotes disjoint union. 

For a local variable mapping \(\rho_{\mathrm{loc}}\) and a variable \(x\), we write
\(\rho_{\mathrm{loc}}[x]\) for the quantum register assigned to \(x\) when \(x \in \dom(\rho_{\mathrm{loc}})\).
In set‑theoretic expressions we overload \(\rho_{\mathrm{loc}}(x)\) to denote the singleton
\(\{\rho_{\mathrm{loc}}[x]\}\) if \(x \in \dom(\rho_{\mathrm{loc}})\), and the empty set \(\emptyset\) otherwise.
For a set of variables \(X\) we lift \(\rho_{\mathrm{loc}}\) pointwise by defining
\[
\rho_{\mathrm{loc}}(X) \triangleq \bigcup_{x \in X} \rho_{\mathrm{loc}}(x).
\]
The same conventions apply to the global mapping \(\rho_{\mathrm{glob}}\).

For each function \(h\), define
\[
X_h \triangleq \{\, x \in \mathsf{Var}(G_h) \mid x.\flag=2 \,\}.
\]
That is, \(X_h\) is the set of variable nodes in the dependency graph of \(h\) whose cleanup flag is \(2\). 

We then define, for any \(h \in \Fun(\ctx.\St)\),
\[
Q_{h} \triangleq \bigcup_{x \in X_h, \var(\idx(\ctx.\rho_{\mathrm{loc},h}(x))) \neq \emptyset} \ctx.\rho_{\mathrm{loc},h}(x).
\]
In the compilation algorithm we arrange that for the linear recursively‑coupled
temporaries (those with \(\flag = 2\)) the condition \(\var(\idx(\ctx.\rho_{\mathrm{loc},h}(x))) \neq \emptyset\)
is always satisfied; thus \(Q_{h}\) collects all currently relevant temporary
quantum variables for \(h\) that are not cleaned.

Similarly, we define
\[
Q_{\succeq f} \triangleq \bigcup_{h \in \Fun(\ctx.\St), \pos(h) \ge \pos(f)} Q_{h}. \] 

Let \(\SCC(f)\) be the strongly connected component of \(f\) in \(G_{\mathrm{call}}\). Define \(f^{*}\) to be the function in \(\SCC(f)\) with the smallest stack position in \(\ctx.\St\), i.e.,
\[
f^{*} \triangleq \arg\min_{h \in \SCC(f)\cap \Fun(\ctx.\St)} \pos(h).
\] 
Furthermore, we define \(\ctx.\Gar_f \triangleq \ctx.Q_{\succeq f^*}[+] \cup \ctx.Q_{\succeq f}\) as the set of garbage variables of $f$. 

% In the main text we briefly introduced the notions of \(cv\) and \(A_f\) to convey the key ideas. For the sake of readability, those descriptions were simplified and omitted some technical details. Now we give their complete formal definitions, which are the ones actually used in the proof. The definitions below supersede the informal descriptions in the main text and should be taken as the definitive ones.

\paragraph{Clean variable pool and accessible variable set.}
The \emph{clean variable pool} of \(\ctx\), written \(\ctx.cv\), is defined by
\[
\begin{aligned}
\ctx.cv \triangleq {} &\, \ctx.\Qpool \;\cup\; \ctx.\Qpool[+] \\
&\cup\;
\Bigl(
\ctx.V_{\activelabel,\succeq (f_{\mathrm{cur}})^{*}}[+]
\;\cup\;
\cod\bigl(\ctx.\rho_{\mathrm{loc},\succeq (f_{\mathrm{cur}})^{*}}\bigr)[+]
\Bigr)
\setminus
\bigl(\ctx.Q_{\succeq f_{\mathrm{cur}}^*}[+]\cap \ctx.V_{\activelabel,f_{\mathrm{cur}}}\bigr).
\end{aligned}
\]
Intuitively, \(\ctx.cv\) consists of all quantum registers that are currently clean and available for allocation: (1) the currently free registers \(\ctx.\Qpool\); (2) their higher-index extensions; and (3) the higher-index extensions induced by the active-variable set and the local variable mappings from \((f_{\mathrm{cur}})^{*}\) onward, excluding those variables that belong to \(\ctx.Q_{\succeq f_{\mathrm{cur}}^*}[+]\) and are already active in \(\ctx.V_{\activelabel,f_{\mathrm{cur}}}\).

The \textit{accessible variable set} $A_f$ for a function $f$ is defined as:
\[
\ctx.A_{f} \triangleq \ctx.cv \cup \ctx.V_{\activelabel,\succeq f} \cup \cod(\ctx.\rho_{\mathrm{loc},f}).
\]
Here, \(V_{\activelabel,\succeq f}\) is understood as the union of all active-variable sets in the corresponding sequence. Intuitively, $A_f$ contains all quantum variables that may be directly or indirectly used by function $f$ and its subsequent call chain.  
% In fact, we guarantee that every element of \(\rho_{l,f}\), except for the images of the formal parameters and the return value, belongs to \(ucv_{\succeq f}\). Hence the property stated in the main text — that the ancilla of \(\Theta(f)\) is contained in \(cv \cup ucv_{\succeq f}\) — is correct. However, because these sets change dynamically during compilation, to facilitate the verification of invariants such as the invariance of \(A_f\) and other related properties, we incorporate \(\rho_{l,f}\) into the definition of \(A_f\). This makes the proof simpler and cleaner. 

Define
\[
\ctx.B \triangleq \{\, q \in \ctx.\Qpool \cup \ctx.V_{\activelabel,\succeq f_{\mathrm{cur}}^*} \cup \cod(\ctx.\rho_{\mathrm{loc},f_{\mathrm{cur}}^*}) \mid \base[q] = q \,\}.
\]
From the definition of \(A_{f_{\mathrm{cur}}^*}\) one immediately sees that
\(\ctx.B = \{\, q \in \ctx.A_{f_{\mathrm{cur}}^*} \mid \base[q] = q \,\}\);
hence, whenever we prove that \(A_{f_{\mathrm{cur}}^*}\) remains invariant during compilation,
the invariance of \(\ctx.B\) follows automatically. 

\paragraph{$k$-th syntactic approximation.}

For a program \(C\) and a function \(f\), let
\[
\anc_f(C) \triangleq \qv(C) \setminus \bigl( \cod(\rho_{\mathrm{glob}}) \cup \rho_{\mathrm{loc},f}(\fp(f)) \cup \rho_{\mathrm{loc},f}(\re(f))) \bigr) 
\]
denote the set of ancilla quantum variables of \(C\) \emph{with respect to~\(f\)}.
That is, \(\anc_f(C)\) consists of those quantum variables occurring in \(C\) except the
images of the global variables, the formal parameters of \(f\), and the  return variables of \(f\).  When the function~\(f\) is clear from context, we simply write \(\anc(C)\) for
\(\anc_f(C)\).

Given a set of declarations $D = \{P_1(\overline{u}) \Leftarrow C_1, \ldots, P_n(\overline{u}) \Leftarrow C_n\}$ of $\RQC^{++}$, the $k$-th syntactic approximation $C^k$ of a well-defined quantum program $C \in \text{RQC}^{++}$ is defined by induction on $k \geq 0$:
\begin{enumerate}
    \item $C^0 \equiv \abort$, where $\abort$ denotes a program such that for any classical state $\sigma$ and quantum state $|\phi\rangle$, we have $(\abort, \sigma, |\phi\rangle) \not\rightarrow$;
    \item $C^{k+1} \equiv C[C_1^k / P_1, \ldots, C_n^k / P_n]$, where $C_i^k$ is the $k$-th approximation of $C_i$, and $C[C_1^k / P_1, \ldots, C_n^k / P_n]$ denotes the program obtained by replacing all procedure calls $P_i$ in $C$ with their bodies $C_i^k$. A more detailed definition can be found in the work~\cite{yingVerificationRecursivelyDefined2024}.
\end{enumerate}
The $k$-th syntactic approximation for the high-level language RQIMP is defined analogously. 

Building on the notations introduced above, we now define the
\(k\)-th syntactic approximation equivalence that relates a source function
to its target implementation.

\paragraph{$k$-th syntactic approximation equivalence of functions under a quantum variable set \(A\)}
Fix a quantum variable set \(A\). Given a target function declaration \(F(\overline{t}) \Leftarrow C\) and a source function \(f : \{\overline{\tau x};\, s;\, \tau_r v; \Gamma'\}\), we say that \(F(\overline{t}) \Leftarrow C\) is \emph{\(k\)-th syntactic- approximation equivalent under \(A\)} to \(f\), written \(F^{k} \approx_{A} f^{k}\), if the following holds.

For any classical states \(\sigma,\sigma'\) and variables \(\overline{u},y\) such that
\[
\langle y \leftarrow f^{k}(\overline{u}),\, \sigma \rangle \to \sigma',
\]
and for any quantum state \(\ket{\varphi}\) and quantum variables \(\overline{p}, r, \overline{\iota}\) satisfying:
\begin{enumerate}
    \item \textbf{Length match.} We have
    \(
    |\overline{p}| = |\mathrm{get\_Q}(\overline{\tau x})|,
    \qquad
    |r| = |v|.
    \)

    \item \textbf{Initialization.} We have \(\ket{\varphi}_{\cod(\rho_{\mathrm{glob}})} = \ket{\sigma(\dom(\rho_{\mathrm{glob}}))}\), 
    \(
    \ket{\varphi}_{\overline{p}} = \ket{\sigma(\mathrm{get\_Q}(\overline{u}))}\), \( 
    \ket{\varphi}_{r} = \ket{0},\) \( 
    \ket{\varphi}_{\anc(F^{k}(\mathrm{get\_C}(\overline{u}),\overline{p},r,\overline{\iota}))} = \ket{0}.
    \)

    \item \textbf{Ancilla correspondence.} There exists a bijection $h$ between the formal ancilla set \(\anc(F^{k}(\overline{t}))\) and the concrete ancilla set \(\anc(F^{k}(\mathrm{get\_C}(\overline{u}),\overline{p},r,\overline{\iota}))\).
\end{enumerate}

Then the execution of the \(k\)-th approximation of the target function satisfies
\[
\bigl\langle F^{k}(\mathrm{get\_C}(\overline{u}),\overline{p},r,\overline{\iota}),\, (\sigma_c, \ket{\varphi}) \bigr\rangle
\to
\bigl\langle \downarrow,\, (\sigma'_c,  \ket{\sigma'(\dom(\rho_{\mathrm{glob}}))} \otimes 
\ket{\sigma'(y)}_{r}
\otimes
\ket{\theta}_{h(A \cap \anc(F^k))} \otimes \ket{\varphi}_{\mathrm{rem}})
\bigr\rangle,
\]
 This definition formalizes the notion of compilation correctness at unfolding depth \(k\): the target approximation \(F^{k}\) must have the same input-output behavior as the corresponding source approximation \(f^{k}\), while preserving the input registers and explicitly accounting for the residual states of special recursive temporaries determined by the current compilation context. 

For any family \(\D\) of target function declarations, let \(\dom(\D) \triangleq \{\, F \mid \exists\, \overline{t}, C.\; F(\overline{t}) \Leftarrow C \in \D \,\}\), and for each \(F \in \dom(\D)\), let \(\D[F]\) denote the unique declaration in \(\D\) whose function name is \(F\). For a declaration family \(\D\) and \(F \in \dom(\D)\), we write \(\D[F]^k\) for the \(k\)-th syntactic approximation of the body of \(F\) as given by \(\D\).  When the family \(\D\) is clear from context, we abbreviate this as \(F^k\).  The same convention applies to derived notations such as \(\anc(F^k)\).

For two target declarations \(\delta_1\) and \(\delta_2\) with the same target name \(F\), let \(f = \Theta^{-1}(F)\) be the corresponding source function. We write \(\delta_1 \equiv_{A} \delta_2\) if for every \(k\),
\[
\delta_1 \approx_{A}^{k} f \;\Longleftrightarrow\; \delta_2 \approx_{A}^{k} f.
\]
For two declaration families \(\D_1\) and \(\D_2\), we write \(\D_1 \equiv_{A} \D_2\) if for every target name \(F \in \dom(\D_1) \cap \dom(\D_2)\),
\[
\D_1[F] \equiv_{A} \D_2[F].
\]

Let \(\D_{\SCC(f_{\mathrm{cur}})}\) be the specification family of target declarations
indexed by \(\{ \Theta(f) \mid f \in \SCC(f_{\mathrm{cur}})\,\}\).
Here \(\SCC(f_{\mathrm{cur}})\) keeps its standard meaning as the strongly connected
component of \(f_{\mathrm{cur}}\) in the call graph.  The auxiliary declaration
family \(\D_{\SCC(f_{\mathrm{cur}})}\) is mainly used for recursive or mutually
recursive components: in the proof, it records the corresponding target
declarations and forms part of the induction hypothesis, stating that these
declarations already satisfy the required semantic correctness at unfolding
depth \(k\).  If the current component is non-recursive, no such recursive
induction hypothesis is needed; in that case, this auxiliary family can be taken
to be empty, and the following state-equivalence definition and subsequent
proofs proceed in the same way.  To avoid repeatedly distinguishing these two
cases, we uniformly use the notation \(\D_{\SCC(f_{\mathrm{cur}})}\) below; the
case where the auxiliary family is empty is analogous to the recursive case.

\paragraph{State equivalence}
  We say that \(\ctx\) is
\emph{equivalent} to a source state \(\sigma\) for \(f\) under
\(\D_{\SCC(f_{\mathrm{cur}})}\) with respect to a target program state \((\sigma_0,\ket{\varphi})\)
and an unfolding depth \(k\), written
\[
\ctx \approx_{(f,(\sigma_0,\ket{\varphi}),k)}^{\D_{\SCC(f_{\mathrm{cur}})}} \sigma,
\]
if
\[
\langle \ctx.C_{f}^{\,k}, \ket{\varphi} \rangle
\to_{\D_{\SCC(f)} \cup \D_{\dom(\D) \setminus \Theta(\SCC(f))}}
\Bigl\langle \downarrow,\,
\mathcal E(\sigma,\rho_{\mathrm{active},f})
\otimes
\ket{\theta}_{\,\ctx.V_{\activelabel,f}\setminus \cod(\rho_{\mathrm{active},f})}
\otimes
\ket{\varphi}_{\mathrm{rem}}
\Bigr\rangle,
\]
where \(\rho_f \triangleq \rho_{\mathrm{glob}} \cup \rho_{\mathrm{loc},f}\),
\[
\rho_{\mathrm{active},f}
\triangleq
\rho_f \upharpoonright
\Bigl(
\dom(\rho_{\mathrm{glob}})
\;\cup\;
\{\, x \in \dom(\rho_{\mathrm{loc},f}) \mid
x=\re(f) \vee \neg(\delta(\node(x))=0 \land \node(x).\flag=0) \,\}
\Bigr),
\]
and let \(\sys\) be the set of all quantum variables; then 
\[
\mathrm{rem}
\triangleq \sys \setminus \bigl(\cod(\rho_{\mathrm{active},f})\cup
(\ctx.V_{\activelabel,f}\setminus \cod(\rho_{\mathrm{active},f}))\bigr).
\]
Intuitively, this equivalence requires agreement only on those variables
whose quantum states are still semantically relevant to the source
state \(\sigma\); local input variables that have already been restored
and are no longer used are excluded from \(\rho_{\mathrm{active},f}\) and
\(\ctx.V_{\activelabel,f}\).  When the specification family \(\D_{\SCC(f_{\mathrm{cur}})}\) is clear
from context, we simply write \(\ctx \approx_{(f,(\sigma_0,\ket{\varphi}),k)} \sigma\).
Furthermore, in the vast majority of cases we instantiate this definition
with the current function \(f_{\mathrm{cur}}\); under this instantiation we further
abbreviate the notation to \(\ctx \approx_{((\sigma_0,\ket{\varphi}),k)} \sigma\).  
Since we only focus on the quantum part in the proof, from this point on we write only the quantum state \(\ket{\phi}\) for the target program state. 

Other notations will be introduced as needed later.

\subsection{Valid Compilation Context} 
\label{app:valid}
In this subsection we present the validity of compilation states. 
We first define static validity for the static input parameters. Since the compilation algorithm only reads these parameters without modifying them, they are invariant throughout compilation. To facilitate the proof, we encapsulate this property in the following lemma. 

Let \(\Xi\) be the source function environment.  
For each \(f \in \dom(\Xi)\), let \(s_f\) be the body of \(f\). Write \(\Call(s_f)\) for the set of function names that occur as direct callees in \(s_f\), and let \(\Succ(f)\) denote the set of direct successors of \(f\) in the call graph. For a statement \(s\) and a variable \(x\), let \(\contuses(s,x)\) denote the total number of (syntactic) read occurrences of \(x\) in \(s\); when \(s\) is a sequence of statements, the counts are summed over the sequence.   

\begin{lemma}[Well-formedness of the static context]
  \label{lem:static-wf}
  The static compilation parameters \(\K = (\Xi, \Gamma, \G=(G_{\mathrm{call}}, G_{\mathrm{fun}}), \Theta, \rho_{\mathrm{glob}}) \)
  satisfy the following properties.
  \begin{enumerate}
   \item \textbf{Well-typedness.}  For every function \(f \in \dom(\Xi)\), the declaration of \(f\) passes the
  function-declaration typing rule under \(\Xi\) and \(\Gamma\).
  Moreover, every global variable is quantum-mode, i.e.,
  \(M(z)=\mathtt{Q}\) for all \(z \in \dom(\Gamma)\). 
    \item \textbf{Flag consistency.} For every \(f \in \dom(\Xi)\),
    \begin{itemize}
      \item for all \(n,w \in \mathsf{Node}(G_f)\),
            if \(n.\flag=1\) and \(w.\flag=1\) then \(n=w\),
            and for all \(x,y \in  \mathsf{Var}(G_f)\),
            if \(x.\flag=2\) and \(y.\flag=2\) then \(x=y\);
      \item moreover,
            \[
            \Bigl(\exists n \in \mathsf{Node}_{\SCC(f)}.\; n.\flag=1\Bigr)
            \Longrightarrow
            \Bigl(\forall x \in \mathsf{Var}_{\SCC(f)}.\; x.\flag\neq 2\Bigr).
            \]
    \end{itemize}
   \item \textbf{Dependence-graph consistency.} For every \(f \in \dom(\Xi)\),
    \begin{itemize}
      \item \(\var(s_f) \subseteq \mathsf{Var}(G_f)\) and
            \(\node(\var(s_f)) \subseteq \mathsf{Node}(G_f)\);
      \item for every \(x \in \mathsf{Var}(G_f)\),
            if \(x \notin \dom(\rho_{\mathrm{glob}})\) then \(\var(\node(x)) \cap \dom(\rho_{\mathrm{glob}}) = \emptyset\), and
            if \(x \notin \re(f)\) then \(\var(\node(x)) \cap \re(f) = \emptyset\).
      \item for every \(u,v \in \mathsf{Node}(G_f)\) such that $u \neq v$, $\var(u) \neq \var(v)$  and if \(u \in \mathrm{pred}^+(v)\) then \(v \notin \mathrm{pred}^*(u)\). 
    \end{itemize}
\item \textbf{Global mapping injectivity.} The global variable mapping \(\rho_{\mathrm{glob}}\) is injective: for any distinct \(x, y \in \dom(\rho_{\mathrm{glob}})\), \(\rho_{\mathrm{glob}}(x) \neq \rho_{\mathrm{glob}}(y)\). 
    \item \textbf{Call-graph consistency.} For every \(f \in \dom(\Xi)\),
          \(\Succ(f) = \Call(s_f)\). In addition, for any \(f \in \dom(\Theta)\)
        and every node \(u \in \mathsf{Node}(G_f)\),
        \[
        \Outd_{G_f}(u) = \contuses(s_f, \var(u)).
        \]

    \item \textbf{Function name-map consistency.} \(\dom(\Xi) = \dom(\Theta)\), and
          \(\Theta\) is injective on \(\dom(\Xi)\) (i.e., for all distinct
          \(f_1,f_2 \in \dom(\Theta)\), \(\Theta(f_1) \neq \Theta(f_2)\)).
  \end{enumerate}
\end{lemma}
  \begin{proof}
All properties follow directly from the static analysis that produces the initial compilation parameters: well‑typedness from type checking; flag consistency, dependence‑graph consistency, and call‑graph consistency from the dependency graph construction (where the part about \(\rho_{\mathrm{glob}}\) in dependence‑graph consistency additionally relies on well‑typedness and the initialization of \(\rho_{\mathrm{glob}}\)); global mapping injectivity and function name‑map consistency from the initialization of \(\rho_{\mathrm{glob}}\) and \(\Theta\). Since the compilation algorithm never modifies any static component, these properties remain invariant throughout compilation. 
\end{proof}

To facilitate the subsequent proof, we add a parameter \(R_{\mathrm{stmt}}\) to the compilation state \(m\), which records the remaining sequence of source statements yet to be compiled in the function body. When compilation of a function begins, this parameter is initialized to the whole function body; after each statement is compiled, it is removed from the sequence. 

Let \(\ctx_r\) denote the completion context obtained by continuing the
compilation from \(\ctx\) until the current function reaches the
declaration-generation point of \(\textsc{Compile\_Fun}\) (the analogue of
\(\ctx_4\) in Notation~\ref{nota:fun}). That is, starting from \(\ctx\), the
compiler finishes compiling \(\ctx.R_{\mathrm{stmt}}\), cleans the remaining
temporary variables, and records the generated target declaration in \(\D\).

For a target declaration \(\D\), let \(\args(\D)\) denote its extra-parameter list, excluding the input and output parameters. Let
\(\D_{\SCC(f_{\mathrm{cur}})}\) be the specification family of target declarations.
We say that the global compilation state of \(\ctx\) is \emph{valid}
with respect to a depth \(k\) and \(\D_{\SCC(f_{\mathrm{cur}})}\) if the following conditions hold.
\begin{enumerate}
    \item \textbf{Declaration consistency.}
        For every \(f \in \SCC(f_{\mathrm{cur}}) \cap \Theta^{-1}(\dom(\ctx.\D))\),
        \[
        \ctx.\D[\Theta(f)] \equiv_{\ctx_r.\Gar_{f^*}} \D_{\SCC(f_{\mathrm{cur}})}[\Theta(f)].
        \]

    \item \textbf{Stack discipline.}
        \(f_{\mathrm{cur}} \in \Fun(\ctx.\St)\) and
        \[
        \Fun(\ctx.\St_{\succ f_{\mathrm{cur}}})
        \subseteq
        \Theta^{-1}(\dom(\ctx.\D)) \cap \SCC(f_{\mathrm{cur}}).
        \]
        Moreover, \(\Fun(\ctx.\St) \subseteq \dom(\ctx.\M) \cup \{f_{\mathrm{cur}}\}\) and,
        for every \(f \in \Fun(\ctx.\St)\),
        \[
        \Fun(\ctx.\St_{\succ f}) \subseteq \{\, g \mid f \mathrel{R^*} g \,\},
        \]
        where \(f_1 \mathrel{R^*} f_2 \triangleq f_2 \in \Succ(f_1)\).

    % \item \textbf{Out‑degree consistency.}
    %     For any \(f \in \dom(\Theta) \setminus
    %         (\Fun(\ctx.\St) \cap \Theta^{-1}(\dom(\ctx.\D)))\)
    %     and every node \(u \in \mathsf{Node}(G_f)\),
    %     \[
    %     \delta(G_f)(u) = \contuses(s_f, \var(u)).
    %     \]

    \item \textbf{Equivalence conditions.}
        \begin{itemize}
            \item[(a)] For all \(f \in \SCC(f_{\mathrm{cur}}) \cap (\Fun(\ctx_r.\St))\),
                \begin{itemize}
                    \item \(\D_{\SCC(f_{\mathrm{cur}})}[\Theta(f)]^k \approx_{\ctx_r.\Gar_f} f^k\);
                    \item \(\ctx_r.E_{\args,f} \preceq \args(\D_{\SCC(f_{\mathrm{cur}})}[\Theta(f)])\); 
                    \item If \(f \in (\Fun(\ctx.\St)) \setminus \Theta^{-1}(\dom(\ctx_r.\D))\), 
                          then
                          \( \anc(\D_{\SCC(f_{\mathrm{cur}})}[\Theta(f)]^k) \subseteq \ctx_r.A_f \); 
                    \item \(\ctx_r.\D \equiv_{\ctx_r.\Gar_{f^*}} \D_{\SCC(f_{\mathrm{cur}})}\).
                          Moreover, if \(f \in \Theta^{-1}(\dom(\ctx_r.\D))\), for any $z$, 
                          \[ 
                          \anc(\D_{\SCC(f_{\mathrm{cur}})}[\Theta(f)]^z) \subseteq \anc(\ctx_r.\D[\Theta(f)]^z) \cup \anc(\ctx_r.\D[\Theta(f)]^z)[+].
                          \]
                \end{itemize}

            \item[(b)] For all \(f \in \SCC(f_{\mathrm{cur}}) \cap  \Fun(\ctx.\St) \cap
                \Theta^{-1}(\dom(\ctx.\D))\),
                we have \(\ctx.R_{\mathrm{stmt},f} = \emptyset\), and
                \[
                \D_{\SCC(f_{\mathrm{cur}})}[\Theta(f)]^{k+1} \approx_{\ctx_r.\Gar_f} f^{k+1}.
                \]

            \item[(c)] For all \(f \in \Theta^{-1}(\dom(\ctx.\D))
                \setminus \Fun(\ctx.\St)\)
                and all \(z\),
                we have \(\ctx.\D[\Theta(f)]^z \approx_{G'} f^z\) for some
                \(G'\) with the following property:
                if \(\forall g \in \SCC(f),\; X_g = \emptyset\) then
                \(G' = \emptyset\); otherwise \(G' = \anc(C)\).
                Additionally,
                \[
                \{\, \base(q) \mid q \in \qv(\anc(C)) \,\}
                \cup
                \{\, \idx(q) \mid q \in \qv(\anc(C)) \,\}
                \subseteq \args(\ctx.\D[\Theta(f)]).
                \]
        \end{itemize}
	Conditions (a) and (b) refer to the specification family \(\D_{\SCC(f_{\mathrm{cur}})}\),
  which remains fixed throughout compilation, while condition (c) refers to the
	implementation family \(\ctx.\D\) for functions whose compilation is already complete.
\end{enumerate}

Given a node \(v\), we write \(\rho_{\mathrm{loc}}(v)\) for the set of registers assigned to the variables in \(v\); i.e., \(\rho_{\mathrm{loc}}(v) =\rho_{\mathrm{loc}}(\var(v))\).  Let \(f\) be a source function, and let \(\rho_{\mathrm{loc}}\), \(C\), \(\kappa\), \(\delta_f\) and 
\(R_{\mathrm{stmt}}\) denote, respectively, the local variable mapping, the compiled 
code, the cleanup-circuit mapping, the residual out‑degree map of \(f\), 
and the sequence of remaining statements to be compiled.
Let \(\Vactive\) be the 
unclean variable set used by the function \(f\) and \(S_A\), \(S_B\) be two sets of variables.  
Let \(\ket{\phi}\) be an initial quantum state, and let \(z \in \mathbb{N}\) be a 
non-negative integer.

We say that the tuple \((\rho_{\mathrm{loc}}, C, \kappa, \delta_f, R_{\mathrm{stmt}})\) is 
\emph{valid} with respect to \(\Vactive\), \(S_A\), \(S_B\), \(\ket{\phi}\) and \(z\) 
if the following conditions hold: 

\begin{enumerate}
    \item \textbf{Injectivity of \(\rho_{\mathrm{loc}}\).} For all \(l,l' \in \dom(\rho_{\mathrm{loc}})\), if \(l \neq l'\), then \(\rho_{\mathrm{loc}}(l) \neq \rho_{\mathrm{loc}}(l')\).

    \item \textbf{Residual-use consistency.}
          \(R_{\mathrm{stmt}}\) is a suffix of \(s_f\) and for every node \(v \in \mathsf{Node}(G_f)\), 
          \[
          \delta_f(v) = \contuses(R_{\mathrm{stmt}}, v),
          \]
          which states that the remaining out‑degree correctly reflects the actual usage of variables in the yet‑to‑be‑compiled statements. 
          Moreover, $\dom(\rho_{\mathrm{glob}}) \cap \dom(\rho_{\mathrm{loc}}) =\emptyset$ and 
          \[
          \dom(\rho_{\mathrm{loc}}) \setminus \bigl( \fp(f) \cup \{\re(f)\} \bigr)
          \subseteq 
          \{\, x \in \mathsf{Var}(G_f) \mid
              \delta(\node(x)) > 0 \;\lor\; x.\flag \neq 0 \,\}.
          \]

    \item \textbf{Correctness of cleanup circuits.} 
For every \(q \in \cod(\rho_{\mathrm{loc}}) \setminus \rho_{\mathrm{loc}}(\re(f))\) such  that $\rho_{\mathrm{loc}}^{-1}(q).\flag \neq 2$, let
\(v \triangleq \node(\rho_{\mathrm{loc}}^{-1}(q))\). We have $v \in \dom(\kappa)$. 
          For any $C'$, define the set of required registers for cleanup as
          \[
          \req_{\rho,v}(C') \triangleq \qv(C') \setminus \rho (\operatorname{pred}^*(v)).
          \]
          The following hold:
          \begin{itemize} 
              \item[(a)] \(\bigl(\rho_{\mathrm{loc}}(\re(f)) \cup \rho_{\mathrm{loc}}(\fp(f))\bigr) \cap \req_{\rho,v}(\kappa(v)^z) = \emptyset\).

              \item[(b)] For every \(k\), \(\qv(\kappa(v)^k) \subseteq \qv(C^k) \cup S\) such that \(S \subseteq S_A\) and \(\base(S) \subseteq S_B\).  

              \item[(c)] Assume \(\langle C^z,\ket{\phi}\rangle \to^{*} \langle \downarrow,\ket{\psi}\rangle\). For any tuple of fresh quantum variables \(\overline{p}\) with \(\ket{\psi} \models \ket{0}_{\overline{p}}\) and \(\overline{p} \cap (\cod(\rho)\cup \Vactive) =\emptyset\), let \(\kappa(v)^z[\req_{\rho,v} \mapsto \overline{p}]\) denote the circuit obtained from \(\kappa(v)^z\) by replacing every register in \(\req_{\rho,v}(\kappa(v)^z)\) with the corresponding fresh registers \(\overline{p}\) (preserving order). Then there exists a quantum state \(\ket{\theta}\) such that
\[
\bigl\langle (C^z;\,\qinv(\kappa(v)^z[\req_{\rho,v} \mapsto \overline{p}])),\; \ket{\phi} \bigr\rangle 
\to^{*}
\bigl\langle \downarrow,\; \ket{\phi}_{\rho_{\mathrm{loc}}(U_v)} \otimes \ket{\theta}_{\rho_{\mathrm{loc}}(v) \setminus \rho_{\mathrm{loc}}(U_v)} \otimes \ket{\psi}_{\qv(C^z)\setminus \rho_{\mathrm{loc}}(v)} \bigr\rangle, 
\]
where \(U_v \triangleq \{\, x \in v \mid x.\flag \neq 2 \,\}\). \end{itemize} 
\end{enumerate} 

We use the notation \(\ket{\psi}\models\ket{q}_{X}\) to mean that the
reduced state of \(\ket{\psi}\) on the subsystem \(X\) coincides with the pure
state \(\ket{q}\); formally,
\(\tr_{\sys\setminus X}(\lvert\psi\rangle\langle\psi\rvert)=\lvert q\rangle\langle q\rvert\).
In particular, \(\ket{\psi}\models\ket{0}_{X}\) signifies that all qubits in \(X\)
are in the \(|0\rangle\) state.

Let \(\Theta\) be the function name mapping and let
\(\D_{\SCC(f_{\mathrm{cur}})}\) be the specification family of target declarations.
For a source function \(f\) we write \(F_f \triangleq \Theta(f)\) for the
corresponding target function name. 
In the sequel, whenever we write \(F_f^k\) or \(\anc(F_f^k)\) without
explicitly mentioning a declaration family, the function is understood to
be resolved in \(\D_{\SCC(f_{\mathrm{cur}})}\). 
When the specification family is clear from context, for a set \(S\) of source functions we define the shorthand
\[
\anc(F_S^{\,z}) \;\triangleq\; \bigcup_{f \in S} \anc\bigl(\D_{\SCC(f_{\mathrm{cur}})}[\Theta(f)]^{\,z}\bigr).
\] 
In particular, \(\anc(F_{\succeq f^*}^{\,z}) = \bigcup_{f \in \Fun(\ctx.\St_{\succeq f^*})} \anc\bigl(\D_{\SCC(f_{\mathrm{cur}})}[\,\Theta(f)\,]^{\,z}\bigr)\). 

% Let \(\mathrm{prev}_{\St}(f)\) be the first function before \(f\) on the stack \(\St\) that is still uncompiled (i.e., not in \(\Theta^{-1}(\dom(\D))\)). If none, \(\mathrm{prev}_{\St}(f)\) is undefined.  

Finally, given a compilation context \(\ctx\), a depth \(k\), a state \(\ket{\phi}\),
and a specification family of target declarations \(\D_{\SCC(f_{\mathrm{cur}})}\),
we say that \(\ctx\) is \emph{dynamic valid} (for short, \emph{valid}) with respect to \((\ket{\phi}, k, \D_{\SCC(f_{\mathrm{cur}})})\), written
\[
  \bigl(\ctx,(\ket{\phi},k,\D_{\SCC(f_{\mathrm{cur}})})\bigr)\in\mathcal V,
\]
if the following conditions hold.
\begin{enumerate}
  \item \textbf{Global validity.}
        The global compilation state of \(\ctx\) is valid with respect to
        \(k\) and \(\D_{\SCC(f_{\mathrm{cur}})}\).

  \item \textbf{Initial-state consistency.}
        \begin{gather*}
          \ket{\phi}\models\ket{0}_{\ctx.A_{f_{\mathrm{cur}}}\setminus\ctx.\rho_{\mathrm{loc},f_{\mathrm{cur}}}(\fp(f_{\mathrm{cur}}))}, \\
          \ctx.cv \;\cap\; \bigl( \ctx.V_{\activelabel,\succeq f_{\mathrm{cur}}^{*}}
               \cup \cod(\ctx.\rho_{\mathrm{loc},\succeq f_{\mathrm{cur}}^{*}})
               \cup \cod(\rho_{\mathrm{glob}}) \bigr)
          = \emptyset, \\
          \ctx.Q_{\succeq f_{\mathrm{cur}}^{*}}[+] \cap \bigl(\cod(\ctx.\rho_{\mathrm{loc},\succeq f_{\mathrm{cur}}^{*}}) \cup \cod(\rho_{\mathrm{glob}}) \bigr)
          = \emptyset, \\
          \ctx.\Qpool\cap (\ctx.\Qpool[+] \cup \bigl( \ctx.V_{\activelabel,\succeq f_{\mathrm{cur}}^{*}}[+]
               \cup \cod(\ctx.\rho_{\mathrm{loc},\succeq f_{\mathrm{cur}}^{*}})[+]
               \bigr))
          = \emptyset, \\
          \forall\, q \in \ctx.\Qpool,\;
                  \var(\idx(q))=\emptyset \Longrightarrow \base(q)=q.
        \end{gather*}

  \item For every \(f\in \{f_{\mathrm{cur}}\} \cup \Fun(\ctx.\St)\cap\SCC(f_{\mathrm{cur}}) \setminus \Theta^{-1}(\dom(\ctx.\D))\), the following hold.
        \begin{enumerate}
          \item \textbf{Local validity.}
                The tuple \((\ctx.\rho_{\mathrm{loc},f}, \ctx.C_f, \ctx.\kappa_f, \ctx.\delta_f, \ctx.R_{\mathrm{stmt},f})\)
                is valid with respect to \(\ctx.V_{\activelabel,f}\),  \(\ctx.\Qpool \cup \ctx.V_{\activelabel,\succeq f} \setminus \ctx.Q_{\succeq f^*}[+]\), \(\ctx.V_{\alloc,f}\), \(\ket{\varphi}\) and \(k\)
                for every state \(\ket{\varphi}\) with
                \(\ket{\varphi}\models\ket{0}_{\ctx.A_f\setminus\ctx.\rho_{\mathrm{loc},f}(\fp(f))}\). 

          \item \textbf{Code preservation.}
                \[
                  \ctx.C_{f}^{\,k}\ket{\varphi}
                  \;\models\;
                  \ket{\varphi}_{\,
                        \ctx.A_f\setminus\bigl( \ctx.V_{\activelabel,f}
                                              \cup \cod(\ctx.\rho_{\mathrm{loc},f}(\fp(f))) \bigr)}
                \]
                for every \(\ket{\varphi}\) with
                \(\ket{\varphi}\models\ket{0}_{\ctx.A_f\setminus\ctx.\rho_{\mathrm{loc},f}(\fp(f))}\).

          \item \textbf{Allocation consistency.}
                \[
                \begin{aligned}
                &\fp(f)\cup \re(f) \subseteq\dom(\ctx.\rho_{\mathrm{loc},f}), \\
                &\ctx.A_f \cap \cod(\rho_{\mathrm{glob}}) = \emptyset, \\
                &\bigl( \ctx.A_f \setminus \ctx.\rho_{\mathrm{loc},f}(\fp(f)) \setminus \ctx.\rho_{\mathrm{loc},f}(\re(f)) \bigr)
                  \cap
                  \bigl( \ctx.V_{\activelabel,[f^{*},f)} \cup \cod(\ctx.\rho_{\mathrm{loc},[f^{*},f)}) \bigr)
                  = \emptyset, \\
                &f \neq f_{\mathrm{cur}}^* \Longrightarrow  \ctx.\rho_{\mathrm{loc},f}(\re(f)) \subseteq \ctx.V_{\activelabel,\mathrm{prev}(f)}  \\ 
                &\ctx.\rho_{\mathrm{loc},f}(\fp(f))\cap \ctx.V_{\activelabel,f} = \emptyset, \\
                &\base\bigl( \cod(\ctx.\rho_{\mathrm{loc},f}) \setminus
                            \ctx.\rho_{\mathrm{loc},f}(\fp(f)) \setminus \ctx.\rho_{\mathrm{loc},f}(\re(f)) \bigr)
                  \subseteq \ctx.V_{\alloc,f} .
                \end{aligned}
                \]
          \end{enumerate}
  \item For every \(f\in\Fun(\ctx.\St)\cap\SCC(f_{\mathrm{cur}})\), the following hold.
        \begin{enumerate}
          \item \textbf{Allocation consistency.}
                \[
                \begin{aligned}
                &\bigl( \cod(\ctx.\rho_{\mathrm{loc},f}) \setminus
                        \ctx.\rho_{\mathrm{loc},f}(\fp(f)) \setminus \ctx.\rho_{\mathrm{loc},f}(\re(f)) \bigr)
                  \subseteq \ctx.V_{\activelabel,f}, \\ 
                & f \in \Theta^{-1}(\dom(\ctx.\D)) \setminus \{f_{\mathrm{cur}}\} \Longrightarrow \cod(\ctx.\rho_{\mathrm{loc},f}) \cap \cod(\ctx.\rho_{\mathrm{loc},[f^{*},f)})=\emptyset \\ 
                &\begin{cases}
                  \forall\, q \in \ctx.V_{\activelabel,\succeq f_{\mathrm{cur}}^*} \setminus (\ctx.V_{\activelabel,f_{\mathrm{cur}}} \cup \ctx.\rho_{\mathrm{loc},f_{\mathrm{cur}}}(\re(f_{\mathrm{cur}}))),\; 
                  \var(\idx(q)) \neq \emptyset,  \\ 
                  \forall\, q \in \ctx.V_{\activelabel,f_{\mathrm{cur}}} \cup \ctx.\rho_{\mathrm{loc},f_{\mathrm{cur}}}(\re(f_{\mathrm{cur}})),\; 
                  \var(\idx(q))=\emptyset \Longrightarrow \base[q]=q,
                \end{cases}  \\   
                &\forall\, q \in \ctx.V_{\activelabel,\succeq f_{\mathrm{cur}}^*} \cap \ctx.\rho_{\mathrm{loc},f_{\mathrm{cur}}}(\re(f_{\mathrm{cur}})),\; 
                  \var(\idx(q))=\emptyset \Longrightarrow  q \in \ctx.\rho_{\mathrm{loc},f}(\re(f))  
                \end{aligned}
                \]
          \item \textbf{Garbage discipline.}
                \[
                \begin{cases}
                  \ctx.V_{\activelabel,f} \setminus \ctx.\Gar_f \subseteq \cod(\ctx.\rho_{\mathrm{loc},f}), \\
                  h \neq f_{\mathrm{cur}} \;\Longrightarrow\; \ctx.Q_{\succeq f_{\mathrm{cur}}^*}[+] \cap \ctx.V_{\activelabel,h} = \emptyset, \\ 
                  \begin{aligned}
                    \ctx.Q_{\succeq f_{\mathrm{cur}}^*}[+] \cap \ctx.V_{\activelabel,f_{\mathrm{cur}}}\neq\emptyset
                    &\;\Longrightarrow\;
                      \exists\, x\in\dom(\ctx.\rho_{\mathrm{loc},f_{\mathrm{cur}}}),\;
                      x.\flag = 2 \;\wedge\;
                      \ctx.\rho_{\mathrm{loc},f_{\mathrm{cur}}}(x) \in \ctx.V_{\activelabel,f_{\mathrm{cur}}} \\
                    &\qquad \wedge\; \ctx.\Gar_{f_{\mathrm{cur}}}\subseteq \ctx.V_{\activelabel,f_{\mathrm{cur}}}.
                  \end{aligned}
                \end{cases}
                \]
          \item \textbf{Ancilla inclusion.}
                \(\anc(\ctx.C_{f}^{\,k}) \subseteq \ctx.A_f\).
          \item \textbf{Ancilla disjointness.}
                \(\bigl(\anc(\ctx.C_{f}^{\,k}) \setminus \bigl( \anc(F_{\succeq f_{\mathrm{cur}}^*}^k) \cup \anc(F_{\succeq f_{\mathrm{cur}}^*}^k)[+] \bigr)\bigr) \cap \ctx.B[+] = \emptyset\).

          \item \textbf{Index containment.}
                \(\var(\idx(\ctx.A_{f_{\mathrm{cur}}^*})) \subseteq \ctx.E_{\args,f_{\mathrm{cur}}^*} = \ctx.E_{\args,f}\).

          \item \textbf{Ancilla footprint.}
                 For every \(z\),
      \(
        \base\bigl(\anc(\ctx.C_f^{\,z})\bigr)
        \subseteq \ctx.V_{\alloc,\succeq f^{*}} \cup \base\bigl( \anc(F_{\succeq f_{\mathrm{cur}}^*}^z) \bigr).
      \)

          \item \textbf{Call closure.}
                \(
                  (\Succ(f) \setminus \Call(R_{\mathrm{stmt},f})) \cap \SCC(f)
                  \;\subseteq\;
                  \Fun(\ctx.\St). 
                \)
        \end{enumerate}
\end{enumerate} 

\paragraph{Proof strategy.}
Although the validity predicate \(\mathcal{V}\) bundles together all the
properties of a compilation context, internally it is organized into
several clauses according to the objects they describe---a structure
that is deliberately chosen to facilitate a layered proof.
Conceptually, these clauses can be grouped into two levels:
\begin{itemize}
  \item \textbf{Structural properties.} This level covers all conditions
        except the semantic ones listed below.  Their preservation proofs
        require only the structural properties of the previous context
        and the algorithmic steps; they never appeal to semantic
        \(k\)-equivalence conditions, the concrete behavior of
        compiled code, or the correctness of cleanup circuits.

  \item \textbf{Semantic properties.} This level consists mainly of the
        following:
          \begin{enumerate}
          \item declaration consistency with the specification family
                \(\D_{\SCC(f_{\mathrm{cur}})}\) (Condition~1(1));
          \item \(k\)-equivalence conditions (Condition~1(4));
          \item the correctness of cleanup circuits, i.e.\ the
                sub‑conditions (a) and (c) within the local validity
                Condition~3(a)(3);
          \item the validity of the compiled code (Condition~3(b)); 
          \item Condition~4(c) (containment of ancilla usage within the declaration).
          \end{enumerate}
        The common feature of these conditions is that their proofs
        rely on the structural properties of the previous context,
        on its semantic properties, and on the properties of the
        specification family.
\end{itemize}
This separation is crucial: it allows us to reuse the structural
invariants even when the semantic invariants have not yet been fully established.  This will be exploited in the proof for compiling a (mutually) recursive function, especially when extracting the
specification family for a recursive SCC; we will elaborate further
at that point.  

\paragraph{Notational convention for the proof.} 
In the subsequent proofs we often need to refer to components of several 
compilation contexts (the current one, its predecessor, or the result of 
a compilation step).  To keep the presentation manageable, we write 
	\(\star_i\) for \(\ctx_i.\star\) and \(\star'\) for \(\ctx'.\star\) whenever the 
	context is clear from the discussion.  The full prefix \(\ctx_i.\star\) or 
	\(\ctx'.\star\) is used only when we need to avoid ambiguity.

Before starting the main correctness proofs, we record one basic fact that will
be used implicitly throughout the subsequent proofs.  Static validity ensures
that the source functions being compiled satisfy the RQIMP typing rules, as
recorded in Lemma~\ref{lem:static-wf}.  As proved in
Theorem~\ref{theo:target-wf}, under these typing rules and the compilation
algorithms, the generated target declarations satisfy the semantic
well-formedness conditions of \(\RQC^{++}\) (corresponding to
Conditions~\ref{wf-rqc-coin}--\ref{wf-rqc-proc}, recalled in
Appendix~\ref{app:tar}) and therefore have well-defined semantics.  Hence, in
the following correctness proofs, we take the generated target declarations to
be semantically well formed and do not repeatedly re-check their
well-formedness conditions.

We first state several theorems that will be used repeatedly in the subsequent subsections. 

\begin{lemma}
\label{lem:curf}
Assume \(\bigl(\ctx,(\ket{\phi},k,\D_{\SCC(f_{\mathrm{cur}})})\bigr)\in \mathcal V\), and let
\[
R \triangleq \ctx.A_{f_{\mathrm{cur}}} \setminus \bigl(\ctx.V_{\activelabel,f_{\mathrm{cur}}} \cup \ctx.\rho_{\mathrm{loc},f_{\mathrm{cur}}}(\fp(f_{\mathrm{cur}}))\bigr).
\]
Then, for any quantum state \(\ket{\varphi}\) such that \(\ket{\varphi}\models \ket{0}_{\ctx.A_{f_{\mathrm{cur}}} \setminus \ctx.\rho_{\mathrm{loc},f_{\mathrm{cur}}}(\fp(f_{\mathrm{cur}}))}\), we have
\[
\ctx.C_{f_{\mathrm{cur}}}^{\,k}\ket{\varphi}\models \ket{0}_R.
\]
\end{lemma}

\begin{proof}
By the assumption, we have \(\ket{\varphi}\models \ket{0}_{\ctx.A_{f_{\mathrm{cur}}} \setminus \ctx.\rho_{\mathrm{loc},f_{\mathrm{cur}}}(\fp(f_{\mathrm{cur}}))}\). Since \(R \subseteq \ctx.A_{f_{\mathrm{cur}}}\), it follows that \(\ket{\varphi}\models \ket{0}_R\). Now by validity Clause~(3).(b), we have \[\ctx.C_{f_{\mathrm{cur}}}^{\,k}\ket{\varphi} \models \ket{\varphi}_{\ctx.A_{f_{\mathrm{cur}}} \setminus \bigl(\ctx.V_{\activelabel,f_{\mathrm{cur}}} \cup (\ctx.\rho_{\mathrm{loc},f_{\mathrm{cur}}}(\fp(f_{\mathrm{cur}})))\bigr)}.\] Therefore \(\ctx.C_{f_{\mathrm{cur}}}^{\,k}\ket{\varphi}\models \ket{0}_R\).
\end{proof}

\begin{lemma}
\label{lem:cv-clean}
Assume \(\bigl(\ctx,(\ket{\phi},k,\D_{\SCC(f_{\mathrm{cur}})})\bigr)\in \mathcal V.\) 
Then, for any quantum state \(\ket{\varphi}\) such that \(\ket{\varphi}\models \ket{0}_{\ctx.A_{f_{\mathrm{cur}}} \setminus (\ctx.\rho_{\mathrm{loc},f_{\mathrm{cur}}}(\fp(f_{\mathrm{cur}})))}\), we have
\[
\ctx.C_{f_{\mathrm{cur}}}^{\,k}\ket{\varphi}\models \ket{0}_{\ctx.cv}.
\]
\end{lemma}
\begin{proof}
By the initial-state consistency condition of \(\ctx\), we have \(\ctx.cv \cap \bigl(\ctx.V_{\activelabel,f_{\mathrm{cur}}} \cup \ctx.\rho_{\mathrm{loc},f_{\mathrm{cur}}}(\fp(f_{\mathrm{cur}}))\bigr)=\emptyset\). By definition, we have \(\ctx.cv \subseteq \ctx.A_{f_{\mathrm{cur}}}\). It follows from Lemma~\ref{lem:curf} that \(\ctx.C_{f_{\mathrm{cur}}}^{\,k}\ket{\varphi}\models \ket{0}_{\ctx.cv}\). 
\end{proof}

\subsection{Correctness Proof for the Cleanup Algorithm}
\label{app:proof-cleanup}
In this subsection we establish the correctness of Algorithm~\ref{clean_temp} for variable cleanup. Since the proof is rather large, we structure it as follows. We first fix the common 
notation and assumptions, then present the overall proof structure
together with the dependencies among the intermediate theorems, and
finally give the detailed proofs of those theorems.

To keep the presentation manageable, in the sequel we abbreviate
\(\req_{\rho_{f_{\mathrm{cur}}},\,v}\) as \(\req_v\) and
\(\req_{\rho_{f_{\mathrm{cur}},i},\,v}\) as \(\req_{v,i}\)
whenever the accompanying context makes the underlying local mapping clear. We also write \(\rho_{\mathrm{loc},\succeq f_{\mathrm{cur}},0} \setminus (-,\rho_{\mathrm{loc},f_{\mathrm{cur}},0}(V))\) for the sequence of maps obtained by removing from each constituent map any entry whose value lies in \(\rho_{\mathrm{loc},f_{\mathrm{cur}},0}(V)\).  

\begin{notation} \label{nota:clean}
  Let \(\ctx_0\) be a compilation context, \(\ket{\phi} \in \mathcal{H}\) a quantum state,
  \(k \in \mathbb{N}^+\) a depth, and \(\D_{\SCC(f_{\mathrm{cur}})}\) a family of
  target declarations. Fix a node \(v \in \mathsf{Node}(G_{f_{\mathrm{cur}}})\)
  and set
  \[
  U_v \triangleq \{\, x \in \operatorname{var}(v) \mid x.\flag \neq 2 \,\}.
  \]
  Define
  \[
  (\ctx, C_{\mathrm{cl}}) \triangleq \textsc{Clean\_Temp}(\K, \ctx_0, v).
  \]
  By definition of the cleanup operation we distinguish two cases
  according to the flag of \(v\).

  \noindent\textbf{Case 1: \(v.\flag = 0 \wedge \exists x \in \var(v), x.\flag \neq 2\).} 
  Then
  \begin{align*}
  \ctx = \ctx_0\bigl[
    &C_{f_{\mathrm{cur}},0} \mapsto C_{f_{\mathrm{cur}},0}; C_{\mathrm{cl}},\;
    V_{\activelabel,\succeq f_{\mathrm{cur}},0} \mapsto V_{\activelabel,\succeq f_{\mathrm{cur}},0} \setminus \rho_{\mathrm{loc},f_{\mathrm{cur}},0}(V),\\ 
    &\rho_{\mathrm{loc},\succeq f_{\mathrm{cur}},0} \mapsto \rho_{\mathrm{loc},\succeq f_{\mathrm{cur}},0} \setminus (-,\rho_{\mathrm{loc},f_{\mathrm{cur}},0}(V)),\; V_{\alloc,f_{\mathrm{cur}},0} \mapsto V_{\alloc,f_{\mathrm{cur}},0} \cup \base(\overline{p}),\\ 
    &\Qpool_0 \mapsto \Qpool,\;
    \kappa_{f_{\mathrm{cur}},0} \mapsto \kappa
  \bigr],
  \end{align*}
  where the updated components are given by
  \[
  \begin{aligned}
  &\overline{p} = \pop\bigl(\Qpool_0,\, |\req_{v,0}(\kappa_{f_{\mathrm{cur}},0}(v))|\bigr), \quad \Qpool'= \Qpool_0 \setminus \overline{p},  
  \quad
    C_{\mathrm{cl}} = \qinv\bigl(\kappa_{f_{\mathrm{cur}},0}(v)[\req_{v,0} \mapsto \overline{p}]\bigr),\\[1mm]
  &\Qpool = \push\bigl(\push(\Qpool', \overline{p}),\, \rho_{\mathrm{loc},f_{\mathrm{cur}},0}(V)\bigr), \ V = \{\, y \in \operatorname{var}(v) \mid
        y.\flag \neq 2 \;\land\; \rho_{\mathrm{loc},f_{\mathrm{cur}},0}(y) \in V_{\activelabel,f_{\mathrm{cur}},0} \,\} \\ 
  &\kappa = \kappa_{f_{\mathrm{cur}},0}\bigl[ 
            w \mapsto \qinv(C_{\mathrm{cl}}); \kappa_{f_{\mathrm{cur}},0}(w); C_{\mathrm{cl}}
            \mid w \in \dom(\kappa_{f_{\mathrm{cur}},0})  \wedge
         v \in \operatorname{pred}^+(w)
           \bigr].
  \end{aligned}
  \]

  \noindent\textbf{Case 2: \(v.\flag = 1\).}
  Then
  \begin{align*}
  \ctx = \ctx_0\bigl[
    &C_{f_{\mathrm{cur}},0} \mapsto C_{f_{\mathrm{cur}},0}; \qinv(\kappa_{f_{\mathrm{cur}},0}(v)),\;
    V_{\activelabel,\succeq f_{\mathrm{cur}},0} \mapsto V_{\activelabel,\succeq f_{\mathrm{cur}},0} \setminus \rho_{\mathrm{loc},f_{\mathrm{cur}},0}(V),\\ 
    &\rho_{\mathrm{loc},\succeq f_{\mathrm{cur}},0} \mapsto \rho_{\mathrm{loc},\succeq f_{\mathrm{cur}},0} \setminus (-,\rho_{\mathrm{loc},f_{\mathrm{cur}},0}(V)), 
    \Qpool_0 \mapsto \push(\Qpool_0, \rho_{\mathrm{loc},f_{\mathrm{cur}},0}(V))
  \bigr],
   \end{align*}
  where $ V = \{\, y \in \operatorname{var}(v) \mid
        \rho_{\mathrm{loc},f_{\mathrm{cur}},0}(y) \in V_{\activelabel,f_{\mathrm{cur}},0} \,\}$.
\end{notation}

\begin{assumption} \label{hyp:clean}
  In addition to Notation~\ref{nota:clean}, assume
  \[
  \bigl(\ctx_0, (\ket{\phi}, k, \D_{\SCC(f_{\mathrm{cur}})})\bigr) \in \mathcal{V}.
  \]
  Moreover, we assume \(v \notin \node(\re(f_{\mathrm{cur}}))\) and $\var(v) \cap \dom(\rho_{\mathrm{loc},f_{\mathrm{cur}},0})\neq \emptyset$; 
  and if \(v.\flag = 1\), assume additionally
  \[
  \req_{v,0}(\kappa_{f_{\mathrm{cur}},0}(v)) \subseteq
  A_{f_{\mathrm{cur}},0} \setminus \bigl( V_{\activelabel,f_{\mathrm{cur}},0} \cup \cod(\rho_{\mathrm{loc},f_{\mathrm{cur}},0}) \bigr).
  \]
\end{assumption}

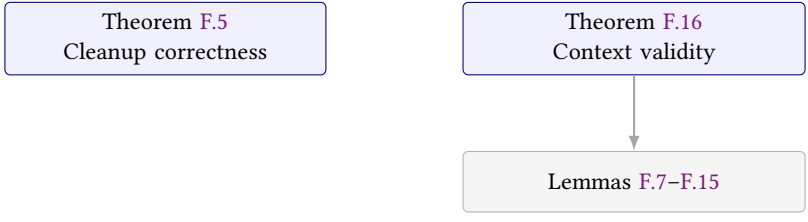
\begin{figure}[t]
\centering
\begin{tikzpicture}[
  node distance=9mm and 15mm,
  proofnode/.style={
    draw,
    rounded corners=2pt,
    align=center,
    inner xsep=5pt,
    inner ysep=4pt,
    minimum height=8mm,
    font=\small
  },
  thmnode/.style={proofnode, fill=blue!6, draw=blue!45!black},
  lemmanode/.style={proofnode, fill=gray!8, draw=gray!60},
  dep/.style={-{Latex[length=2mm]}, thick, draw=gray!70}
]
\node[thmnode, text width=.28\linewidth] (clean)
  {Theorem~\ref{theo:clean}\\
   Cleanup correctness};
\node[thmnode, text width=.30\linewidth, right=18mm of clean] (valid)
  {Theorem~\ref{theo:clean-valid}\\
   Context validity};
\node[lemmanode, text width=.30\linewidth, below=10mm of valid] (lemmas)
  {Lemmas~\ref{lemma:con-2}--\ref{lemma:con-4.f}};
\draw[dep] (valid) -- (lemmas);
\end{tikzpicture}
\caption{Proof dependencies for the correctness of Clean\_Temp; arrows point from a result to the lemmas or theorems it depends on.}
\label{fig:clean-proof-dep}
\Description{A dependency diagram showing that cleanup correctness follows from Theorem clean and Theorem clean-valid, and that Theorem clean-valid depends on subsidiary lemmas.}
\end{figure}

As depicted in Figure~\ref{fig:clean-proof-dep}, the correctness of \textsc{Clean\_Temp} algorithm follows from two constituent theorems,
Theorems~\ref{theo:clean} and~\ref{theo:clean-valid}.  Theorem~\ref{theo:clean} establishes the semantic
correctness of the cleanup algorithm itself; its proof directly appeals
to the cleanup‑circuit correctness condition that is part of the
validity of \(\ctx_0\).  Theorem~\ref{theo:clean-valid} shows that the output compilation
context remains valid.  It verifies each clause of the validity
predicate in turn; several of the more involved conditions are handled
by subsidiary lemmas (Lemmas~\ref{lemma:con-2} to~\ref{lemma:con-4.f}). In particular, Lemmas~\ref{lemma:clean-inv} and~\ref{lemma:Aset} describe the
relation between certain compilation state of the input context
\(\ctx_0\) and that of the output context \(\ctx\).

\begin{lemma}\label{lemma:clean-inv} 
    Under Notation~\ref{nota:clean}, we have \[\ctx.(\D, \delta,f_{\mathrm{cur}},R_{\mathrm{stmt}})=\ctx_0.(\D, \delta, f_{\mathrm{cur}},R_{\mathrm{stmt}}),\] $\Fun(\St)=\Fun(\St_0)$, $\dom(\M)=\dom(\M_0)$, $\M \upharpoonright_{\Fun(\St_{\prec f_{\mathrm{cur}}})}=\M \upharpoonright_{\Fun(\St_{\prec f_{\mathrm{cur}},0})}$ and $\St_{\preceq f_{\mathrm{cur}}}=\St_{\preceq f_{\mathrm{cur}},0}$. Besides, for $h$ such that $\pos(h)\geq \pos(f_{\mathrm{cur}})$ \(\rho_{\mathrm{loc},h}=\rho_{\mathrm{loc},h,0} \upharpoonright_{\{x \in \dom(\rho_{\mathrm{loc},h,0})\mid \rho_{\mathrm{loc},h,0}(x) \not\subseteq \rho_{\mathrm{loc},f_{\mathrm{cur}},0}(U_v) \cap V_{\activelabel,f_{\mathrm{cur}},0} \}}\), $V_{\activelabel,h}=V_{\activelabel,h,0} \setminus \rho_{\mathrm{loc},f_{\mathrm{cur}},0}(U_v)$, $E_{\args,h}=E_{\args,h,0}$.  
\end{lemma}
\begin{proof}
  This conclusion follows directly from the definition of \(\ctx\).
\end{proof}

\begin{theorem}
  \label{theo:clean}
  \label{theo:correctness-cleanup}
  Under the Notation~\ref{nota:clean} and Assumption~\ref{hyp:clean},
  \[
  \bigl\langle (C_{f_{\mathrm{cur}},0};\,C_{\mathrm{cl}})^k,\ket{\phi}\bigr\rangle
  \to^{*}
  \bigl\langle \downarrow,\,
  (C_{f_{\mathrm{cur}},0}^k\ket{\phi})_{\qv(C_{f_{\mathrm{cur}},0}^k)\setminus \rho_{\mathrm{loc},f_{\mathrm{cur}},0}(v)}
  \otimes \ket{\theta}_{\rho_{\mathrm{loc},f_{\mathrm{cur}},0}(\var(v) \setminus U_v)}
  \otimes \ket{\phi}_{\rho_{\mathrm{loc},f_{\mathrm{cur}},0}(U_v)}
  \bigr\rangle,
  \]
  for some state \(\ket{\theta}\).
\end{theorem}

\begin{proof}
  If \(v.\flag = 0\) and \(\forall x \in \var(v),\, x.\flag = 2\), then the algorithm performs no operation, sets \(C_{\mathrm{cl}}\) to \(I\), and the conclusion holds trivially; the validity of the subsequent output state is also satisfied. Hence, without loss of generality, we assume that when \(v.\flag = 0\), there exists some \(x \in \var(v)\) with \(x.\flag \neq 2\).

Since \(\bigl(\ctx_0,(\ket{\phi},k,\D_{\SCC(f_{\mathrm{cur}})})\bigr)\in\mathcal{V}\), the cleanup circuit \(\kappa_0\) is valid.  
From the assumptions \(v \notin \node(\re(f_{\mathrm{cur}}))\) and \(\var(v) \cap \dom(\rho_{\mathrm{loc},f_{\mathrm{cur}},0}) \neq \emptyset\), it follows that there exists a \(q \in \cod(\rho_{\mathrm{loc},f_{\mathrm{cur}},0}) \setminus \rho_{\mathrm{loc},f_{\mathrm{cur}},0}(\re(f_{\mathrm{cur}}))\) such that \(\rho_{\mathrm{loc},f_{\mathrm{cur}},0}^{-1}(q).\flag \neq 2\) and \(v = \node(\rho_{\mathrm{loc},f_{\mathrm{cur}},0}^{-1}(q))\). Consequently, \(v \in \dom(\kappa_0)\). Then, for any quantum state \(\ket{\varphi}\) satisfying  
\( \ket{\varphi} \models \ket{0}_{\ctx_0.A_{f_{\mathrm{cur}}} \setminus \rho_{\mathrm{loc},f_{\mathrm{cur}},0}(\fp(f_{\mathrm{cur}}))}, \)
any state \(\ket{\psi}\) such that  
\(
\langle C_{f_{\mathrm{cur}},0}^{\,k}, \ket{\varphi} \rangle \to^{*} \langle \downarrow, \ket{\psi} \rangle,
\)  
and any tuple of fresh quantum variables \(\overline{q}\) with  
\[
\ket{\psi} \models \ket{0}_{\overline{q}},\qquad
\overline{q} \cap \bigl( \cod(\rho_{f_{\mathrm{cur}},0}) \cup V_{\activelabel,f_{\mathrm{cur}},0} \bigr) = \emptyset,
\]  
we have  
\[
\begin{aligned}
&\bigl\langle \bigl(C_{f_{\mathrm{cur}},0}^k;\, \qinv(\kappa_{f_{\mathrm{cur}},0}(v)^k[\req_{v,0} \mapsto \overline{q}])\bigr),\, \ket{\varphi} \bigr\rangle \\
\to^{*}&\;
\bigl\langle \downarrow,\,
\ket{\varphi}_{\rho_{\mathrm{loc},f_{\mathrm{cur}},0}(U_v)}
\otimes \ket{\theta}_{\rho_{\mathrm{loc},f_{\mathrm{cur}},0}(\var(v)\setminus U_v)}
\otimes \ket{\psi}_{\qv(C_{f_{\mathrm{cur}},0}^k)\setminus \rho_{\mathrm{loc},f_{\mathrm{cur}},0}(v)}
\bigr\rangle .
\end{aligned}
\] 

  Now let \(\ket{\psi}\) satisfy 
  \(\langle C_{f_{\mathrm{cur}},0}^{\,k}, \ket{\phi} \rangle \to^{*} \langle \downarrow, \ket{\psi} \rangle\).
  By validity of \(\ctx_0\) we have
  \(\ket{\phi} \models \ket{0}_{\ctx_0.A_{f_{\mathrm{cur}}} \setminus \rho_{\mathrm{loc},f_{\mathrm{cur}},0}(\fp(f_{\mathrm{cur}}))}\).
  Hence, to apply the cleanup-circuit correctness clause above, it remains
  to exhibit a tuple \(\overline{q}\) such that
  \(\overline{q} \cap \bigl( \cod(\rho_{f_{\mathrm{cur}},0}) \cup V_{\activelabel,f_{\mathrm{cur}},0} \bigr) = \emptyset\)
  and \(\ket{\psi} \models \ket{0}_{\overline{q}}\).

  We distinguish two cases.

  \noindent\textbf{Case 1: \(v.\flag = 0\).}
  Let \(\overline{q} \triangleq \overline{p}\), where
  \(\overline{p} = \qv(C_{\mathrm{cl}}) \setminus \rho_{f_{\mathrm{cur}},0}\bigl(\operatorname{pred}^*(v)\bigr)\). 
  By definition of \(\textsc{Clean\_Temp}\), the tuple \(\overline{p}\) is popped
  from \(\Qpool_0\); hence \(\overline{p} \subseteq \Qpool_0 \subseteq cv_0\).
  From the validity of \(\ctx_0\) and Lemma~\ref{lem:curf} we obtain
  \(cv_0 \cap \bigl( \cod(\rho_{f_{\mathrm{cur}},0}) \cup V_{\activelabel,f_{\mathrm{cur}},0} \bigr) = \emptyset\) and \(\ket{\psi} \models \ket{0}_{cv_0}\).  Consequently, \(\overline{p} \subseteq cv_0\) implies \(\ket{\psi} \models \ket{0}_{\overline{p}}\). 

  \noindent\textbf{Case 2: \(v.\flag = 1\).}
  Let \(\overline{q} \triangleq \req_{v,0}(\kappa_{f_{\mathrm{cur}},0}(v))\).
  By hypothesis,
  \(\req_{v,0}(\kappa_{f_{\mathrm{cur}},0}(v)) \subseteq A_{f_{\mathrm{cur}},0} \setminus
  \bigl( V_{\activelabel,f_{\mathrm{cur}},0} \cup \cod(\rho_{\mathrm{loc},f_{\mathrm{cur}},0}) \bigr)\).
  Since $A_{f_{\mathrm{cur}},0} \cap \cod(\rho_{\mathrm{glob}})=\emptyset$, 
  \(\overline{q} \cap \bigl( \cod(\rho_{f_{\mathrm{cur}},0}) \cup V_{\activelabel,f_{\mathrm{cur}},0} \bigr) = \emptyset\). 
  Again, Lemma~\ref{lem:curf} and the validity of \(\ctx_0\) yield
  \(\ket{\psi} \models \ket{0}_{\overline{q}}\).

  Thus the cleanup-circuit correctness clause applies in both cases.
  Recalling that \[C_{\mathrm{cl}} = \qinv\bigl(\kappa_{f_{\mathrm{cur}},0}(v)[\req_{v,0} \mapsto \overline{p}]\bigr)\]
  in Case~1 and \(C_{\mathrm{cl}} = \qinv(\kappa_{f_{\mathrm{cur}},0}(v))\) in Case~2, we obtain 
  \[
  \bigl\langle (C_{f_{\mathrm{cur}},0};\,C_{\mathrm{cl}})^{\,k},\, \ket{\phi} \bigr\rangle
  \to^{*}
  \bigl\langle \downarrow,\,
  \ket{\phi}_{\rho_{\mathrm{loc},f_{\mathrm{cur}},0}(U_v)}
  \otimes \ket{\theta}_{\rho_{\mathrm{loc},f_{\mathrm{cur}},0}(\var(v)\setminus U_v)}
  \otimes \ket{\psi}_{\qv(C_{f_{\mathrm{cur}},0}^k)\setminus \rho_{\mathrm{loc},f_{\mathrm{cur}},0}(v)} 
  \bigr\rangle.
  \]
\end{proof} 

\begin{lemma}\label{lemma:Aset}
 Under Notation~\ref{nota:clean} and assuming that \(\ctx_0\) satisfies the structural properties required for validity, the following hold:
\begin{itemize}
    \item \(\ctx.cv = \ctx_0.cv \cup \rho_{\mathrm{loc},f_{\mathrm{cur}},0}(V)\).
    \item \(\ctx.A_f = \ctx_0.A_f\) for every
          \(f \in \Fun(\St) \cap \SCC(f_{\mathrm{cur}})\) with \(\pos(f) \le \pos(f_{\mathrm{cur}})\).
    \item \(\ctx.A_f = \ctx_0.A_f \cup \rho_{\mathrm{loc},f_{\mathrm{cur}},0}(V)\) for every
          \(f \in \Fun(\St) \cap \SCC(f_{\mathrm{cur}})\) with \(\pos(f) > \pos(f_{\mathrm{cur}})\).
\end{itemize} 
\end{lemma}

\begin{proof}
  In both cases, by definition of \(\textsc{Clean\_Temp}\) we have
  \(\Qpool = \push(\Qpool_0, \rho_{\mathrm{loc},f_{\mathrm{cur}},0}(V))\),
  $V_{\activelabel,\succeq f_{\mathrm{cur}}} = V_{\activelabel,\succeq f_{\mathrm{cur}},0} \setminus \rho_{\mathrm{loc},f_{\mathrm{cur}},0}(V)$ and $ \rho_{\mathrm{loc},\succeq f_{\mathrm{cur}}} = \rho_{\mathrm{loc},\succeq f_{\mathrm{cur}},0} \setminus (-,\rho_{\mathrm{loc},f_{\mathrm{cur}},0}(V))$. 
  Let
  \(R_v \triangleq \rho_{\mathrm{loc},f_{\mathrm{cur}},0}(V)\),
  where
  \[
  V =
  \begin{cases}
    \{\, y \in \operatorname{var}(v) \mid
       y.\flag \neq 2 \;\land\; \rho_{\mathrm{loc},f_{\mathrm{cur}},0}(y) \in V_{\activelabel,f_{\mathrm{cur}},0} \,\}, & \text{if } v.\flag = 0,\\[2mm]
    \{\, y \in \operatorname{var}(v) \mid
       \rho_{\mathrm{loc},f_{\mathrm{cur}},0}(y) \in V_{\activelabel,f_{\mathrm{cur}},0} \,\}, & \text{if } v.\flag = 1 .
  \end{cases}
  \]
  By construction of \(V\) we have
  \(R_v \subseteq V_{\activelabel,f_{\mathrm{cur}},0} \cap \cod(\rho_{\mathrm{loc},f_{\mathrm{cur}},0})\).
  The validity of \(\ctx_0\) gives \(\cod(\rho_{\mathrm{loc},f_{\mathrm{cur}},0}) \cap Q_{\succeq f_{\mathrm{cur}}^*,0}[+] = \emptyset\), hence \(R_v \cap Q_{\succeq f_{\mathrm{cur}}^*,0}[+] = \emptyset\). On the other hand, by definition, \(Q_{\succeq f_{\mathrm{cur}}^*,0}\) collects the registers corresponding to all variables with \(\flag=2\) in the relevant functions, while \(V\) contains no variable with \(\flag=2\) in either case. Moreover, we can easily show that \(V\) does not contain the formal parameters or the return value:

\begin{itemize}
  \item For \(\fp(f_{\mathrm{cur}})\): validity Condition~3(c) of \(\ctx_0\) gives \(\rho_{\mathrm{loc},f_{\mathrm{cur}},0}(\fp(f_{\mathrm{cur}})) \cap V_{\activelabel,f_{\mathrm{cur}},0} = \emptyset\); together with \(V \subseteq V_{\activelabel,f_{\mathrm{cur}},0}\) we obtain \(V \cap \fp(f_{\mathrm{cur}}) = \emptyset\).  
  \item For \(\re(f_{\mathrm{cur}})\): the assumption \(v \notin \node(\re(f_{\mathrm{cur}}))\) implies \(\re(f_{\mathrm{cur}}) \cap \var(v) = \emptyset\), so \(V \cap \re(f_{\mathrm{cur}}) = \emptyset\).  
\end{itemize} 
By the injectivity of the local mapping, it follows that \(R_v\) does not contain the images of the formal parameters or the return value (i.e., $\rho_{\mathrm{loc},f_{\mathrm{cur}},0}(\fp(f_{\mathrm{cur}}) \cup \re(f_{\mathrm{cur}})) \cap R_v =\emptyset$). Consequently, the validity of \(\ctx_0\) ensures that \(R_v\) is disjoint from the local mappings of any other function; together with the injectivity of the current local mapping, we obtain \(R_v \cap Q_{\succeq f_{\mathrm{cur}}^*,0} = \emptyset\). Note that by definition, both \(R_v\) and \(Q_{\succeq f_{\mathrm{cur}}^*,0}\) are subsets of \(\rho_{\mathrm{loc},\succeq f_{\mathrm{cur}}^*,0}\), and the validity of \(\ctx_0\) guarantees \(\rho_{\mathrm{loc},\succeq f_{\mathrm{cur}}^*,0} \cap \rho_{\mathrm{loc},\succeq f_{\mathrm{cur}}^*,0}[+] = \emptyset\); hence \(R_v[+] \cap Q_{\succeq f_{\mathrm{cur}}^*,0}[+] = \emptyset\). Moreover, from \(R_v \cap Q_{\succeq f_{\mathrm{cur}}^*,0} = \emptyset\) we can further deduce that \(Q_{\succeq f_{\mathrm{cur}}^*}[+] = Q_{\succeq f_{\mathrm{cur}}^*,0}[+]\) holds. 

 Based on this, we analyze the update of the clean variable pool. Recall that
  \[
  cv
  = \Qpool \cup \Qpool[+]
    \cup
    \Bigl(
      \bigl(V_{\activelabel,\succeq f_{\mathrm{cur}}^{*}}[+] \cup \cod(\rho_{\mathrm{loc},\succeq f_{\mathrm{cur}}^{*}})[+]\bigr)
      \setminus
      \bigl(Q_{\succeq f_{\mathrm{cur}}^{*}}[+] \cap V_{\activelabel,f_{\mathrm{cur}}}\bigr)
    \Bigr).
  \] 

From Condition~3(c) in the validity of \(\ctx_0\), together with the fact that \(R_v\) does not contain the images of the formal parameters or the return value (as argued above), it follows that for any \(f\) with \(\pos(f) < \pos(f_{\mathrm{cur}})\),
\[
(V_{\activelabel,f,0} \cup \cod(\rho_{\mathrm{loc},f,0})) \cap R_v = \emptyset.
\]
Therefore we obtain
\[
V_{\activelabel,\succeq f_{\mathrm{cur}}^{*}} = V_{\activelabel,\succeq f_{\mathrm{cur}}^{*},0} \setminus R_v,
\qquad
\cod(\rho_{\mathrm{loc},\succeq f_{\mathrm{cur}}^{*}}) = \cod(\rho_{\mathrm{loc},\succeq f_{\mathrm{cur}}^{*},0}) \setminus R_v. 
\] 
Substituting these updates into the definition of \(cv\) yields
\(cv = cv_0 \cup R_v\), as shown in Figure~\ref{fig:cv-trans-clean}.  Thus the cleanup operation returns exactly the registers in \(R_v\) to the clean pool.  
\begin{figure}[t]
\begin{align}
cv
&= (\Qpool_0 \cup R_v) \cup (\Qpool_0[+] \cup R_v[+]) \notag\\
&\qquad\cup
\Bigl(
\bigl((V_{\activelabel,\succeq f_{\mathrm{cur}}^{*},0}\setminus R_v)[+] \cup (\cod(\rho_{\mathrm{loc},\succeq f_{\mathrm{cur}}^{*},0})\setminus R_v)[+]\bigr)
\setminus
\bigl(Q_{\succeq f_{\mathrm{cur}}^*,0}[+]\cap V_{\activelabel,f_{\mathrm{cur}}}\bigr)
\Bigr) \tag{by the updates of \(\Qpool\), \(\Vactive\), \(\rho\) and \(Q_{\succeq f_{\mathrm{cur}}^*}[+] = Q_{\succeq f_{\mathrm{cur}}^*,0}[+]\)}\\
&= (\Qpool_0 \cup R_v) \cup (\Qpool_0[+] \cup R_v[+]) \notag\\
&\qquad\cup
\Bigl(
\bigl((V_{\activelabel,\succeq f_{\mathrm{cur}}^{*},0}[+] \setminus R_v[+]) \cup (\cod(\rho_{\mathrm{loc},\succeq f_{\mathrm{cur}}^{*},0})[+] \setminus R_v[+])\bigr)
\setminus
\bigl(Q_{\succeq f_{\mathrm{cur}}^*,0}[+]\cap V_{\activelabel,f_{\mathrm{cur}}}\bigr)
\Bigr) \tag{since \((S\setminus R_v)[+] = S[+]\setminus R_v[+]\)}\\
&= (\Qpool_0 \cup R_v) \cup \Qpool_0[+] \cup R_v[+] \notag\\
&\qquad\cup
\Bigl(
\bigl((V_{\activelabel,\succeq f_{\mathrm{cur}}^{*},0}[+] \cup \cod(\rho_{\mathrm{loc},\succeq f_{\mathrm{cur}}^{*},0})[+]) \setminus R_v[+]\bigr)
\setminus
\bigl(Q_{\succeq f_{\mathrm{cur}}^*,0}[+]\cap V_{\activelabel,f_{\mathrm{cur}}}\bigr)
\Bigr) \tag{since \((A\setminus C)\cup(B\setminus C)=(A\cup B)\setminus C\)}\\
&= (\Qpool_0 \cup R_v) \cup \Qpool_0[+] \notag\\
&\qquad\cup
\Bigl(
R_v[+]\cup
\Bigl(
\bigl((V_{\activelabel,\succeq f_{\mathrm{cur}}^{*},0}[+] \cup \cod(\rho_{\mathrm{loc},\succeq f_{\mathrm{cur}}^{*},0})[+]) \setminus R_v[+]\bigr)
\setminus
\bigl(Q_{\succeq f_{\mathrm{cur}}^*,0}[+]\cap V_{\activelabel,f_{\mathrm{cur}}}\bigr)
\Bigr)
\Bigr) \tag{by associativity and commutativity of \(\cup\)}\\
&= (\Qpool_0 \cup R_v) \cup \Qpool_0[+] \notag\\
&\qquad\cup
\Bigl(
\bigl(R_v[+] \cup ((V_{\activelabel,\succeq f_{\mathrm{cur}}^{*},0}[+] \cup \cod(\rho_{\mathrm{loc},\succeq f_{\mathrm{cur}}^{*},0})[+]) \setminus R_v[+])\bigr)
\setminus
\bigl(Q_{\succeq f_{\mathrm{cur}}^*,0}[+]\cap V_{\activelabel,f_{\mathrm{cur}}}\bigr)
\Bigr) \tag{since \(R_v[+] \cap (Q_{\succeq f_{\mathrm{cur}}^*,0}[+]\cap V_{\activelabel,f_{\mathrm{cur}}}) = \emptyset\)}\\
&= (\Qpool_0 \cup R_v) \cup \Qpool_0[+] \notag\\
&\qquad\cup
\Bigl(
(V_{\activelabel,\succeq f_{\mathrm{cur}}^{*},0}[+] \cup \cod(\rho_{\mathrm{loc},\succeq f_{\mathrm{cur}}^{*},0})[+])
\setminus
\bigl(Q_{\succeq f_{\mathrm{cur}}^*,0}[+]\cap V_{\activelabel,f_{\mathrm{cur}}}\bigr)
\Bigr) \tag{since \(R_v[+] \subseteq V_{\activelabel,\succeq f_{\mathrm{cur}}^{*},0}[+] \cup \cod(\rho_{\mathrm{loc},\succeq f_{\mathrm{cur}}^{*},0})[+]\)}\\
&= (\Qpool_0 \cup R_v) \cup \Qpool_0[+] \notag\\
&\qquad\cup
\Bigl(
(V_{\activelabel,\succeq f_{\mathrm{cur}}^{*},0}[+] \cup \cod(\rho_{\mathrm{loc},\succeq f_{\mathrm{cur}}^{*},0})[+])
\setminus
\bigl(Q_{\succeq f_{\mathrm{cur}}^*,0}[+]\cap V_{\activelabel,f_{\mathrm{cur}},0}\bigr)
\Bigr) \tag{since \(R_v \cap Q_{\succeq f_{\mathrm{cur}}^*,0}[+] = \emptyset\)}\\
&= cv_0 \cup R_v. 
\end{align}
\caption{Transformation of the Clean Variable Pool in the \textsc{Clean\_Temp} Algorithm}
\label{fig:cv-trans-clean}
\Description{...}
\end{figure}

  More generally, for any function \(f \in \Fun(\St) \cap \SCC(f_{\mathrm{cur}})\),
  the accessible set \(A_f\) is defined by
  \[
  A_f = cv \cup V_{\activelabel,\succeq f} \cup \cod(\rho_{\mathrm{loc},f}).
  \]
  Using the equalities established above, we obtain:
  \begin{itemize}
    \item If \(\pos(f) \le \pos(f_{\mathrm{cur}})\), then
          \(V_{\activelabel,f_{\mathrm{cur}},0} \subseteq V_{\activelabel,\succeq f,0}\) and therefore
          \[
          \begin{aligned}
          A_f
          = cv \cup V_{\activelabel,\succeq f} \cup \cod(\rho_{\mathrm{loc},f}) &= (cv_0 \cup R_v) \cup (V_{\activelabel,\succeq f,0} \setminus R_v) \cup \cod(\rho_{\mathrm{loc},f,0}) \\
          &= cv_0 \cup V_{\activelabel,\succeq f,0} \cup \cod(\rho_{\mathrm{loc},f,0}) = A_{f,0}. 
          \end{aligned}
          \]
    \item If \(\pos(f) > \pos(f_{\mathrm{cur}})\), then
          \[
          \begin{aligned}
          A_f
          = cv \cup V_{\activelabel,\succeq f} \cup \cod(\rho_{\mathrm{loc},f})
          &= (cv_0 \cup R_v) \cup (V_{\activelabel,\succeq f,0} \setminus R_v) \cup \cod(\rho_{\mathrm{loc},f,0}) \\
          &= cv_0 \cup V_{\activelabel,\succeq f,0} \cup \cod(\rho_{\mathrm{loc},f,0}) \cup R_v = A_{f,0} \cup R_v.
          \end{aligned}
          \]
  \end{itemize}
\end{proof}

\begin{lemma}\label{lemma:con-2}
Under Notation~\ref{nota:clean} and Assumption~\ref{hyp:clean}, Condition~2 in the validity of \(\ctx\) holds.
\end{lemma} 
\begin{proof}
    First of all, by Lemma~\ref{lemma:Aset} the accessible set of \(f_{\mathrm{cur}}\) is
preserved, i.e. \(\ctx.A_{f_{\mathrm{cur}}} = \ctx_0.A_{f_{\mathrm{cur}}}\).
Since \(\rho_{\mathrm{loc},f_{\mathrm{cur}}} = \rho_{\mathrm{loc},f_{\mathrm{cur}},0} \setminus (-, R_v)\) and $\rho_{\mathrm{loc},f_{\mathrm{cur}},0}(\fp(f_{\mathrm{cur}})) \cap R_v =\emptyset$, \(\rho_{\mathrm{loc},f_{\mathrm{cur}}}(\fp(f_{\mathrm{cur}})) = \rho_{\mathrm{loc},f_{\mathrm{cur}},0}(\fp(f_{\mathrm{cur}}))\). Together with \(\ket{\phi} \models \ket{0}_{A_{f_{\mathrm{cur}},0} \setminus \rho_{\mathrm{loc},f_{\mathrm{cur}},0}(\fp(f_{\mathrm{cur}}))}\) 
(which follows from the validity of \(\ctx_0\)), we obtain
\[
  \ket{\phi} \models \ket{0}_{A_{f_{\mathrm{cur}}} \setminus \rho_{\mathrm{loc},f_{\mathrm{cur}}}(\fp(f_{\mathrm{cur}}))}.
\]

Again by Lemma~\ref{lemma:Aset} we have \(cv = cv_0 \cup R_v\).
Validity Condition~2 for \(\ctx_0\) yields
\(cv_0 \cap \bigl( V_{\activelabel,\succeq f_{\mathrm{cur}}^*,0} \cup \cod(\rho_{\mathrm{loc},\succeq f_{\mathrm{cur}}^*,0}) \cup \rho_{\mathrm{glob},0} \bigr) = \emptyset\);
in particular \(R_v \cap cv_0 = \emptyset\) and $R_v \subseteq A_{f_{\mathrm{cur}}}$ since \(R_v \subseteq V_{\activelabel,f_{\mathrm{cur}},0}\). 
Moreover, directly from the construction we obtain the global mapping \(\rho_{\mathrm{glob}}\) remains unchanged and 
\[
  V_{\activelabel,\succeq f_{\mathrm{cur}}^*} = V_{\activelabel,\succeq f_{\mathrm{cur}}^*,0} \setminus R_v,\qquad
  \cod(\rho_{\mathrm{loc},\succeq f_{\mathrm{cur}}^*}) = \cod(\rho_{\mathrm{loc},\succeq f_{\mathrm{cur}}^*,0}) \setminus R_v. 
\]

Consequently,
\[
\begin{aligned}
&cv \cap \bigl( V_{\activelabel,\succeq f_{\mathrm{cur}}^*} \cup \cod(\rho_{\mathrm{loc},\succeq f_{\mathrm{cur}}^*}) \cup \rho_{\mathrm{glob}} \bigr) \\
 = & (cv_0 \cup R_v) \cap \bigl( (V_{\activelabel,\succeq f_{\mathrm{cur}}^*,0}\setminus R_v)
        \cup (\cod(\rho_{\mathrm{loc},\succeq f_{\mathrm{cur}}^*,0})\setminus R_v) \cup \rho_{\mathrm{glob},0} \bigr)
      = \emptyset .
\end{aligned}
\]
Indeed, \(cv_0 \cap \bigl( V_{\activelabel,\succeq f_{\mathrm{cur}}^*,0} \cup \cod(\rho_{\mathrm{loc},\succeq f_{\mathrm{cur}}^*,0}) \cup \rho_{\mathrm{glob},0} \bigr) = \emptyset\)
by validity Condition~2; \(R_v \cap \rho_{\mathrm{glob},0} = \emptyset\) by validity
Condition~3(c); and \(R_v\) is disjoint from \(V_{\activelabel,\succeq f_{\mathrm{cur}}^*,0}\setminus R_v\)
and \(\cod(\rho_{\mathrm{loc},\succeq f_{\mathrm{cur}}^*,0})\setminus R_v\) by construction.

Thirdly, from the validity of \(\ctx_0\) we have
\(Q_{\succeq f_{\mathrm{cur}}^*,0}[+] \cap \bigl( \cod(\rho_{\mathrm{loc},\succeq f_{\mathrm{cur}}^*,0}) \cup \rho_{\mathrm{glob},0} \bigr) = \emptyset\).
Since \(Q_{\succeq f_{\mathrm{cur}}^*}[+] = Q_{\succeq f_{\mathrm{cur}}^*,0}[+]\), \(\rho_{\mathrm{glob}} = \rho_{\mathrm{glob},0}\) and
\(\cod(\rho_{\mathrm{loc},\succeq f_{\mathrm{cur}}^*}) = \cod(\rho_{\mathrm{loc},\succeq f_{\mathrm{cur}}^*,0}) \setminus R_v
    \subseteq \cod(\rho_{\mathrm{loc},\succeq f_{\mathrm{cur}}^*,0})\),
we immediately obtain
\[
  Q_{\succeq f_{\mathrm{cur}}^*}[+] \cap \bigl(\cod(\rho_{\mathrm{loc},\succeq f_{\mathrm{cur}}^*}) \cup \rho_{\mathrm{glob}} \bigr) = \emptyset.
\]

From the above we already have
\[
  \Qpool = \Qpool_0 \cup R_v,\qquad
  V_{\activelabel,\succeq f_{\mathrm{cur}}^*} = V_{\activelabel,\succeq f_{\mathrm{cur}}^*,0} \setminus R_v,\qquad
  \cod(\rho_{\mathrm{loc},\succeq f_{\mathrm{cur}}^*}) = \cod(\rho_{\mathrm{loc},\succeq f_{\mathrm{cur}}^*,0}) \setminus R_v. 
\]
Since \(R_v \subseteq V_{\activelabel,\succeq f_{\mathrm{cur}}^*,0} \cup \cod(\rho_{\mathrm{loc},\succeq f_{\mathrm{cur}}^*,0})\),
the overall union \(\Qpool[+] \cup V_{\activelabel,\succeq f_{\mathrm{cur}}^*}[+] \cup \cod(\rho_{\mathrm{loc},\succeq f_{\mathrm{cur}}^*})[+]\)
remains identical to the corresponding union in \(\ctx_0\).
Moreover \(\Qpool = \Qpool_0 \cup R_v \subseteq \Qpool_0 \cup \cod(\rho_{\mathrm{loc},f_{\mathrm{cur}},0})\).
The validity Condition~2 of \(\ctx_0\) guarantees
\[
  (\Qpool_0 \cup \cod(\rho_{\mathrm{loc},f_{\mathrm{cur}},0})) \cap
  \bigl( \Qpool_0[+] \cup V_{\activelabel,\succeq f_{\mathrm{cur}}^*,0}[+] \cup \cod(\rho_{\mathrm{loc},\succeq f_{\mathrm{cur}}^*,0})[+] \bigr) = \emptyset ,
\]
hence 
\[
  \Qpool \cap \bigl( \Qpool[+] \cup V_{\activelabel,\succeq f_{\mathrm{cur}}^*}[+] \cup \cod(\rho_{\mathrm{loc},\succeq f_{\mathrm{cur}}^*})[+] \bigr) = \emptyset .
\]

Finally, since \(R_v=\rho_{\mathrm{loc},f_{\mathrm{cur}},0}(V) \subseteq V_{\activelabel,f_{\mathrm{cur}},0} \setminus \rho_{\mathrm{loc},f_{\mathrm{cur}},0}(\re(f_{\mathrm{cur}}))\),
we have
\[
  \Qpool = \Qpool_0 \cup R_v 
      \subseteq \Qpool_0 \cup \bigl( V_{\activelabel,f_{\mathrm{cur}},0} \setminus \rho_{\mathrm{loc},f_{\mathrm{cur}},0}(\re(f_{\mathrm{cur}})) \bigr) .
\]
The validity of \(\ctx_0\) tells us that for every
\(q \in \Qpool_0 \cup \bigl( V_{\activelabel,f_{\mathrm{cur}},0} \setminus \rho_{\mathrm{loc},f_{\mathrm{cur}},0}(\re(f_{\mathrm{cur}})) \bigr)\),
if \(\var(\idx(q)) = \emptyset\) then \(\base[q] = q\).
Consequently, for all \(q \in \Qpool\),
\[
  \var(\idx(q)) = \emptyset \;\Longrightarrow\; \base[q] = q .
\]
\end{proof}

\begin{lemma}\label{lemma:con-3.a}
  Under Notation~\ref{nota:clean} and Assumption~\ref{hyp:clean},
  Condition~3(a) in the validity of \(\ctx\) holds.
\end{lemma}

\begin{proof}
Take any \(f \in \{f_{\mathrm{cur}}\} \cup \bigl( \Fun(\St) \cap \SCC(f_{\mathrm{cur}}) \setminus \Theta^{-1}(\dom(\D)) \bigr)\). Note that
\[
\{f_{\mathrm{cur}}\} \cup \bigl( \Fun(\St) \cap \SCC(f_{\mathrm{cur}}) \setminus \Theta^{-1}(\dom(\D)) \bigr)
= \{f_{\mathrm{cur}}\} \cup \bigl( \Fun(\St_0) \cap \SCC(f_{\mathrm{cur}}) \setminus \Theta^{-1}(\dom(\D_0)) \bigr).
\]
Validity Condition~1 in \(\ctx_0\) implies that for every \(h\) with \(\pos(h) > \pos(f_{\mathrm{cur}})\), we have \(h \in \Theta^{-1}(\dom(\D_0))\); hence, no such \(h\) belongs to the set on the right‑hand side. Consequently, besides \(f_{\mathrm{cur}}\), the only functions we need to check are those \(f \neq f_{\mathrm{cur}}\) with \(\pos(f) < \pos(f_{\mathrm{cur}})\). For any such \(f\), the compiled state of \(f\) remains unchanged by the definition of the algorithm. Moreover, it is easy to verify that
\[
\Qpool_0 \cup V_{\activelabel,\succeq f,0} \setminus Q_{\succeq f_{\mathrm{cur}}^*,0}[+]
= (\Qpool \cup R_v) \cup (V_{\activelabel,\succeq f} \setminus R_v) \setminus Q_{\succeq f_{\mathrm{cur}}^*}[+]
= \Qpool \cup V_{\activelabel,\succeq f} \setminus Q_{\succeq f_{\mathrm{cur}}^*}[+].
\]
Therefore, the required property follows directly from validity Condition~3(a) for \(\ctx_0\). Thus, it remains only to consider \(f_{\mathrm{cur}}\) itself. 

First of all, injectivity of \(\rho_{\mathrm{loc},f_{\mathrm{cur}}}\) follows immediately, since it is obtained from the injective mapping \(\rho_{\mathrm{loc},f_{\mathrm{cur}},0}\) by removing the entries whose values lie in \(R_v\). Moreover, the out‑degree mapping \(\delta = \delta_0\) and the remaining statement sequence \(R_{\mathrm{stmt}} = R_{\mathrm{stmt},0}\) are unchanged, and \(\cod(\rho_{\mathrm{loc},f_{\mathrm{cur}}}) \subseteq \cod(\rho_{\mathrm{loc},f_{\mathrm{cur}},0})\), so the corresponding validity clauses are inherited directly from \(\ctx_0\). It therefore remains only to verify the cleanup-circuit correctness conditions for the updated mapping \(\kappa\). 

\noindent\textbf{Cleanup-circuit correctness for \(\kappa\).}
  % Since \(\dom(\kappa) = \dom(\kappa_{f_{\mathrm{cur}},0})\), every node
  % \(u \in \mathsf{Node}(G_{f_{\mathrm{cur}}})\) still belongs to \(\dom(\kappa)\).
  % For any \(u \in \dom(\kappa)\), if \(\rho_{\mathrm{loc},f_{\mathrm{cur}}}(u) = \emptyset\), then
  % either \(\rho_{\mathrm{loc},f_{\mathrm{cur}},0}(u) = \emptyset\) or \(u = v\).  In either case,
  % by the definition of \(ctx\) and the validity of \(\ctx_0\) it is easy to
  % see that \(\kappa(u) = I\). 

  Now let \(r \in \cod(\rho_{\mathrm{loc},f_{\mathrm{cur}}}) \setminus \rho_{\mathrm{loc},f_{\mathrm{cur}}}(\re(f_{\mathrm{cur}}))\) such that \(\rho_{\mathrm{loc},f_{\mathrm{cur}}}^{-1}(r).\flag \neq 2\) and set \(w \triangleq \node(\rho_{\mathrm{loc},f_{\mathrm{cur}}}^{-1}(r))\). It is easy to see that \(r \in \cod(\rho_{\mathrm{loc},f_{\mathrm{cur}},0}) \setminus \rho_{\mathrm{loc},f_{\mathrm{cur}},0}(\re(f_{\mathrm{cur}}))\); hence \(w \in \dom(\kappa_0) = \dom(\kappa)\). Besides, since \(U_v \cap \dom(\rho_{\mathrm{loc},f_{\mathrm{cur}}}) = \emptyset\), we have \(w \neq v\). 

According to the update operation of the cleanup algorithm on \(\kappa\),  if \(v \in \operatorname{pred}^+(w)\), then
    \[
    \kappa_{f_{\mathrm{cur}}}(w) = \qinv(C_{\mathrm{cl}})\,;\,\kappa_{f_{\mathrm{cur}},0}(w)\,;\,C_{\mathrm{cl}},
    \qquad
    C_{\mathrm{cl}} = \qinv\bigl(\kappa_{f_{\mathrm{cur}},0}(v)[\req_{v,0} \mapsto \overline{p}]\bigr).
    \]
Otherwise, \(\kappa_{f_{\mathrm{cur}}}(w) = \kappa_{f_{\mathrm{cur}},0}(w)\). 

We distinguish the two cases.

  \noindent\textbf{Case A: \(\kappa_{f_{\mathrm{cur}}}(w)\) is modified.}
  In this case, \(v \in \operatorname{pred}^+(w)\).
  Since \(v \in \operatorname{pred}^+(w)\), we have \(w \notin \operatorname{pred}^*(v)\) and \(\var(w)\cap \var(\operatorname{pred}^*(v))=\emptyset\) by Lemma~\ref{lem:static-wf}. Moreover, by definition 
\[
\qv(C_{\mathrm{cl}})
= \qv\bigl(\kappa_{f_{\mathrm{cur}},0}(v)[\req_v \mapsto \overline{p}]\bigr)
\subseteq \overline{p} \cup \rho_{f_{\mathrm{cur}},0}\bigl(\operatorname{pred}^*(v)\bigr).
\]
Since \(\overline{p} \subseteq \Qpool_0 \subseteq cv_0\), we have \(\overline{p} \cap \rho_{f_{\mathrm{cur}},0}(w) = \emptyset\).
Furthermore, \(\var(w)\cap \var(\operatorname{pred}^*(v))=\emptyset\) and the map \(\rho_{f_{\mathrm{cur}},0}\) is injective (since \(\rho_{\mathrm{loc},f_{\mathrm{cur}},0}\) and \(\rho_{\mathrm{glob}}\) are injective, with disjoint domains and codomains), hence \(\rho_{f_{\mathrm{cur}},0}(\operatorname{pred}^*(v)) \cap \rho_{f_{\mathrm{cur}},0}(w) = \emptyset\).
Therefore \(\qv(C_{\mathrm{cl}}) \cap \rho_{f_{\mathrm{cur}},0}(w) = \emptyset\). 

    We first check the two auxiliary requirements in Condition~3(a).

  \begin{enumerate}
    \item[(a)] We need to show
    \(\bigl(\rho_{\mathrm{loc},f_{\mathrm{cur}}}(\re(f_{\mathrm{cur}})) \cup \rho_{\mathrm{loc},f_{\mathrm{cur}}}(\fp(f_{\mathrm{cur}}))\bigr)
       \cap \req_w(\kappa_{f_{\mathrm{cur}}}(w)^z) = \emptyset.\)

    Since \(\kappa_{f_{\mathrm{cur}}}(w) = \qinv(C_{\mathrm{cl}}); \kappa_{f_{\mathrm{cur}},0}(w); C_{\mathrm{cl}}\),
 for any \(z\), we have
    \begin{align*}
    &\req_w(\kappa_{f_{\mathrm{cur}}}(w)^z) \\ 
    &= \qv(\kappa_{f_{\mathrm{cur}}}(w)^z) \setminus \rho_{f_{\mathrm{cur}}}\bigl(\operatorname{pred}^*(w)\bigr) \\
    &= \bigl(\qv(\kappa_{f_{\mathrm{cur}},0}(w)^z) \cup \qv(C_{\mathrm{cl}}^z)\bigr)
       \setminus \rho_{f_{\mathrm{cur}}}\bigl(\operatorname{pred}^*(w)\bigr) \\
    &\subseteq \Bigl(\qv(\kappa_{f_{\mathrm{cur}},0}(w)^z)
            \setminus \rho_{f_{\mathrm{cur}},0}\bigl(\operatorname{pred}^*(w)\bigr)\Bigr) 
       \cup
       \Bigl(\qv(C_{\mathrm{cl}}^z)
            \setminus \rho_{f_{\mathrm{cur}},0}\bigl(\operatorname{pred}^*(w)\bigr)\Bigr) \cup \rho_{\mathrm{loc},f_{\mathrm{cur}},0}(V) \\
    &\subseteq \req_{w,0}(\kappa_{f_{\mathrm{cur}},0}(w)^z) 
       \cup \Bigl((\overline{p} \cup \rho_{f_{\mathrm{cur}},0}(\operatorname{pred}^*(v))) \setminus \rho_{f_{\mathrm{cur}},0}\bigl(\operatorname{pred}^*(w)\bigr)\Bigr) \cup \rho_{\mathrm{loc},f_{\mathrm{cur}},0}(V) .
    \end{align*}
    In the inclusion step we use that
    \(\rho_{f_{\mathrm{cur}},0}(\operatorname{pred}^*(w)) = \rho_{f_{\mathrm{cur}}}(\operatorname{pred}^*(w)) \cup \rho_{\mathrm{loc},f_{\mathrm{cur}},0}(V)\) and
    \(\qv(C_{\mathrm{cl}}^z) = \qv(C_{\mathrm{cl}}) \subseteq
     \overline{p} \cup \rho_{f_{\mathrm{cur}},0}(\operatorname{pred}^*(v))\).
    Continuing the computation,
    \begin{align*}
    & \req_w(\kappa_{f_{\mathrm{cur}}}(w)^z) \\ 
    &\subseteq \req_{w,0}(\kappa_{f_{\mathrm{cur}},0}(w)^z)
       \cup \Bigl((\overline{p} \cup \rho_{f_{\mathrm{cur}},0}(\operatorname{pred}^*(v)))
                \setminus \rho_{f_{\mathrm{cur}},0}\bigl(\operatorname{pred}^*(v)\bigr)\Bigr) \cup \rho_{\mathrm{loc},f_{\mathrm{cur}},0}(V)
       \tag{since \(v \in \operatorname{pred}^*(w)\)} \\
    &\subseteq \req_{w,0}(\kappa_{f_{\mathrm{cur}},0}(w)^z)
       \cup \overline{p} \cup \rho_{\mathrm{loc},f_{\mathrm{cur}},0}(V) 
        \tag{since \(\overline{p} \cap \cod(\rho_{f_{\mathrm{cur}},0}) = \emptyset\)}.
    \end{align*} 
    By the arguments already established,
    \(\rho_{\mathrm{loc},f_{\mathrm{cur}},0}(V)\) is disjoint from
    \(\rho_{\mathrm{loc},f_{\mathrm{cur}}}(\fp(f_{\mathrm{cur}})) \cup \rho_{\mathrm{loc},f_{\mathrm{cur}}}(\re(f_{\mathrm{cur}}))\).
    Moreover, \(\overline{p}\) is fresh, hence also disjoint from
    \(\rho_{\mathrm{loc},f_{\mathrm{cur}}}(\fp(f_{\mathrm{cur}})) \cup \rho_{\mathrm{loc},f_{\mathrm{cur}}}(\re(f_{\mathrm{cur}}))\).
    Finally, by the validity of \(\ctx_0\),
    \(\bigl(\rho_{\mathrm{loc},f_{\mathrm{cur}},0}(\re(f_{\mathrm{cur}})) \cup \rho_{\mathrm{loc},f_{\mathrm{cur}},0}(\fp(f_{\mathrm{cur}}))\bigr)
     \cap \req_{w,0}(\kappa_{f_{\mathrm{cur}},0}(w)^z)= \emptyset.\)
   Therefore, since the quantum variables corresponding to the formal parameters and the return value do not change from \(\ctx_0\) to \(\ctx\), we obtain
\[
\bigl(\rho_{\mathrm{loc},f_{\mathrm{cur}}}(\re(f_{\mathrm{cur}})) \cup \rho_{\mathrm{loc},f_{\mathrm{cur}}}(\fp(f_{\mathrm{cur}}))\bigr)
\cap \req_{w}(\kappa_{f_{\mathrm{cur}}}(w)^z) = \emptyset. \]

\item[(b)] We need to show  \(\qv(\kappa_{f_{\mathrm{cur}}}(w)^z) \subseteq \qv(C_{f_{\mathrm{cur}}}^z) \cup S\) for some \(S\) with \(S \subseteq \Qpool \cup V_{\activelabel,\succeq f_{\mathrm{cur}}} \setminus Q_{\succeq f_{\mathrm{cur}}^*}[+]\) and \(\base(S) \subseteq V_{\alloc,f_{\mathrm{cur}}}\). 
By the local validity of \(\ctx_0\), there exists a set \(S_0 \subseteq \Qpool_0 \cup V_{\activelabel,\succeq f_{\mathrm{cur}},0} \setminus Q_{\succeq f_{\mathrm{cur}}^*,0}[+]\) with \(\base(S_0) \subseteq V_{\alloc,f_{\mathrm{cur}},0}\) such that
\(\qv(\kappa_{f_{\mathrm{cur}},0}(v)^z) \cup \qv(\kappa_{f_{\mathrm{cur}},0}(w)^z) \subseteq \qv(C_{f_{\mathrm{cur}},0}^z) \cup S_0. \)
Note that \(\qv(C_{\mathrm{cl}}^z) \subseteq \qv(\kappa_{f_{\mathrm{cur}},0}(v)^z) \cup \overline{p}\).  
Combining this with \(\qv(C_{f_{\mathrm{cur}},0}^z) \subseteq \qv(C_{f_{\mathrm{cur}}}^z)\), we first obtain
\[
\begin{aligned}
\qv(\kappa_{f_{\mathrm{cur}}}(w)^z)
&= \qv(\kappa_{f_{\mathrm{cur}},0}(w)^z) \cup \qv(C_{\mathrm{cl}}^z) \\
&\subseteq \qv(\kappa_{f_{\mathrm{cur}},0}(w)^z) \cup \qv(\kappa_{f_{\mathrm{cur}},0}(v)^z) \cup \overline{p} \\
&\subseteq \qv(C_{f_{\mathrm{cur}}}^z) \cup S_0 \cup \overline{p}.
\end{aligned}
\]
Let \(S = S_0 \cup \overline{p}\).  
Since \(\Qpool_0 \cup V_{\activelabel,\succeq f_{\mathrm{cur}},0} \setminus Q_{\succeq f_{\mathrm{cur}}^*,0}[+] = \Qpool \cup V_{\activelabel,\succeq f_{\mathrm{cur}}} \setminus Q_{\succeq f_{\mathrm{cur}}^*}[+]\) and \(V_{\alloc,f_{\mathrm{cur}},0} \subseteq V_{\alloc,f_{\mathrm{cur}}}\), we have \(S_0 \subseteq \Qpool \cup V_{\activelabel,\succeq f_{\mathrm{cur}}} \setminus Q_{\succeq f_{\mathrm{cur}}^*}[+]\) and \(\base(S_0) \subseteq V_{\alloc,f_{\mathrm{cur}}}\).  
Together with \(\overline{p} \subseteq \Qpool\) and \(\base(\overline{p}) \subseteq V_{\alloc,f_{\mathrm{cur}}}\), we obtain \(S \subseteq \Qpool \cup V_{\activelabel,\succeq f_{\mathrm{cur}}} \setminus Q_{\succeq f_{\mathrm{cur}}^*}[+]\) and \(\base(S) \subseteq V_{\alloc,f_{\mathrm{cur}}}\). 
  \end{enumerate} 

  Now let \(\ket{\varphi}\) be any state such that
  \(\ket{\varphi} \models \ket{0}_{\ctx.A_{f_{\mathrm{cur}}} \setminus \rho_{\mathrm{loc},f_{\mathrm{cur}}}(\fp(f_{\mathrm{cur}}))}\),
  and define
  \[
  \langle C_{f_{\mathrm{cur}},0}^{\,k}, \ket{\varphi}\rangle \to^{*} \langle \downarrow, \ket{\chi}\rangle,
  \qquad
  \langle C_{f_{\mathrm{cur}}}^{\,k}, \ket{\varphi}\rangle \to^{*} \langle \downarrow, \ket{\psi}\rangle.
  \]
  Let \(\overline{q}\) be any tuple of quantum variables such that
  \(\ket{\psi} \models \ket{0}_{\overline{q}}\) and
  \(\overline{q} \cap \bigl( \cod(\rho_{f_{\mathrm{cur}}}) \cup V_{\activelabel,f_{\mathrm{cur}}} \bigr) = \emptyset\).

  First, it is immediate from the definitions that
  \[
  (\overline{q} \setminus \rho_{\mathrm{loc},f_{\mathrm{cur}},0}(V)) \cap
  \bigl( \cod(\rho_{f_{\mathrm{cur}},0}) \cup V_{\activelabel,f_{\mathrm{cur}},0} \bigr) = \emptyset.  
  \]
  In particular, since \(\rho_{\mathrm{loc},f_{\mathrm{cur}},0}(v) \subseteq
  \cod(\rho_{f_{\mathrm{cur}},0}) \cup V_{\activelabel,f_{\mathrm{cur}},0}\), this implies
  \( (\overline{q} \setminus \rho_{\mathrm{loc},f_{\mathrm{cur}},0}(V)) \cap \rho_{\mathrm{loc},f_{\mathrm{cur}},0}(v) = \emptyset. \)
  The cleanup circuit \(C_{\mathrm{cl}}\) only modifies the registers associated with \(\operatorname{var}(v)\); together with the disjointness
  just established, the cleanup argument for \(C_{\mathrm{cl}}\) yields
  \[
  \ket{\chi} \models \ket{0}_{\overline{q} \setminus \rho_{\mathrm{loc},f_{\mathrm{cur}},0}(V)} .
  \]
  Finally, since \(C_{\mathrm{cl}}\) restores the registers \(\rho_{\mathrm{loc},f_{\mathrm{cur}},0}(V)\) to their initial zero state and does not change the values of \(\overline{p}\), we also have
\[
\ket{\psi} \models \ket{0}_{\overline{p} \cup \rho_{\mathrm{loc},f_{\mathrm{cur}},0}(V)} .
\] 

%   Furthermore, because \(v \in \operatorname{pred}^+(w)\) and
%   \(\operatorname{pred}^+(\cdot)\) denotes the full predecessor set,
%   every register in \(\rho_{\mathrm{loc},f_{\mathrm{cur}},0}(\operatorname{var}(v) \setminus V)\)
%   that remains in the local mapping belongs to
%   \(\rho_{\mathrm{loc},f_{\mathrm{cur}}}\bigl(\operatorname{pred}^*(w) \setminus \{w\}\bigr)\).

 We now verify the semantic part of Condition~3(a) by direct reduction:
\begin{align*}
&\bigl\langle (C_{f_{\mathrm{cur}}}; \qinv(\kappa_{f_{\mathrm{cur}}}(w)[\req_w \mapsto \overline{q}]))^{\,k},\,
        \ket{\varphi} \bigr\rangle \\
= &\bigl\langle \bigl(C_{f_{\mathrm{cur}},0}; C_{\mathrm{cl}};
          \qinv\bigl((\qinv(C_{\mathrm{cl}});\kappa_{f_{\mathrm{cur}},0}(w);C_{\mathrm{cl}})
                   [\req_w \mapsto \overline{q}]\bigr)\bigr)^{\,k},\,
        \ket{\varphi} \bigr\rangle \\
&\qquad \tag{by the definitions of \(C_{f_{\mathrm{cur}}}\) and \(\kappa_{f_{\mathrm{cur}}}(w)\)} \\[2mm] 
= &\bigl\langle \bigl(C_{f_{\mathrm{cur}},0}; C_{\mathrm{cl}};
          \qinv\bigl(C_{\mathrm{cl}}[\req_w \mapsto \overline{q}]\bigr);
          \qinv\bigl(\kappa_{f_{\mathrm{cur}},0}(w)[\req_w \mapsto \overline{q}]\bigr);
          C_{\mathrm{cl}}[\req_w \mapsto \overline{q}]\bigr)^{\,k},\,
        \ket{\varphi} \bigr\rangle \\
\qquad &\tag{by distributing the fresh tuple \(\overline{q}\) over the sequential composition} \\
= &\bigl\langle \bigl(C_{f_{\mathrm{cur}},0}; C_{\mathrm{cl}};
          \qinv\bigl(C_{\mathrm{cl}}[\overline{p} \cup \rho_{\mathrm{loc},f_{\mathrm{cur}},0}(V) \mapsto \overline{q}]\bigr);\\ 
 & \qquad \qquad   \qinv\bigl(\kappa_{f_{\mathrm{cur}},0}(w)[\req_{w,0}
                   \cup \rho_{\mathrm{loc},f_{\mathrm{cur}},0}(V) \mapsto \overline{q}]\bigr);
          C_{\mathrm{cl}}[\overline{p} \cup \rho_{\mathrm{loc},f_{\mathrm{cur}},0}(V) \mapsto \overline{q}]\bigr)^{\,k},\,
        \ket{\varphi} \bigr\rangle .
\end{align*}
In the last step we used the key identity
\(\qv(C_{\mathrm{cl}}^z) \setminus \rho_{f_{\mathrm{cur}}}\bigl(\operatorname{pred}^*(w)\bigr)
 = \overline{p} \cup \rho_{\mathrm{loc},f_{\mathrm{cur}},0}(V)\) and \(\qv(\kappa_{f_{\mathrm{cur}},0}(w)^z) \setminus \rho_{f_{\mathrm{cur}}}\bigl(\operatorname{pred}^*(w)\bigr)
 = \req_{w,0}(\kappa_{f_{\mathrm{cur}},0}(w)) \cup \rho_{\mathrm{loc},f_{\mathrm{cur}},0}(V)\)
which tells us precisely which registers are
considered ``fresh'' and need to be substituted by \(\overline{q}\).
Since we have already proved \(\ket{\psi} \models \ket{0}_{\overline{p} \cup \rho_{\mathrm{loc},f_{\mathrm{cur}},0}(V)}\), renaming these registers in \(\kappa_{f_{\mathrm{cur}}}(w)\) is semantically equivalent to omitting them from the substitution and replacing the remaining part with the portion of \(\overline{q}\) that is disjoint from \(\overline{p} \cup \rho_{\mathrm{loc},f_{\mathrm{cur}},0}(V)\).  Hence
\begin{align*}
&\bigl\langle (C_{f_{\mathrm{cur}}}; \qinv(\kappa_{f_{\mathrm{cur}}}(w)[\req_w \mapsto \overline{q}]))^{\,k},\,
        \ket{\varphi} \bigr\rangle \\
= &\bigl\langle \bigl(C_{f_{\mathrm{cur}},0}; C_{\mathrm{cl}};
          \qinv(C_{\mathrm{cl}});
          \qinv\bigl(\kappa_{f_{\mathrm{cur}},0}(w)[\req_w \setminus \overline{p} 
                   \mapsto \overline{q} \setminus (\rho_{\mathrm{loc},f_{\mathrm{cur}},0}(V) \cup \overline{p})]\bigr);
          C_{\mathrm{cl}}\bigr)^{\,k},\,
        \ket{\varphi} \bigr\rangle \\
= &\bigl\langle \bigl(C_{f_{\mathrm{cur}},0};
          \qinv\bigl(\kappa_{f_{\mathrm{cur}},0}(w)[\req_w \setminus \overline{p} \mapsto \overline{q} \setminus (\rho_{\mathrm{loc},f_{\mathrm{cur}},0}(V) \cup \overline{p})]\bigr);
          C_{\mathrm{cl}}\bigr)^{\,k},\,
        \ket{\varphi} \bigr\rangle
        \qquad \tag{cancelling \(C_{\mathrm{cl}}\) with its inverse} \\[2mm]
\rightarrow^{*} &\bigl\langle \bigl(
          \qinv\bigl(\kappa_{f_{\mathrm{cur}},0}(w)[\req_w \setminus \overline{p} \mapsto \overline{q} \setminus (\rho_{\mathrm{loc},f_{\mathrm{cur}},0}(V) \cup \overline{p})]\bigr);
          C_{\mathrm{cl}}\bigr)^{\,k},\,
        \ket{\chi} \bigr\rangle.
        \qquad \tag{by the definition of \(\ket{\chi}\)}
\end{align*}
Now we invoke the already established validity of \(\kappa_{f_{\mathrm{cur}},0}(w)\) in \(\ctx_0\).  Recall that 
\(\ket{\chi} \models \ket{0}_{\overline{q} \setminus \rho_{\mathrm{loc},f_{\mathrm{cur}},0}(V)}\) and
\((\overline{q} \setminus \rho_{\mathrm{loc},f_{\mathrm{cur}},0}(V)) \cap
 (\cod(\rho_{\mathrm{loc},f_{\mathrm{cur}},0}) \cup V_{\activelabel,f_{\mathrm{cur}},0}) = \emptyset\); 
moreover, \(\overline{p}\) itself also satisfies these conditions. 
Therefore we obtain that there exists a state $\ket{\theta}$ such that 
\begin{align*}
&\bigl\langle (C_{f_{\mathrm{cur}}}; \qinv(\kappa_{f_{\mathrm{cur}}}(w)[\req_w \mapsto \overline{q}]))^{\,k},\,
        \ket{\varphi} \bigr\rangle \\
\rightarrow^{*} &\bigl\langle C_{\mathrm{cl}}^{\,k},\,
        \ket{\chi}_{\sys \setminus \rho_{\mathrm{loc},f_{\mathrm{cur}},0}(w)}
        \otimes \ket{\theta}_{\rho_{\mathrm{loc},f_{\mathrm{cur}},0}(w) \setminus U_w}
        \otimes \ket{\varphi}_{\rho_{\mathrm{loc},f_{\mathrm{cur}},0}(U_w)} \bigr\rangle \\
\rightarrow^{*} &\bigl\langle \downarrow,\,
        C_{\mathrm{cl}}^{\,k}\bigl(\ket{\chi}_{\sys \setminus \rho_{\mathrm{loc},f_{\mathrm{cur}},0}(w)}\bigr)
        \otimes \ket{\theta}_{\rho_{\mathrm{loc},f_{\mathrm{cur}},0}(w) \setminus U_w}
        \otimes \ket{\varphi}_{\rho_{\mathrm{loc},f_{\mathrm{cur}},0}(U_w)} \bigr\rangle
        \qquad (\text{since } \qv(C_{\mathrm{cl}}) \cap \rho_{\mathrm{loc},f_{\mathrm{cur}},0}(w) = \emptyset) \\
= \ &\bigl\langle \downarrow,\,
        \ket{\psi}_{\sys \setminus \rho_{\mathrm{loc},f_{\mathrm{cur}}}(w)}
        \otimes \ket{\theta}_{\rho_{\mathrm{loc},f_{\mathrm{cur}}}(w) \setminus U_w}
        \otimes \ket{\varphi}_{\rho_{\mathrm{loc},f_{\mathrm{cur}}}(U_w)} \bigr\rangle. \tag{by the definition of \(\ket{\psi}\) and
          \(\rho_{\mathrm{loc},f_{\mathrm{cur}}}(w) = \rho_{\mathrm{loc},f_{\mathrm{cur}},0}(w)\)}
\end{align*}
This is exactly the required conclusion.

\noindent\textbf{Case B: \(\kappa_{f_{\mathrm{cur}}}(w)\) is unchanged.}
In this case, we have \(v \notin \operatorname{pred}^+(w)\) and \(v \neq w\), and by the update definition of \(\kappa\), \(\kappa_{f_{\mathrm{cur}}}(w) = \kappa_{f_{\mathrm{cur}},0}(w)\). First, Lemma~\ref{lem:static-wf} yields \(\var(v) \cap \var(\operatorname{pred}^*(w)) = \emptyset\). Therefore $\rho_{f_{\mathrm{cur}},0}(w) \cap \rho_{f_{\mathrm{cur}},0}(v) =\emptyset$ by the injection of $\rho_{f_{\mathrm{cur}},0}$. Moreover, since \(\rho_{\mathrm{loc},f_{\mathrm{cur}}}\) is obtained from \(\rho_{\mathrm{loc},f_{\mathrm{cur}},0}\) by deleting \(R_v = \rho_{\mathrm{loc},f_{\mathrm{cur}},0}(V) \subseteq \rho_{\mathrm{loc},f_{\mathrm{cur}},0}(v)\), while \(\rho_{\mathrm{glob}}\) remains unchanged, we have \(\rho_{f_{\mathrm{cur}}}(\operatorname{pred}^*(w)) = \rho_{f_{\mathrm{cur}},0}(\operatorname{pred}^*(w))\). Consequently, 
\[
\req_w = \qv(\kappa_{f_{\mathrm{cur}}}(w)) \setminus \rho_{f_{\mathrm{cur}}}(\operatorname{pred}^*(w))
       = \qv(\kappa_{f_{\mathrm{cur}},0}(w)) \setminus \rho_{f_{\mathrm{cur}},0}(\operatorname{pred}^*(w))
       = \req_{w,0}.
\]
Thus requirements (a) and (b) of Condition~3(a) concerning the correctness of \(\kappa\) follow directly from the validity of \(\ctx_0\) and an argument analogous to that of Case~A.  

Now let \(\ket{\varphi}\), \(\ket{\chi}\), \(\ket{\psi}\), and \(\overline{q}\)
be as above.  Without loss of generality, we may further assume that
\[
\overline{q} \cap \rho_{\mathrm{loc},f_{\mathrm{cur}},0}(V) = \emptyset.
\]
Indeed, if some registers in \(\overline{q}\) overlap with \(\rho_{\mathrm{loc},f_{\mathrm{cur}},0}(V)\), we may replace them by equally fresh zero-valued registers that are disjoint from \(\rho_{\mathrm{loc},f_{\mathrm{cur}},0}(V)\). Since \(\rho_{\mathrm{loc},f_{\mathrm{cur}},0}(V)\) is also in the \(|0\rangle\) state under \(\ket{\psi}\), the first step in the proof of the subsequent circuit execution replaces the occurrences of \(\rho_{\mathrm{loc},f_{\mathrm{cur}},0}(V)\) with other zero-valued systems, which is equivalent to the semantics of the effect original circuit.  
Consequently, we have 
\(\overline{q} \cap \bigl( \cod(\rho_{f_{\mathrm{cur}},0}) \cup V_{\activelabel,f_{\mathrm{cur}},0} \bigr) 
 = \emptyset\) and $\overline{q} \cap \rho_{\mathrm{loc},f_{\mathrm{cur}},0}(v)=\emptyset$. 
And \(\ket{\chi} \models \ket{0}_{\overline{q}}\) also holds by the cleanup argument for \(C_{\mathrm{cl}}\). 

% Combing with \(C_{\mathrm{cl}}\) acts on
% registers disjoint from \(\rho_{\mathrm{loc},f_{\mathrm{cur}}}\bigl(\operatorname{pred}^*(w)\bigr)\), 

We then verify the semantic part by direct reduction:
\begin{align*} 
&\bigl\langle (C_{f_{\mathrm{cur}}}; \qinv(\kappa_{f_{\mathrm{cur}}}(w)[\req_w \mapsto \overline{q}]))^{\,k},\;
        \ket{\varphi} \bigr\rangle \\
= \ &\bigl\langle (C_{f_{\mathrm{cur}},0}; C_{\mathrm{cl}};
          \qinv\bigl(\kappa_{f_{\mathrm{cur}},0}(w)[\req_{w,0} \mapsto \overline{q}]\bigr))^{\,k},\;
        \ket{\varphi} \bigr\rangle \\
\rightarrow^{*} \ &\bigl\langle (C_{\mathrm{cl}};
          \qinv\bigl(\kappa_{f_{\mathrm{cur}},0}(w)[\req_{w,0} \mapsto \overline{q}]\bigr))^{\,k},\;
        \ket{\chi} \bigr\rangle
        \tag{by the definition of \(\ket{\chi}\)} \\
\rightarrow^{*} \ &\bigl\langle \qinv\bigl(\kappa_{f_{\mathrm{cur}},0}(w)[\req_{w,0} \mapsto \overline{q}]\bigr)^{\,k},\;
        \ket{\chi}_{\sys \setminus \rho_{\mathrm{loc},f_{\mathrm{cur}},0}(v)}
        \otimes \ket{\theta}_{\rho_{\mathrm{loc},f_{\mathrm{cur}},0}(v) \setminus U_v}
        \otimes \ket{\varphi}_{\rho_{\mathrm{loc},f_{\mathrm{cur}},0}(U_v)} \bigr\rangle
        \tag{by semantics of $C_{\mathrm{cl}}$} \\
= \ &\bigl\langle \qinv\bigl(\kappa_{f_{\mathrm{cur}},0}(w)[\req_{w,0} \mapsto \overline{q}]\bigr)^{\,k},\;
        \ket{\chi}_{\sys \setminus (\rho_{\mathrm{loc},f_{\mathrm{cur}},0}(v) \cup \rho_{\mathrm{loc},f_{\mathrm{cur}},0}(w))}
        \otimes \ket{\chi}_{\rho_{\mathrm{loc},f_{\mathrm{cur}},0}(w)}
        \otimes \ket{\psi}_{\rho_{\mathrm{loc},f_{\mathrm{cur}},0}(v)} \bigr\rangle
        \tag{by decomposing the register space and definition of $\ket{\psi}$}. 
\end{align*}
Now we apply the validity of \(\kappa_{f_{\mathrm{cur}},0}(w)\) in \(\ctx_0\).  Recall that we already have $\rho_{f_{\mathrm{cur}},0}(w) \cap \rho_{f_{\mathrm{cur}},0}(v) =\emptyset$ and $\overline{q}  \cap \rho_{f_{\mathrm{cur}},0}(v) =\emptyset$, we get that \(\qv(\kappa_{f_{\mathrm{cur}},0}(w)[\req_{w,0} \mapsto \overline{q}]) \cap \rho_{\mathrm{loc},f_{\mathrm{cur}},0}(v) = \emptyset\). Combining with 
\(\overline{q} \cap (\cod(\rho_{f_{\mathrm{cur}},0}) \cup V_{\activelabel,f_{\mathrm{cur}},0}) = \emptyset\), 
and \(\ket{\chi} \models \ket{0}_{\overline{q}}\), we obtain that there exists a state $\ket{\theta'}$ such that 
\begin{align*}
\rightarrow^{*} \ &\bigl\langle \downarrow,\;
        \ket{\chi}_{\sys \setminus (\rho_{\mathrm{loc},f_{\mathrm{cur}},0}(v) \cup \rho_{\mathrm{loc},f_{\mathrm{cur}},0}(w))}
        \otimes \ket{\theta'}_{\rho_{\mathrm{loc},f_{\mathrm{cur}},0}(w) \setminus U_w}
        \otimes \ket{\varphi}_{\rho_{\mathrm{loc},f_{\mathrm{cur}},0}(U_w)}
        \otimes \ket{\psi}_{\rho_{\mathrm{loc},f_{\mathrm{cur}},0}(v)} \bigr\rangle \\
= \ &\bigl\langle \downarrow,\;
        \ket{\psi}_{\sys \setminus \rho_{\mathrm{loc},f_{\mathrm{cur}}}(w)}
        \otimes \ket{\theta'}_{\rho_{\mathrm{loc},f_{\mathrm{cur}}}(w) \setminus U_w}
        \otimes \ket{\varphi}_{\rho_{\mathrm{loc},f_{\mathrm{cur}}}(U_w)} \bigr\rangle
        \tag{by the definition of \(\ket{\psi}\) and
              \(\rho_{\mathrm{loc},f_{\mathrm{cur}}}(w) = \rho_{\mathrm{loc},f_{\mathrm{cur}},0}(w)\)}.
\end{align*}
This again yields the required conclusion.

Since \(w\) was arbitrary, \(\kappa_{f_{\mathrm{cur}}}\) satisfies the corresponding part of Condition~3(a) in the validity definition.
\end{proof}

\begin{lemma}
    \label{lemma:con-3.b}
  Under Notation~\ref{nota:clean} and Assumption~\ref{hyp:clean},
  Condition~3(b) in the validity of \(\ctx\) holds.
\end{lemma}

\begin{proof} 
 For any \(f \in \{f_{\mathrm{cur}}\} \cup \bigl( \Fun(\St) \cap \SCC(f_{\mathrm{cur}}) \setminus \Theta^{-1}(\dom(\D)) \bigr)\),
we must prove, for every state \(\ket{\varphi}\) with
\(\ket{\varphi} \models \ket{0}_{\ctx.A_f \setminus \rho_{\mathrm{loc},f}(\fp(f))}\),
\[
  C_{f}^k\ket{\varphi} \models \ket{\varphi}_{\ctx.A_f \setminus (V_{\activelabel,f} \cup \cod(\rho_{\mathrm{loc},f}(\fp(f))))}.
\]
Similarly, for any such \(f \neq f_{\mathrm{cur}}\), the required property follows directly from validity Condition~3(b) for \(\ctx_0\). 
We now treat \(f = f_{\mathrm{cur}}\).
  Condition~3(b) for \(\ctx_0\) gives
  \[
    C_{f_{\mathrm{cur}},0}^k \ket{\varphi}
    \models
    \ket{\varphi}_{\ctx_0.A_{f_{\mathrm{cur}}} \setminus V_{\activelabel,f_{\mathrm{cur}},0} \setminus \rho_{\mathrm{loc},f_{\mathrm{cur}},0}(\fp(f_{\mathrm{cur}}))}. \tag{1}
  \]

  To apply the cleanup-circuit correctness condition for \(f_{\mathrm{cur}}\), it remains
  to verify the side conditions on the fresh registers used by the cleanup code.
  These are exactly the same conditions established in the preceding argument:
  in both cases the chosen registers satisfy
  \(\overline{q} \subseteq \ctx_0.A_{f_{\mathrm{cur}}} \setminus (V_{\activelabel,f_{\mathrm{cur}},0} \cup \cod(\rho_{\mathrm{loc},f_{\mathrm{cur}},0}))\).
  Hence \(\overline{q} \subseteq \ctx_0.A_{f_{\mathrm{cur}}} \setminus (V_{\activelabel,f_{\mathrm{cur}},0} \cup \rho_{\mathrm{loc},f_{\mathrm{cur}},0}(\fp(f_{\mathrm{cur}})))\)
  and \(\overline{q}\) is disjoint from \(V_{\activelabel,f_{\mathrm{cur}},0} \cup \cod(\rho_{\mathrm{loc},f_{\mathrm{cur}},0})\).
  By the same reasoning as before (using Lemma~\ref{lem:curf}), we obtain
  \(C_{f_{\mathrm{cur}},0}^k\ket{\varphi} \models \ket{0}_{\overline{q}}\).
  Thus the cleanup-circuit correctness condition applies to \(\ket{\varphi}\),
  yielding
  \[
    (C_{f_{\mathrm{cur}},0}; C_{\mathrm{cl}})^k \ket{\varphi}
    \models
    (C_{f_{\mathrm{cur}},0}^k\ket{\varphi})_{\bigl( \ctx_0.A_{f_{\mathrm{cur}}} \setminus
      (V_{\activelabel,f_{\mathrm{cur}},0} \cup \rho_{\mathrm{loc},f_{\mathrm{cur}},0}(\fp(f_{\mathrm{cur}}))) \bigr) \setminus \rho_{\mathrm{loc},f_{\mathrm{cur}},0}(v)} 
    \otimes
    \ket{\varphi}_{R_v}. \tag{2}
  \]

  Since $\rho_{\mathrm{loc},f_{\mathrm{cur}},0}(V) \subseteq (V_{\activelabel,f_{\mathrm{cur}},0} \cup \rho_{\mathrm{loc},f_{\mathrm{cur}},0}(\fp(f_{\mathrm{cur}})))$, combining (1) and (2) we obtain
  \[
    (C_{f_{\mathrm{cur}},0}; C_{\mathrm{cl}})^k \ket{\varphi}
    \models
    \ket{\varphi}_{\ctx_0.A_{f_{\mathrm{cur}}} \setminus
      \bigl( V_{\activelabel,f_{\mathrm{cur}},0} \cup \rho_{\mathrm{loc},f_{\mathrm{cur}},0}(\fp(f_{\mathrm{cur}})) \bigr) \cup R_v}. \tag{3} 
  \]

  By definition of \(\textsc{Clean\_Temp}\),
  \[
    V_{\activelabel,f_{\mathrm{cur}}} = V_{\activelabel,f_{\mathrm{cur}},0} \setminus R_v,\qquad
    \rho_{\mathrm{loc},f_{\mathrm{cur}}}(\fp(f_{\mathrm{cur}})) = \rho_{\mathrm{loc},f_{\mathrm{cur}},0}(\fp(f_{\mathrm{cur}})).
  \]
  Therefore,
  \begin{align*}
    &\ctx.A_{f_{\mathrm{cur}}} \setminus \bigl( V_{\activelabel,f_{\mathrm{cur}}} \cup \rho_{\mathrm{loc},f_{\mathrm{cur}}}(\fp(f_{\mathrm{cur}})) \bigr) \\
    = &\ctx_0.A_{f_{\mathrm{cur}}} \setminus \bigl( (V_{\activelabel,f_{\mathrm{cur}},0} \setminus R_v)
                                      \cup \rho_{\mathrm{loc},f_{\mathrm{cur}},0}(\fp(f_{\mathrm{cur}})) \bigr) \\
    = &\bigl( \ctx_0.A_{f_{\mathrm{cur}}} \setminus
             (V_{\activelabel,f_{\mathrm{cur}},0} \cup \rho_{\mathrm{loc},f_{\mathrm{cur}},0}(\fp(f_{\mathrm{cur}}))) \bigr) \cup R_v .
  \end{align*}
  Together with (3), this yields
  \[
    C_{f_{\mathrm{cur}}}^k \ket{\varphi}
    \models
    \ket{\varphi}_{\ctx.A_{f_{\mathrm{cur}}} \setminus (V_{\activelabel,f_{\mathrm{cur}}} \cup \rho_{\mathrm{loc},f_{\mathrm{cur}}}(\fp(f_{\mathrm{cur}})))}.
  \]

  Thus both parts of Condition~3(b) are satisfied for the updated context \(\ctx\).
\end{proof}

\begin{lemma}\label{lemma:con-3.c}
  Under Notation~\ref{nota:clean} and Assumption~\ref{hyp:clean},
  Condition~3(c) in the validity of \(\ctx\) holds.
\end{lemma}

\begin{proof}
  Take any
  \(f \in \{f_{\mathrm{cur}}\} \cup \bigl( \Fun(\St) \cap \SCC(f_{\mathrm{cur}}) \setminus \Theta^{-1}(\dom(\D)) \bigr)\).
  As before, it suffices to consider the case \(f = f_{\mathrm{cur}}\).

  By definition of \(\textsc{Clean\_Temp}\), in both subcases we have
  \[
  \begin{aligned}
  \rho_{\mathrm{loc},f_{\mathrm{cur}}} &= \rho_{\mathrm{loc},f_{\mathrm{cur}},0} \setminus (-, R_v),\\
  V_{\activelabel,f_{\mathrm{cur}}} &= V_{\activelabel,f_{\mathrm{cur}},0} \setminus R_v,\\
  V_{\alloc,0} &\subseteq \Valloc,\\
  \Qpool &= \push(\Qpool_0,R_v),\qquad
  R_v \triangleq \rho_{\mathrm{loc},f_{\mathrm{cur}},0}(V).
  \end{aligned}
  \]

  \noindent\textbf{Conditions~3(c)(1), 3(c)(4), and 3(c)(5)}
  We need to prove
  \[
  \fp(f_{\mathrm{cur}}) \cup \re(f_{\mathrm{cur}}) \subseteq \dom(\rho_{\mathrm{loc},f_{\mathrm{cur}}}),
  \]
  \[
  \rho_{\mathrm{loc},f_{\mathrm{cur}}}(\fp(f_{\mathrm{cur}})) \cap V_{\activelabel,f_{\mathrm{cur}}} = \emptyset,
  \qquad
  \rho_{\mathrm{loc},f_{\mathrm{cur}}}(\re(f_{\mathrm{cur}}) \cup \fp(f_{\mathrm{cur}})) \subseteq V_{\activelabel,\mathrm{prev}(f_{\mathrm{cur}})}.
  \]

  Since \(\rho_{\mathrm{loc},f_{\mathrm{cur}}} = \rho_{\mathrm{loc},f_{\mathrm{cur}},0} \setminus (-, R_v)\), and we have already shown that neither the formal parameters nor the return variable of \(f_{\mathrm{cur}}\) appear in \(V\) and  $R_v \cap \rho_{\mathrm{loc},f_{\mathrm{cur}},0}(\fp(f_{\mathrm{cur}}) \cup \re(f_{\mathrm{cur}}))=\emptyset$.  Therefore, \(\rho_{\mathrm{loc},f_{\mathrm{cur}}}(\fp(f_{\mathrm{cur}})) = \rho_{\mathrm{loc},f_{\mathrm{cur}},0}(\fp(f_{\mathrm{cur}}))\), \(\rho_{\mathrm{loc},f_{\mathrm{cur}}}(\re(f_{\mathrm{cur}})) = \rho_{\mathrm{loc},f_{\mathrm{cur}},0}(\re(f_{\mathrm{cur}}))\).   
Since \(\fp(f_{\mathrm{cur}}) \cup \re(f_{\mathrm{cur}}) \subseteq \dom(\rho_{\mathrm{loc},f_{\mathrm{cur}},0})\) holds by the validity of \(\ctx_0\), we obtain  
\[
\fp(f_{\mathrm{cur}}) \cup \re(f_{\mathrm{cur}}) \subseteq \dom(\rho_{\mathrm{loc},f_{\mathrm{cur}}}).
\] 
  The validity Condition~3(c) for \(\ctx_0\) gives
  \[
  \rho_{\mathrm{loc},f_{\mathrm{cur}},0}(\fp(f_{\mathrm{cur}})) \cap V_{\activelabel,f_{\mathrm{cur}},0} = \emptyset,\quad
  \rho_{\mathrm{loc},f_{\mathrm{cur}},0}(\re(f_{\mathrm{cur}})) \subseteq
  V_{\activelabel,\mathrm{prev}(f_{\mathrm{cur}}),0}. 
  \]
  Since \(V_{\activelabel,f_{\mathrm{cur}}} \subseteq V_{\activelabel,f_{\mathrm{cur}},0}\), we have
  \(\rho_{\mathrm{loc},f_{\mathrm{cur}}}(\fp(f_{\mathrm{cur}})) \cap V_{\activelabel,f_{\mathrm{cur}}} = \emptyset\).
  Moreover, by definition the stack components below \(f_{\mathrm{cur}}\) are unchanged, i.e.
  \(V_{\activelabel,\mathrm{prev}(f_{\mathrm{cur}})} = V_{\activelabel,\mathrm{prev}(f_{\mathrm{cur}}),0}\).
  Hence
  \[
  \rho_{\mathrm{loc},f_{\mathrm{cur}}}(\re(f_{\mathrm{cur}})) \subseteq V_{\activelabel,\mathrm{prev}(f_{\mathrm{cur}})}.
  \]

  \noindent\textbf{Condition~3(c)(2).}
We need to show
\[
A_{f_{\mathrm{cur}}} \cap \cod(\rho_{\mathrm{glob}}) = \emptyset .
\]
Lemma~\ref{lemma:Aset} gives \(\ctx.A_{f_{\mathrm{cur}}} = \ctx_0.A_{f_{\mathrm{cur}}}\).
The validity of \(\ctx_0\) guarantees
\(\ctx_0.A_{f_{\mathrm{cur}}} \cap \cod(\rho_{\mathrm{glob}}) = \emptyset\); combining this with the fact that \(\rho_{\mathrm{glob}}\) is unchanged
hence the required disjointness follows directly. 

%   We need to show
%   \[
%   A_{f_{\mathrm{cur}}} \cap \cod(\rho_{\mathrm{glob}}) = \emptyset .
%   \]
%   Lemma~1 gives \(\ctx.A_{f_{\mathrm{cur}}} \subseteq \ctx_0.A_{f_{\mathrm{cur}}}\).
%   Because \(R_v \subseteq V_{\activelabel,f_{\mathrm{cur}},0} \subseteq \ctx_0.A_{f_{\mathrm{cur}}}\), the validity of
%   \(\ctx_0\) yields \(R_v \cap \cod(\rho_{\mathrm{glob},0}) = \emptyset\); it also
%   guarantees \(\ctx_0.A_{f_{\mathrm{cur}}} \cap \cod(\rho_{\mathrm{glob},0}) = \emptyset\).
%   The required disjointness follows directly.

\noindent\textbf{Condition~3(c)(3).}
We need to show
\[
\bigl( A_{f_{\mathrm{cur}}} \setminus \rho_{\mathrm{loc},f_{\mathrm{cur}}}(\fp(f_{\mathrm{cur}})) \setminus \rho_{\mathrm{loc},f_{\mathrm{cur}}}(\re(f_{\mathrm{cur}})) \bigr)
\;\cap\;
\bigl( V_{\activelabel,[f_{\mathrm{cur}}^*,f_{\mathrm{cur}})} \cup \cod(\rho_{\mathrm{loc},[f_{\mathrm{cur}}^*,f_{\mathrm{cur}})}) \bigr)
= \emptyset .
\]

Observe that \(V_{\activelabel,[f_{\mathrm{cur}}^*,f_{\mathrm{cur}})}\) and \(\cod(\rho_{\mathrm{loc},[f_{\mathrm{cur}}^*,f_{\mathrm{cur}})})\) are unchanged
from \(\ctx_0\).  From the argument above we already have
\[
\rho_{\mathrm{loc},f_{\mathrm{cur}}}(\fp(f_{\mathrm{cur}})) = \rho_{\mathrm{loc},f_{\mathrm{cur}},0}(\fp(f_{\mathrm{cur}})),\qquad
\rho_{\mathrm{loc},f_{\mathrm{cur}}}(\re(f_{\mathrm{cur}})) = \rho_{\mathrm{loc},f_{\mathrm{cur}},0}(\re(f_{\mathrm{cur}})).
\]
Moreover, Lemma~\ref{lemma:Aset} gives \(A_{f_{\mathrm{cur}}} = A_{f_{\mathrm{cur}},0}\).
Therefore the left‑hand side coincides with the corresponding set for \(\ctx_0\):
\begin{align*}
&\bigl( A_{f_{\mathrm{cur}}} \setminus \rho_{\mathrm{loc},f_{\mathrm{cur}}}(\fp(f_{\mathrm{cur}})) \setminus \rho_{\mathrm{loc},f_{\mathrm{cur}}}(\re(f_{\mathrm{cur}})) \bigr)
  \cap
  \bigl( V_{\activelabel,[f_{\mathrm{cur}}^*,f_{\mathrm{cur}})} \cup \cod(\rho_{\mathrm{loc},[f_{\mathrm{cur}}^*,f_{\mathrm{cur}})}) \bigr) \\
= &\bigl( A_{f_{\mathrm{cur}},0} \setminus \rho_{\mathrm{loc},f_{\mathrm{cur}},0}(\fp(f_{\mathrm{cur}})) \setminus \rho_{\mathrm{loc},f_{\mathrm{cur}},0}(\re(f_{\mathrm{cur}})) \bigr)
  \cap
  \bigl( V_{\activelabel,[f_{\mathrm{cur}}^*,f_{\mathrm{cur}}),0} \cup \cod(\rho_{\mathrm{loc},[f_{\mathrm{cur}}^*,f_{\mathrm{cur}}),0}) \bigr).
\end{align*}
The latter intersection is empty by validity Condition~3(c)(3) of \(\ctx_0\),
which completes the proof of this condition.

  \noindent\textbf{Condition~3(c)(6).}
  We need to show
  \[
  \base\bigl( \cod(\rho_{\mathrm{loc},f_{\mathrm{cur}}}) \setminus
              (\rho_{\mathrm{loc},f_{\mathrm{cur}}}(\fp(f_{\mathrm{cur}})) \cup \rho_{\mathrm{loc},f_{\mathrm{cur}}}(\re(f_{\mathrm{cur}}))) \bigr)
  \subseteq V_{\alloc,f_{\mathrm{cur}}}.
  \]
  From the reasoning above,
  \(\rho_{\mathrm{loc},f_{\mathrm{cur}}}(\fp(f_{\mathrm{cur}})) = \rho_{\mathrm{loc},f_{\mathrm{cur}},0}(\fp(f_{\mathrm{cur}}))\) and
  \(\rho_{\mathrm{loc},f_{\mathrm{cur}}}(\re(f_{\mathrm{cur}})) = \rho_{\mathrm{loc},f_{\mathrm{cur}},0}(\re(f_{\mathrm{cur}}))\).
  Since \(\cod(\rho_{\mathrm{loc},f_{\mathrm{cur}}}) = \cod(\rho_{\mathrm{loc},f_{\mathrm{cur}},0}) \setminus R_v\), we have 
  \[
  \cod(\rho_{\mathrm{loc},f_{\mathrm{cur}}}) \setminus
  \bigl( \rho_{\mathrm{loc},f_{\mathrm{cur}}}(\fp(f_{\mathrm{cur}})) \cup \rho_{\mathrm{loc},f_{\mathrm{cur}}}(\re(f_{\mathrm{cur}})) \bigr)
  \subseteq
  \cod(\rho_{\mathrm{loc},f_{\mathrm{cur}},0}) \setminus
  \bigl( \rho_{\mathrm{loc},f_{\mathrm{cur}},0}(\fp(f_{\mathrm{cur}})) \cup \rho_{\mathrm{loc},f_{\mathrm{cur}},0}(\re(f_{\mathrm{cur}})) \bigr).
  \]
  Hence
  \begin{align*}
  &\base\bigl( \cod(\rho_{\mathrm{loc},f_{\mathrm{cur}}}) \setminus
              (\rho_{\mathrm{loc},f_{\mathrm{cur}}}(\fp(f_{\mathrm{cur}})) \cup \rho_{\mathrm{loc},f_{\mathrm{cur}}}(\re(f_{\mathrm{cur}}))) \bigr)\\ 
  \subseteq
  &\base\bigl( \cod(\rho_{\mathrm{loc},f_{\mathrm{cur}},0}) \setminus
              (\rho_{\mathrm{loc},f_{\mathrm{cur}},0}(\fp(f_{\mathrm{cur}})) \cup \rho_{\mathrm{loc},f_{\mathrm{cur}},0}(\re(f_{\mathrm{cur}}))) \bigr).
   \end{align*}
  By validity Condition~3(c)(5) for \(\ctx_0\), the right‑hand side is contained
  in \(V_{\alloc,0}\).  Since \(V_{\alloc,0} \subseteq \Valloc\), the desired inclusion follows.
\end{proof}

\begin{lemma}\label{lemma:con-4.a}
  Under Notation~\ref{nota:clean} and Assumption~\ref{hyp:clean},
  Condition~4(a) in the validity of \(\ctx\) holds.
\end{lemma}

\begin{proof}
  Take any \(f \in \Fun(\St) \cap \SCC(f_{\mathrm{cur}})\).
  By definition of \(\textsc{Clean\_Temp}\), in both subcases we have
  \[
  \rho_{\mathrm{loc},f_{\mathrm{cur}}} = \rho_{\mathrm{loc},f_{\mathrm{cur}},0} \setminus (-,R_v), \quad
  V_{\activelabel,f_{\mathrm{cur}}} = V_{\activelabel,f_{\mathrm{cur}},0} \setminus R_v,\quad
  \Qpool = \push(\Qpool_0,R_v), \quad R_v \triangleq \rho_{\mathrm{loc},f_{\mathrm{cur}},0}(V)
  \]

  \noindent\textbf{Condition~4(a)(1).}
  We need to show
  \[
  \cod(\rho_{\mathrm{loc},f_{\mathrm{cur}}}) \setminus \bigl( \rho_{\mathrm{loc},f_{\mathrm{cur}}}(\fp(f_{\mathrm{cur}})) \cup \rho_{\mathrm{loc},f_{\mathrm{cur}}}(\re(f_{\mathrm{cur}})) \bigr)
  \subseteq V_{\activelabel,f_{\mathrm{cur}}}.
  \]

  As argued above, \(\rho_{\mathrm{loc},f_{\mathrm{cur}}}(\fp(f_{\mathrm{cur}})) = \rho_{\mathrm{loc},f_{\mathrm{cur}},0}(\fp(f_{\mathrm{cur}}))\) and
  \(\rho_{\mathrm{loc},f_{\mathrm{cur}}}(\re(f_{\mathrm{cur}})) = \rho_{\mathrm{loc},f_{\mathrm{cur}},0}(\re(f_{\mathrm{cur}}))\).  Using the definitions
  of \(\rho_{\mathrm{loc},f_{\mathrm{cur}}}\), \(V_{\activelabel,f_{\mathrm{cur}}}\) and the validity of \(\ctx_0\), we obtain
  \[
  \begin{aligned}
  &\cod(\rho_{\mathrm{loc},f_{\mathrm{cur}}}) \setminus \bigl( \rho_{\mathrm{loc},f_{\mathrm{cur}}}(\fp(f_{\mathrm{cur}})) \cup \rho_{\mathrm{loc},f_{\mathrm{cur}}}(\re(f_{\mathrm{cur}})) \bigr) \\
  = \ &\bigl( \cod(\rho_{\mathrm{loc},f_{\mathrm{cur}},0}) \setminus R_v \bigr)
     \setminus \bigl( \rho_{\mathrm{loc},f_{\mathrm{cur}},0}(\fp(f_{\mathrm{cur}})) \cup \rho_{\mathrm{loc},f_{\mathrm{cur}},0}(\re(f_{\mathrm{cur}})) \bigr) \\[2mm]
  \subseteq \ &\bigl( \cod(\rho_{\mathrm{loc},f_{\mathrm{cur}},0})
     \setminus \bigl( \rho_{\mathrm{loc},f_{\mathrm{cur}},0}(\fp(f_{\mathrm{cur}})) \cup \rho_{\mathrm{loc},f_{\mathrm{cur}},0}(\re(f_{\mathrm{cur}})) \bigr) \bigr)
     \setminus R_v \\[2mm]
  \subseteq \ &\; V_{\activelabel,f_{\mathrm{cur}},0} \setminus R_v
       = V_{\activelabel,f_{\mathrm{cur}}}.
  \end{aligned}
  \]

 For \(f \neq f_{\mathrm{cur}}\), by definition we similarly have
\[
\begin{aligned}
&\cod(\rho_{\mathrm{loc},f}) \setminus \bigl( \rho_{\mathrm{loc},f}(\fp(f)) \cup \rho_{\mathrm{loc},f}(\re(f)) \bigr) \\
= &\bigl( \cod(\rho_{\mathrm{loc},f,0})
   \setminus \bigl( \rho_{\mathrm{loc},f,0}(\fp(f)) \cup \rho_{\mathrm{loc},f,0}(\re(f)) \bigr) \bigr)
   \setminus R_v \\[2mm]
\subseteq &\; V_{\activelabel,f,0} \setminus R_v
     = V_{\activelabel,f}.
\end{aligned}
\] 

 \noindent\textbf{Condition~4(a)(2).} For  
$f \in \Theta^{-1}(\dom(\D)) \setminus \{f_{\mathrm{cur}}\}$,  $\cod(\rho_{\mathrm{loc},f}) \cap \cod(\rho_{\mathrm{loc},[f^{*},f)})=\emptyset$. This holds easily since both \(\cod(\rho_{\mathrm{loc},f})\) and \(\cod(\rho_{\mathrm{loc},[f^{*},f)})\) are obtained by removing \(R_v\) from their counterparts in \(\ctx_0\), which are already disjoint. 

  \noindent\textbf{Condition~4(a)(3).}
  Since $V_{\activelabel,\succeq f_{\mathrm{cur}}^*} \setminus (V_{\activelabel,f_{\mathrm{cur}}} \cup  \rho_{\mathrm{loc},f_{\mathrm{cur}}}(\re(f_{\mathrm{cur}}))) \subseteq V_{\activelabel,\succeq f_{\mathrm{cur}}^*,0} \setminus (V_{\activelabel,f_{\mathrm{cur}},0} \cup  \rho_{\mathrm{loc},f_{\mathrm{cur}},0}(\re(f_{\mathrm{cur}})))$, the required property again follows immediately from 
  the validity of \(\ctx_0\). 
  % For \(f \neq f_{\mathrm{cur}}\), the required property again follows immediately from 
  % the validity of \(\ctx_0\), since \(V_{\activelabel,f} \setminus \rho_{\mathrm{loc},f}(\re(f))
  % \subseteq V_{\activelabel,f,0} \setminus \rho_{\mathrm{loc},f,0}(\re(f))\).

  We now verify the condition: 
  \[
  \forall\, q \in V_{\activelabel,f_{\mathrm{cur}}} \cup  \rho_{\mathrm{loc},f_{\mathrm{cur}}}(\re(f_{\mathrm{cur}})),\;
  \var(\idx(q)) = \emptyset \;\Longrightarrow\; \base[q] = q .
  \]
  The case is similar. Since \(V_{\activelabel,f_{\mathrm{cur}}} = V_{\activelabel,f_{\mathrm{cur}},0} \setminus R_v\),
  \(R_v \cap \rho_{\mathrm{loc},f_{\mathrm{cur}},0}(\re(f_{\mathrm{cur}})) = \emptyset\), and
  \(\rho_{\mathrm{loc},f_{\mathrm{cur}}}(\re(f_{\mathrm{cur}})) = \rho_{\mathrm{loc},f_{\mathrm{cur}},0}(\re(f_{\mathrm{cur}}))\),
  we have
  \[
  V_{\activelabel,f_{\mathrm{cur}}} \cup \rho_{\mathrm{loc},f_{\mathrm{cur}}}(\re(f_{\mathrm{cur}}))
  \subseteq V_{\activelabel,f_{\mathrm{cur}},0} \cup \rho_{\mathrm{loc},f_{\mathrm{cur}},0}(\re(f_{\mathrm{cur}})).
  \] 
  Hence the claim follows directly from validity Condition~3(c) for \(\ctx_0\).

  \noindent\textbf{Condition~4(a)(4).}
  We need to show
  \[
  \forall\, q \in \ctx.V_{\activelabel,\succeq f_{\mathrm{cur}}^*} \cap \ctx.\rho_{\mathrm{loc},f_{\mathrm{cur}}}(\re(f_{\mathrm{cur}})),\;
  \var(\idx(q)) = \emptyset \;\Longrightarrow\; q \in \rho_{\mathrm{loc},f}(\re(f)).  
  \]
  By definition, \(\ctx.V_{\activelabel,\succeq f_{\mathrm{cur}}^*} \subseteq \ctx.V_{\activelabel,\succeq f_{\mathrm{cur},0}^*}\), \(\rho_{\mathrm{loc},f}(\re(f)) = \rho_{\mathrm{loc},f,0}(\re(f))\) and \(\rho_{\mathrm{loc},f_{\mathrm{cur}}}(\re(f_{\mathrm{cur}})) = \rho_{\mathrm{loc},f_{\mathrm{cur}},0}(\re(f_{\mathrm{cur}}))\) hold; these equalities are immediate from the validity of \(\ctx_0\). 

\end{proof} 

\begin{lemma}
    \label{lemma:con-4.b}
  Under Notation~\ref{nota:clean} and Assumption~\ref{hyp:clean},
  Condition~4(b) in the validity of \(\ctx\) holds.
\end{lemma}

\begin{proof}
 We take every \(f \in \SCC(f_{\mathrm{cur}}) \cap \Fun(\St)\). Recall that \(R_v \triangleq \rho_{\mathrm{loc},f_{\mathrm{cur}},0}(V)\). From the previous analysis we have already obtained  
\[
R_v \cap Q_{\succeq f,0} = \emptyset,\qquad
R_v \cap Q_{\succeq f_{\mathrm{cur}}^*,0} = \emptyset,\qquad
Q_{\succeq f_{\mathrm{cur}}^*}[+] = Q_{\succeq f_{\mathrm{cur}}^*,0}[+],
\] 
which indicates that we first show \(R_v \cap Gar_{f,0} = \emptyset\) and \(Gar_f = Gar_{f,0}\) for all such \(f\). 

  \noindent\textbf{Condition~4(b)(1).} 
  We need to show
  \[
  V_{\activelabel,f} \setminus Gar_f \subseteq \cod(\rho_{\mathrm{loc},f}) .
  \]
  The validity of \(\ctx_0\) (Condition~3(c)(4)) gives
  \[
  V_{\activelabel,f,0} \setminus Gar_{f,0} \subseteq \cod(\rho_{\mathrm{loc},f,0}) .
  \]
  By definition of \(\textsc{Clean\_Temp}\),
  \[
  V_{\activelabel,f} = V_{\activelabel,f,0} \setminus R_v,\qquad
  \cod(\rho_{\mathrm{loc},f}) = \cod(\rho_{\mathrm{loc},f,0}) \setminus R_v .
  \]
  Hence, using \(Gar_f = Gar_{f,0}\) and \(R_v \cap Gar_{f,0} = \emptyset\),
  \[
  \begin{aligned}
  V_{\activelabel,f} \setminus Gar_f = \bigl( V_{\activelabel,f,0} \setminus R_v \bigr) \setminus Gar_{f,0} = \bigl( V_{\activelabel,f,0} \setminus Gar_{f,0} \bigr) \setminus R_v \subseteq \cod(\rho_{\mathrm{loc},f,0}) \setminus R_v
   = \cod(\rho_{\mathrm{loc},f}). 
  \end{aligned}
  \]

  \noindent\textbf{Condition~4(b)(2).}
  We need to prove that for every \(h \neq f_{\mathrm{cur}}\),
  \(Q_{\succeq f_{\mathrm{cur}}^*}[+] \cap V_{\activelabel,h} = \emptyset\).
  Since for \(h \neq f_{\mathrm{cur}}\) we have \(V_{\activelabel,h} \subseteq V_{\activelabel,h,0}\), while
  \(Q_{\succeq f_{\mathrm{cur}}^*}[+]\) is unchanged by the cleanup step, the condition
  follows immediately from the validity of \(\ctx_0\).

  \noindent\textbf{Condition~4(b)(3).}
  We need to show
  \[
  Gar_{f_{\mathrm{cur}}} \cap V_{\activelabel,f_{\mathrm{cur}}} \neq \emptyset
  \;\Longrightarrow\;
  \exists\, x \in \dom(\rho_{\mathrm{loc},f_{\mathrm{cur}}}),\;
  x.\flag = 2 \;\wedge\;
  \rho_{\mathrm{loc},f_{\mathrm{cur}}}(x) \in V_{\activelabel,f_{\mathrm{cur}}} \;\wedge\;
  Gar_{f_{\mathrm{cur}}} \subseteq V_{\activelabel,f_{\mathrm{cur}}}.
  \]

Assume \(Gar_{f_{\mathrm{cur}}} \cap V_{\activelabel,f_{\mathrm{cur}}} \neq \emptyset\). Since \(Gar_{f_{\mathrm{cur}}} = Gar_{f_{\mathrm{cur}},0}\) and \(V_{\activelabel,f_{\mathrm{cur}}} \subseteq V_{\activelabel,f_{\mathrm{cur}},0}\), we have \(Gar_{f_{\mathrm{cur}},0} \cap V_{\activelabel,f_{\mathrm{cur}},0} \neq \emptyset\). 
The validity of \(\ctx_0\) then supplies
\[
\exists\, x \in \dom(\rho_{\mathrm{loc},f_{\mathrm{cur}},0}),\;
x.\flag = 2 \;\wedge\;
\rho_{\mathrm{loc},f_{\mathrm{cur}},0}(x) \in V_{\activelabel,f_{\mathrm{cur}},0} \;\wedge\;
Gar_{f_{\mathrm{cur}},0} \subseteq V_{\activelabel,f_{\mathrm{cur}},0}.
\]
Since \(x\) carries \(\flag = 2\), we have \(x \notin R_v\); still, \(x\) remains in both \(\rho_{\mathrm{loc},f_{\mathrm{cur}}}\) and \(V_{\activelabel,f_{\mathrm{cur}}}\). Combining this with \(R_v \cap Gar_{f_{\mathrm{cur}},0} = \emptyset\), which implies that \(Gar_{f_{\mathrm{cur}},0}\) also stays in \(V_{\activelabel,f_{\mathrm{cur}}}\), we obtain
\[
\exists\, x \in \dom(\rho_{\mathrm{loc},f_{\mathrm{cur}}}),\;
x.\flag = 2 \;\wedge\;
\rho_{\mathrm{loc},f_{\mathrm{cur}}}(x) \in V_{\activelabel,f_{\mathrm{cur}}} \;\wedge\;
Gar_{f_{\mathrm{cur}}} \subseteq V_{\activelabel,f_{\mathrm{cur}}},
\]
which completes the proof.  
\end{proof} 

\begin{lemma}
    \label{lemma:con-4.c}
  Under Notation~\ref{nota:clean} and Assumption~\ref{hyp:clean},
  Condition~4(c) in the validity of \(\ctx\) holds.
\end{lemma}
\begin{proof}
    Take any \(f \in \Fun(\St) \cap \SCC(f_{\mathrm{cur}})\).
For \(f \neq f_{\mathrm{cur}}\), the code \(C_f\) is unchanged by the cleanup step,
and Lemma~\ref{lemma:Aset} gives \(\ctx_0.A_f \subseteq \ctx.A_f\).
Since the validity of \(\ctx_0\) yields \(\anc(C_{f,0}^{\,k}) \subseteq A_{f,0}\),
we obtain \(\anc(C_f^{\,k}) \subseteq A_f\) directly.
Hence it suffices to consider \(f = f_{\mathrm{cur}}\).

\noindent\textbf{Case 1: \(v.\flag = 0\).}
Here \(C_{f_{\mathrm{cur}}} = C_{f_{\mathrm{cur}},0} ; C_{\mathrm{cl}}\), where
\(C_{\mathrm{cl}} = \qinv\bigl(\kappa_{f_{\mathrm{cur}},0}(v)[\req_v \mapsto \overline{p}]\bigr)\).
Since inversion does not change quantum variables,
\[
\anc(C_{\mathrm{cl}}^{\,k}) \subseteq \qv(C_{\mathrm{cl}}^{\,k})
   = \qv\bigl(\kappa_{f_{\mathrm{cur}},0}(v)^k[\req_v \mapsto \overline{p}]\bigr)
   \subseteq \rho_{\mathrm{loc},f_{\mathrm{cur}},0}(\operatorname{pred}^*(v)) \cup \overline{p}
   \subseteq \cod(\rho_{\mathrm{loc},f_{\mathrm{cur}},0}) \cup \overline{p}.
\]
By validity Condition~4(c) for \(\ctx_0\), we have
\(\anc(C_{f_{\mathrm{cur}},0}^{\,k}) \subseteq A_{f_{\mathrm{cur}},0}\).
Moreover, \(\overline{p} \subseteq \Qpool_0 \subseteq cv_0 \subseteq A_{f_{\mathrm{cur}},0}\),
and \(\cod(\rho_{\mathrm{loc},f_{\mathrm{cur}},0}) \subseteq A_{f_{\mathrm{cur}},0}\) by the definition of \(A_{f_{\mathrm{cur}},0}\).
Since Lemma~\ref{lemma:Aset} gives \(A_{f_{\mathrm{cur}}} = A_{f_{\mathrm{cur}},0}\), we obtain
\[
\anc(C_{f_{\mathrm{cur}}}^{\,k})
   \subseteq \anc(C_{f_{\mathrm{cur}},0}^{\,k}) \cup \anc(C_{\mathrm{cl}}^{\,k})
   \subseteq A_{f_{\mathrm{cur}}}.
\]

\noindent\textbf{Case 2: \(v.\flag = 1\).}
Here \(C_{f_{\mathrm{cur}}} = C_{f_{\mathrm{cur}},0} ; \qinv(\kappa_{f_{\mathrm{cur}},0}(v))\).
Again, inversion preserves quantum variables, so
\[
\anc(\qinv(\kappa_{f_{\mathrm{cur}},0}(v))^k) = \anc(\kappa_{f_{\mathrm{cur}},0}(v)^k).
\]
By the local validity of \(\ctx_0\),
\(\qv(\kappa_{f_{\mathrm{cur}},0}(v)^k) \subseteq \qv(C_{f_{\mathrm{cur}},0}^k)\),
hence \(\anc(\kappa_{f_{\mathrm{cur}},0}(v)^k) \subseteq \anc(C_{f_{\mathrm{cur}},0}^k) \cup S\)
with \(S \subseteq \Qpool_0 \cup V_{\activelabel,\succeq f_{\mathrm{cur}},0}\setminus Q_{\succeq f_{\mathrm{cur}}^*,0}[+]\).
Using validity Condition~4(c) for \(\ctx_0\), we have
\(\anc(C_{f_{\mathrm{cur}},0}^k) \subseteq A_{f_{\mathrm{cur}},0}\).
Since \(\Qpool_0 \cup V_{\activelabel,\succeq f_{\mathrm{cur}},0}\setminus Q_{\succeq f_{\mathrm{cur}}^*,0}[+] \subseteq A_{f_{\mathrm{cur}},0}\) and
\(A_{f_{\mathrm{cur}}} = A_{f_{\mathrm{cur}},0}\) (Lemma~\ref{lemma:Aset}), it follows that
\[
\anc(C_{f_{\mathrm{cur}}}^{\,k})
   \subseteq \anc(C_{f_{\mathrm{cur}},0}^{\,k}) \cup \anc(\kappa_{f_{\mathrm{cur}},0}(v)^k)
   \subseteq A_{f_{\mathrm{cur}}}.
\]

In both cases the required inclusion \(\anc(C_f^{\,k}) \subseteq A_f\) holds for all
\(f \in \Fun(\St) \cap \SCC(f_{\mathrm{cur}})\), which establishes Condition~4(c).
\end{proof}

\begin{lemma}
  \label{lemma:con-4.d}
  Under Notation~\ref{nota:clean} and Assumption~\ref{hyp:clean},
  Condition~4(d) in the validity of \(\ctx\) holds.
\end{lemma}

\begin{proof}
  Take any \(f \in \Fun(\St) \cap \SCC(f_{\mathrm{cur}})\).
  For \(f \neq f_{\mathrm{cur}}\), the code fragment \(C_f\) is unchanged by the cleanup step.
  Moreover, Lemma~\ref{lemma:Aset} gives
  \(\ctx.A_{f_{\mathrm{cur}}^*} = \ctx_0.A_{f_{\mathrm{cur}}^*}\), which implies \(\ctx.B = \ctx_0.B\).
  Hence for those \(f\) the condition follows directly from the validity of
  \(\ctx_0\), and it suffices to verify the case \(f = f_{\mathrm{cur}}\).

  By definition,
  \[
  \anc(C_{f_{\mathrm{cur}}}^k) = \anc(C_{f_{\mathrm{cur}},0}^k) \cup \anc(C_{\mathrm{cl}}^k),
  \]
  hence
  \begin{align*}
   &\anc(C_{f_{\mathrm{cur}}}^k) \setminus \bigl( \anc(F_{\succeq f_{\mathrm{cur}}^*}^k) \cup \anc(F_{\succeq f_{\mathrm{cur}}^*}^k)[+] \bigr) \\ 
  = &\Bigl( \anc(C_{f_{\mathrm{cur}},0}^k) \setminus \bigl( \anc(F_{\succeq f_{\mathrm{cur}}^*}^k) \cup \anc(F_{\succeq f_{\mathrm{cur}}^*}^k)[+] \bigr) \Bigr) 
    \cup \Bigl( \anc(C_{\mathrm{cl}}^k) \setminus \bigl( \anc(F_{\succeq f_{\mathrm{cur}}^*}^k) \cup \anc(F_{\succeq f_{\mathrm{cur}}^*}^k)[+] \bigr) \Bigr).
   \end{align*}
  The validity of \(\ctx_0\) already gives
  \[
  \Bigl( \anc(C_{f_{\mathrm{cur}},0}^k) \setminus \bigl( \anc(F_{\succeq f_{\mathrm{cur}}^*}^k) \cup \anc(F_{\succeq f_{\mathrm{cur}}^*}^k)[+] \bigr) \Bigr) \cap \ctx_0.B[+] = \emptyset .
  \]

  Observe that, by construction,
  \[
  \anc(C_{\mathrm{cl}}^k) \subseteq \anc(\kappa_{f_{\mathrm{cur}},0}(v)) \cup \Qpool_0 \cup \cod(\rho_{\mathrm{loc},f_{\mathrm{cur}},0}),
  \]
  and therefore
  \begin{align*} 
    &\anc(C_{\mathrm{cl}}^k) \setminus \bigl( \anc(F_{\succeq f_{\mathrm{cur}}^*}^k) \cup \anc(F_{\succeq f_{\mathrm{cur}}^*}^k)[+] \bigr) \\ 
  \subseteq &\Bigl( \anc(\kappa_{f_{\mathrm{cur}},0}(v)) \setminus \bigl( \anc(F_{\succeq f_{\mathrm{cur}}^*}^k) \cup \anc(F_{\succeq f_{\mathrm{cur}}^*}^k)[+] \bigr) \Bigr)
            \cup \Qpool_0 \cup \cod(\rho_{\mathrm{loc},f_{\mathrm{cur}},0}).
  \end{align*} 
  Again, the validity of \(\ctx_0\) (specifically the condition concerning
  the cleanup circuit \(\kappa_{f_{\mathrm{cur}},0}(v)\)) yields
  \[
  \anc(\kappa_{f_{\mathrm{cur}},0}(v)) \subseteq \anc(C_{f_{\mathrm{cur}},0}^k) \cup \Qpool_0 \cup (V_{\activelabel,\succeq f_{\mathrm{cur}},0}\setminus Q_{\succeq f_{\mathrm{cur}}^*,0}[+]) \subseteq \anc(C_{f_{\mathrm{cur}},0}^k) \cup \Qpool_0 \cup (\cod(\rho_{\mathrm{loc},\succeq f_{\mathrm{cur}},0})).
  \]

  Thus it remains to prove
  \[
  \bigl( \Qpool_0 \cup \cod(\rho_{\mathrm{loc},\succeq f_{\mathrm{cur}},0}) \bigr) \cap \ctx_0.B[+] = \emptyset  .
  \]
  Take any \(q \in \Qpool_0 \cup \cod(\rho_{\mathrm{loc},\succeq f_{\mathrm{cur}},0})\).
  If \(\base[q] = q\), then clearly \(q \notin \ctx_0.B[+]\).
  If \(\base[q] \neq q\), we show that \(\base(q) \not\subseteq \ctx_0.B\);
  otherwise, assume for contradiction that \(\base(q) \subseteq \ctx_0.B\).
  By definition of \(\ctx_0.B\),
  \[
  \base(q) \subseteq \Qpool_0 \cup V_{\activelabel,\succeq f_{\mathrm{cur}}^*,0} \cup \cod(\rho_{\mathrm{loc},f_{\mathrm{cur}}^*,0}).
  \]
  \begin{itemize}
    \item If \(\base(q) \subseteq \Qpool_0\), then \(q \in \Qpool_0[+]\).
          But the validity of \(\ctx_0\) demands
          \(\Qpool_0[+] \cap (\Qpool_0 \cup \cod(\rho_{\mathrm{loc},\succeq f_{\mathrm{cur}},0})) = \emptyset\),
          contradicting \(q \in \Qpool_0 \cup \cod(\rho_{\mathrm{loc},f_{\mathrm{cur}},0})\).
    \item If \(\base(q) \subseteq V_{\activelabel,\succeq f_{\mathrm{cur}}^*,0} \cup \cod(\rho_{\mathrm{loc},f_{\mathrm{cur}}^*,0})\),
          then \(q \in (V_{\activelabel,\succeq f_{\mathrm{cur}}^*,0} \cup \cod(\rho_{\mathrm{loc},f_{\mathrm{cur}}^*,0}))[+]\).
          The validity of \(\ctx_0\) gives
          \((V_{\activelabel,\succeq f_{\mathrm{cur}}^*,0} \cup \cod(\rho_{\mathrm{loc},f_{\mathrm{cur}}^*,0}))[+] \cap (\Qpool_0 \cup \cod(\rho_{\mathrm{loc},\succeq f_{\mathrm{cur}},0})) = \emptyset\),
          again a contradiction.
  \end{itemize}
  Hence \(\base(q) \not\subseteq \ctx_0.B\), which implies \(q \notin \ctx_0.B[+]\).
  Since \(q\) was arbitrary, we conclude
  \[
  \bigl( \Qpool_0 \cup \cod(\rho_{\mathrm{loc},\succeq f_{\mathrm{cur}},0}) \bigr) \cap \ctx_0.B[+] = \emptyset .
  \]

  Finally, since \(\ctx.B = \ctx_0.B\), the required disjointness for
  Condition~4(d) follows immediately.
\end{proof} 

\begin{lemma}
     \label{lemma:con-4.f}
Under Notation~\ref{nota:clean} and Assumption~\ref{hyp:clean}, Condition~4(f) in the validity of \(\ctx\) holds.
\end{lemma}
\begin{proof}
Similarly, it suffices to verify the case \(f = f_{\mathrm{cur}}\): for every depth \(z\),
\[
\base(\anc(C_{f_{\mathrm{cur}}}^{\,z})) \subseteq V_{\alloc,\succeq f_{\mathrm{cur}}^{*}} \cup \base(\anc(F_{\succeq f_{\mathrm{cur}}^{*}}^{\,z})). 
\]

\noindent\textbf{Case 1: \(v.\flag = 0\).}
In this case the compiled code is \(C_{f_{\mathrm{cur}}} = C_{f_{\mathrm{cur}},0}\,;\,C_{\mathrm{cl}}\),
where \(C_{\mathrm{cl}} \triangleq \qinv(\kappa_{f_{\mathrm{cur}},0}(v)[\req_v \mapsto \overline{p}])\).
Hence
\[
\anc(C_{f_{\mathrm{cur}}}^{\,z}) \subseteq \anc(C_{f_{\mathrm{cur}},0}^{\,z}) \cup \anc(C_{\mathrm{cl}}^{\,z}) .
\]

For the cleanup part we have
\begin{align*}
\anc(C_{\mathrm{cl}}^{\,z})
&= \qv(C_{\mathrm{cl}}^{\,z}) \setminus \bigl( \rho_{\mathrm{loc},f_{\mathrm{cur}}}(\fp(f_{\mathrm{cur}})) \cup \rho_{\mathrm{loc},f_{\mathrm{cur}}}(\operatorname{var}(v)) \bigr) \\
&\subseteq \bigl( \rho_{\mathrm{loc},f_{\mathrm{cur}},0}(\mathrm{pred}^*(v)) \cup \overline{p} \bigr)
   \setminus \bigl( \rho_{\mathrm{loc},f_{\mathrm{cur}},0}(\fp(f_{\mathrm{cur}})) \cup \rho_{\mathrm{loc},f_{\mathrm{cur}},0}(\operatorname{var}(v)) \bigr) \\
&\subseteq \Bigl( \cod(\rho_{\mathrm{loc},f_{\mathrm{cur}},0}) \setminus
                \bigl( \rho_{\mathrm{loc},f_{\mathrm{cur}},0}(\fp(f_{\mathrm{cur}})) \cup \rho_{\mathrm{loc},f_{\mathrm{cur}},0}(\operatorname{var}(v)) \bigr) \Bigr)
          \cup \overline{p} .
\end{align*}
By the local validity of \(\ctx_0\), we know
\(\base(\cod(\rho_{\mathrm{loc},f_{\mathrm{cur}},0}) \setminus
 \bigl( \rho_{\mathrm{loc},f_{\mathrm{cur}},0}(\fp(f_{\mathrm{cur}})) \cup \rho_{\mathrm{loc},f_{\mathrm{cur}},0}(\operatorname{var}(v))\bigr) ) 
 \subseteq V_{\alloc,f_{\mathrm{cur}},0}\);
thus
\[
\base(\anc(C_{\mathrm{cl}}^{\,z})) \subseteq V_{\alloc,f_{\mathrm{cur}},0} \cup \base(\overline{p}).
\]

Now,
\begin{align*}
\base\bigl(\anc(C_{f_{\mathrm{cur}}}^{\,z})\bigr)
&= \base\bigl(\anc(C_{f_{\mathrm{cur}},0}^{\,z})\bigr) \cup \base(\anc(C_{\mathrm{cl}}^{\,z})) 
   \tag{since \(\base\) distributes over unions} \\
&\subseteq \base\bigl(\anc(C_{f_{\mathrm{cur}},0}^{\,z})\bigr) \cup V_{\alloc,f_{\mathrm{cur}},0} \cup \base(\overline{p}) \\
&\subseteq \Bigl( V_{\alloc,\succeq f_{\mathrm{cur}}^*,0} \cup
                  \base\bigl(\anc(F_{\succeq f_{\mathrm{cur}}^*}^{\,z})\bigr) \Bigr) 
          \cup \base(\overline{p})
   \tag{by validity Condition~4(c) for \(\ctx_0\)} \\
&= V_{\alloc,\succeq f_{\mathrm{cur}}^*,0} \cup \base(\overline{p})
   \cup \base\bigl(\anc(F_{\succeq f_{\mathrm{cur}}^*}^{\,z})\bigr) .
\end{align*}
By the definition of the cleanup step,
\(V_{\alloc,\succeq f_{\mathrm{cur}}^*} = V_{\alloc,\succeq f_{\mathrm{cur}}^*,0} \cup \base(\overline{p})\).
Therefore
\[
\base\bigl(\anc(C_{f_{\mathrm{cur}}}^{\,z})\bigr)
\subseteq V_{\alloc,\succeq f_{\mathrm{cur}}^*} \cup \base\bigl(\anc(F_{\succeq f_{\mathrm{cur}}^*}^{\,z})\bigr) ,
\]
which is exactly the required inclusion for Case~1. 

\noindent\textbf{Case 2: \(v.\flag = 1\).}
In this case, \(C_{f_{\mathrm{cur}}} = C_{f_{\mathrm{cur}},0} ; \qinv(\kappa_{f_{\mathrm{cur}},0}(v))\).
Since inversion preserves quantum variables,
\[
\anc(C_{f_{\mathrm{cur}}}^z) \subseteq \anc(C_{f_{\mathrm{cur}},0}^z) \cup \anc(\kappa_{f_{\mathrm{cur}},0}(v)^z).
\]
By the validity of \(\ctx_0\) (cleanup-circuit correctness),
we have \(\anc(\kappa_{f_{\mathrm{cur}},0}(v)^z) \subseteq \anc(C_{f_{\mathrm{cur}},0}^z) \cup S\)
with \(\base(S) \subseteq V_{\alloc,f_{\mathrm{cur}},0}\).
Notice that in this case the cleanup step does not add new base names to
\(\Valloc\); hence \(V_{\alloc,\succeq f_{\mathrm{cur}}^*} = V_{\alloc,\succeq f_{\mathrm{cur}}^*,0}\).
Therefore
\[
\begin{aligned}
\base\bigl(\anc(C_{f_{\mathrm{cur}}}^{\,z})\bigr)
&\subseteq \base\bigl(\anc(C_{f_{\mathrm{cur}},0}^{\,z})\bigr) \cup V_{\alloc,f_{\mathrm{cur}},0} \\
&\subseteq \Bigl( V_{\alloc,\succeq f_{\mathrm{cur}}^*,0} \cup
                  \base\bigl(\anc(F_{\succeq f_{\mathrm{cur}}^*}^{\,z})\bigr) \Bigr) \\
&= V_{\alloc,\succeq f_{\mathrm{cur}}^*} \cup \base\bigl(\anc(F_{\succeq f_{\mathrm{cur}}^*}^{\,z})\bigr) .
\end{aligned}
\]
The required inclusion follows. 
\end{proof}

\begin{theorem}
\label{theo:clean-valid} 
Under Notation~\ref{nota:clean} and Assumption~\ref{hyp:clean}, \(\ctx\) preserves validity with respect to \((\ket{\phi},k,\D_{\SCC(f_{\mathrm{cur}})})\): if \(v.\flag=0\), then
\[
\bigl(\ctx,(\ket{\phi},k,\D_{\SCC(f_{\mathrm{cur}})})\bigr)\in\mathcal V; 
\]
if \(v.\flag=1\), then all validity conditions are preserved except that the Condition~3(a) clause for \(f=f_{\mathrm{cur}}\) needs not hold.
\end{theorem}

\begin{proof}

Recall that \(\textsc{Clean\_Temp}\) modifies only the components
\(\Qpool\), \(C\), \(\Vactive\), \(\Valloc\), \(\rho_{\mathrm{loc}}\), and \(\kappa\)
associated with \(f_{\mathrm{cur}}\), together with the local mappings \(\rho_{\mathrm{loc}}\) and unclean sets \(\Vactive\) stored in the compilation stack; the function name 
stack \(\St\) and all other parts of the context remain unchanged.
In particular, none of the variables appearing in Condition~1 are
modified by this step.  Hence Condition~1 carries over directly from
\(\ctx_0\), and it remains to verify Conditions~2, 3, and 4. 

\begin{itemize}
\item Condition~2: By Lemma~\ref{lemma:con-2}. 

\item Condition~3(a): By Lemma~\ref{lemma:con-3.a}.

\item Condition~3(b): By Lemma~\ref{lemma:con-3.b}.

\item Condition~3(c): By Lemma~\ref{lemma:con-3.c}. 
 
\item Condition~4(a): By Lemma~\ref{lemma:con-4.a}.

\item Condition~4(b): By Lemma~\ref{lemma:con-4.b}.  
\item Condition~4(c): By Lemma~\ref{lemma:con-4.c}.

\item Condition~4(d): By Lemma~\ref{lemma:con-4.d}.

\item Condition~4(e). 
      We need to show \(\var(\idx(A_{f_{\mathrm{cur}}^*})) \subseteq E_{\args,f_{\mathrm{cur}}^*} = E_{\args,f}\)
      for every \(f\).
      Since \(\textsc{Clean\_Temp}\) does not modify the argument set of any function,
      we have \(E_{\args,f} = E_{\args,f,0}\) and \(E_{\args,f_{\mathrm{cur}}^*} = E_{\args,f_{\mathrm{cur}}^*,0}\).
      Moreover, we already proved \(A_{f_{\mathrm{cur}}^*} = A_{f_{\mathrm{cur}}^*,0}\) by Lemma~\ref{lemma:Aset}.
      Hence the claim follows immediately from the validity of \(\ctx_0\)
      (Condition~4(e)). 
\item Condition~4(f): By Lemma~\ref{lemma:con-4.f}.
\item Condition~4(g). 
We need to show
\( \bigl(\Succ(f)\setminus \Call(R_{\mathrm{stmt},f})\bigr) \cap \SCC(f_{\mathrm{cur}}) \subseteq \Fun(\St). \) 
Since \(\textsc{Clean\_Temp}\) does not modify \(R_{\mathrm{stmt},f}\), both sides of the inclusion remain unchanged.
Hence the claim follows immediately from the validity of \(\ctx_0\)
(Condition~4(g)). 
\end{itemize}

\end{proof}

\subsection{Correctness Proof for Compilation of Bodies}
\label{app:proof-body}
In this subsection we prove the correctness theorem of Algorithm~\ref{com_body} for the compilation of function‑bodies.  The proof follows a similar
structure: we first fix the common notation and assumptions, then
present the overall proof architecture together with the dependencies
among the theorems (lemmas), and finally give their detailed proofs.

\begin{notation}
    \label{nota:body}
   Suppose we are given a state \(\ket{\phi} \in \mathcal{H}\), a depth \(k \in \mathbb{N}^+\), and a set of procedure declarations \(\D_{\SCC(f_{\mathrm{cur}})}\). Let \(s\) be a source statement, and let \(\sigma_0\), \(\sigma\) be source language states such that \(\langle s^{\,k}, \sigma_0 \rangle \rightarrow_{\Xi}^{*} \langle \downarrow, \sigma \rangle\). Let \(\ctx_0\) be a compilation context, and \(\ctx = \textsc{Compile\_Body}(\K, \ctx_0, s)\). If \(s\) is not a sequential statement (\(s \neq s_1; s_2\)), then by definition, the compiled context is further denoted as:  
\[
\ctx=\ctx_{(1+\vert V_b\vert)'}[\delta_{(1+\vert V_b\vert)'} \mapsto \delta_a, R_{\mathrm{stmt},(1+\vert V_b\vert)'} \mapsto \mathrm{tail}(R_{\mathrm{stmt},(1+\vert V_b\vert)'})],  
\]
where
\begin{align*}
&(\ctx_1, C_a) = \textsc{Compile\_Com}(\K, \ctx_0, s), \\
&\kappa_a = \kappa_1[w \mapsto \kappa_1(w); C_a, 
               v \mapsto \operatorname{qinv}(C_a); \kappa_1(v); C_a \mid v \in V_1], \\
&\qquad \text{for } w = \node(\mvq(s)) \wedge \neg(\flag(\mvq(s)) = 2 \wedge \Call(s)\neq \emptyset) \wedge \mvq(s) \not\subseteq \dom(\rho_{\mathrm{glob}}); \\ 
& \qquad \quad  V_1 =\{ v \in \dom(\kappa_1) \mid w\in \operatorname{pred}^+(v) \wedge  \neg(\flag(\mvq(s)) = 2 \wedge \Call(s)\neq \emptyset) \} \\ 
% & \quad \text{and} \  V_2=\{v \in \dom(\kappa_1) \mid (w \notin \dom(\rho_{0}) \wedge  \neg (\flag(\mvq(s)) = 2 \wedge \Call(s)\neq \emptyset) \wedge w \in  \operatorname{pred}^+(v)) \\ 
% & \qquad \qquad \qquad \qquad \qquad  \qquad  \vee ((\flag(\mvq(s)) = 2 \wedge \Call(s)\neq \emptyset)  \wedge w \in (\operatorname{pred}^*(v))) \}; \\
& \ctx_{1'}= \ctx_1[\kappa \mapsto \kappa_a]; \\ 
% & ctx_{1'}= ctx_1[\kappa \mapsto \kappa_a][allv \mapsto allv \cup \{q\} \mid V_2 \neq \emptyset]; \\ 
&(\ctx_{(i+1)'},C_{i+1}) = \textsc{Clean\_Temp}\bigl(\K, \ctx_{i'}, u_i \bigr) \ \text{for} \ u_i \in V_b, i=1,2,\cdots \vert V_b\vert \\ 
&V_a = \node(\rvq(s)); \quad V_c = \{u \in V_a \mid \delta_{a}(u) = 0 \wedge u.\flag=0 \}\\  
&V_{b} =\{ u \in V_c \mid \var(u) \cap (\dom(\rho_{\mathrm{glob}}) \cup \re(f_{\mathrm{cur}})) =\emptyset \};\\  
& \delta_a = \delta_1[u \mapsto \delta_1(u) - 1 \mid u \in V_a],
\end{align*}
\end{notation}

\begin{assumption} \label{hpy:body}
    In addition to Notation~\ref{nota:body}, assume that $s$ is a prefix of \(\ctx_0.R_{\mathrm{stmt}}\), $\ctx_0 \approx_{(\ket{\phi}, k)} \sigma_0$  and \[(\ctx_0, (\ket{\phi}, k, \D_{\SCC(f_{\mathrm{cur}})})) \in \mathcal{V}. \]  
\end{assumption}

\begin{figure}[t]
\centering
\begin{tikzpicture}[
  x=1cm,
  y=1cm,
  proofnode/.style={
    draw,
    rounded corners=2pt,
    align=center,
    inner xsep=3pt,
    inner ysep=3pt,
    minimum height=7mm,
    font=\scriptsize
  },
  thmnode/.style={proofnode, fill=blue!6, draw=blue!45!black},
  lemmanode/.style={proofnode, fill=gray!8, draw=gray!60},
  dep/.style={-{Latex[length=1.8mm]}, thick, draw=gray!70}
]
\node[thmnode, text width=.18\linewidth] (body) at (-1.05,1.35)
  {Theorem~\ref{theo:body}\\
   Body correctness};
\node[lemmanode, text width=.11\linewidth] (validone) at (-4.2,0)
  {Lemma~\ref{lemma:valid-1}};
\node[thmnode, text width=.11\linewidth] (clean) at (-2.1,0)
  {Theorem~\ref{theo:clean}};
\node[thmnode, text width=.11\linewidth] (cleanvalid) at (0,0)
  {Theorem~\ref{theo:clean-valid}};
\node[lemmanode, text width=.11\linewidth] (cleaninv) at (2.1,0)
  {Lemma~\ref{lemma:clean-inv}};
\node[thmnode, text width=.11\linewidth] (stmt) at (-4.2,-1.15)
  {Theorem~\ref{theo_com}};
\node[lemmanode, text width=.11\linewidth] (bodyinv) at (3.65,1.35)
  {Lemma~\ref{lem:body-inv}};
\node[lemmanode, text width=.11\linewidth] (stateinv) at (5.2,0)
  {Lemma~\ref{lem:state-inv-2}};
\draw[dep] (body) -- (validone);
\draw[dep] (body) -- (clean);
\draw[dep] (body) -- (cleanvalid);
\draw[dep] (body) -- (cleaninv);
\draw[dep] (validone) -- (stmt);
\draw[dep] (bodyinv) -- (cleaninv);
\draw[dep] (bodyinv) -- (stateinv);
\end{tikzpicture}
\caption{Proof dependencies for the correctness of \textsc{Compile\_Body}; arrows point from a result to the lemmas or theorems it depends on.}
\label{fig:body-proof-dep}
\Description{A dependency diagram showing that Theorem body depends on Lemma body-inv, Lemma valid-1, and cleanup correctness results; Lemma valid-1 depends on statement correctness, and Lemma body-inv depends on state-invariant and cleanup-invariant lemmas.}
\end{figure}
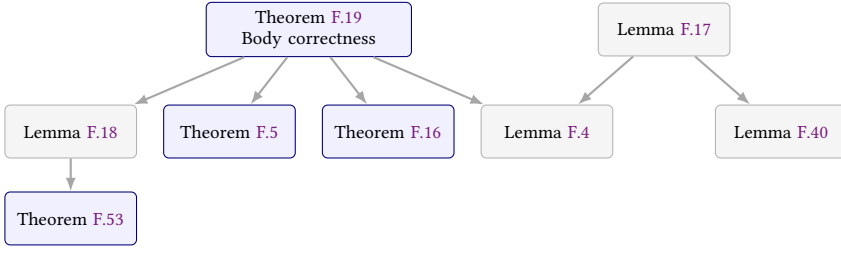

As illustrated in Figure~\ref{fig:body-proof-dep}, the correctness of Algorithm~\ref{com_body} is established by Theorem~\ref{theo:body}. Its proof proceeds as follows. Using the correctness theorem for statement compilation (Theorem~\ref{theo_com}), Lemma~\ref{lemma:valid-1} first builds the validity of the intermediate context \(\ctx_{1'}\) together with the compilation correctness of the generated code. From the validity of \(\ctx_{1'}\), Theorem~\ref{theo:body} then inductively applies the correctness theorems of the cleanup algorithm (Lemma~\ref{lemma:clean-inv}, Theorem~\ref{theo:clean}, Theorem~\ref{theo:clean-valid} of the previous subsection), finally obtaining the validity of \(\ctx_{(1+\vert V_b\vert)'}\) and the compilation correctness of the generated code, thereby completing the proof. Similarly, Lemma~\ref{lem:body-inv} gives a relation between certain compilation states of the input context \(\ctx_0\) and that of the output context \(\ctx\), which depends on Lemmas~\ref{lemma:clean-inv} and~\ref{lem:state-inv-2} (giving the corresponding properties for the cleanup algorithm and the statement compilation algorithm, respectively).

\begin{lemma}
\label{lem:body-inv}
Under Notation~\ref{nota:body} and Assumption that $\ctx_0$ satisfies the structural properties of validity, the following holds:
\begin{enumerate}
    \item For $\D$, 
    \begin{itemize}
        \item $\dom(\D_0) \subseteq \dom(\D)$; 
        \item $\dom(\D) \setminus \dom(\D_0) \subseteq \Theta(\Succ^*(f_{\mathrm{cur}}) \setminus \Fun(\St_0))$; 
        \item $\D\upharpoonright_{\dom(\D_0) \setminus \Theta(\SCC(f_{\mathrm{cur}}))} = \D_0\upharpoonright_{\dom(\D_0) \setminus \Theta(\SCC(f_{\mathrm{cur}}))}$;
        \item $\D\upharpoonright_{\dom(\D_0) \cap \Theta(\SCC(f_{\mathrm{cur}}))} \approx_{\ctx.Gar_{f_{\mathrm{cur}}^*}} \D_0\upharpoonright_{\dom(\D_0) \cap \Theta(\SCC(f_{\mathrm{cur}}))}$; and for any $h \in \SCC(f_{\mathrm{cur}}) \cap \Theta^{-1}(\dom(\D_0))$, $\anc(\D[\Theta(h)]) \subseteq \anc(\D_0[\Theta(h)]) \cup \anc(\D_0[\Theta(h)])[+]$. 
    \end{itemize} 
    \item For $\Fun(\St)$, 
    \begin{itemize}
        \item $\Fun(\St_0) \subseteq \Fun(\St)$;
        \item $\Fun(\St_{\preceq h }) = \Fun(\St_{\preceq h,0})$ for any $h \in \Fun(\St_{\preceq f_{\mathrm{cur}},0})$;
        \item $\Fun(\St_{\succ f_{\mathrm{cur}} }) \setminus \Fun(\St_{\succ f_{\mathrm{cur}},0}) \subseteq \Succ^*(\Succ(f_{\mathrm{cur}}) \setminus \Fun(\St_{0})) \cap \SCC(f_{\mathrm{cur}})$. 
    \end{itemize}
    \item \(\St_{\preceq f_{\mathrm{cur}}^*}=\St_{\preceq f_{\mathrm{cur}}^*,0}\) and \(\M_{\preceq f_{\mathrm{cur}}^*}=\M_{\preceq f_{\mathrm{cur}}^*,0}\). 
    \item For any $h \in \SCC(f_{\mathrm{cur}}) \cap \Fun(\St_0)$, 
     \begin{itemize}
        \item \(E_{\args,h,0} \preceq E_{\args,h}\);
        \item $\ctx.A_h \subseteq \ctx_0.A_h$ if $\pos(h) \leq \pos (f_{\mathrm{cur}})$;
        % \item $\rho_{\mathrm{loc},h}=\rho_{\mathrm{loc},h,0}$, $V_{\activelabel,h}=V_{\activelabel,h,0}$ and $V_{\activelabel,[f_{\mathrm{cur}}^*,h]}=V_{\activelabel,[f_{\mathrm{cur}}^*,h],0}$ if $h < f_{\mathrm{cur}}$; 
        \item $\dom(\rho_{\mathrm{loc},h}) \subseteq \dom(\rho_{\mathrm{loc},h,0})$ if $h \neq f_{\mathrm{cur}}$;
        \item $\rho_{\mathrm{loc},h}(x)=\rho_{\mathrm{loc},h,0}(x)$, if $x \in \dom(\rho_{\mathrm{loc},h,0})\cap \dom(\rho_{\mathrm{loc},h})$ and $\var(\idx(\rho_{\mathrm{loc},h,0}(x)))\neq \emptyset$. 
    \end{itemize}
\end{enumerate}
\end{lemma}
\begin{proof}
    For a single statement, the conclusion follows directly from the construction of $\ctx$ together with Lemmas~\ref{lemma:clean-inv} and~\ref{lem:state-inv-2}. For sequential statements, the result is obtained by induction. 
\end{proof}

\begin{lemma}\label{lemma:valid-1}
    Under Notation~\ref{nota:body} and Assumption~\ref{hpy:body}, we have that  \((\ctx_{1'}, (\ket{\phi}, k, \D_{\SCC(f_{\mathrm{cur}})})) \in \mathcal{V}\).  
\end{lemma}

\begin{proof}
Since $s$  is a prefix of $\ctx_0.R_{\mathrm{stmt}}$, 
\((\ctx_0, (\ket{\phi}, k, \D_{\SCC(f_{\mathrm{cur}})})) \in \mathcal{V}\) and
\(\ctx_0 \approx_{(\ket{\phi}, k)} \sigma_0\), 
by Theorem~\ref{theo_com}, there exists a compilation context \(\ctx_{0'}\) such that
\[
\ctx_{0'} \approx_{(\ket{\phi}, k)} \sigma_0,\qquad
(\ctx_{0'}, (\ket{\phi}, k, \D_{\SCC(f_{\mathrm{cur}})})) \in \mathcal{V},
\]
and there exist sets \(D_1, D_2\) with the following properties: \(D_1 = \rho_{1}(\mvq(s))\) if \(\neg\bigl(\operatorname{flag}(\mvq(s))=2 \land \Call(s)\neq\emptyset\bigr)\);
        otherwise \(D_1 = \rho_{1}(\mvq(s)) \cup Gar_{f_{\mathrm{cur}}}\).   
% \begin{itemize}
%   \item \(D_1 = \rho_{l,1}(\mvq(s))\) if \(\neg\bigl(\operatorname{flag}(\mvq(s))=2 \land \Call(s)\neq\emptyset\bigr)\);
%         otherwise \(D_1 = \rho_{l,1}(\mvq(s)) \cup Gar_{\curf}\);
%   \item \(\rho_{l,1}(x) = \rho_{l,0'}(x)\) for $x \in \mvq(s) \cap \dom(\rho_{l,0'})$.
% \end{itemize}
Moreover, \(\ctx_1\) can be written as
\begin{align*}
\ctx_1 = \ctx_{0'}[\, &C_{0'} \mapsto C_{0'}; C_a,\;
                     V_{\activelabel,0'} \mapsto V_{\activelabel,0'} \cup \bigl(D_1 \setminus \rho_{\mathrm{loc},0'}(\fp(f_{\mathrm{cur}})) \cup \cod(\rho_{\mathrm{glob}})\bigr),\; \\
                    &V_{\alloc,0'} \mapsto V_{\alloc,0'} \cup D_2,\;
                     \rho_{\mathrm{loc},0'} \mapsto \rho_{\mathrm{loc},0'}[\mvq(s) \mapsto \rho_{\mathrm{loc},1}(\mvq(s)) \mid \mvq(s) \not\subseteq \dom(\rho_{0'})],\\ 
                     & \Qpool_{0'} \mapsto \Qpool_{0'} \setminus \rho_{\mathrm{loc},1}(\mvq(s))\,], 
\end{align*}
and the code \(C_a\) satisfies the following conditions:
\begin{itemize}
  \item \(\qv(C_a^{k}) \setminus \rho_{0'}(\mvq(s) \cup \rvq(s))
        \subseteq \ctx_{0'}.A_{f_{\mathrm{cur}}} \setminus \bigl( \ctx_{0'}.V_{\activelabel,f_{\mathrm{cur}}} \cup \cod(\ctx_{0'}.\rho_{\mathrm{loc}}) \bigr)\);
  \item \(C_{0'}^k \ket{\varphi} \models \ket{0}_{\rho_{\mathrm{loc},1}(\mvq(s) \setminus \dom(\rho_{0}))}\)
        for any \(\ket{\varphi} \models \ket{0}_{\ctx_{0'}.A_{f_{\mathrm{cur}}} \setminus \rho_{\mathrm{loc},0'}(\fp(f_{\mathrm{cur}}))}\);
  \item set \(E \triangleq \qv(C_a^k) \setminus \rho_{0'}(\mvq(s) \cup \rvq(s)) 
        \cup \rho_{\mathrm{loc},1}(\mvq(s) \setminus \dom(\rho_{0}))\);
        then for every \(\ket{\psi}\) with \(\ket{\psi} \models \ket{0}_{E}\),
        \(\tr_{D_1}(C_a^{k} \ket{\psi}) = \tr_{D_1}(\ket{\psi})\).
\end{itemize}
Furthermore, we have
\[
\ctx_1 \approx_{(\ket{\phi}, k)} \sigma
\quad\text{and}\quad
(\ctx_1, (\ket{\phi}, k, \D_{\SCC(f_{\mathrm{cur}})})) \in \mathcal{V},
\]
where the validity condition holds except possibly for the correctness of
the cleanup circuit in Condition~3(a) for \(f_{\mathrm{cur}}\).

We now prove \((\ctx_{1'}, (\ket{\phi}, k, \D_{\SCC(f_{\mathrm{cur}})})) \in \mathcal{V}\).
Notice that \(\ctx_1\) already satisfies all validity conditions except the correctness of
the cleanup circuit in Condition~3(a) for \(f_{\mathrm{cur}}\), and the step from \(\ctx_1\) to \(\ctx_{1'}\)
modifies only the cleanup circuit \(\kappa\) in the buffer.   
% Because \(allv_{1'} \subseteq allv_1\), any condition involving \(allv\) is
% trivially preserved.
Thus it remains to verify the cleanup-circuit correctness of \(\kappa\) for $f_{\mathrm{cur}}$. 
Observe that \(\kappa_{0'}\) is known to be valid (it satisfies the
cleanup-circuit correctness condition in \(\ctx_{0'}\)), and
\(\kappa_{1} = \kappa_{0'}\).
Consequently, we can establish the required property for \(\ctx_{1'}\) by
relating it directly to \(\ctx_{0'}\); the argument is as follows.

First, let \(\ket{\varphi}\) be any state satisfying
\(\ket{\varphi} \models \ket{0}_{\ctx_{1'}.A_{f_{\mathrm{cur}}} \setminus \rho_{\mathrm{loc},1'}(\fp(f_{\mathrm{cur}}))}\),
and define 
\[
\langle \ctx_{0'}.C_{f_{\mathrm{cur}}}^{\,k},\ket{\varphi}\rangle \to^{*} \langle \downarrow,\ket{\chi}\rangle,
\qquad
\langle \ctx_{1'}.C_{f_{\mathrm{cur}}}^{\,k},\ket{\varphi}\rangle \to^{*} \langle \downarrow,\ket{\psi}\rangle .
\]
Set \(\req_{w,1'}(C_a^k) \triangleq \qv(C_a^k) \setminus \rho_{1'}(\operatorname{pred}^*(w)) \).
Since 
\(\rho_{0'}(\rvq(s)) \cup \rho_{0'}(\mvq(s))
 \subseteq \rho_{1'}(\rvq(s)) \cup \rho_{1'}(\mvq(s))
 \subseteq \rho_{1'}(\operatorname{pred}^*(w))\),
we have 
\begin{align*}
&\req_{w,1'}(C_a^k) \subseteq \bigl(\qv(C_a^{k}) \setminus \rho_{1'}(\mvq(s)) \setminus \rho_{1'}(\rvq(s))\bigr) \\ 
\subseteq &\bigl(\qv(C_a^{k}) \setminus \rho_{0'}(\mvq(s)) \setminus \rho_{0'}(\rvq(s))\bigr)
\subseteq \bigl( \ctx_{0'}.A_{f_{\mathrm{cur}}} \setminus \ctx_{0'}.V_{\activelabel,f_{\mathrm{cur}}} \cup \rho_{\mathrm{loc},0'}(\fp(f_{\mathrm{cur}})) \bigr).
\end{align*}
Then Lemma~\ref{lem:curf} gives \(\ket{\chi} \models \ket{0}_{\req_{w,1'}(C_a^k)}\).

Similarly, by the properties of \(\ctx_{0'}\) established above and
Lemma~\ref{lem:curf}, we obtain \(\ket{\chi} \models \ket{0}_{E}\), and therefore
\begin{equation}\label{eqn:body-1}
  \tr_{D_1}(C_a^{k} \ket{\chi}) = \tr_{D_1}(\ket{\chi}) .
\end{equation}
If \(\neg \bigl( \operatorname{flag}(\mvq(s)) = 2 \land \Call(s) \neq \emptyset \bigr)\),
then by definition \(D_1 = \rho_{1'}(\mvq(s))\); hence
\(D_1 \subseteq \rho_{1'}(w)\) and we further obtain 
\begin{equation}\label{equ:body-2} 
    \tr_{\rho_{1'}(w)}(C_a^{k} \ket{\chi})
    = \tr_{\rho_{1'}(w)}(\ket{\chi}),
\end{equation} 
which implies
\(\tr_{\rho_{1'}(w)}(\ket{\psi})
 = \tr_{\rho_{1'}(w)}(\ket{\chi})\).
Note that when \(\mvq(s) \notin \dom(\rho_{0})\) or
\(\mvq(s) \notin \dom(\rho_{0'})\),
the definition of \(E\) yields \(\rho_{1'}(\mvq(s)) \subseteq E\);
consequently \(\ket{\chi} \models \ket{0}_{\rho_{1'}(\mvq(s))}\).

Next, we prove the correctness of \(\kappa_{1'}\).  For any \(q \in \cod(\rho_{\mathrm{loc},1'}) \setminus \rho_{\mathrm{loc},1'}(\re(f_{\mathrm{cur}}))\) with \(\rho_{\mathrm{loc},1}^{-1}(q).\flag \neq 2\), we distinguish two cases:

\begin{itemize}
    \item \textbf{If \(q \in \rho_{\mathrm{loc},1'}(\mvq(s))\):} Since \(\mvq(s)\) contains exactly one element, by definition we have \(\mvq(s) \subseteq \dom(\rho_{\mathrm{loc},1})\) and \(\mvq(s).\flag \neq 2\). Moreover, since \(\dom(\rho_{\mathrm{loc},1}) \cap \dom(\rho_{\mathrm{glob}}) = \emptyset\), it further follows that \(\mvq(s) \cap \dom(\rho_{\mathrm{glob}}) = \emptyset\). Therefore, by definition, \(\node(\mvq(s)) \subseteq \dom(\kappa_{1'})\).

    \item \textbf{If \(q \notin \rho_{\mathrm{loc},1'}(\mvq(s))\):} From the definition of \(\rho_{\mathrm{loc},1'}\), we obtain \(q \in \cod(\rho_{\mathrm{loc},0'}) \setminus \rho_{\mathrm{loc},0'}(\re(f_{\mathrm{cur}}))\) and \(\rho_{\mathrm{loc},0'}^{-1}(q).\flag = 2\). Then, by the validity of \(\kappa_{0'}\), we have \(\node(\rho_{\mathrm{loc},0'}^{-1}(q)) \in \dom(\kappa_{0'}) \subseteq \dom(\kappa_{1'})\).
\end{itemize}

Based on the above analysis, we only need to consider the correctness of the cleanup circuits corresponding to the node \(w=\node(\mvq(s))\) and the nodes in \(\dom(\kappa_{0'})\). We divide their proof into three cases as follows. 

\noindent
\textbf{Case 1:} Let \(w = \node(\mvq(s))\) and assume
\(\neg \bigl( \operatorname{flag}(\mvq(s)) = 2 \land \Call(s) \neq \emptyset \bigr)\). Assume \(\mvq(s) \cap \dom(\rho_{\mathrm{glob}}) \neq \emptyset\). Then \(\var(w) \cap \dom(\rho_{\mathrm{loc},1'}) = \emptyset\). Indeed, if the conclusion were false, there would exist \(x \in w\) such that \(x \in \dom(\rho_{\mathrm{loc},1'})\), which implies \(x \notin \dom(\rho_{\mathrm{glob}})\). By Lemma~\ref{lem:static-wf}, we obtain \(\var(w) \cap \dom(\rho_{\mathrm{glob}}) = \emptyset\), contradicting the condition. Hence, we only need to consider the case where \(\mvq(s) \cap \dom(\rho_{\mathrm{glob}}) = \emptyset\). 
Then \(\kappa_{1'}(w) = \kappa_1(w); C_a = \kappa_{0'}(w); C_a\) by definition.  

We first verify conditions~(a) and~(b).

\begin{enumerate}
    \item[(a)]  Note that \( \req_{w,1'}(\kappa_{1'}(w)^{k}) =
\req_{w,1'}(\kappa_{0'}(w)^{k}) \cup \req_{w,1'}(C_a^{k}). \)
Since 
\(
\req_{w,1'}(C_a^{k}) 
\subseteq \qv(C_a^{k}) \setminus \rho_{0'}(\mvq(s)) \setminus \rho_{0'}(\rvq(s)), 
\)
and 
\(
\bigl( \qv(C_a^{k}) \setminus \rho_{0'}(\mvq(s)) \setminus \rho_{0'}(\rvq(s)) \bigr)
\cap \bigl( \rho_{\mathrm{loc},0'}(\re(f_{\mathrm{cur}})) \cup \rho_{\mathrm{loc},0'}(\fp(f_{\mathrm{cur}})) \bigr) = \emptyset,
\)
we obtain
\[
\req_{w,1'}(C_a^{k}) \cap \bigl( \rho_{\mathrm{loc},0'}(\re(f_{\mathrm{cur}})) \cup \rho_{\mathrm{loc},0'}(\fp(f_{\mathrm{cur}})) \bigr) = \emptyset.
\]

On the other hand, by the validity of \(\ctx_{0'}\),
\(
\bigl( \rho_{\mathrm{loc},0'}(\re(f_{\mathrm{cur}})) \cup \rho_{\mathrm{loc},0'}(\fp(f_{\mathrm{cur}})) \bigr)
\cap \req_{w,0'}(\kappa_{0'}(w)^{k}) = \emptyset.
\) 
Since \(\req_{w,1'}(\kappa_{0'}(w)^{k}) \subseteq \req_{w,0'}(\kappa_{0'}(w)^{k})\), it follows that
\[
\bigl( \rho_{\mathrm{loc},0'}(\re(f_{\mathrm{cur}})) \cup \rho_{\mathrm{loc},0'}(\fp(f_{\mathrm{cur}})) \bigr)
\cap \req_{w,1'}(\kappa_{0'}(w)^{k}) = \emptyset.
\]
Combining with \(\rho_{\mathrm{loc},1'}(\re(f_{\mathrm{cur}})) \cup \rho_{\mathrm{loc},1'}(\fp(f_{\mathrm{cur}})) = \rho_{\mathrm{loc},0'}(\re(f_{\mathrm{cur}})) \cup \rho_{\mathrm{loc},0'}(\fp(f_{\mathrm{cur}}))\), we finally obtain 
\[
\bigl( \rho_{\mathrm{loc},1'}(\re(f_{\mathrm{cur}})) \cup \rho_{\mathrm{loc},1'}(\fp(f_{\mathrm{cur}})) \bigr)
\cap \req_{w,1'}(\kappa_{1'}(w)^{k}) = \emptyset.
\] 

    \item[(b)] For any $z$, we have $\qv(\kappa_{0'}(w)^z) \subseteq \qv(\ctx_{0'}.C_{f_{\mathrm{cur}}}^z) \cup S$ for some $S$ such that $S \in \Qpool_{0'} \cup V_{\activelabel,\succeq f_{\mathrm{cur}},0'} \setminus Q_{\succeq f_{\mathrm{cur}}^*,0'}[+]$, with $\base(S) \subseteq V_{\alloc,0'}$. Combining this with the definitions of $\kappa_{1'}(w) = \kappa_{0'}(w); C_a$ and $\ctx_{1'}.C_{f_{\mathrm{cur}}} = \ctx_{0'}.C_{f_{\mathrm{cur}}}; C_a$, we obtain
\[ 
\qv(\kappa_{1'}(w)^z) = \qv(\kappa_{0'}(w)^z) \cup \qv(C_a^z)
\subseteq \qv(\ctx_{0'}.C_{f_{\mathrm{cur}}}^z) \cup S \cup \qv(C_a^z)
= \qv(\ctx_{1'}.C_{f_{\mathrm{cur}}}^z) \cup S.
\] 

It is easy to prove that $\Qpool_{0'} \cup V_{\activelabel,\succeq f_{\mathrm{cur}},0'} \setminus Q_{\succeq f_{\mathrm{cur}}^*,0'}[+] \subseteq \Qpool_{1'} \cup V_{\activelabel,\succeq f_{\mathrm{cur}},1'} \setminus Q_{\succeq f_{\mathrm{cur}}^*,1'}[+]$ and $V_{\alloc,0'} \subseteq V_{\alloc,1'}$. Hence the required condition follows immediately. 
\end{enumerate}

Now let \(\overline{p}\) be such that
\(\ket{\psi} \models \ket{0}_{\overline{p}}\) and
\(\overline{p} \cap \bigl( \cod(\rho_{1'}) \cup V_{\activelabel,f_{\mathrm{cur}},1'} \bigr) = \emptyset\).
Since \(\cod(\rho_{0'}) \cup V_{\activelabel,f_{\mathrm{cur}},0'} \subseteq
      \cod(\rho_{1'}) \cup V_{\activelabel,f_{\mathrm{cur}},1'}\),
we also have
\(\overline{p} \cap \bigl( \cod(\rho_{0'}) \cup V_{\activelabel,f_{\mathrm{cur}},0'} \bigr) = \emptyset\).
Furthermore, \(\mvq(s) \in \var(\operatorname{pred}^*(w))\) implies \(\req_{w,1'}(C_a^k) \cap \rho_{1'}(\mvq(s)) = \emptyset\). Also, \(\overline{p} \cap \rho_{1'}(\mvq(s)) = \emptyset\). Hence, by Equation~\ref{eqn:body-1}, we have \(\ket{\chi} \models \ket{0}_{\overline{p}}\) and \(\ket{\psi} \models \ket{0}_{\req_{w,1'}(C_a^k)}\) since $D_1=\rho_{1'}(\mvq(s))$ in this case.  

Consequently,
\begin{align*}
&\bigl\langle \ctx_{1'}.C_{f_{\mathrm{cur}}}^k;
        \qinv\bigl(\kappa_{1'}(w)[\req_{w,1'} \mapsto \overline{p}]\bigr)^k,
        \ket{\varphi} \bigr\rangle \\
= \ &\bigl\langle \ctx_{0'}.C_{f_{\mathrm{cur}}}^k; C_a^k;
        \qinv\bigl(\kappa_{1'}(w)[\req_{w,1'} \mapsto \overline{p}]\bigr)^k,
        \ket{\varphi} \bigr\rangle \\
=\  &\bigl\langle \ctx_{0'}.C_{f_{\mathrm{cur}}}^k; C_a^k;
        \bigl(\qinv(C_a)[\req_{w,1'} \mapsto \overline{p}]\bigr)^k;
        \bigl(\kappa_{0'}(w)[\req_{w,1'} \mapsto \overline{p}]\bigr)^k,
        \ket{\varphi} \bigr\rangle \\
= \ &\bigl\langle \ctx_{0'}.C_{f_{\mathrm{cur}}}^k; C_a^k; \bigl(\qinv(C_a)\bigr)^k;
        \bigl(\kappa_{0'}(w)[(\req_{w,1'} \setminus \req_{w,1'}(C_a^k)) \mapsto (\overline{p}  \setminus \req_{w,1'}(C_a^k)) ]\bigr)^k,
        \ket{\varphi} \bigr\rangle \tag{ \(\ket{\psi} \models \ket{0}_{\req_{w,1'}(C_a^k)}\)} \\ 
= 
\ &\bigl\langle \ctx_{0'}.C_{f_{\mathrm{cur}}}^k;
        \qinv\bigl( \kappa_{0'}(w)[(\req_{w,1'} \setminus \req_{w,1'}(C_a^k)) \mapsto (\overline{p}  \setminus \req_{w,1'}(C_a^k))]\bigr),
        \ket{\varphi} \bigr\rangle. 
\end{align*} 
To apply the validity of \(\kappa_{0'}\), it remains to verify that the quantum bits indexed by \(\req_{w,0'}\) are replaced by those in the zero state in \(\ket{\chi}\) and are disjoint from \(\cod(\rho_{0'}) \cup V_{\activelabel,f_{\mathrm{cur}},0'}\). 

Firstly, we consider $\kappa_{0'}(w)[\req_{w,1'} \mapsto \overline{p}]$.  If \(\mvq(s) \in \dom(\rho_{0'})\), then by definition \(\rho_{1'} = \rho_{0'}\), 
hence \(\req_{w,1'}(\kappa_{0'}(w)) = \req_{w,0'}(\kappa_{0'}(w))\) and the condition holds trivially.  

Otherwise, \(\req_{w,0'}(\kappa_{0'}(w)) = \req_{w,1'}(\kappa_{0'}(w)) \cup \{\rho_{1'}(\mvq(s))\}\). 
We have \(\ket{\chi} \models \ket{0}_{\rho_{1'}(\mvq(s))}\) and
\(\rho_{1'}(\mvq(s)) \cap \bigl( \cod(\rho_{0'}) \cup V_{\activelabel,f_{\mathrm{cur}},0'} \bigr) = \emptyset\).
Thus the additional qubits in \(\req_{w,0'} \setminus \req_{w,1'}\) are already
in the zero state under \(\ket{\chi}\), and they do not interfere with the
freshness requirement.  Hence, under \(\ket{\chi}\), substituting \(\overline{p}\)
for \(\req_{w,1'}\) has the same effect as substituting \((\overline{p}, \overline{0})\)
for \(\req_{w,0'}\), where \(\overline{0}\) corresponds to the registers \(\rho_{1'}(\mvq(s))\).

Now, we consider $\kappa_{0'}(w)[(\req_{w,1'} \setminus \req_{w,1'}(C_a^k)) \mapsto (\overline{p}  \setminus \req_{w,1'}(C_a^k))]$. 
Note that \(\req_{w,1'}(C_a^k) \subseteq
 \qv(C_a^{k}) \setminus \rho_{0'}(\mvq(s)) \setminus \rho_{0'}(\rvq(s))\) 
and
\(\bigl( \qv(C_a^{k}) \setminus \rho_{0'}(\mvq(s)) \setminus \rho_{0'}(\rvq(s)) \bigr) \cap \bigl(\cod(\rho_{0'}) \cup V_{\activelabel,f_{\mathrm{cur}},0'} \bigr) = \emptyset\). 
% Together with \(\ket{\chi} \models \ket{0}_{\req_{w,1'}(C_a^k)}\),  we get $\req_{w,1'}(C_a^k)$. Therefore it suffices to
% substitute only for \(\req_{w,1'} \setminus \req_{w,1'}(C_a^k)\).
Together with \(\ket{\chi} \models \ket{0}_{\req_{w,1'}(C_a^k)}\), we obtain that \(\req_{w,1'}(C_a^k)\) itself already satisfies the required conditions. Hence, it suffices to substitute only for \(\req_{w,1'} \setminus \req_{w,1'}(C_a^k)\). 

Therefore, since \(\ket{\chi} \models \ket{0}_{\overline{p}}\) and
\(\overline{p} \cap \bigl( \cod(\rho_{0'}) \cup V_{\activelabel,f_{\mathrm{cur}},0'} \bigr) = \emptyset\), 
the validity of \(\kappa_{0'}\) yields
\begin{align*}
&\bigl\langle \ctx_{0'}.C_{f_{\mathrm{cur}}}^k;
        \qinv\bigl(\kappa_{0'}(w)[(\req_{w,1'} \setminus \req_{w,1'}(C_a^k)) \mapsto (\overline{p}  \setminus \req_{w,1'}(C_a^k))]\bigr),
        \ket{\varphi} \bigr\rangle \\
\rightarrow^{*} \ &\bigl\langle \downarrow,
        \bigl(\ket{\chi}\bigr)_{\sys \setminus \rho_{\mathrm{loc},0'}(w)}
        \otimes \ket{\theta}_{\rho_{\mathrm{loc},0'}(w) \setminus \rho_{\mathrm{loc},0'}(U_{w})}
        \otimes \ket{\varphi}_{\rho_{\mathrm{loc},0'}(U_{w})} \bigr\rangle .
\end{align*}
Since in this case \(\mvq(s) \cap \dom(\rho_{\mathrm{glob}}) = \emptyset\), we have \(\mvq(s) \in \dom(\rho_{\mathrm{loc},1'})\). Thus \(D_1 = \rho_{\mathrm{loc},1'}(\mvq(s))\) and Equations~\ref{eqn:body-1} and~\ref{equ:body-2} can be applied as follows. 

If \(\mvq(s) \in \dom(\rho_{\mathrm{loc},0'})\), then \(\rho_{\mathrm{loc},1'} = \rho_{\mathrm{loc},0'}\) and we obtain 
\begin{align*}
&\bigl\langle \downarrow,
        \bigl(\ket{\chi}\bigr)_{\sys \setminus \rho_{\mathrm{loc},0'}(w)}
        \otimes \ket{\theta}_{\rho_{\mathrm{loc},0'}(w) \setminus \rho_{\mathrm{loc},0'}(U_{w})}
        \otimes \ket{\varphi}_{\rho_{\mathrm{loc},0'}(U_{w})} \bigr\rangle \\
= \ &\bigl\langle \downarrow,
        \bigl(\ket{\psi}\bigr)_{\sys \setminus \rho_{\mathrm{loc},1'}(w)}
        \otimes \ket{\theta}_{\rho_{\mathrm{loc},1'}(w) \setminus \rho_{\mathrm{loc},1'}(U_{w})}
        \otimes \ket{\varphi}_{\rho_{\mathrm{loc},1'}(U_{w})} \bigr\rangle. 
        \qquad \tag{by Equation~\ref{eqn:body-1}}  
\end{align*}

If \(\mvq(s) \notin \dom(\rho_{\mathrm{loc},0'})\), then
\(\rho_{\mathrm{loc},1'}(\mvq(s)) \in \qv(C_a^k) \setminus \cod(\rho_{\mathrm{loc},0'})\) and
\(\ket{\chi} \models \ket{0}_{\rho_{\mathrm{loc},1'}(\mvq(s))} =
 \ket{\varphi}_{\rho_{\mathrm{loc},1'}(\mvq(s))}\).
Therefore,
\begin{align*}
&\bigl\langle \downarrow,
        \bigl(\ket{\chi}\bigr)_{\sys \setminus \rho_{\mathrm{loc},0'}(w)}
        \otimes \ket{\theta}_{\rho_{\mathrm{loc},0'}(w) \setminus \rho_{\mathrm{loc},0'}(U_{w})}
        \otimes \ket{\varphi}_{\rho_{\mathrm{loc},0'}(U_{w})} \bigr\rangle \\
= &\bigl\langle \downarrow,
        \bigl(\ket{\chi}\bigr)_{\sys \setminus (\rho_{\mathrm{loc},0'}(w) \cup \rho_{\mathrm{loc},1'}(\mvq(s)))} 
        \otimes \bigl(\ket{\chi}\bigr)_{\rho_{\mathrm{loc},1'}(\mvq(s))}
        \otimes \ket{\theta}_{\rho_{\mathrm{loc},0'}(w) \setminus \rho_{\mathrm{loc},0'}(U_{w})}
        \otimes \ket{\varphi}_{\rho_{\mathrm{loc},0'}(U_{w})} \bigr\rangle \\
= &\bigl\langle \downarrow,
        \bigl(\ket{\psi}\bigr)_{\sys \setminus \rho_{\mathrm{loc},1'}(w)}
        \otimes \ket{\varphi}_{\rho_{\mathrm{loc},1'}(\mvq(s))}
        \otimes \ket{\theta}_{\rho_{\mathrm{loc},0'}(w) \setminus \rho_{\mathrm{loc},0'}(U_{w})}
        \otimes \ket{\varphi}_{\rho_{\mathrm{loc},0'}(U_{w})} \bigr\rangle \tag{by Equation~\ref{equ:body-2} and 
        \(\rho_{\mathrm{loc},1'}(w) = \rho_{\mathrm{loc},0'}(w) \cup \rho_{\mathrm{loc},1'}(\mvq(s))\)} \\
= &\bigl\langle \downarrow,
        \ket{\psi}_{\sys \setminus \rho_{\mathrm{loc},1'}(w)}
        \otimes \ket{\theta}_{\rho_{\mathrm{loc},1'}(w) \setminus \rho_{\mathrm{loc},1'}(U_{w})}
        \otimes \ket{\varphi}_{\rho_{\mathrm{loc},1'}(U_{w})} \bigr\rangle.
\end{align*}
The last equality holds since \(\operatorname{flag}(\mvq(s)) \neq 2\), hence \(\mvq(s)\) is merged with \(U_w\).

\textbf{Case 2:} Consider any node \(v \in \dom(\kappa_{0'})\) such that \(w \in \operatorname{pred}^+(v)\), where \(w = \node(\mvq(s))\) (as in Case~1). Then 
\(
\kappa_{1'}(v) = \qinv(C_a); \kappa_1(v); C_a = \qinv(C_a); \kappa_{0'}(v); C_a .
\)

Let \(\req_{v,1'}(C_a^k) \triangleq \qv(C_a^k) \setminus \rho_{1'}(\operatorname{pred}^*(v))\). Since \(w \in \operatorname{pred}^+(v)\), we have \(\operatorname{pred}^*(w) \subseteq \operatorname{pred}^*(v)\) and thus  
\[
\req_{v,1'}(C_a^k) = \qv(C_a^k) \setminus \rho_{1'}(\operatorname{pred}^*(v))
\subseteq \qv(C_a^k) \setminus \rho_{1'}(\operatorname{pred}^*(w))
\subseteq \req_{w,1'}(C_a^k).
\]
The arguments for conditions~(a) and~(b) are analogous to Case~1 and are omitted here.

Now let \(\overline{p}\) be a tuple such that \(\ket{\psi} \models \ket{0}_{\overline{p}}\) and \(\overline{p} \cap \bigl( \cod(\rho_{1'}) \cup V_{\activelabel,f_{\mathrm{cur}},1'} \bigr) = \emptyset\). The condition \(\cod(\rho_{0'}) \cup V_{\activelabel,f_{\mathrm{cur}},0'} \subseteq \cod(\rho_{1'}) \cup V_{\activelabel,f_{\mathrm{cur}},1'}\) implies \(\overline{p} \cap \bigl( \cod(\rho_{0'}) \cup V_{\activelabel,f_{\mathrm{cur}},0'} \bigr) = \emptyset\).  Since we also have \(\overline{p} \cap \rho_{1'}(\mvq(s)) = \emptyset\), \(\ket{\chi} \models \ket{0}_{\overline{p}}\) by Equation~\ref{eqn:body-1}.  

Similarly, we have \(\ket{\chi} \models \ket{0}_{\req_{v,1'}(C_a^k)}\), \(\ket{\psi} \models \ket{0}_{\req_{v,1'}(C_a^k)}\), and \(\req_{v,1'}(C_a^k) \cap \bigl( \cod(\rho_{0'}) \cup V_{\activelabel,f_{\mathrm{cur}},0'} \bigr) = \emptyset\), since \(\req_{v,1'}(C_a^k) \subseteq \req_{w,1'}(C_a^k)\) and \(\req_{w,1'}(C_a^k)\) already satisfies these conditions. 

Then we obtain:
\begin{align*}
&\bigl\langle \ctx_{1'}.C_{f_{\mathrm{cur}}}^k; \qinv\bigl(\kappa_{1'}(v)[\req_{v,1'} \mapsto \overline{p}]\bigr)^k, \ket{\varphi} \bigr\rangle \\
= \ &\bigl\langle \ctx_{0'}.C_{f_{\mathrm{cur}}}^k; C_a^k; \qinv\bigl(\kappa_{1'}(v)[\req_{v,1'} \mapsto \overline{p}]\bigr)^k, \ket{\varphi} \bigr\rangle \\
= \ &\bigl\langle \ctx_{0'}.C_{f_{\mathrm{cur}}}^k; C_a^k;
      \bigl(\qinv(C_a)[\req_{v,1'} \mapsto \overline{p}]\bigr)^k;
      \bigl(\kappa_{0'}(v)[\req_{v,1'} \mapsto \overline{p}]\bigr)^k;
      \bigl(C_a[\req_{v,1'} \mapsto \overline{p}]\bigr)^k,
      \ket{\varphi} \bigr\rangle \\
= \ &\bigl\langle \ctx_{0'}.C_{f_{\mathrm{cur}}}^k; C_a^k; \qinv(C_a)^k;
      \bigl(\kappa_{0'}(v)[(\req_{v,1'} \setminus \req_{v,1'}(C_a^k)) \mapsto (\overline{p} \setminus \req_{v,1'}(C_a^k)) ]\bigr)^k;
      C_a^k, \ket{\varphi} \bigr\rangle
      \tag{since \(\ket{\psi} \models \ket{0}_{\req_{v,1'}(C_a^k)}\)} \\
= \ &\bigl\langle \ctx_{0'}.C_{f_{\mathrm{cur}}}^k;
      \qinv\bigl(\kappa_{0'}(v)[(\req_{v,1'} \setminus \req_{v,1'}(C_a^k)) \mapsto (\overline{p} \setminus \req_{v,1'}(C_a^k)) ]\bigr)^k; 
      C_a^k, \ket{\varphi} \bigr\rangle .
\end{align*}
As in Case~1, if \(\req_{v,1'} \neq \req_{v,0'}\), we have \(\ket{\chi} \models \ket{0}_{\rho_{1'}(\mvq(s))}\)  and \(\rho_{1'}(\mvq(s)) \cap \bigl( \cod(\rho_{0'}) \cup V_{\activelabel,f_{\mathrm{cur}},0'} \bigr) = \emptyset\).  Combining this with \(\overline{p}\) and \(\req_{v,1'}(C_a^k)\) which also satisfy these conditions as analysed above, we apply the validity of \(\kappa_{0'}(v)\) to obtain 
\begin{align*}
&\bigl\langle \ctx_{0'}.C_{f_{\mathrm{cur}}}^k;
      \qinv\bigl(\kappa_{0'}(v)[(\req_{v,1'} \setminus \req_{v,1'}(C_a^k)) \mapsto (\overline{p}  \setminus \req_{v,1'}(C_a^k))]\bigr)^k;
      C_a^k, \ket{\varphi} \bigr\rangle \\
\rightarrow^{*} \ &\bigl\langle C_a^k,
      \bigl(\ket{\chi}\bigr)_{\sys \setminus \rho_{\mathrm{loc},0'}(v)}
      \otimes \ket{\theta}_{\rho_{\mathrm{loc},0'}(v) \setminus \rho_{\mathrm{loc},0'}(U_{v})}
      \otimes \ket{\varphi}_{\rho_{\mathrm{loc},0'}(U_{v})} \bigr\rangle \\
= \ &\bigl\langle \downarrow,
      \ket{\psi}_{\sys \setminus \rho_{\mathrm{loc},1'}(v)}
      \otimes \ket{\theta}_{\rho_{\mathrm{loc},1'}(v) \setminus \rho_{\mathrm{loc},1'}(U_{v})}
      \otimes \ket{\varphi}_{\rho_{\mathrm{loc},1'}(U_{v})} \bigr\rangle
      \tag{by \(\rho_{\mathrm{loc},1'}(v) = \rho_{\mathrm{loc},0'}(v)\)}. 
\end{align*} 

The key to establishing the last equality is to prove that \(\qv(C_a^k) \cap \rho_{\mathrm{loc},0'}(v) = \emptyset\). We proceed as follows. Since
\(
\qv(C_a^{k}) \setminus \rho_{0'}(\mvq(s) \cup \rvq(s)) \subseteq \ctx_{0'}.A_{f_{\mathrm{cur}}} \setminus \bigl( \ctx_{0'}.V_{\activelabel,f_{\mathrm{cur}}} \cup \cod(\ctx_{0'}.\rho_{\mathrm{loc}}) \bigr),
\)
we immediately obtain
\(
\rho_{\mathrm{loc},0'}(v) \cap \bigl( \qv(C_a^{k}) \setminus \rho_{0'}(\mvq(s) \cup \rvq(s)) \bigr) = \emptyset.
\)
Thus it suffices to prove \(\rho_{\mathrm{loc},0'}(v) \cap \rho_{0'}(\mvq(s) \cup \rvq(s)) = \emptyset\).

By the injectivity of \(\rho_{\mathrm{loc},0'}\) and \(\rho_{\mathrm{glob}}\), together with \(\cod(\rho_{\mathrm{loc},0'}) \cap \cod(\rho_{\mathrm{glob}}) = \emptyset\), we only need to show
\(
\var(v) \cap (\mvq(s) \cup \rvq(s)) = \emptyset.
\) Since  \(w \in \operatorname{pred}^+(v)\), we have \(v \neq w\) and \(v \notin \operatorname{pred}^+(w)\). 
\begin{itemize}
  \item  \(v \neq w\) implies \(\var(v) \cap \var(w) = \emptyset\), hence \(\var(v) \cap \mvq(s) = \emptyset\).  
\item  \(v \notin \operatorname{pred}^+(w)\) implies \(\var(v) \cap \rvq(s) = \emptyset\).
\end{itemize} 
Therefore, the required equality holds. 

\textbf{Case 3:} For any node \(v\) not yet covered by the previous cases, we have
\[
\bigl( \neg (\operatorname{flag}(\mvq(s)) = 2 \land \Call(s) \neq \emptyset) \land w \notin \operatorname{pred}^*(v) \bigr)
\;\lor\;
\bigl( \operatorname{flag}(\mvq(s)) = 2 \land \Call(s) \neq \emptyset \bigr).
\]
By definition, for these \(v\) we have \(\kappa_{1'}(v) = \kappa_{1}(v) = \kappa_{0'}(v)\).

First consider the subcase \(w \notin \operatorname{pred}^*(v)\). From the definition of \(\rho_{1'}\) we have \(\rho_{0'}(\operatorname{pred}^*(v)) = \rho_{1'}(\operatorname{pred}^*(v))\). Then, it follows that 
\[
\qv\bigl(\kappa_{0'}(v)\bigr)
\setminus \rho_{1'}(\operatorname{pred}^*(v))
= \qv\bigl(\kappa_{0'}(v)\bigr)
\setminus \rho_{0'}(\operatorname{pred}^*(v)).
\] Since \(v \neq w\), Lemma~\ref{lem:static-wf} guarantees \(\var(v) \cap \var(w) = \emptyset\); together with the injectivity of \(\rho_{1'}\), we obtain that \(\rho_{1'}(\operatorname{pred}^*(v))\) and \(\rho_{1'}(\mvq(s))\) are disjoint. Consequently,
\[
\qv(\kappa_{0'}(v)) \cap \rho_{1'}(\operatorname{pred}^*(v)) \cap \rho_{1'}(\mvq(s)) = \emptyset. 
\]

Next consider the subcase \(w \in \operatorname{pred}^*(v)\). Here we have \(\operatorname{flag}(\mvq(s)) = 2 \land \Call(s) \neq \emptyset\). The source type system ensures that \(\mvq(s)\) is written for the first time, hence \(\mvq(s) \notin \dom(\rho_{0})\). In this situation,
\[
\rho_{1'}(\operatorname{pred}^*(v)) = \rho_{0'}(\operatorname{pred}^*(v)) \cup \rho_{1'}(\mvq(s)).
\]
As argued in the previous subsection, we may always assume \(\qv(\kappa_{0'}(v)) \cap \rho_{1'}(\mvq(s)) = \emptyset\).   We  also have 
\[
\qv\bigl(\kappa_{0'}(v)\bigr)
\setminus \rho_{\mathrm{loc},1'}(\operatorname{pred}^*(v))
= \qv\bigl(\kappa_{0'}(v)\bigr)
  \setminus \rho_{\mathrm{loc},0'}(\operatorname{pred}^*(v)). 
\]

%   by definition,
% \[
% \kappa_{1'}(v) = \kappa_{1}(v)[\rho_{l,1'}(\mvq(s)) \mapsto q]
%             = \kappa_{0'}(v)[\rho_{l,1'}(\mvq(s)) \mapsto q],
% \qquad q \in \xi_{1} = \xi_{1'}.
% \]
% Therefore
% \[
% \qv\bigl(\kappa_{0'}(v)[\rho_{l,1'}(\mvq(s)) \mapsto q]\bigr)
% \cap \rho_{l,1'}(\operatorname{pred}^*(v))
% \cap \{\rho_{l,1'}(\mvq(s))\}
% = \emptyset .
% \]

% In all subcases we have
% \[
% \qv(\kappa_{1'}(v)) \cap \rho_{l,1'}(\operatorname{pred}^*(v))
% \cap \{\rho_{l,1'}(\mvq(s))\}
% = \emptyset .
% \]
% We now prove the claim for the situation \(w \in \operatorname{pred}^+(v)\);
% the situation \(w \notin \operatorname{pred}^+(v)\) is analogous.

% Since \(q \in \xi_{1'}\) and \(\base(q) \subseteq allv_{1'}\),
% conditions~(a) and~(b) follow directly from the validity of \(\kappa_{0'}(v)\).

Now let \(\overline{p}\) be a tuple such that \(\ket{\psi} \models \ket{0}_{\overline{p}}\) and \(\overline{p} \cap \bigl( \cod(\rho_{1'}) \cup V_{\activelabel,f_{\mathrm{cur}},1'} \bigr) = \emptyset\). Since \(D_1 \subseteq V_{\activelabel,f_{\mathrm{cur}},1'}\), we have \(\overline{p} \cap D_1 = \emptyset\), and consequently \(\ket{\chi} \models \ket{0}_{\overline{p}}\) by Equation~\ref{eqn:body-1}. Moreover, since \(\cod(\rho_{0'}) \cup V_{\activelabel,f_{\mathrm{cur}},0'} \subseteq \cod(\rho_{1'}) \cup V_{\activelabel,f_{\mathrm{cur}},1'}\), it follows that \(\overline{p} \cap \bigl( \cod(\rho_{0'}) \cup V_{\activelabel,f_{\mathrm{cur}},0'} \bigr) = \emptyset\).

Since \(
\qv(\kappa_{0'}(v)) \cap \rho_{1'}(\operatorname{pred}^*(v)) \cap \rho_{1'}(\mvq(s)) = \emptyset, \)
and by the definition of \(D_1\) together with the other validity conditions of \(\ctx_{1'}\) we can ensure that
\(
\rho_{1'}(\operatorname{pred}^*(v)) \cap \bigl( D_1 \setminus \rho_{1'}(\mvq(s)) \bigr) = \emptyset, \)
it follows that
\(
\bigl( \qv(\kappa_{0'}(v)) \cap \rho_{1'}(\operatorname{pred}^*(v)) \bigr) \cap D_1 = \emptyset.
\)
Together with \(\overline{p} \cap D_1 = \emptyset\), we obtain  
\[
\bigl( \qv(\kappa_{0'}(v))[\req_{v,1'} \mapsto \overline{p}] \bigr) \cap D_1 = \emptyset. \] 

% Since \(\bigl( \qv(\kappa_{0'}(v))
%      \cap \rho_{1'}(\operatorname{pred}^*(v))
%      \bigr)
% \cap \{\rho_{1'}(\mvq(s))\}
% = \emptyset, \) and 
% % we obtain
% % \( \bigl( \qv(\kappa_{0'}(v)[\req_{v,1'} \mapsto \overline{p}]) \bigr) \cap \{\rho_{1'}(\mvq(s))\} = \emptyset. \)
% by the definition of \(D_1\),
% \(
% \bigl( \rho_{1'}(\operatorname{pred}^*(v)) \bigr)
% \cap \bigl( D_1 \setminus \{\rho_{1'}(\mvq(s)) \} \bigr)
% = \emptyset , 
% \)
% hence
% \[
% \bigl( \qv(\kappa_{0'}(v))[\req_{v,1'} \mapsto \overline{p}] \bigr)
% \cap D_1
% = \emptyset.
% \] 

Then we obtain:
\begin{align*}
&\bigl\langle \ctx_{1'}.C_{f_{\mathrm{cur}}}^k;
        \qinv\bigl(\kappa_{1'}(v)[\req_{v,1'} \mapsto \overline{p}]\bigr)^k,
        \ket{\varphi} \bigr\rangle \\
% = &\bigl\langle ctx_{0'}.C_{\curf}^k; C_a^k;
%         \qinv\bigl(\kappa_{1}(v)[\req_{v,1'} \mapsto \overline{p}]\bigr)^k,
%         \ket{\varphi} \bigr\rangle \\ 
= \ &\bigl\langle \ctx_{0'}.C_{f_{\mathrm{cur}}}^k; C_a^k;
        \qinv\bigl(\kappa_{0'}(v) 
                 [\req_{v,1'} \mapsto \overline{p}] \bigr)^k,
        \ket{\varphi} \bigr\rangle \\
\rightarrow^{*} \ &\bigl\langle C_a^k;
        \qinv\bigl(\kappa_{0'}(v)
                 [\req_{v,1'} \mapsto \overline{p}]\bigr),
        \bigl(\ket{\chi}\bigr)_{D_1} \otimes
        \bigl(\ket{\chi}\bigr)_{\sys \setminus D_1} \bigr\rangle \\
\rightarrow^{*} \ &\bigl\langle
        \qinv\bigl(\kappa_{0'}(v)
                 [\req_{v,1'} \mapsto \overline{p}]\bigr),
        \bigl(\ket{\psi}\bigr)_{D_1} \otimes
        \bigl(\ket{\chi}\bigr)_{\sys \setminus (D_1)} \bigr\rangle
        \tag{since \(\tr_{D_1}(\ket{\chi}) = \tr_{D_1}(\ket{\psi})\)} \\ 
= \ &\bigl\langle
  \qinv\bigl(\kappa_{0'}(v)[\req_{v,0'} \mapsto \overline{p}]\bigr), 
  \bigl(\ket{\psi}\bigr)_{D_1} \otimes
  \bigl(\ket{\chi}\bigr)_{\sys \setminus D_1} \bigr\rangle \tag{by \(\req_{v,1'} = \req_{v,0'}\)} \\ 
\end{align*}
Since $\qv(\kappa_{0'}(v)[\req_{v,0'} \mapsto \overline{p}]) \cap D_1=\emptyset$, \(\ket{\chi} \models \ket{0}_{\overline{p}}\) and 
              \(\overline{p} \cap (\cod(\rho_{0'}) \cup V_{\activelabel,f_{\mathrm{cur}},0'}) = \emptyset\), we can apply the validity of \(\kappa_{0'}\) to yield 
\begin{align*}
    \rightarrow^{*} &\bigl\langle \downarrow,
  \bigl(\ket{\psi}\bigr)_{D_1} \otimes
  \bigl(\ket{\chi}\bigr)_{\sys \setminus D_1 \setminus \rho_{\mathrm{loc},0'}(v)}
  \otimes \ket{\theta}_{\rho_{\mathrm{loc},0'}(v) \setminus \rho_{\mathrm{loc},0'}(U_{v})}
  \otimes \ket{\varphi}_{\rho_{\mathrm{loc},0'}(U_{v})} \bigr\rangle \\
= &\bigl\langle \downarrow,
  \bigl(\ket{\psi}\bigr)_{\sys \setminus \rho_{\mathrm{loc},1'}(v)}
  \otimes \ket{\theta}_{\rho_{\mathrm{loc},1'}(v) \setminus \rho_{\mathrm{loc},1'}(U_{v})}
  \otimes \ket{\varphi}_{\rho_{\mathrm{loc},1'}(U_{v})} \bigr\rangle .
  \tag{by \(\rho_{\mathrm{loc},1'}(v) = \rho_{\mathrm{loc},0'}(v)\) and
        \(\tr_{D_1}(\ket{\chi}) = \tr_{D_1}(\ket{\psi})\)} 
\end{align*} 

Hence \(\kappa_{1'}\) is valid, and we obtain
\(
\bigl(\ctx_{1'}, (\ket{\phi}, k, \D_{\SCC(f_{\mathrm{cur}})})\bigr) \in \mathcal{V}. 
\)
\end{proof}

\begin{theorem} 
    \label{theo:body}
    \label{theo:correctness-body}
    Under the Notation~\ref{nota:body} and Assumption~\ref{hpy:body}, we have that \[(\ctx, (\ket{\phi}, k, \D_{\SCC(f_{\mathrm{cur}})})) \in \mathcal{V}\] and $\ctx\approx_{(\ket{\phi}, k)} \sigma$. 
\end{theorem}

\begin{proof}
We proceed by induction on the structure of \(s\).

\textbf{Case 1: \(s\) is not a sequential statement (i.e., \(s \neq s_1; s_2\)).}  
First, by Lemma~\ref{lemma:valid-1} we directly obtain \((\ctx_{1'}, (\ket{\phi}, k, \D_{\SCC(f_{\mathrm{cur}})})) \in \mathcal{V}\).
 
Now assume inductively that \(\ctx_{i'}\) is valid, i.e., 
\(\bigl(\ctx_{i'}, (\ket{\phi}, k, \D_{\SCC(f_{\mathrm{cur}})})\bigr) \in \mathcal{V}\).  
Under this induction hypothesis, consider any \(u_i \in V_b\). Since \(\var(u_i) \cap \rvq(s) \neq \emptyset\) and \(\rvq(s) \subseteq \dom(\rho_{0'})\), we obtain \(\var(u_i) \cap \dom(\rho_{0'}) \neq \emptyset\). Moreover, \(\var(u_i) \cap (\dom(\rho_{\mathrm{glob}}) \cup \re(f_{\mathrm{cur}})) = \emptyset\). Therefore, \(\var(u_i) \cap \dom(\rho_{\mathrm{loc},0'}) \neq \emptyset\) and \(u_i \notin \node(\re(f_{\mathrm{cur}}))\). Hence we may successively apply Lemma~\ref{lemma:clean-inv}, Theorem~\ref{theo:clean-valid}, and Theorem~\ref{theo:clean} to obtain  
\[
\bigl(\ctx_{(i+1)'}, (\ket{\phi}, k, \D_{\SCC(f_{\mathrm{cur}})})\bigr) \in \mathcal{V},
\qquad
\rho_{\mathrm{loc},(i+1)'} = \rho_{\mathrm{loc},i'} \setminus (U_{u_i} \setminus \fp(f_{\mathrm{cur}})),
\]  
and  
\[
\bigl\langle \ctx_{(i+1)'}.C_{f_{\mathrm{cur}}},\; \ket{\phi} \bigr\rangle
\rightarrow^{*}
\bigl\langle \downarrow,\;
(\ctx_{i'}.C_{f_{\mathrm{cur}}}\ket{\phi})_{\sys \setminus \rho_{\mathrm{loc},i'}(u_i)}
\otimes \ket{\theta}_{\rho_{\mathrm{loc},i'}(u_i) \setminus \rho_{\mathrm{loc},i'}(U_{u_i})}
\otimes \ket{\phi}_{\rho_{\mathrm{loc},i'}(U_{u_i})} \bigr\rangle,
\]  
where \(U_{u_i} \subseteq \{\, x \in \operatorname{var}(u_i) \mid x.\flag \neq 2 \,\}\).

Thus, by induction on \(i\), we obtain  
\[
\bigl(\ctx_{(1+\lvert V_b\rvert)'}, (\ket{\phi}, k, \D_{\SCC(f_{\mathrm{cur}})})\bigr) \in \mathcal{V},
\quad
\rho_{\mathrm{loc},(1+\lvert V_b\rvert)'}
= \rho_{\mathrm{loc},1'} \setminus (U_{V_b} \setminus \fp(f_{\mathrm{cur}}))
= \rho_{\mathrm{loc},1} \setminus (U_{V_b} \setminus \fp(f_{\mathrm{cur}})),
\]
and  
\begin{align}\label{eqn:body-5}
&\bigl\langle \ctx_{(1+\lvert V_b\rvert)'}.C_{f_{\mathrm{cur}}},\; \ket{\phi} \bigr\rangle \notag \\
\rightarrow^{*} \ &
\bigl\langle \downarrow,\;
(\ctx_{1'}.C_{f_{\mathrm{cur}}}\ket{\phi})_{\sys \setminus \rho_{\mathrm{loc},1'}(V_b)}
\otimes \ket{\theta}_{\rho_{\mathrm{loc},1'}(V_b) \setminus \rho_{\mathrm{loc},1'}(U_{V_b})}
\otimes \ket{\phi}_{\rho_{\mathrm{loc},1'}(U_{V_b})} \bigr\rangle \notag \\
= \ &\bigl\langle \downarrow,\;
(\ctx_{1}.C_{f_{\mathrm{cur}}}\ket{\phi})_{\sys \setminus \rho_{\mathrm{loc},1}(V_b)}
\otimes \ket{\theta}_{\rho_{\mathrm{loc},1}(V_b) \setminus \rho_{\mathrm{loc},1}(U_{V_b})}
\otimes \ket{\phi}_{\rho_{\mathrm{loc},1}(U_{V_b})} \bigr\rangle,
\end{align}  
where \(U_{V_b} \subseteq \{\, x \in \operatorname{var}(u) \mid u \in V_b,\; x.\flag \neq 2 \,\}\).  

Next, we prove that \((\ctx, (\ket{\phi}, k, \D_{\SCC(f_{\mathrm{cur}})}))\) is valid. The modifications from \(\ctx_{(1+\lvert V_b\rvert)'}\) to \(\ctx\) only affect \(\delta\) and \(R_{\mathrm{stmt},f_{\mathrm{cur}}}\); therefore, it suffices to verify Conditions~3(a)(2) and~4(g) in the validity of \(\ctx\) for $f_{\mathrm{cur}}$. For convenience, let \(n = (1+\lvert V_b\rvert)'\). 

\noindent\textbf{Condition~3(a).}
Note that \(R_{\mathrm{stmt}} = \operatorname{tail}(R_{\mathrm{stmt},n})\). Since \(R_{\mathrm{stmt},n}\) is a suffix of \(s_f\) by the condition in the validity of $\ctx_n$, its tail \(\operatorname{tail}(R_{\mathrm{stmt},n})\) is also a suffix of \(s_f\).

Now consider the update to the out-degree map \(\delta\). Recall that
\[
\delta= \delta_1\bigl[
            u \mapsto \delta_1(u) - 1
            \mid u \in \node(\rvq(s))
           \bigr].
\]
By Theorem~\ref{lemma:clean-inv}, we have \(\delta_n = \delta_1\); consequently, 
\[
\delta= \delta_n\bigl[
            u \mapsto \delta_n(u) - 1
            \mid u \in \node(\rvq(s))
           \bigr].
\]
From the validity of \(\ctx_n\), we know that for every node \(u \in \mathsf{Node}(G_{f_{\mathrm{cur}}})\),
\(\delta_n(u) = \contuses(R_{\mathrm{stmt},n}, u)\).
It remains to show that after the update,
\(\delta(u) = \contuses(R_{\mathrm{stmt}}, u)\) holds for all
\(u \in \mathsf{Node}(G_{f_{\mathrm{cur}}})\). 

We consider two cases:
\begin{itemize}
    \item If \(u \in \node(\rvq(s))\), then by definition \(\contuses(s, u) = 1\).
          Hence,
          \[
          \begin{aligned}
          \delta(u) &= \delta_n(u) - 1 = \contuses(R_{\mathrm{stmt},n}, u) - 1 \\
          &= \contuses(s; R_{\mathrm{stmt}}, u) - 1 = \bigl(\contuses(s, u) + \contuses(R_{\mathrm{stmt}}, u)\bigr) - 1 \\ 
          &= (1 + \contuses(R_{\mathrm{stmt}}, u)) - 1 = \contuses(R_{\mathrm{stmt}}, u).
          \end{aligned}
          \]
    \item For any other \(u\), \(\contuses(s, u) = 0\) by definition, so
          \[
          \begin{aligned}
          \delta(u)
          &= \delta_n(u) = \contuses(R_{\mathrm{stmt},n}, u) = \contuses(s; R_{\mathrm{stmt}}, u) \\
          &= \contuses(s, u) + \contuses(R_{\mathrm{stmt}}, u) = \contuses(R_{\mathrm{stmt}}, u).
          \end{aligned}
          \]
\end{itemize}
Thus, the updated out-degree map for \(G_{f_{\mathrm{cur}}}\) correctly records the continuation usage for the remaining statements \(R_{\mathrm{stmt}}\). 

Consider any node \(u \in V_b \). If \(\var(u) \cap \bigl(\dom(\rho_{\mathrm{glob}}) \cup \re(f_{\mathrm{cur}})\bigr) \neq \emptyset\), then \(\var(u) \cap \bigl(\dom(\rho_{\mathrm{loc},f_{\mathrm{cur}},1}) \setminus \re(f_{\mathrm{cur}})\bigr) = \emptyset\). Indeed, suppose to the contrary that there exists \(x \in \var(u)\) with \(x \in \dom(\rho_{\mathrm{loc},f_{\mathrm{cur}},1}) \setminus \re(f_{\mathrm{cur}})\). Since $\dom(\rho_{\mathrm{loc},f_{\mathrm{cur}},1}) \cap \dom(\rho_{\mathrm{glob}})=\emptyset$, \(x \notin \dom(\rho_{\mathrm{glob}})\). By Lemma~\ref{lem:static-wf} we obtain \(\var(u) \cap \dom(\rho_{\mathrm{glob}}) = \emptyset\) and \(\var(u) \cap \re(f_{\mathrm{cur}}) = \emptyset \), contradicting the assumption.

Observe that \[
V_c = \{\, u \in V_a \mid \delta(u)=0 \;\wedge\; u.\flag=0 \,\}.
\]  Moreover,  
\[
V_b = \{\, u \in V_c \mid \var(u) \cap \bigl(\dom(\rho_{\mathrm{glob}}) \cup \re(f_{\mathrm{cur}})\bigr) = \emptyset \,\}. 
\]

Finally, we obtain the following derivation:
\begin{align*}
&\dom(\rho_{\mathrm{loc},f_{\mathrm{cur}}}) \setminus \bigl( \fp(f_{\mathrm{cur}}) \cup \re(f_{\mathrm{cur}}) \bigr) \\
=&\; \dom(\rho_{\mathrm{loc},f_{\mathrm{cur}},1}) \setminus (U_{V_b}  \setminus \fp(f_{\mathrm{cur}})) \setminus \bigl( \fp(f_{\mathrm{cur}}) \cup \re(f_{\mathrm{cur}}) \bigr) \\
=&\; \dom(\rho_{\mathrm{loc},f_{\mathrm{cur}},1}) \setminus \bigl( \fp(f_{\mathrm{cur}}) \cup \re(f_{\mathrm{cur}}) \bigr) \setminus U_{V_b} \\
=&\; \dom(\rho_{\mathrm{loc},f_{\mathrm{cur}},1}) \setminus \bigl( \fp(f_{\mathrm{cur}}) \cup \re(f_{\mathrm{cur}}) \bigr) \setminus \Bigl( U_{V_c} \setminus U_{\{\, u \in V_c \mid \var(u) \cap (\dom(\rho_{\mathrm{glob}})\cup \re(f_{\mathrm{cur}}) ) \ne \emptyset \,\}} \Bigr) \\ 
=&\; \dom(\rho_{\mathrm{loc},f_{\mathrm{cur}},1}) \setminus \bigl( \fp(f_{\mathrm{cur}}) \cup \re(f_{\mathrm{cur}}) \bigr) \setminus U_{V_c} \\ 
\subseteq&\; \{\, x \in \mathsf{Node}(G_{f_{\mathrm{cur}}}) \mid \delta_1(\node(x)) > 0 \;\lor\; x.\flag \neq 0 \,\} \setminus U_{V_c} \\
=&\; \{\, x \in \mathsf{Node}(G_{f_{\mathrm{cur}}}) \mid \delta_n(\node(x)) > 0 \;\lor\; x.\flag \neq 0 \,\} \setminus U_{V_c} \\
=&\; \{\, x \in \mathsf{Node}(G_{f_{\mathrm{cur}}}) \mid \delta(\node(x)) > 0 \;\lor\; x.\flag \neq 0 \,\}.  
\end{align*}

\noindent\textbf{Condition~4(g).}
We need to prove
\[
\bigl( \Succ(f_{\mathrm{cur}}) \setminus \Call(R_{\mathrm{stmt},f_{\mathrm{cur}}}) \bigr) \cap \SCC(f_{\mathrm{cur}}) \subseteq \Fun(\St).
\] 
From the validity of \(\ctx_1\) we have
\[
\bigl( \Succ(f_{\mathrm{cur}}) \setminus \Call(R_{\mathrm{stmt},f_{\mathrm{cur}},1}) \bigr) \cap \SCC(f_{\mathrm{cur}})
\subseteq \Fun(\St_1). 
\]
It is straightforward to prove that \(\Call(s) \cap \SCC(f_{\mathrm{cur}}) \subseteq \Fun(\St_1)\). 
By Theorem~\ref{lemma:clean-inv}, we have \(R_{\mathrm{stmt},f_{\mathrm{cur}},n} = R_{\mathrm{stmt},f_{\mathrm{cur}},1}\) and \(\Fun(\St_n)=\Fun(\St_1)\). Consequently,
\begin{align*}
    &\bigl( \Succ(f_{\mathrm{cur}}) \setminus \Call(R_{\mathrm{stmt},f_{\mathrm{cur}}}) \bigr) \cap \SCC(f_{\mathrm{cur}}) \\
    =&\; \bigl( \Succ(f_{\mathrm{cur}}) \setminus \bigl( \Call(R_{\mathrm{stmt},f_{\mathrm{cur}},n}) \setminus \Call(s) \bigr) \bigr) \cap \SCC(f_{\mathrm{cur}}) \\
    =&\; \bigl( (\Succ(f_{\mathrm{cur}}) \setminus \Call(R_{\mathrm{stmt},f_{\mathrm{cur}},n})) \cup \Call(s) \bigr) \cap \SCC(f_{\mathrm{cur}}) \\
    =&\; \bigl( (\Succ(f_{\mathrm{cur}}) \setminus \Call(R_{\mathrm{stmt},f_{\mathrm{cur}},1})) \cap \SCC(f_{\mathrm{cur}}) \bigr)
        \cup \bigl( \Call(s) \cap \SCC(f_{\mathrm{cur}}) \bigr) \\
    \subseteq&\; \Fun(\St_1) = \Fun(\St_n) = \Fun(\St).
\end{align*} 
Thus we conclude that \((\ctx, (\ket{\phi}, k, \D_{\SCC(f_{\mathrm{cur}})})) \in \mathcal{V}\). 

Finally, we prove that \(\ctx \approx_{(\ket{\phi}, k)} \sigma\). 
Since \(\ctx_1 \approx_{(\ket{\phi}, k)} \sigma\), by definition of state equivalence we have
\begin{equation} \label{eqn:body-6}
\bigl\langle \ctx_1.C_{f_{\mathrm{cur}}}^{\,k},\; \ket{\phi} \bigr\rangle
\rightarrow^{*}
\bigl\langle \downarrow,\; \mathcal{E}(\sigma, \rho_{\mathrm{active},1}) \otimes \ket{\theta'}_{V_{\activelabel,f_{\mathrm{cur}},1}\setminus \cod(\rho_{\mathrm{active},1})} \otimes \ket{\phi}_{\mathrm{rem}} \bigr\rangle. 
\end{equation}

Since \(\rho_{\mathrm{loc},1}(U_{V_b}) \subseteq \rho_{\mathrm{loc},1}(V_b)\) and
\(\rho_{\mathrm{active}} = \rho_{\mathrm{active},1} \setminus \var(V_b)\) (proved below), 
together with Lemma~\ref{lemma:clean-inv} which gives
\(V_{\activelabel,f_{\mathrm{cur}}} = V_{\activelabel,f_{\mathrm{cur}},1}\setminus \rho_{\mathrm{loc},1}(U_{V_b})\),
a straightforward set calculation yields
\[
\bigl( V_{\activelabel,f_{\mathrm{cur}},1}\setminus \cod(\rho_{\mathrm{active},1}) \bigr)
\cup \bigl( \rho_{\mathrm{loc},1}(V_b) \setminus \rho_{\mathrm{loc},1}(U_{V_b}) \bigr)
= V_{\activelabel,f_{\mathrm{cur}}}\setminus \cod(\rho_{\mathrm{active}}).
\]

Combining Equations~\ref{eqn:body-5} and~\ref{eqn:body-6}, there exists $\ket{\theta''}=\ket{\theta'} \otimes \ket{\theta}$ such that 
\begin{align} \label{eqn:body-7} 
    \bigl\langle \ctx.C_{f_{\mathrm{cur}}}^{\,k},\; \ket{\phi} \bigr\rangle = \ & \bigl\langle \ctx_n.C_{f_{\mathrm{cur}}}^{\,k},\; \ket{\phi} \bigr\rangle \notag \\
    \rightarrow^{*}\  & \bigl\langle \downarrow,\;
(\ctx_{1}.C_{f_{\mathrm{cur}}}\ket{\phi})_{\sys \setminus \rho_{\mathrm{loc},1}(V_b)}
\otimes \ket{\theta}_{\rho_{\mathrm{loc},1}(V_b) \setminus \rho_{\mathrm{loc},1}(U_{V_b})}
\otimes \ket{\phi}_{\rho_{\mathrm{loc},1}(U_{V_b})} \bigr\rangle \notag \\ 
    = \ & 
    \bigl\langle \downarrow,\;
        \mathcal{E}(\sigma, \rho_{\mathrm{active},1})_{\sys \setminus \rho_{\mathrm{loc},1}(V_b)}
        \otimes \ket{\theta''}_{(V_{\activelabel,f_{\mathrm{cur}},1}\setminus \cod(\rho_{\mathrm{active},1}))
                                 \cup (\rho_{\mathrm{loc},1}(V_b) \setminus \rho_{\mathrm{loc},1}(U_{V_b}))}
        \otimes \ket{\phi}_{\mathrm{rem}} \bigr\rangle  \notag \\
    = \ & \bigl\langle \downarrow,\;
        \mathcal{E}(\sigma, \rho_{\mathrm{active},1})_{\sys \setminus \rho_{\mathrm{loc},1}(V_b)}
        \otimes \ket{\theta''}_{V_{\activelabel,f_{\mathrm{cur}}}\setminus \cod(\rho_{\mathrm{active}})}
        \otimes \ket{\phi}_{\mathrm{rem}} \bigr\rangle.
\end{align}
The third equality holds since \(\rho_{\mathrm{loc},1}(V_b) \subseteq \rho_{\mathrm{active},1}\). 

It remains to verify \(\rho_{\mathrm{active}} = \rho_{\mathrm{active},1} \setminus \var(V_b)\). 
Recall that \(\rho_{\mathrm{loc},f_{\mathrm{cur}}} = \rho_{\mathrm{loc},f_{\mathrm{cur}},1} \setminus (U_{V_b} \setminus \fp(f_{\mathrm{cur}}))\).
\begin{align*}
&\rho_{\mathrm{loc},f_{\mathrm{cur}}} \setminus \{ x \in \dom(\rho_{\mathrm{loc},f_{\mathrm{cur}}}) \setminus (\re(f_{\mathrm{cur}})) \mid \delta(\node(x))=0 \wedge (\node(x).\flag=0) \}  \\ 
=&\rho_{\mathrm{loc},f_{\mathrm{cur}},1} \setminus (U_{V_b} \setminus \fp(f_{\mathrm{cur}})) \\ &\setminus \{x \in \dom(\rho_{\mathrm{loc},f_{\mathrm{cur}},1}) \setminus (U_{V_b} \setminus \fp(f_{\mathrm{cur}})) \setminus (\re(f_{\mathrm{cur}})) \mid \delta(\node(x))=0 \wedge (\node(x).\flag=0) \} \\ 
=&\rho_{\mathrm{loc},f_{\mathrm{cur}},1} \setminus (U_{V_b} \setminus \fp(f_{\mathrm{cur}})) \\ &\setminus \{x \in \dom(\rho_{\mathrm{loc},f_{\mathrm{cur}},1}) \setminus (\re(f_{\mathrm{cur}})) \mid \delta(\node(x))=0 \wedge (\node(x).\flag=0) \} \setminus (U_{V_b} \setminus \fp(f_{\mathrm{cur}})) \\ 
=&\rho_{\mathrm{loc},f_{\mathrm{cur}},1} \setminus \{ x \in \dom(\rho_{\mathrm{loc},f_{\mathrm{cur}},1}) \setminus (\re(f_{\mathrm{cur}})) \mid \delta(\node(x))=0 \wedge (\node(x).\flag=0) \} \\ 
=&\rho_{\mathrm{loc},f_{\mathrm{cur}},1}  \setminus (\var(V_b) \cup \{ x \in \dom(\rho_{\mathrm{loc},f_{\mathrm{cur}},1}) \setminus (\re(f_{\mathrm{cur}})) \mid \delta_1(\node(x))=0 \wedge (\node(x).\flag=0) \}) \\  
&\qquad \tag{by definition of \(\delta\)}\\  
=&\rho_{\mathrm{loc},f_{\mathrm{cur}},1} \setminus \{ x \in \dom(\rho_{\mathrm{loc},f_{\mathrm{cur}},1}) \setminus (\re(f_{\mathrm{cur}}))  \mid \delta_1(\node(x))=0 \wedge \node(x).\flag=0 \} \setminus \var(V_b) \\  
\end{align*} 

Since $\var(V_b) \cap \dom(\rho_{\mathrm{glob}})=\emptyset$, 
\begin{align*}
&\rho_{\mathrm{active}}\\  
=&\rho_{\mathrm{glob}} \cup \rho_{\mathrm{loc},f_{\mathrm{cur}}} \setminus \{ x \in \dom(\rho_{\mathrm{loc},f_{\mathrm{cur}}}) \setminus (\re(f_{\mathrm{cur}})) \mid \delta(\node(x))=0 \wedge (\node(x).\flag=0) \} \\ 
=&\rho_{\mathrm{glob}} \cup (\rho_{\mathrm{loc},f_{\mathrm{cur}},1} \setminus \{ x \in \dom(\rho_{\mathrm{loc},f_{\mathrm{cur}},1}) \setminus (\re(f_{\mathrm{cur}})) \mid \delta_1(\node(x))=0 \wedge \node(x).\flag=0 \} \setminus \var(V_b))  \\ 
=& \rho_{\mathrm{active},1} \setminus \var(V_b).  
\end{align*} 

Consequently, Equation~\ref{eqn:body-7} rewrites to
\[
\bigl\langle \ctx.C_{f_{\mathrm{cur}}}^{\,k},\; \ket{\phi} \bigr\rangle
\rightarrow^{*}
\bigl\langle \downarrow,\;
    \mathcal{E}(\sigma, \rho_{\mathrm{active}})
    \otimes \ket{\theta''}_{V_{\activelabel,f_{\mathrm{cur}}}\setminus \cod(\rho_{\mathrm{active}})}
    \otimes \ket{\phi}_{\mathrm{rem}} \bigr\rangle,
\]
which is exactly \(\ctx \approx_{(\ket{\phi}, k)} \sigma\).

\textbf{Case~2: Sequential composition (i.e., $s=s_1;s_2$).} By definition,
we have \[\ctx_1 = \textsc{Compile\_Body}(\K, \ctx_0, s_1)\]  and 
\(\ctx = \textsc{Compile\_Body}(\K, \ctx_1, s_2)\). Since \(\langle s^{\,k}, \sigma_0 \rangle \rightarrow_{\Xi}^{*} \langle \downarrow, \sigma \rangle\), 
by the operational semantics, there exists an intermediate classical state 
\(\sigma_1\) such that
\[
\langle s_1^{k}, \sigma_0 \rangle \rightarrow_{\Xi}^{*} \langle \downarrow, \sigma_1 \rangle,
\qquad
\langle s_2^{k}, \sigma_1 \rangle \rightarrow_{\Xi}^{*} \langle \downarrow, \sigma  \rangle .
\]

Applying the induction hypothesis to \(s_1\) yields
\[
\ctx_1 \approx_{(\ket{\phi}, k)} \sigma_1
\quad\text{and}\quad
(\ctx_1, (\ket{\phi}, k, \D_{\SCC(f_{\mathrm{cur}})})) \in \mathcal{V}.
\]

Since \(s = s_1; s_2\) is a prefix of \(\ctx_0.R_{\mathrm{stmt},f_{\mathrm{cur}}}\), by the definition of how \(R_{\mathrm{stmt},f_{\mathrm{cur}}}\) is updated during compilation (i.e., after compiling the first statement \(s_1\) it is removed from the sequence), it follows directly that \(s_2\) is a suffix of $\ctx_1.R_{\mathrm{stmt},f_{\mathrm{cur}}}$. 
Now apply the induction hypothesis to \(s_2\) with initial context \(\ctx_1\)
and state \(\sigma_1\).  We obtain
\[
\ctx \approx_{(\ket{\phi}, k)} \sigma
\quad\text{and}\quad
(\ctx, (\ket{\phi}, k, \D_{\SCC(f_{\mathrm{cur}})})) \in \mathcal{V}.
\]
Thus the theorem holds for the sequential case. 

\end{proof}

\subsection{Correctness Proof for Compilation of Functions} 
\label{app:proof-fun}

This subsection presents the correctness proof of Algorithm~\ref{com_fun} for compilation of functions.
As before, we first introduce some notations and assumptions. 

\begin{notation}\label{nota:fun}
Let \(\ctx_0\) be a compilation context, \(f\) be a function name, and \(\overline{p}\) and \(\overline{r}\) be quantum variables. Define \(\ctx = \textsc{Compile\_Fun}(\K, \ctx_0, f, \{\overline{p}, \overline{r}\})\). 
According to the definition of \(\textsc{Compile\_Fun}\), the computation proceeds through a sequence of intermediate compilation contexts; for clarity in the subsequent proofs, we denote these intermediate states explicitly as follows:
\[
\begin{aligned}
& \{\overline{\tau x}, s; \tau_r v; \Gamma'\} \gets \Xi(f); \qquad  \Qpool_{\mathrm{temp}} \gets \{\, q \in \Qpool_{0} \mid \base[q] \neq q \,\}; \qquad g \gets f_{\mathrm{cur}}; \\[2pt]
& \ctx_1 =
\begin{cases}
 \ctx_0\bigl[\,\St \mapsto \push(f, \push((g,d_0), \St_0)),\; 
      \M \mapsto \M_0[f_{\mathrm{cur}} \mapsto m_0],\ f_{\mathrm{cur}} \mapsto f,\  \Vactive \mapsto \emptyset, \\ 
      \qquad  m \mapsto (I, \{x \mapsto \overline{p},\; v \mapsto \overline{r}\},\kappa_I,\; \Eargs = E_{\args,g},\; \Valloc = \emptyset,\; R_{\mathrm{stmt}} = s)\bigr]; f \in \SCC(g)\\[4pt]
\ctx_0\bigl[\,\St \mapsto \push(f, \push((g,d_0), \St_0)),\; 
      \M \mapsto \M_0[f_{\mathrm{cur}} \mapsto m_0],\;  
 \ f_{\mathrm{cur}} \mapsto f,\; 
      \Vactive \mapsto \emptyset,\; \Qpool \mapsto \Qpool_{1}\\ 
    \qquad m \mapsto (I, \{x \mapsto \overline{p}',\; v\mapsto \overline{r}'\},\kappa_I,\; \Eargs = \emptyset,\; \Valloc = \emptyset,\; R_{\mathrm{stmt}} = s) \bigr]; f\notin \SCC(g), \\[4pt]
\end{cases} \\
& \qquad \text{where } \ \overline{p}'\cup \overline{r}'= (\pop(\Qpool_{0} \setminus \Qpool_{\mathrm{temp}}, \vert (x, v) \vert )) \text{ and } \Qpool_{1}=\Qpool_{0} \setminus (\overline{p}'\cup \overline{r}')\\ 
& \ctx_2 = \textsc{Compile\_Body}(\K[\Gamma \mapsto \Gamma'], \ctx_1, s); \\[4pt]
& \ctx_3 = \textsc{Clean\_Temp}(\ctx_2, w), \\ 
& \qquad  \text{where } w \in \{\, \node(\rho_{\mathrm{loc},2}^{-1}(q)) \mid \; q \in \cod(\rho_{\mathrm{loc},2}) \setminus \rho_{\mathrm{loc},2}(\re(f)) \wedge \node(\rho_{\mathrm{loc},2}^{-1}(q)).\flag=1 \,\}; \\[4pt]
& \ctx_4 = \ctx_3\Bigl[\,\D \mapsto \bigl(
        \D_3 \cup \{\, F_f(\getC(\overline{\tau x}), \overline{p}, \overline{r}, \ctx_3.E_{\args,f})
        \Leftarrow \ctx_3.C_f \,\}\bigr)\Bigr]; \\[4pt]
& \ctx_5 = \begin{cases}
    \ctx_4\bigl[\,\D \mapsto \D_5,\; \St \mapsto \St_5,\; \M \mapsto \M_5,\\ 
    \qquad  \Qpool \mapsto \Qpool_{\mathrm{temp}} \cup \{ \base[q'] \mid q' \in \Qpool_{4} \cup \cod(\rho_{\succeq f^*,4}) \cup V_{\activelabel,\succeq f^*,4} \}\bigr],
        & f \notin \SCC(g), \\[6pt]
    \ctx_4\bigl[\,\St \mapsto \pop(\push(\St_4,\ctx_4.\Vactive), (\ctx_4.V_{\activelabel,f} \cap Q_{\succeq f^*,4}))\; 
                 \M \mapsto \M_4[f \mapsto m_4]\;\bigr],
        & f \in \SCC(g), 
\end{cases} \\[4pt]
& \qquad \text{where } E_{\args,h} \triangleq \operatorname{add}(E_{\args,h}, V_{\alloc,\succeq f}) 
        \quad \text{for all } h \in \Fun(\St_4) \text{ with } \pos(h) \ge \pos(f), \\
& \qquad\qquad \ \D_5 = \operatorname{update}(E_{\args,h}, h, \D_4), \qquad \St_5 = \St_{\prec f,4}, \qquad 
     \M_5 = \M_4 \upharpoonright_{\{h \mid \pos(h) < \pos(f)\}}; \\[6pt]
& \ctx = \ctx_5\bigl[\,\Vactive \mapsto \emptyset,\; m \mapsto \M_5(g),\; f_{\mathrm{cur}} \mapsto g\, \bigr]. \\ 
\end{aligned}
\]
\end{notation}

\begin{assumption} \label{hyp:fun_1}
Under Notation~\ref{nota:fun}, assume that there exist a state
\(\ket{\phi} \in \mathcal{H}\), a depth
\(k \in \mathbb{N}^+\), and a family of target declarations 
\(\D_{\SCC(g)}\) such that
\[
(\ctx_0, (\ket{\phi}, k, \D_{\SCC(g)})) \in \mathcal{V}.
\]
\end{assumption} 

\begin{assumption} \label{hpy:fun_2}
  Under Notation~\ref{nota:fun}, assume that if \(\ctx_0\)
  satisfies the structural properties of the validity predicate,
  then the following also hold.
  \begin{itemize}
    \item \(\overline{p} \cap \overline{r} = \emptyset\), \(\overline{p} \cup \overline{r}  \subseteq \ctx_0.\rho_{\mathrm{loc},g}\) and $\overline{r} \subseteq V_{\activelabel,g,0}$.   
    \item \(f \in \Succ(g)\).
    \item $f \notin (\Theta^{-1}(\dom(\D_0)) \cup \Fun(\St_0))$. 
    \item \(\ctx_{r,0} = \ctx_r\). 
    \item If \(f \in \SCC(g)\), then
           \(\ctx_0.Q_{\succeq g^*}[+] \cap \ctx_0.V_{\activelabel,g} = \emptyset\) and 
          \(\forall\, q \in \ctx_0.V_{\activelabel,g} \setminus (\rho_{\mathrm{loc},g,0}(\re(g)) \cap \overline{r}),\; \var(\idx(q)) \neq \emptyset\). 
          Moreover, if \(\ctx\) satisfies the
          structural properties of validity,
          we have \(\ctx_r.Gar_h \subseteq \ctx.Gar_h\) for every \(h \in \SCC(g) \cap \Fun(\St)\).  
  \end{itemize}
\end{assumption}

\begin{assumption}\label{hpy:fun_3} 
Under Notation~\ref{nota:fun}, assume that there exists a family of target declarations \(\D'_{\SCC(f)}\) such that, if \(f \in \SCC(g)\), then \(\D'_{\SCC(f)} = \D_{\SCC(g)}\). Furthermore, suppose the following hold:
 For every \(h \in \SCC(f) \cap \Fun(\ctx_4.\St)\) and every repetition count \(z > 0\) (where \(z = k \) whenever \(f \in \SCC(g)\)):
    \begin{itemize}
        \item $\Theta(h)^z \approx_{\ctx_4.Gar_h} h^z$; 
        \item \(\ctx_4.E_{\args,h}\) is a prefix of \(\args(\D'_{\SCC(f)}(h))\);
        \item  if $h \in \Fun(\ctx_1.\St) \setminus \Theta^{-1}(\dom(\ctx_4.\D))$, $\anc(\Theta(h)^z) \subseteq \ctx_4.A_h$; 
        \item \(\ctx_4.\D \approx_{\ctx_4.Gar_{f^*}} \D'_{\SCC(f)}\) and if $h \in \Theta^{-1}(\dom(\ctx_4.\D))$, \[\anc(\D'_{\SCC(f)}[\Theta(h)]) \subseteq \anc(\ctx_4.\D[\Theta(h)]) \cup \anc(\ctx_4.\D[\Theta(h)])[+].\]  
    \end{itemize} 
\end{assumption}

\begin{figure}[t]
\centering
\resizebox{\linewidth}{!}{%
\begin{tikzpicture}[
  x=1cm,
  y=1cm,
  proofnode/.style={
    draw,
    rounded corners=2pt,
    align=center,
    inner xsep=3pt,
    inner ysep=3pt,
    minimum height=7mm,
    font=\scriptsize
  },
  thmnode/.style={proofnode, fill=blue!6, draw=blue!45!black},
  lemmanode/.style={proofnode, fill=gray!8, draw=gray!60},
  dep/.style={-{Latex[length=1.8mm]}, thick, draw=gray!70}
]
\node[lemmanode, text width=.10\linewidth] (f4) at (-2.7,1.55)
  {Lemma~\ref{lemma:clean-inv}};
\node[lemmanode, text width=.10\linewidth] (f17) at (-0.6,1.55)
  {Lemma~\ref{lem:body-inv}};
\node[lemmanode, text width=.10\linewidth] (f20) at (-1.65,0.45)
  {Lemma~\ref{lem:fstar-fixed}};

\node[lemmanode, text width=.10\linewidth] (f29) at (-0.25,-0.95)
  {Lemma~\ref{lemma:ctx-A}};
\node[lemmanode, text width=.10\linewidth] (f30) at (0.35,-2.00)
  {Lemma~\ref{lem:fun-inv}};

\node[lemmanode, text width=.10\linewidth] (f21) at (1.55,3.60)
  {Lemma~\ref{lemma:cv-A-1}};
\node[lemmanode, text width=.10\linewidth] (f22) at (3.60,3.60)
  {Lemma~\ref{lemma:Gar-4}};
\node[lemmanode, text width=.10\linewidth] (f23) at (2.55,2.55)
  {Lemma~\ref{lemma:step_1}};
\node[lemmanode, text width=.10\linewidth] (f24) at (2.55,1.50)
  {Lemma~\ref{lemma:step_2}};
\node[lemmanode, text width=.10\linewidth] (f25) at (2.55,0.45)
  {Lemma~\ref{lemma:valid-4}};
\node[thmnode, text width=.11\linewidth] (f19) at (4.55,2.25)
  {Theorem~\ref{theo:body}};
\node[thmnode, text width=.11\linewidth] (f5) at (4.65,1.50)
  {Theorem~\ref{theo:clean}};
\node[thmnode, text width=.11\linewidth] (f16) at (4.55,0.75)
  {Theorem~\ref{theo:clean-valid}};

\node[lemmanode, text width=.10\linewidth] (f26) at (6.85,1.50)
  {Lemma~\ref{lem:R2-closure}};
\node[lemmanode, text width=.10\linewidth] (f27) at (6.85,0.45)
  {Lemma~\ref{lem:SCC-in-stack-domD}};
\node[lemmanode, text width=.10\linewidth] (f28) at (8.80,0.45)
  {Lemma~\ref{lemma:cv-A-4}};

\node[thmnode, text width=.12\linewidth] (f32) at (5.40,-1.15)
  {Theorem~\ref{theo_fun}\\
   Function correctness};

\draw[dep] (f20) -- (f4);
\draw[dep] (f20) -- (f17);
\draw[dep] (f29) -- (f20);
\draw[dep] (f30) -- (f29);
\draw[dep] (f23) -- (f20);
\draw[dep] (f23) -- (f21);
\draw[dep] (f23) -- (f22);
\draw[dep] (f24) -- (f23);
\draw[dep] (f24) -- (f19);
\draw[dep] (f24) -- (f5);
\draw[dep] (f24) -- (f16);
\draw[dep] (f25) -- (f24);
\draw[dep] (f25) -- (f20);
\draw[dep] (f27) -- (f26);
\draw[dep] (f32) -- (f20);
\draw[dep] (f32) -- (f29);
\draw[dep] (f30) -- (f32);
\draw[dep] (f32) -- (f25);
\draw[dep] (f32) -- (f27);
\draw[dep] (f32) -- (f28);
\draw[dep] (f29) -- (f25);
\end{tikzpicture}%
}
\caption{Proof dependencies for the correctness of \textsc{Compile\_Fun}; arrows point from a result to the lemmas or theorems it depends on.}
\label{fig:fun-proof-dep}
\Description{A dependency diagram showing the main lemmas used in the proof of function compilation correctness.}
\end{figure}
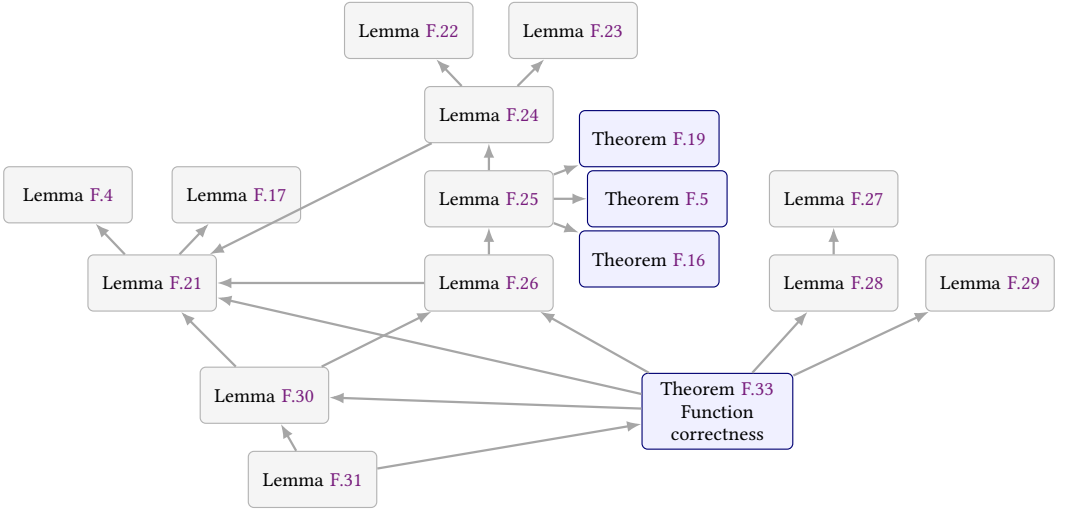

The correctness of Algorithm~\ref{com_fun} is established by Theorem~\ref{theo_fun}. Since the argument is rather involved, we first outline how it uses the existing hypotheses and auxiliary
lemmas under the notation; the dependencies among the theorems are shown in Figure~\ref{fig:fun-proof-dep} (which depicts only the main dependencies).

Under Notation~\ref{nota:fun} and Assumptions~\ref{hyp:fun_1}
and~\ref{hpy:fun_2}, we establish the main theorem of this section
(Theorem~\ref{theo_fun}), i.e., the validity of the compilation context
\(\ctx\) together with the correctness of the generated code.  The proof
verifies each clause of the validity predicate \(\mathcal{V}\) in turn.
Along the way we need to construct a specification family that satisfies
Assumption~\ref{hpy:fun_3}, so that Lemmas~\ref{lemma:step_1},
\ref{lemma:step_2} and~\ref{lemma:valid-4}---which successively
establish the validity of \(\ctx_1\), \(\ctx_3\) and \(\ctx_4\)---can be
applied (these lemmas require Assumption~\ref{hpy:fun_3} to hold).

As noted earlier in the Proof Strategy, the structural properties
within validity do not depend on the semantic conditions nor on the
extra specification family.  This observation leads to the following
proof outline.

\begin{itemize}
\item \textbf{Starting point.} Notation~\ref{nota:fun} and
      Assumptions~\ref{hyp:fun_1} and~\ref{hpy:fun_2} are given.

\item \textbf{Step~1 (structural transfer).}  Using the notation,
      the structural part of the validity of \(\ctx_0\) contained in
      Assumption~\ref{hyp:fun_1}, and Assumption~\ref{hpy:fun_2}
      (which itself depends only on structural properties), we follow
      the proof chain established in the Proof Strategy---a chain that
      never invokes semantic conditions---to first obtain the
      structural properties of $\ctx_1$ to \(\ctx_4\) (see the structural parts of Lemmas~\ref{lemma:step_1}, \ref{lemma:step_2}
      and~\ref{lemma:valid-4}).  Once the structural properties of
     $\ctx_1$ to \(\ctx_4\) are in place, we further employ Lemmas~\ref{lem:fstar-fixed},~\ref{lem:SCC-in-stack-domD},~\ref{lemma:cv-A-4} and~\ref{lemma:ctx-A} (among them Lemma~~\ref{lem:fstar-fixed} requires only Condition~1(2) of the structural properties of the relevant context,
      while the remaining lemmas presuppose that the structural
      properties already hold); with these conclusions
we are then able to establish the structural properties of the final context \(\ctx\) (see the corresponding parts of Theorem~\ref{theo_fun}).  This step needs neither the semantic
      properties of \(\ctx_0\) nor a specification family satisfying Assumption~\ref{hpy:fun_3}; the
      latter serves only the subsequent semantic part.

\item \textbf{Step~2 (semantic recovery).}  Once the structural
      properties of \(\ctx_4\) and \(\ctx\) as well as the properties of
      Lemmas~\ref{lem:fstar-fixed},~\ref{lem:SCC-in-stack-domD},~\ref{lemma:cv-A-4} and~\ref{lemma:ctx-A} are available, we construct, 
      depending on the concrete situation (e.g., whether
      \(f \in \SCC(g)\)), a specification family
      \(\D'_{\SCC(f)}\) that meets Assumption~\ref{hpy:fun_3}.
      Returning to \(\ctx_0\) and using its semantic properties (in
      particular the \(k\)-equivalence with respect to
      \(\D_{\SCC(g)}\)), we re‑apply
      Lemmas~\ref{lemma:step_1}, \ref{lemma:step_2}
      and~\ref{lemma:valid-4} to transfer the semantic conditions of
      \(\ctx_0\) to \(\ctx_4\), from which the semantic properties of
      \(\ctx\) are then derived.

\item \textbf{Closing the circle.}  The structural properties are laid
      down first, then the semantic properties are proved on top of
      them, yielding the full validity of \(\ctx\).  In the remainder of
      the proof we carefully ensure that the structural reasoning never
      uses the semantic properties of \(\ctx_0\) or any specification
      family satisfying Assumption~\ref{hpy:fun_3}, while the semantic
      part explicitly shows how it depends on the semantic properties
      of the original compilation state, on the already established
      structural properties, and on the construction of a suitable
      specification family that satisfies Assumption~\ref{hpy:fun_3}
      to invoke the preceding lemmas.  Consequently, whenever we prove
      a semantic property we may always assume that the structural
      properties from \(\ctx_0\) up to \(\ctx\) are already in place.
\end{itemize}

Furthermore, Lemma~\ref{lem:fun-inv} describes the relation between compilation
variables of the input context \(\ctx_0\) and those of the output
context \(\ctx\); it relies only on the structural properties of the
validity of \(\ctx\) established by Theorem~\ref{theo_fun}.  Lemmas~\ref{lemma:clean-inv} and~\ref{lem:body-inv} 
concern the relation between the input and output contexts of the
function‑body compilation algorithm and of the cleanup algorithm; they are directly used by Lemmas~\ref{lemma:step_1}, \ref{lemma:ctx-A}, \ref{lem:fun-inv} and by Theorem~\ref{theo_fun}.  Since these dependencies are already reflected indirectly in the diagram, they are not drawn separately.

\begin{lemma}[Residual completion invariant]
\label{lem:completion-inv} Let \(\ctx\) be any completion context. Let \(\ctx_r\) be the completion context obtained from \(\ctx\), and assume that
\(\ctx\) satisfies the structural properties of validity. Then we have the following
properties. 
\begin{enumerate}
    \item For \(\D\),
    \begin{itemize}
        \item \(\dom(\ctx.\D) \subseteq \dom(\ctx_r.\D)\);
        \item \(\dom(\ctx_r.\D) \setminus \dom(\ctx.\D)
        \subseteq \Theta(\Succ^*(f_{\mathrm{cur}}) \setminus \Fun(\ctx.\St))\);
        \item
        \(
        \ctx_r.\D\upharpoonright_{\dom(\ctx.\D) \setminus \Theta(\SCC(f_{\mathrm{cur}}))}
        =
        \ctx.\D\upharpoonright_{\dom(\ctx.\D) \setminus \Theta(\SCC(f_{\mathrm{cur}}))};
        \)
        \item
        \(
        \ctx_r.\D\upharpoonright_{\dom(\ctx.\D) \cap \Theta(\SCC(f_{\mathrm{cur}}))}
        \approx_{\ctx_r.\Gar_{f_{\mathrm{cur}}^*}}
        \ctx.\D\upharpoonright_{\dom(\ctx.\D) \cap \Theta(\SCC(f_{\mathrm{cur}}))};
        \)
        and for any \(h \in \SCC(f_{\mathrm{cur}}) \cap
        \Theta^{-1}(\dom(\ctx.\D))\),
        \[
        \anc(\ctx_r.\D[\Theta(h)])
        \subseteq
        \anc(\ctx.\D[\Theta(h)]) \cup \anc(\ctx.\D[\Theta(h)])[+].
        \]
    \end{itemize}

    \item For \(\Fun(\St)\),
    \begin{itemize}
        \item \(\Fun(\ctx.\St) \subseteq \Fun(\ctx_r.\St)\);
        \item \(\Fun(\ctx_r.\St_{\preceq h}) =
        \Fun(\ctx.\St_{\preceq h})\) for any
        \(h \in \Fun(\ctx.\St_{\preceq f_{\mathrm{cur}}})\);
        \item
        \(
        \Fun(\ctx_r.\St_{\succ f_{\mathrm{cur}}})
        \setminus
        \Fun(\ctx.\St_{\succ f_{\mathrm{cur}}})
        \subseteq
        \Succ^*(\Succ(f_{\mathrm{cur}}) \setminus \Fun(\ctx.\St))
        \cap \SCC(f_{\mathrm{cur}}).
        \)
    \end{itemize}

    \item
    \(
    \ctx_r.\St_{\preceq f_{\mathrm{cur}}^*}
    =
    \ctx.\St_{\preceq f_{\mathrm{cur}}^*},
    \qquad
    \ctx_r.\M_{\preceq f_{\mathrm{cur}}^*}
    =
    \ctx.\M_{\preceq f_{\mathrm{cur}}^*}.
    \)

    \item For any \(h \in \SCC(f_{\mathrm{cur}}) \cap \Fun(\ctx.\St)\),
    \begin{itemize}
        \item \(\ctx.E_{\args,h} \preceq \ctx_r.E_{\args,h}\);
        \item \(\ctx_r.A_h \subseteq \ctx.A_h\) if
        \(\pos(h) \leq \pos(f_{\mathrm{cur}})\);
        % \item \(\ctx_r.\rho_{\mathrm{loc},h}=\ctx.\rho_{\mathrm{loc},h}\),
        % \(\ctx_r.V_{\activelabel,h}=\ctx.V_{\activelabel,h}\), and
        % \(\ctx_r.V_{\activelabel,[f_{\mathrm{cur}}^*,h]}
        % =\ctx.V_{\activelabel,[f_{\mathrm{cur}}^*,h]}\) if \(h < f_{\mathrm{cur}}\);
        \item \(\dom(\ctx_r.\rho_{\mathrm{loc},h})
        \subseteq \dom(\ctx.\rho_{\mathrm{loc},h})\) if \(h \neq f_{\mathrm{cur}}\);
        \item \(\ctx_r.\rho_{\mathrm{loc},h}(x)=\ctx.\rho_{\mathrm{loc},h}(x)\), if
        % \(\pos(h) \geq \pos(f_{\mathrm{cur}})\),
        \(x \in \dom(\ctx.\rho_{\mathrm{loc},h})\cap
        \dom(\ctx_r.\rho_{\mathrm{loc},h})\), and
        \[\var(\idx(\ctx.\rho_{\mathrm{loc},h}(x)))\neq \emptyset.\]
    \end{itemize}
\end{enumerate}
\end{lemma}

\begin{proof}
The residual completion from \(\ctx\) to \(\ctx_r\) is a suffix of the execution
of the compilation algorithm for the current function. We prove the statement by
induction on this residual execution. Each step is one of the already verified
compilation transitions: statement compilation, body compilation, cleanup of
temporary variables, compilation of an uncompiled callee, or the final
declaration-generation step. The corresponding preservation properties are given
by the invariants for statement compilation, Lemma~\ref{lem:body-inv},
Lemma~\ref{lemma:clean-inv}, and Lemma~\ref{lem:fun-inv}; the last one covers
the case where the residual execution invokes the compilation of an uncompiled
callee. The final
declaration-generation step only records the generated target declaration and
preserves the listed prefix, stack, ancilla-containment, and
declaration-restriction properties. Therefore all listed properties hold for the
completion context \(\ctx_r\).
\end{proof}

\begin{lemma}
  \label{lem:fstar-fixed}
  Under Notation~\ref{nota:fun}, assume that \(\ctx_0\) satisfies
  Condition~1(2) (stack discipline) of the validity predicate.
  Let \(\ctx'\) be any of \(\ctx_1\)–\(\ctx_4\) and assume that \(\ctx'\)
  also satisfies Condition~1(2).  If \(f \notin \SCC(g)\), then
  \(\ctx'.f^* = f\).
\end{lemma}

\begin{proof}
  We first extract some consequences of the stack discipline in
  \(\ctx_0\).  Lemmas~\ref{lem:body-inv} and~\ref{lemma:clean-inv} give
  \[
    \Fun(\St_{\preceq h',3}) = \Fun(\St_{\preceq h',2})
    = \Fun(\St_{\preceq h',1})
  \]
  for every \(h' \in \Fun(\St_{\preceq f,1})\).  By construction,
  \(\St_3 = \St_4\).

  Since \(g = \ctx_0.f_{\mathrm{cur}} \in \Fun(\St_0)\) and
  \(\St_1 = \push(\push(\St_0,(g,\ctx_0.d)),f)\), we have
  \(f,g \in \Fun(\St_{\preceq f,1})\) and
  \(\pos(g) < \pos(f)\) in \(\St_1\).  Hence
  \(f,g \in \Fun(\St')\) and \(\pos(g) < \pos(f)\) also hold in
  \(\St'\).  Since \(\ctx_0.f_{\mathrm{cur}} = g\), Condition~1(2) for \(\ctx_0\)
  implies that every \(h'\) with \(\pos(h') > \pos(g)\) belongs to
  \(\SCC(g) \cap \Fun(\St_0)\); together with the equalities above,
  this yields
  \(h' \in \SCC(g) \cap \Fun(\St')\) for every \(h'\) with
  \(\pos(f) > \pos(h') > \pos(g)\) in \(\St'\).

  Now suppose, for contradiction, that \(\ctx'.f^* \neq f\).
  By definition of \(f^*\) there then exists
  \(h \in \Fun(\St'_{\prec f})\) such that \(\pos(h) < \pos(f)\) and
  \(h \in \SCC(f)\); in particular \(h \in \Succ^*(f)\).

  We distinguish two cases according to the position of \(h\).

  \begin{itemize}
    \item \textbf{Case \(\pos(h) > \pos(g)\).}
          Since \(\pos(h) < \pos(f)\), the observation above forces
          \(f \in \SCC(g)\).

    \item \textbf{Case \(\pos(h) < \pos(g)\).}
          Condition~1(2) for \(\ctx'\) gives
          \(\Fun(\St'_{\succ h}) \subseteq \{\, h' \mid h \mathrel{R_1^*} h' \,\}\).
          Since \(\pos(h) < \pos(g)\), we have
          \(g \in \Fun(\St'_{\succ h})\), hence
          \(g \in \{\, h' \mid h \mathrel{R_1^*} h' \,\}\); therefore
          \(g \in \Succ^*(h)\).
          Together with \(h \in \Succ^*(f)\) this implies
          \(g \in \Succ^*(f)\).
          On the other hand, \(\pos(g) < \pos(f)\) yields
          \(f \in \Succ^*(g)\) by the stack discipline.
          Consequently \(f \in \SCC(g)\).
  \end{itemize}
  In either case we obtain \(f \in \SCC(g)\), which contradicts the
  hypothesis that \(f \notin \SCC(g)\).  Hence \(\ctx'.f^* = f\).
\end{proof}

\begin{lemma} \label{lemma:cv-A-1}
  Under Notation~\ref{nota:fun} and Assumption~\ref{hpy:fun_2}, assume that $\ctx_0$ satisfies the structural properties of validity. If \(f \in \SCC(g)\),
  then \(cv_1 = cv_0\) and
  \(\ctx_1.A_h = \ctx_0.A_h\) for all \(h \in \SCC(g) \cap \Fun(\St_0)\).
\end{lemma}

\begin{proof}
Since $f\in \SCC(g)$ and \(f,g \in \Fun(\St_1)\) we have $\ctx_1.f^*=\ctx_1.g^*$.  
By definition, \(\Qpool_{1} = \Qpool_{0}\),
\[
  V_{\activelabel,\succeq f^{*},1} =V_{\activelabel,\succeq g^{*},1}
  = V_{\activelabel,\succeq g^{*},0} \cup V_{\activelabel,1}
  = V_{\activelabel,\succeq g^{*},0} \cup \emptyset
  = V_{\activelabel,\succeq g^{*},0}
\]
and
\[
  \cod(\rho_{\mathrm{loc},\succeq f^{*},1})
  = \cod(\rho_{\mathrm{loc},\succeq g^{*},0}) \cup \{\overline{p}, \overline{r}\}.
\]
Since \(\{\overline{p}, \overline{r}\} \subseteq \ctx_0.\rho_{g}\) by Assumption~\ref{hpy:fun_2}, 
the last equality simplifies to
\(\cod(\rho_{\mathrm{loc},\succeq f^{*},1}) = \cod(\rho_{\mathrm{loc},\succeq g^{*},0})\),
which also yields \(Q_{\succeq f^{*},1} = Q_{\succeq g^{*},0}\).
Assumption~\ref{hpy:fun_2} already gives \(Q_{\succeq g^{*},0}[+] \cap V_{\activelabel,g,0} = \emptyset\).
Moreover, since \(V_{\activelabel,f,1} = \emptyset\), we have
\begin{align*}
  cv_1
  &= \Qpool_{1} \cup \Qpool_{1}[+] \cup
     \Bigl( V_{\activelabel,\succeq f^{*},1}[+] \cup \cod(\rho_{\mathrm{loc},\succeq f^{*},1})[+]
            \setminus \bigl( Q_{\succeq f^{*},1}[+] \cap V_{\activelabel,f,1} \bigr) \Bigr) \\
  &= \Qpool_{0} \cup \Qpool_{0}[+] \cup
     \Bigl( V_{\activelabel,\succeq g^{*},0}[+] \cup \cod(\rho_{\mathrm{loc},\succeq g^{*},0})[+]
            \setminus \bigl( Q_{\succeq g^{*},0}[+] \cap V_{\activelabel,g,0} \bigr) \Bigr)
    = cv_0 .
\end{align*}
Thus \(cv_1 = cv_0\). 

  For every \(h \in \SCC(g) \cap \Fun(\St_0)\), by construction
  we have
  \(V_{\activelabel,\succeq h,1} \cup \cod(\rho_{\mathrm{loc},h,1})
   = V_{\activelabel,\succeq h,0} \cup \cod(\rho_{\mathrm{loc},h,0})\);
  the definition of \(A_h\) then gives \(\ctx_1.A_h = \ctx_0.A_h\).
\end{proof}

\begin{lemma}\label{lemma:Gar-4}
  Under Notation~\ref{nota:fun}, if \(f \in \SCC(g)\), then
  \(Gar_{h,4} = Gar_h\) for every \(h \in \SCC(g) \cap \Fun(\St_4)\).
\end{lemma}

\begin{proof}
  Since \(f \in \SCC(g)\) and \(f,g \in \Fun(\St_{\succeq f^{*},4})\), we have \(\ctx_4.f^{*} = \ctx_4.g^{*}\). 
  By the construction of \(\ctx\) from \(\ctx_4\) we have
  \(\Fun(\St_{\succeq f^{*},4}) = \Fun(\St_{\succeq g^{*}})\),
  \(\M = \M_4[f \mapsto m_4]\), and \(m = \M_4(g)\).

 Therefore, we obtain
  \[
    \rho_{\succeq g^{*}} = \rho_{\succeq g^{*},4} = \rho_{\succeq f^{*},4},
    \qquad
    \rho_{\succeq h}=\rho_{\succeq h,4} 
  \]
  for every \(h \in \SCC(f) \cap \Fun(\St_4)\).
  Together with the definition of \(Gar\)
  (\(Gar_h = Q_{\succeq h^{*}}[+] \cup Q_{\succeq h}\)),
  this directly yields
  \(Gar_{h,4} = Gar_h\) for all \(h \in \SCC(g) \cap \Fun(\St_4)\).
\end{proof}

\begin{lemma}\label{lemma:step_1}
    Under Notation~\ref{nota:fun} and Assumptions~\ref{hyp:fun_1},~\ref{hpy:fun_2} and~\ref{hpy:fun_3},  let \(\ket{\phi'}\) be such that $\ket{\phi'}\models \ket{0}_{{\ctx_1.A_f\setminus\rho_{\mathrm{loc},f,1}(\fp(f))}}$. Then we have 
\[
(\ctx_1, (\ket{\phi'}, z, \D'_{\SCC(f)})) \in \mathcal{V}. 
\] 
\end{lemma}

\begin{proof}
     We verify that \((\ctx_1, (\ket{\phi'}, z, \D'_{\SCC(f)})) \in \mathcal{V}\) by checking each validity condition.

    \begin{enumerate}
          \item \textbf{Condition 1.} We first prove that the global compilation state of $\ctx_1$ is valid with respect to \(z\) and \(\D'_{\SCC(f)}\). 
          
          Since \((\ctx_0, (\ket{\phi}, k, \D_{\SCC(g)})) \in \mathcal{V}\), we have that the global compilation state of $\ctx_0$ is valid with respect to \(k\) and \(\D_{\SCC(g)}\), which indicates that \(\Fun(\St_{\succ g,0}) \subseteq \dom(\M_0)\cup \{g\}\). 
          
          % Note that \((\D_1, G_{\mathrm{fun},1})=(\D_0, G_{\mathrm{fun},0})\) and $\Fun(\St_1) =\Fun(\St_0)\cup \{f\}$. Then, for any $h \in \dom(\Theta) \setminus (\Fun(\St_1) \cap \Theta^{-1}(\dom(\D_1)))$, we have $h \in \dom(\Theta) \setminus (\Fun(\St_0) \cap \Theta^{-1}(\dom(\D_0)))$, so Condition~3 follows directly from the validity of \(\ctx_0\). 
          
          The validity of \(\ctx_0\) gives that $\Fun(\St_0) \subseteq \dom(\M_0)\cup \{g\}$. 
          Since \(\Fun(\St_1)=\Fun(\St_0) \cup \{ f \}\), \(\dom(\M_1)=\dom(\M_0)\cup \{ g \}\). Then we have \(f \in \Fun(\St_1)\) and \(\Fun(\St_1) =\Fun(\St_0)\cup \{f\} \subseteq \dom(\M_0)\cup \{g\} \cup \{f\} =\dom(\M_1)\cup \{f\}\). Besides, since \(\Fun(\St_{\succ f, 1})=\emptyset\) and  thus \(\Fun(\St_{\succ f, 1})\subseteq \Theta^{-1}(\dom(\D_1)) \cap \SCC(f)\). We then prove that for every \(h \in \Fun(\St_1)\),
\[
\Fun(\St_{\succ h,1}) \subseteq \{\, h' \mid h \mathrel{R_1^*} h' \,\}.
\]
Observe that
\(
\Fun(\St_{\succ h,1}) = \Fun(\St_{\succ h,0}) \cup \{f\}.
\)
By the validity of \(\ctx_0\), we already have
\(
\Fun(\St_{\succ h,0}) \subseteq \{\, h' \mid h \mathrel{R_1^*} h' \,\}.
\)
It therefore remains to show that
\(
f \in \{\, h' \mid h \mathrel{R_1^*} h' \,\}.
\)

Since \(f \in \Succ(g)\), it suffices to show that
\(
g \in \{\, h' \mid h \mathrel{R_1^*} h' \,\}.
\)

We distinguish two cases.

If \(\pos(h) < \pos(g)\), then by the validity of \(\ctx_0\), we have
\(
g \in \{\, h' \mid h \mathrel{R_1^*} h' \,\}.
\)

If \(\pos(h) \ge \pos(g)\), then by the validity of \(\ctx_0\), we have
\(
h \in \SCC(g).
\)
This implies that
\(
g \in \{\, h' \mid h \mathrel{R_1^*} h' \,\}.
\)
	In summary, since \(f \in \Succ(g)\), we obtain
\(
f \in \{\, h' \mid h \mathrel{R_1^*} h' \,\}.
\)

Therefore,
\[
\Fun(\St_{\succ h,1}) \subseteq \{\, h' \mid h \mathrel{R_1^*} h' \,\}.
\]
This completes the proof. Thus Condition~1(2) holds. 

Since \(f \notin \Theta^{-1}(\dom(\D_0))\), we have \(\Theta^{-1}(\dom(\D_1)) \setminus \Fun(\St_1)=\Theta^{-1}(\dom(\D_0)) \setminus \Fun(\St_0)\). Combining this with \(\D_1=\D_0\), Condition~1(3)(c) follows from the validity of $\ctx_0$. 

Since \(R_{\mathrm{stmt},1,f}=s_f\), the completion context from \(\ctx_1\) is \(\ctx_4\), i.e.,
  \(\ctx_{r,1}=\ctx_4\). 
If $f \notin \SCC(g)$, then $\SCC(f) \cap \Fun(\St_1) \cap \Theta^{-1}(\dom(\D_0)) =\emptyset$, and thus Condition~1(3)(b) holds. Otherwise, we have  
\(\SCC(f) \cap \Fun(\St_1) \cap \Theta^{-1}(\dom(\D_1))=\SCC(g)\cap \Fun(\St_0) \cap \Theta^{-1}(\dom(\D_0))\). First, by the construction, we have for any $h \in \Fun(\St_1)\setminus \{f\}$, $R_{\mathrm{stmt},h}$ unchanged.  By Lemma~\ref{lemma:Gar-4}, we have $\ctx_4.Gar_{f_{\mathrm{cur}}^*}=\ctx.Gar_{f_{\mathrm{cur}}^*}$ and by the assumption, we have $\ctx.Gar_{f_{\mathrm{cur}}^*}=\ctx_r.Gar_{f_{\mathrm{cur}}^*}$ since we can prove that $\ctx$ satisfies structural properties. Then $\ctx_{r,1}.Gar_{f_{\mathrm{cur}}^*}=\ctx_{4}.Gar_{f_{\mathrm{cur}}^*}=\ctx.Gar_{f_{\mathrm{cur}}^*}=\ctx_{r}.Gar_{f_{\mathrm{cur}}^*}=\ctx_{r,0}.Gar_{f_{\mathrm{cur}}^*}$, we get Condition~1(3)(b). It remains to prove that 
          \begin{itemize}
                \item Condition~1(1): \(\D_1 \approx_{\ctx_{r,1}.Gar_{f^*}} \D'_{\SCC(f)}\);
              \item Condition~1(3)(a): for any \(h \in \SCC(f)\), \(\Theta(h)^{z} \approx_{\ctx_{r,1}.Gar_{h}} h^{z}\); \\ 
            \(\anc(\Theta(h)^{z}) \subseteq \ctx_{r,1}.A_h\), \quad  
        \(\ctx_{r,1}.E_{\args,h} \preceq \args(\D'_{\SCC(f)}(h))\), \quad 
        \(\ctx_{r,1}.\D \approx_{\ctx_{r,1}.Gar_{f^*}} \D'_{\SCC(f)}\).
          \end{itemize}
        Using the identity \(\ctx_{r,1}=\ctx_4\), 
        % By definition, we have $ctx_{r,1}=ctx_4$,
        %  \textcolor{red}{ We have \(\ctx_{r,1}=\ctx_2\). 
        %  Note that \(\ctx_2.E_{\args,f}=\ctx_4.E_{\args,f}\) and \(\ctx_2.A_f=\ctx_4.A_f\). Together with \(\ctx_2.Gar_{h}=\ctx_4.Gar_{h}\) for any $h$ and \(\D_2 = \D_4\)}, 
         Condition~1(3)(a) holds by the properties of \(\D'_{\SCC(f)}\) in Assumption~\ref{hpy:fun_3}. Finally, by Lemma~\ref{lem:completion-inv}, we have \(\ctx_{r,1}.\D\upharpoonright_{\dom(\D_1)} \approx_{\ctx_{r,1}.Gar_{f^*}} \D_1\) and thus \(\D_1 \approx_{\ctx_{r,1}.Gar_{f^*}} \D'_{\SCC(f)}\).  

\item \textbf{Condition 2.} 
By the premise we immediately obtain
\(\ket{\phi'} \models \ket{0}_{\ctx_1.A_f \setminus \rho_{\mathrm{loc},f,1}(\fp(f))}\).

We now prove
\(cv_1 \cap \bigl( V_{\activelabel,\succeq f^*,1} \cup \cod(\rho_{\mathrm{loc},\succeq f^*,1}) \cup \cod(\rho_{\mathrm{glob},1}) \bigr) = \emptyset\).

If \(f \in \SCC(g)\), Lemma~\ref{lemma:cv-A-1} gives \(cv_1 = cv_0\) together with
\(V_{\activelabel,\succeq f^*,1} \cup \cod(\rho_{\mathrm{loc},\succeq f^*,1})
= V_{\activelabel,\succeq g^*,0} \cup \cod(\rho_{\mathrm{loc},\succeq g^*,0})\).
Hence, together with the fact that \(\rho_{\mathrm{glob}}\) is unchanged,
\[
cv_1 \cap \bigl( V_{\activelabel,\succeq f^*,1} \cup \cod(\rho_{\mathrm{loc},\succeq f^*,1}) \cup \cod(\rho_{\mathrm{glob}}) \bigr) = \emptyset
\]
directly follows from the validity of \(\ctx_0\). 

If \(f \notin \SCC(g)\), we have \(cv_1 = \Qpool_{1}\) and
\(V_{\activelabel,\succeq f^*,1} \cup \cod(\rho_{\mathrm{loc},\succeq f^*,1}) \cup \cod(\rho_{\mathrm{glob}})
= \{\overline{p}, \overline{r}\} \cup \cod(\rho_{\mathrm{glob}})\). 
Since \(\Qpool_{1} = \Qpool_{0} \setminus \{\overline{p}, \overline{r}\}\) and the validity of \(\ctx_0\) gives
\(\Qpool_{0} \cap \cod(\rho_{\mathrm{glob}}) = \emptyset\), we obtain
\(\Qpool_{1} \cap \bigl( \{\overline{p}, \overline{r}\} \cup \cod(\rho_{\mathrm{glob}}) \bigr) = \emptyset\).  

\noindent
Next we prove
\(Q_{\succeq f^*}[+] \cap \bigl( \rho_{\mathrm{loc},\succeq f^*,1} \cup \cod(\rho_{\mathrm{glob}}) \bigr) = \emptyset\).
Observe that \(\rho_{\mathrm{glob}}\) is unchanged and consider two cases.

If \(f \in \SCC(g)\), then \(\rho_{\mathrm{loc},\succeq f^*,1} = \rho_{\mathrm{loc},\succeq g^*,0}\) and
\(Q_{\succeq f^*}[+] = Q_{\succeq g^*,0}[+]\); the condition holds directly
from the validity of \(\ctx_0\).

If \(f \notin \SCC(g)\), by Lemma~\ref{lem:fstar-fixed} we have $f^*=f$, then
\(\rho_{\mathrm{loc},\succeq f^*,1} = \rho_{\mathrm{loc},f,1} = \{\overline{p}, \overline{r}\} \subseteq \Qpool_{0} \setminus \Qpool_{\mathrm{temp}}\).
For every \(q \in \Qpool_{0} \setminus \Qpool_{\mathrm{temp}}\) we have \(\base[q] = q\),
hence \(Q_{\succeq f^*,1} = Q_{f,1} = \emptyset\).
The validity of \(\ctx_0\) guarantees \(\Qpool_{0} \cap \cod(\rho_{\mathrm{glob}}) = \emptyset\),
so the required disjointness follows.  

\noindent
We also prove
\(\Qpool_{1} \cap \bigl( \Qpool_{1}[+] \cup (V_{\activelabel,\succeq f^*,1})[+] \cup \cod(\rho_{\mathrm{loc},\succeq f^*,1})[+] \bigr) = \emptyset\).

If \(f \in \SCC(g)\), we already have
\(V_{\activelabel,\succeq f^*,1} \cup \cod(\rho_{\mathrm{loc},\succeq f^*,1})
= V_{\activelabel,\succeq g^*,0} \cup \cod(\rho_{\mathrm{loc},\succeq g^*,0})\)
and \(\Qpool_{1} = \Qpool_{0}\); hence the claim follows directly from the validity of \(\ctx_0\).

If \(f \notin \SCC(g)\), we have
\(\Qpool_{1} = \Qpool_{0} \setminus \Qpool_{\mathrm{temp}} \setminus \{\overline{p}, \overline{r}\}\),
\(V_{\activelabel,\succeq f^*,1} = V_{\activelabel,f^*,1} = \emptyset\), and
\(\cod(\rho_{\mathrm{loc},\succeq f^*,1}) = \cod(\rho_{\mathrm{loc},f^*,1})
= \{\overline{p}, \overline{r}\} \subseteq \Qpool_{0} \setminus \Qpool_{\mathrm{temp}}\). 
Consequently,
\[
\Qpool_{1}[+] \cup (V_{\activelabel,\succeq f^*,1})[+] \cup \cod(\rho_{\mathrm{loc},\succeq f^*,1})[+]
= (\Qpool_{0} \setminus \Qpool_{\mathrm{temp}} \setminus \{\overline{p}, r\})[+] \cup \{\overline{p}, \overline{r}\}[+]
= (\Qpool_{0} \setminus \Qpool_{\mathrm{temp}})[+]
\subseteq \Qpool_{0}[+],
\]
and \(\Qpool_{1} \subseteq \Qpool_{0}\).
Since the validity of \(\ctx_0\) gives
\(\Qpool_{0} \cap \Qpool_{0}[+] = \emptyset\),
the desired disjointness follows.

\noindent
Finally, \(\Qpool_{1} \subseteq \Qpool_{0}\) also implies that the condition
\(\forall\, q \in \Qpool_{1},\; \var(\idx(q)) = \emptyset \Longrightarrow \base[q] = q\)
is directly inherited from the validity of \(\ctx_0\).

\item \textbf{Condition 3.} For any \(h \in \{f\} \cup \bigl( \Fun(\ctx_1.\St) \cap \SCC(f) \setminus \Theta^{-1}(\dom(\ctx_1.\D)) \bigr)\):
note that \(\Fun(\ctx_1.\St) = \Fun(\ctx_0.\St) \cup \{f\}\).
\begin{itemize}
    \item If \(f \in \SCC(g)\), then \(\Fun(\ctx_1.\St) \cap \SCC(f) = (\Fun(\ctx_0.\St) \cap \SCC(g)) \cup \{f\}\);
    \item otherwise \(\Fun(\ctx_1.\St) \cap \SCC(f) = \{f\}\).
\end{itemize}
Moreover, \(\Theta^{-1}(\dom(\ctx_1.\D)) = \Theta^{-1}(\dom(\ctx_0.\D))\).
Hence, whenever \(h \neq f\) exists, we obtain \(h \in \Fun(\ctx_0.\St) \cap \SCC(g) \setminus \Theta^{-1}(\dom(\ctx_0.\D))\).
In this step only the compilation state of \(f\) is modified. Moreover, we already have \(\ctx_1.A_h = \ctx_0.A_h\) if \(f \in \SCC(g)\) by Lemma~\ref{lemma:cv-A-1} and $\Qpool_{1} \cup V_{\activelabel,h,1}\setminus Q_{\succeq f^*,1}[+]= \Qpool_{0} \cup V_{\activelabel,h,0}\setminus Q_{\succeq f^*,0}[+]$; therefore, for any such \(h \neq f\),
the correctness of the conditions follows directly from the validity of \(\ctx_0\).
Consequently, when verifying these conditions it suffices to check the case \(h = f = f_{\mathrm{cur}}\).

\begin{enumerate}
\item[3(a)] Observe that
\(
(\rho_{\mathrm{loc},f,1},\, \ctx_1.C_f^{\,z},\, \kappa_{f,1},\, \delta_{f,1},\, R_{\mathrm{stmt},f,1})
= \bigl( \{\fp(f) \mapsto \overline{p},\; \re(f) \mapsto \overline{r}\},\; I,\; \{\},\; \delta_{f,0},\; s \bigr).
\)
Since \(\overline{p} \cap \overline{r} =\emptyset\), \(\rho_{\mathrm{loc},f,1}\) is injective. Note that \(R_{\mathrm{stmt},f,1} = s_f\), so \(R_{\mathrm{stmt},f,1}\) is the suffix of \(s_f\). It is easy to get $\dom(\rho_{\mathrm{loc},f,1}) \cap \rho_{\mathrm{glob}} =\emptyset$. 
Moreover, \(\dom(\rho_{\mathrm{loc},f,1}) \setminus \bigl( \rho_{\mathrm{loc},f,1}(\fp(f)) \cup \rho_{\mathrm{loc},f,1}(\re(f)) \bigr) = \emptyset\); hence
\[
\dom(\rho_{\mathrm{loc},f,1}) \setminus \bigl( \rho_{\mathrm{loc},f,1}(\fp(f)) \cup \rho_{\mathrm{loc},f,1}(\re(f)) \bigr)
\subseteq \{\, x \in \mathsf{Var}(G_{f}) \mid \delta_{f,1}(\node(x)) > 0 \;\lor\; x.\flag \neq 0 \,\}.
\]
By Lemma~\ref{lem:static-wf}, 
for every node \(u \in \mathsf{Node}(G_{f})\),
\[
\Outd_{G_{f}}(u) = \contuses(s_f, \var(u)).
\]
Together with \(R_{\mathrm{stmt},f,1} = s_f\) and \(\delta_{f,1} = \Outd_{G_{f}}\), 
this shows that Condition~3(a)(2) holds for \(\delta_{f,1}\).

Finally, since \(\ctx_1.C_f^{\,z} = I\), \(\dom(\kappa_{f,1}) = \{\}\) and $\rho_{\mathrm{loc},f,1}=\{\fp(f) \mapsto \overline{p},\; \re(f) \mapsto \overline{r}\}$, the cleanup circuit \(\kappa_{f,1}\) is trivially valid.
Therefore, the tuple \((\rho_{\mathrm{loc},f,1},\, \ctx_1.C_f^{\,z},\, \kappa_{f,1},\, \delta_{f,1}, R_{\mathrm{stmt},f,1})\) is valid
for $V_{\activelabel,f,1}$, $\Qpool_1 \cup (V_{\activelabel,\succeq f,1} \setminus Q_{\succeq f^*}[+])$, $V_{\alloc, f,1}$, $z$ and every quantum state \(\ket{\varphi}\) with \(\ket{\varphi} \models \ket{0}_{\ctx_1.A_f \setminus \rho_{\mathrm{loc},f,1}(\fp(f))}\). 

\item[3(b)]
  Since \(\ctx_1.C_f = I\), for any \(\ket{\varphi}\) with
  \(\ket{\varphi} \models \ket{0}_{\ctx_1.A_f \setminus \rho_{\mathrm{loc},f,1}(\fp(f))}\),
  we have \(\ctx_1.C_f^{\,k}\ket{\varphi} = \ket{\varphi}\).
  Consequently,
  \[
    \ctx_1.C_f^{\,k}\ket{\varphi}
    \models \ket{\varphi}_{\ctx_1.A_f \setminus \bigl( V_{\activelabel,f,1} \cup \rho_{\mathrm{loc},f,1}(\fp(f)) \bigr)}
  \]
  holds trivially, since  the reduced state of \(\ket{\varphi}\) on any
  subsystem equals itself.

\item[3(c)] The following items collectively verify the sub-conditions of
  Condition~3(c) (allocation consistency).

  From the construction of \(\ctx_1\) we have
  \(V_{\activelabel,\succeq f,1} = V_{\activelabel,1} = \emptyset\) and \(\cod(\rho_{\mathrm{glob}})\) is unchanged.
  If \(f \in \SCC(g)\),
  \(V_{\activelabel,\succeq f^*,1} = V_{\activelabel,\succeq g^*,0}\),
  \(\cod(\rho_{\succeq f^*,1}) = \cod(\rho_{\mathrm{loc},\succeq g^*,0})\),
  \(\cod(\rho_{\mathrm{loc},f,1}) = \rho_{\mathrm{loc},f,1}(\fp(f) \cup \re(f)) \subseteq \cod(\rho_{\mathrm{loc},g,0})\), $ \rho_{\mathrm{loc},f,1}(\re(f)) \subseteq   V_{\activelabel,g,0}$, 
  and \(cv_1 = cv_0\).
  If \(f \notin \SCC(g)\), \(V_{\activelabel,\succeq f^*,1} = \emptyset\),
  \(\cod(\rho_{\mathrm{loc},\succeq f^*,1}) = \cod(\rho_{\mathrm{loc},f,1}) = \rho_{\mathrm{loc},f,1}(\fp(f) \cup \re(f))\) and
  \(cv_1 = \Qpool_{1} \subseteq \Qpool_{0} \setminus \Qpool_{\mathrm{temp}}\).

\begin{itemize} 
    \item
      Since \(V_{\activelabel,1} = \emptyset\) and
      \(\cod(\rho_{\mathrm{loc},f,1}) = \rho_{\mathrm{loc},f,1}(\fp(f) \cup \re(f)) \subseteq \cod(\rho_{\mathrm{loc},g,0}) \subseteq \ctx_0.A_g\),
      \[
        \ctx_1.A_f \setminus \rho_{\mathrm{loc},f,1}(\fp(f) \cup \re(f))
        \subseteq cv_1.
      \]
      If \(f \in \SCC(g)\), then \(cv_1 = cv_0\).
      The validity of \(\ctx_0\) gives 
      \[
        cv_0 \cap \bigl( V_{\activelabel,\succeq g^*,0} \cup \cod(\rho_{\succeq g^*,0}) \cup \cod(\rho_{\mathrm{glob}}) \bigr) = \emptyset,
        \qquad \ctx_0.A_g \cap \rho_{\mathrm{glob}} = \emptyset, 
      \]
      and therefore \(\ctx_1.A_f \cap \cod(\rho_{\mathrm{glob}}) = \emptyset\) and
      \[
        \bigl( \ctx_1.A_f \setminus \rho_{\mathrm{loc},f,1}(\fp(f) \cup \{\re(f)\}) \bigr)
        \cap \bigl( V_{\activelabel,[f^*, f),1} \cup \cod(\rho_{[f^*, f),1}) \bigr)
        = \emptyset .
      \]
      If \(f \notin \SCC(g)\), then \(cv_1 \subseteq \Qpool_{0} \subseteq cv_0\) and
      \(V_{\activelabel,[f^*, f),1} \cup \cod(\rho_{[f^*, f),1}) = \emptyset\), so the conditions hold trivially.

    \item Since \(V_{\activelabel,g,0}=V_{\activelabel,g,1}\) and \(\mathrm{pred}(f)=g\), the inclusion \(\rho_{\mathrm{loc},f,1}(\re(f)) \subseteq V_{\activelabel,g,0}\) immediately gives \(\rho_{\mathrm{loc},f,1}(\re(f)) \subseteq V_{\activelabel,\mathrm{pred}(f),1}\). 
    From \(V_{\activelabel,f,1} = \emptyset\) we immediately obtain
      \(\rho_{\mathrm{loc},f,1}(\fp(f)) \cap V_{\activelabel,f,1} = \emptyset\).
      The definition \(\rho_{\mathrm{loc},f,1} = \{\overline{\tau x} \mapsto \overline{p},\; v \mapsto r\}\) implies
      \(\fp(f) \cup \re(f) \subseteq \dom(\rho_{\mathrm{loc},f,1})\) and
      \(\cod(\rho_{\mathrm{loc},f,1}) \setminus \rho_{\mathrm{loc},f,1}(\overline{\tau x}) \setminus \rho_{\mathrm{loc},f,1}(v) = \emptyset\);
      hence
      \[
      \base\bigl( \cod(\rho_{\mathrm{loc},f,1}) \setminus \rho_{\mathrm{loc},f,1}(\overline{\tau x}) \setminus \rho_{\mathrm{loc},f,1}(v) \bigr)
      = \emptyset \subseteq \base(V_{\alloc,f}).
      \]

    %   For the return variable we check that
    %   \[
    %     \rho_{f,1}(\re(f))\cap\bigl(ucv_{[f^{*},f),1}\cup
    %       \rho_{f^{*},1}(\fp(f^{*}))\cup\cod(\rho_{\mathrm{glob},1})\bigr)=\emptyset .
    %   \]
    %   \begin{itemize}
    %     \item If \(f\in\SCC(g)\), then Assumption~\ref{hyp_fun} directly
    %       supplies this equality.
    %     \item If \(f\notin\SCC(g)\), then Lemma~\ref{lem:fstar-fixed}
    %       gives \(f^{*}=f\); consequently \(ucv_{[f^{*},f),1}=\emptyset\).
    %       Moreover, injectivity of \(\rho_{f,1}\) and
    %       \(\fp(f)\cap\{\re(f)\}=\emptyset\) imply
    %       \(\rho_{f,1}(\re(f))\cap\rho_{f,1}(\fp(f))=\emptyset\).
    %       Finally, \(\rho_{f,1}(\re(f))\in\xi_0\) yields
    %       \(\rho_{f,1}(\re(f))\cap\cod(\rho_{\mathrm{glob},1})=\emptyset\).
    %       The three empty intersections together establish the required
    %       equality.
    %   \end{itemize}

    % \item
    %   If \(f^{*}\neq f\), then necessarily \(f\in\SCC(g)\).
    %   Using Assumption~\ref{hyp_fun} and the construction of \(ctx_1\),
    %   \[
    %     \rho_{f,1}(\fp(f))\subseteq ucv_{g,0}\subseteq ucv_{[f^{*},f),1},
    %     \qquad
    %     \rho_{f,1}(\re(f))\in\cod(\rho_{g,0})\subseteq\cod(\rho_{[f^{*},f),1}).
    %   \]

  \end{itemize}

\end{enumerate}

\item \textbf{Condition~4.}
For any \(h \in \Fun(\St_1) \cap \SCC(f)\), 
as in the previous item, if \(h \neq f\) then \(f \in \SCC(g)\),
\(h \in \Fun(\St_0) \cap \SCC(g)\), and Lemma~\ref{lemma:cv-A-1} yields
\(\ctx_1.A_h = \ctx_0.A_h\) and further gives
\(Q_{\succeq f^*,1}[+] = Q_{\succeq g^*,0}[+]\) and \(Gar_{h,1} = Gar_{h,0}\).
In this step only the compilation state of \(f\) is altered;
moreover, the stack and the domain of target declarations are extended
monotonically (\(\Fun(\St_0) \subseteq \Fun(\St_1)\)).
When \(f \in \SCC(g)\) the premise also gives \(\D_{\SCC(f)} = \D_{\SCC(g)}\).
Consequently, for these conditions the required
properties for \(h \neq f\) follow directly from the validity of \(\ctx_0\);
in the remainder of the proof we concentrate on the case \(h = f\)
and will only discuss the other cases when extra care is needed. 
    \begin{enumerate}
        \item [4(a)] 
      \begin{itemize}
        \item \(\cod(\rho_{\mathrm{loc},f,1}) \setminus \rho_{\mathrm{loc},f,1}(\overline{\tau x}) \setminus \rho_{\mathrm{loc},f,1}(v) = \emptyset\); hence
\(\cod(\rho_{\mathrm{loc},f,1}) \setminus \rho_{\mathrm{loc},f,1}(\overline{\tau x}) \setminus \rho_{\mathrm{loc},f,1}(v) \subseteq V_{\activelabel,f,1}\). 
\item The condition \(h \in \Theta^{-1}(\dom(\D_1))\setminus \{f\}\), \(\cod(\rho_{\mathrm{loc},h}) \cap \cod(\rho_{\mathrm{loc},[f^*,h]}) =\emptyset\), follows directly from the validity of \(\ctx_0\). 
    \item  We must prove
\[
  \forall\, q \in V_{\activelabel,\succeq f^*,1} \setminus (V_{\activelabel,f} \cup  \rho_{\mathrm{loc},h,1}(\re(h))),\;  \var(\idx(q)) \neq \emptyset. 
\]

If \(f \notin \SCC(g)\), Lemma~\ref{lem:fstar-fixed} gives \(f^{*}=f\);
hence the condition is vacuously
satisfied. 
If \(f \in \SCC(g)\), then \(f^{*} = g^{*}\).  Besides what is already 
provided by the validity of \(\ctx_0\) for the functions different from 
\(g\), we additionally need to verify the property for \(g\) itself:
\[
  \forall\, q \in V_{\activelabel,g,1} \setminus (V_{\activelabel,f} \cup  \rho_{\mathrm{loc},h,1}(\re(h))),\; \var(\idx(q)) \neq \emptyset.  
\]
Assumption~\ref{hpy:fun_2} supplies the same condition for \(\ctx_0\):
\(\forall\, q \in V_{\activelabel,g,0} \setminus (\rho_{\mathrm{loc},g,0}(\re(g)) \cap \overline{r}),\; \var(\idx(q)) \neq \emptyset\). Since $(V_{\activelabel,f} \cup  \rho_{\mathrm{loc},h,1}(\re(h))) =\overline{r}$, 
together with construction of \(\ctx_1\) leaves both \(V_{\activelabel,g}\) 
unchanged; therefore
\(V_{\activelabel,g,1} \setminus (V_{\activelabel,f} \cup  \rho_{\mathrm{loc},h,1}(\re(h))) 
= V_{\activelabel,g,0} \setminus (\rho_{\mathrm{loc},g,0}(\re(g)) \cap \overline{r}) \), 
and the desired property follows immediately. 

    %   Otherwise (i.e., \(f\in\SCC(g)\)), we need prove that $Q_{\succeq f^*,1}[+] \cap ucv_g=\emptyset$.  
    %   By the assumption, we have for any
    %   \(q\in ucv_{\succeq f^{*},1}\setminus ucv_{f,1}\subseteq
    %   ctx_0.ucv_{\succeq f^{*}}\cup ctx_0.ucv_g\),
    %   the validity of \(ctx_0\) (and the extra condition
    %   \(\forall q\in ctx_0.ucv_g,\;\base(q)\neq q\) from
    %   Assumption~\ref{hyp_fun}) immediately yields \(\base(q)\neq q\). 

     Since \(V_{\activelabel,f,1} = \emptyset\) and $\rho_{\mathrm{loc},f,1}(\re(f))=\overline{r} \subseteq V_{\activelabel,g,0}$; hence the condition   
\[ 
  \forall\, q \in V_{\activelabel,f,1} \cup \rho_{\mathrm{loc},f,1}(\re(f)),\;
  \var(\idx(q)) = \emptyset \;\Longrightarrow\; \base[q] = q
\]
holds from the validity of $\ctx_0$.  
  \item We need to prove
  \[
  \forall\, q \in \ctx.V_{\activelabel,\succeq f^*} \cap \rho_{\mathrm{loc},f,1}(\re(f)),\;
  \var(\idx(q)) = \emptyset \;\Longrightarrow\; q \in \rho_{\mathrm{loc},h,1}(\re(h)). 
\]

If \(f \notin \SCC(g)\), Lemma~\ref{lem:fstar-fixed} gives \(f^* = f\) and the conclusion holds immediately.
If \(f \in \SCC(g)\), then \(f^* = g^*\).  
Note that \(\rho_{\mathrm{loc},f,1}(\re(f)) = \overline{r}\). Assumption~\ref{hpy:fun_2} states two conditions: 
\(
\forall\, q \in \ctx_0.V_{\activelabel,g} \setminus (\rho_{\mathrm{loc},g,0}(\re(g)) \cap \overline{r}),\quad \var(\idx(q)) \neq \emptyset,
\)
and \(\overline{r} \subseteq V_{\activelabel,g,0}\). These conditions imply that for any \(q \in \overline{r}\), if \(\var(\idx(q)) = \emptyset\), then \(q \in \rho_{\mathrm{loc},g,0}(\re(g))\).  
  Consequently the conclusion follows from the validity of \(\ctx_0\).  
   \end{itemize}  
  \item[4(b)]
  We need to prove the following three conditions:
  \begin{enumerate}
    \item \(V_{\activelabel,f,1}\setminus Gar_{f,1}\subseteq\cod(\rho_{\mathrm{loc},f,1})\);
    \item \(Q_{\succeq f^*,1}[+] \cap V_{\activelabel,h,1} = \emptyset\) for all \(h \neq f\);
    \item \(Q_{\succeq f^*,1}[+] \cap V_{\activelabel,f,1}\neq\emptyset
          \;\Longrightarrow\;
          \exists\,x\in\dom(\rho_{\mathrm{loc},f,1}),\; x.\flag=2 \;\wedge\;
          \rho_{\mathrm{loc},f,1}(x)\in V_{\activelabel,f,1} \;\wedge\; Gar_{f,1}\subseteq V_{\activelabel,f,1}\).
  \end{enumerate}

  The first condition follows immediately from \(V_{\activelabel,f,1}=\emptyset\).

  For the second condition, if \(f\notin\SCC(g)\), Lemma~\ref{lem:fstar-fixed}
  gives \(f^{*}=f\); thus \(\SCC(f)\cap\Fun(\St_1)=\{f\}\) and the condition
  is vacuously satisfied for all \(h \neq f\).
  If \(f\in\SCC(g)\), then \(f^{*}=g^{*}\) and, in addition to what is
  already guaranteed by the validity of \(\ctx_0\) for functions different
  from \(g\), we must also verify the property for \(g\) itself, i.e.\
  \(Q_{\succeq f^*,1}[+] \cap V_{\activelabel,g,1} = \emptyset\). 
  Assumption~\ref{hpy:fun_2} yields \(Q_{\succeq g^*,0}[+] \cap V_{\activelabel,g,0} = \emptyset\).
  Since \(Q_{\succeq g^*,0} = Q_{\succeq f^*,1}\) and
  \(V_{\activelabel,g,0} = V_{\activelabel,g,1}\), the condition holds.

  The third condition is vacuously true since \(V_{\activelabel,f,1}=\emptyset\)
  makes the antecedent false.

      \item[4(c)]
        Since  \(\ctx_1.C_f^{\,z} = I\), we have
        \(\anc(\ctx_1.C_f^{\,z}) = \emptyset \subseteq \ctx_1.A_f\).

\item[4(d)]
        From \(\ctx_1.C_f^{\,z} = I\) it follows that
        \(\anc(\ctx_1.C_f^{\,z}) = \emptyset\); consequently
        \[
          \anc(\ctx_1.C_f^{\,z}) \setminus \bigl( \anc(F_{\succeq f_{\mathrm{cur}}^*}^k) \cup \anc(F_{\succeq f_{\mathrm{cur}}^*}^k)[+] \bigr)
          \cap \ctx_1.B = \emptyset .
        \] 
\item[4(e)] We verify \(\var(\idx(\ctx_1.A_{f^*})) \subseteq E_{\args,f^*,1} = E_{\args,f,1}\).
  \begin{itemize}
    \item If \(f \notin \SCC(g)\), then \(f^* = f\).
          Here \(\ctx_1.A_f = \Qpool_{1} \cup \{\overline{p}, r\} \subseteq \Qpool_{0} \setminus \Qpool_{\mathrm{temp}}\).
          By definition \(\Qpool_{\mathrm{temp}} = \{\, q \in \Qpool_{0} \mid \base[q] \neq q \,\}\);
          hence every variable \(q\) in \(\Qpool_{0} \setminus \Qpool_{\mathrm{temp}}\) satisfies \(\base[q] = q\). For such a base variable the index term is a constant (or \(0\)), so \(\var(\idx(q)) = \emptyset\).   
          Consequently \(\var(\idx(\ctx_1.A_f)) = \emptyset \subseteq E_{\args,f,1}\).
    \item If \(f \in \SCC(g)\), then \(f^* = g^*\).
          We have \(\ctx_1.A_{g^*} = \ctx_0.A_{g^*}\),
          \(E_{\args,g^*,1} = E_{\args,g^*,0}\), and
          \(E_{\args,f,1} = E_{\args,g,0}\).
          The validity of \(\ctx_0\) gives \(\var(\idx(\ctx_0.A_{g^*})) \subseteq E_{\args,g^*,0}\);
          therefore \(\var(\idx(\ctx_1.A_{f^*})) \subseteq E_{\args,f^*,1} = E_{\args,f,1}\).
  \end{itemize} 

 \item[4(f)]
        Since \(\anc(\ctx_1.C_f^{\,z}) = \emptyset\), we directly obtain
        \[
          \base\bigl(\anc(\ctx_1.C_f^{\,z})\bigr)
          \subseteq V_{\alloc,\succeq f^*} \cup \base\bigl(\anc(F_{\succeq f^*}^z)\bigr).
        \]

\item[4(g)]
        By Lemma~\ref{lem:static-wf}, \(\Succ(f) = \Call(s_f)\).
        Since \(R_{\mathrm{stmt},f,1} = s_f\),
        \[
          (\Succ(f) \setminus \Call(R_{\mathrm{stmt},f,1})) \cap \SCC(f)
          = \emptyset \subseteq \Fun(\St_1).
        \]
         
    \end{enumerate}
      \end{enumerate}
      
      Hence \((\ctx_1, (\ket{\phi'}, z, \D'_{\SCC(f)})) \in \mathcal{V}\).
\end{proof}

\begin{lemma}\label{lemma:step_2}
   Under Notation~\ref{nota:fun} and Assumptions~\ref{hyp:fun_1},~\ref{hpy:fun_2} and~\ref{hpy:fun_3},  there exists \(\ket{\phi'}: \mathcal{H}\)
such that
\[
(\ctx_3, (\ket{\phi'}, z, \D'_{\SCC(f)})) \in \mathcal{V}. 
\]
and \(\Theta(f)^{z+1}\approx_{\ctx_4.Gar_f} f^{z+1}\), where \(\Theta(f)\) is defined as
\[
F_f\bigl(\getC(\overline{x}), \rho_{\mathrm{loc},3}(\getQ(\overline{x})), \rho_{\mathrm{loc},3}(v), \ctx_3.E_{\args,f}\bigr) \Leftarrow \ctx_3.C_f. 
\] 
\end{lemma}

\begin{proof}
By the definition of function equivalence, for \(\Theta(f)^{z+1}\approx_{\ctx_4.Gar_f} f^{z+1}\) we must show that for any classical states \(\sigma_0,\sigma\) and variables \(y,\overline{u}\) such that  
\[
\langle y \gets f^{\,z+1}(\overline{u}), \sigma_0 \rangle \rightarrow^{*} \langle \downarrow, \sigma \rangle,
\]
and for any quantum state \(\ket{\psi}\), lists \(\overline{q},\overline{t},\overline{\alpha}\) with  
\(\vert  \overline{q} \vert = \vert \overline{\tau y} \vert\),  
\(\vert \overline{t} \vert = \vert y \vert\),  
satisfying  
\[
\ket{\psi}_{\cod(\rho_{\mathrm{glob}})} = \ket{\sigma_0(\dom(\rho_{\mathrm{glob}}))}, \quad 
\ket{\psi}_{\overline{q}} = \ket{\sigma_0(\getQ(\overline{u}))},\quad 
\ket{\psi}_{\overline{t}} = \ket{0},\quad 
\ket{\psi}_{\anc\bigl(F_f(\getC(\overline{u}),\overline{q},\overline{t},\overline{\iota})\bigr)} = \ket{0}, 
\]
and such that there exists a bijection from  
\(\anc(F_f^{z+1}(\getC(\overline{x}), \rho_{\mathrm{loc},3}(\getQ(\overline{x})), \rho_{\mathrm{loc},3}(v), \ctx_3.E_{\args,f}))\) to  
\(\anc(F_f^{z+1}(\getC(\overline{u}), \overline{q}, \overline{t}, \overline{\alpha}))\),
the following reduction holds. There exists some state $\ket{\theta}$, 
\[
\bigl\langle F_f^{\,z+1}\bigl(\getC(\overline{u}), \overline{q}, \overline{t}, \overline{\alpha}\bigr), \ket{\psi} \bigr\rangle
\rightarrow^{*} \bigl\langle \downarrow,\;
    \ket{\sigma(\dom(\rho_{\mathrm{glob}}))} \otimes 
    \ket{\sigma(y)}_{\overline{t}} 
    \otimes \ket{\theta}_{Gar_{f,4}}
    \otimes \ket{\psi}_{\mathrm{rem}} \bigr\rangle. 
\]

The proof proceeds by examining each compilation stage \(\ctx_1\)–\(\ctx_3\) step by step. Define  
\(\displaystyle 
      \ket{\phi'} = \ket{\sigma_0(\overline{u})}_{\overline{p}} \otimes \ket{0}_{\sys \setminus \overline{p}}.
      \) Note that $\ket{\phi'} \models \ket{0}_{{\ctx_1.A_f\setminus(\rho_{\mathrm{loc},f,1}(\fp(f)))}}$. 
By Lemma~\ref{lemma:step_1}, we get 
\[ (\ctx_1, (\ket{\phi'}, k, \D'_{\SCC(f)})) \in \mathcal{V}. 
\]

 Furthermore, since \(\dom(\rho_1) = \{\overline{p}, \overline{r}\}\), \(\ctx_1.C_f = I\) and \(\rho_{\mathrm{active},1} = \rho_{\mathrm{loc},f,1} \cup \rho_{\mathrm{glob}}\), we obtain 
      \[
      \ctx_1.C_f \ket{\phi'} = \ket{\phi'}
        = \ket{\sigma_0(\overline{u})}_{\overline{p}} \otimes \ket{0}_{r} \otimes \ket{\phi'}_{\mathrm{rem}}
        = \mathcal{E}\bigl(\sigma_0[\overline{ x}\mapsto\overline{u},\;  v \mapsto \overline{0}],\; \rho_{\mathrm{active},1}\bigr) 
          \otimes \ket{\phi'}_{\mathrm{rem}}.
      \]
      Consequently, \(\ctx_1 \approx_{(\ket{\phi'},z)} \sigma_0[\overline{x}\mapsto\overline{u},\; v  \mapsto \overline{0}]\).

   Now  let \(\sigma'\) be such that  
      \(\langle s^{\,z},\; \sigma_0[\overline{x}\mapsto\overline{u},\; v \mapsto \overline{0}] \rangle \rightarrow^{*} \langle \downarrow, \sigma'\rangle\).  
      By the semantics of the function call we have \(\sigma = \sigma_0[y \mapsto \sigma'(v), \dom(\rho_{\mathrm{glob}}) \mapsto \sigma'(\dom(\rho_{\mathrm{glob}}))]\) and thus $\sigma(y)=\sigma'(v)$.
      Applying Theorem~\ref{theo:body} we obtain 
\[
(\ctx_2, (\ket{\phi'}, z, \D'_{\SCC(f)})) \in \mathcal{V}
\quad\text{and}\quad
\ctx_2 \approx_{(\phi', z)} \sigma', 
\]
which indicates that 
\[
      \begin{aligned}
      \bigl\langle \ctx_2.C_f^{\,z},\; \ket{\phi'} \bigr\rangle & \rightarrow^{*} \bigl\langle \downarrow,\; 
         \mathcal{E}(\sigma', \rho_{\mathrm{active},2}) 
         \otimes \ket{\theta}_{V_{\activelabel,f,2} \setminus \cod(\rho_{\mathrm{active},2})} 
         \otimes \bigl(\ket{\phi}\bigr)_{\mathrm{rem}} \bigr\rangle \\
      &= \bigl\langle \downarrow,\; 
         \ket{\sigma'(\dom(\rho_{\mathrm{glob}}))} \otimes  
         \ket{\sigma'(v)}_{\rho_{\mathrm{loc},2}(v)}  
         \otimes \ket{\sigma'(y')}_{\rho_{\mathrm{loc},2}(y')} 
         \otimes \ket{\theta}_{V_{\activelabel,f,2} \setminus \cod(\rho_{\mathrm{active},2})}  
         \otimes \ket{\phi}_{\mathrm{rem}} \bigr\rangle,
      \end{aligned}
      \]
where $y' \in \dom(\rho_{\mathrm{active},2})$ and $y' \neq v$. 
If \(w = \emptyset\), then no variable originates from a node with flag \(1\).
By Lemma~\ref{lem:body-inv}, \(\ctx_2.R_{\mathrm{stmt},f} = \ctx_1.R_{\mathrm{stmt},f} \setminus \{s_f\}\),
and since \(\ctx_1.R_{\mathrm{stmt},f} = s_f\) we obtain \(\ctx_2.R_{\mathrm{stmt},f} = \emptyset\). 
The dependency-graph validity in Condition~3(a) requires
\(\delta_2(u) = \contuses(R_{\mathrm{stmt},f}, u)\) for every node \(u\);
with \(R_{\mathrm{stmt},f} = \emptyset\) this forces every node in \(G_{f}\) to
have out-degree \(0\).
Moreover, \(\dom(\rho_{\mathrm{loc},2}) \subseteq \mathsf{Var}(G_f)\), hence all nodes 
corresponding to variables in \(\dom(\rho_{\mathrm{loc},2})\) also have out-degree \(0\).
By the definition of \(\rho_{\mathrm{active}}\), any variable appearing in
\(\rho_{\mathrm{active}}\) that is not the return variable \(\re(f)\) must
have a flag different from \(0\); together with the assumption that no node
carries flag \(1\), we conclude that for every local variable \(y'\) retained in 
\(\rho_{\mathrm{active}}\), \(y'.\flag = 2\). Thus we have \(\ctx_3 = \ctx_2\). The equation above then becomes:
\begin{align*}
\bigl\langle & \ctx_3.C_f^{\,z},\; \ket{\phi'} \bigr\rangle \\ 
 \rightarrow^{*} \ & \bigl\langle \downarrow,\; 
    \ket{\sigma'(\dom(\rho_{\mathrm{glob}}))} \otimes  
         \ket{\sigma'(v)}_{\rho_{\mathrm{loc},3}(v)} 
         \otimes \ket{\sigma'(y')}_{\rho_{\mathrm{loc},3}(y')} 
         \otimes \ket{\theta}_{V_{\activelabel,f,3} \setminus \cod(\rho_{\mathrm{active},3})} 
         \otimes \ket{\phi}_{\mathrm{rem}} \bigr\rangle\\ 
 = \ & \bigl\langle \downarrow,\;
 \ket{\sigma'(\dom(\rho_{\mathrm{glob}}))} \otimes  
  \ket{\sigma'(v)}_{\rho_{\mathrm{loc},3}(v)} \otimes
  \ket{\theta}_{V_{\activelabel,f,3} \setminus \cod(\rho_{\mathrm{active},3}) \cup \rho_{\mathrm{loc},3}(y')} \otimes
  \ket{\phi}_{\text{rem}} \bigr\rangle \\ 
= \ & \bigl\langle \downarrow,\;
\ket{\sigma'(\dom(\rho_{\mathrm{glob}}))} \otimes   
  \ket{\sigma'(v)}_{\rho_{\mathrm{loc},3}(v)} \otimes
  \ket{\theta}_{Gar_{f,3} \cup \rho_{\mathrm{loc},3}(y')} \otimes
  \ket{\phi}_{\text{rem}} \bigr\rangle \tag{$V_{\activelabel,f,3} \setminus \cod(\rho_{\mathrm{active},3}) \subseteq Gar_{f,3}$}.
\end{align*}

If \(w \neq \emptyset\), then there exists a node with flag \(1\).
By Lemma~\ref{lem:static-wf}, we have
\[
\forall x \in \mathsf{Var}_{\SCC(f)},\; x.\flag \neq 2,
\qquad\text{and}\qquad
\forall n \in \mathsf{Node}(G_f),\; n \neq w \implies n.\flag \neq 1.
\]
Hence \(\lvert w \rvert = 1\).

Recall \(Gar_{f,2} \triangleq \rho_{\mathrm{loc},\succeq f^*,2}(\overline{x'})[+] \cup \rho_{\mathrm{loc}, \succeq f,2}(\overline{x'})\),
where \(\overline{x'}.\flag=2\).
Condition~1 in the validity of \(\ctx_2\) ensures that every function in
\(\Fun(\St_{\succeq f^*,2})\) belongs to \(\SCC(f)\); together with
the fact that the set \(\{\overline{x'}\}\) of flag‑2 variables within
\(\SCC(f)\) is empty, we obtain \(Gar_{f,2} = \emptyset\).

The validity requirement \(V_{\activelabel,f,2} \setminus Gar_{f,2}
\subseteq \cod(\rho_{\mathrm{loc},f,2})\) then simplifies to
\(V_{\activelabel,f,2} \subseteq \cod(\rho_{\mathrm{loc},f,2})\), hence
\(V_{\activelabel,f,2} \setminus \cod(\rho_{\mathrm{loc},f,2}) = \emptyset\).

Similarly, \(R_{\mathrm{stmt},2} = \emptyset\); therefore every local variable
\(y'\) retained in \(\rho_{\mathrm{active}}\) with \(y' \neq v\) satisfies
\(y'.\flag \neq 0\).
Since no variable in this SCC carries flag \(2\), it follows that
\(y'.\flag = 1\) and consequently \(y' = w\).
The equation thus becomes
\[
\bigl\langle \downarrow,\;
  \ket{\sigma'(\dom(\rho_{\mathrm{glob}}))} \otimes  
   \ket{\sigma'(v)}_{\rho_{\mathrm{loc},2}(v)} \otimes
   \ket{\sigma'(w)}_{\rho_{\mathrm{loc},2}(w)} \otimes
   \ket{\phi}_{\text{rem}} \bigr\rangle. 
\]

To apply the properties of the cleanup algorithm, we first establish the following facts.

First, the validity of \(\ctx_2\) gives
\[
\dom(\rho_{\mathrm{loc},2}) \setminus (\overline{x} \cup v) \subseteq
\{\, x' \in \mathsf{Var}(G_{f}) \mid \delta_{f,2}(\node(x')) > 0 \;\lor\; x'.\flag \neq 0 \,\}.  
\]
We have already shown that every node in \(G_{f}\) has out-degree \(0\) and that
no variable in \(\SCC(f)\) carries flag \(2\).  By the same reasoning as for
\(\rho_{\mathrm{active}}\), any variable \(x'\) in the above set with \(x'.\flag \neq 0\)
must have flag \(1\), hence \(\node(x') = w\).  Consequently,
\[
\cod(\rho_{\mathrm{loc},2}) \subseteq \{\rho_{\mathrm{loc},2}(v),\; \rho_{\mathrm{loc},2}(\overline{\tau x}),\; \rho_{\mathrm{loc},2}(w)\},
\]
and therefore
\[
V_{\activelabel,f,2} \cup \cod(\rho_{\mathrm{loc},2}) \subseteq \{\rho_{\mathrm{loc},2}(v),\; \rho_{\mathrm{loc},2}(\overline{\tau x}),\; \rho_{\mathrm{loc},2}(w)\}.
\] 

Second, from the cleanup-circuit correctness (Condition~3(b) in the validity of \(\ctx_2\)), we have:
\begin{itemize}
    \item \((\rho_{\mathrm{loc},2}(\re(f)) \cup \rho_{\mathrm{loc},2}(\fp(f))) \cap \req_{w,2}(\kappa_2(w)^z) = \emptyset\);
    \item \(\qv(\kappa_2(w)^z) \subseteq \ctx_2.A_{f}\) for any \(z\).
\end{itemize}
By the definition of \(\req_{w,2}\), \(\req_{w,2}(\kappa_2(w)) = \qv(\kappa_2(w)) \setminus \rho_2(\operatorname{pred}^*(w))\),
which implies \(\rho_2(w) \notin \req_{w,2}(\kappa_2(w))\).
Moreover, \(V_{\activelabel,f,2} \subseteq \cod(\rho_{\mathrm{loc},2})\) together with the first item gives
\[
\req_{w,2}(\kappa_2(w)) \cap \bigl( V_{\activelabel,f,2} \cup \cod(\rho_{\mathrm{loc},2}) \bigr) = \emptyset, 
\]
and consequently
\[
\req_{w,2}(\kappa_2(w)) \subseteq \ctx_2.A_{f} \setminus (V_{\activelabel,f,2} \cup \cod(\rho_{\mathrm{loc},2})). 
\]

By Condition~3(c) of the validity of \(\ctx_2\), we have
\[
\bigl( \ctx_2.A_{f} \setminus \rho_{\mathrm{loc},2}(\fp(f) \cup \re(f))\bigr)
\cap \bigl( V_{\activelabel,[f^*,f],2} \cup \cod(\rho_{\mathrm{loc},[f^*,f],2})  \bigr) = \emptyset.
\]
Combining this with the inclusion above yields
\[
\req(\kappa_2(w)) \subseteq \ctx_2.A_{f} \setminus (V_{\activelabel,f,2} \cup \cod(\rho_{\mathrm{loc},2})). 
\] 
Then, applying Theorem~\ref{theo:clean} and Theorem~\ref{theo:clean-valid} yields:
\[
\bigl\langle \ctx_3.C_f^{\,z},\; \ket{\phi} \bigr\rangle \rightarrow^{*}  \bigl\langle \downarrow,\;
\ket{\sigma'(\dom(\rho_{\mathrm{glob}}))} \otimes   
   \ket{\sigma'(v)}_{\rho_{\mathrm{loc},3}(v)} \otimes
   \ket{\phi}_{\text{rem}} \bigr\rangle
\]
and
\[
(\ctx_3, (\ket{\phi'}, z, \D'_{\SCC(f)})) \in \mathcal{V},
\]
except for Condition~3(b) for \(f_{\mathrm{cur}}=f\). 

Thus we conclude that
\[
(\ctx_3, (\ket{\phi'}, z, \D'_{\SCC(f)})) \in \mathcal{V},
\]
except for Condition~3(b) for \(f_{\mathrm{cur}}=f\)
 and \[
\bigl\langle \ctx_3.C_f^{\,z},\; \ket{\phi} \bigr\rangle \rightarrow^{*} \bigl\langle \downarrow,\; 
 \ket{\sigma'(\dom(\rho_{\mathrm{glob}}))} \otimes  
  \ket{\sigma'(v)}_{\rho_{\mathrm{loc},3}(v)} \otimes
  \ket{\theta}_{Gar_{f,3} \cup \rho_{\mathrm{loc},3}(y')} \otimes
  \ket{\phi}_{\mathrm{rem}} \bigr\rangle, 
\]
where \(y.\flag=2\). 

\item Note that \(\anc(\ctx_3.C_f^{\,z}) \subseteq \ctx_3.A_f \subseteq \ctx_1.A_f\). 
% By the validity of \(\ctx_1\), we obtain \((\ctx_1.A_f \setminus \rho(\overline{x}\cup v )) \cap (V_{\activelabel,[f^*,f],1} \cup \cod(\rho_{\mathrm{loc},[f^*,f],1})) = \emptyset\). Moreover, since \(\ctx_1.V_{\activelabel,\succeq f} = \emptyset\), it follows that \((\ctx_1.A_f \setminus \rho(\overline{x} \cup v)) \cap (\ctx_1.V_{\activelabel,\all} \cup \rho_{\mathrm{glob}}) = \emptyset\).
 By premises, we have \(\ket{\phi'} \models \ket{0}_{\ctx_1.A_f \setminus \rho_{\mathrm{loc},1}(\overline{x})}\). Therefore \(\ket{\phi'} \models \ket{0}_{\anc(\ctx_3.C_f^{\,z})}\). 
      Consequently,
      \begin{align*}
        &\bigl\langle \ctx_3.C_f^{\,z},\; 
        \ket{\phi'}_{qv(\ctx_3.C_f^{\,z})}  \bigr\rangle \\ 
      =&\bigl\langle \ctx_3.C_f^{\,z},\; 
      \ket{\sigma_0(\dom(\rho_{\mathrm{glob}}))} \otimes  
        \ket{\sigma_0(\overline{u})}_{\rho_{\mathrm{loc},3}(\getQ(\overline{x}))} 
        \otimes \ket{0}_{\rho_{\mathrm{loc},3}(v)} 
        \otimes \ket{0}_{\anc(\ctx_3.C_f^{\,z})} \bigr\rangle  \tag{\(\rho_{\mathrm{loc},3}(\getQ(\overline{x})) = \rho_{\mathrm{loc},1}(\getQ(\overline{x}))= \overline{p}, \rho_{\mathrm{loc},3}(v) \subseteq \rho_{\mathrm{loc},1}(v)=\overline{r}\) } \\
      \rightarrow^{*} & 
        \bigl\langle \downarrow,\;
        \ket{\sigma'(\dom(\rho_{\mathrm{glob}}))} 
          \otimes \ket{\sigma'(v)}_{\rho_{\mathrm{loc},3}(v)}
         \otimes \ket{\theta}_{\rho_{\mathrm{loc},3}(y') \cup Gar_{f,3}} 
         \otimes \ket{0}_{\mathrm{rem}} \bigr\rangle.
      \end{align*}

      Now, by the semantics of function call in \(\text{RQC}^{++}\),  
      consider arbitrary \(\ket{\psi}\), lists \(\overline{q}, \overline{t}, \overline{\alpha}\) with  
      \(\vert \overline{q} \vert =\vert \getQ(\overline{x}) \vert \), \(\vert \overline{t} \vert =\vert v \vert \),  \(\ket{\psi}_{\cod(\rho_{\mathrm{glob}})} = \ket{\sigma_0(\dom(\rho_{\mathrm{glob}}))}\), 
      \(\ket{\psi}_{\overline{q}} = \ket{\sigma_0(\overline{u})}\), \(\ket{\psi}_{\overline{t}} = \ket{0}\),  
      and \(\ket{\psi} \models \ket{0}_{\anc(F_f^{\,z+1}(\getC(\overline{x}),\overline{q},\overline{t},\overline{\alpha}))}\).  
Assume there exists a bijection \(h'\) from \[\anc(F_f^{\,z+1}(\getC(\overline{x}),\rho_{\mathrm{loc},3}(\getQ(\overline{x})), \rho_{\mathrm{loc},3}(v),E_{\args,f}))\] to \(\anc(F_f^{\,z+1}(C,\overline{q},\overline{t},\overline{\alpha}))\). Then we can construct the following mapping. 

Let \(h\) be a mapping from  
\[
X = \qv\bigl(\anc(F_f^{\,z+1}(\getC(\overline{x}),\rho_{\mathrm{loc},3}(\getQ(\overline{x})), \rho_{\mathrm{loc},3}(v),E_{\args,f}))\bigr)
\]
to  
\[
Y = \qv\bigl(\anc(F_f^{\,z+1}(\getC(\overline{u}),\overline{q},\overline{t},\overline{\iota}))\bigr)
\]
such that \(h(\rho_{\mathrm{loc},3}(\getQ(\overline{x}))) = \overline{q}\), \(h(\rho_{\mathrm{loc},3}(v)) = \overline{t}\), and  
\[
\begin{aligned}
 &h(\anc(F_f^{\,z+1}(\getC(\overline{x}),\rho_{\mathrm{loc},3}(\getQ(\overline{x})), \rho_{\mathrm{loc},3}(v),E_{\args,f}))) \\ 
= \ & h'(\anc(F_f^{\,z+1}(\getC(\overline{x}), \rho_{\mathrm{loc},3}(\getQ(\overline{x})), \rho_{\mathrm{loc},3}(v),E_{\args,f}))).
\end{aligned}
\] 
Note that $\anc(F_f^{\,z+1}(\getC(\overline{u}),\overline{q},\overline{t},\overline{\alpha}))$ is disjoint with $\overline{q}$ and $\overline{t}$ by definition and \(h'\) is a bijection.  
Moreover, from the state \(\ket{\psi}\), we have \(\ket{\phi}_{X} = \ket{\psi}_{Y}\).
      Then by Lemma~\ref{lemma:rename} we have the following reduction:
      \[
      \begin{aligned}
      &\bigl\langle F_f^{\,z+1}(\getC(\overline{u}),\overline{q},\overline{t},\overline{\alpha}),\; \ket{\psi} \bigr\rangle \\[2pt]
      = \ &\bigl\langle F_f^{\,z+1}(\getC(\overline{u}),\overline{q},\overline{t},\overline{\alpha}),\; 
        \ket{\sigma_0(\dom(\rho_{\mathrm{glob}}))} \otimes \ket{\sigma_0(\overline{u})}_{\overline{q}} \otimes \ket{0}_{\overline{t}} 
         \otimes \ket{0}_{\anc} \otimes \ket{\psi}_{\mathrm{rem}} \bigr\rangle \\[2pt]
      \rightarrow^{*} \  &\bigl\langle \getC(\overline{x}) \mapsto \getC(\overline{u});\;
                     \getC(E_{\args,f}) \mapsto \getC(\overline{\alpha}); \\ 
                     &\qquad \ctx_3.C_f^{\,z}\bigl[\rho_{\mathrm{loc},3}(\overline{x})\mapsto\overline{q},\; \rho_{\mathrm{loc},3}(v)\mapsto \overline{t},\; 
                     \anc(\ctx_3.C_f^{\,z})[\getQ(E_{\args,f})\mapsto \getQ(\overline{\alpha})]\bigr], \\
                     &\qquad \ket{\sigma_0(\dom(\rho_{\mathrm{glob}}))} \otimes \ket{\sigma_0(\overline{u})}_{\overline{q}} \otimes \ket{0}_{\overline{t}} 
                     \otimes \ket{0}_{\anc} \otimes \ket{\psi}_{\mathrm{rem}} \bigr\rangle \\[2pt]
      \rightarrow^{*} \ &\bigl\langle \ctx_3.C_f^{\,z}\bigl[\getQ(E_{\args,f})\mapsto \getQ(\overline{\alpha}),\; 
                     \anc(\ctx_3.C_f^{\,z})[\getQ(E_{\args,f})\mapsto \getQ(\overline{\alpha})]\bigr], \\
                     &\qquad \sigma_c\bigl[\getC(\overline{\tau x})\mapsto \getC(\overline{u}),\; 
                               \getC(E_{\args,f}) \mapsto \getC(\overline{\alpha})\bigr], \\
                     &\qquad \ket{\sigma_0(\dom(\rho_{\mathrm{glob}}))} \otimes \ket{\sigma_0(\overline{u})}_{\overline{q}} \otimes \ket{0}_{\overline{t}} \otimes \ket{0}_{\anc} \otimes \ket{\psi}_{\mathrm{rem}} \bigr\rangle \\[2pt]
      \rightarrow^{*} \ &\bigl\langle \downarrow\;,  
         \ctx_3.C_f^{\,z} (\sigma_c\bigl[\getC(\overline{x})\mapsto \getC(\overline{u})\bigr],\; \\ 
         & \qquad 
         (\ket{\sigma_0(\dom(\rho_{\mathrm{glob}}))} \otimes \ket{\sigma_0(\overline{u})}_{\rho_{\mathrm{loc},3}(\getQ(\overline{x}))} \otimes \ket{0}_{\rho_{\mathrm{loc},3}(v)} 
         \otimes \ket{0}_{\anc} \otimes \ket{\psi}_{\mathrm{rem}})_{X \rightarrow Y}) \bigr\rangle \\[2pt]
      = \ &\bigl\langle \downarrow,\;
         \ket{\sigma'(\dom(\rho_{\mathrm{glob}}))}  
         \otimes \ket{\sigma'(v)}_{\overline{t}} 
         \otimes \ket{\theta}_{h(\rho_{\mathrm{loc},3}(y') \cup Gar_{f,3})} 
         \otimes \ket{0}_{\anc(\ctx_3.C_f^{\,z})\setminus h(\rho_{\mathrm{loc},3}(y') \cup Gar_{f,3})} 
         \otimes \ket{\psi}_{\mathrm{rem}} \bigr\rangle \\[2pt]
   = \ &\bigl\langle \downarrow,\;
         \ket{\sigma(\dom(\rho_{\mathrm{glob}}))}
         \otimes \ket{\sigma(y)}_{\overline{t}}
         \otimes \ket{\theta}_{h(\rho_{\mathrm{loc},3}(y') \cup Gar_{f,3})} 
         \otimes \ket{0}_{\anc(\ctx_3.C_f^{\,z})\setminus h(\rho_{\mathrm{loc},3}(y') \cup Gar_{f,3})} 
         \otimes \ket{\psi}_{\mathrm{rem}} \bigr\rangle \\[2pt]
      = \ &\bigl\langle \downarrow,\;
         \ket{\sigma(\dom(\rho_{\mathrm{glob}}))}
         \otimes \ket{\sigma(y)}_{\overline{t}} 
         \otimes \ket{\theta}_{h(\rho_{\mathrm{loc},3}(y') \cup Gar_{f,3})} 
         \otimes \ket{\psi}_{\mathrm{rem}} \bigr\rangle.
      \end{aligned} 
      \] 

      In the above, \(\sigma_c\) denotes the classical state during the execution of the compiled code. Since our proof concerns only the quantum part, we omit \(\sigma_c\) in the subsequent derivation steps; its evolution is standard.  
      The renaming \(h\) is applied consistently, and Lemma~\ref{lemma:rename} justifies the variable replacement.

      Note that by the definition $\rho_{\mathrm{loc},3}(y')\subseteq Gar_{f,3}$ and $\ctx_3.Gar_f=\ctx_4.Gar_f$. If there is no \(x'\) with \(x'.\flag=2\), then \(\ctx_4.Gar_f = \rho_{\mathrm{loc},3}(y') \cup Gar_{f,3} = \emptyset\) and the output simplifies to
      \[
      \bigl\langle F_f^{\,z+1}(\getC(\overline{u}),\overline{q},\overline{t},\overline{\iota}),\; \ket{\psi} \bigr\rangle
      \rightarrow^{*} \bigl\langle \downarrow,\;
       \ket{\sigma(\dom(\rho_{\mathrm{glob}}))} 
        \otimes \ket{\sigma(y)}_{\overline{t}} 
        \otimes \ket{\psi}_{\mathrm{rem}} \bigr\rangle .
      \]

      This reduction matches exactly the behaviour of \(f^{\,z+1}\); therefore we have shown  
      \(F_f^{\,z+1} \approx_{\ctx_4.Gar_f} f^{\,z+1}\) under the given hypotheses. 
\end{proof}

\begin{lemma} \label{lemma:valid-4}
    Under Notation~\ref{nota:fun} and Assumptions~\ref{hyp:fun_1},~\ref{hpy:fun_2} and~\ref{hpy:fun_3},  there exists \(\ket{\phi'}: \mathcal{H}\)
such that
\[
(\ctx_4, (\ket{\phi'}, z, \D'_{\SCC(f)})) \in \mathcal{V}. 
\] 
\end{lemma}
\begin{proof}

Since \(\ctx_4\) only adds the declaration of \(\Theta(f)\) to the context of \(\ctx_3\), \(\ctx_4.f_{\mathrm{cur}}=\ctx_3.f_{\mathrm{cur}}=f\), \(\M_3=\M_4\), \(\St_3=\St_4\), $\ctx_4.G_{fun}=\ctx_3.G_{fun}$,  $\ctx_4.R_{\mathrm{stmt}}=R_{\mathrm{stmt},3}=\emptyset$, $\ctx_{4,r}=\ctx_4$ and using the hypothesis \(\ctx_4.\D \approx_{\ctx_4.Gar_{f^*}} \D'_{\SCC(f)}\), the main task reduces to proving that \((\ctx_3, (\ket{\phi'}, k, \D'_{\SCC(f)})) \in \mathcal{V}\) except that Condition~3(b) for \(f\), for some \(\ket{\phi'}: \mathcal{H}\) and \(\Theta(f)^{\,z+1}\approx_{\ctx_4.Gar_{f}} f^{\,z+1}\), where \(\Theta(f)\) is defined as 
\[
F_f\bigl(\getC(\overline{\tau x}), \rho_{\mathrm{loc},3}(\getQ(\overline{x})), \rho_{\mathrm{loc},3}(v), \ctx_3.E_{\args,f}\bigr) \Leftarrow \ctx_3.C_f.
\]
This follows directly from Lemma~\ref{lemma:step_2}.  
Consequently, \((\ctx_4,(\ket{\phi'},k,\D'_{\SCC(f)}))\in\mathcal{V}\). It remains to prove \(\ctx_4.\D \approx_{\ctx_4.Gar_{f^*}} \D'_{\SCC(f)}\). Note that \(\dom(\ctx_4.\D) =\dom(\ctx_3.\D) \cup \Theta(f)\). From the validity of $\ctx_3$, we have \(\ctx_3.\D \approx_{\ctx_{r,3}.Gar_{f^*}} \D'_{\SCC(f)}\). Since \(\ctx_{r,3}=\ctx_4\), we have
  \(\ctx_3.\D \approx_{\ctx_4.Gar_{f^*}} \D'_{\SCC(f)}\). It remains to prove that \(\Theta(f)^{\,z+1}\approx_{\ctx_4.Gar_{f^*}} f^{\,z+1}\). Note that $\ctx_4.Gar_{f}\subseteq \ctx_4.Gar_{f^*}$; from \(\Theta(f)^{\,z+1}\approx_{\ctx_4.Gar_{f}} f^{\,z+1}\), we immediately get \(\Theta(f)^{\,z+1}\approx_{\ctx_4.Gar_{f^*}} f^{\,z+1}\). 
\end{proof}

\paragraph{Auxiliary relation.}
Define a binary relation \(R_2\) on source functions by
\[
f_1 \mathrel{R_2} f_2
\quad\triangleq\quad
f_2 \in \Succ(f_1)\cap \SCC(f_1).
\]
Let
\[
C(f_{\mathrm{cur}})\triangleq \{\, h \mid f_{\mathrm{cur}} \mathrel{R_2^{*}} h \,\},
\]
where \(R_2^{*}\) denotes the reflexive-transitive closure of \(R_2\).

\begin{lemma}
\label{lem:R2-closure}
Assume that \(\ctx'\) satisfies the structural properties of validity and that 
\(\ctx'.f_{\mathrm{cur}} = (\ctx'.f_{\mathrm{cur}})^*\) and
\(
\ctx'.f_{\mathrm{cur}} \in \Fun(\ctx'.\St)\cap \Theta^{-1}(\dom(\ctx'.\D)).
\)
Then
\[
\SCC(\ctx'.f_{\mathrm{cur}})\subseteq \Fun(\ctx'.\St)\cap \Theta^{-1}(\dom(\ctx'.\D)).
\]
\end{lemma}

\begin{proof}
We first establish the following one-step closure property:
\[
h \in \Fun(\ctx'.\St)\cap \SCC(\ctx'.f_{\mathrm{cur}})\cap \Theta^{-1}(\dom(\ctx'.\D))
\;\Longrightarrow\;
\Succ(h)\cap \SCC(h)
\subseteq
\Fun(\ctx'.\St)\cap \Theta^{-1}(\dom(\ctx'.\D)).
\]
Indeed, since \(h \in \Fun(\ctx'.\St)\cap \SCC(\ctx'.f_{\mathrm{cur}})\), 
Condition~4(d) of the validity of \(\ctx'\) implies
\[
(\Succ(h)\setminus \Call(\ctx'.R_{\mathrm{stmt},h}))\cap \SCC(h)
\subseteq
\Fun(\ctx'.\St).
\]
On the other hand, since \(h \in \Theta^{-1}(\dom(\ctx'.\D))\), 
Condition~1(b) of the validity of \(\ctx'\) yields
\(
\ctx'.R_{\mathrm{stmt},h}=\emptyset.
\)
Hence
\[
\Succ(h)\cap \SCC(h)
=
(\Succ(h)\setminus \Call(\ctx'.R_{\mathrm{stmt},h}))\cap \SCC(h)
\subseteq \Fun(\ctx'.\St).
\]

Besides, since \(h \in \Fun(\ctx'.\St)\cap \SCC(\ctx'.f_{\mathrm{cur}})\), we have 
\(\pos(h) \geq \pos((\ctx'.f_{\mathrm{cur}})^*)\), and together with 
\(\ctx'.f_{\mathrm{cur}} = (\ctx'.f_{\mathrm{cur}})^*\) this gives \(\pos(h) \geq \pos(\ctx'.f_{\mathrm{cur}})\).  
Then Condition~1(2) in the validity of \(\ctx'\) further yields 
\(h \in \Theta^{-1}(\dom(\ctx'.\D))\).  Thus 
\[
\Succ(h)\cap \SCC(h) \subseteq \Fun(\ctx'.\St) \cap \Theta^{-1}(\dom(\ctx'.\D)). 
\]

Now we prove the lemma by induction on the length of an \(R_2^*\)-derivation 
from \(f_{\mathrm{cur}} = \ctx'.f_{\mathrm{cur}}\).  
The base case is immediate from the assumption
\[
\ctx'.f_{\mathrm{cur}} \in \Fun(\ctx'.\St)\cap \Theta^{-1}(\dom(\ctx'.\D)).
\]
For the induction step, suppose that \(\ctx'.f_{\mathrm{cur}} \mathrel{R_2^i} h\) and
\(
h \in \Fun(\ctx'.\St)\cap \Theta^{-1}(\dom(\ctx'.\D)),
\)
and let \(g\) satisfy \(h \mathrel{R_2} g\). Then \(h \in \SCC(\ctx'.f_{\mathrm{cur}})\) 
and \(g \in \Succ(h)\cap \SCC(h)\) by the definition of \(R_2\).  
By the one-step closure property above,
\[
g \in \Fun(\ctx'.\St)\cap \Theta^{-1}(\dom(\ctx'.\D)).
\]
Therefore, every \(R_2^{i+1}\)-successor of \(\ctx'.f_{\mathrm{cur}}\) also belongs to 
\(\Fun(\ctx'.\St)\cap \Theta^{-1}(\dom(\ctx'.\D))\).  Hence
\[
\SCC(\ctx'.f_{\mathrm{cur}})\subseteq \Fun(\ctx'.\St)\cap \Theta^{-1}(\dom(\ctx'.\D)).
\]
\end{proof}

\begin{lemma}
\label{lem:SCC-in-stack-domD}
Let \(\ctx'\) be a compilation context.  
Assume that \(\ctx'\) satisfies the structural properties, 
\(\ctx'.f_{\mathrm{cur}} = (\ctx'.f_{\mathrm{cur}})^*\), and
\[
\ctx'.f_{\mathrm{cur}} \in \Fun(\ctx'.\St) \cap \Theta^{-1}(\dom(\ctx'.\D)).
\]
Then
\[
\SCC(\ctx'.f_{\mathrm{cur}}) \subseteq \Fun(\ctx'.\St) \cap \Theta^{-1}(\dom(\ctx'.\D)).
\]
\end{lemma}

\begin{proof}
It suffices to prove
\[
\SCC(\ctx'.f_{\mathrm{cur}}) \subseteq C(\ctx'.f_{\mathrm{cur}}).
\]
Let \(h \in \SCC(\ctx'.f_{\mathrm{cur}})\).  By the definition of strongly connected
components, there exists a path
\[
\ctx'.f_{\mathrm{cur}} = h_0 \to h_1 \to \cdots \to h_n = h
\]
such that \(h_i \in \SCC(\ctx'.f_{\mathrm{cur}})\) for all \(0 \le i \le n\).
Since each edge \(h_i \to h_{i+1}\) stays inside \(\SCC(\ctx'.f_{\mathrm{cur}})\),
we have \(h_i \mathrel{R_2} h_{i+1}\) for all \(0 \le i < n\).
Therefore \(\ctx'.f_{\mathrm{cur}} \mathrel{R_2^*} h\), which means
\(h \in C(\ctx'.f_{\mathrm{cur}})\).  Hence
\[
\SCC(\ctx'.f_{\mathrm{cur}}) \subseteq C(\ctx'.f_{\mathrm{cur}}).
\]
Combining this with Lemma~\ref{lem:R2-closure} yields
\[
\SCC(\ctx'.f_{\mathrm{cur}}) \subseteq \Fun(\ctx'.\St) \cap \Theta^{-1}(\dom(\ctx'.\D)).
\]
\end{proof} 

\begin{lemma}\label{lemma:cv-A-4}
  Under Notation~\ref{nota:fun}, assume that \(\ctx_4\) satisfies the
  structural properties of validity.
  If \(f \in \SCC(g)\), then
  \[
    cv = cv_4 \cup \bigl(Q_{\succeq f^*,4}[+] \cap V_{\activelabel,f,4}\bigr),
  \]
  and for all
  \(h \in \SCC(g) \cap \Fun(\St_4)\), if $\pos(h)< \pos(f)$,
  \(\ctx.A_h = \ctx_4.A_h\); otherwise, \(\ctx_4.A_h \subseteq \ctx.A_h\). 
\end{lemma}

\begin{proof}
  Since \(f \in \SCC(g)\) and $f, g \in \Fun(\St_4)$, we have \(\ctx_4.f^{*} = \ctx_4.g^{*}\). 
  By construction,
  \(\Qpool = \Qpool_{4}\),
  \(\M = \M_4[f \mapsto m_4]\),
  \(\St = \pop(\St_4, (C_{f,4},\, V_{\activelabel,f,4} \cap Q_{\succeq f^*,4}[+]))\),
  \(m = \M_4(g)\), and \(d = (I, V_{\activelabel,4})\). Therefore, we have $\Fun(\St_{\succeq f^*,4}) = \Fun(\St_{\succeq g^{*},4})=\Fun(\St_{\succeq g^{*}})$ and 
  we obtain
  \[
    \rho_{\mathrm{loc},\succeq g^{*}} = \rho_{\mathrm{loc},\succeq f^*,4},
    \qquad
    \rho_{\mathrm{loc},\succeq h,4} = \rho_{\mathrm{loc},\succeq h}
  \]
  for every \(h \in \SCC(f) \cap \Fun(\St_4)\), which also indicates that $Q_{\succeq f^*,4}=Q_{\succeq g^*}$. 

  For any \(h \neq f\), Condition~4(b) of the validity of \(\ctx_4\)
  gives \(V_{\activelabel,h,4} \cap Q_{\succeq f^*,4}[+] = \emptyset\); hence
  \[
    V_{\activelabel,\succeq g^{*}}
    = V_{\activelabel,\succeq f^*,4} \setminus \bigl( V_{\activelabel,f,4} \cap Q_{\succeq f^*,4}[+] \bigr).
  \]
  In particular, \(Q_{\succeq f^*,4}[+] \cap V_{\activelabel,g,4} = \emptyset\).

  Now we compute \(cv\):
  \begin{align*}
    cv
    &= \Qpool \cup \Qpool[+] \cup
       \Bigl( \cod(\rho_{\mathrm{loc}, \succeq g^{*}})[+] \cup V_{\activelabel,\succeq g^{*}}[+]
              \setminus \bigl( Q_{\succeq g^{*}}[+] \cap V_{\activelabel,g} \bigr) \Bigr) \\
    &= \Qpool_{4} \cup \Qpool_{4}[+] \cup
       \Bigl( \cod(\rho_{\mathrm{loc}, \succeq f^*,4})[+] \cup (V_{\activelabel,\succeq f^*,4}[+] \setminus V_{\activelabel,f,4} \cap Q_{\succeq f^*,4}[+])  \Bigr) \tag{since $Q_{\succeq g^{*}}[+] \cap V_{\activelabel,g}=Q_{\succeq f^*,4}[+] \cap V_{\activelabel,g,4} = \emptyset$} \\ 
    &= \Qpool_{4} \cup \Qpool_{4}[+] \cup
       \Bigl( \cod(\rho_{\mathrm{loc},\succeq f^*,4})[+] \cup V_{\activelabel,\succeq f^*,4}[+] \Bigr) \tag{since $Q_{\succeq f^*,4}[+] \subseteq \cod(\rho_{\succeq f^*,4})[+]$} \\ 
    &= \Qpool_{4} \cup \Qpool_{4}[+] \cup
       \Bigl( \bigl( \cod(\rho_{\mathrm{loc},\succeq f^*,4})[+] \cup V_{\activelabel,\succeq f^*,4}[+] \bigr)
              \setminus \bigl( Q_{\succeq f^*,4}[+] \cap V_{\activelabel,f,4} \bigr)  \cup \bigl( Q_{\succeq f^*,4}[+] \cap V_{\activelabel,f,4} \bigr)  \Bigr)  \\
     &= \Qpool_{4} \cup \Qpool_{4}[+] \cup
       \Bigl( \bigl( \cod(\rho_{\mathrm{loc}, \succeq f^*,4})[+] \cup V_{\activelabel,\succeq f^*,4}[+] \bigr)
              \setminus \bigl( Q_{\succeq f^*,4}[+] \cap V_{\activelabel,f,4} \bigr) \Bigr) \cup \bigl( Q_{\succeq f^*,4}[+] \cap V_{\activelabel,f,4} \bigr) \\ 
    &= cv_4 \cup \bigl( Q_{\succeq f^*,4}[+] \cap V_{\activelabel,f,4} \bigr).
  \end{align*}

  Now let \(h \in \SCC(g) \cap \Fun(\St_4)\).
  Observe that \(V_{\activelabel,f} = V_{\activelabel,f,4} \setminus Q_{\succeq f^*,4}[+]\);
  for all \(h\), \(\rho_{\mathrm{loc},h}\) is unchanged.
  By Condition~2 of \(\ctx_4\) we also have
  \(Q_{\succeq f^*,4}[+] \cap \cod(\rho_{\mathrm{loc},h,4}) = \emptyset\). Combining this with  
   \(h \neq f\) and \(V_{\activelabel,h,4} \cap Q_{\succeq f^*,4}[+] = \emptyset\), we obtain:
  \begin{align*}
    V_{\activelabel,\succeq h} \cup \cod(\rho_{\mathrm{loc},h})
    &= \bigl( V_{\activelabel,\succeq h} \setminus
             ( Q_{\succeq f^*,4}[+] \cap V_{\activelabel,f,4} ) \bigr)
       \cup \cod(\rho_{\mathrm{loc},h,4}) \\
    &= \bigl( V_{\activelabel,\succeq h,4} \cup \cod(\rho_{\mathrm{loc},h,4}) \bigr)
       \setminus \bigl( Q_{\succeq f^*,4}[+] \cap V_{\activelabel,f,4} \bigr).
  \end{align*}

  We distinguish two cases.
  \begin{itemize}
    \item If \(\pos(h) \le \pos(f)\), then
          \(Q_{\succeq f^*,4}[+] \cap V_{\activelabel,f,4} \subseteq V_{\activelabel,f,4}
           \subseteq V_{\activelabel,\succeq h,4}\).
          Together with \(cv = cv_4 \cup (Q_{\succeq f^*,4}[+] \cap V_{\activelabel,f,4})\),
          we obtain \(\ctx.A_h = \ctx_4.A_h\).
    \item If \(\pos(h) > \pos(f)\), then
          \(cv = cv_4 \cup (Q_{\succeq f^*,4}[+] \cap V_{\activelabel,f,4})\) implies
          \(\ctx_4.A_h \subseteq \ctx.A_h\).
  \end{itemize}
\end{proof}  

\begin{lemma}\label{lemma:ctx-A}
  Under Notation~\ref{nota:fun} and Assumption~\ref{hpy:fun_2}, assume that \(\ctx_0\) satisfies the
  structural properties of validity.  Then for every
  \(h \in \SCC(g) \cap \Fun(\St_0)\) with \(\pos(h) \le \pos(g)\),
  \[
    \ctx.A_h \subseteq \ctx_0.A_h .
  \]
\end{lemma}

\begin{proof}
  Since \(\ctx_0\) satisfies the structural properties of validity and Assumption~\ref{hpy:fun_2}, the
  reasoning of Lemmas~\ref{lemma:step_1},~\ref{lemma:step_2} and~\ref{lemma:valid-4}
  shows that each of \(\ctx_1\)--\(\ctx_4\) also satisfies the same structural
  properties.

  Lemma~\ref{lem:body-inv} gives
  \(\ctx_2.A_h \subseteq \ctx_1.A_h\) for every
  \(h \in \SCC(f) \cap \Fun(\St_1)\) with \(\pos(h) \le \pos(f)\).
  Lemma~\ref{lemma:clean-inv} gives
  \(\ctx_3.A_h \subseteq \ctx_2.A_h\) for every
  \(h \in \SCC(f) \cap \Fun(\St_2)\) with \(\pos(h) \le \pos(f)\).
  From the construction of \(\ctx_4\) we directly obtain
  \(\ctx_4.A_h \subseteq \ctx_3.A_h\) for every
  \(h \in \SCC(f) \cap \Fun(\St_3)\).
  Moreover, Lemma~\ref{lem:body-inv}, Lemma~\ref{lemma:clean-inv}, and the
  construction together imply
  \(\Fun(\St_1) \subseteq \Fun(\St_2) = \Fun(\St_3)\).
  These facts together yield
  \[
    \ctx_4.A_h \subseteq \ctx_1.A_h
    \quad\text{for all } h \in \SCC(f) \cap \Fun(\St_1)
    \text{ with } \pos(h) \le \pos(f).
  \]

  We now distinguish two cases.

  \noindent
  \textit{Case \(f \in \SCC(g)\).}
  Since Assumption~\ref{hpy:fun_2} holds and \(\ctx_0\) satisfies the structural properties of validity,
  applying Lemma~\ref{lemma:cv-A-1}, we obtain
  \(\ctx_1.A_h = \ctx_0.A_h\) for any $h \in \SCC(g) \cap \Fun(\St_0)$.  
  Since \(\ctx_4\) also satisfies the structural properties,
  Lemma~\ref{lemma:cv-A-4} yields
  \(\ctx.A_h = \ctx_4.A_h\) for every
  \(h \in \SCC(f) \cap \Fun(\St_4)\) with \(\pos(h) \le \pos(f)\).
  By construction, \(\Fun(\St_3) = \Fun(\St_4)\) and
  \(\Fun(\St_0) \subseteq \Fun(\St_1)\).
  Together with \(\pos(g) \le \pos(f)\) and \(\SCC(f) = \SCC(g)\),
  we conclude
  \(\ctx.A_h \subseteq \ctx_0.A_h\) for all
  \(h \in \SCC(g) \cap \Fun(\St_0)\) with \(\pos(h) \le \pos(g)\).

  \noindent
  \textit{Case \(f \notin \SCC(g)\).}
  We first prove \(\ctx_5.A_f \subseteq \ctx_4.A_f\):
  \begin{align*}
    \ctx_5.A_f
    &= \Qpool_{5} \cup \Qpool_{5}[+] \\
    &= \bigl( \Qpool_{4} \cup \cod(\rho_{\mathrm{loc}, \succeq f^*,4}) \cup V_{\activelabel,\succeq f^*,4} \bigr)
       \cup \bigl( \Qpool_{4}[+] \cup \cod(\rho_{\mathrm{loc}, \succeq f^*,4})[+] \cup V_{\activelabel,\succeq f^*,4}[+] \bigr) \\
    &= \Qpool_{4} \cup \Qpool_{4}[+] \cup
       \bigl( \cod(\rho_{\succeq f^*,4})[+] \cup V_{\activelabel,\succeq f^*,4}[+] \bigr)
       \cup \bigl( \cod(\rho_{\mathrm{loc},\succeq f^*,4}) \cup V_{\activelabel,\succeq f^*,4} \bigr) \\
    &= \Qpool_{4} \cup \Qpool_{4}[+] \cup
       \Bigl( \cod(\rho_{\mathrm{loc}, \succeq f^*,4})[+] \cup V_{\activelabel,\succeq f^*,4}[+]
              \setminus \bigl( Q_{\succeq f^*,4}[+] \cap V_{\activelabel,f,4} \bigr) \Bigr) \\ 
       & \qquad \cup \bigl( \cod(\rho_{\mathrm{loc}, \succeq f^*,4}) \cup V_{\activelabel,\succeq f^*,4} \bigr) \\
    &\qquad (\text{since } V_{\activelabel,f,4} \subseteq V_{\activelabel,\succeq f^*,4}) \\
    &= \ctx_4.A_f .
  \end{align*}
  Hence \(\ctx_5.A_f \subseteq \ctx_1.A_f\).
  Observe that \(\ctx_1.A_f = \Qpool_{1} \cup \{\overline{p},\overline{r}\}\) and
  \(
    \ctx.\Qpool = \base(\ctx_5.\Qpool) \cup \Qpool_{\mathrm{temp}}
            = \base(\ctx_5.A_f) \cup \Qpool_{\mathrm{temp}}
            \subseteq \base(\ctx_1.A_f) \cup \Qpool_{\mathrm{temp}}
            = \base(\Qpool_{1} \cup \{\overline{p}, \overline{r}\}) \cup \Qpool_{\mathrm{temp}}
            = \base(\Qpool_{0} \setminus \Qpool_{\mathrm{temp}}) \cup \Qpool_{\mathrm{temp}}
            = \Qpool_{0} \setminus \Qpool_{\mathrm{temp}} \cup \Qpool_{\mathrm{temp}}
            = \Qpool_{0}. 
  \)

  By construction, \(\St = \St_{\prec f,4} = \St_{\prec f,3}\) and
  \(\St_{\prec f,1} = \St_0 \cup \{(g, \ctx_0.d)\}\). By Lemma~\ref{lem:fstar-fixed} we have \(f = f^*\) throughout
\(\ctx_1\)--\(\ctx_4\); with this fact,
Lemmas~\ref{lem:body-inv} and~\ref{lemma:clean-inv} yield
\(\St_{\prec f,3} = \St_{\prec f,1}\); therefore
  \(\St = \St_0 \cup \{(g, \ctx_0.d)\}\).
  Similarly, \(\M = \M_0[g \mapsto \ctx_0.m]\).
  Consequently, the sets \(V_{\activelabel,\succeq h}\), \(V_{\activelabel,\succeq g^*}\),
  \(\rho_{\succeq h}\), \(\rho_{\succeq h^*}\), and \(Q_{\succeq g^*}\)
  are all unchanged from their counterparts in \(\ctx_0\).
  Hence \(\ctx.A_h = \ctx_0.A_h\) for all such \(h\).
\end{proof}

\begin{lemma}
\label{lem:fun-inv}
Under Notation~\ref{nota:fun}, Assumption~\ref{hpy:fun_2}, and the assumption that $\ctx_0$ satisfies the structural properties of validity, the following properties hold. 
\begin{enumerate}
    \item For $\D$,
    \begin{itemize}
        \item $\dom(\D_0) \subseteq \dom(\D)$; 
        \item $\dom(\D) \setminus \dom(\D_0) \subseteq \Theta(\Succ^*(g) \setminus \Fun(\St_0))$; 
        \item $\D\upharpoonright_{\dom(\D_0) \setminus \Theta(\SCC(g))} = \D_0\upharpoonright_{\dom(\D_0) \setminus \Theta(\SCC(g))}$; 
        \item $\D\upharpoonright_{\dom(\D_0) \cap \Theta(\SCC(g))} \approx_{\ctx_r.Gar_{g^*}} \D_0\upharpoonright_{\dom(\D_0) \cap \Theta(\SCC(g))}$; and for any $h \in \SCC(g) \cap \Theta^{-1}(\dom(\D_0))$, $\anc(\D[\Theta(h)]) \subseteq \anc(\D_0[\Theta(h)]) \cup \anc(\D_0[\Theta(h)])[+]$. 
    \end{itemize} 
    \item For $\Fun(\St)$, 
    \begin{itemize}
        \item $\Fun(\St_0) \subseteq \Fun(\St)$;
        \item $\Fun(\St_{\preceq h}) = \Fun(\St_{\preceq h,0})$ for any $h \in \Fun(\St_{\preceq g,0})$;
        \item $\Fun(\St_{\succ g}) \setminus \Fun(\St_{\succ g,0}) \subseteq \Succ^*(\Succ(g) \setminus \Fun(\St_{0})) \cap \SCC(g)$.  
    \end{itemize}
    \item \(\St_{\preceq g^*}=\St_{\preceq g^*,0}\) and \(\M_{\preceq g^*}=\M_{\preceq g^*,0}\)  
    \item For any $h \in \SCC(g) \cap \Fun(\St_0)$, 
     \begin{itemize}
        \item \(\ctx_0.E_{\args,h} \preceq \ctx.E_{\args,h}$;
        \item $\ctx.A_h \subseteq \ctx_0.A_h$ if $\pos(h) \leq \pos (g)$;
        % \item $\rho_{\mathrm{loc},h}=\rho_{\mathrm{loc},h,0}$, $V_{\activelabel,h}=V_{\activelabel,h,0}$ and $V_{\activelabel,[g^*,h]}=V_{\activelabel,[g^*,h],0}$ if $h \leq g$ 
        \item $\dom(\rho_{\mathrm{loc},h}) \subseteq \dom(\rho_{\mathrm{loc},h,0})$ if $h \neq  g$; 
        \item $\rho_{\mathrm{loc},h}(x)=\rho_{\mathrm{loc},h,0}(x)$, if $x \in \dom(\rho_{\mathrm{loc},h,0})\cap \dom(\rho_{\mathrm{loc},h})$ and $\var(\idx(\rho_{\mathrm{loc},h,0}(x)))\neq \emptyset$. 
    \end{itemize}
\end{enumerate}
\end{lemma}
\begin{proof}
    By Lemma~\ref{lemma:ctx-A}, we already have for every
  \(h \in \SCC(g) \cap \Fun(\St_0)\) with \(\pos(h) \le \pos(g)\), \( \ctx.A_h \subseteq \ctx_0.A_h \). 
    Firstly, we prove the required statements for the declaration family \(\D\) (i.e., the first item).
By the construction of the algorithm we have \(\D_0 = \D_1\) and \(\dom(\D_4) = \dom(\D)\),
together with \(\dom(\D_4) = \dom(\D_3) \cup \{\Theta(f)\}\).
Lemmas~\ref{lem:body-inv} and~\ref{lemma:clean-inv} further yield
\(\D_3 = \D_2\) and
\[
\dom(\D_1) \subseteq \dom(\D_2),\qquad
\dom(\D_2) \setminus \dom(\D_1) \subseteq
\Theta\bigl(\operatorname{succ}^*(f) \setminus \Fun(\St_1)\bigr).
\]
From these we obtain \(\dom(\D_0) \subseteq \dom(\D)\) and
\begin{align*}
\dom(\D) \setminus \dom(\D_0)
&\subseteq \Theta\bigl(\operatorname{succ}^*(f) \setminus \Fun(\St_1)\bigr) \cup \{\Theta(f)\} \\
&= \Theta\bigl(\operatorname{succ}^*(f) \setminus (\Fun(\St_0) \cup \{f\})\bigr) \cup \{\Theta(f)\} \\
&\subseteq \Theta\bigl(\operatorname{succ}^*(f) \setminus \Fun(\St_0)\bigr) \cup \{\Theta(f)\} \\
&\subseteq \Theta\bigl(\operatorname{succ}^*(g) \setminus \Fun(\St_0)\bigr) \tag{$f \in \Succ(g)$}. 
\end{align*}

We already have \(\D_0 = \D_1\).
Again by the construction of the algorithm, \(\D_4 \upharpoonright_{\dom(\D_3)} = \D_3\).
For the step from \(\D_4\) to \(\D\) we distinguish two cases:
\begin{itemize}
  \item if \(f \in \SCC(g)\), then \(\D_4 = \D\);
  \item if \(f \notin \SCC(g)\), then \(\D \upharpoonright_{\dom(\D_4) \setminus \SCC(f)} = \D_4\).
\end{itemize}

Lemma~\ref{lem:body-inv} and~\ref{lemma:clean-inv} also give
\[
\D_3 = \D_2,\qquad
\D_2 \upharpoonright_{\dom(\D_1) \setminus \Theta(\SCC(f))} =
\D_1 \upharpoonright_{\dom(\D_1) \setminus \Theta(\SCC(f))},
\]
\[
\D_2 \upharpoonright_{\dom(\D_1) \cap \Theta(\SCC(f))} \approx_{\ctx_2.\Gar_{f^*}}
\D_1 \upharpoonright_{\dom(\D_1) \cap \Theta(\SCC(f))},
\]
and for every \(h \in \SCC(f)\) with \(\Theta(h) \in \dom(\D_1)\),
\[
\anc(\D_2[\Theta(h)]) \subseteq
\anc(\D_1[\Theta(h)]) \cup \anc(\D_1[\Theta(h)])[+].
\]

Putting these facts together, when \(f \in \SCC(g)\) we obtain
\[
\D \upharpoonright_{\dom(\D_0) \setminus \Theta(\SCC(f))} =
\D_0 \upharpoonright_{\dom(\D_0) \setminus \Theta(\SCC(f))},
\]
\[
\D \upharpoonright_{\dom(\D_0) \cap \Theta(\SCC(f))} \approx_{\ctx_2.\Gar_{f^*}}
\D_0 \upharpoonright_{\dom(\D_0) \cap \Theta(\SCC(f))},
\]
and for every \(h \in \SCC(f)\) with \(\Theta(h) \in \dom(\D_0)\),
\[
\anc(\D[\Theta(h)]) \subseteq
\anc(\D_0[\Theta(h)]) \cup \anc(\D_0[\Theta(h)])[+].
\]

Since  \(\SCC(f) = \SCC(g)\), \(\ctx_2.f^* = \ctx_2.g^*\), and for every
\(h \in \SCC(f) = \SCC(g)\) we have
\(\Gar_{h,2} = \Gar_{h,4} = \Gar_h = \ctx_r.\Gar_h\),
the above implies
\[
\D \upharpoonright_{\dom(\D_0) \setminus \Theta(\SCC(g))} =
\D_0 \upharpoonright_{\dom(\D_0) \setminus \Theta(\SCC(g))},
\]
\[
\D \upharpoonright_{\dom(\D_0) \cap \Theta(\SCC(g))} \approx_{\ctx_r.\Gar_{f^*}}
\D_0 \upharpoonright_{\dom(\D_0) \cap \Theta(\SCC(g))},
\]
and for every \(h \in \SCC(g)\) with \(\Theta(h) \in \dom(\D_0)\),
\[
\anc(\D[\Theta(h)]) \subseteq
\anc(\D_0[\Theta(h)]) \cup \anc(\D_0[\Theta(h)])[+].
\]

If \(f \notin \SCC(g)\), then \(\SCC(f) \cap \Theta^{-1}(\dom(\D_0)) = \emptyset\).
In this case we directly obtain
\[
\D \upharpoonright_{\dom(\D_0) \cap \Theta(\SCC(g))} =
\D_0 \upharpoonright_{\dom(\D_0) \cap \Theta(\SCC(g))},\qquad
\D \upharpoonright_{\dom(\D_0) \setminus \Theta(\SCC(g))} =
\D_0 \upharpoonright_{\dom(\D_0) \setminus \Theta(\SCC(g))},
\]
which also entails the required properties. Thus all the claims concerning \(\D\) are established.   

For the remaining claims, if \(f \notin \SCC(g)\),
similarly to the proof of Lemma~\ref{lemma:ctx-A}, we have  \(\St = \St_0 \cup \{(g,\ctx_0.d)\}\) and 
\(\M = \M_0 \cup \{(g,\ctx_0.m)\}\).
Therefore the remaining conditions in items~2,~3 and~4 are immediate when \(f \notin \SCC(g)\).

We now treat the case \(f \in \SCC(g)\).

\noindent\textbf{Item~2 (concerning \(\Fun(\St)\)).}
By construction,
\[
\Fun(\St_1) = \Fun(\push(\push(\St_0,\{(g,\ctx_0.d)\}),f)),\qquad
\Fun(\St_3) = \Fun(\St_4) = \Fun(\St).
\]
Lemmas~\ref{lem:body-inv} and~\ref{lemma:clean-inv} give
\[
\Fun(\St_2) = \Fun(\St_3),\quad
\Fun(\St_{\preceq h,1}) = \Fun(\St_{\preceq h,2}) \text{ for } h \in \Fun(\St_{\preceq f,1}), \quad 
\Fun(\St_{\succ f,1}) \subseteq \Fun(\St_{\succ f,2}),
\]
\[
\Fun(\St_{\succ f,2}) \setminus \Fun(\St_{\succ f,1})
\subseteq \operatorname{succ}^*(\operatorname{succ}(f) \setminus \Fun(\St_1)) \cap \SCC(f).
\]
Moreover, \(\pos(g) < \pos(f)\).  From these facts we obtain
\[
\Fun(\St_0) \subseteq \Fun(\St),\qquad
\Fun(\St_{\preceq h}) = \Fun(\St_{\preceq h,0}) \text{ for } h \in \Fun(\St_{\preceq g,1}) =\Fun(\St_{\preceq g,0}) 
\]
and
\begin{align*}
\Fun(\St_{\succ g}) \setminus \Fun(\St_{\succ g,0})
&\subseteq \{f\} \cup \bigl( \operatorname{succ}^*(\operatorname{succ}(f) \setminus \Fun(\St_1)) \cap \SCC(f) \bigr) \\
&= \{f\} \cup \bigl( \operatorname{succ}^*(\operatorname{succ}(f) \setminus \Fun(\St_0) \setminus \{f\}) \cap \SCC(g) \bigr) \\
&= \operatorname{succ}^*(\operatorname{succ}(g) \setminus \Fun(\St_0)) \cap \SCC(g). \tag{$f \in \Succ(g)$ and $f \notin \Fun(\St_0)$} 
\end{align*}

\noindent\textbf{Item~3.}
The compilation variables involved are unchanged from \(\ctx_0\) to \(\ctx_1\)
and from \(\ctx_3\) to \(\ctx\).  Together with
Lemmas~\ref{lem:body-inv},~\ref{lemma:clean-inv} and $f^*=g^*$, the claim follows.

\noindent\textbf{Item~4.}
Similarly, the steps from \(\ctx_0\) to \(\ctx_1\) and from \(\ctx_3\) to \(\ctx\)
modify only the compilation data of \(f\); since \(f \notin \Fun(\St_0)\),
the variables appearing in these conditions are in fact untouched. Together with the relations among
\(\ctx_1\)–\(\ctx_3\) established by
Lemmas~\ref{lem:body-inv} and~\ref{lemma:clean-inv}, as well as the fact \(\pos(g) < \pos(f)\), the desired property follows. 
\end{proof} 

\begin{lemma}
\label{lem:A_f-disjointness}
Under Notation~\ref{nota:fun}, Assumption~\ref{hpy:fun_2}, and the assumption that $\ctx_0$ satisfies the structural properties of validity, if $f \in \SCC(g)$, then $\ctx.A_f \setminus \rho_{\mathrm{loc},f}(\fp(f)\cup \re(f)) \cap (V_{\activelabel,[f^*,f)} \cup \cod(\rho_{[f^*,f)}))=\emptyset$.
%  and $\cod(\rho_{\mathrm{loc},g}) \cup V_{\activelabel,g} \subseteq (V_{\activelabel,[f^*,f)} \cup \cod(\rho_{[f^*,f)}))$.  
\end{lemma}
\begin{proof}
  Since \(\ctx_0\) satisfies the structural properties of validity and Assumption~\ref{hpy:fun_2}, the
  reasoning of Lemmas~\ref{lemma:step_1},~\ref{lemma:step_2}
  shows that $\ctx_3$ also satisfies the same structural
  properties. Since $\ctx_3.f_{\mathrm{cur}}=f$, we have $\ctx_3.A_f \setminus \rho_{\mathrm{loc},f,3}(\fp(f)\cup \re(f)) \cap (V_{\activelabel,[f^*,f),3} \cup \cod(\rho_{[f^*,f),3}))=\emptyset$. By construction and Lemma~\ref{lemma:cv-A-4}, we have $\ctx.A_f=\ctx_4.A_f$; together with the fact that the compilation states in this condition are all unchanged from $\ctx_3$ to $\ctx$ by the construction, we have $\ctx.A_f \setminus \rho_{\mathrm{loc},f}(\fp(f)\cup \re(f)) \cap (V_{\activelabel,[f^*,f)} \cup \cod(\rho_{[f^*,f)}))=\emptyset$. 
  %  By the construction of $\ctx_1$, we have $\cod(\rho_{\mathrm{loc},g,1}) \cup V_{\activelabel,g,1} \subseteq (V_{\activelabel,[f^*,f),1} \cup \cod(\rho_{[f^*,f),1}))$. 
  % As in the proof of Lemma~\ref{lem:fun-inv}, we have $\cod(\rho_{\mathrm{loc},g}) \cup V_{\activelabel,g}=\cod(\rho_{\mathrm{loc},g,1}) \cup V_{\activelabel,g,1}$ and $(V_{\activelabel,[f^*,f)} \cup \cod(\rho_{[f^*,f)}))=(V_{\activelabel,[f^*,f),1} \cup \cod(\rho_{[f^*,f),1}))$, the desired condition holds. 
\end{proof}

\begin{theorem}
\label{theo_fun}
\label{theo:correctness-fun}
Under Notation~\ref{nota:fun} and Assumptions~\ref{hyp:fun_1} and~\ref{hpy:fun_2}, we have 
\[
(\ctx, (\ket{\phi}, k, \D_{\SCC(g)})) \in \mathcal{V}
\quad \text{and} \quad
\ctx \approx_{(\ket{\phi}, k)} \sigma.
\]
\end{theorem}

\begin{proof}

First of all, 
the validity of \(\ctx_0\) gives, for every
\(h \in \SCC(g) \cap \Fun(\St_{r,0})\),
\begin{enumerate}
    \item \(F_h^{\,k} \approx_{\ctx_{r,0}.Gar_h} h^{k}\);
    \item \(E_{\args,h,r,0}\) is a prefix of \(\args(\D_{\SCC(g)}(h))\);
    \item if \(h \in \Fun(\St_0) \setminus \Theta^{-1}(\dom(\D_{r,0}))\),
          then \(\anc(\Theta(h)^{k}) \subseteq \ctx_{r,0}.A_h\);
    \item \(\ctx_{r,0}.\D \approx_{\ctx_{r,0}.Gar_{g^*}} \D_{\SCC(g)}\),
          and if \(h \in \Theta^{-1}(\dom(\D_{r,0}))\),
          then
          \(\anc(\D_{\SCC(g)}[\Theta(h)]) \subseteq
           \anc(\ctx_{r,0}.\D[\Theta(h)]) \cup
           \anc(\ctx_{r,0}.\D[\Theta(h)])[+]\).
\end{enumerate}

% Since \(f_{\mathrm{cur}} = g = \ctx_0.f_{\mathrm{cur}}\),
% \(\ctx.R_{\mathrm{stmt},g} = \ctx_0.R_{\mathrm{stmt},g}\), we have
% \[
% \textsc{Compile\_Body}(\ctx, \ctx.R_{\mathrm{stmt},g}) =\textsc{Compile\_Body}(\ctx, \ctx_0.R_{\mathrm{stmt},g})
%  = \textsc{Compile\_Body}(\ctx_0, \ctx_0.R_{\mathrm{stmt},g}),
% \]
Since \(\ctx\) is reached from \(\ctx_0\) during the residual compilation of the
  current function \(g\), completing the current function from \(\ctx\) is the
  suffix of completing it from \(\ctx_0\). Hence \(\ctx_r=\ctx_{r,0}\). 
Moreover, \(f \in \Theta^{-1}(\dom(\ctx.\D))\) implies
\[
\Fun(\St_1) \setminus \Theta^{-1}(\dom(\ctx.\D))
 = \Fun(\St_0) \setminus \Theta^{-1}(\dom(\ctx.\D)).
\]

We also need the following relations between \(\ctx_r\) and \(\ctx\), which
hold whenever \(\ctx\) satisfies the structural properties of the
validity predicate.  For every \(h \in \SCC(g) \cap \Fun(\St)\):
\begin{enumerate}
    \item \(\ctx_r.Gar_{h} \subseteq \ctx.Gar_{h}\) (by assumption);
    \item \(E_{\args,h}\) is a prefix of \(E_{\args,h,r}\) (by Lemma~\ref{lem:completion-inv});
    \item if \(h \notin \Theta^{-1}(\dom(\ctx.\D))\),
          then \(\ctx_r.A_h \subseteq \ctx.A_h\) (by Lemma~\ref{lem:completion-inv});
    \item \(\dom(\ctx.\D) \subseteq \dom(\ctx_r.\D)\),
          \(\ctx_r.\D \upharpoonright_{\dom(\ctx.\D)}
           \approx_{\ctx_r.Gar_{g^*}} \ctx.\D\),
          and if \(h \in \Theta^{-1}(\dom(\ctx.\D))\),
          then
          \(\anc(\ctx_r.\D[\Theta(h)]) \subseteq
           \anc(\ctx.\D[\Theta(h)]) \cup \anc(\ctx.\D[\Theta(h)])[+]\)
          (by Lemma~\ref{lem:completion-inv}). 
\end{enumerate}

Note that \(\ctx_0\) satisfies the structural properties of the validity
predicate.  By the Proof Strategy, this is already sufficient --- using
the very same argument that established the validity of \(\ctx_4\) --- to
conclude that \(\ctx_4\) also satisfies the structural properties,
without ever invoking the semantic conditions of \(\ctx_0\) or
Assumption~\ref{hpy:fun_3}.  Likewise, the structural properties of \(\ctx\) follow
directly from those of \(\ctx_4\), exactly as in the subsequent validity
proof from \(\ctx_4\) to \(\ctx\).

Combining the above facts, we obtain for every
\(h \in \SCC(g) \cap \Fun(\St)\):
\begin{enumerate}
    \item \(F_h^{\,k} \approx_{\ctx.Gar_h} h^{k}\);
    \item \(E_{\args,h}\) is a prefix of \(\args(\D_{\SCC(g)}(h))\);
    \item if \(h \in \Fun(\St_1) \setminus \Theta^{-1}(\dom(\ctx.\D))\),
          then \(\anc(F_h^{\,k}) \subseteq \ctx.A_h\);
    \item \(\ctx.\D \approx_{\ctx.Gar_{g^*}} \D_{\SCC(g)}\),
          and if \(h \in \Theta^{-1}(\dom(\ctx.\D))\),
          then
          \(\anc(\D_{\SCC(g)}[\Theta(h)]) \subseteq
           \anc(\ctx.\D[\Theta(h)]) \cup
           \anc(\ctx.\D[\Theta(h)])[+]\).
\end{enumerate}

Now consider two cases.  

\textbf{Case 1: \(f\in \SCC(g)\)}.
Since \(\D'_{\SCC(f)}=\D_{\SCC(g)}\), by the construction of the algorithm, we have $\ctx_4.\D=\ctx.\D$. 
And for any $h \in \Fun(\St_4) \cap \SCC(f)$, we have \(\ctx_{4}.Gar_h = \ctx.Gar_h\) by Lemma~\ref{lemma:Gar-4}, \(\ctx_{4}.A_h = \ctx.A_h\) if $h\notin \dom(\D_4)$ by Lemma~\ref{lemma:cv-A-4} and \(E_{\args,h}=E_{\args,h,4}\) by the construction. 

% By lemma~\ref{lemma:Gar-4}, we have \(ctx_{4}.Gar_h = ctx.Gar_h\) for any $h \in \Fun(\St_4) \cap SCC(f)$.  Besides, $\dom(D_0)=\dom(D_1)$ and $ctx_4.(\Fun(\St)\cup \Theta^{-1}\dom(\D)) \subseteq ctx.(\Fun(\St)\cup \Theta^{-1}\dom(\D))$, $SCC(\curf)\cap ctx.\dom(\D)=SCC(\curf)\cap ctx_4.\dom(\D)$, $\anc(ctx_4.\D[\Theta(h)])=\anc(ctx.\D[\Theta(h)])$.
Therefore, by the above conclusions we obtain for any \(h \in \SCC(f) \cap \Fun(\St_4)\): 
\begin{enumerate}
    \item $F_h^k \approx_{\ctx_4.Gar_h} h^k$;
    \item \(\ctx_4.E_{\args,h}\) is a prefix of \(\args(\D'_{\SCC(f)}(h))\);
    \item If $h \in (\Fun(\St_1) \setminus \Theta^{-1}(\dom(\ctx_4.\D)))$, 
\(\anc(F_h^{k}) \subseteq \ctx_4.A_h\);
    \item \(\ctx_4.\D \approx_{\ctx_4.Gar_{g^*}} \D'_{\SCC(f)}\). For any $h \in \SCC(f)\cap \Theta^{-1}(\dom(\ctx_4.\D))$, $\anc(\D'_{\SCC(f)}[\Theta(h)]) \subseteq \anc(\ctx_4.\D[\Theta(h)]) \cup \anc(\ctx_4.\D[\Theta(h)])[+]$.  
\end{enumerate}
Taking \(z=k\) in Lemma~\ref{lemma:valid-4} yields \(F_f^{k+1}\approx_{\ctx_4.Gar_f}f^{k+1}\) and there exists $\ket{\phi'}$ such that 
\((\ctx_4,(\ket{\phi'}, k,\D_{\SCC(g)}))\in\mathcal{V}\).   

We finally prove the validity of \(\ctx\) for \((\ket{\phi}, k ,\D_{\SCC(g)})\) step by step. Based on the definitions of \(\ctx_5\) and \(\ctx\), we have:
\[
\ctx = \ctx_4\bigl[\St \mapsto \pop(\St_4, (C_f,V_{\activelabel,f})),\; \M \mapsto \M_4[f \mapsto m_4],\; d \mapsto (I,\emptyset),\; m \mapsto \M(g),\; f_{\mathrm{cur}} \mapsto g \bigr].
\]

\begin{enumerate}
    \item  Note that  \(\SCC(g)=\SCC(f), (\D, G_{fun}) = (\D_4, G_{fun_4})\), \(\Fun(\St) = \Fun(\St_4)\) and \(\ctx.R_{\mathrm{stmt},h}=\ctx_4.R_{\mathrm{stmt},h}\) for any $h \in \Fun(\St_4) \cap \Theta^{-1}(\dom(\D_4))$.  Together with $\ctx_{4,r}.Gar_h=\ctx_4.Gar_h=\ctx.Gar_h=\ctx_r.Gar_h$, Conditions~1(1), 1(3)(b), and 1(3)(c) follow directly from the fact that \(\ctx_4, (\ket{\phi'},k,\D_{\SCC(g)})\) is valid.    
    
For Condition~1(2) (stack discipline),
from the construction of \(\ctx\) we recall the key relations with \(\ctx_4\):
\(
\Fun(\St) = \Fun(\St_4),\;
\dom(\M) = \dom(\M_4) \cup \{f\},\;
f_{\mathrm{cur}} = g,
\)
and the stack order between \(g\) and \(f\) satisfies
\(
\Fun(\St_{(g,f)}) = \Fun(\St_{(g,f),0})=\Fun(\St_{\succ g,0})
\) (Lemma~\ref{lem:fun-inv} gives \(\Fun(\St_{\prec f}) = \Fun(\St_{\prec f,0})\)).
Moreover, the target declaration domains are monotone:
\(\dom(\D_0) \subseteq \dom(\D)\) and \(\dom(\D_4) \subseteq \dom(\D)\).

\begin{itemize}
\item[3(a)] \(f_{\mathrm{cur}} \in \Fun(\St)\).
      Since \(g \in \Fun(\St_0)\) (validity of \(\ctx_0\)) and
      Lemma~\ref{lem:fun-inv} gives \(\Fun(\St_{\prec f}) = \Fun(\St_{\prec f,0})\),
      we have \(g \in \Fun(\St_{\prec f,0}) = \Fun(\St_{\prec f})
      \subseteq  \Fun(\St) \subseteq \Fun(\St_4)\).  With \(f_{\mathrm{cur}} = g\) the claim follows.

\item[3(b)] \(\Fun(\St_{\succ f_{\mathrm{cur}}}) \subseteq \Theta^{-1}(\dom(\D)) \cap \SCC(f_{\mathrm{cur}})\).
      Since \(f_{\mathrm{cur}} = g\) and \(\SCC(f_{\mathrm{cur}}) = \SCC(g) = \SCC(f)\), we decompose $\Fun(\St_{\succ g})$ as follows. 
      \[
        \Fun(\St_{\succ g}) =
        \Fun(\St_{(g,f)}) \;\cup\; \Fun(\St_{\succ f,4}) \;\cup\; \{f\}.
      \]
      Now \(\Fun(\St_{(g,f)}) = \Fun(\St_{\succ g,0}) \subseteq
      \Theta^{-1}(\dom(\D_0)) \cap \SCC(g)\) by validity of \(\ctx_0\).
      Likewise, \(\Fun(\St_{\succ f,4}) \subseteq
      \Theta^{-1}(\dom(\D_4)) \cap \SCC(f)\) by validity of \(\ctx_4\).
      Finally, \(f\) itself belongs to \(\Theta^{-1}(\dom(\D))\) since
      its declaration is added during the current compilation step.
      Using \(\dom(\D_0),\dom(\D_4) \subseteq \dom(\D)\) we obtain the
      required inclusion.

\item[3(c)] \(\Fun(\St) \subseteq \dom(\M) \cup \{f_{\mathrm{cur}}\}\).
      The validity of \(\ctx_4\) gives \(\Fun(\St_4) \subseteq \dom(\M_4) \cup \{f\}\)
      (since \(\ctx_4.f_{\mathrm{cur}} = f\)).  Therefore,
      \[
        \Fun(\St) = \Fun(\St_4) \subseteq \dom(\M_4) \cup \{f\}
        \subseteq \dom(\M) \cup \{g\} = \dom(\M) \cup \{f_{\mathrm{cur}}\},
      \]
      where we used \(\dom(\M) = \dom(\M_4) \cup \{f\}\) and \(f_{\mathrm{cur}} = g\).

\item[3(d)] \(\forall\, h \in \Fun(\St),\; \Fun(\St_{\succ h}) \subseteq \{\, j \mid h \mathrel{R^*} j \,\}\)
      (with \(h \mathrel{R^*} j \triangleq j \in \Succ(h)\)).
      The stack modification from \(\ctx_4\) to \(\ctx\) only affects the
      code and clean-up components but leaves the set of function names
      and their order unchanged; hence \(\Fun(\St_{\succ h}) =
      \Fun(\St_{\succ h,4})\) for every \(h\).  The desired inclusion
      therefore follows directly from the validity of \(\ctx_4\).
\end{itemize}
    
    % For Condition~1(3)(a): Since 
    
%     \(f_{\mathrm{cur}}=g=\ctx_0.f_{\mathrm{cur}}\), \(\ctx.R_{\mathrm{stmt},g}=\ctx_0.R_{\mathrm{stmt},g}\) and 
% \begin{align*}
% \textsc{Compile\_Body}(\ctx, \ctx.R_{\mathrm{stmt},g})=\textsc{Compile\_Body}(\ctx, \ctx_0.R_{\mathrm{stmt},g})=\textsc{Compile\_Body}(\ctx_0, \ctx_0.R_{\mathrm{stmt},g}),
% \end{align*} 
Since $\ctx_{r}=\ctx_{r,0}$, Condition~1(3)(a) follows directly from the validity of \(\ctx_0\) with respect to \(\D_{\SCC(g)}\). 

    \item Since we can prove that \(\ctx.A_g = \ctx_4.A_g \subseteq \ctx_1.A_g \) and 
\(\rho_{\mathrm{loc},g}(\fp(g)) = \rho_{\mathrm{loc},g,1}(\fp(g))\).  Hence
\[
  \ket{\phi} \models \ket{0}_{\ctx.A_g \setminus \rho_{\mathrm{loc},g}(\fp(g))}
\]
is directly inherited from the validity of \(\ctx_1\).

Since \(f \in \SCC(g)\), we have \(\ctx.g^* = \ctx_4.f^*\).
Again by Lemma~\ref{lemma:cv-A-4},
\(cv \subseteq cv_4 \cup \bigl(Q_{\succeq f^*,4}[+] \cap V_{\activelabel,f,4}\bigr)\).
From the definition \(\St = \pop(\St_4,(C_f,V_{\activelabel,f}))\) we obtain
\(\rho_{\mathrm{loc},\succeq g^*} = \rho_{\mathrm{loc},\succeq f^*,4}\), and
\(V_{\activelabel,\succeq g^*} = V_{\activelabel,\succeq f^*,4}
   \setminus \bigl(Q_{\succeq f^*,4}[+] \cap V_{\activelabel,f,4}\bigr)\).
The validity of \(\ctx_4\) gives
\(cv_4 \cap \bigl( \rho_{\mathrm{loc}, \succeq f^*,4} \cup V_{\activelabel,\succeq f^*,4} \cup \rho_{\mathrm{glob}}\bigr)
   = \emptyset\);
therefore
\(cv \cap \bigl( \rho_{\mathrm{loc}, \succeq g^*} \cup V_{\activelabel,\succeq g^*} \cup \rho_{\mathrm{glob}}\bigr)
   = \emptyset\)
follows.

Next, by definition \(\rho_{\mathrm{glob}}\), \(\rho_{\mathrm{loc}, \succeq f_{\mathrm{cur}}^*}\) and thus 
\(Q_{\succeq f_{\mathrm{cur}}^*}\) are also unchanged. 
The validity of \(\ctx_4\) supplies
\(Q_{\succeq f_{\mathrm{cur}}^*,4}[+] \cap \bigl(\rho_{\mathrm{loc}, \succeq f_{\mathrm{cur}}^*,4} \cup \rho_{\mathrm{glob}}\bigr)
   = \emptyset\);
hence the condition continues to hold.

We now prove
\(\Qpool \cap \bigl(\Qpool[+] \cup (V_{\activelabel,\succeq f_{\mathrm{cur}}^*})[+] \cup \cod(\rho_{\mathrm{loc}, \succeq f_{\mathrm{cur}}^*})[+]\bigr)
   = \emptyset\).
Observe that \(\Qpool\) remains unchanged, so \(\Qpool[+] = \Qpool_{4}[+]\).
From the construction, \(\cod(\rho_{\mathrm{loc}, \succeq f_{\mathrm{cur}}^*}) = \cod(\rho_{\mathrm{loc},\succeq f_{\mathrm{cur}}^*,4})\).
Only \(V_{\activelabel,f}\) is modified among the \(\Vactive\) components:
\(V_{\activelabel,f} = V_{\activelabel,f,4} \setminus Q_{\succeq f_{\mathrm{cur}}^*,4}[+]\).
Hence \(V_{\activelabel,\succeq f_{\mathrm{cur}}^*} \subseteq V_{\activelabel,\succeq f_{\mathrm{cur}}^*,4}\).
Consequently,
\[
  \Qpool[+] \cup (V_{\activelabel,\succeq f_{\mathrm{cur}}^*})[+] \cup \cod(\rho_{\mathrm{loc}, \succeq f_{\mathrm{cur}}^*})[+]
  \subseteq
  \Qpool_{4}[+] \cup (V_{\activelabel,\succeq f_{\mathrm{cur}}^*,4})[+] \cup \cod(\rho_{\mathrm{loc}, \succeq f_{\mathrm{cur}}^*,4})[+].
\]
The desired disjointness now follows from the corresponding condition in
the validity of \(\ctx_4\).

Finally, since \(\Qpool\) is unchanged, the last required property carries
over automatically from \(\ctx_4\). 

\item Let \(h \in \{g\} \cup \bigl( \SCC(g) \cap \Fun(\St) \setminus \Theta^{-1}(\dom(\D)) \bigr)\). 
Since \(\SCC(g) = \SCC(f)\), \(\Fun(\St) = \Fun(\St_4)\) and
\(\dom(\D) = \dom(\D_4)\), we actually have
\(h \in \bigl( \SCC(f) \cap \Fun(\St_4) \bigr) \setminus \dom(\D_4)\).

By the validity of \(\ctx_4\), all the required conditions hold for such an
\(h\) in the context \(\ctx_4\).  The current step only modifies the
compilation data associated with \(f\), and \(f\) already belongs to
\(\dom(\D)\); therefore \(h \neq f\) and the compilation state of \(h\)
remains unchanged.  In particular, the sets
\(V_{\activelabel,[g^*,h)}\) and \(\cod(\rho_{[g^*,h)})\), as well as
\(\Qpool \cup \bigl(V_{\activelabel,\succeq h} \setminus Q_{\succeq g^*}[+]\bigr)\), are
identical to their counterparts in \(\ctx_4\).

Moreover, for any \(h \in \bigl( \SCC(g) \cap \Fun(\St_4) \bigr)
   \setminus \dom(\D_4)\) we have \(\pos(h) < \pos(f)\).
Lemma~\ref{lemma:cv-A-4} then yields
\(\ctx.A_h = \ctx_4.A_h\) for every such \(h\).
Consequently, Condition~3 follows directly from the validity of \(\ctx_4\).

\item Let \(h \in \bigl(\SCC(f_{\mathrm{cur}}) \cap \Fun(\St)\bigr)\).  Similarly,
\(h \in \bigl(\SCC(f_{\mathrm{cur}}) \cap \Fun(\St_4)\bigr)\).  This step only modifies the
compilation data associated with \(f\).  By Lemma~\ref{lemma:cv-A-4}, 
we have \(\ctx_4.A_h \subseteq \ctx.A_h\) for every such \(h\) with
\(\pos(h) \leq \pos(f)\) (the accessible set can only grow).
Moreover, \(\D_{\SCC(g)} = \D_{\SCC(f)}\) and
\(\Fun(\St) = \Fun(\St_4)\). For \(h \neq f\) all required conditions follow directly from the  validity of \(\ctx_4\); we therefore concentrate on the case \(h = f\) below, and will only discuss the other cases when extra care is needed. 

\begin{itemize}
\item[(a)]
  By definition,
\[
  \cod(\rho_{\mathrm{loc},f}) = \cod(\rho_{\mathrm{loc},f,4}), \qquad
  V_{\activelabel,f} = V_{\activelabel,f,4} \setminus Q_{\succeq f_{\mathrm{cur}}^*,4}[+] .
\]

From the validity of \(\ctx_4\) we obtain
\[
  \cod(\rho_{\mathrm{loc},f}) \setminus \rho_{\mathrm{loc},f}(\fp(f)\cup \re(f))
  = \cod(\rho_{\mathrm{loc},f,4}) \setminus \rho_{\mathrm{loc},f,4}(\fp(f)\cup \re(f))
  \subseteq V_{\activelabel,f,4}.
\]
On the other hand, \(\cod(\rho_{\mathrm{loc},f,4})\cap Q_{\succeq f_{\mathrm{cur}}^*,4}[+]=\emptyset\)
together with the definition of \(V_{\activelabel,f}\) implies
\[
  \cod(\rho_{\mathrm{loc},f}) \setminus \rho_{\mathrm{loc},f}(\fp(f)\cup \re(f)) \subseteq V_{\activelabel,f} .
\]

The validity of \(\ctx_4\) also gives
\[
  V_{\activelabel,f,4} \subseteq Q_{\succeq f,4} \cup Q_{\succeq f_{\mathrm{cur}}^*,4}[+] \cup \cod(\rho_{\mathrm{loc},f,4}) .
\]
From the earlier argument we already know
\[
  \cod(\rho_{\mathrm{loc},f,4}) = \rho_{\mathrm{loc},f,4}(\fp(f)\cup \re(f)) \cup Q_{f,4}.
\]
The local mapping \(\rho_{\mathrm{loc},h}\) is unchanged for every \(h \in \SCC(f)\);
hence \(Q_{\succeq f,4}=Q_{\succeq f}\).
Combining this with the expression for \(V_{\activelabel,f}\) yields
\[
  V_{\activelabel,f} \subseteq Q_{\succeq f,4} \cup \rho_{\mathrm{loc},f,4}(\re(f))
        = Q_{\succeq f} \cup \rho_{\mathrm{loc},f}(\re(f)).
\]

For any \(q \in V_{\activelabel,f} \setminus \rho_{\mathrm{loc},f}(\re(f))\) we thus have
\(q \in Q_{\succeq f}\); by the definition of \(Q_{\succeq f}\), this implies 
\(\var(\idx(q)) \neq \emptyset\).

Moreover, from the construction of \(\ctx_0\) and Lemma~\ref{lem:completion-inv}, it immediately follows that for any \(q \in \rho_{\mathrm{loc},f}(\re(f))\), if \(\var(\idx(q)) = \emptyset\), then \(q \in \rho_{\mathrm{loc},g}(\re(g))\), and vice versa. Since both \(\rho_{\mathrm{loc},g}\) and \(\rho_{\mathrm{loc},f}\) remain unchanged from \(\ctx_4\) to \(\ctx\), and \(\ctx_4\) is valid, the remaining conditions are satisfied.

\item[(b)] 
We verify the three sub‑conditions.

\begin{itemize}
  \item \(V_{\activelabel,h} \setminus Gar_h \subseteq \cod(\rho_{\mathrm{loc},h})\).
    Since \(Q_{\succeq f_{\mathrm{cur}}^*}\) and \(Q_{\succeq h^*}\) are unchanged,
    we have \(Gar_h = Gar_{h,4}\).
    For \(h \neq f\) the sets \(V_{\activelabel,h} = V_{\activelabel,h,4}\) and
    \(\rho_{\mathrm{loc},h} = \rho_{\mathrm{loc},h,4}\) remain unmodified;
    for \(h = f\) we have \(V_{\activelabel,f} \subseteq V_{\activelabel,f,4}\) and
    \(\rho_{\mathrm{loc},f} = \rho_{\mathrm{loc},f,4}\).
    In both cases the required inclusion follows from the
    corresponding property in \(\ctx_4\).

  \item For \(h \neq g\), \(Q_{\succeq g^*}[+] \cap V_{\activelabel,h} = \emptyset\).
    The validity of \(\ctx_4\) already guarantees this for all
    \(h \neq f\).
    The only new obligation after the return is the case \(h = f\):
    we must verify \(Q_{\succeq g^*}[+] \cap V_{\activelabel,f} = \emptyset\).
    By definition \(V_{\activelabel,f} = V_{\activelabel,f,4} \setminus Q_{\succeq f^*,4}[+]\).
    Since \(Q_{\succeq f^*,4} = Q_{\succeq g^*,4}\),
    the condition follows from the validity of \(\ctx_4\).

  \item For \(f_{\mathrm{cur}} = g\):
    \(Q_{\succeq g^*}[+] \cap V_{\activelabel,g} \neq \emptyset
      \;\Longrightarrow\;
      \exists\,x\in\dom(\rho_{\mathrm{loc},g}),\; x.\flag=2 \;\wedge\;
      \rho_{\mathrm{loc},g}(x)\in V_{\activelabel,g} \;\wedge\; Gar_g \subseteq V_{\activelabel,g}\).
    The validity of \(\ctx_4\) provides
    \(Q_{\succeq f^*,4}[+] \cap V_{\activelabel,g,4} = \emptyset\).
    The step from \(\ctx_4\) to \(\ctx\) leaves the sets \(Q_{\succeq f^*}\)
    and \(V_{\activelabel,g}\) unchanged, i.e.\
    \( Q_{\succeq g^*}=Q_{\succeq f^*} = Q_{\succeq f^*,4}\) and
    \(V_{\activelabel,g} = V_{\activelabel,g,4}\).
    Hence the required implication follows directly from the
    validity of \(\ctx_4\).
\end{itemize} 

\item[(c)]
  By Lemma~\ref{lemma:cv-A-4} we have \(\ctx_4.A_f \subseteq \ctx.A_f\).
  The validity of \(\ctx_4\) together with the fact that
  \(\ctx.C_f = \ctx_4.C_f\) immediately yields
  \[
    \anc(\ctx.C_f) = \anc(\ctx_4.C_f)
    \subseteq \ctx_4.A_f \setminus \rho_{\mathrm{loc},f,4}(\fp(f)\cup\re(f))
    \subseteq \ctx.A_f .
  \]

\item[(d)]
  The ancilla sets \(\anc(C_f^k)\) are unchanged since
  \(\ctx.C_f = \ctx_4.C_f\).
  Lemma~\ref{lemma:cv-A-4} gives \(\ctx_4.A_{f^*} = \ctx.A_{f^*}\),
  hence \(\ctx.B = \ctx_4.B\).
  Together with \(\D_{\SCC(g)} = \D_{\SCC(f)}\), the condition
  follows directly from the validity of \(\ctx_4\).

\item[(e)--(g)]
  None of the variables appearing in these conditions is modified
  during this step; therefore they are inherited directly from
  the validity of \(\ctx_4\).

\end{itemize}
\end{enumerate}

Therefore, \(\ctx\) satisfies all validity conditions, i.e., \(\ctx\) is valid. 

\textbf{Case 2: \(f\notin \SCC(g)\).} From the proofs above, under Notation~\ref{nota:fun} and Assumptions~\ref{hyp:fun_1},~\ref{hpy:fun_2} and~\ref{hpy:fun_3}, all
compilation contexts from \(\ctx_1\) to \(\ctx_4\) are valid. 

However, as already noted in the Proof Strategy, the \textbf{structural
properties} within validity --- such as the stack discipline
(Condition~1(2)) and the call closure (Condition~4(g)) --- do not
rely on the full semantic assumptions. More precisely, the same chain of proofs that establishes them requires
only the notation, Assumption~\ref{hyp:fun_1} (minus the semantic conditions in the
validity of \(\ctx_0\)) and Assumption~\ref{hpy:fun_2}; the semantic conditions themselves and
Assumption~\ref{hpy:fun_3} are never used. 
Consequently, under these weaker hypotheses we still obtain that
\(\ctx_4\) satisfies the structural properties. With the structural properties of \(\ctx_4\) in hand, we can now derive
the following key properties.

First, observe that \(\ctx_4.f_{\mathrm{cur}} = f\).  Lemma~\ref{lem:fstar-fixed}
gives \(\ctx_4.f^{*}=f\).  Since \(f \in \Fun(\St_{1})\),
we also have \(f \in \Fun(\St_{4})\) as in the proof of Lemma~\ref{lem:fstar-fixed}.  Moreover, by definition 
\(f \in \Theta^{-1}(\dom(\D_4))\) holds, hence
\(f \in \Fun(\St_4) \cap \Theta^{-1}(\dom(\D_4))\).
Applying Lemma~\ref{lem:SCC-in-stack-domD} yields
\[
  \SCC(f) \subseteq \Fun(\St_{4}) \cap \Theta^{-1}(\dom(\D_4)) .
\]
Together with the stack discipline of \(\ctx_4\), which guarantees
\(\Fun(\St_{\succ f, 4}) \subseteq \SCC(f)\), and the fact \(f^{*}=f\),
we obtain the equality
\[
  \SCC(f) = \Fun(\St_{\succeq f, 4}) ,
\]
and for every \(h \in \Fun(\St_{\succeq f, 4})\) we have
\(h \in \Theta^{-1}(\dom(\D_4))\).  Thus the functions at or above \(f\) in the stack
form exactly one complete SCC, and every one of them already possesses a
declaration in \(\ctx_4.\D\).

With this guarantee we can safely build the required specification
family \(\D'_{\SCC(f)}\): for each function \(h\) in \(\St_{4}\) whose
position is at least that of \(f\), take its declaration from
\(\ctx_4.\D\). 
% and update its argument list to \(args_{h,5}\) (as
% constructed in \(ctx_5\)).  
By construction,
\(\dom(\D'_{\SCC(f)}) = \Theta(\SCC(f))\), and for every
\(h \in \SCC(f)\) the following properties hold:
\begin{itemize}
  \item \(E_{\args,h,4}\) is a prefix of \(\args(\D'_{\SCC(f)}(h))\);
  \item \(\ctx_4.\D \approx_{\ctx_4.\Gar_{f^*}} \D'_{\SCC(f)}\) and for any $h \in \SCC(f) \cap \Theta^{-1}(\dom(\D_4))$, $\anc(\D'_{\SCC(f)}[\Theta(h)]) \subseteq \anc(\D_4[\Theta(h)]) \cup \anc(\D_4[\Theta(h)])[+]$.  
\end{itemize}

We prove by induction on \(z\) that for every \(h \in \SCC(f)\),
the family \(\D'_{\SCC(f)}\) defined above satisfies
\[
F_h^{\,z}\approx_{\ctx_4.\Gar_h} h^{\,z}
\qquad\text{and}\qquad
\anc(F_h^{\,z})\subseteq \ctx_4.A_h .
\]

 \emph{Base case \(z=0\).}  Trivial.

 \emph{Inductive step.}  Suppose the statement holds for some \(z\). Note that by construction and induction hypotheses, \(\D'_{\SCC(f)}\) meets the properties of Assumption~\ref{hpy:fun_3}; 
hence Lemma~\ref{lemma:valid-4} (applied with the current \(z\) and the
SCC \(\SCC(f)\)) yields
\[
F_f^{\,z+1}\approx_{\ctx_4.\Gar_f} f^{\,z+1}
\qquad\text{and}\qquad
(\ctx_4, (\ket{\phi'}, z, \D'_{\SCC(f)})) \in \mathcal{V}
\]
for some state \(\ket{\phi'}\).

Every \(h \in \SCC(f)\) satisfies \(h \in \Fun(\St_{\succ f,4}) \cap \Theta^{-1}(\dom(\D_4))\).
Using Condition~3(a) in the validity of \(\ctx_4\) together with the fact
that \(\ctx_{r,4}=\ctx_4\), we obtain for all such \(h\),
\[
F_h^{\,z+1}\approx_{\ctx_4.\Gar_h} h^{\,z+1}
\qquad\text{and}\qquad
\anc(F_h^{\,z+1})\subseteq \ctx_4.A_h .
\]

Thus the statement holds for \(z+1\).  By induction, it holds for every
depth \(z\), i.e.,
\begin{equation}
    \label{eqn:equiv}
\forall\, h\in \SCC(f):\;
F_h^{\,z}\approx_{\ctx_4.\Gar_h} h^{\,z}
\quad\text{and}\quad
\anc(F_h^{\,z})\subseteq \ctx_4.A_h . 
\end{equation}

Consequently, there are a state \(\ket{\phi'}\) and a depth \(z\) such that
\[
(\ctx_4, (\ket{\phi'}, z, \D'_{\SCC(f)})) \in \mathcal{V}.
\]

This completes the recursive proof.  Finally we prove that \[
(\ctx, (\ket{\phi}, k, \D_{\SCC(g)})) \in \mathcal{V}. 
\]

\noindent 
\begin{itemize}
    \item[(1)] 
By construction, we have \(\St = \St_{\prec f, 4} = \St_{\prec f, 3}\) and
\(\St_{\prec f, 1} = \St_0 \cup \{(g, \ctx_0.d)\}\).  Since \(f^* = f\),
Lemmas~\ref{lem:body-inv} and~\ref{lemma:clean-inv} give
\(\St_{\prec f, 3} = \St_{\prec f, 1}\); hence
\(\St = \St_0 \cup \{(g, \ctx_0.d)\}\).  Similarly,
\(\M = \M_0 \cup \{(g, \ctx_0.m)\}\). 

\textbf{(Declaration consistency).} 
    Since \(\ctx.f_{\mathrm{cur}} = g = \ctx_0.f_{\mathrm{cur}}\), the validity of \(\ctx_0\) yields
    \(\D_{\SCC(g)} \approx_{\ctx_{r,0}} \D_0\) for every
    \(h \in \SCC(g)\) with \(\Theta(h) \in \dom(\D_0)\).

    As in the proof of Lemma~\ref{lem:fun-inv}, we have 
    \(\dom(\D) \setminus \dom(\D_0) \subseteq \Theta(\Succ^*(f) \setminus \Fun(\St_0))\).
    We first show that this implies
    \(\SCC(g) \cap \Theta^{-1}(\dom(\D)) = \SCC(g) \cap \Theta^{-1}(\dom(\D_0))\).
    Indeed, suppose for contradiction that
    \(h \in \SCC(g) \cap \Theta^{-1}(\dom(\D) \setminus \dom(\D_0))\).
    Then \(h \in (\Succ^*(f) \setminus \Fun(\St_0))\) by the above inclusion.  Since
    \(h \in \SCC(g)\), there is a path from \(h\) to \(g\) in the call
    graph; together with the edge \(f \rightarrow h\) this gives a path
    from \(f\) to \(g\).  On the other hand, \(f \in \Succ(g)\) means
    \(g \rightarrow f\).  Hence \(f\) and \(g\) belong to the same SCC,
    contradicting \(f \notin \SCC(g)\) (if \(f \in \SCC(g)\) the
    statement is trivial).  Thus no such \(h\) exists and the two
    intersections coincide.

     Note that \(\SCC(f) \cap \SCC(g) = \emptyset\).  
    Moreover, again as in the proof of Lemma~\ref{lem:fun-inv}, we have 
    \(\D \upharpoonright_{\dom(\D_0) \setminus \Theta(\SCC(f))} = \D_0 \upharpoonright_{\dom(\D_0) \setminus \Theta(\SCC(f))}\). 
  Consequently, 
    \begin{align*}
      &\D \upharpoonright_{\dom(\D_0) \cap \Theta(\SCC(g))} \\ 
    = \ &(\D \upharpoonright_{\dom(\D_0) \setminus \Theta(\SCC(f))}) \upharpoonright_{\dom(\D_0) \cap \Theta(\SCC(g))} \\ 
    = \ &(\D_0 \upharpoonright_{\dom(\D_0) \setminus \Theta(\SCC(f))}) \upharpoonright_{\dom(\D_0) \cap \Theta(\SCC(g))} \\ 
    =\  &\D_0 \upharpoonright_{\dom(\D_0) \cap \Theta(\SCC(g))} .
     \end{align*}
    Thus the restrictions of \(\D\) and \(\D_0\) to \(\dom(\D_0) \cap \Theta(\SCC(g))\) are
    identical.  Together with
    \(\D_{\SCC(g)} \approx_{\ctx_{r,0}} \D_0\),
    \(\SCC(g) \cap \Theta^{-1}(\dom(\D)) = \SCC(g) \cap \Theta^{-1}(\dom(\D_0))\), and
    \(\ctx_{r,0} = \ctx_{r}\), we obtain
    \[
    \D_{\SCC(g)} \approx_{\ctx_{r}} \D .
    \]

    \textbf{(Stack discipline).}
By construction, \(\St = \St_0 \cup \{(g, \ctx_0.d)\}\),
\(\M = \M_0\), and \(f_{\mathrm{cur}} = \ctx_0.f_{\mathrm{cur}}\).
Moreover, Lemma~\ref{lem:fun-inv} also gives that 
\(\dom(\D_0) \subseteq \dom(\D)\).
Therefore all the relevant components coincide with those in \(\ctx_0\),
and the condition follows directly from the validity of \(\ctx_0\).

% \textbf{(Dependency-graph consistency).}
% Since \(G_{\mathrm{fun}} = G_{\mathrm{fun},4}\) and
% \(\Fun(\St) \cup \Theta^{-1}(\dom(\D)) = \Fun(\St_4) \cup \Theta^{-1}(\dom(\D_4))\),
% the set of functions to which this condition applies is unchanged.
% Moreover, the validity of \(\ctx_4\) directly yields the required property
% for every function in \(\Fun(\St_{\succeq f_{\mathrm{cur}}, 4}) \cap \Theta^{-1}(\dom(\D_4))\).
% Hence the condition follows immediately.
    
\textbf{(Equivalence conditions).} 
\begin{itemize}
    \item[(a)] As in Case 1, using \(\ctx_{r,0}=\ctx_r\), Condition~1(3)(a) follows directly
  from the validity of \(\ctx_0\) with respect to \(\D_{\SCC(g)}\).  

\item[(b)]
Since \(\SCC(g) \cap \Theta^{-1}(\dom(\D)) = \SCC(g) \cap \Theta^{-1}(\dom(\D_0))\)
and \(\Fun(\St) = \Fun(\ctx_0.\St)\),
for any \(h \in \SCC(g) \cap \Fun(\St) \cap \Theta^{-1}(\dom(\D))\)
we actually have
\(h \in \SCC(g) \cap \Fun(\St_0) \cap \Theta^{-1}(\dom(\D_0))\).
Moreover \(\M = \M_0 \cup \{(g, \ctx_0.m)\}\) implies
\(R_{\mathrm{stmt},h} = R_{\mathrm{stmt},h,0}\).
Together with \(\ctx_{r,0} = \ctx_{r}\), the condition follows directly
from the validity of \(\ctx_0\).

\item[(c)]
By definition,
\[
  \Theta^{-1}(\dom(\D)) \setminus \Fun(\St)
  = \bigl( \Theta^{-1}(\dom(\D_4)) \setminus \Fun(\St_4) \bigr)
    \cup \SCC(f).
\]
For every \(h\) belonging to the first part on the right‑hand side,
the required property is guaranteed by the validity of \(\ctx_4\):
this uses \(\dom(\D) = \dom(\D_4)\),
\(\D \upharpoonright_{\dom(\D)\setminus \SCC(f)}= \D_4 \upharpoonright_{\dom(\D_4)\setminus \SCC(f)}\),
and \(\SCC(f) \subseteq \Fun(\St_4)\).

It remains to prove that for every \(h \in \SCC(f)\),
\[
\begin{aligned}
  &\forall z,\; \ctx.\D[\Theta(h)]^z \approx_{G'} h^z, \\ 
  & \var(\base(\anc(\ctx.\D[\Theta(h)]))) \;\cup\;
  \var(\idx(\anc(\ctx.\D[\Theta(h)])))
  \subseteq \args(\ctx.\D[\Theta(h)]),
\end{aligned}
\]
where \(G'\) satisfies: if \(\forall g \in \SCC(f),\; X_g = \emptyset\)
then \(G' = \emptyset\); otherwise \(G' = \anc(C_h)\).

By construction of \(\D'_{\SCC(f)}\) we have
\(\D'_{\SCC(f)} = \D_4 \upharpoonright_{\Theta(\SCC(f))}\).
From the equivalence~\eqref{eqn:equiv} proved above, we have 
\(\forall z,\; \ctx_4.\D[\Theta(h)]^z \approx_{\ctx_4.\Gar_{h}} h^z\). The transition from \(\ctx_4\) to \(\ctx_5\) lambda‑lifts the 
 affected functions: the free ancilla variables that already
appear in the function bodies are collected in \(V_{\alloc,\succeq f}\) and promoted to explicit formal parameters; accordingly, every call site (including recursive calls) is
supplied with the corresponding actual arguments.  The
function bodies themselves are otherwise unchanged, and the
lifted variables remain bound to the same quantum registers.
 Hence the updated declarations are semantically equivalent to
 their original versions in \(\ctx_4.\D\). Therefore, the first part
\(\forall z,\; \ctx_4.\D[\Theta(h)]^z \approx_{G'} h^z\). Since \(\ctx_4.\Gar_{h}\) contains only registers arising from
variables with \(\flag = 2\), if no such variable exists in
\(\SCC(f)\) (i.e., \(X_g = \emptyset\) for all \(g \in \SCC(f)\)),
we may indeed take \(G' = \emptyset\). 

We now verify the second part (ancilla base and index inclusion).
By the definition of \(\ctx\),
\[
  \args(\ctx.\D[\Theta(h)]) = \args(\ctx_5.\D[\Theta(h)])
  = E_{\args,h,4} \cup \ctx_4.V_{\alloc,\succeq f},
\]
and
\[
  \anc(\ctx.\D[\Theta(h)]) = \anc(\ctx_5.\D[\Theta(h)])
  = \anc(\ctx_4.\D[\Theta(h)]) \cup \ctx_4.V_{\alloc,\succeq f}
  = \anc(\ctx_4.C_h) \cup \ctx_4.V_{\alloc,\succeq f}.
\]

Since \(\ctx_4\) is valid, for every
\(h \in \SCC(f) \cap \Fun(\St_4) = \SCC(f)\) we have
\[
  \base(\anc(\ctx_4.C_h)) \subseteq \ctx_4.V_{\alloc,\succeq f},\ 
  \var(\idx(\anc(\ctx_4.C_h))) \subseteq \var(\idx(\ctx_4.A_{f^*}))
  \subseteq E_{\args,h,4}.
\]

Putting everything together,
\[
\begin{aligned}
&\var(\base(\anc(\ctx.\D[\Theta(h)]))) \cup
 \var(\idx(\anc(\ctx.\D[\Theta(h)]))) \\
= &\var(\base(\anc(\ctx_4.C_h))) \cup
   \ctx_4.V_{\alloc,\succeq f}\cup
   \var(\idx(\anc(\ctx_4.C_h))) \cup \var(\idx(\ctx_4.V_{\alloc,\succeq f})) \\
\subseteq &\ctx_4.V_{\alloc,\succeq f}\cup E_{\args,h,4}
   = \args(\ctx.\D[\Theta(h)]).
\end{aligned}
\] 
\end{itemize}

    \item[2] The construction above yields
\[
  \St = \St_0 \cup \{(g, \ctx_0.d)\}, \qquad
  \M = \M_0[g \mapsto m_0],
\]
while the current buffer is reset to \(d = (I,\emptyset)\) and therefore
\(m = \M(g) = m_0\).  Consequently every component that appears in the
initial-state consistency condition remains unchanged: $V_{\activelabel,g}=V_{\activelabel,g,0}$, \(\rho_g(\fp(g)) = \rho_{0,g}(\fp(g))\) and 
\begin{align*}
  V_{\activelabel,\succeq f_{\mathrm{cur}}^*} &= \ctx_0.V_{\activelabel,\succeq f_{\mathrm{cur}}^*},\\
  \cod(\rho_{\mathrm{loc}, \succeq f_{\mathrm{cur}}^*}) &= \cod(\ctx_0.\rho_{\mathrm{loc}, \succeq f_{\mathrm{cur}}^*}),\\
  \cod(\rho_{\mathrm{glob}}) &= \cod(\rho_{\mathrm{glob},0}).
\end{align*}
Lemma~\ref{lemma:ctx-A} further gives \(\Qpool = \Qpool_{0}\); together with the
equalities above this implies \(cv = cv_0\).  Hence, 
\[
  cv \cap \bigl( V_{\activelabel,\succeq f_{\mathrm{cur}}^*} \cup \cod(\rho_{\mathrm{loc}, \succeq f_{\mathrm{cur}}^*}) \cup \cod(\rho_{\mathrm{glob}}) \bigr)
  = cv_0 \cap \bigl( \ctx_0.V_{\activelabel,\succeq f_{\mathrm{cur}}^*} \cup \cod(\ctx_0.\rho_{\mathrm{loc}, \succeq f_{\mathrm{cur}}^*}) \cup \cod(\rho_{\mathrm{glob},0}) \bigr)
  = \emptyset,
\]
where the last equality follows from the validity of \(\ctx_0\).
Applying Lemma~\ref{lemma:ctx-A} again we obtain \(\ctx.A_g = \ctx_0.A_g\);
together with \(\rho_g(\fp(g)) = \rho_{0,g}(\fp(g))\) this yields
\[
  \ket{\phi} \models \ket{0}_{\ctx.A_g \setminus \rho_{\mathrm{loc},g}(\fp(g))}
  = \ket{0}_{\ctx_0.A_g \setminus \rho_{\mathrm{loc},g,0}(\fp(g))}.
\]
Similarly, all remaining conditions follow directly from the validity of \(\ctx_0\). 

\item[3] Consider any 
  \(h \in \{g \} \cup \bigl((\Fun(\St) \cap \SCC(g)) \setminus \Theta^{-1}(\dom(\D))\bigr)\).
Since \(\Fun(\St) = \Fun(\St_0)\) and
  \(\Theta^{-1}(\dom(\D)) \cap \SCC(g) = \Theta^{-1}(\dom(\D_0)) \cap \SCC(g)\),
  this set coincides with
  \((\Fun(\St_0) \cap \SCC(g)) \setminus \Theta^{-1}(\dom(\D_0))\).
  Observe that the construction of \(\ctx\) only adds an entry
  for \(g\) to the stack and to \(\M\), while leaving all other
  functions untouched:
  \[
    \St = \St_0 \cup \{(g, \ctx_0.d)\}, \qquad
    \M = \M_0[g \mapsto m_0].
  \]
   Lemma~\ref{lemma:ctx-A} further gives $\Qpool=\Qpool_{0}$ and 
  \(\ctx.A_h = \ctx_0.A_h\) for all \(h \in \Fun(\St_0) \cap \SCC(g) = \Fun(\St) \cap \SCC(g)\).  Consequently, all conditions required
  for \(h\) in the validity definition follow directly from the validity of \(\ctx_0\).

\item[4] Consider any \(h \in \Fun(\St) \cap \SCC(g)\).
  As shown above, the stack, the local mappings, and the persistent
  compilation state of every function are the same as in \(\ctx_0\);
  in particular, \(\Fun(\St) = \Fun(\St_0)\) and
  \(\ctx.A_h = \ctx_0.A_h\).  Therefore all conditions that must be
  satisfied by \(h\)
  are immediate consequences of the validity of \(\ctx_0\).

\end{itemize}
Thus in both cases, we conclude that 
\[
(\ctx, (\ket{\phi}, k, \D_{\SCC(g)})) \in \mathcal{V}.
\]
Finally, if \(f \notin \SCC(g)\), it is straightforward to get \(\ctx \approx_{(\ket{\phi},k)} \sigma_0\) since the compilation state of 
  \(g\) is restored by \(\St = \St_0 \cup \{(g, \ctx_0.d)\}\) and  \(\M = \M_0[g \mapsto m_0]\). Otherwise, \(\ctx \approx_{(\ket{\phi},k)} \sigma_0\) by Lemma~\ref{lem:state-pre}.  
\end{proof}

\subsection{Correctness Proof for Compilation of Statements}
\label{app:proof-com}

This subsection gives a correctness proof of Algorithm~\ref{Com_state} for the compilation of statements. As before, we first introduce some notations and assumptions. 

\begin{notation}\label{note:state_1}
  Suppose we are given a state \(\ket{\phi} \in \mathcal{H}\), a depth \(k \in \mathbb{N}^+\), and a set of procedure declarations \(\D_{\SCC(f_{\mathrm{cur}})}\). Let \(s\) be a source statement, and let \(\sigma_0\), \(\sigma\) be source language states such that \(\langle s^{\,k}, \sigma_0 \rangle \rightarrow_{\Xi}^{*} \langle \downarrow, \sigma \rangle\). Let \(\ctx_0\) be a compilation context, and \(\ctx = \textsc{Compile\_Com}(\K, \ctx_0, s)\).  
\end{notation}

\begin{notation}\label{note:state_2}
Under Notation~\ref{note:state_1}, if $s = \tau x \gets f(\overline{u})$ for some variables $x$, $\overline{u}$ and a function \(f\) with \(f \in \SCC(f_{\mathrm{cur}})\), we further define $\ctx_1$ and $\ctx_2$ as follows:
    \[
\ctx_1 = 
\begin{cases}
\ctx_0[\Qpool_0 \mapsto \Qpool_1,\;  
    \rho_{\mathrm{loc},0} \mapsto \rho_{\mathrm{loc},0}[x \mapsto \overline{p}]\\ 
    \qquad V_{\alloc,0} \mapsto V_{\alloc,0} \cup \base(\overline{p}),  V_{\activelabel,0} \mapsto V_{\activelabel,0} \cup \overline{p}], &x \notin \dom(\rho_{0}); \\
\ctx_0[V_{\activelabel,0} \mapsto V_{\activelabel,0} \cup \rho_{\mathrm{loc},0}(x) \setminus \rho_{\mathrm{loc},0}(\fp(f_{\mathrm{cur}}))], & x \in \dom(\rho_{0}),
\end{cases} 
\]
where $\overline{p}=\pop(\Qpool_0, \vert x \vert)$ and $\Qpool_1 =\Qpool_0 \setminus \overline{p}$. 

Now define \(\ctx_2\): Let $S=\{ q \in (V_{\activelabel,f_{\mathrm{cur}},1}) \setminus \rho_{\mathrm{loc},1}(\{x\} \cap \re(f_{\mathrm{cur}})) \mid \base[q]=q\}$. 

Let \(i = \iota\) and define \(\ctx_2 = \ctx_1[\Bbuf \mapsto \Bbuf_2,\; \St \mapsto \St_2,\; \M \mapsto \M_2, \D \mapsto \D_2]\), where
\begin{align*}
\Bbuf_2 &= \Bbuf_1\bigl[\Eargs \mapsto \Eargs \mathbin{+\!+} [i],\; 
        (C, \Vactive, \rho_{\mathrm{loc}}, \kappa) \mapsto (C, \Vactive, \rho_{\mathrm{loc}}, \kappa)[q \mapsto q[i]]\bigr], \\
\St_2 &= \St_1[(g, \Vactive) \mapsto (g, \Vactive)[q \mapsto q[i]]], \quad \D_2 =\D_1[\Theta(h) \mapsto \D_1[\Theta(h)][q \mapsto q[i]]]\\
\M_2 &= \M_1\bigl[g \mapsto \M_1(g)[\Eargs \mapsto \Eargs \mathbin{+\!+} [i], C \mapsto C[q \mapsto q[i]]], \rho_{\mathrm{loc}} \mapsto \rho_{\mathrm{loc}}[q \mapsto q[i]], \kappa \mapsto \kappa[q \mapsto q[i]]\bigr],\\  
& \quad \text{for all } q \in S, g \in \SCC(f_{\mathrm{cur}}) \cap \Fun(\St_1) \text{and } h \in (\SCC(f_{\mathrm{cur}}) \cap \Theta^{-1}(\dom(\D)));. 
\end{align*}
\end{notation}

\begin{assumption}\label{hpy:com}
In addition to Notations~\ref{note:state_1} and~\ref{note:state_2}, suppose that $s$ is a prefix of \(\ctx_0.R_{\mathrm{stmt}}\), \(\ctx_0 \approx_{(\ket{\phi}, k)} \sigma_0\) and \[(\ctx_0, (\ket{\phi}, k, \D_{\SCC(f_{\mathrm{cur}})})) \in \mathcal{V}.\]  
\end{assumption}

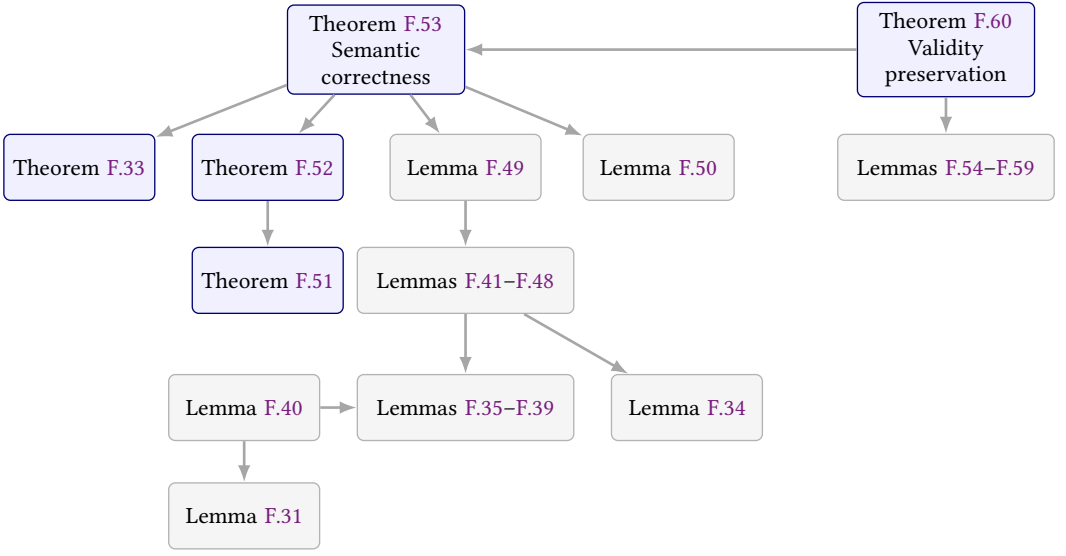
\begin{figure}[t]
\centering
\resizebox{\linewidth}{!}{%
\begin{tikzpicture}[
  x=1cm,
  y=1cm,
  proofnode/.style={
    draw,
    rounded corners=2pt,
    align=center,
    inner xsep=3pt,
    inner ysep=3pt,
    minimum height=7mm,
    font=\scriptsize
  },
  thmnode/.style={proofnode, fill=blue!6, draw=blue!45!black},
  lemmanode/.style={proofnode, fill=gray!8, draw=gray!60},
  dep/.style={-{Latex[length=1.8mm]}, thick, draw=gray!70}
]
\node[thmnode, text width=.12\linewidth] (f52) at (-1.2,3.20)
  {Theorem~\ref{theo_com}\\
   Semantic correctness};
\node[thmnode, text width=.10\linewidth] (f32) at (-4.35,1.95)
  {Theorem~\ref{theo_fun}};
\node[thmnode, text width=.10\linewidth] (f51) at (-2.35,1.95)
  {Theorem~\ref{theo_assgn}};
\node[lemmanode, text width=.10\linewidth] (f48) at (-0.25,1.95)
  {Lemma~\ref{lem:state-valid}};
\node[lemmanode, text width=.10\linewidth] (f49) at (1.80,1.95)
  {Lemma~\ref{lem:state-pre}};
\node[thmnode, text width=.10\linewidth] (f50) at (-2.35,0.75)
  {Theorem~\ref{theo_exp}};
\node[lemmanode, text width=.15\linewidth] (f40to47) at (-0.25,0.75)
  {Lemmas~\ref{lem:state-valid-3.a}--\ref{lem:state-valid-2}};
\node[lemmanode, text width=.15\linewidth] (f34to38) at (-0.25,-0.60)
  {Lemmas~\ref{lem:S-plus}--\ref{lem:state-inv-1}};
\node[lemmanode, text width=.10\linewidth] (f33) at (2.10,-0.60)
  {Lemma~\ref{lemma:rename}};
\node[lemmanode, text width=.10\linewidth] (f39) at (-2.60,-0.60)
  {Lemma~\ref{lem:state-inv-2}};
\node[lemmanode, text width=.10\linewidth] (f30) at (-2.60,-1.75)
  {Lemma~\ref{lem:fun-inv}};

\node[thmnode, text width=.12\linewidth] (f59) at (4.85,3.20)
  {Theorem~\ref{theo:com_valid}\\
   Validity preservation};
\node[lemmanode, text width=.15\linewidth] (f53to58) at (4.85,1.95)
  {Lemmas~\ref{state_valid_2}--\ref{state_valid_4.f}};

\draw[dep] (f52) -- (f32);
\draw[dep] (f52) -- (f51);
\draw[dep] (f52) -- (f48);
\draw[dep] (f52) -- (f49);
\draw[dep] (f51) -- (f50);
\draw[dep] (f48) -- (f40to47);
\draw[dep] (f40to47) -- (f34to38);
\draw[dep] (f40to47) -- (f33);
\draw[dep] (f39) -- (f34to38);
\draw[dep] (f39) -- (f30);
\draw[dep] (f59) -- (f52);
\draw[dep] (f59) -- (f53to58);
\end{tikzpicture}%
}
\caption{Proof dependencies for the correctness of \textsc{Compile\_Com}; arrows point from a result to the lemmas or theorems it depends on.}
\label{fig:stmt-proof-dep}
\Description{A dependency diagram showing the main proof dependencies for statement compilation correctness and validity preservation.}
\end{figure}

As illustrated in Figure~\ref{fig:stmt-proof-dep} (which depicts only the main dependencies), the correctness of Algorithm~\ref{Com_state} comprises Theorems~\ref{theo_com} and~\ref{theo:com_valid}.  

\begin{itemize}
  \item \textbf{Theorem~\ref{theo_com} (Semantic correctness of compiled code).}  
    The proof proceeds by induction on the structure of \(s\).  
    For an assignment statement, the theorem invokes Theorem~\ref{theo_assgn}, which establishes the correctness of compiling assignment statements; Theorem~\ref{theo_assgn} itself depends on Theorem~\ref{theo_exp}, the correctness of compiling expressions. 

    For a recursive call, the proof first relies on Lemmas~\ref{lem:state-valid} and~\ref{lem:state-pre}. These lemmas respectively guarantee the preservation of the validity of the resulting context \(\ctx_2\) and the correctness of the compiled code after renaming the variables in \(\Vactive\). Lemma~\ref{lem:state-valid} checks each condition of the validity predicate individually. The verification of some complex conditions requires Lemmas~\ref{lem:state-valid-3.a}–\ref{lem:state-valid-2}, which in turn repeatedly depend on Lemma~\ref{lemma:rename} for renaming and the auxiliary properties established in Lemmas~\ref{lem:S-plus}–\ref{lem:state-inv-1}. Finally, based on the validity of \(\ctx_2\) and the semantic correctness of the code, we establish the semantic correctness of \(\ctx\) for this case.

    For a call to an uncompiled function (i.e., a function not yet compiled), the proof recursively applies Theorem~\ref{theo_fun}, which states the correctness of function compilation.

  \item \textbf{Theorem~\ref{theo:com_valid} (Preservation of validity states).}  
    The proof relies on several additional properties that Theorem~\ref{theo_com} yields as a byproduct when proving semantic correctness. It then checks each condition of the validity predicate in turn. The proofs of some complex conditions are provided by Lemmas~\ref{state_valid_2}–\ref{state_valid_4.f}.  
\end{itemize}

Similarly, Lemma~\ref{lem:state-inv-2} establishes a relationship between the partial state of the input and output compilation contexts for the function compilation algorithm. This lemma depends on Theorem~\ref{lem:fun-inv} (a theorem analogous to function compilation) and Lemma~\ref{lem:state-inv-1}.

\begin{lemma}
\label{lemma:rename}
Let \(C\) be a quantum circuit, and let \(g : \qv(C) \to \mathcal{Q}\) be
an injection that renames the registers occurring in \(C\).  Denote by
\(g(C)\) the circuit obtained by replacing every register \(x\) in \(C\)
with \(g(x)\).

Let \(\ket{\varphi}\) and \(\ket{\psi}\) be quantum states on a system
that contains at least the registers \(\qv(C) \cup g(\qv(C))\), and
assume that they agree on the corresponding registers under \(g\):
\[
\ket{\varphi}_{g(\qv(C))} = \ket{\psi}_{\qv(C)} .
\]

Then the following equality of reduced states holds:
\[
\bigl( g(C) \ket{\varphi} \bigr)_{g(\qv(C))}
= \bigl( C \ket{\psi} \bigr)_{\qv(C)} .
\]
\end{lemma}

\begin{lemma} \label{lem:S-plus}
  Under Notations~\ref{note:state_1},~\ref{note:state_2}, if
  \(s = \tau x \gets f(\overline{u})\) for some variables \(x\), \(\overline{u}\) and a function \(f\) with \(f \in \SCC(f_{\mathrm{cur}})\), and Condition~2 in the
  validity of \(\ctx_1\) holds, then 
  \[
    S[+] \subseteq V_{\activelabel,f_{\mathrm{cur}},1}[+],\qquad
    S[+] \cap Q_{\succeq f_{\mathrm{cur}}^*}[+] = \emptyset,
  \]
  and consequently \(S[+] \subseteq cv_1\).
\end{lemma}

\begin{proof}
  From \(S \subseteq V_{\activelabel,f_{\mathrm{cur}},1}\) we directly obtain
  \(S[+] \subseteq V_{\activelabel,f_{\mathrm{cur}},1}[+]\) by the definition of higher-index extension. 
  By the definition of \(Q_{\succeq f_{\mathrm{cur}}^*}\), we have
  \(S \cap Q_{\succeq f_{\mathrm{cur}}^*} = \emptyset\).

  Condition~2 in the validity of \(\ctx_1\) gives
  \[
    \cod(\rho_{\mathrm{loc},\succeq f_{\mathrm{cur}}^*}) \cap \bigl( cv_1 \cup Q_{\succeq f_{\mathrm{cur}}^*}[+] \bigr) = \emptyset .
  \]
  The definition of \(cv_1\) yields the inclusion
  \[
    V_{\activelabel,f_{\mathrm{cur}},1}[+] \subseteq cv_1 \cup Q_{\succeq f_{\mathrm{cur}}^*}[+],
  \]
  and together with \(S[+] \subseteq V_{\activelabel,f_{\mathrm{cur}},1}[+]\) we obtain
  \(S[+] \subseteq cv_1 \cup Q_{\succeq f_{\mathrm{cur}}^*}[+]\).

  By the definition of \(Q_{\succeq f_{\mathrm{cur}}^*}\), we have
  \(Q_{\succeq f_{\mathrm{cur}}^*} \subseteq \cod(\rho_{\mathrm{loc},\succeq f_{\mathrm{cur}}^*})\).
  Combining this with the disjointness from Condition~2 and the inclusion
  \(S[+] \subseteq cv_1 \cup Q_{\succeq f_{\mathrm{cur}}^*}[+]\) immediately yields
  \(S[+] \cap Q_{\succeq f_{\mathrm{cur}}^*} = \emptyset\). Together with \(S \cap Q_{\succeq f_{\mathrm{cur}}^*} = \emptyset\), we have \(S[+] \cap Q_{\succeq f_{\mathrm{cur}}^*}[+] = \emptyset\). 

  Finally, from \(S[+] \subseteq cv_1 \cup Q_{\succeq f_{\mathrm{cur}}^*}[+]\) and
  \(S[+] \cap Q_{\succeq f_{\mathrm{cur}}^*}[+] = \emptyset\) we conclude
  \(S[+] \subseteq cv_1\).
\end{proof}

\begin{lemma}\label{lem:Q-plus-holds-ctx2}
  Under Notations~\ref{note:state_1},~\ref{note:state_2}, if
  \(s = \tau x \gets f(\overline{u})\) for some variables \(x\), \(\overline{u}\) and a function \(f\) with \(f \in \SCC(f_{\mathrm{cur}})\), and Condition~2 in the
  validity of \(\ctx_1\) and \(\ctx_0\) holds, then \[ Q_{\succeq f_{\mathrm{cur}}^*,2}[+] \cap V_{\activelabel,f_{\mathrm{cur}},2}= 
Q_{\succeq f_{\mathrm{cur}}^*,1}[+] \cap V_{\activelabel,f_{\mathrm{cur}},1}
= Q_{\succeq f_{\mathrm{cur}}^*,1}[+] \cap V_{\activelabel,f_{\mathrm{cur}},0}. 
\]
\end{lemma}

\begin{proof}
For $\ctx_1$,
if \(x \in \dom(\rho_0)\), then by definition
\[
Q_{\succeq f_{\mathrm{cur}}^*,1}[+] = Q_{\succeq f_{\mathrm{cur}}^*,0}[+],\qquad
V_{\activelabel,f_{\mathrm{cur}},1} = V_{\activelabel,f_{\mathrm{cur}},0},
\]
and the equality holds trivially.

Otherwise, \(x \notin \dom(\rho_0)\) and thus
\(\rho_1(x) = \rho_{\mathrm{loc},1}(x) \subseteq \Qpool_0\).  In this case
\[
\begin{aligned}
Q_{\succeq f_{\mathrm{cur}}^*,1}[+]
   &= Q_{\succeq f_{\mathrm{cur}}^*,0}[+] \cup \rho_{\mathrm{loc},1}(x)[+]
    \subseteq Q_{\succeq f_{\mathrm{cur}}^*,0}[+] \cup \Qpool_0[+],\\
V_{\activelabel,f_{\mathrm{cur}},1}
   &= V_{\activelabel,f_{\mathrm{cur}},0} \cup \rho_{\mathrm{loc},1}(x)
    \subseteq V_{\activelabel,f_{\mathrm{cur}},0} \cup \Qpool_0 .
\end{aligned}
\]
The validity of \(\ctx_0\) supplies the disjointness relations
\[
\Qpool_0 \cap \bigl( \Qpool_0[+] \cup Q_{\succeq f_{\mathrm{cur}}^*,0}[+] \bigr) = \emptyset,
\qquad
\Qpool_0[+] \cap V_{\activelabel,f_{\mathrm{cur}},0} = \emptyset .
\]
Using these, one immediately obtains the desired equality. 

For $\ctx_2$, 
 by the definition of  $Q_{\succeq f_{\mathrm{cur}}^*,2}$, we have
\[
  Q_{\succeq f_{\mathrm{cur}}^*,2}[+]
  \subseteq Q_{\succeq f_{\mathrm{cur}}^*,1}[+] \cup S[i][+].
\]
% By the definition of \(ctx_2\), only \(ucv_{\succeq \curf,1}\) and
% \(\rho_{l,\succeq \curf,1}\) are renamed, while for $\pos(f) < \pos(\curf)$
% \(ucv_{f,1} \cup \cod(\rho_{l,f,1})\) is left unchanged.
% Condition~3.(c) in the validity of \(ctx_1\) gives for
% \[
% \bigl( ctx_1.A_f \setminus \rho_{l,\curf,1}(\fp(\curf) \cup \re(\curf)) \bigr)
% \cap
% \bigl( ucv_{[\curf^*,\curf),1} \cup \cod(\rho_{l,[\curf^*,\curf),1}) \bigr)
% = \emptyset
% \]
% and
% \[
% \rho_{l,\curf,1}(\fp(\curf)) \cap ucv_{\curf,1} = \emptyset .
% \]
% Together with the definition of \(S\), we have
% \[
% S \subseteq ucv_{\curf,1} \setminus \rho_{l,\curf,1}(\re(\curf))
% \subseteq ctx_1.A_f \setminus \rho_{l,\curf,1}(\fp(\curf) \cup \re(\curf)) ,
% \]
% hence
% \[
% S \cap \bigl( ucv_{[\curf^*,\curf),1} \cup \cod(\rho_{l,[\curf^*,\curf),1}) \bigr)
% = \emptyset .
% \]
% % For $f$ such that $\pos(f) > \pos(\curf)$, we have $f\in \Theta^{-1}(\dom(\D_1))$. Condition~3.(c) in the validity of \(ctx_1\) gives for such $f$, $\re(f)\cup\fp(f) \not\subseteq \dom(\rho_{f,1})$ and thus $ctx_1.A_f \cap  \bigl( ucv_{[\curf^*, f),1} \cup \cod(\rho_{[\curf^*,f),1}) \bigr)$. So we have \[
% % S \cap \bigl( ucv_{\succ \curf,1} \cup \cod(\rho_{\succ \curf,1}) \bigr)
% % = \emptyset.
% % \]
% Therefore, for any component \(X\) built from
% \(ucv_{\succeq \curf^*,1} \cup \cod(\rho_{l,\succeq \curf^*,1})\),
% its counterpart in \(ctx_2\) satisfies \(X_2 = X_1[q \mapsto q[i]]=X_1\setminus S \cup S[i]\).
% In the sequel we use this observation without repeating it. 
By definition, 
\[
  V_{\activelabel,f_{\mathrm{cur}},2}
  = \bigl( V_{\activelabel,f_{\mathrm{cur}},1} \setminus S \bigr) \cup S[i].
\]

First observe that
\[
  S[i][+] \cap S[i] = \emptyset.
\]
Moreover,
\[
  S[i][+] \cup S[i] \subseteq S[+].
\]
Applying Lemma~\ref{lem:S-plus} together with Condition~2 in the validity of \(\ctx_1\),
we obtain \(S[+] \subseteq cv_1\) and
\(S[+] \cap Q_{\succeq f_{\mathrm{cur}}^*,1}[+] = \emptyset\).
Hence
\[
  S[i][+] \subseteq cv_1, \qquad
  S[i] \cap Q_{\succeq f_{\mathrm{cur}}^*,1}[+] = \emptyset.
\]

By Condition~2 in the validity of \(\ctx_1\), we further have
\[
  cv_1 \cap \bigl( V_{\activelabel,f_{\mathrm{cur}},1} \cup Q_{\succeq f_{\mathrm{cur}}^*,1}[+] \bigr) = \emptyset.
\]
Therefore
\[
  S[i][+] \cap \bigl( V_{\activelabel,f_{\mathrm{cur}},1} \setminus S \bigr) = \emptyset.
\]
Consequently, together with $S \cap Q_{\succeq f_{\mathrm{cur}}^*,1}[+]=\emptyset$, the required equality follows. 
\end{proof}

\begin{lemma}
\label{lem:com-Aset-1} 
Under Notations~\ref{note:state_1},~\ref{note:state_2} and the assumption that Condition~2 in the
  validity of \(\ctx_0\) holds, if \(s = \tau x \gets f(\overline{u})\) for some variables \(x\), \(\overline{u}\) and a function \(f\) with \(f \in \SCC(f_{\mathrm{cur}})\), we have for every \(h \in \SCC(f) \cap \Fun(\St_1)\), if \(h =f_{\mathrm{cur}} \vee h \notin \Theta^{-1}(\dom(\D_0))\) then \(\ctx_1.A_h = \ctx_0.A_h\); otherwise \(\ctx_0.A_h \subseteq \ctx_1.A_h\). 
\end{lemma}
\begin{proof}

If \(x \in \dom(\rho_{f_{\mathrm{cur}},0})\), by Lemma~\ref{lem:Q-plus-holds-ctx2} we first obtain
\[
Q_{\succeq f_{\mathrm{cur}}^*,1}[+] \cap V_{\activelabel,f_{\mathrm{cur}},1}
   = Q_{\succeq f_{\mathrm{cur}}^*,0}[+] \cap V_{\activelabel,f_{\mathrm{cur}},0} .
\]

By definition \(\Qpool_1 = \Qpool_0\).
Since \(\rho_{\mathrm{loc},0}(x) \setminus \rho_{\mathrm{loc},0}(\fp(f_{\mathrm{cur}})) \subseteq \cod(\rho_{\mathrm{loc},0})
      \subseteq \cod(\rho_{\mathrm{loc},\succeq f_{\mathrm{cur}}^*,0})\),
we have
\[
\begin{aligned}
V_{\activelabel,\succeq f_{\mathrm{cur}}^*,1} \cup \cod(\rho_{\mathrm{loc},\succeq f_{\mathrm{cur}}^*,1})
&= V_{\activelabel,\succeq f_{\mathrm{cur}}^*,0}
   \cup \bigl(\rho_{\mathrm{loc},0}(x) \setminus \rho_{\mathrm{loc},0}(\fp(f_{\mathrm{cur}}))\bigr)
   \cup \cod(\rho_{\mathrm{loc},\succeq f_{\mathrm{cur}}^*,0}) \\
&= V_{\activelabel,\succeq f_{\mathrm{cur}}^*,0} \cup \cod(\rho_{\mathrm{loc},\succeq f_{\mathrm{cur}}^*,0}) .
\end{aligned}
\]
From the definition of \(cv\) it thus follows that \(cv_1= cv_0\).

Now take any \(h \in \Fun(\St_1) \cap \SCC(f_{\mathrm{cur}})\).
Since \(V_{\activelabel,f_{\mathrm{cur}},1} = V_{\activelabel,f_{\mathrm{cur}},0} \cup \bigl(\rho_{\mathrm{loc},0}(x) \setminus \rho_{\mathrm{loc},0}(\fp(f_{\mathrm{cur}}))\bigr)\),
the definition gives
\[
V_{\activelabel,\succeq h,1} \cup \cod(\rho_{\mathrm{loc},h,1})
   = V_{\activelabel,\succeq h,0}
     \cup \bigl(\rho_{\mathrm{loc},0}(x) \setminus \rho_{\mathrm{loc},0}(\fp(f_{\mathrm{cur}}))\bigr)
     \cup \cod(\rho_{\mathrm{loc},h,0}).
\]

For \(h = f_{\mathrm{cur}}\) the same reasoning as above yields
\[
V_{\activelabel,\succeq f_{\mathrm{cur}},1} \cup \cod(\rho_{\mathrm{loc},f_{\mathrm{cur}},1})
   = V_{\activelabel,\succeq f_{\mathrm{cur}},0} \cup \cod(\rho_{\mathrm{loc},f_{\mathrm{cur}},0}) .
\]

For \(h \neq f_{\mathrm{cur}}\) with \(h \notin \Theta^{-1}(\dom(\D))\) we argue as follows.
Condition~1 in the validity of \(\ctx_0\) gives \(\pos(h) < \pos(f_{\mathrm{cur}})\).
The validity condition 
\(\cod(\rho_{\mathrm{loc},0}) \setminus (\rho_{\mathrm{loc},0}(\fp(f_{\mathrm{cur}})) \cup \rho_{\mathrm{loc},0}(\re(f_{\mathrm{cur}}))) \subseteq V_{\activelabel,f_{\mathrm{cur}},0}\)
implies
\[
\rho_{\mathrm{loc},0}(x) \setminus \rho_{\mathrm{loc},0}(\fp(f_{\mathrm{cur}}))
   \subseteq \cod(\rho_{\mathrm{loc},0}) \setminus \rho_{\mathrm{loc},0}(\fp(f_{\mathrm{cur}}))
   \subseteq V_{\activelabel,f_{\mathrm{cur}},0} \cup \rho_{\mathrm{loc},0}(\re(f_{\mathrm{cur}})).
\]
Condition~3(c) in the validity of \(\ctx_0\) ensures
\(\rho_{\mathrm{loc},0}(\re(f_{\mathrm{cur}})) \subseteq V_{\activelabel,\mathrm{pred}(f_{\mathrm{cur}}),0}\);
consequently,
\[
\rho_{\mathrm{loc},0}(x) \setminus \rho_{\mathrm{loc},0}(\fp(f_{\mathrm{cur}}))
   \subseteq V_{\activelabel,f_{\mathrm{cur}},0} \cup V_{\activelabel,\mathrm{pred}(f_{\mathrm{cur}}),0}
   \subseteq V_{\activelabel,\succeq h,0}.
\]

Putting these inclusions together we obtain
\[
V_{\activelabel,\succeq h} \cup \cod(\rho_{\mathrm{loc},h})
   = V_{\activelabel,\succeq h,0} \cup \cod(\rho_{\mathrm{loc},h,0}) .
\]

Summarising, for every \(h \in \Fun(\St) \cap \SCC(f_{\mathrm{cur}})\):
\begin{itemize}
  \item if \(h = f_{\mathrm{cur}}\) or \(h \notin \Theta^{-1}(\dom(\D))\), then
        \(\ctx.A_h = \ctx_0.A_h\);
  \item otherwise, \(\ctx_0.A_h \subseteq \ctx.A_h\).
\end{itemize}
If \(x \notin \dom(\rho_{f_{\mathrm{cur}},0})\): 
   The new variable \(\rho_{\mathrm{loc},f_{\mathrm{cur}},1}(x)\) is taken from \(\Qpool_0 \subseteq cv_0\).
   We have \(\Qpool_1 = \Qpool_0 \setminus \rho_{\mathrm{loc},f_{\mathrm{cur}},1}(x) \) and
   \(\Qpool_1[+] = \Qpool_0[+] \setminus \rho_{\mathrm{loc},f_{\mathrm{cur}},1}(x)[+]\).
   Since \(V_{\activelabel,f_{\mathrm{cur}},1} = V_{\activelabel,f_{\mathrm{cur}},0} \cup \rho_{\mathrm{loc},f_{\mathrm{cur}},1}(x) \) and
   \(\rho_{\mathrm{loc},f_{\mathrm{cur}},1} = \rho_{\mathrm{loc},f_{\mathrm{cur}},0} \cup \rho_{\mathrm{loc},f_{\mathrm{cur}},1}(x) \),
   we have for any \(h\), \[
   V_{\activelabel,\succeq h,1} \cup \cod(\rho_{\mathrm{loc},h,1})
   = V_{\activelabel,\succeq h,0} \cup \cod(\rho_{\mathrm{loc},h,0}) \cup \rho_{\mathrm{loc},f_{\mathrm{cur}},1}(x),
   \] and 
   \begin{align*}
   &V_{\activelabel,\succeq f_{\mathrm{cur}}^*,1}[+] \cup \cod(\rho_{\mathrm{loc},\succeq f_{\mathrm{cur}}^*,1})[+] \\ 
   =&\bigl(V_{\activelabel,\succeq f_{\mathrm{cur}}^*,0} \cup \rho_{\mathrm{loc},f_{\mathrm{cur}},1}(x) \bigr)[+]
     \cup \bigl(\cod(\rho_{\mathrm{loc},\succeq f_{\mathrm{cur}}^*,0}) \cup \rho_{\mathrm{loc},f_{\mathrm{cur}},1}(x) \bigr)[+] \\ 
   =& V_{\activelabel,\succeq f_{\mathrm{cur}}^*,0}[+] \cup \cod(\rho_{\mathrm{loc},\succeq f_{\mathrm{cur}}^*,0})[+] \cup \rho_{\mathrm{loc},f_{\mathrm{cur}},1}(x)[+]. 
   \end{align*}
   Combining with
   \((Q_{\succeq f_{\mathrm{cur}}^*,1}[+] \cap V_{\activelabel,f_{\mathrm{cur}},1})
   = (Q_{\succeq f_{\mathrm{cur}}^*,0}[+] \cap V_{\activelabel,f_{\mathrm{cur}},0})\)
   by Lemma~\ref{lem:Q-plus-holds-ctx2}, we obtain $cv_0 \setminus \rho_{\mathrm{loc},f_{\mathrm{cur}},1}(x)$ as presented in Figure~\ref{fig:cv-com}.

   \begin{figure}[t] 
   \begin{align*}
   cv_1
   &=\Qpool_1 \cup \Qpool_1[+] \cup \Bigl( \bigl(V_{\activelabel,\succeq f_{\mathrm{cur}}^*,1}[+] \cup \cod(\rho_{\mathrm{loc},\succeq f_{\mathrm{cur}}^*,1})[+]\bigr)
          \setminus \bigl(Q_{\succeq f_{\mathrm{cur}}^*,1}[+] \cap V_{\activelabel,f_{\mathrm{cur}},1}\bigr) \Bigr) \\ 
   &=\bigl(\Qpool_0 \setminus \rho_{\mathrm{loc},f_{\mathrm{cur}},1}(x)\bigr)
     \cup \bigl(\Qpool_0[+] \setminus \rho_{\mathrm{loc},f_{\mathrm{cur}},1}(x)[+]\bigr) \\ 
   &\qquad \cup \Bigl( \bigl(V_{\activelabel,\succeq f_{\mathrm{cur}}^*,0}[+] \cup \cod(\rho_{\mathrm{loc},\succeq f_{\mathrm{cur}}^*,0})[+] \cup \rho_{\mathrm{loc},f_{\mathrm{cur}},1}(x)[+]\bigr) \\
   &\qquad\qquad \setminus \bigl(Q_{\succeq f_{\mathrm{cur}}^*,0}[+] \cap V_{\activelabel,f_{\mathrm{cur}},0}\bigr) \Bigr) \\
   &=\bigl(\Qpool_0 \setminus \rho_{\mathrm{loc},f_{\mathrm{cur}},1}(x)\bigr)
     \cup \bigl(\Qpool_0[+] \setminus \rho_{\mathrm{loc},f_{\mathrm{cur}},1}(x)[+]\bigr) \\ 
   &\qquad \cup \Bigl( \bigl(V_{\activelabel,\succeq f_{\mathrm{cur}}^*,0}[+] \cup \cod(\rho_{\mathrm{loc},\succeq f_{\mathrm{cur}}^*,0})[+]\bigr)
          \setminus \bigl(Q_{\succeq f_{\mathrm{cur}}^*,0}[+] \cap V_{\activelabel,f_{\mathrm{cur}},0}\bigr)
          \cup \rho_{\mathrm{loc},f_{\mathrm{cur}},1}(x)[+] \Bigr) \\
   &\qquad\qquad \tag{since \(\rho_{\mathrm{loc},f_{\mathrm{cur}},1}(x) \cap V_{\activelabel,f_{\mathrm{cur}},0} = \emptyset\)} \\ 
   &=\bigl(\Qpool_0 \setminus \rho_{\mathrm{loc},f_{\mathrm{cur}},1}(x) \bigr) \cup \Qpool_0[+] \\ 
   &\qquad \cup \Bigl( \bigl(V_{\activelabel,\succeq f_{\mathrm{cur}}^*,0}[+] \cup \cod(\rho_{\mathrm{loc},\succeq f_{\mathrm{cur}}^*,0})[+]\bigr)
          \setminus \bigl(Q_{\succeq f_{\mathrm{cur}}^*,0}[+] \cap V_{\activelabel,f_{\mathrm{cur}},0}\bigr) \Bigr) \\
   &\qquad\qquad \tag{since \(\rho_{\mathrm{loc},f_{\mathrm{cur}},1}(x)[+] \subseteq \Qpool_0[+]\)} \\ 
   &=\Qpool_0 \cup \Qpool_0[+] \cup \Bigl( \bigl(V_{\activelabel,\succeq f_{\mathrm{cur}}^*,0}[+] \cup \cod(\rho_{\mathrm{loc},\succeq f_{\mathrm{cur}}^*,0})[+]\bigr)
          \setminus \bigl(Q_{\succeq f_{\mathrm{cur}}^*,0}[+] \cap V_{\activelabel,f_{\mathrm{cur}},0}\bigr) \Bigr)
          \setminus \rho_{\mathrm{loc},f_{\mathrm{cur}},1}(x) \\
   &\qquad\qquad \tag{\(\rho_{\mathrm{loc},f_{\mathrm{cur}},1}(x) \subseteq \Qpool_0\) and
          \(\Qpool_0 \cap \bigl(\Qpool_0[+] \cup V_{\activelabel,\succeq f_{\mathrm{cur}}^*,0}[+] \cup \cod(\rho_{\mathrm{loc},\succeq f_{\mathrm{cur}}^*,0})[+]\bigr) = \emptyset\)} \\ 
   &=cv_0 \setminus \rho_{\mathrm{loc},f_{\mathrm{cur}},1}(x).  
   \end{align*} 
 \caption{Transformation of the clean variable pool \(cv\) during the transition from \(\ctx_0\) to \(\ctx_1\)} 
 \label{fig:cv-com} 
 \Description{...}
\end{figure}

   Note that for any \(h\), \(\rho_{\mathrm{loc},f_{\mathrm{cur}},1}(x)\) is merely moved from \(cv\) to
   \(V_{\activelabel,\succeq h,1} \cup \cod(\rho_{\mathrm{loc},h,1})\); therefore the accessible sets
   remain unchanged for every \(h\), i.e., \(\ctx_1.A_h = \ctx_0.A_h\).  

\end{proof}

\begin{lemma}
\label{lem:com-Aset-2} 
Under Notations~\ref{note:state_1},~\ref{note:state_2} and the assumption that Condition~2 in the
  validity of \(\ctx_0\) holds, if \(s = \tau x \gets f(\overline{u})\) for some variables \(x\), \(\overline{u}\) and a function \(f\) with \(f \in \SCC(f_{\mathrm{cur}})\), we have for every \(g \in \SCC(f) \cap \Fun(\St_2)\), \(\ctx_2.A_g \subseteq \ctx_1.A_g\).   
\end{lemma}
\begin{proof}
Recall that the accessible set for a function \(g\) is defined as
\[
A_{g} = cv \cup V_{\activelabel,\succeq g} \cup \cod(\rho_{\mathrm{loc},g}) .
\]

By the definition, for any component \(X\) built from
\(V_{\activelabel,\succeq f_{\mathrm{cur}}^*,1} \cup \cod(\rho_{\mathrm{loc},\succeq f_{\mathrm{cur}}^*,1})\),
its counterpart in \(\ctx_2\) satisfies \(X_2 = X_1[q \mapsto q[i]]=X_1\setminus S \cup S[i]\).
In the sequel we use this observation without repeating it. 
Observe that
\((A \cup B)[q \mapsto q[i]] = A[q \mapsto q[i]] \cup B[q \mapsto q[i]]\). 

Applying this to the relevant components yields
\begin{align*}
V_{\activelabel,\succeq g,2} \cup \cod(\rho_{\mathrm{loc},g,2})
&= V_{\activelabel,\succeq g,1}[q \mapsto q[i]] \cup \cod(\rho_{\mathrm{loc},g,1})[q \mapsto q[i]] \\
&= \bigl( V_{\activelabel,\succeq g,1} \cup \cod(\rho_{\mathrm{loc},g,1}) \bigr)[q \mapsto q[i]] \\
&\subseteq \bigl( V_{\activelabel,\succeq g,1} \cup \cod(\rho_{\mathrm{loc},g,1}) \bigr) \setminus S \;\cup\; S[i]. \tag{1}
\end{align*}

Moreover, the validity of \(\ctx_1\) ensures
\((\Qpool_1 \cup \rho_{\mathrm{glob},1}) \cap V_{\activelabel,f_{\mathrm{cur}},1} = \emptyset\); consequently
\[
\Qpool_2 = \Qpool_1, \qquad \rho_{\mathrm{glob},2} = \rho_{\mathrm{glob},1}. \tag{2}
\]

Since the index \(i\) is fresh, the substitution commutes with the higher-index extension as follows:
for any set \(X\),
\(
X[q \mapsto q[i]][+] = X[+] \setminus \{q[1{:}i]\}.
\)
Moreover, \((A \cup B)[+] = A[+] \cup B[+]\) by the definition of
future extension. 

Using these properties we obtain
\begin{align*}
&V_{\activelabel,\succeq f_{\mathrm{cur}}^*,2}[+] \cup \cod(\rho_{\mathrm{loc},\succeq f_{\mathrm{cur}}^*,2})[+] \\
 = \ &V_{\activelabel,\succeq f_{\mathrm{cur}}^*,1}[q \mapsto q[i]][+]
      \cup \cod(\rho_{\mathrm{loc},\succeq f_{\mathrm{cur}}^*,1})[q \mapsto q[i]][+] \\
= \ &\bigl( (V_{\activelabel,\succeq f_{\mathrm{cur}}^*,1} \cup \cod(\rho_{\mathrm{loc},\succeq f_{\mathrm{cur}}^*,1}))[q \mapsto q[i]] \bigr)[+] \\
 = \ &\bigl( V_{\activelabel,\succeq f_{\mathrm{cur}}^*,1} \cup \cod(\rho_{\mathrm{loc},\succeq f_{\mathrm{cur}}^*,1}) \bigr)[+] \setminus S[1{:}i]  \\
 = \ & \bigl( V_{\activelabel,\succeq f_{\mathrm{cur}}^*,1}[+] \cup \cod(\rho_{\mathrm{loc},\succeq f_{\mathrm{cur}}^*,1})[+] \bigr) \setminus S[1{:}i]. \tag{3}
\end{align*}

By Lemma~\ref{lem:Q-plus-holds-ctx2} we have
\[
Q_{\succeq f_{\mathrm{cur}}^*,2}[+] \cap V_{\activelabel,f_{\mathrm{cur}},2}
= Q_{\succeq f_{\mathrm{cur}}^*,1}[+] \cap V_{\activelabel,f_{\mathrm{cur}},1}, \tag{4}
\]

Recall that
\begin{align}
cv_2 = \Qpool_2 \cup \Qpool_2[+] \cup \Bigl( \bigl( V_{\activelabel,\succeq f_{\mathrm{cur}}^*,2}[+] \cup \cod(\rho_{\mathrm{loc},\succeq f_{\mathrm{cur}}^*,2})[+] \bigr)
                 \setminus \bigl( Q_{\succeq f_{\mathrm{cur}}^*,2}[+] \cap V_{\activelabel,f_{\mathrm{cur}},2} \bigr) \Bigr). 
\end{align}

Using (2)-(4) we deduce
\begin{align*}
cv_2 = &\Qpool_1 \cup \Qpool_1[+] \cup \Bigl( \bigl( V_{\activelabel,\succeq f_{\mathrm{cur}}^*,1}[+] \cup \cod(\rho_{\mathrm{loc},\succeq f_{\mathrm{cur}}^*,1})[+] \bigr) \setminus S[1{:}i]
                 \setminus \bigl( Q_{\succeq f_{\mathrm{cur}}^*,1}[+] \cap V_{\activelabel,f_{\mathrm{cur}},1} \bigr) \Bigr) \\[2mm]
= &\Qpool_1 \cup \Qpool_1[+] \cup \Bigl( \bigl( V_{\activelabel,\succeq f_{\mathrm{cur}}^*,1}[+] \cup \cod(\rho_{\mathrm{loc},\succeq f_{\mathrm{cur}}^*,1})[+] \bigr)
                 \setminus \bigl( Q_{\succeq f_{\mathrm{cur}}^*,1}[+] \cap V_{\activelabel,f_{\mathrm{cur}},1} \bigr)
                 \setminus S[1{:}i] \Bigr).
\end{align*}

By Lemma~\ref{lem:S-plus},
\(S[1{:}i] \subseteq S[+] \subseteq
   (V_{\activelabel,\succeq f_{\mathrm{cur}}^*,1}[+] \cup \cod(\rho_{\succeq f_{\mathrm{cur}}^*,1})[+])
   \setminus Q_{\succeq f_{\mathrm{cur}}^*,1}[+] \subseteq cv_1\);
hence \(S[1{:}i] \cap (\Qpool_1 \cup \Qpool_1[+]) = \emptyset\) by Condition~2 in the validity of \(\ctx_1\).
We obtain
\begin{align}
cv_2 = cv_1 \setminus S[1{:}i], \qquad cv_1 = cv_2 \cup S[1{:}i]. \tag{6}
\end{align}

Combining (6) with (1), together with
\(S[1{:}i] \subseteq cv_1\), gives
\begin{align*}
\ctx_2.A_{g}
&= cv_2 \cup V_{\activelabel,\succeq g,2} \cup \cod(\rho_{\mathrm{loc},g,2})  \\
&\subseteq \bigl( cv_1 \setminus S[1{:}i] \bigr)
 \cup \bigl( V_{\activelabel,\succeq g,1} \cup \cod(\rho_{\mathrm{loc},g,1}) \setminus S \cup S[i] \bigr) \\
&= cv_1 \cup \bigl( V_{\activelabel,\succeq g,1} \cup \cod(\rho_{\mathrm{loc},g,1}) \setminus S \bigr) \setminus S[1{:}i]  \cup S[i] \\
&= cv_1 \cup \bigl( V_{\activelabel,\succeq g,1} \cup \cod(\rho_{\mathrm{loc},g,1}) \bigr)  \setminus S  \setminus S[1{:}i]  \cup S[i] \\
&= cv_1 \cup \bigl( V_{\activelabel,\succeq g,1} \cup \cod(\rho_{\mathrm{loc},g,1}) \bigr)  \setminus (S  \cup S[1{:}i]) \cup S[i] \\
&= \ctx_1.A_{g} \setminus (S[0:i-1]). 
\end{align*}

Thus we have \(\ctx_2.A_g \subseteq \ctx_1.A_g = \ctx_0.A_g\) for any \(g\).
\end{proof}

\begin{lemma}
\label{lem:state-inv-1}
Under Notations~\ref{note:state_1},~\ref{note:state_2} and the assumption that $\ctx_0$ satisfies the structural properties of validity, if
\(s = \tau x \gets f(\overline{u})\) for some variables \(x\), \(\overline{u}\) and a function \(f\) with \(f \in \SCC(f_{\mathrm{cur}})\), then the following properties hold: 
\begin{enumerate}
    \item For $\D$,
    \begin{itemize}
        \item $\dom(\D_0) \subseteq \dom(\D_2)$; 
        \item $\dom(\D_2) \setminus \dom(\D_0) \subseteq \Theta(\Succ^*(f_{\mathrm{cur}}) \setminus \Fun(\St_0))$; 
        \item $\D_2\upharpoonright_{\dom(\D_0) \setminus \Theta(\SCC(f_{\mathrm{cur}}))} = \D_0\upharpoonright_{\dom(\D_0) \setminus \Theta(\SCC(f_{\mathrm{cur}}))}$;
        \item $\D_2\upharpoonright_{\dom(\D_0) \cap \Theta(\SCC(f_{\mathrm{cur}}))} \approx_{\ctx_2.Gar_{f_{\mathrm{cur}}^*}} \D_0\upharpoonright_{\dom(\D_0) \cap \Theta(\SCC(f_{\mathrm{cur}}))}$; and for any $h \in \SCC(f_{\mathrm{cur}}) \cap \Theta^{-1}(\dom(\D_0))$, $\anc(\D_2[\Theta(h)]) \subseteq \anc(\D_0[\Theta(h)]) \cup \anc(\D_0[\Theta(h)])[+]$. 
    \end{itemize} 
    \item For $\Fun(\St)$, 
    \begin{itemize}
        \item $\Fun(\St_0) \subseteq \Fun(\St_2)$;
        \item $\Fun(\St_{\preceq h,2}) = \Fun(\St_{\preceq h,0})$ for any $h \in \Fun(\St_{\preceq f_{\mathrm{cur}},0})$;
        \item $\Fun(\St_{\succ f_{\mathrm{cur}},2}) \setminus \Fun(\St_{\succ f_{\mathrm{cur}},0}) \subseteq \Succ^*(\Succ(f_{\mathrm{cur}}) \setminus \Fun(\St_{0})) \cap \SCC(f_{\mathrm{cur}})$. 
    \end{itemize}
    \item \(\St_{\preceq f_{\mathrm{cur}}^*,2}=\St_{\preceq f_{\mathrm{cur}}^*,0}\) and \(\M\upharpoonright_{\Fun(\St_{\prec f_{\mathrm{cur}}^*,2})} = \M\upharpoonright_{\Fun(\St_{\prec f_{\mathrm{cur}}^*,0})}\) 
    \item For any $h \in \SCC(f_{\mathrm{cur}}) \cap \Fun(\St_0)$,  
     \begin{itemize} 
        \item \(\ctx_0.E_{\args,h} \preceq \ctx_2.E_{\args,h}\);
        \item $\ctx_2.A_h \subseteq \ctx_0.A_h$ if $\pos(h) \leq \pos (f_{\mathrm{cur}})$;
        % \item $\rho_{\mathrm{loc},h,2}=\rho_{\mathrm{loc},h,0}$, $V_{\activelabel,h,2}=V_{\activelabel,h,0}$ if $h < f_{\mathrm{cur}}$; 
        \item $\dom(\rho_{\mathrm{loc},h,2}) \subseteq \dom(\rho_{\mathrm{loc},h,0})$ if $h \neq f_{\mathrm{cur}}$;
        \item $\rho_{\mathrm{loc},h,2}(x)=\rho_{\mathrm{loc},h,0}(x)$, if $x \in \dom(\rho_{\mathrm{loc},h,0})\cap \dom(\rho_{\mathrm{loc},h,2})$ and $\var(\idx(\rho_{\mathrm{loc},h,0}(x)))\neq \emptyset$. 
    \end{itemize}
\end{enumerate}
\end{lemma} 
\begin{proof}
    These properties mostly follow directly from the definition of the algorithm; we outline them briefly.

By definition, \(\Fun(\St_2) = \Fun(\St_0)\), hence the claim concerning
\(\Fun(\St)\) holds immediately.  The updates to \(\St\) and \(\M\) from
\(\ctx_0\) to \(\ctx_2\) only affect functions that belong to
\(\SCC(f_{\mathrm{cur}})\).  Consequently, by the definition of \(f_{\mathrm{cur}}^*\),
\[
  \St_{\prec f_{\mathrm{cur}}^*,2} = \St_{\prec f_{\mathrm{cur}}^*,0},
  \qquad
  \M \upharpoonright_{\Fun(\St_{\prec f_{\mathrm{cur}}^*,2})}
  = \M \upharpoonright_{\Fun(\St_{\prec f_{\mathrm{cur}}^*,0})},
\]
so the third claim also follows.

Moreover, \(\ctx_1\) only adds new mappings to the local mapping of the current function; updates to \(\rho_{\mathrm{loc}}\) for functions in \(\SCC(f_{\mathrm{cur}})\) in \(\ctx_2\) only affect the parts intersecting \(S\) and are restricted to renaming. Note that the set \(S\) itself collects 
quantum variables whose index terms contain no classical variables;
therefore the condition concerning \(\rho_{\mathrm{loc}}\) in the fourth
item holds naturally.  For each \(f \in \SCC(f_{\mathrm{cur}})\), the step adds
the index \(i\) to the argument list, hence
\[
  E_{\args,f,0} \text{ is a prefix of } E_{\args,f,2}.
\]

By Lemma~\ref{lem:com-Aset-2}, \(\ctx_2.A_h \subseteq \ctx_0.A_h\) for every \(h\) with \(\pos(h) \le \pos(f_{\mathrm{cur}})\).

Concerning the declaration family, \(\dom(\D_2) = \dom(\D_0)\) by
definition.  The update to \(\D\) again only affects functions in
\(\SCC(f_{\mathrm{cur}})\); thus
\[
  \D_2 \upharpoonright_{\dom(\D_0) \setminus \Theta(\SCC(f_{\mathrm{cur}}))}
  = \D_0 \upharpoonright_{\dom(\D_0) \setminus \Theta(\SCC(f_{\mathrm{cur}}))}.
\]
It remains to prove the last clauses concerning \(\D\), namely
\(\D_2 \approx_{\ctx_{r,0}.Gar_{f_{\mathrm{cur}}^*}} \D_0\) and, for every
\(h \in \SCC(f_{\mathrm{cur}}) \cap \Theta^{-1}(\dom(\D_0))\),
\[
  \anc(\D_2[\Theta(h)]) \subseteq \anc(\D_0[\Theta(h)]) \cup \anc(\D_0[\Theta(h)])[+].
\]
The second one follows directly from the construction since we update \(\D_2[\Theta(g)] = \D_1[\Theta(g)][q \mapsto q[i]]
= \D_0[\Theta(g)][q \mapsto q[i]]\), where \(q \in S\) for any \(g \in \SCC(f_{\mathrm{cur}})\). Since \(\ctx_2\) is reached from \(\ctx_0\) during the residual compilation of the current function, completing the current function from \(\ctx_2\) is the suffix of completing it from \(\ctx_0\). Hence \(\ctx_{r,0}=\ctx_{r,2}\), 
it suffices to show, for each \(g \in \SCC(f_{\mathrm{cur}}) \cap \Theta^{-1}(\dom(\D_0))\),
\[
  \D_2[\Theta(g)] \approx_{\ctx_{r,0}.Gar_{f_{\mathrm{cur}}^*}} \D_0[\Theta(g)].
\]
By construction, \(\D_2[\Theta(g)] = \D_1[\Theta(g)][q \mapsto q[i]]
= \D_0[\Theta(g)][q \mapsto q[i]]\) where \(q \in S\).
Define a mapping \(h\) by \(h(p) = p[i]\) if \(p \in S\), and
\(h(p) = p\) otherwise.  Since 
\(\qv(\D_0[\Theta(g)]) \setminus S \cap S[+] = \emptyset\),
\(h\) is a bijection from \(\qv(\D_0[\Theta(g)])\) onto \(\qv(\D_2[\Theta(g)])\).
A straightforward renaming argument yields
\(\D_0[\Theta(g)][q \mapsto q[i]] \approx_{\ctx_{r,0}.Gar_{f_{\mathrm{cur}}^*}} \D_0[\Theta(g)]\),
hence \(\D_2[\Theta(g)] \approx_{\ctx_{r,0}.Gar_{f_{\mathrm{cur}}^*}} \D_0[\Theta(g)]\). Since $\ctx_{r,0}=\ctx_{r,2}$, we have \(\D_2[\Theta(g)] \approx_{\ctx_{r,2}.Gar_{f_{\mathrm{cur}}^*}} \D_0[\Theta(g)]\). 
\end{proof}

\begin{lemma}
\label{lem:state-inv-2}
Under Notations~\ref{note:state_1}, ~\ref{note:state_2} and the assumption that $\ctx_0$ satisfies the structural properties of validity, the following  properties hold. 
\begin{enumerate}
    \item For $\D$, 
    \begin{itemize}
        \item $\dom(\D_0) \subseteq \dom(\D)$; 
        \item $\dom(\D) \setminus \dom(\D_0) \subseteq \Theta(\Succ^*(f_{\mathrm{cur}}) \setminus \Fun(\St_0))$; 
        \item $\D\upharpoonright_{\dom(\D_0) \setminus \Theta(\SCC(f_{\mathrm{cur}}))} = \D_0\upharpoonright_{\dom(\D_0) \setminus \Theta(\SCC(f_{\mathrm{cur}}))}$;
        \item $\D\upharpoonright_{\dom(\D_0) \cap \Theta(\SCC(f_{\mathrm{cur}}))} \approx_{\ctx_r.Gar_{f_{\mathrm{cur}}^*}} \D_0\upharpoonright_{\dom(\D_0) \cap \Theta(\SCC(f_{\mathrm{cur}}))}$; and for any $h \in \SCC(f_{\mathrm{cur}}) \cap \Theta^{-1}(\dom(\D_0))$, $\anc(\D[\Theta(h)]) \subseteq \anc(\D_0[\Theta(h)]) \cup \anc(\D_0[\Theta(h)])[+]$. 
    \end{itemize} 
    \item For $\Fun(\St)$, 
    \begin{itemize}
        \item $\Fun(\St_0) \subseteq \Fun(\St)$;
        \item $\Fun(\St_{\preceq h }) = \Fun(\St_{\preceq h,0})$ for any $h \in \Fun(\St_{\preceq f_{\mathrm{cur}},0})$;
        \item $\Fun(\St_{\succ f_{\mathrm{cur}} }) \setminus \Fun(\St_{\succ f_{\mathrm{cur}},0}) \subseteq \Succ^*(\Succ(f_{\mathrm{cur}}) \setminus \Fun(\St_{0})) \cap \SCC(f_{\mathrm{cur}})$. 
    \end{itemize}
    \item \(\St_{\preceq f_{\mathrm{cur}}^*}=\St_{\preceq f_{\mathrm{cur}}^*,0}\) and \(\M_{\preceq f_{\mathrm{cur}}^*}=\M_{\preceq f_{\mathrm{cur}}^*,0}\) 
    \item For any $h \in \SCC(f_{\mathrm{cur}}) \cap \Fun(\St_0)$, 
     \begin{itemize}
        \item \(\ctx_0.E_{\args,h} \preceq \ctx.E_{\args,h}\); 
        \item $\ctx.A_h \subseteq \ctx_0.A_h$ if $\pos(h) \leq \pos (f_{\mathrm{cur}})$;
        % \item $\rho_{\mathrm{loc},h}=\rho_{\mathrm{loc},h,0}$, $V_{\activelabel,h}=V_{\activelabel,h,0}$ and $V_{\activelabel,[f_{\mathrm{cur}}^*,h]}=V_{\activelabel,[f_{\mathrm{cur}}^*,h],0}$ if $h < f_{\mathrm{cur}}$; 
        \item $\dom(\rho_{\mathrm{loc},h}) \subseteq \dom(\rho_{\mathrm{loc},h,0})$ if $h \neq  f_{\mathrm{cur}}$;
        \item $\rho_{\mathrm{loc},h}(x)=\rho_{\mathrm{loc},h,0}(x)$, if $x \in \dom(\rho_{\mathrm{loc},h,0})\cap \dom(\rho_{\mathrm{loc},h})$ and $\var(\idx(\rho_{\mathrm{loc},h,0}(x)))\neq \emptyset$. 
    \end{itemize}
\end{enumerate}
\end{lemma}
\begin{proof} 
If $s = \tau x \gets f(\overline{u})$ with $f \in \SCC(f_{\mathrm{cur}})$, then by Lemma~\ref{lem:state-inv-1}, the condition holds for $\ctx_2$.  
If $f \notin \Fun(\St_0)$, the conclusion for $\ctx$ follows by combining Lemma~\ref{lem:fun-inv} for function compilation with the construction of $\ctx$ from $\ctx_3$.  
If $f \in \Fun(\St_0)$, the conclusion for $\ctx$ follows from the construction of $\ctx$ from $\ctx_2$.  
If $f \notin \SCC(f_{\mathrm{cur}})$, the conclusion follows directly from the construction of $\ctx$.
\end{proof}

\begin{lemma}\label{lem:state-valid-3.a}
Under Notations~\ref{note:state_1},~\ref{note:state_2} and Assumption~\ref{hpy:com}, if \(s = \tau x \gets f(\overline{u})\) for some variables \(x\), \(\overline{u}\) and a function \(f\) with \(f \in \SCC(f_{\mathrm{cur}})\), then Condition~3(a) of the validity of \(\ctx_2\) holds. 
\end{lemma}
\begin{proof}
For any \(g \in \{\ctx_2.f_{\mathrm{cur}}\} \cup \bigl( \Fun(\St_2) \cap \SCC(f_{\mathrm{cur}}) \setminus \Theta^{-1}(\dom(\D_2)) \bigr)\),
we have
\(
g \in \{\ctx_1.f_{\mathrm{cur}}\} \cup \bigl( \Fun(\St_1) \cap \SCC(f_{\mathrm{cur}}) \setminus \Theta^{-1}(\dom(\D_1)) \bigr)
\)
and
\(
g \in \{\ctx_0.f_{\mathrm{cur}}\} \cup \bigl( \Fun(\St_0) \cap \SCC(f_{\mathrm{cur}}) \setminus \Theta^{-1}(\dom(\D_0)) \bigr),
\)
since $\ctx_2.f_{\mathrm{cur}}=\ctx_1.f_{\mathrm{cur}}=\ctx_0.f_{\mathrm{cur}}$, \(\Fun(\St_2) = \Fun(\St_1) = \Fun(\St_0)\) and
\(\Theta^{-1}(\dom(\D_2)) = \Theta^{-1}(\dom(\D_1)) = \Theta^{-1}(\dom(\D_0))\).

\textbf{Part I: Condition~3(a) for \(\ctx_1\).} 
\begin{itemize}
    \item If \(g \neq f_{\mathrm{cur}}\) or \(g=f_{\mathrm{cur}} \wedge x \in \dom(\rho_0)\), then  
          \[\ctx_0.(\rho_{\mathrm{loc},g},C_{g},\kappa_{g}, \delta_g, R_{\mathrm{stmt},g})
           = \ctx_1.(\rho_{\mathrm{loc},g},C_{g},\kappa_{g}, \delta_g, R_{\mathrm{stmt},g}),\] $(V_{\activelabel,g,0} \cup \cod(\rho_{g,0}))= (V_{\activelabel,g,1} \cup \cod(\rho_{g,1}))$, $\Qpool_0 \cup (V_{\activelabel,\succeq g,0} \setminus Q_{\succeq f_{\mathrm{cur}}^*,0}[+]) \subseteq \Qpool_1 \cup (V_{\activelabel,\succeq g,1} \setminus Q_{\succeq f_{\mathrm{cur}}^*,1}[+])$ and $V_{\alloc,g,0}=V_{\alloc,g,1}$ by definition.   
          Validity follows directly from the validity of \(\ctx_0\).
    \item If \(g = f_{\mathrm{cur}}\) and \(x \notin \dom(\rho_0)\), we have 
          \[
          \begin{aligned}
          &\ctx_1.(\rho_{\mathrm{loc},f_{\mathrm{cur}}}\, C_{f_{\mathrm{cur}}},\,\kappa_{f_{\mathrm{cur}}},\, \delta_{f_{\mathrm{cur}}},\,R_{\mathrm{stmt},f_{\mathrm{cur}}}) \\ 
          = &\ctx_0.(\rho_{\mathrm{loc},f_{\mathrm{cur}}} \cup \{ x \mapsto \ctx_1.\rho_{\mathrm{loc},f_{\mathrm{cur}}}(x)\},\, C_{f_{\mathrm{cur}}},\,\kappa_{f_{\mathrm{cur}}},\, \delta_{f_{\mathrm{cur}}},\,R_{\mathrm{stmt},f_{\mathrm{cur}}}).
            \end{aligned}  
          \]
\end{itemize}

We check the three sub‑conditions of validity for this case:
\begin{enumerate}
    \item \textbf{Injectivity of \(\rho_{\mathrm{loc},f_{\mathrm{cur}},1}\).} 
          Since \(\ctx_0\) is valid, \(\rho_{\mathrm{loc},f_{\mathrm{cur}},0}\) is injective.  
          Moreover \(cv_0 \cap \cod(\rho_{\mathrm{loc},f_{\mathrm{cur}},0}) = \emptyset\) and 
          \(\rho_{\mathrm{loc},f_{\mathrm{cur}},1}(x) \in \Qpool_0 \subseteq cv_0\); hence 
          \(\rho_{\mathrm{loc},f_{\mathrm{cur}},1}(x) \notin \cod(\rho_{\mathrm{loc},f_{\mathrm{cur}},0})\).  
          Therefore \(\rho_{\mathrm{loc},f_{\mathrm{cur}},1} = \rho_{\mathrm{loc},f_{\mathrm{cur}},0} \cup \{x \mapsto \rho_{\mathrm{loc},f_{\mathrm{cur}},1}(x)\}\)
          remains injective.

    \item \textbf{Residual-use consistency.}  
         Since \(R_{\mathrm{stmt},f_{\mathrm{cur}},1}=R_{\mathrm{stmt},f_{\mathrm{cur}},0}\) and \(\delta_{f_{\mathrm{cur}},1} = \delta_{f_{\mathrm{cur}},0}\), we have that \(R_{\mathrm{stmt},f_{\mathrm{cur}},1}\) is a suffix of \(s_{f_{\mathrm{cur}}}\) and, for every node \(v \in \mathsf{Node}(G_{f_{\mathrm{cur}}})\), \[
\delta_{f_{\mathrm{cur}},1}(v) = \contuses(R_{\mathrm{stmt},f_{\mathrm{cur}},1}, v),
\]
both of which follow directly from the validity of \(\ctx_0\). 
 Moreover, \(\dom(\rho_{\mathrm{loc},f_{\mathrm{cur}},1}) = \dom(\rho_{\mathrm{loc},f_{\mathrm{cur}},0}) \cup \{x\}\). Since \(x \notin \dom(\rho_{\mathrm{glob}})\) and \(\dom(\rho_{\mathrm{loc},f_{\mathrm{cur}},0}) \cap \dom(\rho_{\mathrm{glob}}) = \emptyset\), we obtain \(\dom(\rho_{\mathrm{loc},f_{\mathrm{cur}},1}) \cap \dom(\rho_{\mathrm{glob}}) = \emptyset\). 
        Since \(s\) is a prefix of \(R_{\mathrm{stmt},f_{\mathrm{cur}},0}\), and \(R_{\mathrm{stmt},f_{\mathrm{cur}},0}\) is a suffix of \(s_{f_{\mathrm{cur}}}\), it follows that \(s\) is a subsequence of \(s_{f_{\mathrm{cur}}}\). Since \(x \in \var(s)\), \(x \in \var(s_{f_{\mathrm{cur}}})\), and by Lemma~\ref{lem:static-wf} we have \(\var(s_{f_{\mathrm{cur}}}) \subseteq \mathsf{Var}(G_{f_{\mathrm{cur}},0})\); thus \(x \in \mathsf{Var}(G_{f_{\mathrm{cur}},0})\). 
          Without loss of generality, since \(x\) is declared for the first time,
          we may assume that it will be used by the remaining statements, hence 
          \(\delta_{f_{\mathrm{cur}},0}(\node(x)) > 0\).  Combining this with the
          validity of \(\ctx_0\) we obtain
          \[
          \begin{aligned}
          & \dom(\rho_{\mathrm{loc},f_{\mathrm{cur}},1}) \setminus (\fp(f_{\mathrm{cur}}) \cup \re(f_{\mathrm{cur}})) =  \bigl(\dom(\rho_{\mathrm{loc}, f_{\mathrm{cur}},0}) \setminus (\fp(f_{\mathrm{cur}}) \cup \re(f_{\mathrm{cur}}))\bigr)
             \cup \{x\} \\
         \subseteq \ & \{\, y \in \mathsf{Var}(G_{f_{\mathrm{cur}},0}) \mid
                       \delta_{f_{\mathrm{cur}},0}(\node(y)) > 0 \lor y.\flag \neq 0 \,\} 
          = \{\, y \in \mathsf{Var}(G_{f_{\mathrm{cur}},1}) \mid
                       \delta_{f_{\mathrm{cur}},1}(\node(y)) > 0 \lor y.\flag \neq 0 \,\}.
          \end{aligned}
          \]

    \item \textbf{Correctness of \(\kappa_{f_{\mathrm{cur}},1}\).}
          Let \(q \in \cod(\rho_{\mathrm{loc},f_{\mathrm{cur}},1}) \setminus \rho_{\mathrm{loc},f_{\mathrm{cur}},1}(\re(f_{\mathrm{cur}}))\) such that $\rho_{f_{\mathrm{cur}},1}^{-1}(q).\flag \neq 2$ and \(v = \rho_{f_{\mathrm{cur}},1}^{-1}(q)\).  
        First, we easily obtain the following inclusions:
\(
(V_{\activelabel,f_{\mathrm{cur}},0} \cup \cod(\rho_{f_{\mathrm{cur}},0})) \subseteq (V_{\activelabel,f_{\mathrm{cur}},1} \cup \cod(\rho_{f_{\mathrm{cur}},1})),\) 
\(\Qpool_0 \cup (V_{\activelabel,\succeq f_{\mathrm{cur}},0} \setminus Q_{\succeq f_{\mathrm{cur}}^*,0}[+]) = \Qpool_1 \cup (V_{\activelabel,\succeq f_{\mathrm{cur}},1} \setminus Q_{\succeq f_{\mathrm{cur}}^*,1}[+]),
\) 
and \(V_{\alloc,f_{\mathrm{cur}},0} \subseteq V_{\alloc,f_{\mathrm{cur}},1}\).
          \begin{itemize}
              \item If \(v \neq \node(x)\), then it follows that \(v \in \dom(\kappa_0)\) and, by definition,
\( \rho_{\mathrm{loc},f_{\mathrm{cur}},1}(v) = \rho_{\mathrm{loc},f_{\mathrm{cur}},0}(v),\)  
\(\kappa_{f_{\mathrm{cur}},1}(v) = \kappa_{f_{\mathrm{cur}},0}(v), \) and 
\(\ctx_1.C_{f_{\mathrm{cur}}} = \ctx_0.C_{f_{\mathrm{cur}}}.
\)
Together with the inclusions above, since \(\ctx_0\) is valid, the three required properties for \(\kappa_{f_{\mathrm{cur}},0}(v)\) are satisfied; they transfer immediately to \(\kappa_{f_{\mathrm{cur}},1}(v)\). 

\item If \(v = \node(x)\). 
       \begin{itemize}
           \item If \(\var(v) \cap \dom(\rho_{\mathrm{loc},f_{\mathrm{cur}},0}) \neq \emptyset\): Note that \(\kappa_{f_{\mathrm{cur}},1}(v)= \kappa_{f_{\mathrm{cur}},0}(v)\).  
By the validity conditions of \(\ctx_0\) we have
\( \bigl(\rho_{f_{\mathrm{cur}},0}(\re(f_{\mathrm{cur}}) \cup \fp(f_{\mathrm{cur}}))\bigr) \cap \req_{v,0}(\kappa_{f_{\mathrm{cur}},0}^k(v)) = \emptyset, \)
and for any \(z\), there exists a set \(S_0\) with
\(
\qv(\kappa_{f_{\mathrm{cur}},0}(v)^z) \subseteq \qv(C_{f_{\mathrm{cur}},0}^z) \cup S_0,
\)
where \(S_0 \subseteq \Qpool_0 \cup (V_{\activelabel,\succeq f_{\mathrm{cur}},0} \setminus Q_{\succeq f_{\mathrm{cur}}^*,0}[+])\) and \(\base(S_0) \subseteq V_{\alloc,f_{\mathrm{cur}},0}\).  
Lemma~\ref{lem:state-valid-3.c} gives
\(
\rho_{\mathrm{loc},f_{\mathrm{cur}},1}(\re(f_{\mathrm{cur}}) \cup \fp(f_{\mathrm{cur}})) = \rho_{\mathrm{loc},f_{\mathrm{cur}},0}(\re(f_{\mathrm{cur}}) \cup \fp(f_{\mathrm{cur}})).
\)
Combining this with \(\req_{v,1}(\kappa_{f_{\mathrm{cur}},1}(v)) \subseteq \req_{v,0}(\kappa_{f_{\mathrm{cur}},0}(v))\), \(C_{f_{\mathrm{cur}},1}=C_{f_{\mathrm{cur}},0}\), and the inclusions mentioned above, we conclude that the same properties hold for \(\kappa_{f_{\mathrm{cur}},1}(v)\). 
\item If \(\var(v) \cap \dom(\rho_{\mathrm{loc},f_{\mathrm{cur}},0}) = \emptyset\):
By definition $\rho_{\mathrm{loc},f_{\mathrm{cur}},1}(v)=\rho_{\mathrm{loc},f_{\mathrm{cur}},1}(x)$ and \(\kappa_{f_{\mathrm{cur}},1}(v) = \kappa_{f_{\mathrm{cur}},0}(v)= I\).  
We trivially have
\(
\bigl(\rho_{f_{\mathrm{cur}},1}(\re(f_{\mathrm{cur}}) \cup \fp(f_{\mathrm{cur}}))\bigr) \cap \req_{v,1}(\kappa_{f_{\mathrm{cur}},1}(v)^k) = \emptyset,
\)
and for any \(z\), there exists a set \(S_1\) such that
\(
\qv(\kappa_{f_{\mathrm{cur}},0}(v)^z) \subseteq \qv(C_{f_{\mathrm{cur}},0}^k) \cup S_1,
\)
with \(S_1 \subseteq \Qpool_1 \cup (V_{\activelabel,\succeq f_{\mathrm{cur}},1} \setminus Q_{\succeq f_{\mathrm{cur}}^*,1}[+])\) and \(\base(S_1) \subseteq V_{\alloc,f_{\mathrm{cur}},1}\). 
                    \end{itemize}
          \end{itemize}

        Now let \(\ket{\varphi}\) be a state with 
          \(\ket{\varphi} \models \ket{0}_{\ctx_1.A_{f_{\mathrm{cur}}} \setminus \rho_{\mathrm{loc},f_{\mathrm{cur}},1}(\fp(f_{\mathrm{cur}}))}\). 
          By the proof of Lemma~\ref{lem:state-valid-2} we also have 
          \(\ket{\varphi} \models \ket{0}_{\ctx_1.A_{f_{\mathrm{cur}}} \setminus \rho_{\mathrm{loc},f_{\mathrm{cur}},0}(\fp(f_{\mathrm{cur}}))}\). 

          Now take any tuple \(\overline{p}\) of fresh quantum variables such that
          \(\ctx_1.C_{f_{\mathrm{cur}}}^k\ket{\varphi} \models \ket{0}_{\overline{p}}\) and 
          \(\overline{p} \cap \bigl(\cod(\rho_{f_{\mathrm{cur}},1}) \cup V_{\activelabel,f_{\mathrm{cur}},1}\bigr) = \emptyset\).
         Since  \(\ctx_0.C_{f_{\mathrm{cur}}} = \ctx_1.C_{f_{\mathrm{cur}}}\) and 
          \(\cod(\rho_{f_{\mathrm{cur}},0}) \cup V_{\activelabel,f_{\mathrm{cur}},0} \subseteq
           \cod(\rho_{f_{\mathrm{cur}},1}) \cup V_{\activelabel,f_{\mathrm{cur}},1}\),
          we also have 
          \(\ctx_0.C_{f_{\mathrm{cur}}}^k\ket{\varphi} = \ctx_1.C_{f_{\mathrm{cur}}}^k\ket{\varphi}
           \models \ket{0}_{\overline{p}}\) and 
          \(\overline{p} \cap \bigl(\cod(\rho_{f_{\mathrm{cur}},0}) \cup V_{\activelabel,f_{\mathrm{cur}},0}\bigr) = \emptyset\).
          Moreover, \(\rho_{\mathrm{loc},f_{\mathrm{cur}},1}(x) \in \Qpool_0 \subseteq cv_0\) implies
          \(\rho_{\mathrm{loc},f_{\mathrm{cur}},1}(x) \cap \bigl(\cod(\rho_{f_{\mathrm{cur}},0}) \cup V_{\activelabel,f_{\mathrm{cur}},0}\bigr) = \emptyset\);
          together with the fact that 
          \(\ctx_0.C_{f_{\mathrm{cur}}}^k\ket{\varphi} \models \ket{0}_{\rho_{\mathrm{loc},f_{\mathrm{cur}},1}(x)}\) by Lemma~\ref{lem:curf} this yields
          \begin{align*}
          &\langle \ctx_1.C_{f_{\mathrm{cur}}}^k;\,
            \kappa_{f_{\mathrm{cur}},1}(v)[\req_{v,1}\mapsto\overline{p}],\;
            \ket{\varphi}\rangle \\
          =\ &\langle \ctx_0.C_{f_{\mathrm{cur}}}^k;\,
            \kappa_{f_{\mathrm{cur}},0}(v)[(\req_{v,0}\setminus\rho_{\mathrm{loc},f_{\mathrm{cur}},1}(x))\mapsto\overline{p}],\;
            \ket{\varphi}\rangle \\
          =\ &\langle \ctx_0.C_{f_{\mathrm{cur}}}^k;\,
            \kappa_{f_{\mathrm{cur}},0}(v)[\req_{v,0}\setminus\rho_{\mathrm{loc},f_{\mathrm{cur}},1}(x)\mapsto\overline{p},\,
            \rho_{f_{\mathrm{cur}},1}(x)\mapsto\rho_{\mathrm{loc},f_{\mathrm{cur}},1}(x)],\;
            \ket{\varphi}\rangle \\
          \rightarrow^{*} \ &\bigl\langle \downarrow,\;
            \ket{\varphi}_{\rho_{\mathrm{loc},f_{\mathrm{cur}},0}(U_v)} \otimes \ket{\theta}_{\rho_{\mathrm{loc},f_{\mathrm{cur}},0}(v) \setminus \rho_{\mathrm{loc},f_{\mathrm{cur}},0}(U_v)}
            \otimes (\ctx_0.C_{f_{\mathrm{cur}}}^k\ket{\varphi})_{\sys\setminus\rho_{\mathrm{loc},f_{\mathrm{cur}},0}(v)}
            \bigr\rangle \\
          = \ &\bigl\langle \downarrow,\;
            \ket{\varphi}_{\rho_{\mathrm{loc},f_{\mathrm{cur}},0}(U_v)} \otimes \ket{\theta}_{\rho_{\mathrm{loc},f_{\mathrm{cur}},0}(v) \setminus \rho_{\mathrm{loc},f_{\mathrm{cur}},0}(U_v)}  
            \otimes \ket{\varphi}_{\rho_{\mathrm{loc},f_{\mathrm{cur}},1}(x)}
            \otimes (\ctx_0.C_{f_{\mathrm{cur}}}^k\ket{\varphi})_{\mathrm{rem}}
            \bigr\rangle \\
          = \ &\bigl\langle \downarrow,\;
            \ket{\varphi}_{\rho_{\mathrm{loc},f_{\mathrm{cur}},1}(U_v)} \otimes \ket{\theta}_{\rho_{\mathrm{loc},f_{\mathrm{cur}},1}(v) \setminus \rho_{\mathrm{loc},f_{\mathrm{cur}},1}(U_v)} 
            \otimes (\ctx_1.C_{f_{\mathrm{cur}}}^k\ket{\varphi})_{\sys\setminus\rho_{\mathrm{loc},f_{\mathrm{cur}},1}(v)}
            \bigr\rangle.  \tag{$\rho_{\mathrm{loc},f_{\mathrm{cur}},1}(v)=\rho_{\mathrm{loc},f_{\mathrm{cur}},0}(v)\cup \rho_{\mathrm{loc},f_{\mathrm{cur}},0}(x)$} 
          \end{align*} 
      Here we processed the subcase where \(\var(v) \cap \dom(\rho_{f_{\mathrm{cur}},0}) \neq \emptyset\) and \(x.\flag \neq 2\); the subcase where \(\var(v) \cap \dom(\rho_{f_{\mathrm{cur}},0}) = \emptyset\) or \(x.\flag = 2\) is analogous and simpler. 
\end{enumerate}
Thus \(\ctx_1\) satisfies Condition~3(a).

\textbf{Part II: Condition~3(a) for \(\ctx_2\).}
Let \(\ket{\varphi}\) be a state with 
\(\ket{\varphi} \models \ket{0}_{\ctx_2.A_g \setminus \rho_{\mathrm{loc},g,2}(\fp(g))}\).
Let \(\ket{\psi}\) be a state with 
\(\ket{\psi} \models \ket{0}_{\ctx_1.A_g \setminus \rho_{\mathrm{loc},g,1}(\fp(g))}\) and 
\(\ket{\psi}_{\rho_{\mathrm{loc},g,1}(\fp(g))} = \ket{\varphi}_{\rho_{\mathrm{loc},g,2}(\fp(g))}\).

For any \(g \in \SCC(f_{\mathrm{cur}})\), the components are obtained by renaming:
\[
(\rho_{\mathrm{loc},g,2}, C_{g,2}, \kappa_{g,2}) = (\rho_{\mathrm{loc},g,1}, C_{g,1}, \kappa_{g,1})[q \mapsto q[i]], \qquad
(\delta_{g,2}, R_{\mathrm{stmt},g,2}) = (\delta_{g,1}, R_{\mathrm{stmt},g,1}).
\]

\begin{enumerate}
    \item \textbf{Injectivity of \(\rho_{\mathrm{loc},g,2}\).}
          The renaming \(q \mapsto q[i]\) is a bijection on the involved quantum variables,
          hence \(\rho_{g,2}\) remains injective.

    \item \textbf{Residual-use consistency.}
          Since \(\delta_{g,2} = \delta_{g,1}\), \(R_{\mathrm{stmt},g,2} = R_{\mathrm{stmt},g,1}\),
          and \(\dom(\rho_{\mathrm{loc},g,2}) = \dom(\rho_{\mathrm{loc},g,1})\), 
          the required properties are unchanged from the validity of \(\ctx_1\).  

    \item \textbf{Correctness of \(\kappa_{g,2}\).}
          Let \(q \in \cod(\rho_{\mathrm{loc},g,2}) \setminus \rho_{\mathrm{loc},g,2}(\re(g))\) such that $\rho_{\mathrm{loc},g,2}^{-1}(q).\flag \neq 2$  and \(v = \rho_{\mathrm{loc},g,2}^{-1}(q)\).
          There exists a unique \(q' \in \cod(\rho_{\mathrm{loc},g,1}) \setminus \rho_{\mathrm{loc},g,1}(\re(f_{\mathrm{cur}}))\) with 
          \(q'[S \mapsto S[i]] = q\) and \(v = \rho_{\mathrm{loc},g,1}^{-1}(q')\). Then $v \in \dom(\kappa_1) =\dom(\kappa_2)$. 

          From the validity of \(\ctx_1\) we have for \(v\):
          \begin{itemize}
              \item \(\rho_{\mathrm{loc},g,1}(\re(g) \cup \fp(g)) \cap \req_{v,1}(\kappa_{g,1}(v)^k) = \emptyset\); 
              \item for any $z$, \(\qv(\kappa_{g,1}(v)^z) \subseteq \qv(C_{g,1}^z) \cup S_1\) for some $S_1$ such that $S_1\subseteq \Qpool_1 \cup  (V_{\activelabel,\succeq f_{\mathrm{cur}},1} \setminus Q_{\succeq f_{\mathrm{cur}}^*,1}[+])$ and $\base(S_1) \subseteq V_{\alloc,f_{\mathrm{cur}},1}$.  
              \item Let \(\ket{\psi'} = \ctx_1.C_{g}^k\ket{\psi}\). For
                    any \(\overline{p}\) such that \(\ket{\psi'} \models \ket{0}_{\overline{p}}\) and
                    \(\overline{p} \cap (\cod(\rho_{g,1}) \cup V_{\activelabel,g,1}) = \emptyset\), 
                    \[
                    \langle \ctx_1.C_{g}^k; \kappa_{g,1}(v)[\req_{v,1} \mapsto \overline{p}], \ket{\psi} \rangle
                    \rightarrow^{*} \langle \downarrow,\;
                        \ket{\psi}_{\rho_{\mathrm{loc},g,1}(v)} \otimes \ket{\psi'}_{\sys \setminus \rho_{\mathrm{loc},g,1}(v)} \rangle .
                    \]
          \end{itemize}

          Applying the renaming we obtain:
          \begin{itemize}
              \item Note that
                    \(\rho_{\mathrm{loc},g,2}(\re(g) \cup \fp(g)) = (\rho_{\mathrm{loc},g,1}(\re(g) \cup \fp(g)))[q \mapsto q[i]]\) and
                    \(\req_{v,2}(\kappa_{g,2}(v)^k) = \req_{v,2}(\kappa_{g,1}(v)^k[q \mapsto q[i]])
                       = (\req_{v,1}(\kappa_{g,1}(v)^k))[q \mapsto q[i]]\).  
                    Since \(\rho_{\mathrm{loc},g,1}(\re(g) \cup \fp(g)) \cap \req_{v,1}(\kappa_{g,1}(v)^k) = \emptyset\), 
                    we have
                    \begin{align*}
                    \bigl(\rho_{\mathrm{loc},g,1}(\re(g) \cup \fp(g))\bigr)[q \mapsto q[i]]
                    \cap \req_{v,1}(\kappa_{g,1}(v)^k)[q \mapsto q[i]] = \emptyset,
                    \end{align*}
                    and thus
                    \(\rho_{\mathrm{loc},g,2}(\re(g) \cup \fp(g)) \cap \req_{v,2}(\kappa_{g,2}(v)^k) = \emptyset\).
              \item For any \(z\), we have
\(
\qv(\kappa_{g,2}(v)^z) = \qv(\kappa_{g,1}(v)^z[q \mapsto q[i]]) = \bigl(\qv(\kappa_{g,1}(v)^z)\bigr)[q \mapsto q[i]] \subseteq \bigl(\qv(C_{g,1}^z) \cup S_1\bigr)[q \mapsto q[i]].
\)
Moreover,
\(
\bigl(\qv(C_{g,1}^z) \cup S_1\bigr)[q \mapsto q[i]] = \qv(C_{g,1}^z)[q \mapsto q[i]] \cup S_1[q \mapsto q[i]] = \qv(C_{g,2}^z) \cup S_1[q \mapsto q[i]].
\)
Let \(S_2 = S_1[q \mapsto q[i]]\). Note that
\(
S_1[q \mapsto q[i]] \subseteq \bigl(\Qpool_1 \cup (V_{\activelabel,\succeq f_{\mathrm{cur}},1} \setminus Q_{\succeq f_{\mathrm{cur}}^*,1}[+])\bigr)[q \mapsto q[i]] = \Qpool_2 \cup (V_{\activelabel,\succeq f_{\mathrm{cur}},2} \setminus Q_{\succeq f_{\mathrm{cur}}^*,2}[+]) 
\) and $\base(S_1[q \mapsto q[i]])=\base(S_1) \subseteq V_{\alloc,f_{\mathrm{cur}},1}=V_{\alloc,f_{\mathrm{cur}},2}$. 
Thus the desired conclusion follows naturally.
          \end{itemize} 

          Now let \(\ket{\varphi'} = \ctx_2.C_{g}^k\ket{\varphi}\). Take
          any \(\overline{p}\) such that \(\ket{\varphi'} \models \ket{0}_{\overline{p}}\)
          and \(\overline{p} \cap (\cod(\rho_{g,1}) \cup V_{\activelabel,g,1}) = \emptyset\). We need to prove that
          \[
          \langle \ctx_2.C_{g}^k; \kappa_{g,2}(v)[\req_{v,2} \mapsto \overline{p}], \ket{\varphi} \rangle
          \rightarrow^{*} \langle \downarrow,\;
              \ket{\varphi}_{\rho_{\mathrm{loc},g,2}(v)} \otimes \ket{\varphi'}_{\sys \setminus \rho_{\mathrm{loc},g,2}(v)} \rangle .
          \]

          First, choose \(\overline{p}' \in \Qpool_1\) such that
          \(\overline{p}' \cap (\overline{p} \cup S \cup S[i]) = \emptyset\).
          Note that \(\Qpool_2 = \Qpool_1\); then by Condition~2 in the validity of \(\ctx_1\) and \(\ctx_2\),
          we have
          \(\overline{p}' \cap (\cod(\rho_{g,1}) \cup V_{\activelabel,g,1}) = \emptyset\) and 
          \(\overline{p}' \cap (\cod(\rho_{g,2}) \cup V_{\activelabel,g,2}) = \emptyset\).
          Moreover, Condition~3(b) in the validity of \(\ctx_1\) and \(\ctx_2\) yields
          \(\ket{\psi'} \models \ket{0}_{\overline{p}'}\) and
          \(\ket{\varphi'} \models \ket{0}_{\overline{p}'}\).

          We now first prove that
          \[
          \langle \ctx_2.C_{g}^k; \kappa_{g,2}(v)[\req_{v,2} \mapsto \overline{p}'], \ket{\varphi} \rangle
          \rightarrow^{*} \langle \downarrow,\;
              \ket{\varphi}_{\rho_{\mathrm{loc},g,2}(v)} \otimes \ket{\varphi'}_{\sys \setminus \rho_{\mathrm{loc},g,2}(v)} \rangle .
          \]

          Let \(C_a = (\ctx_1.C_{g}^k; \kappa_{g,1}(v)[\req_{v,1} \mapsto \overline{p}'])\) and
          \(C_b = \ctx_2.C_{g}^k; \kappa_{g,2}(v)[\req_{v,1} \mapsto \overline{p}']\).
          Since \(\overline{p}' \cap S = \emptyset\), we have
          \begin{align*}
          C_b &= \ctx_1.C_{g}^k[q \mapsto q[i]];
                \kappa_{g,1}(v)[q \mapsto q[i]][\req_{v,2} \mapsto \overline{p}'] \\
              &= \bigl( \ctx_1.C_{g}^k; \kappa_{g,1}(v) \bigr)[q \mapsto q[i]]
                 [\req_{v,2} \mapsto \overline{p}'] \\
              &= \bigl( \ctx_1.C_{g}^k; \kappa_{g,1}(v)[\req_{v,1} \mapsto \overline{p}'] \bigr)[q \mapsto q[i]]
               = C_a[q \mapsto q[i]] .
          \end{align*}

          Define \(h_1 : \qv(C_a) \rightarrow \mathcal{Q}\) by
          \(h_1(q) = q[i]\) for \(q \in S\) and \(h_1(q) = q\) otherwise.
          Since \(\overline{p}' \cap S[i] = \emptyset\), Condition~4(d) in the validity
          of \(\ctx_1\) implies \(\bigl(\qv(C_a) \setminus S\bigr) \cap S[i] = \emptyset\).
          Thus \(h_1\) is a bijection onto its image.
         Since \(\qv(C_a) \subseteq \ctx_1.A_g\) and
          \(h_1(\qv(C_a)) = \qv(h_1(C_a)) = \qv(C_a[q \mapsto q[i]]) = \qv(C_b) \subseteq \ctx_2.A_g\)
          (by Lemma~\ref{lem:state-valid-4.c}),
          we have \(\ket{\varphi}_{h_1(\qv(C_a))} = \ket{\psi}_{\qv(C_a)}\).

          Applying Lemma~\ref{lemma:rename},
          \begin{align*}
          (C_b \ket{\varphi})_{\qv(C_b)}
          = (h_1(C_a) \ket{\varphi})_{h_1(\qv(C_a))}
          = \bigl(C_a \ket{\psi}\bigr)_{\qv(C_a)} .
          \end{align*}

          Similarly, since \(\ctx_1.C_{g}^k \preceq C_a\) and \(\ctx_2.C_{g}^k \preceq C_b\),
          \(h_1\) restricts to a bijection from \(\qv(\ctx_1.C_{g}^k)\) to its image, and
          \(\ket{\varphi}_{h_1(\qv(\ctx_1.C_{g}^k))} = \ket{\psi}_{\qv(\ctx_1.C_{g}^k)}\).
          Applying Lemma~\ref{lemma:rename} again,
          \begin{align*}
          (\ctx_2.C_{g}^k \ket{\varphi})_{\qv(\ctx_2.C_{g}^k)}
          &= (h_1(\ctx_1.C_{g}^k) \ket{\varphi})_{h_1(\qv(\ctx_1.C_{g}^k))}
           = \bigl(\ctx_1.C_{g}^k \ket{\psi}\bigr)_{\qv(\ctx_1.C_{g}^k)} .
          \end{align*}
          Thus \(\ket{\varphi'}_{h_1(\qv(\ctx_1.C_{g}^k))} = \ket{\psi'}_{\qv(\ctx_1.C_{g}^k)}\).

          From the validity for \(\ctx_1\) we know
          \(C_a \ket{\psi} = \ket{\psi}_{\rho_{\mathrm{loc},g,1}(v)} \otimes \ket{\psi'}_{\sys \setminus \rho_{\mathrm{loc},g,1}(v)}\).
          Since \(h_1(\rho_{g,1}(v)) = \rho_{g,2}(v)\), we obtain
          \begin{align*}
          (C_b \ket{\varphi})_{\rho_{\mathrm{loc},g,2}(v)}
          &= \ket{\psi}_{\rho_{\mathrm{loc},g,1}(v)}
           = \ket{\varphi}_{\rho_{\mathrm{loc},g,2}(v)}, \\
          (C_b \ket{\varphi})_{(\sys \setminus \rho_{\mathrm{loc},g,2}(v)) \cap \qv(C_b)}
          &= \ket{\psi'}_{(\sys \setminus \rho_{\mathrm{loc},g,1}(v)) \cap \qv(C_a)}
           = \ket{\varphi'}_{(\sys \setminus \rho_{\mathrm{loc},g,2}(v)) \cap \qv(C_b)}. 
          \end{align*}
          Hence
          \(C_b \ket{\varphi}
          = \ket{\varphi}_{\rho_{\mathrm{loc},g,2}(v)}
            \otimes \ket{\varphi'}_{(\sys \setminus \rho_{\mathrm{loc},g,2}(v)) \cap \qv(C_b)}
            \otimes \ket{\varphi'}_{\sys \setminus \rho_{\mathrm{loc},g,2}(v) \setminus \qv(C_b)}
          = \ket{\varphi}_{\rho_{\mathrm{loc},g,2}(v)}
            \otimes \ket{\varphi'}_{\sys \setminus \rho_{\mathrm{loc},g,2}(v)}\), 
          which is exactly the desired reduction for \(\overline{p}'\).

          Finally, let \(h'\) be a bijection from \(\overline{p}\) to \(\overline{p}'\),
          and define \(h_2 : \qv(\kappa_{g,2}(v)[\req_{v,2} \mapsto \overline{p}]) \to \mathcal{Q}\) by
          \(h_2(q) = h'(q)\) for \(q \in \overline{p}\) and \(h_2(q) = q\) otherwise.
         Since \(h'\) is a bijection and \(\overline{p}' \cap \cod(\rho_{g,2}) = \emptyset\),
          \(h_2\) is also a bijection.
          Moreover, again by \(\overline{p}' \cap \cod(\rho_{g,2}) = \emptyset\)
          \[
          \kappa_{g,2}(v)[\req_{v,2} \mapsto \overline{p}]
          = \kappa_{g,2}(v)[\req_{v,2} \mapsto \overline{p}'][\overline{p}' \mapsto \overline{p}]
          = h_2\bigl( \kappa_{g,2}(v)[\req_{v,2} \mapsto \overline{p}'] \bigr).
          \]
          Note that \(\ket{\varphi'} \models \ket{0}_{\overline{p} \cup \overline{p}'}\).
          Applying Lemma~\ref{lemma:rename} once more,
          \begin{align*}
          \bigl\langle \ctx_2.C_{g}^k; \kappa_{g,2}(v)[\req_{v,2} \mapsto \overline{p}], \ket{\varphi} \bigr\rangle
          &= \bigl\langle \kappa_{g,2}(v)[\req_{v,2} \mapsto \overline{p}], \ket{\varphi'} \bigr\rangle \\
          &= \bigl\langle h_2\bigl( \kappa_{g,2}(v)[\req_{v,2} \mapsto \overline{p}'] \bigr), \ket{\varphi'} \bigr\rangle \\
          &\rightarrow^{*} \bigl\langle \downarrow,\;
              \ket{\varphi}_{\rho_{\mathrm{loc},g,2}(v)} \otimes \ket{\varphi'}_{\sys \setminus \rho_{\mathrm{loc},g,2}(v)} \bigr\rangle
              \tag{since \(\overline{p} \cup \overline{p}' \subseteq \sys \setminus \rho_{g,2}(v)\)}.
          \end{align*}

          Thus \(\kappa_{g,2}\) satisfies the required reduction property.
\end{enumerate}

Therefore \(\ctx_2\) also satisfies Condition~3(a).  Together with the previous arguments, 
Condition~3(a) holds for both \(\ctx_1\) and \(\ctx_2\).
\end{proof}

\begin{lemma}\label{lem:state-valid-3.b}
  Under Notations~\ref{note:state_1},~\ref{note:state_2} and
  Assumption~\ref{hpy:com}, if \(s = \tau x \gets f(\overline{u})\) for some
  variables \(x\), \(\overline{u}\) and a function \(f\) with \(f \in \SCC(f_{\mathrm{cur}})\), then 
  Condition~3(b) of the validity of \(\ctx_2\) holds.
\end{lemma} 

\begin{proof}
Consider any \(g \in \{\ctx_2.f_{\mathrm{cur}}\} \cup \bigl( \Fun(\St_2) \cap \SCC(f_{\mathrm{cur}}) \setminus \Theta^{-1}(\dom(\D_2)) \bigr)= \{\ctx_1.f_{\mathrm{cur}}\} \cup \bigl( \Fun(\St_1) \cap \SCC(f_{\mathrm{cur}}) \setminus \Theta^{-1}(\dom(\D_1)) \bigr)= \{\ctx_0.f_{\mathrm{cur}}\} \cup \bigl( \Fun(\St_0) \cap \SCC(f_{\mathrm{cur}}) \setminus \Theta^{-1}(\dom(\D_0)) \bigr)\).

  Let \(\ket{\varphi}\) be a state satisfying
  \(\ket{\varphi} \models \ket{0}_{\ctx_2.A_g \setminus \rho_{\mathrm{loc},g,2}(\fp(g))}\).
  We must show
  \[
    \ctx_2.C_{g}^k \ket{\varphi}
    \models \ket{\varphi}_{\ctx_2.A_g \setminus \rho_{\mathrm{loc},g,2}(\fp(g)) \setminus V_{\activelabel,g,2}} .
  \]

  Define a mapping \(h\) with domain \(\qv(\ctx_1.C_{g})\) by
  \(h(q)=q[i]\) for \(q \in S\) and \(h(q)=q\) otherwise.
  Condition~4(d) in the validity of \(\ctx_2\) gives
  \(\bigl( \qv(\ctx_1.C_{g}) \setminus S \bigr) \cap S[i] = \emptyset\);
  thus \(h\) is injective, hence a bijection onto its image.

  Choose a state \(\ket{\psi}\) such that
  \(\ket{\psi} \models \ket{0}_{\ctx_1.A_g \setminus \rho_{\mathrm{loc},g,1}(\fp(g))}\)
  and
  \(\ket{\psi}_{\rho_{g,1}(\fp(g))} = \ket{\varphi}_{\rho_{\mathrm{loc},g,2}(\fp(g))}\).
  Since 
  \(\qv(C_{g,1}^k) \subseteq \ctx_1.A_g\)
  and
  \(h(\qv(C_{g,1}^k)) = \qv(h(C_{g,1}^k))
    = \qv(C_{g,1}^k[q \mapsto q[i]]) = \qv(C_{g,2}^k) \subseteq \ctx_2.A_g\) (by Lemma~\ref{lem:state-valid-4.c}),
  we have
  \(\ket{\varphi}_{h(\qv(C_{g,1}^k))} = \ket{\psi}_{\qv(C_{g,1}^k)}\).

  Applying Lemma~\ref{lemma:rename} yields
  \[
    (\ctx_2.C_{g}^k \ket{\varphi})_{\qv(\ctx_2.C_{g}^k)}
    = (h(\ctx_1.C_{g})^k \ket{\varphi})_{h(\qv(\ctx_1.C_{g}^k))}
    = (\ctx_1.C_{g}^k \ket{\psi})_{\qv(\ctx_1.C_{g}^k)} .
  \]

  From \(\ctx_1.A_g = \ctx_0.A_g\) and
  \(\rho_{g,0}(\fp(g)) = \rho_{\mathrm{loc},g,1}(\fp(g))\) we obtain
  \(\ket{\psi} \models \ket{0}_{\ctx_0.A_g \setminus \rho_{\mathrm{loc},g,0}(\fp(g))}\) (see Lemma~\ref{lem:state-valid-2}).
  The validity of \(\ctx_0\) then gives
  \[
    \ctx_0.C_{g}^k \ket{\psi}
    \models \ket{\psi}_{\ctx_0.A_g \setminus \rho_{\mathrm{loc},g,0}(\fp(g)) \setminus V_{\activelabel,g,0}} .
  \]
  Since \(\ctx_1\) does not modify the code of \(g\) nor its formal parameters
  (i.e., \(\ctx_1.C_{g} = \ctx_0.C_{g}\),
  \(\rho_{\mathrm{loc},g,1}(\fp(g)) = \rho_{\mathrm{loc},g,0}(\fp(g))\))
  and \(V_{\activelabel,g,0} \subseteq V_{\activelabel,g,1}\), we obtain
  \[
    \ctx_1.C_{g}^k \ket{\psi}
    \models \ket{\psi}_{\ctx_1.A_g \setminus \rho_{\mathrm{loc},g,1}(\fp(g)) \setminus V_{\activelabel,g,1}} . \tag{1}
  \]

  Now observe that
  \(h(\rho_{\mathrm{loc},g,1}(\fp(g)) \cup V_{\activelabel,g,1})
    = \rho_{\mathrm{loc},g,2}(\fp(g)) \cup V_{\activelabel,g,2}\).
  Therefore
  \begin{align*}
    &\bigl( \ctx_2.C_{g}^k \ket{\varphi}
      \bigr)_{\qv(\ctx_2.C_{g}^k) \cap (\ctx_2.A_g 
             \setminus \rho_{\mathrm{loc},g,2}(\fp(g)) \setminus V_{\activelabel,g,2})} \\
    = \ & \bigl( \ctx_1.C_{g}^k \ket{\psi}
      \bigr)_{\qv(\ctx_1.C_{g}^k) \cap (\ctx_1.A_g
             \setminus \rho_{\mathrm{loc},g,1}(\fp(g)) \setminus V_{\activelabel,g,1})} \\
    = \ &  \ket{\psi}_{\qv(\ctx_1.C_{g}^k) \cap
          (\ctx_1.A_g \setminus \rho_{\mathrm{loc},g,1}(\fp(g)) \setminus V_{\activelabel,g,1})} \\
    = \ & \ket{\varphi}_{\qv(\ctx_2.C_{g}^k) \cap
          (\ctx_2.A_g \setminus \rho_{\mathrm{loc},g,2}(\fp(g)) \setminus V_{\activelabel,g,2})}. 
  \end{align*}
  Hence
  \[
    \ctx_2.C_{g}^k \ket{\varphi}
    \models \ket{\varphi}_{\ctx_2.A_g \setminus \rho_{\mathrm{loc},g,2}(\fp(g)) \setminus V_{\activelabel,g,2}} .
  \]
  This establishes Condition~3(b) for \(\ctx_2\).
\end{proof} 

\begin{lemma} \label{lem:state-valid-3.c}
Under Notations~\ref{note:state_1},~\ref{note:state_2} and Assumption~\ref{hpy:com}, if \(s = \tau x \gets f(\overline{u})\) for some variables \(x\), \(\overline{u}\) and a function \(f\) with \(f \in \SCC(f_{\mathrm{cur}})\), then Condition~3(c) of the validity of \(\ctx_2\) holds.
\end{lemma}
\begin{proof}
Similarly, consider any
\(h \in \{f_{\mathrm{cur}}\} \cup \SCC(f_{\mathrm{cur}}) \cap \Fun(\St_2) \setminus \Theta^{-1}(\dom(\D_2))= \{\ctx_1.f_{\mathrm{cur}}\} \cup \bigl( \Fun(\St_1) \cap \SCC(f_{\mathrm{cur}}) \setminus \Theta^{-1}(\dom(\D_1)) \bigr)= \{\ctx_0.f_{\mathrm{cur}}\} \cup \bigl( \Fun(\St_0) \cap \SCC(f_{\mathrm{cur}}) \setminus \Theta^{-1}(\dom(\D_0)) \bigr).\) 
Since Condition~1 in the validity of \(\ctx_0\) implies that for any \(g\) with \(\pos(g) > \pos(f_{\mathrm{cur}})\) we have \(g \in \Theta^{-1}(\dom(\D_0)) = \Theta^{-1}(\dom(\D_2))\), it follows that for every \(g\) under consideration, \(\pos(g) \le \pos(f_{\mathrm{cur}})\). We shall use these facts repeatedly in the sequel without further mention.   
\begin{enumerate}
     \item \(\fp(h)\cup \re(h) \subseteq \dom(\rho_{\mathrm{loc},h,2})\):  
For any \(h\), we have \(\dom(\rho_{\mathrm{loc},h,0}) \subseteq \dom(\rho_{\mathrm{loc},h,1}) = \dom(\rho_{\mathrm{loc},h,2})\) by definition.  
Since \(\fp(h)\cup \re(h) \subseteq \dom(\rho_{\mathrm{loc},h,0})\) holds by the validity of \(\ctx_0\), the claim follows directly.

\item Lemmas~\ref{lem:com-Aset-1} and~\ref{lem:com-Aset-2} give \(\ctx_2.A_h \subseteq \ctx_0.A_h\).
By definition \(\cod(\rho_{\mathrm{glob}})\) is unchanged.
Since the validity of \(\ctx_0\) yields
\(\bigl( \ctx_0.A_h \cap \cod(\rho_{\mathrm{glob}}) \bigr) = \emptyset\),
we immediately obtain
\[
\bigl( \ctx_2.A_h \cap \cod(\rho_{\mathrm{glob}}) \bigr) = \emptyset .
\] 
\item We now prove
\[
\bigl( \ctx_2.A_h \setminus \rho_{\mathrm{loc},h,2}(\fp(h) \cup \re(h)) \bigr)
\cap
\bigl( V_{\activelabel,[f_{\mathrm{cur}}^*, h),2} \cup \cod(\rho_{\mathrm{loc},[f_{\mathrm{cur}}^*, h),2}) \bigr)
= \emptyset .
\]

% The construction of \(\ctx_1\) and \(\ctx_2\) does not modify the local mapping
% nor the \(\Vactive\) sets for any function whose stack position lies below
% \(f_{\mathrm{cur}}\); the formal parameters and return variable of \(f_{\mathrm{cur}}\) are also
% left unchanged.  Together with \(\pos(h) \le \pos(f_{\mathrm{cur}})\) we therefore have
% \begin{align*}
% \rho_{\mathrm{loc},h,2}(\fp(h) \cup \re(h)) &= \rho_{\mathrm{loc},h,0}(\fp(h) \cup \re(h)), \\
% V_{\activelabel,[f_{\mathrm{cur}}^*, h),2} \cup \cod(\rho_{\mathrm{loc},[f_{\mathrm{cur}}^*, h),2})
% &= V_{\activelabel,[f_{\mathrm{cur}}^*, h),0} \cup \cod(\rho_{\mathrm{loc},[f_{\mathrm{cur}}^*, h),0}).
% \end{align*}
% Lemmas~\ref{lem:com-Aset-1} and~\ref{lem:com-Aset-2} give \(\ctx_2.A_h \subseteq \ctx_0.A_h\); 
% hence
% \[
% \ctx_2.A_h \setminus \rho_{\mathrm{loc},h,2}(\fp(h) \cup \re(h))
% \subseteq \ctx_0.A_h \setminus \rho_{\mathrm{loc},h,0}(\fp(h) \cup \re(h)).
% \]

% The validity of \(\ctx_0\) supplies
% \[
% \bigl( \ctx_0.A_h \setminus \rho_{\mathrm{loc},h,0}(\fp(h) \cup \re(h)) \bigr)
% \cap
% \bigl( V_{\activelabel,[f_{\mathrm{cur}}^*, h),0} \cup \cod(\rho_{\mathrm{loc},[f_{\mathrm{cur}}^*, h),0}) \bigr)
% = \emptyset,
% \]
% and the required conclusion follows immediately.  

By construction we have
  \(
  \rho_{\mathrm{loc},1,h}(\fp(h) \cup \re(h))
  =
  \rho_{\mathrm{loc},0,h}(\fp(h) \cup \re(h))\),
  and \(
  V_{\activelabel,1,[f_{\mathrm{cur}}^*, h)}
   \cup \cod(\rho_{\mathrm{loc},1,[f_{\mathrm{cur}}^*, h)})
  =
  V_{\activelabel,0,[f_{\mathrm{cur}}^*, h)}
   \cup \cod(\rho_{\mathrm{loc},0,[f_{\mathrm{cur}}^*, h)}).
  \)
  Using the definition of \(\ctx_2\) we obtain the following chain of inclusions.
  For the left-hand side, let \(q \in S\); then
  \[
  \begin{aligned}
    &\ctx_2.A_h \setminus \rho_{\mathrm{loc},2,h}(\fp(h) \cup \re(h)) \\
    =  \ & \ctx_2.A_h \setminus
       \bigl( \rho_{\mathrm{loc},1,h}(\fp(h) \cup \re(h))[q \mapsto q[i]] \bigr) \\
    \subseteq \ & \ctx_1.A_h \setminus
       \bigl( (\rho_{\mathrm{loc},1,h}(\fp(h) \cup \re(h)) \setminus S) \cup S[i] \bigr) \\
    \subseteq \ & \bigl( \ctx_1.A_h \setminus \rho_{\mathrm{loc},1,h}(\fp(h) \cup \re(h)) \bigr)
       \setminus S[i] \;\cup\; S \\
    = \ & \bigl( \ctx_0.A_h \setminus \rho_{\mathrm{loc},0,h}(\fp(h) \cup \{\re(h)\}) \bigr)
       \setminus S[i] \;\cup\; S .
  \end{aligned}
  \]
  The second step relies on \(\ctx_2.A_h \subseteq \ctx_1.A_h\), which follows from Lemma~\ref{lem:state-inv-1} together with the definition of \(\ctx_2\).
  The third step uses the general set-theoretic inclusion
  \(A \setminus ((B \setminus S) \cup S[i]) \subseteq (A \setminus B) \setminus S[i] \cup S\).
  The last equality is justified by \(\ctx_1.A_h = \ctx_0.A_h\), also obtained from Lemma~\ref{lem:state-inv-1}.

  For the right-hand side,
  \[
  \begin{aligned}
    &V_{\activelabel,2,[f_{\mathrm{cur}}^*, h)}
      \cup \cod(\rho_{\mathrm{loc},2,[f_{\mathrm{cur}}^*, h)}) \\
    = \ & \bigl(
        V_{\activelabel,1,[f_{\mathrm{cur}}^*, h)}
        \cup \cod(\rho_{\mathrm{loc},1,[f_{\mathrm{cur}}^*, h)})
       \bigr)[q \mapsto q[i]] \\
    = \ & \bigl(
        V_{\activelabel,1,[f_{\mathrm{cur}}^*, h)}
        \cup \cod(\rho_{\mathrm{loc},1,[f_{\mathrm{cur}}^*, h)})
        \setminus S
       \bigr) \cup S[i] \\
    = \ & \bigl(
        V_{\activelabel,0,[f_{\mathrm{cur}}^*, h)}
        \cup \cod(\rho_{\mathrm{loc},0,[f_{\mathrm{cur}}^*, h)})
        \setminus S
       \bigr) \cup S[i].
  \end{aligned}
  \]

  The validity of \(\ctx_0\) supplies
  \[
  \bigl( \ctx_0.A_h \setminus
  \rho_{\mathrm{loc},0,h}(\fp(h) \cup \re(h)) \bigr)
  \cap
  \bigl(
  V_{\activelabel,0,[f_{\mathrm{cur}}^*, h)}
  \cup \cod(\rho_{\mathrm{loc},0,[f_{\mathrm{cur}}^*, h)})
  \bigr)
  = \emptyset .
  \]

  Now compare the two sides. The left-hand side is contained in
  \(\bigl( (A_0 \setminus B_0) \setminus S[i] \bigr) \cup S\), where
  \(A_0 = \ctx_0.A_h\) and
  \(B_0 = \rho_{\mathrm{loc},0,h}(\fp(h) \cup \re(h))\).
  The right-hand side equals \((U_0 \setminus S) \cup S[i]\), where
  \[
  U_0 =
  V_{\activelabel,0,[f_{\mathrm{cur}}^*, h)}
  \cup \cod(\rho_{\mathrm{loc},0,[f_{\mathrm{cur}}^*, h)}).
  \]
  Their intersection therefore satisfies
  \[
  \begin{aligned}
  &\bigl( ((A_0 \setminus B_0) \setminus S[i]) \cup S \bigr)
    \cap \bigl( (U_0 \setminus S) \cup S[i] \bigr) \\
  =&\bigl( ((A_0 \setminus B_0) \setminus S[i]) \cap (U_0 \setminus S) \bigr)
     \cup \bigl( ((A_0 \setminus B_0) \setminus S[i]) \cap S[i] \bigr) \\
  &\quad \cup \bigl( S \cap (U_0 \setminus S) \bigr)
     \cup \bigl( S \cap S[i] \bigr).
  \end{aligned}
  \]
  The first term is contained in
  \((A_0 \setminus B_0) \cap U_0 = \emptyset\) by the validity of \(\ctx_0\);
  the second and third terms are obviously empty;
  \(S \cap S[i] = \emptyset\) because \(S\) and \(S[i]\) use distinct indices.
  Thus the whole intersection is empty, which proves the required disjointness. 

\item \(\rho_{\mathrm{loc},h,2}(\fp(h)) \cap V_{\activelabel,h,2} = \emptyset\):

% For any \(h \neq f_{\mathrm{cur}}\), the condition \(\pos(h) \le \pos(f_{\mathrm{cur}})\) established above 
% actually yields \(\pos(h) < \pos(f_{\mathrm{cur}})\). 
% Thus, as before, the construction of \(\ctx_1\) and \(\ctx_2\) leaves the local
% mapping of \(h\) and its \(\Vactive\) set unchanged;
% hence \(\rho_{\mathrm{loc},h,2}(\fp(h)) = \rho_{\mathrm{loc},h,0}(\fp(h))\) and \(V_{\activelabel,h,2} = V_{\activelabel,h,0}\).
% The required condition for such \(h\) follows directly from the validity of \(\ctx_0\).

% We now treat the case \(h = f_{\mathrm{cur}}\).

First consider \(\ctx_1\). We only need to treat the case \(h = f_{\mathrm{cur}}\). 
Assume \(x \notin \dom(\rho_{f_{\mathrm{cur}},0})\).
Since \(\fp(f_{\mathrm{cur}}) \subseteq \dom(\rho_{\mathrm{loc},f_{\mathrm{cur}},0})\), the construction gives
\[
\rho_{\mathrm{loc},f_{\mathrm{cur}},1}(\fp(f_{\mathrm{cur}})) = \rho_{\mathrm{loc},f_{\mathrm{cur}},0}(\fp(f_{\mathrm{cur}})), \qquad
V_{\activelabel,f_{\mathrm{cur}},1} = V_{\activelabel,f_{\mathrm{cur}},0} \cup \rho_{\mathrm{loc},f_{\mathrm{cur}},1}(x),
\]
with \(\rho_{\mathrm{loc},f_{\mathrm{cur}},1}(x) \subseteq \Qpool_0\).
The validity of \(\ctx_0\) gives
\[
\rho_{\mathrm{loc},f_{\mathrm{cur}},0}(\fp(f_{\mathrm{cur}})) \cap V_{\activelabel,f_{\mathrm{cur}},0} = \emptyset, \qquad
\Qpool_0 \cap \rho_{\mathrm{loc},f_{\mathrm{cur}},0}(\fp(f_{\mathrm{cur}})) = \emptyset.
\]
Consequently,
\[
\rho_{\mathrm{loc},f_{\mathrm{cur}},1}(\fp(f_{\mathrm{cur}})) \cap V_{\activelabel,f_{\mathrm{cur}},1}
= \rho_{\mathrm{loc},f_{\mathrm{cur}},0}(\fp(f_{\mathrm{cur}})) \cap \bigl( V_{\activelabel,f_{\mathrm{cur}},0} \cup \rho_{\mathrm{loc},f_{\mathrm{cur}},1}(x) \bigr) = \emptyset. 
\]

If \(x \in \dom(\rho_{f_{\mathrm{cur}},0})\), then the construction gives
\(\rho_{\mathrm{loc},f_{\mathrm{cur}},1}(\fp(f_{\mathrm{cur}})) = \rho_{\mathrm{loc},f_{\mathrm{cur}},0}(\fp(f_{\mathrm{cur}}))\) and
\(V_{\activelabel,f_{\mathrm{cur}},1} = V_{\activelabel,f_{\mathrm{cur}},0} \cup (\rho_{\mathrm{loc},f_{\mathrm{cur}},0}(x) \setminus \rho_{\mathrm{loc},f_{\mathrm{cur}},0}(\fp(f_{\mathrm{cur}})))\).
It immediately follows from the validity of \(\ctx_0\) that
\[
\rho_{\mathrm{loc},f_{\mathrm{cur}},1}(\fp(f_{\mathrm{cur}})) \cap V_{\activelabel,f_{\mathrm{cur}},1}
= \rho_{\mathrm{loc},f_{\mathrm{cur}},0}(\fp(f_{\mathrm{cur}})) \cap \bigl( V_{\activelabel,f_{\mathrm{cur}},0} \cup (\rho_{\mathrm{loc},f_{\mathrm{cur}},0}(x) \setminus \rho_{\mathrm{loc},f_{\mathrm{cur}},0}(\fp(f_{\mathrm{cur}}))) \bigr)
= \emptyset .
\]

Now turn to \(\ctx_2\).  
Note that \(\rho_{\mathrm{loc},h,1}(\fp(h)) \cap V_{\activelabel,h,1} = \emptyset\).

By definition,
\[
\begin{aligned}
\rho_{\mathrm{loc},h,2}(\fp(h)) &= \rho_{\mathrm{loc},h,1}(\fp(h))[q \mapsto q[i]] \\ 
V_{\activelabel,h,2} &= V_{\activelabel,h,1}[q \mapsto q[i]].
\end{aligned}
\]
Since the renaming is bijective, it preserves disjointness of sets. Therefore,
\[
\rho_{\mathrm{loc},h,2}(\fp(h)) \cap V_{\activelabel,h,2} = \emptyset.
\]
% Note that \(\rho_{\mathrm{loc},f_{\mathrm{cur}},1}(\fp(f_{\mathrm{cur}})) \cap V_{\activelabel,f_{\mathrm{cur}},1} = \emptyset\). 

% % the renaming
% % operation does not affect the image of the formal parameters, so
% By definition,
% \[
% \begin{aligned}
% \rho_{\mathrm{loc},f_{\mathrm{cur}},2}(\fp(f_{\mathrm{cur}})) &= \rho_{\mathrm{loc},f_{\mathrm{cur}},1}(\fp(f_{\mathrm{cur}}))[q \mapsto q[i]] \\ 
% V_{\activelabel,f_{\mathrm{cur}},2} &= V_{\activelabel,f_{\mathrm{cur}},1}[q \mapsto q[i]].
% \end{aligned}
% \]
% Since the renaming is bijective, it preserves disjointness of sets. Therefore,
% \[
% \rho_{\mathrm{loc},f_{\mathrm{cur}},2}(\fp(f_{\mathrm{cur}})) \cap V_{\activelabel,f_{\mathrm{cur}},2} = \emptyset.
% \]
% We have \(S[i] \subseteq S[+] \subseteq cv_1\) by Lemma~\ref{lem:S-plus},
% and the validity of \(\ctx_1\) implies \(cv_1 \cap \rho_{\mathrm{loc},f_{\mathrm{cur}},1}(\fp(f_{\mathrm{cur}})) = \emptyset\).  
% Therefore
% \[
% \rho_{\mathrm{loc},f_{\mathrm{cur}},2}(\fp(f_{\mathrm{cur}})) \cap V_{\activelabel,f_{\mathrm{cur}},2}
% = \rho_{\mathrm{loc},f_{\mathrm{cur}},1}(\fp(f_{\mathrm{cur}})) \cap \bigl( (V_{\activelabel,f_{\mathrm{cur}},1} \setminus S) \cup S[i] \bigr) = \emptyset .
% \]

% Thus \(\rho_{\mathrm{loc},h,2}(\fp(h)) \cap V_{\activelabel,h,2} = \emptyset\) holds for all \(h\).

\item We need to show that
\(
\rho_{\mathrm{loc},h,2}(\re(h)) \subseteq V_{\activelabel,\mathrm{pred}(h),2}.
\)

First, for any \(g\) with \(\pos(g) < \pos(f_{\mathrm{cur}})\), the construction of \(\ctx_1\) leaves both the local mapping \(\rho_{\mathrm{loc}}\) and the active-variable set \(\Vactive\) of \(g\) unchanged. For the current function \(f_{\mathrm{cur}}\), the construction of \(\ctx_1\) also preserves the quantum variables corresponding to its return value. Hence, from the validity of \(\ctx_0\), it immediately follows that \(\ctx_1\) satisfies the same condition for any $h$ considered here. 

Moreover, the transition from \(\ctx_1\) to \(\ctx_2\) only involves a bijective renaming of variables. Since bijective substitutions preserve subset relations, the property remains valid in \(\ctx_2\):
\[
\rho_{\mathrm{loc},h,2}(\re(h)) \subseteq V_{\activelabel,\mathrm{pred}(h),2}.
\] 
% By the same reasoning as before, since \(\pos(h) \le \pos(f_{\mathrm{cur}})\), the construction of \(\ctx_1\) does not alter the mapping of the formal parameters and the return variable of \(h\); hence
% \(
% \rho_{\mathrm{loc},h,2}(\re(h)) = \rho_{\mathrm{loc},h,0}(\re(h)). 
% \) 
% Furthermore, the \(\Vactive\) sets for any function whose stack position lies below \(f_{\mathrm{cur}}\) remain unchanged; therefore, by definition, \(V_{\activelabel,\mathrm{pred}(h),2}= V_{\activelabel,\mathrm{pred}(h),0}\). 
% The corresponding condition holds for \(\ctx_0\) by its validity, therefore it follows immediately for \(\ctx_2\). 

% For \(ctx_2\): the renaming operation gives
% \[
% \rho_{h,2}(\{\re(h)\} \cup \fp(h))
% = \rho_{h,1}(\{\re(h)\} \cup \fp(h))[q \mapsto q[i]]
% \]
% and, similarly,
% \[
% ucv_{[\curf^*,h),2} \cup \cod(\rho_{[\curf^*,h),2})
% = \bigl( ucv_{[\curf^*,h),1} \cup \cod(\rho_{[\curf^*,h),1}) \bigr)[q \mapsto q[i]].
% \]
% Because we already have
% \(\rho_{h,1}(\{\re(h)\} \cup \fp(h))
% \subseteq ucv_{[\curf^*,h),1} \cup \cod(\rho_{[\curf^*,h),1})\),
% applying the same renaming to both sides yields the required inclusion for \(ctx_2\).

\item \(\base(\cod(\rho_{\mathrm{loc},h,2}) \setminus \rho_{\mathrm{loc},h,2}(\fp(h)) \setminus \rho_{\mathrm{loc},h,2}(\re(h))) \subseteq V_{\alloc,h,2}\).

For \(\ctx_1\), if \(h \neq f_{\mathrm{cur}}\) or \((h = f_{\mathrm{cur}} \land x \in \dom(\rho_{f_{\mathrm{cur}},0}))\), then
\(\rho_{\mathrm{loc},h,1} = \rho_{\mathrm{loc},h,0}\) and \(V_{\alloc,h,1} = V_{\alloc,h,0}\); the inclusion follows
directly from the validity of \(\ctx_0\).

If \(h = f_{\mathrm{cur}} \land x \notin \dom(\rho_{f_{\mathrm{cur}},0})\), then
\begin{align*}
    &\base\bigl( \cod(\rho_{\mathrm{loc},f_{\mathrm{cur}},1}) \setminus \rho_{\mathrm{loc},f_{\mathrm{cur}},1}(\re(f_{\mathrm{cur}})) \setminus \rho_{\mathrm{loc},f_{\mathrm{cur}},1}(\fp(f_{\mathrm{cur}})) \bigr) \\
\subseteq \  & \base\!\Bigl( \bigl( \cod(\rho_{\mathrm{loc},f_{\mathrm{cur}},0}) \cup \rho_{\mathrm{loc},f_{\mathrm{cur}},1}(x) \bigr)
                      \setminus \rho_{\mathrm{loc},f_{\mathrm{cur}},0}(\re(f_{\mathrm{cur}}))
                      \setminus \rho_{\mathrm{loc},f_{\mathrm{cur}},0}(\fp(f_{\mathrm{cur}})) \Bigr) \\
 \subseteq \  & \base\!\Bigl( \bigl( \cod(\rho_{\mathrm{loc},f_{\mathrm{cur}},0})
                      \setminus \rho_{\mathrm{loc},f_{\mathrm{cur}},0}(\re(f_{\mathrm{cur}}))
                      \setminus \rho_{\mathrm{loc},f_{\mathrm{cur}},0}(\fp(f_{\mathrm{cur}})) \bigr)
                      \cup \rho_{\mathrm{loc},f_{\mathrm{cur}},1}(x) \Bigr) \\
    &\qquad \tag{by \((A \cup B) \setminus C \subseteq (A \setminus C) \cup B\)} \\
 =  \  & \base\bigl( \cod(\rho_{\mathrm{loc},f_{\mathrm{cur}},0})
                      \setminus \rho_{\mathrm{loc},f_{\mathrm{cur}},0}(\re(f_{\mathrm{cur}}))
                      \setminus \rho_{\mathrm{loc},f_{\mathrm{cur}},0}(\fp(f_{\mathrm{cur}})) \bigr)
            \cup \base(\rho_{\mathrm{loc},f_{\mathrm{cur}},1}(x)) \\
    &\qquad \tag{since \(\base(A \cup B) = \base(A) \cup \base(B)\)} \\
 \subseteq \  & V_{\alloc,f_{\mathrm{cur}},0} \cup \rho_{\mathrm{loc},f_{\mathrm{cur}},1}(x)
      = V_{\alloc,f_{\mathrm{cur}},1}. 
\end{align*}

For \(\ctx_2\) the conclusion follows from the derivation below.
\begin{align*}
&\base\bigl( \cod(\rho_{\mathrm{loc},h,2}) \setminus \rho_{\mathrm{loc},h,2}(\fp(h)) \setminus \rho_{\mathrm{loc},h,2}(\re(h)) \bigr) \\
= &\base\!\Bigl( \cod(\rho_{\mathrm{loc},h,1})[q \mapsto q[i]] \Bigr)
   \setminus \Bigl( \rho_{\mathrm{loc},h,1}(\re(h))[q \mapsto q[i]] \Bigr)
   \setminus \Bigl( \rho_{\mathrm{loc},h,1}(\fp(h))[q \mapsto q[i]] \Bigr) \\
= &\base\!\Bigl( \bigl( \cod(\rho_{\mathrm{loc},h,1}) \setminus \rho_{\mathrm{loc},h,1}(\re(h)) \setminus \rho_{\mathrm{loc},h,1}(\fp(h)) \bigr)[q \mapsto q[i]] \Bigr) \\
  &\qquad \tag{since both \(\rho_{\mathrm{loc},h,1}\), \(\rho_{\mathrm{loc},h,2}\) and the renaming are injective} \\
= &\base\bigl( \cod(\rho_{\mathrm{loc},h,1}) \setminus \rho_{\mathrm{loc},h,1}(\fp(h)) \setminus \rho_{\mathrm{loc},h,1}(\re(h)) \bigr) \\
= &\,V_{\alloc,h,1} = V_{\alloc,h,2}.
  \tag{by the validity of \(\ctx_1\) and the definition of \(V_{\alloc,h,2}\)}
\end{align*}

\end{enumerate}
\end{proof}

\begin{lemma} \label{lem:state-valid-4.a}
Under Notations~\ref{note:state_1},~\ref{note:state_2} and Assumption~\ref{hpy:com}, if \(s = \tau x \gets f(\overline{u})\) for some variables \(x\), \(\overline{u}\) and a function \(f\) with \(f \in \SCC(f_{\mathrm{cur}})\), then Condition~4(a) of the validity of \(\ctx_2\) holds.
\end{lemma}
\begin{proof}
Now consider \(h \in \SCC(f_{\mathrm{cur}}) \cap \Fun(\St_2) \). 
\begin{enumerate}

\item \(\cod(\rho_{\mathrm{loc},h,2}) \setminus \rho_{\mathrm{loc},h,2}(\re(h)) \setminus \rho_{\mathrm{loc},h,2}(\fp(h)) \subseteq V_{\activelabel,h,2}\):

For \(\ctx_1\), if \(h \neq f_{\mathrm{cur}}\), then \(\rho_{\mathrm{loc},h,1} = \rho_{\mathrm{loc},h,0}\) and
\(V_{\activelabel,h,1} = V_{\activelabel,h,0}\); the inclusion follows directly from the validity
of \(\ctx_0\).
For \(h = f_{\mathrm{cur}}\), if \(x \in \dom(\rho_{f_{\mathrm{cur}},0})\), then
\(\rho_{\mathrm{loc},f_{\mathrm{cur}},1} = \rho_{\mathrm{loc},f_{\mathrm{cur}},0}\) and
\(V_{\activelabel,f_{\mathrm{cur}},0} \subseteq V_{\activelabel,f_{\mathrm{cur}},1}\); again the inclusion follows
directly from the validity of \(\ctx_0\).
If \(x \notin \dom(\rho_{f_{\mathrm{cur}},0})\),
\begin{align*}
    &\cod(\rho_{\mathrm{loc},f_{\mathrm{cur}},1}) \setminus \rho_{\mathrm{loc},f_{\mathrm{cur}},1}(\re(f_{\mathrm{cur}})) \setminus \rho_{\mathrm{loc},f_{\mathrm{cur}},1}(\fp(f_{\mathrm{cur}})) \\
 \subseteq \  & \bigl( \cod(\rho_{\mathrm{loc},f_{\mathrm{cur}},0}) \cup \rho_{\mathrm{loc},f_{\mathrm{cur}},1}(x) \bigr)
                      \setminus \rho_{\mathrm{loc},f_{\mathrm{cur}},0}(\re(f_{\mathrm{cur}}))
                      \setminus \rho_{\mathrm{loc},f_{\mathrm{cur}},0}(\fp(f_{\mathrm{cur}})) \\
 \subseteq \  & \bigl( \cod(\rho_{\mathrm{loc},f_{\mathrm{cur}},0})
                      \setminus \rho_{\mathrm{loc},f_{\mathrm{cur}},0}(\re(f_{\mathrm{cur}}))
                      \setminus \rho_{\mathrm{loc},f_{\mathrm{cur}},0}(\fp(f_{\mathrm{cur}})) \bigr)
                  \cup \rho_{\mathrm{loc},f_{\mathrm{cur}},1}(x) \\
 \subseteq \  & V_{\activelabel,f_{\mathrm{cur}},0} \cup \rho_{\mathrm{loc},f_{\mathrm{cur}},1}(x)
      = V_{\activelabel,f_{\mathrm{cur}},1}.
\end{align*}

For \(\ctx_2\):
\[
\begin{aligned}
&\cod(\rho_{\mathrm{loc},h,2}) \setminus \rho_{\mathrm{loc},h,2}(\re(h)) \setminus \rho_{\mathrm{loc},h,2}(\fp(h)) \\
= \  & \bigl( \cod(\rho_{\mathrm{loc},h,1}) \setminus \rho_{\mathrm{loc},h,1}(\re(h)) \setminus \rho_{\mathrm{loc},h,1}(\fp(h)) \bigr)[q \mapsto q[i]] \\
\subseteq \  & V_{\activelabel,h,1}[q \mapsto q[i]] \\
= \  & V_{\activelabel,h,2}.
\end{aligned}
\] 

\item We need to prove that for any \(h \in \Theta^{-1}(\dom(\D_2)) \setminus \{f_{\mathrm{cur}}\}\),
\[
\cod(\rho_{\mathrm{loc},h,2}) \cap \cod(\rho_{\mathrm{loc},[f_{\mathrm{cur}}^*,h],2}) = \emptyset.
\]
The transition from \(\ctx_0\) to \(\ctx_1\) leaves the relevant compilation variables unchanged, and the step from \(\ctx_1\) to \(\ctx_2\) merely performs a synchronous renaming. Hence the property follows directly from the validity of \(\ctx_0\). 
\item We need to show that for every \(q \in V_{\activelabel,\succeq f_{\mathrm{cur}}^* ,2} \setminus \bigl(V_{\activelabel,f_{\mathrm{cur}},2} \cup \rho_{\mathrm{loc},f_{\mathrm{cur}},2}(\re(f_{\mathrm{cur}}))\bigr)\), we have \(\var(\idx[q]) \neq \emptyset\). By construction, it follows that
\[
V_{\activelabel,\succeq f_{\mathrm{cur}}^* ,2} \setminus \bigl(V_{\activelabel,f_{\mathrm{cur}},2} \cup \rho_{\mathrm{loc},f_{\mathrm{cur}},2}(\re(f_{\mathrm{cur}}))\bigr)
\subseteq V_{\activelabel,\succeq f_{\mathrm{cur}}^* ,0} \setminus \bigl(V_{\activelabel,f_{\mathrm{cur}},0} \cup \rho_{\mathrm{loc},f_{\mathrm{cur}},0}(\re(f_{\mathrm{cur}}))\bigr),
\]
hence the condition for \(\ctx_2\) is inherited directly from the validity of \(\ctx_0\).  

% The validity of \(ctx_1\) implies that for every
% \(q \in ucv_{h,1} \setminus \rho_{l,h,1}(\re(h))\) we already have
% \(\var(\idx(q)) \neq \emptyset\).  This also yields
% \(\bigl( ucv_{h,1} \setminus \rho_{l,h,1}(\re(h)) \bigr) \cap S = \emptyset\).
% Consequently,
% \[
% ucv_{h,2} \setminus \rho_{l,h,2}(\re(h))
% = ucv_{h,1} \setminus \rho_{l,h,1}(\re(h)),
% \]
% and the condition therefore remains valid in \(ctx_2\). 

We need to show that for every \(q \in V_{\activelabel,f_{\mathrm{cur}},2} \cup \rho_{\mathrm{loc},f_{\mathrm{cur}},2}(\re(f_{\mathrm{cur}}))\), if \(\var(\idx[q]) = \emptyset\) then \(\base[q] = q\).

We first analyse \(\ctx_1\). If \(x \in \dom(\rho_{f_{\mathrm{cur}},0})\), then
\[
V_{\activelabel,f_{\mathrm{cur}},1} \cup \rho_{\mathrm{loc},f_{\mathrm{cur}},1}(\re(f_{\mathrm{cur}}))
= V_{\activelabel,f_{\mathrm{cur}},0} \cup (\rho_{\mathrm{loc},f_{\mathrm{cur}},0}(x) \setminus \rho_{\mathrm{loc},f_{\mathrm{cur}},0}(\fp(f_{\mathrm{cur}}))) \cup \rho_{\mathrm{loc},f_{\mathrm{cur}},0}(\re(f_{\mathrm{cur}})).
\]
The validity of \(\ctx_0\) already guarantees that for every
\(q \in V_{\activelabel,f_{\mathrm{cur}},0} \cup \rho_{\mathrm{loc},f_{\mathrm{cur}},0}(\re(f_{\mathrm{cur}}))\),
\(\var(\idx[q]) = \emptyset\) implies \(\base[q] = q\).
Moreover, the validity of \(\ctx_0\) also gives
\[
\rho_{\mathrm{loc},f_{\mathrm{cur}},0}(x) \setminus \rho_{\mathrm{loc},f_{\mathrm{cur}},0}(\fp(f_{\mathrm{cur}}))
\subseteq V_{\activelabel,f_{\mathrm{cur}},0} \cup \rho_{\mathrm{loc},f_{\mathrm{cur}},0}(\re(f_{\mathrm{cur}})),
\]
hence the required property holds for the newly added part as well.

If \(x \notin \dom(\rho_{f_{\mathrm{cur}},0})\), then
\[
V_{\activelabel,f_{\mathrm{cur}},1} \cup \rho_{\mathrm{loc},f_{\mathrm{cur}},1}(\re(f_{\mathrm{cur}}))
= V_{\activelabel,f_{\mathrm{cur}},0} \cup \rho_{\mathrm{loc},f_{\mathrm{cur}},1}(x) \cup \rho_{\mathrm{loc},f_{\mathrm{cur}},0}(\re(f_{\mathrm{cur}})),
\]
with \(\rho_{\mathrm{loc},f_{\mathrm{cur}},1}(x) \in \Qpool_0\).
The validity of \(\ctx_0\) tells us that for every \(r \in \Qpool_0\) with \(\var(\idx[r]) = \emptyset\), we have \(\base[r] = r\).
Together with the property for \(V_{\activelabel,f_{\mathrm{cur}},0}\) inherited from \(\ctx_0\), the condition follows for \(\ctx_1\).

The step from \(\ctx_1\) to \(\ctx_2\) renames variables by \(q' \mapsto q'[i]\) for $q' \in S$. 
By the construction of \(\ctx_2\), for every
\(q \in V_{\activelabel,f_{\mathrm{cur}},2} \cup \rho_{f_{\mathrm{cur}},2}(\re(f_{\mathrm{cur}}))\):
\begin{itemize}
    \item If \(q \in V_{\activelabel,f_{\mathrm{cur}},1} \cup \rho_{f_{\mathrm{cur}},1}(\re(f_{\mathrm{cur}}))\), then the condition holds directly from the validity of \(\ctx_1\).
    \item Otherwise, we have \(q = q'[i]\) for some \(q' \in S\). By construction, \(\var(\idx[q]) = \{i\} \neq \emptyset\); therefore the condition holds trivially for \(\ctx_2\).
\end{itemize} 

\item For every \(q \in V_{\activelabel,\succeq f_{\mathrm{cur}}^* ,2} \cap \rho_{\mathrm{loc},f_{\mathrm{cur}},2}(\re(f_{\mathrm{cur}}))\), if \(\var(\idx[q]) = \emptyset\) then \(q \in \rho_{\mathrm{loc},h,2}(\re(h))\).

The construction of \(\ctx_1\) does not modify the mapping of return values, so the condition also holds for \(\ctx_1\).

Now assume some \(q \in V_{\activelabel,\succeq f_{\mathrm{cur}}^* ,2} \cap \rho_{\mathrm{loc},f_{\mathrm{cur}},2}(\re(f_{\mathrm{cur}}))\) with \(\var(\idx[q]) = \emptyset\). By definition, we have
\[
\rho_{\mathrm{loc},f_{\mathrm{cur}},2}(\re(f_{\mathrm{cur}})) = \rho_{\mathrm{loc},f_{\mathrm{cur}},1}(\re(f_{\mathrm{cur}}))[q' \mapsto q'[i]] \quad\text{for } q' \in S.
\]
Thus \(q\) falls into one of the following two cases.
\begin{itemize}
    \item If \(q \in V_{\activelabel,\succeq f_{\mathrm{cur}}^*, 1} \cap \rho_{\mathrm{loc},f_{\mathrm{cur}},1}(\re(f_{\mathrm{cur}}))\), then \(q \cap S = \emptyset\). By the validity of \(\ctx_1\), we have \(q \in \rho_{\mathrm{loc},h,1}(\re(h))\). Since \(q \cap S = \emptyset\), the definition yields \(q \in \rho_{\mathrm{loc},h,2}(\re(h))\) since \(\rho_{\mathrm{loc},h,2}(\re(h)) = \rho_{\mathrm{loc},h,1}(\re(h))[q' \mapsto q'[i]]\) for $q' \in S$. 

    \item Otherwise, \(q = q'[i]\) for some \(q' \in S\). This implies \(\var(\idx[q]) \neq \emptyset\); hence the premise of the implication is false, and the implication holds trivially.
\end{itemize} 
% \(\rho_{l,h,2}(\re(h)) = \rho_{l,h,1}(\re(h))\) and the condition holds.
% If \(\rho_{l,h,1}(\re(h)) \cap S \neq \emptyset\), then by construction
% \(\rho_{l,h,2}(\re(h)) = \rho_{l,h,1}(\re(h))[q \mapsto q[i]]\), which implies
% \(\idx(\rho_{l,h,2}(\re(h))) \neq \emptyset\); thus the premise is false and
% the implication holds trivially. 

% \item For $f \in \Theta^{-1}(\dom(\D_2))$, $\fp(f)\cup \re(f) \notin \dom(\rho_{f,2})$ and $ucv_{f,2} \subseteq Q_{\succeq f,2}$. 
% Similarly, for all such functions, the local mapping  and $V_{\activelabel,f}$ remains unchanged throughout the
% transformation and $Q_{\succeq f,0}\subseteq Q_{\succeq f,2}$; hence the required condition continues to hold. 
\end{enumerate}

Thus all required containments and disjointness properties are satisfied.
\end{proof}

\begin{lemma}
    \label{lem:Gar_0-empty}
  Under Notations~\ref{note:state_1},~\ref{note:state_2} and the assumptions,
  if \(s = \tau x \gets f(\overline{u})\) for some variables \(x\), \(\overline{u}\)
  and a function \(f\) with \(f \in \SCC(f_{\mathrm{cur}})\), then
  \(\ctx_0.\Gar_{f_{\mathrm{cur}}} \cap V_{\activelabel,f_{\mathrm{cur}},0} = \emptyset\).
\end{lemma}

\begin{proof}
  Note that \(x \in \var(s)\). Since \(s\) is a prefix of \(R_{\mathrm{stmt}}\) and the validity of \(\ctx_0\) guarantees that \(R_{\mathrm{stmt}}\) is a suffix of \(s_{f_{\mathrm{cur}}}\), it follows that \(s\) is a subsequence of \(s_{f_{\mathrm{cur}}}\). Thus $x \in \var(s_{f_{\mathrm{cur}}})$. Since \(f \in \SCC(f_{\mathrm{cur}})\), by the definition of flags we obtain \(x.\flag \neq 0\).  

  \begin{itemize}
    \item If \(x.\flag = 1\), Lemma~\ref{lem:static-wf} implies that
          no variable in \(\SCC(f_{\mathrm{cur}})\) has flag \(2\).
          Since \(\dom(\rho_{\mathrm{loc},h,0}) \subseteq \mathsf{Var}(G_{h,0})\)
          for every \(h \in \Fun(\St_0) \cap \SCC(f_{\mathrm{cur}})\),
          we obtain \(\Gar_{f_{\mathrm{cur}}} = \emptyset\).

    \item If \(x.\flag = 2\), Lemma~\ref{lem:static-wf} tells us that
          apart from \(x\), no variable in \(f_{\mathrm{cur}}\) carries flag \(2\).
          Moreover, \(x \notin \dom(\rho_{\mathrm{loc},f_{\mathrm{cur}},0})\) and
          \(x \notin \fp(f_{\mathrm{cur}}) \cup \re(f_{\mathrm{cur}})\);
          the validity of \(\ctx_0\) gives
          \(\dom(\rho_{\mathrm{loc},f_{\mathrm{cur}},0}) \setminus (\fp(f_{\mathrm{cur}}) \cup \re(f_{\mathrm{cur}}))
           \subseteq V_{\activelabel,f_{\mathrm{cur}},0}\),
          hence \(x \notin V_{\activelabel,f_{\mathrm{cur}},0}\).
          Since \(\rho_{\mathrm{loc},f_{\mathrm{cur}},0}(x) \notin V_{\activelabel,f_{\mathrm{cur}},0}\) and no other variable
          can have flag \(2\), there exists no variable with flag \(2\)
          whose image lies in \(V_{\activelabel,f_{\mathrm{cur}},0}\).
          The validity of \(\ctx_0\) therefore yields
          \(\ctx_0.\Gar_{f_{\mathrm{cur}}} \cap V_{\activelabel,f_{\mathrm{cur}},0} = \emptyset\).
  \end{itemize}
\end{proof} 

\begin{lemma} \label{lem:state-valid-4.b}
Under Notations~\ref{note:state_1},~\ref{note:state_2} and Assumption~\ref{hpy:com}, if \(s = \tau x \gets f(\overline{u})\) for some variables \(x\), \(\overline{u}\) and a function \(f\) with \(f \in \SCC(f_{\mathrm{cur}})\), then Condition~4(b) of the validity of \(\ctx_2\) holds.
\end{lemma}
\begin{proof}
 Now consider any
\(h \in \SCC(f_{\mathrm{cur}}) \cap \Fun(\St_2)\). We check each condition one by one. 
    \begin{itemize}
    \item \(V_{\activelabel,h,2} \setminus Gar_{h,2} \subseteq \cod(\rho_{\mathrm{loc},h,2})\).

Observe that \(Gar_{h,0} \subseteq Gar_{h,1} \subseteq Gar_{h,2}\).
For \(\ctx_1\), if \(h \neq f_{\mathrm{cur}}\), the involved compilation variables
remain unchanged, so the condition follows immediately from the validity
of \(\ctx_0\).
For \(h = f_{\mathrm{cur}}\), if \(x \notin \dom(\rho_{f_{\mathrm{cur}},0})\), we have
\(V_{\activelabel,f_{\mathrm{cur}},1} = V_{\activelabel,f_{\mathrm{cur}},0} \cup \rho_{\mathrm{loc},f_{\mathrm{cur}},1}(x)\) and
\(\cod(\rho_{\mathrm{loc},f_{\mathrm{cur}},1}) = \cod(\rho_{\mathrm{loc},f_{\mathrm{cur}},0}) \cup \rho_{\mathrm{loc},f_{\mathrm{cur}},1}(x)\);
thus the property follows directly from the validity of \(\ctx_0\).
If \(x \in \dom(\rho_{f_{\mathrm{cur}},0})\), we have
\(V_{\activelabel,f_{\mathrm{cur}},1} = V_{\activelabel,f_{\mathrm{cur}},0} \cup (\rho_{\mathrm{loc},f_{\mathrm{cur}},0}(x) \setminus \rho_{\mathrm{loc},f_{\mathrm{cur}},0}(\fp(f_{\mathrm{cur}})))\)
and \(\cod(\rho_{\mathrm{loc},f_{\mathrm{cur}},1}) = \cod(\rho_{\mathrm{loc},f_{\mathrm{cur}},0})\);
since \(\rho_{\mathrm{loc},f_{\mathrm{cur}},0}(x) \in \cod(\rho_{\mathrm{loc},f_{\mathrm{cur}},0})\),
the property again carries over from the validity of \(\ctx_0\).

For \(\ctx_2\) we have the following derivation, where \(q \in S\), 
\begin{align*}
V_{\activelabel,h,2} \setminus Gar_{h,2}
&\subseteq V_{\activelabel,h,1}[q \mapsto q[i]] \setminus Gar_{h,1}
  \tag{since \(Gar_{h,1} \subseteq Gar_{h,2}\)}\\
&\subseteq (V_{\activelabel,h,1} \setminus Gar_{h,1})[q \mapsto q[i]]
  \tag{since \(S \cap Gar_{h,1} = \emptyset\)}\\
&\subseteq \cod(\rho_{\mathrm{loc},h,1})[q \mapsto q[i]] \\
&= \cod(\rho_{\mathrm{loc},h,2}).
\end{align*} 

\item For \(h \neq f_{\mathrm{cur}}\), we prove \(Q_{\succeq f_{\mathrm{cur}}^*,2}[+] \cap V_{\activelabel,h,2} = \emptyset\).

Consider $\ctx_1$. 
The validity of \(\ctx_0\) gives
\(Q_{\succeq f_{\mathrm{cur}}^*,0}[+] \cap V_{\activelabel,h,0} = \emptyset\).
By definition \(V_{\activelabel,h,1} = V_{\activelabel,h,0}\).
If \(x \in \dom(\rho_{f_{\mathrm{cur}},0})\), then
\(Q_{\succeq f_{\mathrm{cur}}^*,1}[+] = Q_{\succeq f_{\mathrm{cur}}^*,0}[+]\), and the claim holds trivially.
Otherwise, \(x \notin \dom(\rho_{f_{\mathrm{cur}},0})\) and therefore
\(\cod(\rho_{\mathrm{loc},f_{\mathrm{cur}},1}) = \cod(\rho_{\mathrm{loc},f_{\mathrm{cur}},0}) \cup \rho_{\mathrm{loc},f_{\mathrm{cur}},1}(x) \).
From the definition of \(Q\) we obtain
\[
Q_{\succeq f_{\mathrm{cur}}^*,1}[+] \subseteq Q_{\succeq f_{\mathrm{cur}}^*,0}[+] \cup \rho_{\mathrm{loc},f_{\mathrm{cur}},1}(x)[+] .
\]
Since \(\rho_{\mathrm{loc},f_{\mathrm{cur}},1}(x) \subseteq \Qpool_0\), we have
\(\rho_{\mathrm{loc},f_{\mathrm{cur}},1}(x)[+] \subseteq \Qpool_0[+]\).
The validity of \(\ctx_0\) ensures \(\Qpool_0[+] \cap V_{\activelabel,h,0} = \emptyset\); consequently
\[
Q_{\succeq f_{\mathrm{cur}}^*,1}[+] \cap V_{\activelabel,h,1}
\subseteq \bigl( Q_{\succeq f_{\mathrm{cur}}^*,0}[+] \cup \Qpool_0[+] \bigr) \cap V_{\activelabel,h,0}
= \emptyset .
\]
Thus the required property holds for \(\ctx_1\). 
For \(\ctx_2\), by definition,
\[
Q_{\succeq f_{\mathrm{cur}}^*,2}[+] \subseteq Q_{\succeq f_{\mathrm{cur}}^*,1}[+] \cup S[i][+], \quad
V_{\activelabel,h,2} = V_{\activelabel,h,1}[q \mapsto q[i]] \subseteq (V_{\activelabel,h,1} \setminus S) \cup S[i].
\]
Since \(Q_{\succeq f_{\mathrm{cur}}^*,1}[+] \cap V_{\activelabel,h,1} = \emptyset\) by the validity of \(\ctx_1\)
and \(S[i] \cap S[i][+] = \emptyset\), it remains to show
\(S[i][+] \cap V_{\activelabel,h,1} = \emptyset\) and \(S[i] \cap Q_{\succeq f_{\mathrm{cur}}^*,1}[+] = \emptyset\).
By Lemma~\ref{lem:S-plus}, we have \(S[+] \subseteq cv_1\) and
\(S[+] \cap Q_{\succeq f_{\mathrm{cur}}^*,1}[+] = \emptyset\).
Clearly \(S[i] \subseteq S[+]\), hence \(S[i] \cap Q_{\succeq f_{\mathrm{cur}}^*,1}[+] = \emptyset\).
The validity of \(\ctx_1\) also gives \(cv_1 \cap V_{\activelabel,h,1} = \emptyset\).
Since \(S[i][+] \subseteq S[+] \subseteq cv_1\), we obtain \(S[i][+] \cap V_{\activelabel,h,1} = \emptyset\),
which completes the proof. 

\item Lemma~\ref{lem:Gar_0-empty} gives \(Q_{\succeq f_{\mathrm{cur}}^*,0}[+] \cap V_{\activelabel,f_{\mathrm{cur}},0} = \emptyset\).
From Lemma~\ref{lem:Q-plus-holds-ctx2}, we have
\[
Q_{\succeq f_{\mathrm{cur}}^*,2}[+] \cap V_{\activelabel,f_{\mathrm{cur}},2}
= Q_{\succeq f_{\mathrm{cur}}^*,1}[+] \cap V_{\activelabel,f_{\mathrm{cur}},1}
= Q_{\succeq f_{\mathrm{cur}}^*,0}[+] \cap V_{\activelabel,f_{\mathrm{cur}},0}=\emptyset.
\]
Therefore, the required conclusion follows directly.
\end{itemize}
\end{proof}

\begin{lemma} \label{lem:state-valid-4.c}
Under Notations~\ref{note:state_1},~\ref{note:state_2} and Assumption~\ref{hpy:com}, if \(s = \tau x \gets f(\overline{u})\) for some variables \(x\), \(\overline{u}\) and a function \(f\) with \(f \in \SCC(f_{\mathrm{cur}})\), then Condition~4(c) of the validity of \(\ctx_2\) holds. 
\end{lemma}
\begin{proof}
 Now consider any \(h \in \SCC(f_{\mathrm{cur}}) \cap \Fun(\St_2)\). We need to show \(\anc(\ctx_2.C_{h}^{k}) \subseteq \ctx_2.A_h\). 
 By definition, \(\ctx_1.C_{h} = \ctx_0.C_{h}\), and by Lemma~\ref{lem:com-Aset-1},
\(\ctx_0.A_h \subseteq \ctx_1.A_h\).  Hence the required property for \(\ctx_1\) follows
directly from the validity of \(\ctx_0\).  We therefore focus on \(\ctx_2\) in the
sequel.
Firstly, since renaming only affects variables in \(q \in S \subseteq V_{\activelabel,f_{\mathrm{cur}},1}\),
\begin{align*}
\ctx_1.A_h[q\mapsto q[i]]
   = &cv_1 \cup \bigl(V_{\activelabel,\succeq h,1} \cup \cod(\rho_{1, h})\bigr)[q\mapsto q[i]] \\ 
= &(cv_2 \cup S[1{:}i]) \cup \bigl(V_{\activelabel,\succeq h,2} \cup \cod(\rho_{2, h})\bigr) \\ 
= &\ctx_2.A_h \cup S[1{:}i]. 
\end{align*}

By definition, let
\(\mathcal{F}_{\succeq}^k \triangleq \anc(F_{\succeq f_{\mathrm{cur}}^*}^k) \cup \anc(F_{\succeq f_{\mathrm{cur}}^*}^k)[+]\),
\[
\anc(C_{h,2}^k) = \anc(C_{h,1}^k[q \mapsto q[i]])
                \triangleq \bigl(\anc(C_{h,1}^k) \setminus \mathcal{F}_{\succeq}^k\bigr)[q \mapsto q[i]] \;\cup\; \mathcal{F}_{\succeq}^k.
\]
We first show
\(\bigl(\anc(C_{h,1}^k) \setminus \mathcal{F}_{\succeq}^k\bigr)[q \mapsto q[i]] \subseteq \ctx_2.A_h\).
The validity of \(\ctx_1\) gives \(\anc(C_{h,1}^k) \subseteq \ctx_1.A_h\);
together with the definition this yields
\[
\bigl(\anc(C_{h,1}^k) \setminus \mathcal{F}_{\succeq}^k\bigr)[q \mapsto q[i]]
   \subseteq \ctx_1.A_h[q \mapsto q[i]]
   = \ctx_2.A_h \cup S[1{:}i].
\]
On the other hand, Condition~4(d) in the validity of \(\ctx_1\) also gives 
\((\anc(C_{h,1}^k) \setminus \mathcal{F}_{\succeq}^k) \cap \ctx_1.B[+] = \emptyset\),
and by definition \(S[1{:}i] \subseteq \ctx_1.B[+]\).
These two facts together imply 
\[
\bigl(\anc(C_{h,1}^k) \setminus \mathcal{F}_{\succeq}^k\bigr)[q \mapsto q[i]] \cap S[1{:}i] = \emptyset .
\]
Hence \(\bigl(\anc(C_{h,1}^k) \setminus \mathcal{F}_{\succeq}^k\bigr)[q \mapsto q[i]] \subseteq \ctx_2.A_h\).

Next we prove \(\mathcal{F}_{\succeq}^k \subseteq \ctx_2.A_h\).
From the definition of \(\mathcal{F}_{\succeq}^k\) and the fact that if 
\(\anc(F_h^z) \subseteq \ctx_2.A_h\) for any \(z\), then \(\anc(F_h^z)[+] \subseteq \ctx_2.A_h\).
The problem thus reduces to proving
\[
\anc(F_h^k) \subseteq \ctx_2.A_h
\quad\text{for every } h \in \Fun(\St_{\succeq f_{\mathrm{cur}}^*,2}).
\]
We distinguish two cases.

\noindent\textbf{Case 1:} \(F_h \notin \dom(\D_2)\).
Condition~1 of the validity of \(\ctx_2\) gives
\(\anc(F_h^k) \subseteq \ctx_{r,2}.A_h\).
Together with \(\ctx_{r,2}.A_h \subseteq \ctx_2.A_h\) we obtain
\(\anc(F_h^k) \subseteq \ctx_2.A_h\).

\noindent\textbf{Case 2:} \(F_h \in \dom(\D_2)\).
We proceed by induction on \(z\).
For \(z = 0\) we have
\(\anc(F_h^0) = \emptyset \subseteq \ctx_2.A_h\) for all such \(h\).
Assume inductively that \(\anc(F_h^z) \subseteq \ctx_2.A_h\)
holds for all such \(h\) and all \(z < k\).
Now consider \(z + 1 \le k\).  By definition,
\[
\anc(\D_2[F_h^{z+1}]) \triangleq \anc(C_{h,2}^z)
   = \bigl(\anc(C_{h,1}^z) \setminus \mathcal{F}_{\succeq}^z\bigr)[q \mapsto q[i]] \;\cup\; \mathcal{F}_{\succeq}^z .
\]
Using the already established inclusion
\(\bigl(\anc(C_{h,1}^k) \setminus \mathcal{F}_{\succeq}^k\bigr)[q \mapsto q[i]] \subseteq \ctx_2.A_h\) (noting that
\(\anc(C_{h,1}^z) \subseteq \anc(C_{h,1}^k)\)) 
and the induction hypothesis, we obtain
\[
\anc(\D_2[F_h^{z+1}]) \subseteq \ctx_2.A_h .
\]
Condition~1 of the validity of \(\ctx_2\) further requires
\[
\anc(F_h^{z+1}) \subseteq \anc(\ctx_{r,2}.\D[F_h^{z+1}]) \cup \anc(\ctx_{r,2}.\D[F_h^{z+1}])[+].
\]
Together with \(\anc(\ctx_{r,2}.\D[F_h^{z+1}]) \subseteq \anc(\D_2[F_h^{z+1}])\cup \anc(\D_2[F_h^{z+1}])[+]\),
we have
\[
\anc(F_h^{z+1}) \subseteq \anc(\D_2[F_h^{z+1}]) \cup \anc(\D_2[F_h^{z+1}])[+].
\]
Hence \(\anc(F_h^{z+1}) \subseteq \ctx_2.A_h\), completing the induction.
Thus for every \(h \in \Fun(\St_{\succeq f_{\mathrm{cur}}^*,2})\) we have
\(\anc(F_h^k) \subseteq \ctx_2.A_h\), which yields the desired conclusion. 
\end{proof}

\begin{lemma} \label{lem:state-valid-2}
Under Notations~\ref{note:state_1},~\ref{note:state_2} and Assumption~\ref{hpy:com}, if \(s = \tau x \gets f(\overline{u})\) for some variables \(x\), \(\overline{u}\) and a function \(f\) with \(f \in \SCC(f_{\mathrm{cur}})\), then Condition~2 of the validity of \(\ctx_2\) holds. 
\end{lemma}

\begin{proof}
   We first prove that \(\ctx_1\) satisfies Condition~2. Two cases arise based on whether \(x\) is already mapped in \(\rho_{f_{\mathrm{cur}},0}\). 
\begin{itemize}
  \item 

 \(x \in \dom(\rho_{f_{\mathrm{cur}},0})\).   
   Note that \(\Qpool_1=\Qpool_0\), \(\rho_{\mathrm{loc},f_{\mathrm{cur}},1}=\rho_{\mathrm{loc},f_{\mathrm{cur}},0}\) and  \(\rho_{\mathrm{loc},\succeq f_{\mathrm{cur}}^*,1}=\rho_{\mathrm{loc},\succeq f_{\mathrm{cur}}^*,0}\).  Since \(\rho_{\mathrm{loc},\succeq f_{\mathrm{cur}}^*,1}=\rho_{\mathrm{loc},\succeq f_{\mathrm{cur}}^*,0}\), we also have \(Q_{\succeq f_{\mathrm{cur}}^*,1}[+]=Q_{\succeq f_{\mathrm{cur}}^*,0}[+]\). Moreover, by Lemma~\ref{lem:com-Aset-1}, we have  $cv_1=cv_0$, \(\ctx_1.A_{f_{\mathrm{cur}}}=\ctx_0.A_{f_{\mathrm{cur}}}\) and \(V_{\activelabel,\succeq f_{\mathrm{cur}}^*,1} \cup \cod(\rho_{\mathrm{loc},\succeq f_{\mathrm{cur}}^*,1})=V_{\activelabel,\succeq f_{\mathrm{cur}}^*,0} \cup \cod(\rho_{\mathrm{loc},\succeq f_{\mathrm{cur}}^*,0})\). 
   Hence all conditions follow from the validity of \(\ctx_0\).
   
\item  \(x \notin \dom(\rho_{f_{\mathrm{cur}},0})\). 
By the validity of \(\ctx_0\) we have \(\fp(f_{\mathrm{cur}}) \in \dom(\rho_{\mathrm{loc},0})\);
moreover, the construction gives \(\rho_{\mathrm{loc},f_{\mathrm{cur}},1} = \rho_{\mathrm{loc},f_{\mathrm{cur}},0}[x \mapsto \rho_{\mathrm{loc},f_{\mathrm{cur}},1}(x)]\).
Hence \(\rho_{\mathrm{loc},f_{\mathrm{cur}},1}(\fp(f_{\mathrm{cur}})) = \rho_{\mathrm{loc},f_{\mathrm{cur}},0}(\fp(f_{\mathrm{cur}}))\).
Since Lemma~\ref{lem:com-Aset-1} yields \(\ctx_1.A_{f_{\mathrm{cur}}} = \ctx_0.A_{f_{\mathrm{cur}}}\),
we obtain
\[
\ket{\phi} \models \ket{0}_{\ctx_1.A_{f_{\mathrm{cur}}} \setminus \rho_{\mathrm{loc},f_{\mathrm{cur}},1}(\fp(f_{\mathrm{cur}}))} .
\]

From the proof of Lemma~\ref{lem:com-Aset-1} we already have
\[
\begin{aligned}
cv_1 &= cv_0 \setminus \rho_{\mathrm{loc},f_{\mathrm{cur}},1}(x), \\
V_{\activelabel,\succeq f_{\mathrm{cur}}^*,1} \cup \cod(\rho_{\mathrm{loc},\succeq f_{\mathrm{cur}}^*,1})
     &= V_{\activelabel,\succeq f_{\mathrm{cur}}^*,0} \cup \cod(\rho_{\mathrm{loc},\succeq f_{\mathrm{cur}}^*,0}) \cup \rho_{\mathrm{loc},f_{\mathrm{cur}},1}(x). 
\end{aligned}
\]
The validity of \(\ctx_0\) yields
\(cv_0 \cap \bigl( V_{\activelabel,\succeq f_{\mathrm{cur}}^*,0} \cup \cod(\rho_{\mathrm{loc},\succeq f_{\mathrm{cur}}^*,0}) \cup \cod(\rho_{\mathrm{glob}}) \bigr) = \emptyset\);
together with the fact that \(\rho_{\mathrm{glob}}\) is unchanged, therefore
\[
cv_1 \cap \bigl( V_{\activelabel,\succeq f_{\mathrm{cur}}^*,1} \cup \cod(\rho_{\mathrm{loc},\succeq f_{\mathrm{cur}}^*,1}) \cup \cod(\rho_{\mathrm{glob}}) \bigr) = \emptyset .
\]

Since \(\rho_{\mathrm{loc},f_{\mathrm{cur}},1}(x) \subseteq \Qpool_0\),
using the definition of \(\rho_{\mathrm{loc},f_{\mathrm{cur}},1}\) we obtain 
\[
\begin{aligned}
 \rho_{\mathrm{loc},\succeq f_{\mathrm{cur}}^*,1} \cup \cod(\rho_{\mathrm{glob}})
&= \rho_{\mathrm{loc},\succeq f_{\mathrm{cur}}^*,0} \cup \{\rho_{\mathrm{loc},f_{\mathrm{cur}},1}(x)\}
   \cup \cod(\rho_{\mathrm{glob}}) \\
&\subseteq \rho_{\mathrm{loc},\succeq f_{\mathrm{cur}}^*,0} \cup \cod(\rho_{\mathrm{glob}}) \cup \Qpool_0.
\end{aligned}
\]
The validity of \(\ctx_0\) gives
\(Q_{\succeq f_{\mathrm{cur}}^*,0}[+] \cap \bigl( \rho_{\mathrm{loc},\succeq f_{\mathrm{cur}}^*,0} \cup \cod(\rho_{\mathrm{glob}}) \bigr) = \emptyset\) and \(Q_{\succeq f_{\mathrm{cur}}^*,0}[+] \cap \Qpool_0 = \emptyset\),
hence
\[
Q_{\succeq f_{\mathrm{cur}}^*,0}[+] \cap \bigl(\rho_{\mathrm{loc},\succeq f_{\mathrm{cur}}^*,1} \cup \cod(\rho_{\mathrm{glob}}) \bigr) = \emptyset .
\]

Now observe that
\(Q_{\succeq f_{\mathrm{cur}}^*,1}[+] \subseteq Q_{\succeq f_{\mathrm{cur}}^*,0}[+] \cup \rho_{\mathrm{loc},f_{\mathrm{cur}},1}(x)[+]\).
Since \(\rho_{\mathrm{loc},f_{\mathrm{cur}},1}(x) \in \Qpool_0\), we have \(\rho_{\mathrm{loc},f_{\mathrm{cur}},1}(x)[+] \subseteq \Qpool_0[+]\).
The validity of \(\ctx_0\) also ensures
\(\Qpool_0[+] \cap \bigl( \Qpool_0 \cup \rho_{\mathrm{loc},\succeq f_{\mathrm{cur}}^*,0} \cup \cod(\rho_{\mathrm{glob}}) \bigr) = \emptyset\).
Consequently,
\[
Q_{\succeq f_{\mathrm{cur}}^*,1}[+] \cap \bigl( \rho_{\mathrm{loc},\succeq f_{\mathrm{cur}}^*,1} \cup \cod(\rho_{\mathrm{glob}}) \bigr) = \emptyset .
\]

Since \(\Qpool_1 = \Qpool_0 \setminus \rho_{\mathrm{loc},f_{\mathrm{cur}},1}(x)\), we have
\[
\Qpool_1 \subseteq \Qpool_0,\qquad
\Qpool_1[+] = \Qpool_0[+] \setminus \rho_{\mathrm{loc},f_{\mathrm{cur}},1}(x)[+] .
\]
From the proof of Lemma~\ref{lem:state-inv-1} we already obtain
\[
V_{\activelabel,\succeq f_{\mathrm{cur}}^*,1}[+] \cup \cod(\rho_{\mathrm{loc},\succeq f_{\mathrm{cur}}^*,1})[+]
= V_{\activelabel,\succeq f_{\mathrm{cur}}^*,0}[+] \cup \cod(\rho_{\mathrm{loc},\succeq f_{\mathrm{cur}}^*,0})[+] \cup \rho_{\mathrm{loc},f_{\mathrm{cur}},1}(x)[+].
\]
Since \(\rho_{\mathrm{loc},f_{\mathrm{cur}},1}(x) \subseteq \Qpool_0\), we have 
\[
\Qpool_1[+] \cup V_{\activelabel,\succeq f_{\mathrm{cur}}^*,1}[+] \cup \cod(\rho_{\mathrm{loc},\succeq f_{\mathrm{cur}}^*,1})[+]
= \Qpool_0[+] \cup V_{\activelabel,\succeq f_{\mathrm{cur}}^*,0}[+] \cup \cod(\rho_{\mathrm{loc},\succeq f_{\mathrm{cur}}^*,0})[+] .
\]
The validity of \(\ctx_0\) gives
\[
\Qpool_0 \cap \bigl( \Qpool_0[+] \cup V_{\activelabel,\succeq f_{\mathrm{cur}}^*,0}[+] \cup \cod(\rho_{\mathrm{loc},\succeq f_{\mathrm{cur}}^*,0})[+] \bigr) = \emptyset .
\]
Therefore,
\[
\Qpool_1 \cap \bigl( \Qpool_1[+] \cup V_{\activelabel,\succeq f_{\mathrm{cur}}^*,1}[+] \cup \cod(\rho_{\mathrm{loc},\succeq f_{\mathrm{cur}}^*,1})[+] \bigr)
\subseteq \Qpool_0 \cap \bigl( \Qpool_0[+] \cup V_{\activelabel,\succeq f_{\mathrm{cur}}^*,0}[+] \cup \cod(\rho_{\mathrm{loc},\succeq f_{\mathrm{cur}}^*,0})[+] \bigr)
= \emptyset .
\]

 Since \(\Qpool_1 \subseteq \Qpool_0\), it follows directly from the validity of \(\ctx_0\) that \(\forall q \in \Qpool_1,\; \var(\idx[q])=\emptyset \rightarrow \base(q)=q\). 
Thus Condition~2 is satisfied by \(\ctx_1\).
\end{itemize}

We now verify Condition~2 for \(\ctx_2\).

The validity of \(\ctx_1\) gives \(\rho_{\mathrm{loc},f_{\mathrm{cur}},1}(\fp(f_{\mathrm{cur}})) \cap V_{\activelabel,f_{\mathrm{cur}},1} = \emptyset\); 
hence, by construction, \(\rho_{\mathrm{loc},f_{\mathrm{cur}},2}(\fp(f_{\mathrm{cur}})) = \rho_{\mathrm{loc},f_{\mathrm{cur}},1}(\fp(f_{\mathrm{cur}}))\).
Since the validity of \(\ctx_1\) gives 
\(\ket{\phi} \models \ket{0}_{\ctx_1.A_{f_{\mathrm{cur}}} \setminus \rho_{\mathrm{loc},f_{\mathrm{cur}},1}(\fp(f_{\mathrm{cur}}))}\) and by Lemma~\ref{lem:com-Aset-2} \(\ctx_2.A_{f_{\mathrm{cur}}} \subseteq \ctx_1.A_{f_{\mathrm{cur}}}\),
it follows that
\[
\ket{\phi} \models \ket{0}_{\ctx_2.A_{f_{\mathrm{cur}}} \setminus \rho_{\mathrm{loc},f_{\mathrm{cur}},2}(\fp(f_{\mathrm{cur}}))}.
\]

\noindent
Next we prove
\[
cv_2 \cap \bigl( V_{\activelabel,\succeq f_{\mathrm{cur}}^*,2} \cup \cod(\rho_{\mathrm{loc},\succeq f_{\mathrm{cur}}^*,2}) \cup \cod(\rho_{\mathrm{glob}}) \bigr) = \emptyset .
\]

The proof in Lemma~\ref{lem:com-Aset-2} gives \(cv_2 \subseteq cv_1 \setminus S[1{:}i]\).
Moreover,
\[
\begin{aligned}
&V_{\activelabel,\succeq f_{\mathrm{cur}}^*,2} \cup \cod(\rho_{\mathrm{loc},\succeq f_{\mathrm{cur}}^*,2}) \cup \cod(\rho_{\mathrm{glob}})\\ 
= \ &\bigl(V_{\activelabel,\succeq f_{\mathrm{cur}}^*,1} \cup \cod(\rho_{\mathrm{loc},\succeq f_{\mathrm{cur}}^*,1})\bigr)[q \mapsto q[i]] \cup \cod(\rho_{\mathrm{glob}}) \\
= \ &\bigl( (V_{\activelabel,\succeq f_{\mathrm{cur}}^*,1} \cup \cod(\rho_{\mathrm{loc},\succeq f_{\mathrm{cur}}^*,1})) \setminus S \bigr)
   \cup S[i] \cup \cod(\rho_{\mathrm{glob}}).
\end{aligned}
\]

The validity of \(\ctx_1\) yields
\[
cv_1 \cap \bigl( V_{\activelabel,\succeq f_{\mathrm{cur}}^*,1} \cup \cod(\rho_{\mathrm{loc},\succeq f_{\mathrm{cur}}^*,1}) \cup \cod(\rho_{\mathrm{glob}}) \bigr) = \emptyset .
\]
Since \(cv_2 \subseteq cv_1 \setminus S[1{:}i]\), we obtain
\[
cv_2 \cap \bigl( \bigl( (V_{\activelabel,\succeq f_{\mathrm{cur}}^*,1} \cup \cod(\rho_{\mathrm{loc},\succeq f_{\mathrm{cur}}^*,1})) \setminus S \bigr) \cup \cod(\rho_{\mathrm{glob}}) \bigr) = \emptyset .
\]
Now note that \(S[i] \subseteq S[1{:}i]\), hence
\(S[i] \cap (cv_1 \setminus S[1{:}i]) = \emptyset\).
Consequently \(S[i] \cap cv_2 = \emptyset\).  Together with the above emptiness,
this yields
\[
cv_2 \cap \bigl( V_{\activelabel,\succeq f_{\mathrm{cur}}^*,2} \cup \cod(\rho_{\mathrm{loc},\succeq f_{\mathrm{cur}}^*,2}) \cup \cod(\rho_{\mathrm{glob}}) \bigr) = \emptyset .
\]

Next, we verify 
\[
Q_{\succeq f_{\mathrm{cur}}^*,2}[+] \cap \bigl(\rho_{\mathrm{loc},\succeq f_{\mathrm{cur}}^*,2} \cup \cod(\rho_{\mathrm{glob}}) \bigr) = \emptyset. 
\]

We have the inclusions
\[
\begin{aligned}
Q_{\succeq f_{\mathrm{cur}}^*,2}[+] &\subseteq Q_{\succeq f_{\mathrm{cur}}^*,1}[+] \cup S[i][+], \\
 \rho_{\mathrm{loc},\succeq f_{\mathrm{cur}}^*,2} \cup \cod(\rho_{\mathrm{glob}})
&= (\rho_{\mathrm{loc},\succeq f_{\mathrm{cur}}^*,1} \setminus S) \cup S[i] \cup \cod(\rho_{\mathrm{glob}}) .
\end{aligned}
\]
From the validity of \(\ctx_1\) we already have
\[
Q_{\succeq f_{\mathrm{cur}}^*,1}[+] \cap \bigl(\rho_{\mathrm{loc},\succeq f_{\mathrm{cur}}^*,1} \cup \cod(\rho_{\mathrm{glob}}) \bigr) = \emptyset .
\]

It is obvious that \(S[i][+] \cap S[i] = \emptyset\).
Lemma~\ref{lem:S-plus} gives \(S[+] \cap Q_{\succeq f_{\mathrm{cur}}^*,1}[+] = \emptyset\); thus
\(S[i] \cap Q_{\succeq f_{\mathrm{cur}}^*,1}[+] = \emptyset\) since \(S[i] \subseteq S[+]\).
Moreover, we have that \(S[i][+] \subseteq S[+]\) and, by Lemma~\ref{lem:state-inv-1}, \(S[+] \subseteq cv_1\).

Using Condition~2 in the validity of \(\ctx_1\) we obtain
\[
cv_1 \cap \bigl( \rho_{\mathrm{loc},\succeq f_{\mathrm{cur}}^*,1} \cup \cod(\rho_{\mathrm{glob}}) \bigr) = \emptyset ,
\]
which implies
\(S[i][+] \cap \bigl( \rho_{\mathrm{loc},\succeq f_{\mathrm{cur}}^*,1} \cup \cod(\rho_{\mathrm{glob}}) \bigr) = \emptyset\). Thus 
\[
S[i][+] \cap \bigl((\rho_{\mathrm{loc},\succeq f_{\mathrm{cur}}^*,1} \setminus S) \cup \cod(\rho_{\mathrm{glob}}) \bigr) = \emptyset .
\]

Putting all pieces together yields
\[
Q_{\succeq f_{\mathrm{cur}}^*,2}[+] \cap \bigl(\rho_{\mathrm{loc},\succeq f_{\mathrm{cur}}^*,2} \cup \cod(\rho_{\mathrm{glob}}) \bigr) = \emptyset .
\]

Since \(\Qpool_2 = \Qpool_1\), we have \(\Qpool_2[+] = \Qpool_1[+]\).
From the proof of Lemma~\ref{lem:com-Aset-2} we already obtain 
\[
V_{\activelabel,\succeq f_{\mathrm{cur}}^*,2}[+] \cup \cod(\rho_{\mathrm{loc},\succeq f_{\mathrm{cur}}^*,2})[+]
= \bigl( (V_{\activelabel,\succeq f_{\mathrm{cur}}^*,1}[+] \cup \cod(\rho_{\mathrm{loc},\succeq f_{\mathrm{cur}}^*,1})[+] ) \setminus S[1{:}i] \bigr).
\]
The validity of \(\ctx_1\) gives the disjointness properties
\[
\Qpool_1 \cap \bigl( \Qpool_1[+] \cup V_{\activelabel,\succeq f_{\mathrm{cur}}^*,1}[+] \cup \cod(\rho_{\mathrm{loc},\succeq f_{\mathrm{cur}}^*,1})[+] \bigr) = \emptyset .
\]
Hence
\[
\begin{aligned}
&\Qpool_2 \cap \bigl( \Qpool_2[+] \cup V_{\activelabel,\succeq f_{\mathrm{cur}}^*,2}[+] \cup \cod(\rho_{\mathrm{loc},\succeq f_{\mathrm{cur}}^*,2})[+] \bigr)\\ 
= \ & \Qpool_1 \cap \Bigl( \Qpool_1[+] \cup \bigl( (V_{\activelabel,\succeq f_{\mathrm{cur}}^*,1}[+] \cup \cod(\rho_{\mathrm{loc},\succeq f_{\mathrm{cur}}^*,1})[+] ) \setminus S[1{:}i]  \bigr)\Bigr) \\
\subseteq \ & \Qpool_1 \cap \bigl( \Qpool_1[+] \cup V_{\activelabel,\succeq f_{\mathrm{cur}}^*,1}[+] \cup \cod(\rho_{\mathrm{loc},\succeq f_{\mathrm{cur}}^*,1})[+] \bigr) = \emptyset. 
\end{aligned}
\]

Similarly, since \(\Qpool_2 = \Qpool_1\), the condition that \(\forall q \in \Qpool_2,\; \var(\idx[q])=\emptyset \rightarrow \base[q]=q\) follows directly from the validity of \(\ctx_1\). Thus both requirements of Condition~2 are satisfied by \(\ctx_2\).
\end{proof}

\begin{lemma}\label{lem:state-valid}
Under Notations~\ref{note:state_1},~\ref{note:state_2} and Assumption~\ref{hpy:com}, if \(s = \tau x \gets f(\overline{u})\) for some variables \(x\), \(\overline{u}\) and a function \(f\) with \(f \in \SCC(f_{\mathrm{cur}})\), then the validity of \(\ctx_2\) holds. 
\end{lemma}
\begin{proof}
For every \(h \in \SCC(f_{\mathrm{cur}})\), the transition from \(\ctx_0\) to \(\ctx_2\) modifies only the components \(\rho\), \(\Qpool\), \(\Vactive\), \(C\), \(\kappa\), \(\Eargs\), $\D$, $\St$, and $\M$. Moreover, $\Fun(\St_2)=\Fun(\St_0)$, $\dom(\D_2)=\dom(\D)$, and $\dom(\M_2)=\dom(\M)$. We proceed to verify the validity conditions for \(\ctx_2\).  
\begin{itemize}
\item Condition~1: By definition of the algorithm we have
\[
\begin{aligned}
&\ctx_2.(\dom(\D), \Fun(\St), \dom(\M), \delta, f_{\mathrm{cur}}, R_{\mathrm{stmt},f_{\mathrm{cur}}}) \\
=&\;\ctx_0.(\dom(\D), \Fun(\St), \dom(\M), \delta, f_{\mathrm{cur}}, R_{\mathrm{stmt},f_{\mathrm{cur}}}),
\end{aligned}
\]
and \(\textsc{Compile\_Body}(\ctx_0, R_{\mathrm{stmt},f_{\mathrm{cur}},0}) = \textsc{Compile\_Body}(\ctx_2, R_{\mathrm{stmt},f_{\mathrm{cur}},2})\), and the algorithm does not alter function declarations in \(\D_0\) outside \(\SCC(f_{\mathrm{cur}})\). Therefore, to prove that Condition~1 holds with respect to \(k\) and \(\D_{\SCC(f_{\mathrm{cur}})}\) in this compilation context, it suffices to show
\[
\D_2\upharpoonright_{\dom(\D_0) \cap \SCC(f_{\mathrm{cur}})} \;\approx_{\ctx_{r,2}.Gar_{f_{\mathrm{cur}}^*}}\; \D_0\upharpoonright_{\dom(\D_0) \cap \SCC(f_{\mathrm{cur}})}.
\]
This follows from Lemma~\ref{lem:state-inv-1}.
\item Condition~2: By Lemma~\ref{lem:state-valid-2}. 
\item Condition~3(a): By Lemma~\ref{lem:state-valid-3.a}.
\item Condition~3(b): By Lemma~\ref{lem:state-valid-3.b}.
\item Condition~3(c): By Lemma~\ref{lem:state-valid-3.c}.
\item Condition~4(a): By Lemma~\ref{lem:state-valid-4.a}.
\item Condition~4(b): By Lemma~\ref{lem:state-valid-4.b}.
\item Condition~4(c): By Lemma~\ref{lem:state-valid-4.c}.  
\item Condition~4(d). Similarly, \(\ctx_1.C_h = \ctx_0.C_h\) and \(\ctx_1.B = \ctx_0.B\); hence the property for \(\ctx_1\) follows directly
from the validity of \(\ctx_0\). For \(\ctx_2\), we have 
\[
\begin{aligned}
&\anc(C_{h,2}^k) \setminus \bigl( \anc(F_{\succeq f_{\mathrm{cur}}^*}^k) \cup \anc(F_{\succeq f_{\mathrm{cur}}^*}^k)[+] \bigr) \\ 
= &\bigl( \anc(C_{h,1}^k) \setminus \bigl( \anc(F_{\succeq f_{\mathrm{cur}}^*}^k) \cup \anc(F_{\succeq f_{\mathrm{cur}}^*}^k)[+] \bigr) \bigr)[q \mapsto q[i]] \\ 
\subseteq &\bigl( \anc(C_{h,1}^k) \setminus \bigl( \anc(F_{\succeq f_{\mathrm{cur}}^*}^k) \cup \anc(F_{\succeq f_{\mathrm{cur}}^*}^k)[+] \bigr) \bigr) \setminus S \;\cup\; S[i].
\end{aligned} 
\]
By the definition of \(B\),
\(\ctx_2.B = \ctx_1.B \setminus S\):
\begin{align*} 
\ctx_2.B
&= \{q \in \Qpool_2 \cup V_{\activelabel,\succeq f_{\mathrm{cur}}^*,2} \cup \cod(\rho_{\mathrm{loc},f_{\mathrm{cur}}^*,2}) \mid \base[q]=q\} \\
&= \{q \in \Qpool_2 \mid \base[q]=q\}
   \cup \{q \in V_{\activelabel,\succeq f_{\mathrm{cur}}^*,2} \cup \cod(\rho_{\mathrm{loc},f_{\mathrm{cur}}^*,2}) \mid \base[q]=q\} \\
&= \{q \in \Qpool_1 \mid \base[q]=q\}
   \cup \{q \in (V_{\activelabel,\succeq f_{\mathrm{cur}}^*,1} \cup \cod(\rho_{\mathrm{loc},f_{\mathrm{cur}}^*,1}))[q \mapsto q[i]] \mid \base[q]=q\} \\
&= \{q \in \Qpool_1 \mid \base[q]=q\}
   \cup \{q \in V_{\activelabel,\succeq f_{\mathrm{cur}}^*,1} \cup \cod(\rho_{\mathrm{loc},f_{\mathrm{cur}}^*,1}) \mid \base[q]=q\} \setminus S ;
\end{align*}
hence \(\ctx_2.B[+] = \ctx_1.B[+] \setminus S[+]\).
The validity of \(\ctx_1\) gives
\[
\bigl( \anc(C_{h,1}^k) \setminus \bigl( \anc(F_{\succeq f_{\mathrm{cur}}^*}^k) \cup \anc(F_{\succeq f_{\mathrm{cur}}^*}^k)[+] \bigr) \bigr) \cap \ctx_1.B[+] = \emptyset .
\]
Putting everything together, we have that 
\begin{align*}
&\bigl( \anc(C_{h,2}^k) \setminus \bigl( \anc(F_{\succeq f_{\mathrm{cur}}^*}^k) \cup \anc(F_{\succeq f_{\mathrm{cur}}^*}^k)[+] \bigr) \bigr) \cap \ctx_2.B[+] \\
\subseteq \ & \Bigl( \bigl( \anc(C_{h,1}^k) \setminus \bigl( \anc(F_{\succeq f_{\mathrm{cur}}^*}^k) \cup \anc(F_{\succeq f_{\mathrm{cur}}^*}^k)[+] \bigr) \bigr) \setminus S \;\cup\; S[i] \Bigr)
   \cap \bigl( \ctx_1.B[+] \setminus S[+] \bigr) \\
= \ & \Bigl( \bigl( \anc(C_{h,1}^k) \setminus \bigl( \anc(F_{\succeq f_{\mathrm{cur}}^*}^k) \cup \anc(F_{\succeq f_{\mathrm{cur}}^*}^k)[+] \bigr) \bigr) \setminus S \Bigr)
   \cap \bigl( \ctx_1.B[+] \setminus S[+] \bigr) \\
&\qquad \cup\; \bigl( S[i] \cap (\ctx_1.B[+] \setminus S[+]) \bigr)
   = \emptyset \qquad (\text{since } S[i] \subseteq S[+]).
\end{align*}

\item Condition~4(e):
\(\var(\idx(\ctx_2.A_{f_{\mathrm{cur}}^*})) \subseteq E_{\args,f_{\mathrm{cur}}^*} = E_{\args,h}\).
By construction,
\(\var(\idx(\ctx_2.A_{f_{\mathrm{cur}}^*})) = \var(\idx(\ctx_0.A_{f_{\mathrm{cur}}^*})) \cup \{i\}\)
and \(\ctx_2.E_{\args,h} = \ctx_0.E_{\args,h} \cup \{i\}\).
The validity of \(\ctx_0\) gives
\(\var(\idx(\ctx_0.A_{f_{\mathrm{cur}}^*})) \subseteq \ctx_0.E_{\args,f_{\mathrm{cur}}^*} = \ctx_0.E_{\args,h}\); 
hence the required inclusion follows immediately. 
\item Condition~4(f): We need to show \(\base(\anc(\ctx_2.C_h^z)) \subseteq \ctx_2.V_{\alloc,\succeq f_{\mathrm{cur}}^*} \cup \anc(F_{\succeq f_{\mathrm{cur}}^*}^z)\) for any $z$. 
Since the renaming \(q \mapsto q[i]\) only introduces new indices,
\(\base(\anc(\ctx_2.C_h^z)) = \base(\anc(\ctx_0.C_h^z))\).
Also, by construction, \(\ctx_2.V_{\alloc,\succeq f_{\mathrm{cur}}^*} = \ctx_0.V_{\alloc,\succeq f_{\mathrm{cur}}^*} \cup \base(\rho_{\mathrm{loc},f_{\mathrm{cur}},1}(x))\). 
The validity of \(\ctx_0\) gives the corresponding inclusion for \(\ctx_0\); hence it holds for \(\ctx_2\) as well. 
\item Condition~4(g): The required inclusion
\((\Succ(h) \setminus \Call(R_{\mathrm{stmt},h,2})) \cap \SCC(f_{\mathrm{cur}}) \subseteq \Fun(\St_2)\) 
follows directly, since none of the involved components are modified in this step. 
\end{itemize} 

\end{proof}

\begin{lemma}
    \label{lem:state-pre}
Under Notations~\ref{note:state_1},~\ref{note:state_2} and Assumption~\ref{hpy:com}, 
if \(s = \tau x \gets f(\overline{u})\) for some variables \(x\), \(\overline{u}\) and a function \(f\) with \(f \in \SCC(f_{\mathrm{cur}})\), then for any \(g \in \{f_{\mathrm{cur}}\} \cup \SCC(f_{\mathrm{cur}}) \cap \Fun(\St_0) \setminus \dom(\D_0)\)
and for any state \(\ket{\varphi}\) such that
\(\ket{\varphi} \models \ket{0}_{\ctx_0.A_g \setminus \rho_{g,0}(\fp(g))}\),
if \(\ctx_0 \approx_{(g,\ket{\varphi},k)} \sigma'\) for some \(\sigma'\),
then \(\ctx_2 \approx_{(g,\ket{\varphi},k)} \sigma'\).
\end{lemma}

\begin{proof}
    Since \(\ctx_1.C_{g}^{k} = \ctx_0.C_{g}^{k}\) and \(G_{g,1} = G_{g,0}\), we distinguish two cases:
\begin{itemize}
    \item If \(x \in \dom(\rho_0)\) or \(g \neq f_{\mathrm{cur}}\), then
          \(\rho_{g,1} = \rho_{g,0}\) and \(V_{\activelabel,g,1} = V_{\activelabel,g,0}\).
          Together with \(\ctx_1.C_{g}^{k} = \ctx_0.C_{g}^{k}\) and
          \(G_{g,1} = G_{g,0}\), the hypothesis
          \(\ctx_0 \approx_{(g,\ket{\varphi},k)} \sigma'\) immediately yields
          \(\ctx_1 \approx_{(g,\ket{\varphi},k)} \sigma'\).

    \item If \(x \notin \dom(\rho_0)\) and \(g = f_{\mathrm{cur}}\), then
          \(\rho_{f_{\mathrm{cur}},1} = \rho_{f_{\mathrm{cur}},0} \cup \{x \mapsto \rho_{f_{\mathrm{cur}},1}(x)\}\)
          and \(V_{\activelabel,f_{\mathrm{cur}},1} = V_{\activelabel,f_{\mathrm{cur}},0} \cup \{\rho_{f_{\mathrm{cur}},1}(x)\}\)
          with \(\rho_{f_{\mathrm{cur}},1}(x) \in \Qpool_0 \subseteq cv_0\).
         Since \(\rho_{f_{\mathrm{cur}},1}(x)\) is clean,
          \(\ket{\varphi} \models \ket{0}_{\rho_{f_{\mathrm{cur}},1}(x)}\).

          Without loss of generality, since \(x\) is declared for the first time,
          we can always assume that \(x\) will be used by the remaining statements,
          hence \(\delta_{f_{\mathrm{cur}},1}(\node(x)) > 0\).
          Thus \(\rho_{\mathrm{active},1} = \rho_{\mathrm{active},0} \cup \rho_{f_{\mathrm{cur}},1}(x)\).
          Consequently,
          \(V_{\activelabel,f_{\mathrm{cur}},1} \setminus \cod(\rho_{\mathrm{active},1})
           = V_{\activelabel,f_{\mathrm{cur}},0} \setminus \cod(\rho_{\mathrm{active},0})\).

          The execution therefore proceeds as follows:
          \[
          \begin{aligned}
          \langle \ctx_1.C_{f_{\mathrm{cur}}}^{k}, \ket{\varphi} \rangle
          = \ & \langle \ctx_0.C_{f_{\mathrm{cur}}}^{k}, \ket{\varphi} \rangle \\
          \rightarrow^{*}  \ & \bigl\langle \downarrow,\; 
               \mathcal{E}(\sigma', \rho_{\mathrm{active},0}) 
               \otimes \ket{\theta}_{V_{\activelabel,f_{\mathrm{cur}},0}\setminus \cod(\rho_{\mathrm{active},0})}  
               \otimes \ket{\varphi}_{\text{rem}} \bigr\rangle \\
          = \ & \bigl\langle \downarrow,\;
               \mathcal{E}(\sigma', \rho_{\mathrm{active},0}) 
               \otimes \ket{\theta}_{V_{\activelabel,f_{\mathrm{cur}},0}\setminus \cod(\rho_{\mathrm{active},0})} 
               \otimes \ket{0}_{\rho_{f_{\mathrm{cur}},1}(x)} 
               \otimes \ket{\varphi}_{\text{rem}} \bigr\rangle \\
          = \ &  \bigl\langle \downarrow,\;
               \mathcal{E}(\sigma', \rho_{\mathrm{active},1}) 
               \otimes \ket{\theta}_{V_{\activelabel,f_{\mathrm{cur}},1}\setminus \cod(\rho_{\mathrm{active},1})} 
               \otimes \ket{\varphi}_{\text{rem}} \bigr\rangle .
          \end{aligned}
          \]
          The last equality uses the fact that
          \(\mathcal{E}(\sigma', \rho_{\mathrm{active},0}) \otimes \ket{0}_{\rho_{f_{\mathrm{cur}},1}(x)}
           = \mathcal{E}(\sigma', \rho_{\mathrm{active},1})\)
        since \(\sigma'(x)=0\).
\end{itemize}

Thus in both cases we obtain \(\ctx_1 \approx_{(g,\ket{\varphi},k)} \sigma'\).

Now consider \(\ctx_2\). Firstly we have \(\ket{\varphi} \models \ket{0}_{\ctx_1.A_g \setminus \rho_{g,1}(\fp(g))}\) by Lemma~\ref{lem:state-valid-2}. Let \(h:\qv(\ctx_1.C_{g}^{k}) \rightarrow \mathcal{Q}\) be the mapping defined by
\(h(q)=q[i]\) for \(q \in S\) and \(h(q)=q\) otherwise.
Since \(\bigl(\qv(\ctx_1.C_{g}^{k}) \setminus S\bigr) \cap S[i] = \emptyset\),
\(h\) is a bijection on its image.
The renaming affects only variables in \(S \subseteq V_{\activelabel,f_{\mathrm{cur}},1}\).
Since \(g \in \SCC(f_{\mathrm{cur}}) \cap \Fun(\St_0) \setminus \dom(\D_0)\),
we have \(\pos(g) \leq \pos(f_{\mathrm{cur}})\) and therefore
\(S \subseteq \ctx_1.A_g \setminus \rho_{g,1}(\fp(g))\).
Moreover, by Lemma~\ref{lem:S-plus},
\(S[i] \subseteq cv_1 \subseteq \ctx_1.A_g \setminus \rho_{g,1}(\fp(g))\).
Thus \(\ket{\varphi} \models \ket{0}_{S \cup S[i]}\).

Applying Lemma~\ref{lemma:rename} yields
\[
\bigl( \ctx_2.C_{g}^{k} \ket{\varphi} \bigr)_{\qv(\ctx_2.C_{g}^{k})}
= \bigl( \ctx_1.C_{g}^{k} \ket{\varphi} \bigr)_{\qv(\ctx_1.C_{g}^{k})}.
\]

From the result for \(\ctx_1\) we have
\[
\langle \ctx_1.C_{g}^{k}, \ket{\varphi} \rangle 
   \rightarrow^{*} \bigl\langle \downarrow,\; 
        \mathcal{E}(\sigma', \rho_{\mathrm{active},1}) 
        \otimes \ket{\theta}_{V_{\activelabel,g,1}\setminus \cod(\rho_{\mathrm{active},1})} 
        \otimes \ket{\varphi}_{\text{rem}} \bigr\rangle .
\]
Since \(h(V_{\activelabel,g,1})=V_{\activelabel,g,2}\) and \(h(\rho_{\mathrm{active},1})=\rho_{\mathrm{active},2}\),
we obtain
\[
\begin{aligned}
\langle \ctx_2.C_{g}^{k}, \ket{\varphi} \rangle
\rightarrow^{*} \bigl\langle \downarrow,\; 
        \mathcal{E}(\sigma', \rho_{\mathrm{active},2}) 
        \otimes \ket{\theta}_{V_{\activelabel,g,2}\setminus \cod(\rho_{\mathrm{active},2})} 
        \otimes \ket{\varphi}_{\text{rem}} \bigr\rangle .
\end{aligned}
\]

Therefore \(\ctx_2 \approx_{(g,\ket{\varphi},k)} \sigma'\) as well.
\end{proof}

For the next two theorems we fix
a typing context \( \Gamma' \), a variable‑to‑qubit mapping \( \rho \), and a
pool of fresh quantum variables \( \Qpool \).

\begin{theorem}[Correctness of compiled expressions]
  \label{theo_exp}
  Let \( e \) be an expression with \( \Gamma' \vdash e \) and let
  \( p \) be a quantum variable.  Suppose
  \( \Gamma';\rho;\Qpool \vdash (e,p) \leadsto C;\Qpool' \) and
  \( \var(e) \subseteq \dom(\rho) \).
  Then the generated code \( C \) satisfies:
  \begin{enumerate}
    \item \( \qv(C) = p \cup \rho(\var(e)) \cup S \)
          for some set \( S \subseteq \Qpool \), and \( \Qpool' = \Qpool \);
    \item for any classical state \( \sigma \), any value \( v \) with 
          \( \langle e,\sigma\rangle \rightarrow v \),
          \[
          \begin{aligned}
            &\Bigl\langle C,\;
                \ket{\sigma(\var(e))}_{\rho(\var(e))}
                \otimes \ket{0}_{p}
                \otimes \ket{0}_{\qv(C)\setminus(\rho(\var(e))\cup\{p\})}
            \Bigr\rangle \\
            \rightarrow^{*}
            &\Bigl\langle \downarrow,\;
                \ket{\sigma(\var(e))}_{\rho(\var(e))}
                \otimes \ket{v}_{p}
                \otimes \ket{0}_{\qv(C)\setminus(\rho(\var(e))\cup\{p\})}
            \Bigr\rangle .
          \end{aligned}
          \]
  \end{enumerate}
\end{theorem}

\begin{proof}
  Both parts are proved by induction on the structure of \( e \).  Specifically, the proof has two layers. First, for the predefined target implementations
  directly invoked by the expression-compilation rules, including constant
  initialization, copying, Boolean operations, comparisons, arithmetic operations
  (addition, subtraction, multiplication, division, and modulo), array-location
  routines, and the corresponding inverse programs, we prove their specified
  total-correctness Hoare triples in the \(\RQC^{++}\) proof system described in
  Subsection~\ref{app:tar}. 

  On top of these verified target implementations, the expression-compilation
  theorem is proved by induction on the structure of \(e\). For each expression
  construct, the generated code is composed according to the corresponding
  compilation rule from the target implementations proved correct above. Hence,
  using the sequential-composition, frame/invariance, and inverse-program rules of
  the \(\RQC^{++}\) proof system, we combine these basic triples to obtain the
  total-correctness Hoare triple for the whole generated code \(C\). This triple
  implies the operational statement required by the theorem: after executing
  \(C\), the target register \(p\) contains \(v\), the registers
  corresponding to source variables are preserved, and all auxiliary registers
  used during expression compilation are restored to \(\ket{0}\). 
  
  Since these derivations are routine but lengthy applications of the
  \(\RQC^{++}\) proof rules, we omit the low-level proof details.
%  Then for any $\sigma'$ such that  $\sigma' \models A$, we have 
%  \[\begin{aligned}
%           \Bigl\langle C,\; (\sigma',
%                 \ket{\sigma(\var(e))}_{\rho(\var(e))}
%                 \otimes \ket{c}_{p}
%                 \otimes \ket{0}_{\qv(C)\setminus(\rho(\var(e))\cup\{p\})})
%             \Bigr\rangle 
%             \rightarrow^{*} \Bigl\langle \downarrow,\;
%                (\sigma'', \ket{\phi})
%             \Bigr\rangle,
%           \end{aligned}
%           \]
% where $\sigma'' \models A$ and $ \ket{\phi} = (\ket{\sigma(\var(e))}_{\rho(\var(e))}
%       \otimes \ket{c\oplus v}_{p}
%              \otimes \ket{0}_{\qv(C)\setminus(\rho(\var(e))\cup\{p\})})(\sigma'').$ Given the arbitrariness of $A$, we obtain $\sigma'' =\sigma'$. Additionally, applying the rules (Frame) and (Invariance-Con) to following triple \[
%         \{\ket{\sigma_q(\overline{x})}_{qv(u)} \otimes \ket{0}_{anc} \} \ u \ \{\ket{\sigma_q(\overline{x})}_{qv(u)} \otimes \ket{v}_p \otimes \ket{0}_{anc} \}
%     \]  can yield equation~\ref{(2)}, where $qv(u)=\rho(\overline{x})$.  
\end{proof}

\begin{theorem}[Correctness of compiled assignments]
  \label{theo_assgn}
  Suppose \(\Gamma' \vdash \tau x \gets e \dashv \Gamma'' \) and
  \( \Gamma';\rho;\Qpool \vdash (\tau x \gets e) \leadsto C;\Qpool' \).
  Assume \( \var(e) \cup \{x\} \subseteq \dom(\rho) \).
  Then the generated code \( C \) satisfies:
  \begin{enumerate}
    \item \( \qv(C) = \rho(x) \cup \rho(\var(e)) \cup S \)
          for some set \( S \subseteq \Qpool \), and \( \Qpool' = \Qpool \);
    \item for any classical state \( \sigma \),
          \[
          \begin{aligned}
            &\Bigl\langle C,\;
                \ket{\sigma(\var(e))}_{\rho(\var(e))}
                \otimes \ket{0}_{\rho(x)}
                \otimes \ket{0}_{\qv(C)\setminus(\rho(\var(e))\cup\rho(x))}
            \Bigr\rangle \\
            \rightarrow^{*}
            &\Bigl\langle \downarrow,\;
                \ket{\sigma(\var(e))}_{\rho(\var(e))}
                \otimes \ket{\sigma(e)}_{\rho(x)}
                \otimes \ket{0}_{\qv(C)\setminus(\rho(\var(e))\cup\rho(x))}
            \Bigr\rangle.
          \end{aligned}
          \]
  \end{enumerate}
\end{theorem}

\begin{proof}
  Both parts follow directly from Theorem~\ref{theo_exp} and the definition of compilation for assignments. 
\end{proof}

We next prove the correctness of statement compilation; along the way, we establish several auxiliary properties of the resulting compilation context, which will be used later in the validity preservation proof of Theorem~\ref{theo:com_valid}. Since the compiled statement \(s\) yields a single element in \(\mvq(s)\), we henceforth write \(\mvq(s).\flag\) to refer to the flag of that element. For clarity, throughout Appendix~\ref{app:veri}, the quantification over
\(\ket{\psi}\) in every auxiliary trace-preservation condition is restricted
to statement-boundary configurations reachable from computational-basis encodings of source states. This basis-state formulation suffices for the auxiliary correctness arguments
  in Subsections~\ref{app:proof-cleanup}--\ref{app:proof-com}, since Theorem~\ref{theo:coherent-oracle}
  subsequently extends whole-program correctness to coherent superpositions by
  unitarity and linearity. For simplicity, the auxiliary set \(E\) in the following theorem does not
  explicitly include the clean-register requirement for a return register. When
  the modified register is a return register, the required \(\ket{0}\)
  precondition instead follows from its initialization at function entry and the
  reachable-pre-execution-state convention above: the register remains clean
  until its unique defining assignment. 
\begin{theorem}
  \label{theo_com}
  \label{theo:correctness-com}
  Under Notations~\ref{note:state_1},~\ref{note:state_2} and
  Assumption~\ref{hpy:com}, there exists a compilation context \(\ctx'\)
  and sets \(D_1, D_2\) such that
  \[
    \ctx' \approx_{(\ket{\phi}, k)} \sigma_0,\qquad
    \bigl( \ctx', (\ket{\phi}, k, \D_{\SCC(f_{\mathrm{cur}})}) \bigr) \in \mathcal{V},
  \]
  and the following conditions are satisfied.
  \begin{itemize}
    \item \(\rho(\mvq(s))=\rho_{\mathrm{loc}}(\mvq(s)) \subseteq \Qpool'\) if \(\mvq(s) \not\subseteq \dom(\rho')\). 
    \item \(D_1 = \rho(\mvq(s))\) if \(\neg (\mvq(s).\flag = 2 \wedge \Call(s)\neq \emptyset)\);
          otherwise \(D_1 = \rho(\mvq(s)) \cup Gar_{f_{\mathrm{cur}}}\).
    \item \(\base\bigl(\cod(\rho_{\mathrm{loc}}) \setminus \cod(\rho'_{\mathrm{loc}})\bigr) \subseteq D_2\). 
  \end{itemize}
  Moreover, \(\ctx\) can be expressed as
  \begin{align*}
    \ctx = \ctx'[\, &C \mapsto C; C_s,\;
                   \Vactive \mapsto \Vactive \cup \bigl(D_1 \setminus (\rho'_{\mathrm{loc}}(\fp(f_{\mathrm{cur}})) \cup \cod(\rho_{\mathrm{glob}})) \bigr),\;
                   \Valloc \mapsto \Valloc \cup D_2,\; \\
                 &\rho_{\mathrm{loc}} \mapsto \rho'_{\mathrm{loc}}[\mvq(s) \mapsto \rho_{\mathrm{loc}}(\mvq(s)) \mid \mvq(s) \not\subseteq  \dom(\rho')],\; 
                   \Qpool \mapsto \Qpool' \setminus \rho_{\mathrm{loc}}(\mvq(s)) \,],
  \end{align*}
  and the following properties hold.
  \begin{itemize}
    \item \(cv = cv' \setminus \bigl(D_1 \setminus (\rho'_{\mathrm{loc}}(\fp(f_{\mathrm{cur}})) \cup \cod(\rho_{\mathrm{glob}})) \bigr)\).
    \item For every \(h \in \SCC(f) \cap \Fun(\St')\),
          if \(h \notin \Theta^{-1}(\dom(\D'))\) then \(\ctx.A_h = \ctx'.A_h\);
          otherwise \(\ctx'.A_h \subseteq \ctx.A_h\).
    \item \(\qv(C_s^{k}) \setminus \rho'\bigl(\mvq(s) \cup \rvq(s)\bigr)
          \subseteq \ctx'.A_{f_{\mathrm{cur}}} \setminus \bigl(\ctx'.V_{\activelabel,f_{\mathrm{cur}}} \cup \cod(\rho'_{\mathrm{loc}})\bigr)\).
    \item For any \(\ket{\varphi}\) with
          \(\ket{\varphi} \models \ket{0}_{\ctx'.A_{f_{\mathrm{cur}}} \setminus \rho'_{\mathrm{loc}}(\fp(f_{\mathrm{cur}}))}\),
          \(\ctx'.C_{f_{\mathrm{cur}}}^k \ket{\varphi} \models \ket{0}_{\rho_{\mathrm{loc}}(\mvq(s) \setminus \dom(\rho_{0}))}\).
	    \item For any \(z\),
	          \(\base\bigl(\qv(C_s^z) \setminus \cod(\rho')\bigr)
	          \subseteq D_2 \cup \anc(F_{\succeq f_{\mathrm{cur}}^*}^z)\).
	    \item \(\anc(C_s^k) \setminus \bigl(\anc(F_{\succeq f_{\mathrm{cur}}^*}^k) \cup \anc(F_{\succeq f_{\mathrm{cur}}^*}^k)[+]\bigr) \subseteq \Qpool' \cup \cod(\rho'_{\mathrm{loc}})\).
    \item Set \(E \triangleq \qv(C_s^{k}) \setminus \rho'\bigl(\mvq(s) \cup \rvq(s)\bigr)
                \cup \rho_{\mathrm{loc}}\bigl( \mvq(s)\setminus \dom(\rho_{0})\bigr)\). 
          Then for every \(\ket{\psi}\) with \(\ket{\psi} \models \ket{0}_{E}\),
          \(\tr_{D_1}(C_s^{k} \ket{\psi}) = \tr_{D_1}(\ket{\psi})\).
  \end{itemize}
  Finally, we have \(\ctx \approx_{(\ket{\phi}, k)} \sigma\).
\end{theorem}
\begin{proof}
For readability, write
\[
\mathcal{F}_{\succeq}^{k}
\triangleq
\anc(F_{\succeq f_{\mathrm{cur}}^*}^{k})
\cup
\anc(F_{\succeq f_{\mathrm{cur}}^*}^{k})[+].
\]
We proceed by induction on the structure of \(s\). 

\textbf{Case~1.} We now consider the case \(s \equiv \tau x \gets e\). According to the definition of \(\textsc{Compile\_Com}\), we distinguish two subcases.

If \(x \notin \dom(\rho_0)\), then a fresh clean qubit \(\overline{p}\) is taken from the pool \(\Qpool_0\):
\[
\begin{aligned}
\ctx = \ctx_0[ &C \mapsto C;C_s,\; \Vactive \mapsto \Vactive \cup \{\overline{p}\},\; \Valloc \mapsto \Valloc \cup \base(\qv(C_s) \setminus \cod(\rho_0)) \cup \base(\overline{p}), \\
             &\rho_{\mathrm{loc}} \mapsto \rho_{\mathrm{loc}}[x \mapsto \overline{p}],\; \Qpool \mapsto \Qpool_b],
\end{aligned}
\]
where $\overline{p}=\pop(\Qpool_0)$, $\Qpool_a=\Qpool_0 \setminus \overline{p}$ and \(\Gamma;\rho; \Qpool_a \vdash (\tau x \gets e) \leadsto C_s;\Qpool_b\). 

If \(x \in \dom(\rho_0)\), then \(\rho_0(x)\) is already mapped; we simply add it to the current‑use set and update the all‑variables set accordingly:
\[
\begin{aligned}
\ctx = &\ctx_0[ C \mapsto C;C_s,\; \Vactive \mapsto \Vactive \cup (\rho_{\mathrm{loc},0}(x) \setminus \rho_{\mathrm{loc},0} (\fp(f_{\mathrm{cur}}))),\; \\ 
 &\qquad \Valloc \mapsto V_{\alloc,0} \cup \base(\qv(C_s) \setminus \cod(\rho_0)), \Qpool \mapsto \Qpool_b], 
\end{aligned}
\]
where \(\Gamma;\rho_0; \Qpool_0 \vdash (\tau x \gets e) \leadsto C_s;\Qpool_b\). 
In both cases \(C_s\) is the compiled command obtained from the assignment statement via the translation judgment, and note that \(\mvq(s)=x\). By Lemma~\ref{theo_assgn}, we have \(\rho(x) \subseteq \qv(C_s)\), $\Qpool_b=\Qpool_a$ if $x \notin \dom(\rho_0)$ and $\Qpool_b=\Qpool_0$ if $x \in \dom(\rho_0)$. 

Now set \(\ctx' = \ctx_0\), \(D_1 = \rho(x)\), and \(D_2 = \base(\qv(C_s) \setminus \cod(\rho'))\). We have \(\ctx' \approx_{(\ket{\phi},k)} \sigma_0\) and \((\ctx', (\ket{\phi}, k, \D_{\SCC(f_{\mathrm{cur}})})) \in \mathcal{V}\).
Since \(\Call(s)=\emptyset\), the following properties follow easily from the above definitions: 
\begin{itemize}
    \item If \(x \notin \dom(\rho')\) then \(\rho(x)=\rho_{\mathrm{loc}}(x) \subseteq \Qpool'\);
    \item \(D_1= \rho(x)\) if \(\neg (x.\flag = 2 \wedge \Call(s)\neq \emptyset)\);  otherwise, \(D_1 =\rho(x) \cup Gar_{f_{\mathrm{cur}}}\).
    \item \(\base(\cod(\rho_{\mathrm{loc}}) \setminus \cod(\rho'_{\mathrm{loc}})) \subseteq \base(\rho_{\mathrm{loc}}(x) \setminus \cod(\rho'_{\mathrm{loc}})) \subseteq \base(\qv(C_s) \setminus \cod(\rho')) \subseteq D_2\);
    \item for any repetition count \(z\), \(\base(\qv(C_s^z) \setminus \cod(\rho')) = \base((\qv(C_s) \setminus \cod(\rho'))) = D_2 \subseteq D_2 \cup \anc(F_{\succeq f_{\mathrm{cur}}^*}^z)\); 
\end{itemize}
For the subcase \(x \notin \dom(\rho')\), we have \(\overline{p} = \rho(x) \subseteq \qv(C_s)\) and thus \(\base(\qv(C_s) \setminus \cod(\rho')) \cup \base(\overline{p}) = \base(\qv(C_s) \setminus \cod(\rho'))\). Hence \(\Valloc = \Valloc' \cup D_2\). Besides, if \(x \notin \dom(\rho')\), then \(\rho_{\mathrm{loc}}(x) \subseteq \Qpool'\) and therefore \(\rho_{\mathrm{loc}}(x) \cap (\rho'_{\mathrm{loc}}(\fp(f_{\mathrm{cur}})) \cup \cod(\rho_{\mathrm{glob}}))) = \emptyset\) since \(\fp(f_{\mathrm{cur}}) \subseteq  \dom(\rho'_{\mathrm{loc}})\) and \(\Qpool' \cap (\cod(\rho'_{\mathrm{loc}}) \cup \cod(\rho_{\mathrm{glob}})) = \emptyset\) by Condition~3(c) and Condition~2 in the validity of \(\ctx'\). Moreover, \(\rho'_{\mathrm{loc}}(x) \cap \cod(\rho_{\mathrm{glob}}) = \emptyset\) by Condition~3(c) in the validity of \(\ctx'\). Thus \(\Vactive = \Vactive' \cup (D_1 \setminus (\rho'(\fp(f_{\mathrm{cur}})) \cup \cod(\rho_{\mathrm{glob}})))\). Finally, if \(x \in \dom(\rho')\), then \(\cod(\rho'_{\mathrm{loc}}) \cap \Qpool' = \emptyset\) and $\rho_{\mathrm{loc}}(x)=\rho'_{\mathrm{loc}}(x)$ gives \(\Qpool = \Qpool' \setminus \rho_{\mathrm{loc}}(x)\).  

Therefore, the new context can be written uniformly as
\begin{align*}
\ctx = \ctx'[ &C \mapsto C;C_s,\; \Vactive' \mapsto \Vactive' \cup (D_1 \setminus (\rho'_{\mathrm{loc}}(\fp(f_{\mathrm{cur}})) \cup \cod(\rho_{\mathrm{glob}}))),\; \Valloc' \mapsto \Valloc' \cup D_2,\\
            &\rho'_{\mathrm{loc}} \mapsto \rho'_{\mathrm{loc}}[x \mapsto \rho_{\mathrm{loc}}(x) \mid x \notin \dom(\rho')],\; \Qpool' \mapsto \Qpool' \setminus \rho_{\mathrm{loc}}(x) ].
\end{align*}

For every \(h \in \SCC(f) \cap \Fun(\St')\), if \(h \notin \Theta^{-1}(\dom(\D'))\) then \(\ctx.A_h = \ctx'.A_h\); otherwise \(\ctx'.A_h \subseteq \ctx.A_h\). The proof of this fact is analogous to that of Lemma~\ref{lem:com-Aset-1}.

We then verify \(cv = cv' \setminus (D_1 \setminus (\rho'_{\mathrm{loc}}(\fp(f_{\mathrm{cur}})) \cup \cod(\rho_{\mathrm{glob}})))\):
\begin{itemize}
  \item If \(x \in \dom(\rho')\), then \(D_1 = \rho'(x)\) and
        \(\rho'(x) \cap cv' = \emptyset\). As in Lemma~\ref{lem:com-Aset-1}, we have \(cv=cv'\), and therefore we can rewrite
        \[
        cv = cv' \setminus D_1
           = cv' \setminus \bigl(D_1 \setminus (\rho'_{\mathrm{loc}}(\fp(f_{\mathrm{cur}}))
                                           \cup \cod(\rho_{\mathrm{glob}}))\bigr).
        \]
  \item If \(x \notin \dom(\rho')\), then \(D_1 = \rho(x) =\rho_{\mathrm{loc}}(x)\) and
        \(D_1 \setminus (\rho'_{\mathrm{loc}}(\fp(f_{\mathrm{cur}})) \cup \cod(\rho_{\mathrm{glob}}))
         = \rho_{\mathrm{loc}}(x)\) since $\rho_{\mathrm{loc}}(x)\subseteq \Qpool'$ and $\Qpool' \cap (\rho'_{\mathrm{loc}}(\fp(f_{\mathrm{cur}})) \cup \cod(\rho_{\mathrm{glob}}))=\emptyset$. As in Lemma~\ref{lem:com-Aset-1}, we have \(cv=cv' \setminus \rho_{\mathrm{loc}}(x)\). Therefore, we again obtain
        \(cv = cv' \setminus (D_1 \setminus (\rho'_{\mathrm{loc}}(\fp(f_{\mathrm{cur}})) \cup \cod(\rho_{\mathrm{glob}})))\).
\end{itemize} 

% Finally, from the analysis above we easily deduce
% \[
% B \setminus \bigl(\rho'_l(\fp(\curf)) \cup \rho'_{\mathrm{glob}}\bigr)
%    \subseteq \rho_l(x)
%    \subseteq cv' \cup ucv'_{\curf} \cup ucv'_{\mathrm{pred}(\curf)}
%    \subseteq cv' \cup ucv'_{\succeq \curf^*} .
% \]

% Thus the required inclusions hold.

By Lemma~\ref{theo_assgn} and $\Qpool_a \subseteq \Qpool'$, we know that \(\qv(C_s^k) \setminus (\rho(x) \cup \rho(\rvq(s))) \subseteq \Qpool'\). Hence \(\anc(C_s^k) \subseteq \Qpool' \cup \cod(\rho'_{\mathrm{loc}})\). Thus \(\anc(C_s^k) \setminus (\mathcal{F}_{\succeq}^k) \subseteq \Qpool' \cup \cod(\rho'_{\mathrm{loc}})\).

Note that if \(x \notin \dom(\rho_0)\) then \(\rho_{\mathrm{loc}}(x) \subseteq \Qpool'\). Hence \(\qv(C_s) \setminus (\rho'(x) \cup \rho'(\rvq(s))) \subseteq \Qpool' \subseteq cv'\). Since \(cv'\) is disjoint from \(V'_{\activelabel,f_{\mathrm{cur}}} \cup \cod(\rho'_{\mathrm{loc}})\), we obtain
\[
\qv(C_s) \setminus (\rho'(x) \cup \rho'(\rvq(s))) \subseteq E \subseteq \ctx'.A_{f_{\mathrm{cur}}} \setminus (\ctx'.V_{\activelabel,f_{\mathrm{cur}}} \cup \cod(\rho'_{\mathrm{loc}})).
\]

By Lemma~\ref{lem:curf} and the validity of \(\ctx'\), for any \(\ket{\varphi}\) with \(\ket{\varphi} \models \ket{0}_{\ctx'.A_{f_{\mathrm{cur}}} \setminus \rho'_{\mathrm{loc}}(\fp(f_{\mathrm{cur}}))}\), we have
\[
\ctx'.C_{f_{\mathrm{cur}}}^k \ket{\varphi} \models \ket{0}_{E}.
\]

Moreover, Theorem~\ref{theo_assgn} applied to \(\rho\) yields, for any \(\sigma'\),
\begin{align*}
&\langle C_s, \ket{\sigma'(\mathrm{var}(e))}_{\rho(\mathrm{var}(e))} \otimes \ket{0}_{\rho(x)} \otimes \ket{0}_{\qv(C_s)\setminus (\rho(\mathrm{var}(e)) \cup \rho(x)) } \rangle \\
\rightarrow^{*} \ &\langle \downarrow, \ket{\sigma'(\mathrm{var}(e))}_{\rho(\mathrm{var}(e))} \otimes \ket{\sigma'(e)}_{\rho(x)}\otimes \ket{0}_{\qv(C_s)\setminus (\rho(\mathrm{var}(e)) \cup \rho(x)) } \rangle.
\end{align*}

Since \(\sigma'\) is arbitrary,
  \(\qv(C_s)\setminus(\rho(\mathrm{var}(e))\cup\rho(x))\subseteq E\),
  and every reachable pre-execution state of \(C_s\) has \(\rho(x)\) clean by
  the preceding conventions, this implies the trace condition
  \[
  \tr_{D_1}\bigl(C_s^{k}\ket{\psi}\bigr)
  =
  \tr_{D_1}\bigl(\ket{\psi}\bigr)
  \]
  for any \(\ket{\psi}\) reachable immediately before \(C_s\) and satisfying
  \(\ket{\psi}\models\ket{0}_{E}\).  

\noindent \textbf{Equivalence \(\ctx \approx_{(\ket{\phi},k)} \sigma\).}
Since \(\ctx' \approx_{(\ket{\phi},k)} \sigma_0\), we have
\[
\langle \ctx'.C_{f_{\mathrm{cur}}}^k,\; \ket{\phi} \rangle \rightarrow^{*}
\langle \downarrow,\; \mathcal{E}(\sigma_0, \rho'_{\mathrm{active}}) \otimes \ket{\theta}_{V'_{\activelabel,f_{\mathrm{cur}}} \setminus \cod(\rho'_{\mathrm{active}})} \otimes \ket{\phi}_{\mathrm{rem}} \rangle.
\]
Since \(\ket{\phi} \models \ket{0}_{\ctx'.A_{f_{\mathrm{cur}}} \setminus \rho'_{\mathrm{loc}}(\fp(f_{\mathrm{cur}}))}\) by Condition~2 in the validity of \(\ctx'\), Lemma~\ref{lem:curf} yields \(\ctx'.C_{f_{\mathrm{cur}}}^k\ket{\phi} \models \ket{0}_{E}\). By assumption, \(\langle s^k,\sigma_0\rangle \rightarrow^{*} \langle \downarrow, \sigma \rangle\). By the semantics of the assignment \(s\), we have \(\sigma = \sigma_0[x \mapsto \sigma_0(e)]\).

We then consider two cases for \(\rho(x)\):

\begin{itemize}
    \item If \(x \in \dom(\rho')\) (where \(\rho=\rho'=\rho_0\) and thus \(E=\qv(C_s)\setminus (\rho(\mathrm{var}(e)) \cup \rho(x))\)), then we can write
    \begin{align*}
    &\mathcal{E}(\sigma_0, \rho'_{\mathrm{active}}) \otimes \ket{\theta}_{V'_{\activelabel,f_{\mathrm{cur}}} \setminus \cod(\rho'_{\mathrm{active}})} \otimes \ket{\phi}_{\mathrm{rem}} \\
    =&\; \mathcal{E}(\sigma_0, \rho'_{\mathrm{active}}) \otimes  \ket{0}_{E} \otimes \ket{\theta}_{V'_{\activelabel,f_{\mathrm{cur}}} \setminus \cod(\rho'_{\mathrm{active}})} \otimes \ket{\phi}_{\mathrm{rem}}\\
    =&\; \bigl(\otimes_{y \in \dom(\rho'_{\mathrm{active}})}\ket{\sigma_0(y)}_{\rho'(y)}\bigr) \otimes  \ket{0}_{E} \otimes \ket{\theta}_{V'_{\activelabel,f_{\mathrm{cur}}} \setminus \cod(\rho'_{\mathrm{active}})} \otimes \ket{\phi}_{\mathrm{rem}}\\
    =&\; \ket{\sigma_0(\mathrm{var}(e))}_{\rho'(\mathrm{var}(e))} \otimes \ket{\sigma_0(x)}_{\rho'(x)} \otimes \ket{0}_{E} \otimes \mathcal{E}(\sigma_0, \rho'_{\mathrm{active}}\setminus (\{x\} \cup \mathrm{var}(e))) \otimes \ket{\theta}_{V'_{\activelabel,f_{\mathrm{cur}}} \setminus \cod(\rho'_{\mathrm{active}})} \otimes \ket{\phi}_{\mathrm{rem}}\\
    =&\; \ket{\sigma_0(\mathrm{var}(e))}_{\rho'(\mathrm{var}(e))} \otimes \ket{0}_{\rho(x)} \otimes  \ket{0}_{E} \otimes \mathcal{E}(\sigma_0, \rho'_{\mathrm{active}}\setminus (\{x\} \cup \mathrm{var}(e))) \otimes \ket{\theta}_{V'_{\activelabel,f_{\mathrm{cur}}} \setminus \cod(\rho'_{\mathrm{active}})} \otimes \ket{\phi}_{\mathrm{rem}} \tag{since \(\sigma_0(x)=0\)}
    \end{align*}

    \item If \(x \notin \dom(\rho_0)\) (i.e., \(\rho(x)\) is freshly taken from \(\Qpool'\)), then \(\rho(x) \not\subseteq \dom(\rho'_{\mathrm{active}})\) and \(E=\qv(C_s)\setminus (\rho(\mathrm{var}(e)) \cup \rho(x)) \cup \rho(x)\), and we have
    \begin{align*}
    &\mathcal{E}(\sigma_0, \rho'_{\mathrm{active}}) \otimes \ket{\theta}_{V'_{\activelabel,f_{\mathrm{cur}}} \setminus \cod(\rho'_{\mathrm{active}})} \otimes \ket{\phi}_{\mathrm{rem}}\\
    =&\; \mathcal{E}(\sigma_0, \rho'_{\mathrm{active}}) \otimes \ket{0}_{\rho(x)} \otimes \ket{0}_{E \setminus \rho(x)} \otimes \ket{\theta}_{V'_{\activelabel,f_{\mathrm{cur}}} \setminus \cod(\rho'_{\mathrm{active}})} \otimes \ket{\phi}_{\mathrm{rem}}\\
    =&\; \ket{\sigma_0(\mathrm{var}(e))}_{\rho'(\mathrm{var}(e))} \otimes \ket{0}_{\rho(x)} \otimes \ket{0}_{E \setminus \rho(x)} \otimes \mathcal{E}(\sigma_0, \rho'_{\mathrm{active}} \setminus (\{x\} \cup \mathrm{var}(e))) \otimes \ket{\theta}_{V'_{\activelabel,f_{\mathrm{cur}}} \setminus \cod(\rho'_{\mathrm{active}})} \otimes \ket{\phi}_{\mathrm{rem}} .
    \end{align*}
\end{itemize}

Then
\begin{align*}
& \langle \ctx.C_{f_{\mathrm{cur}}}^k,\; \ket{\phi} \rangle = \langle \ctx'.C_{f_{\mathrm{cur}}}^k; C_s^k,\; \ket{\phi} \rangle \\
\rightarrow^{*} \ &\langle C_s^k,\;  \ket{\sigma_0(\mathrm{var}(e))}_{\rho'(\mathrm{var}(e))} \otimes \ket{0}_{\rho(x)} \otimes \ket{0}_{E}\\ 
  & \qquad \qquad  \otimes \mathcal{E}(\sigma_0, \rho'_{\mathrm{active}} \setminus (\{x\} \cup \mathrm{var}(e))) \otimes  \ket{\theta}_{V'_{\activelabel,f_{\mathrm{cur}}} \setminus \cod(\rho'_{\mathrm{active}})} \otimes \ket{\phi}_{\mathrm{rem}}\rangle \\
\rightarrow^{*}  \ &\langle \downarrow,\;  \ket{\sigma_0(\mathrm{var}(e))}_{\rho'(\mathrm{var}(e))} \otimes \ket{\sigma_0(e)}_{\rho(x)} \otimes \ket{0}_{E} \\ 
  & \qquad \qquad \otimes \mathcal{E}(\sigma_0, \rho'_{\mathrm{active}} \setminus (\{x\} \cup \mathrm{var}(e))) \otimes  \ket{\theta}_{V'_{\activelabel,f_{\mathrm{cur}}} \setminus \cod(\rho'_{\mathrm{active}})} \otimes \ket{\phi}_{\mathrm{rem}}\rangle \\
= \ &\langle \downarrow,\; \mathcal{E}(\sigma_0, \rho'_{\mathrm{active}} \setminus \{x\}) \otimes \ket{\sigma_0(e)}_{\rho(x)} \otimes \ket{0}_{E} \otimes \ket{\theta}_{V'_{\activelabel,f_{\mathrm{cur}}} \setminus \cod(\rho'_{\mathrm{active}})} \otimes \ket{\phi}_{\mathrm{rem}} \rangle \\
= \ &\langle \downarrow,\; \mathcal{E}(\sigma, \rho_{\mathrm{active}}) \otimes \ket{0}_{E} \otimes \ket{\theta}_{V'_{\activelabel,f_{\mathrm{cur}}} \setminus \cod(\rho'_{\mathrm{active}})} \otimes \ket{\phi}_{\mathrm{rem}} \rangle  \tag{\(\sigma = \sigma_0[x \mapsto \sigma_0(e)]\)}\\
= \ &\langle \downarrow,\; \mathcal{E}(\sigma, \rho_{\mathrm{active}}) \otimes \ket{\theta}_{V_{\activelabel,f_{\mathrm{cur}}} \setminus \cod(\rho_{\mathrm{active}})} \otimes \ket{\phi}_{\mathrm{rem}} \rangle,
\end{align*}
which is exactly \(\ctx \approx_{(\ket{\phi},k)} \sigma\). 

\textbf{Case~2.} \(s \equiv l \gets \operatorname{aop} e\) or \(s \equiv \tau x \gets f(\overline{u})\) with \(f \in \Theta^{-1}(\dom(\D_0))\) and \(f \notin \Fun(\St_0)\):  
These cases follow a similar pattern to the assignment case above. For the external function call, the correctness relies on the validity of \(\D_0\), where validity states that for any \(\Theta(f)\in \D_0\) with \(f \notin \Fun(\St_0)\), we have \(\Theta(f)^z \approx_{G'} f^z\) for any \(z\) and some set $G'$. The details are analogous and thus omitted here. 

\textbf{Case~3.} \(s \equiv \mathtt{if} \ x \ \{s_1\}\) with \(M(x)=\mathtt{Q}\).
We have \(C_s = (\mathbf{qif} \ \rho(x) \ \ket{1} \rightarrow C_1 \ \square \ \ket{0} \rightarrow \mathbf{skip})\) and
\(\ctx = \ctx_1[C_1 \mapsto C_s]\), where \(\ctx_1\) and \(C_1\) satisfy
\[
  (\ctx_1, C_1) = \mathrm{Compile\_Com}(\K, \ctx_0, s_1).
\]
Let \(\sigma_1\) be such that \(\langle s_1, \sigma_0 \rangle \rightarrow^{*} \langle \downarrow, \sigma_1 \rangle\).
From the semantics of the conditional we have \(\sigma = \sigma_1\) when \(\sigma_0(x) = \mathtt{true}\)
and \(\sigma = \sigma_0\) when \(\sigma_0(x) = \mathtt{false}\).

Since \(\ctx_0 \approx_{(\ket{\phi},k)} \sigma_0\) and \(\ctx_0\) is valid, the induction hypothesis
provides a compilation context \(\ctx'\) together with sets \(D_1, D_2\) such that
\(\ctx' \approx_{(\ket{\phi},k)} \sigma_0\) and
\((\ctx', (\ket{\phi}, k, \D_{\SCC(f_{\mathrm{cur}})})) \in \mathcal{V}\), satisfying the following conditions.
\begin{itemize}
  \item \(\rho_1(\mvq(s))=\rho_{\mathrm{loc},1}(\mvq(s)) \in \Qpool'\) if \(\mvq(s) \not\subseteq \dom(\rho')\).  
  \item \(D_1 = \rho_1(\mvq(s))\) if \(\neg((\mvq(s)).\flag = 2 \wedge \Call(s)\neq \emptyset)\); otherwise \(D_1 = \rho_1(\mvq(s)) \cup Gar_{f_{\mathrm{cur}}}\). 
  \item \(\base(\cod(\rho_{\mathrm{loc},1}) \setminus \cod(\rho'_{\mathrm{loc}})) \subseteq D_2\).
\end{itemize}
Moreover, \(\ctx_1\) can be written as
\begin{align*}
  \ctx_1 = \ctx'[&C \mapsto C; C_1,\; \Vactive \mapsto \Vactive \cup (D_1 \setminus (\rho'_{\mathrm{loc}}(\fp(f_{\mathrm{cur}})) \cup \cod(\rho_{\mathrm{glob}}))),\; \Valloc \mapsto \Valloc \cup D_2,\;\\ 
                 &\rho_{\mathrm{loc}} \mapsto \rho'_{\mathrm{loc}}[\mvq(s) \mapsto \rho_{\mathrm{loc},1}(\mvq(s)) \mid \mvq(s) \not\subseteq \dom(\rho')],\; \Qpool \mapsto \Qpool' \setminus \rho_{\mathrm{loc},1}(\mvq(s)) \,],
\end{align*}
and the following properties hold.
\begin{itemize}
  \item \(cv = cv' \setminus (D_1 \setminus (\rho'_{\mathrm{loc}}(\fp(f_{\mathrm{cur}})) \cup \cod(\rho_{\mathrm{glob}})))\).
  \item For every \(h \in \SCC(f) \cap \Fun(\St')\), if \(h \notin \Theta^{-1}(\dom(\D'))\) then \(\ctx_1.A_h = \ctx'.A_h\); otherwise \(\ctx'.A_h \subseteq \ctx_1.A_h\). 
  % \item \(D_1 \setminus (\rho'_{\mathrm{loc}}(\fp(f_{\mathrm{cur}})) \cup \cod(\rho_{\mathrm{glob}})) \subseteq cv' \cup V'_{\activelabel,\succeq f_{\mathrm{cur}}^*}\). 
  \item \(\qv(C_1^{k}) \setminus \rho'(\mvq(s) \cup \rvq(s)) \subseteq \ctx'.A_{f_{\mathrm{cur}}} \setminus (\ctx'.V_{\activelabel,f_{\mathrm{cur}}} \cup \cod(\rho'_{\mathrm{loc}}))\).
  \item For any \(\ket{\varphi}\) with \(\ket{\varphi} \models \ket{0}_{\ctx'.A_{f_{\mathrm{cur}}} \setminus \rho'(\fp(f_{\mathrm{cur}}))}\),
        \(\ctx'.C_{f_{\mathrm{cur}}}^k \ket{\varphi} \models \ket{0}_{\rho_{\mathrm{loc},1}(\mvq(s) \setminus \dom(\rho_0))}\).
  \item For any repetition count \(z\), \(\base(\qv(C_1^z) \setminus \cod(\rho')) \subseteq D_2 \cup \anc(F_{\succeq f_{\mathrm{cur}}^*}^z)\). 
  \item \(\anc(C_1^k) \setminus \mathcal{F}_{\succeq}^k \subseteq \Qpool' \cup \cod(\rho'_{\mathrm{loc}})\).
  \item Set \(E \triangleq \qv(C_1^{k}) \setminus \rho'(\mvq(s) \cup \rvq(s)) \cup \rho_{\mathrm{loc},1}(\mvq(s) \setminus \dom(\rho_0))\).
        Then for every \(\ket{\psi}\) with \(\ket{\psi} \models \ket{0}_{E}\),
        \(\tr_{D_1}(C_1^{k} \ket{\psi}) = \tr_{D_1}(\ket{\psi})\).
\end{itemize}
Furthermore, \(\ctx_1 \approx_{(\ket{\phi}, k)} \sigma_1\).

Observe that the step from \(\ctx_1\) to \(\ctx\) only replaces \(C_1\) by \(C_s\); all other compilation components remain unchanged. By definition, \(\mvq(s) = \mvq(s_1)\). For the code fragments we have the following relations.
\begin{enumerate}
  \item Since the guard \(x \in \dom(\rho')\), for any \(z\),
        \[
        \qv(C_s^z) \setminus \cod(\rho')
        = \qv(C_1^z \cup \rho(x)) \setminus \cod(\rho')
        = \qv(C_1^z \cup \rho'(x)) \setminus \cod(\rho')
        = \qv(C_1^z) \setminus \cod(\rho'). 
        \]
  \item Similarly,
        \[
        \begin{aligned}
        \qv(C_s^{k}) \setminus (\rho'(\mvq(s)) \cup \rho'(\rvq(s)))
        = &\qv(C_1^{k} \cup \rho'(x)) \setminus (\rho'(\mvq(s_1)) \cup (\rho'(\rvq(s_1)) \cup \rho'(x))) \\ 
        = &\qv(C_1^{k}) \setminus (\rho'(\mvq(s_1)) \cup \rho'(\rvq(s_1))).
         \end{aligned}
        \]
  \item \(\anc(C_s^k) \setminus \mathcal{F}_{\succeq}^k \subseteq (\anc(C_1^k) \setminus \mathcal{F}_{\succeq}^k) \cup \rho'_{\mathrm{loc}}(x) \subseteq \Qpool' \cup  \cod(\rho'_{\mathrm{loc}})\). 
  \item For any \(\ket{\psi}\), \(\tr_{\rho_1(\mvq(s))}(C_s \ket{\psi}) = \tr_{\rho_1(\mvq(s))}(C_1 \ket{\psi})\), $\rho(\mvq(s))=\rho_1(\mvq(s))$ and \(\rho_1(\mvq(s)) \subseteq D_1\).
\end{enumerate}
These equalities, together with the definition and the fact that \(\ctx_1.A_h = \ctx.A_h\) for any \(h \in \SCC(f) \cap \Fun(\St')\), immediately imply that all the properties listed above for \(\ctx_1\) remain valid for \(\ctx\). 

It remains to prove \(\ctx \approx_{(\ket{\phi}, k)} \sigma\).
From \(\ctx_1 \approx_{(\ket{\phi}, k)} \sigma_1\) we have
\begin{align*}
    \langle \ctx_1.C_{f_{\mathrm{cur}}}^k, \ket{\phi} \rangle
    &= \langle \ctx'.C_{f_{\mathrm{cur}}}^k; C_1, \ket{\phi} \rangle \\
    &\rightarrow^{*} \langle C_1,\;
         \mathcal{E}(\sigma_0, \rho'_{\mathrm{active}}) \otimes \ket{\theta}_{V'_{\activelabel,f_{\mathrm{cur}}} \setminus \cod(\rho'_{\mathrm{active}})} \otimes \ket{\phi}_{\mathrm{rem}} \rangle \\
    &\rightarrow^{*} \langle \downarrow,\;
         \mathcal{E}(\sigma_1, \rho_{\mathrm{active},1}) \otimes \ket{\theta}_{V_{\activelabel,f_{\mathrm{cur}},1} \setminus \cod(\rho_{\mathrm{active},1})} \otimes \ket{\phi}_{\mathrm{rem}} \rangle .
\end{align*}
We now expand the execution of \(\ctx.C_{f_{\mathrm{cur}}}^k\) with \(\rho(x)=\rho_1(x)=\rho'(x)\).  
\begin{align*}
      &\langle \ctx.C_{f_{\mathrm{cur}}}^k, \ket{\phi} \rangle \\ 
    = \ &\langle \ctx'.C_{f_{\mathrm{cur}}}^k; (\mathbf{qif} \ \rho'(x) \ \ket{1} \rightarrow C_1 \ \square \ \ket{0} \rightarrow \mathbf{skip}), \ket{\phi} \rangle \\
    \rightarrow^{*} \ &\langle (\mathbf{qif} \ \rho'(x) \ \ket{1} \rightarrow C_1 \ \square \ \ket{0} \rightarrow \mathbf{skip}),\;
         \mathcal{E}(\sigma_0, \rho'_{\mathrm{active}}) \otimes \ket{\theta}_{V'_{\activelabel,f_{\mathrm{cur}}} \setminus \cod(\rho'_{\mathrm{active}})} \otimes \ket{\phi}_{\mathrm{rem}} \rangle \\
    = \ & \Bigl\langle (\mathbf{qif} \ \rho'(x) \ \ket{1} \rightarrow C_1 \ \square \ \ket{0} \rightarrow \mathbf{skip}),\; \\ 
    & \qquad 
         \mathcal{E}(\sigma_0, \rho'_{\mathrm{active}} \setminus \{x\} )
         \otimes (\alpha\ket{1}+\beta\ket{0})_{\rho'(x)}
         \otimes \ket{\theta}_{V'_{\activelabel,f_{\mathrm{cur}}} \setminus \cod(\rho'_{\mathrm{active}})}
         \otimes \ket{\phi}_{\mathrm{rem}} \Bigr\rangle \\
    \rightarrow \ &\Bigl\langle \downarrow,\;
         \alpha \bigl( \mathcal{E}(\sigma_1, \rho_{\mathrm{active},1} \setminus \{x\})
         \otimes \ket{1}_{\rho_1(x)}
         \otimes \ket{\theta}_{V_{\activelabel,f_{\mathrm{cur}},1} \setminus \cod(\rho_{\mathrm{active},1})}
         \otimes \ket{\phi}_{\mathrm{rem}} \bigr) \\
         &\qquad\quad + \beta \bigl( \mathcal{E}(\sigma_0, \rho'_{\mathrm{active}} \setminus \{x\}) 
         \otimes \ket{0}_{\rho'(x)}
         \otimes \ket{\theta}_{V'_{\activelabel,f_{\mathrm{cur}}} \setminus \cod(\rho'_{\mathrm{active}})}
         \otimes \ket{\phi}_{\mathrm{rem}} \bigr) \Bigr\rangle \\
    = \ &\Bigl\langle \downarrow,\;
         \alpha \bigl( \mathcal{E}(\sigma_1, \rho_{\mathrm{active}} \setminus \rho(x))
         \otimes \ket{1}_{\rho(x)}
         \otimes \ket{\theta}_{V_{\activelabel,f_{\mathrm{cur}}} \setminus \cod(\rho_{\mathrm{active}})}
         \otimes \ket{\phi}_{\mathrm{rem}} \bigr) \\
         &\qquad\quad + \beta \bigl( \mathcal{E}(\sigma_0, \rho'_{\mathrm{active}} \setminus \rho'(x))
         \otimes \ket{0}_{\rho'(x)}
         \otimes \ket{\theta}_{V'_{\activelabel,f_{\mathrm{cur}}} \setminus \cod(\rho'_{\mathrm{active}})}
         \otimes \ket{\phi}_{\mathrm{rem}} \bigr) \Bigr\rangle .
\end{align*}
Here \(\alpha\) and \(\beta\) are real numbers in \([0,1]\). 
The last equality uses the identification \(\rho_{\mathrm{active},1} = \rho_{\mathrm{active}}\) etc.\ from the construction of \(\ctx_1\). 

We now distinguish the two possible outcomes of the conditional test.
\begin{itemize}
  \item If \(\sigma_0(x) = \mathbf{true}\), then from \(\ctx' \approx_{(\ket{\phi},k)} \sigma_0\) the control qubit is in state \(\ket{1}\); hence \(\alpha=1,\ \beta=0\).
        The final state becomes
        \[
        \mathcal{E}(\sigma_1, \rho_{\mathrm{active}} \setminus \rho(x))
        \otimes \ket{1}_{\rho(x)}
        \otimes \ket{\theta}_{V_{\activelabel,f_{\mathrm{cur}}} \setminus \cod(\rho_{\mathrm{active}})}
        \otimes \ket{\phi}_{\mathrm{rem}},
        \]
        which equals \(\mathcal{E}(\sigma, \rho_{\mathrm{active}}) \otimes \ket{\theta}_{V_{\activelabel,f_{\mathrm{cur}}} \setminus \cod(\rho_{\mathrm{active}})} \otimes \ket{\phi}_{\mathrm{rem}}\)
        since \(\sigma = \sigma_1\) and \(\rho(x)\) is already part of \(\rho_{\mathrm{active}}\).

  \item If \(\sigma_0(x) = \mathbf{false}\), then \(\ket{0}_{\rho'(x)}\) implies \(\alpha=0,\ \beta=1\).  We simplify the \(\beta\)-component.
        \begin{align*}
            &\mathcal{E}(\sigma_0, \rho'_{\mathrm{active}} \setminus \rho'(x) )
            \otimes \ket{0}_{\rho'(x)}
            \otimes \ket{\theta}_{V'_{\activelabel,f_{\mathrm{cur}}} \setminus \cod(\rho'_{\mathrm{active}})}
            \otimes \ket{\phi}_{\mathrm{rem}} \\
             =&\; \mathcal{E}(\sigma_0, \rho'_{\mathrm{active}})
            \otimes \ket{\theta}_{V'_{\activelabel,f_{\mathrm{cur}}} \setminus \cod(\rho'_{\mathrm{active}})}
            \otimes \ket{\phi}_{\mathrm{rem}} \\ 
            =&\; \mathcal{E}(\sigma, \rho_{\mathrm{active}})
            \otimes \ket{\theta}_{V_{\activelabel,f_{\mathrm{cur}}} \setminus \cod(\rho_{\mathrm{active}})}
            \otimes \ket{\phi}_{\mathrm{rem}},
        \end{align*}
          where we used \(\sigma = \sigma_0\) and the fact that the compilation data for \(\ctx\) matches that of \(\ctx'\). 
\end{itemize}
In either case we obtain exactly the configuration required for
\(\ctx \approx_{(\ket{\phi},k)} \sigma\).  Hence the equivalence holds.

\textbf{Case~4.} $s \equiv \tau x \gets f(\overline{u}) \ \operatorname{and} \  f \in \Fun(\St_0)$: In this case, $f \in \SCC(f_{\mathrm{cur}})$. 
we define \(\ctx\) based on two mutually exclusive conditions:

\textit{Subcase 3.1.} If \(x.\flag = 2\), then
\[
\ctx = \ctx_2\bigl[C \mapsto C; C_s,\;  
        \Vactive \mapsto V_{\activelabel,2} \cup Gar_{f_{\mathrm{cur}}} \bigr],
\]
where
\[
\begin{aligned}
&C_s = \Theta(f)\Bigl(\getC(\overline{u}),\; \rho_2(\getQ(\overline{u})),\; \rho_2(x),\; 
          (E_{\args,f}[c \mapsto c+1])\Bigr),\\
&c = \{ \var(\idx(q)) \mid q \in V'_{\activelabel,\succeq f} \}
\end{aligned}
\]

\noindent
\textit{Subcase 3.2.} If $x.\flag \neq 2$, then
\[
\ctx = \ctx_2\bigl[C \mapsto C; C_s \bigr],
\]
where \(C_s\) is defined as in Subcase 3.1. 

Finally, the compiled context is \(\ctx\) and the compiled command is \(C_s\). We first have that \(\ctx_2\) is valid  and $\ctx_2 \approx_{\ket{\phi},k} \sigma_0$ by Lemma~\ref{lem:state-valid}.  

We set \(\ctx' = \ctx_2\).  Observe that in this case \(\mvq(s) = x\)
and \(\rho = \rho'\).  
We distinguish two cases:
\[
D_1 \triangleq
\begin{cases}
\rho(x) \cup \ctx.\Gar_{f_{\mathrm{cur}}}, & \text{if } x.\flag = 2 ,\\[2mm]
\rho(x),                        & \text{otherwise},
\end{cases}
\]
and let \(D_2 \triangleq \emptyset\).

From the construction of \(\ctx'\) and \(\ctx\) we obtain
\begin{itemize}
  \item if \(x.\flag = 2\) then \(D_1 = \rho(x) \cup \Gar_{f_{\mathrm{cur}}}\);
        otherwise \(D_1 = \rho(x)\);
  \item \(\cod(\rho_{\mathrm{loc}}) \setminus \cod(\rho'_{\mathrm{loc}}) = \emptyset \subseteq D_2\);
  \item if \(x \notin \dom(\rho')\) then \(\rho(x)=\rho_{\mathrm{loc}}(x)\in \Qpool'\); 
\end{itemize} 
% Because \(x.\flag = 2\), we have  
% \(x \notin \fp(\curf) \cup \dom(\rho_{\mathrm{glob}})\). 
% Since \(ctx.\Gar_{\curf} = ctx'.\Gar_{\curf}\) and this set is disjoint 
% from \(\rho'_l(\fp(\curf) \cup \dom(\rho_{\mathrm{glob}}))\).
By definition, we already have \(\rho(x) = \rho'(x) \subseteq V'_{\activelabel}\). Hence $V_{\activelabel} = V'_{\activelabel} \cup D_1$ by the definition. 
The type system of the source language implies \(x \notin \dom(\rho_{\mathrm{glob}}) \cup \fp(f_{\mathrm{cur}})\). Together with the injectivity of \(\rho'_{\mathrm{loc}}\) and $\cod(\rho'_{\mathrm{loc}}) \cap \cod(\rho_{\mathrm{glob}})=\emptyset$, we obtain
 \[
    \rho_{\mathrm{loc}}(x) \cap \rho'_{\mathrm{loc}}(\fp(f_{\mathrm{cur}})) \cup \cod(\rho_{\mathrm{glob}})=\emptyset.\] 
Moreover, if $x.\flag=2$, then \(D_1 = \rho_{\mathrm{loc},f_{\mathrm{cur}}}(x) \cup Gar_{f_{\mathrm{cur}}}\).  
By definition we have 
          \[
          Gar_{f_{\mathrm{cur}}} = Gar'_{f_{\mathrm{cur}}} \cup \bigl( \rho_{\mathrm{loc}}(x) \cup \rho_{\mathrm{loc}}(x)[+] \bigr).
          \]

          By definition, \(\rho_{\mathrm{loc}}(x) \cup \rho_{\mathrm{loc}}(x)[+]\) is unchanged from \(\ctx'\). 
          The validity of \(\ctx'\) further guarantees
          \[
          \bigl( \rho_{\mathrm{loc}}(x)[+] \bigr)
          \cap \bigl( \rho'_{\mathrm{loc}}(\fp(f_{\mathrm{cur}})) \cup \cod(\rho_{\mathrm{glob}}) \bigr) = \emptyset,
          \]
          and therefore
          \[
          \bigl( \rho_{\mathrm{loc}}(x) \cup \rho_{\mathrm{loc}}(x)[+] \bigr)
          \cap \bigl( \rho'_{\mathrm{loc}}(\fp(f_{\mathrm{cur}})) \cup \cod(\rho_{\mathrm{glob}}) \bigr) = \emptyset .
          \]

          Concerning \(Gar'_{f_{\mathrm{cur}}} = Q'_{\succeq f_{\mathrm{cur}}^*}[+] \cup Q'_{\succeq f_{\mathrm{cur}}}\):
          the validity of \(\ctx'\) ensures
          \[
          Q'_{\succeq f_{\mathrm{cur}}^*}[+] \cap \bigl( \rho'_{\mathrm{loc}}(\fp(f_{\mathrm{cur}})) \cup \cod(\rho_{\mathrm{glob}}) \bigr) = \emptyset .
          \]
          The set \(Q'_{\succeq f_{\mathrm{cur}}}\) collects the quantum variables of all \(\flag = 2\)
          variables of \(f_{\mathrm{cur}}\) and the functions that follow it.
          By the validity conditions of \(\ctx'\), the variable mappings of every function
          are disjoint from \(\rho_{\mathrm{glob}}\), and apart from the images of the formal
          parameters and return values they are also disjoint from the mappings of
          preceding functions.  Since \(\flag = 2\) variables are never formal parameters
          or return variables, we obtain
          \[
          Q'_{\succeq f_{\mathrm{cur}}} \cap \bigl( \rho'_{\mathrm{loc}}(\fp(f_{\mathrm{cur}})) \cup \cod(\rho_{\mathrm{glob}}) \bigr) = \emptyset .
          \] 
It follows that
\(D_1 \cap (\rho'_{\mathrm{loc}}(\fp(f_{\mathrm{cur}})) \cup \cod(\rho_{\mathrm{glob}})) = \emptyset\).  Hence, we obtain that \[
    V_{\activelabel} = V'_{\activelabel} \cup \bigl(D_1 \setminus ( \rho'_{\mathrm{loc}}(\fp(f_{\mathrm{cur}})) \cup \cod(\rho_{\mathrm{glob}})) \bigr). 
          \]
          Moreover, by definition, \(\rho_{\mathrm{loc}}(x) = \rho'_{\mathrm{loc}}(x)\) already belongs to 
\(\cod(\rho'_{\mathrm{loc}})\) and to \(V'_{\activelabel,f_{\mathrm{cur}}}\), but not to \(\Qpool'\). Consequently the update of the compilation context can be written as
\[ 
\begin{aligned}
\ctx = &\;\ctx'[\,C \mapsto C';C_s,\;
             \rho_{\mathrm{loc}} \mapsto \rho'_{\mathrm{loc}}[x \mapsto \rho_{\mathrm{loc}}(x) \mid x \notin \dom(\rho')],\;
             \Qpool \mapsto \Qpool' \setminus \rho_{\mathrm{loc}}(x),\; \\
        &\; \Vactive \mapsto \Vactive' \cup (D_1 \setminus (\rho'_{\mathrm{loc}}(\fp(f_{\mathrm{cur}})) \cup \cod(\rho_{\mathrm{glob}}))),\;
             \Valloc \mapsto \Valloc' \cup D_2\,]. 
\end{aligned}
\] 

Next we analyse the clean variable pool \(cv\).
In subcase~3.2, since only the compiled code \(C\) is modified and all other compilation variables stay unchanged, we immediately obtain \(cv = cv'\) and \(\ctx.A_h = \ctx'.A_h\). In subcase~3.1, we first prove \(cv = cv' \setminus (Q'_{\succeq f_{\mathrm{cur}}^*}[+])\) as follows.
\[
\begin{aligned}
cv &= \Qpool \cup \Qpool[+] \cup
     \bigl((V_{\activelabel,\succeq f_{\mathrm{cur}}^*}[+] \cup \cod(\rho_{\mathrm{loc},\succeq f_{\mathrm{cur}}^*})[+])
            \setminus (Q_{f_{\mathrm{cur}}}[+] \cap V_{\activelabel, f_{\mathrm{cur}}})\bigr) \\
   &= \Qpool' \cup \Qpool'[+] \cup
     \bigl((V'_{\activelabel,\succeq f_{\mathrm{cur}}^*}[+] \cup \cod(\rho'_{\mathrm{loc},\succeq f_{\mathrm{cur}}^*})[+])
            \setminus (Q'_{\succeq f_{\mathrm{cur}}^*}[+] \cap V_{\activelabel,f_{\mathrm{cur}}})\bigr) \\
   &\qquad (\text{since } \ctx'.Gar_{f_{\mathrm{cur}}}[+] \subseteq \cod(\rho'_{\mathrm{loc},\succeq f_{\mathrm{cur}}^*})[+]
            \text{ and } Q_{\succeq f_{\mathrm{cur}}^*} = Q'_{\succeq f_{\mathrm{cur}}^*}) \\[2mm]
   &= \Qpool' \cup \Qpool'[+] \cup
      \bigl((V'_{\activelabel,\succeq f_{\mathrm{cur}}^*}[+] \cup \cod(\rho'_{\mathrm{loc},\succeq f_{\mathrm{cur}}^*})[+])
            \setminus Q'_{\succeq f_{\mathrm{cur}}^*}[+]\bigr) \\[2mm]
   &= \Qpool' \cup \Qpool'[+] \cup
      \bigl((V'_{\activelabel,\succeq f_{\mathrm{cur}}^*}[+] \cup \cod(\rho'_{\succeq f_{\mathrm{cur}}^*})[+])
            \setminus (Q'_{\succeq f_{\mathrm{cur}}^*}[+] \cap V_{\activelabel, f_{\mathrm{cur}}})\bigr)
      \setminus Q'_{\succeq f_{\mathrm{cur}}^*}[+] \\
   &\qquad (\text{by Condition~2 in the validity of } \ctx,\;
            (\Qpool' \cup \Qpool'[+]) \cap Q'_{\succeq f_{\mathrm{cur}}^*}[+] = \emptyset) \\[2mm]
   &= cv' \setminus Q'_{\succeq f_{\mathrm{cur}}^*}[+]. \\
\end{aligned} 
\]
Since \(\Vactive = \Vactive' \cup Gar'_{f_{\mathrm{cur}}}\), by definition we have
\[
V_{\activelabel,\succeq h} \cup \cod(\rho_{\mathrm{loc},h}) = V'_{\activelabel,\succeq h} \cup Gar'_{f_{\mathrm{cur}}} \cup \cod(\rho'_{\mathrm{loc},h})
= V'_{\activelabel,\succeq h} \cup (Q'_{\succeq f_{\mathrm{cur}}^*}[+]) \cup Q'_{\succeq f_{\mathrm{cur}}} \cup \cod(\rho'_{\mathrm{loc},h}).
\]

If \(\pos(f) \le \pos(f_{\mathrm{cur}})\), the validity of \(\ctx'\) implies
\(Q'_{\succeq f_{\mathrm{cur}}} \subseteq V'_{\activelabel,\succeq f_{\mathrm{cur}}} \subseteq V'_{\activelabel,\succeq h}\); hence
\[
V_{\activelabel,\succeq h} \cup \cod(\rho_{\mathrm{loc},h}) = V'_{\activelabel,\succeq h} \cup (Q'_{\succeq f_{\mathrm{cur}}^*}[+]) \cup \cod(\rho'_{\mathrm{loc},h}).
\]
Since \(Q'_{\succeq f_{\mathrm{cur}}^*}[+] \subseteq cv' \cup V'_{\activelabel,f_{\mathrm{cur}}} \subseteq cv' \cup V'_{\activelabel,\succeq h}\)
by the definition of \(cv'\), combining with the change of \(cv\) we obtain
\(\ctx.A_h = \ctx'.A_h\).
If \(\pos(f) > \pos(f_{\mathrm{cur}})\), we have \(\ctx'.A_h \subseteq \ctx.A_h\).

For every \(f = f_{\mathrm{cur}}\) or \(f \notin \Theta^{-1}(\dom(\D)) =
\Theta^{-1}(\dom(\D'))\), the validity of \(\ctx'\) gives
\(\pos(f) \le \pos(f_{\mathrm{cur}})\); therefore the corresponding conclusion on
\(A_h\) holds.

Next we rewrite \(cv\) in the form required by the theorem.
In Subcase~3.2, we have \(cv = cv'\).  By definition \(D_1 = \rho(x) = \rho'(x)\),
and the validity of \(\ctx'\) guarantees \(\rho'(x) \cap cv' = \emptyset\);
consequently
\[
cv = cv' \setminus (D_1 \setminus (\rho'_{\mathrm{loc}}(\fp(f_{\mathrm{cur}})) \cup \cod(\rho_{\mathrm{glob}})).
\]

In Subcase~3.1, we have \(cv = cv' \setminus (Q'_{\succeq f_{\mathrm{cur}}^*}[+])\).
By definition \(D_1 = \rho'(x) \cup (Q'_{\succeq f_{\mathrm{cur}}^*}[+]) \cup Q'_{\succeq f_{\mathrm{cur}}}\).
Similarly \(\bigl(\rho'(x) \cup Q'_{\succeq f_{\mathrm{cur}}}\bigr) \cap cv' = \emptyset\), so we can again write
\[
cv = cv' \setminus (D_1 \setminus (\rho'_{\mathrm{loc}}(\fp(f_{\mathrm{cur}})) \cup \cod(\rho_{\mathrm{glob}}))). 
\]

% Finally, by definition
% \( D_1 \setminus (\rho'_{\mathrm{loc}}(\fp(f_{\mathrm{cur}})) \cup \cod(\rho_{\mathrm{glob}})) \subseteq \rho_{\mathrm{loc}}(x) \cup Gar_{f_{\mathrm{cur}}}
% \subseteq \rho_{\mathrm{loc}}(x) \cup Q'_{\succeq f_{\mathrm{cur}}^*}[+] \cup Q'_{\succeq f_{\mathrm{cur}}}\).
% From the analysis above we already have
% \(Q'_{\succeq f_{\mathrm{cur}}^*}[+] \subseteq cv' \cup V'_{\activelabel,f_{\mathrm{cur}}}\) and
% \(\rho'_{\mathrm{loc}}(x) \cup Q'_{\succeq f_{\mathrm{cur}}} \subseteq \cod(\rho'_{\mathrm{loc}}) \cup V_{\activelabel,\succeq f_{\mathrm{cur}}}\).
% Altogether,
% \[
% D_1 \setminus (\rho'_{\mathrm{loc}}(\fp(f_{\mathrm{cur}})) \cup \cod(\rho_{\mathrm{glob}})) \subseteq cv' \cup V'_{\activelabel,f_{\mathrm{cur}}^*} \cup \cod(\rho'_{\mathrm{loc}}). 
% \]

The next goal is to prove the following five facts:
\begin{enumerate}
  \item \(\base(\qv(C_s^z) \setminus \cod(\rho'))
        \subseteq D_2 \cup \anc(F_{\succeq f_{\mathrm{cur}}^*}^z)\);
  \item \(\anc(C_s^k) \setminus \mathcal{F}_{\succeq}^k \subseteq \Qpool' \cup \cod(\rho'_{\mathrm{loc}})\);  
  \item \(\qv(C_s^k) \setminus (\rho'(\rvq(s)) \cup \rho'(\mvq(s)))
        \subseteq \ctx'.A_f \setminus (\ctx'.V_{\activelabel,f_{\mathrm{cur}}} \cup \rho'_{\mathrm{loc}})\);
  \item \(\tr_{D_1}(C_s^k\ket{\psi}) = \tr_{D_1}(\ket{\psi})\) for every
        \(\ket{\psi}\) with \(\ket{\psi} \models \ket{0}_{E}\). 
 \item For any \(\ket{\varphi}\) with
          \(\ket{\varphi} \models \ket{0}_{\ctx'.A_{f_{\mathrm{cur}}} \setminus \rho'_{\mathrm{loc}}(\fp(f_{\mathrm{cur}}))}\),
          \(\ctx'.C_{f_{\mathrm{cur}}}^k \ket{\varphi} \models \ket{0}_{\rho_{\mathrm{loc}}(\mvq(s) \setminus \dom(\rho_{0}))}\).
\end{enumerate}

Since \(\ctx'\) is valid, Condition~1 supplies, for all \(h \in \SCC(f_{\mathrm{cur}}) \cap \Fun(\ctx'_r.\St)\), 
\begin{itemize}
  \item \(\D_{\SCC(f_{\mathrm{cur}})}[\Theta(h)]^k \approx_{\ctx'_r.Gar_h} h^k\);
  \item \(\ctx'_r.E_{\args,f}\) is a prefix of
        \(\args(\D_{\SCC(f_{\mathrm{cur}})}[\Theta(f)])\);
  \item if \(h \in (\Fun(\ctx'.\St))
          \setminus \Theta^{-1}(\dom(\ctx'_r.\D))\), 
        then \(\anc(\D_{\SCC(f_{\mathrm{cur}})}[\Theta(h)]^k) \subseteq \ctx'_r.A_h\);
  \item if \(h \in \Theta^{-1}(\dom(\ctx'_r.\D))\), 
        then \[\anc(\D_{\SCC(f_{\mathrm{cur}})}[\Theta(h)]^k)
        \subseteq \anc(\ctx'_r.\D[\Theta(h)]^k) \cup \anc(\ctx'_r.\D[\Theta(h)]^k)[+].\]
\end{itemize} 

Observe that \(f \in \Fun(\ctx'.\St)\).  Since \(\Fun(\ctx'.\St)\subseteq \Fun(\ctx'_r.\St)\) by Lemma~\ref{lem:completion-inv}, we have \(f \in \Fun(\ctx'_r.\St)\). 
If \(\pos(f) > \pos(f_{\mathrm{cur}})\), the stack discipline immediately gives
\(f \in \SCC(f_{\mathrm{cur}})\).
If \(\pos(f) < \pos(f_{\mathrm{cur}})\), the stack discipline yields
\(f_{\mathrm{cur}} \in \Succ(f)\).  Together with \(f \in \Call(s)\),
\(s \preceq ctx'.R_{\mathrm{stmt}}\) and the validity of \(\ctx'\)
(which guarantees \(ctx'.R_{\mathrm{stmt}}\) is a suffix of  \(s_{f_{\mathrm{cur}}}\)),
Lemma~\ref{lem:static-wf} (\(\Call(s_{f_{\mathrm{cur}}}) = \Succ(f_{\mathrm{cur}})\)) implies
\(f \in \Succ(f_{\mathrm{cur}})\); consequently \(f \in \SCC(f_{\mathrm{cur}})\) as well.
Thus the above properties hold for \(h = f\).

\textbf{The prefix and ancilla conditions for \(f\).}
First, it is easy to get \(\ctx'.E_{\args,f} \preceq \ctx'_r.E_{\args,f}\) by Lemma~\ref{lem:completion-inv} and thus the second property 
above directly gives that \(\ctx'.E_{\args,f}\) is a prefix of
\(\args(\D_{\SCC(f_{\mathrm{cur}})}[\Theta(f)])\).

We now prove \(\anc(\D_{\SCC(f_{\mathrm{cur}})}[\Theta(f)]^k) \subseteq \ctx'.A_f\)
by case analysis.

\begin{itemize}
  \item \(f \notin \Theta^{-1}(\dom(\ctx'.\D))\). Since $f \in (\Succ(f_{\mathrm{cur}}) \cap \Fun(\St'))$, we can prove that $f \notin \Theta^{-1}(\dom(\ctx'_r.\D))$ by Lemma~\ref{lem:completion-inv}; therefore,
        by the third property, 
        \(\anc(\D_{\SCC(f_{\mathrm{cur}})}[\Theta(f)]^k) \subseteq \ctx'_r.A_f\).
        Since \(\ctx'_r.A_f \subseteq \ctx'.A_f\) by Lemma~\ref{lem:completion-inv}, 
        we obtain \(\anc(\D_{\SCC(f_{\mathrm{cur}})}[\Theta(f)]^k) \subseteq \ctx'.A_f\).

  \item \(f \in \Theta^{-1}(\dom(\ctx'.\D))\).  
        Condition~4(c) of the validity of \(\ctx'\) gives
        \(\anc(\ctx'.\D[\Theta(f)]^{k+1}) \subseteq \ctx'.A_f\).
        Since \(\anc(\ctx'.\D[\Theta(f)]^k) \subseteq
        \anc(\ctx'.\D[\Theta(f)]^{k+1})\), we get
        \(\anc(\ctx'.\D[\Theta(f)]^k) \subseteq \ctx'.A_f\). Since \(\dom(\ctx'.\D) \subseteq \dom(\ctx'_r.\D)\) by Lemma~\ref{lem:completion-inv}, we have \(f \in \Theta^{-1}(\dom(\ctx'_r.\D))\).  
        Then the fourth property yields
        \[\anc(\D_{\SCC(f_{\mathrm{cur}})}[\Theta(f)]^k)
        \subseteq \anc(\ctx'_r.\D[\Theta(f)]^k) \cup \anc(\ctx'_r.\D[\Theta(f)]^k)[+].\]
        Since \(\anc(\ctx'_r.\D[\Theta(f)]^k) \subseteq \anc(\ctx'.\D[\Theta(f)]^k)\cup \anc(\ctx'.\D[\Theta(f)]^k)[+]\) by Lemma~\ref{lem:completion-inv},
        we conclude \(\anc(\D_{\SCC(f_{\mathrm{cur}})}[\Theta(f)]^k)
        \subseteq \anc(\ctx'.\D[\Theta(f)]^k) \subseteq \ctx'.A_f\).
\end{itemize}
In both subcases we obtain \(\anc(\D_{\SCC(f_{\mathrm{cur}})}[\Theta(f)]^k) \subseteq \ctx'.A_f\).

We interpret \(F_f\) with respect to the declaration family
\(\D_{\SCC(f_{\mathrm{cur}})}\), i.e., through the declaration
\(\D_{\SCC(f_{\mathrm{cur}})}[\Theta(f)]\). Note that  \(c \subseteq \var(idx(V'_{\activelabel,\succeq f})) \subseteq \var(idx(\ctx'.A_{f^*})) \subseteq E'_{\args,f}\) and \(E'_{\args,f}\) is a prefix of \(\args(F_f)\). Combining this with the definition of \(C_s\), we obtain for any \(z\),
\[
\begin{aligned}
&\qv(C_s^z) \setminus \rho'(\rvq(s) \cup \mvq(s)) \\ 
= \ & \bigl(\qv(F_f^z[\mathrm{in} \mapsto \rho'(\getQ(\overline{u})),\;
                  \mathrm{out} \mapsto \rho'(x),\;
                  c \mapsto c+1])\bigr)
   \setminus \rho'(\rvq(s) \cup \mvq(s)) \\
\subseteq \ & \anc(F_f^z)[c \mapsto c+1]. \\
\end{aligned}
\]
From these we immediately obtain, for any \(z\),
\[
\begin{aligned}
\base(\qv(C_s^z) \setminus \cod(\rho'))
&\subseteq \base\bigl(\qv(C_s^z) \setminus \rho'(\rvq(s)) \setminus \rho'(\mvq(s))\bigr) \\
&= \base(\anc(F_f^z))[c \mapsto c+1]
 = \base(\anc(F_f^z))
 \subseteq \base(\anc(F_{\succeq f_{\mathrm{cur}}^*}^z)),
\end{aligned}
\]
which is fact~(1). 

Moreover, since \(\anc(C_s^k) \subseteq \anc(F_f^k[c\mapsto c+1]) \cup \cod(\rho')\), we obtain
\[
\anc(C_s^k) \setminus \bigl(\anc(F_{\succeq f_{\mathrm{cur}}^*}^k) \cup \anc(F_{\succeq f_{\mathrm{cur}}^*}^k)[+]\bigr)
\subseteq \Qpool' \cup \cod(\rho'),
\]
hence the required condition (fact~(2)) holds.

\noindent 
\textbf{Ancilla shift is contained in \(cv'\).}
We want to show \(\anc(F_f^k)[c \mapsto c+1] \subseteq \ctx'.cv\). 
We first prove \(\ctx'.Q_{\succeq f_{\mathrm{cur}}^*}[+] \cap \ctx'.V_{\activelabel,f_{\mathrm{cur}}} = \emptyset\);
this will later simplify the expression of \(cv'\). By Lemma~\ref{lem:Gar_0-empty}, we have \(\ctx_0.\Gar_{f_{\mathrm{cur}}} \cap V_{\activelabel,f_{\mathrm{cur}},0} = \emptyset\), which indicates that \(Q_{\succeq f_{\mathrm{cur}}^*,0}[+] \cap \ctx_0.V_{\activelabel,f_{\mathrm{cur}}} = \emptyset\). By Lemma~\ref{lem:Q-plus-holds-ctx2}, we have \(\ctx'.Q_{\succeq f_{\mathrm{cur}}^*}[+] \cap \ctx'.V_{\activelabel,f_{\mathrm{cur}}} = \ctx_0.Q_{\succeq f_{\mathrm{cur}}^*}[+] \cap \ctx_0.V_{\activelabel,f_{\mathrm{cur}}} \), and thus is also empty. Thus \(cv'\) simplifies to
\[
cv' = \Qpool \cup \Qpool[+] \cup V'_{\activelabel,\succeq f_{\mathrm{cur}}^*}[+] \cup \cod(\rho'_{\mathrm{loc},\succeq f_{\mathrm{cur}}^*})[+].
\]

Expand the definition:
\[
\begin{aligned}
\anc(F_f^k)[c \mapsto c+1]
&= \bigl(\anc(F_f^k) \setminus \{q[i] \in \anc(F_f^k) \mid i \in c\}\bigr)
   \;\cup\; \{q[i] \in \anc(F_f^k) \mid i \in c\}[+] \\
&= \bigl(\anc(F_f^k) \setminus \{q[i] \mid i \in c\}\bigr)
   \;\cup\; \{q[i] \in \anc(F_f^k) \mid i \in c\}[+].
\end{aligned}
\]

From Condition (4)(a) of the validity of \(\ctx'\), it follows that for every
\[
q \in V'_{\activelabel,\succeq f} \setminus (V'_{\activelabel,f_{\mathrm{cur}}} \cup \rho'_{\mathrm{loc},f_{\mathrm{cur}}}(\re(f_{\mathrm{cur}}))),
\]
we have \(\var(\idx(q)) \neq \emptyset\). Moreover, the construction of \(\ctx'\) ensures the same property for every
\[
q \in V'_{\activelabel,f_{\mathrm{cur}}} \setminus \rho'_{\mathrm{loc},f_{\mathrm{cur}}}(\{x\} \cap \re(f_{\mathrm{cur}})).
\]
Additionally, the validity of \(\ctx'\) gives that for any \(q \in  V'_{\activelabel,\succeq f_{\mathrm{cur}}^*} \cap \rho'_{\mathrm{loc},f_{\mathrm{cur}}}(\re(f_{\mathrm{cur}}))\), if \(\var(\idx(q)) = \emptyset\), then \(q \in \rho'_{\mathrm{loc},f}(\re(f))\).  
Combining the above three observations, we conclude that all variables in
\[
\ctx'.V_{\activelabel,\succeq f} \setminus \rho'_{\mathrm{loc},f}(\re(f))
\]
are subscripted quantum variables. 
Recalling that \(c\) collects exactly the indices of the arrays in $\ctx'.V_{\activelabel,\succeq f}$, we obtain
\(\ctx'.V_{\activelabel,\succeq f} \setminus \rho'_{\mathrm{loc},f}(\re(f)) \subseteq \{q[i] \mid i \in c\}\).  

We already proved \(\anc(F_f^k) \subseteq \ctx'.A_f\) and, by definition,
\(\anc(F_f^k) \cap (\rho_{\mathrm{loc},f}(\fp(f)) \cup \rho_{\mathrm{loc},f}(\re(f))) = \emptyset\). 
Hence
\[
\begin{aligned}
&\bigl(\anc(F_f^k) \setminus \{q[i] \in \anc(F_f^k) \mid i \in c\}\bigr) \\
 \subseteq \ &
   \bigl(\ctx'.A_f \setminus (\rho'_{\mathrm{loc},f}(\fp(f) \cup \re(f)))\bigr)
   \setminus \ctx'.V_{\activelabel,\succeq f}  \\
\subseteq \ & cv' \;
        \cup \; \bigl((\cod(\rho'_{\mathrm{loc},f}) \setminus (\rho'_{\mathrm{loc},f}(\fp(f) \cup \re(f)))) \setminus V'_{\activelabel,\succeq f} \bigr) .
\end{aligned}
\]

The validity of \(\ctx'\) tells us that \(\cod(\rho'_{\mathrm{loc},f}) \setminus (\rho'_{\mathrm{loc},f}(\fp(f) \cup \re(f))) \subseteq V'_{\activelabel,\succeq f}\), so the left‑hand side is a subset of \(cv'\). 

For the second part,
\[
\begin{aligned}
&\{q[i] \in \anc(F_f^k) \mid i \in c\}[+] \\ 
\subseteq \ & \ctx'.A_f[+] \\
= \ & \bigl(\Qpool' \cup \Qpool'[+] \cup V'_{\activelabel,\succeq f_{\mathrm{cur}}^*}[+] \cup \cod(\rho'_{\mathrm{loc},\succeq f_{\mathrm{cur}}^*})[+]
        \cup V'_{\activelabel,\succeq f} \cup \cod(\rho'_{\mathrm{loc},f})\bigr)[+] \\
\subseteq \ & \Qpool' \cup \Qpool'[+] \cup V'_{\activelabel,\succeq f_{\mathrm{cur}}^*}[+] \cup \cod(\rho'_{\mathrm{loc},\succeq f_{\mathrm{cur}}^*})[+]
        \cup V'_{\activelabel,\succeq f}[+] \cup \cod(\rho'_{\mathrm{loc},f})[+] \\
\subseteq \ & \Qpool' \cup \Qpool'[+] \cup V'_{\activelabel,\succeq f_{\mathrm{cur}}^*}[+] \cup \cod(\rho'_{\mathrm{loc},\succeq f_{\mathrm{cur}}^*})[+]
        \quad (\text{since } f \in \SCC(f_{\mathrm{cur}}) \subseteq \Fun(\St_{\succeq f_{\mathrm{cur}}^*})) \\
=  \ & cv'.
\end{aligned}
\]

Both pieces together yield
\(\anc(F_f^k)[c \mapsto c+1] \subseteq cv' \subseteq
 \ctx'.A_{f_{\mathrm{cur}}} \setminus (V'_{\activelabel,f_{\mathrm{cur}}} \cup \cod(\rho'_{\mathrm{loc}}))\).

% 
% \noindent
% \textbf{Quantum variables of \(C_s^k\) satisfy the same inclusion.}
% Recall that \(\anc(P^k) = \qv(P^k) \setminus \rho_f(\fp(f)) \setminus \rho_f(\re(f))\)
% and, using the property just established,
% \[
% \bigl(\qv(P^k) \setminus \rho_f(\fp(f)) \setminus \rho_f(\re(f))\bigr)[c \mapsto c+1]
% \cap \rho'(\rvq(s) \cup \mvq(s)) = \emptyset.
% \]

% Because \(ctx'.args_f\) is a prefix of \(args(\D_{\SCC(\curf)}(\Theta(f)))\),
% the quantum variables of \(C^k\) are obtained by substitution:
% \[
% \qv(C_s^k) = \qv(P^k)\bigl[\rho_f(\fp(f)) \mapsto \rho'(\getQ),\;
%                            \rho_f(\re(f)) \mapsto \rho'(x),\;
%                            c \mapsto c+1\bigr].
% \]
Consequently,
\[
\begin{aligned}
\qv(C_s^k) \setminus \rho'(\rvq(s) \cup \mvq(s)) \subseteq \anc(F_f^k)[c \mapsto c+1] \subseteq \ctx'.A_{f_{\mathrm{cur}}} \setminus (\ctx'.V_{\activelabel,f_{\mathrm{cur}}} \cup \cod(\rho'_{\mathrm{loc}})). 
\end{aligned}
\] 
This proves fact~(3).

\noindent
\textbf{Trace equality on \(D_1\).}
Since \(x \notin \dom(\rho_0)\), we can set
\(E \triangleq \qv(C_s^k) \setminus \rho'(\rvq(s)) \setminus \rho'(\mvq(s)) \cup \rho_{\mathrm{loc}}(x)\).
We shall prove
\[
\tr_{D_1}(C_s^k\ket{\psi}) = \tr_{D_1}(\ket{\psi})
\quad\text{for all } \ket{\psi} \models \ket{0}_E.
\]

Define a bijection \(g : \anc(F_f^k) \to \anc(C_s^k)\) by 
\[
g(q[i]) = q[i+1] \text{ if } i \in c, \qquad g(v) = v \text{ otherwise}.
\]
The length restriction holds:
\[
\operatorname{length}(\rho(\getQ(\overline{u}))) = \operatorname{length}(\overline{u}),\qquad
\operatorname{length}(\rho(x)) = \operatorname{length}(x).
\]

As in the proof of the next case, we have \(\ctx'_r.\Gar_f \subseteq \ctx'.\Gar_f\).
Then the equivalence \(F_f^k \approx_{\ctx'.\Gar_f} f^k\) (which follows from
\(F_f^k \approx_{\ctx'_r.\Gar_f} f^k\)) guarantees the following properties.
For any classical states \(\sigma_0, \sigma\) with
\(\langle x \gets f^k(\overline{u}), \sigma_0 \rangle \to^{*} \langle \downarrow, \sigma \rangle\)
and any quantum state \(\ket{\psi}\) satisfying
\begin{itemize}
\item \(\ket{\psi}_{\cod(\rho_{\mathrm{glob}})} = \ket{\sigma_0(\dom(\rho_{\mathrm{glob}}))}\),  
  \item \(\ket{\psi}_{\rho(\getQ(\overline{u}))} = \ket{\sigma_0(\overline{u})}\),
  \item \(\ket{\psi}_{\rho(x)} = \ket{0}\),
  \item \(\ket{\psi}_{\anc(C_s^k)} = \ket{0}\), 
\end{itemize}
the computation yields
\[
\langle C_s^k, \ket{\psi} \rangle \to^{*}
\bigl\langle \downarrow,\;  
\ket{\sigma(\dom(\rho_{\mathrm{glob}}))}_{\cod(\rho_{\mathrm{glob}})} \otimes \ket{\sigma(x)}_{\rho(x)}
\otimes \ket{\theta}_{g(\ctx'.Gar_f \cap \anc(F_f^k))}
\otimes \ket{\psi}_{\mathrm{rem}} \bigr\rangle.
\]

If \(x.\flag \neq 2\) (i.e., \(x.\flag = 1\)),
Lemma~\ref{lem:static-wf} implies that \(\SCC(f_{\mathrm{cur}})\) contains no variable with flag~\(2\);
together with Condition~1 of \(\ctx'\) this forces \(\ctx'.Gar_f = \emptyset\).
In that case the reduction simplifies to
\[
\langle C_s^k, \ket{\psi} \rangle \to^{*}
\bigl\langle \downarrow,\;
\ket{\sigma(\dom(\rho_{\mathrm{glob}}))}_{\cod(\rho_{\mathrm{glob}})} \otimes 
 \ket{\sigma(x)}_{\rho(x)}
\otimes \ket{\psi}_{\mathrm{rem}}  \bigr\rangle.
\]

Observe that for every \(h \in \Fun(\St) \cap \SCC(f_{\mathrm{cur}})\),
\(Q_h \subseteq \rho_h \setminus (\rho_h(\fp(h)) \cup \rho_h(\re(f))) \subseteq V_{\activelabel,\succeq h}\);
hence \(Q_{\succeq f} \subseteq V_{\activelabel,\succeq f}\).
Thus all these variables \(Q_{\succeq f}\) are arrays and their indices belong to \(c\),
we obtain
\[
g(\ctx'.Gar_f \cap \anc(F_f^k))
\subseteq \bigl(\ctx'.Q_{\succeq f_{\mathrm{cur}}^*}[+] \cup \ctx'.Q_{\succeq f}\bigr)[c \mapsto c+1]
\subseteq \ctx'.Q_{\succeq f_{\mathrm{cur}}^*}[+]
\subseteq Gar'_{f_{\mathrm{cur}}} \setminus \rho'(x). 
\]
Therefore the reduction can be rewritten uniformly as
\[
\langle C_s^k, \ket{\psi} \rangle \to^{*}
\bigl\langle \downarrow,\; 
\ket{\sigma(\dom(\rho_{\mathrm{glob}}))}_{\cod(\rho_{\mathrm{glob}})} \otimes \ket{\sigma(x)}_{\rho(x)}
\otimes \ket{\theta}_{(Gar'_{f_{\mathrm{cur}}} \setminus \rho'(x))}
\otimes \ket{\psi}_{\mathrm{rem}} \bigr\rangle. 
\]

Recall that in Subcase~3.1 we set \(D_1 = \rho(x) \cup \ctx.Gar_{f_{\mathrm{cur}}}\),
while in SubCase~3.2 \(D_1 = \rho(x)\).  From the definition of \(\ctx\)
one easily checks that \(\ctx'.Gar_{f_{\mathrm{cur}}} = \ctx.Gar_{f_{\mathrm{cur}}}\) and
\(\rho(x) = \rho'(x)\).  Since the construction of \(\sigma_0\) is
arbitrary and $\sigma(\dom(\rho_{\mathrm{glob}}))=\sigma_0(\dom(\rho_{\mathrm{glob}}))$, the above reduction implies the desired trace equality
\[
\tr_{D_1}(C_s^k\ket{\psi}) = \tr_{D_1}(\ket{\psi})
\qquad (\ket{\psi} \models \ket{0}_E).
\] 

Finally, if \(x \notin \dom(\rho_{0})\), we have
\(\rho_{\mathrm{loc},1}(x) \in \Qpool_0\).  Together with
\(\ctx_1.C_{f_{\mathrm{cur}}} = \ctx_0.C_{f_{\mathrm{cur}}}\), $\ctx_1.A_{f_{\mathrm{cur}}}=\ctx_0.A_{f_{\mathrm{cur}}}$ and $\rho_{\mathrm{loc},1}(\fp(f_{\mathrm{cur}}))=\rho_{\mathrm{loc},0}(\fp(f_{\mathrm{cur}}))$, this implies that for any
\(\ket{\varphi}\) satisfying
\(\ket{\varphi} \models \ket{0}_{\ctx_1.A_{f_{\mathrm{cur}}} \setminus \rho_{\mathrm{loc},1}(\fp(f_{\mathrm{cur}}))}\),
\[
  \ctx_1.C_{f_{\mathrm{cur}}}^k \ket{\varphi} \models \ket{0}_{\rho_{\mathrm{loc},1}(x)}.
\]
The same reasoning as for renaming in Lemma~\ref{lem:state-valid-3.b} shows that
\[
  \ctx_2.C_{f_{\mathrm{cur}}}^k \ket{\varphi} \models \ket{0}_{\rho_{\mathrm{loc},2}(x)}.
\]
Since \(\ctx' = \ctx_2\), the required conclusion follows.

\noindent
\textbf{Establishing the state equivalence \(\ctx \approx_{(\ket{\phi},k)} \sigma\).}
From the validity of \(\ctx'\) we know
\(\ket{\phi} \models \ket{0}_{\ctx'.A_{f_{\mathrm{cur}}} \setminus \rho'_{\mathrm{loc},f_{\mathrm{cur}}}(\fp(f_{\mathrm{cur}}))}\).
Moreover \(\anc(C_s^k) \subseteq \ctx'.A_{f_{\mathrm{cur}}} \setminus \cod(\rho'_{\mathrm{loc},f_{\mathrm{cur}}})\); by
Lemma~\ref{lem:curf} this yields
\[
\ctx'.C_{f_{\mathrm{cur}}}^k \ket{\phi} \models \ket{0}_{\anc(C_s^k)}.
\]
Additionally one can show
\(\ctx'.C_{f_{\mathrm{cur}}}^k \ket{\phi} \models \ket{0}_{\rho(x)}\);
hence \(\ctx'.C_{f_{\mathrm{cur}}}^k \ket{\phi} \models \ket{0}_E\).

Now \(\ctx' \approx_{(\ket{\phi},k)} \sigma_0\) gives
\[
\langle \ctx'.C_{f_{\mathrm{cur}}}^k, \ket{\phi} \rangle \to^{*}
\bigl\langle \downarrow,\;
\mathcal{E}(\sigma_0, \rho'_{\mathrm{active}})
\otimes \ket{\theta'}_{V'_{\activelabel,f_{\mathrm{cur}}} \setminus \cod(\rho'_{\mathrm{active}})}
\otimes \ket{\phi}_{\mathrm{rem}} \bigr\rangle.
\]

Observe that
\[
\begin{aligned}
V_{\activelabel,f_{\mathrm{cur}}} \setminus \cod(\rho_{\mathrm{active}})
&= (\Vactive' \cup Gar'_{f_{\mathrm{cur}}}) \setminus \cod(\rho_{\mathrm{active}}) \\
&= (Gar'_{f_{\mathrm{cur}}} \setminus \cod(\rho_{\mathrm{active}}))
   \cup (V'_{\activelabel,f_{\mathrm{cur}}} \setminus \cod(\rho'_{\mathrm{active}})) \\
&= (Gar'_{f_{\mathrm{cur}}} \setminus \rho(x))
   \cup (V'_{\activelabel,f_{\mathrm{cur}}} \setminus \cod(\rho'_{\mathrm{active}})).
\end{aligned}
\]

Putting everything together we obtain the chain of reductions
(in the following we let \(\theta''\) be a suitable intermediate state):
\[
\begin{aligned}
&\langle \ctx_3.C_{f_{\mathrm{cur}}}^k, \ket{\phi} \rangle \\
= \ & \langle \ctx'.C_{f_{\mathrm{cur}}}^k ; C_s^k, \ket{\phi} \rangle \\
\to^{*} \ & \bigl\langle C_s^k,\; 
      \mathcal{E}(\sigma_0, \rho'_{\mathrm{active}})
      \otimes \ket{\theta'}_{V'_{\activelabel,f_{\mathrm{cur}}} \setminus \cod(\rho'_{\mathrm{active}})}
      \otimes \ket{\phi}_{\mathrm{rem}} \bigr\rangle \\
\to^{*} \ & \bigl\langle \downarrow,\;
     \mathcal{E}(\sigma, \rho_{\mathrm{active}}) \otimes \ket{\theta}_{(Gar'_{f_{\mathrm{cur}}} \setminus \rho'(x))}
      \otimes \ket{\theta'}_{V'_{\activelabel,f_{\mathrm{cur}}} \setminus \cod(\rho'_{\mathrm{active}})}
      \otimes \ket{\phi}_{\mathrm{rem}} \bigr\rangle \\
= \ & \bigl\langle \downarrow,\;
       \mathcal{E}(\sigma, \rho_{\mathrm{active}}) \otimes  \ket{\theta''}_{V_{\activelabel,f_{\mathrm{cur}}} \setminus \cod(\rho_{\mathrm{active}})}
      \otimes \ket{\phi}_{\mathrm{rem}} \bigr\rangle. 
\end{aligned}
\]

Thus we have shown \(\ctx \approx_{(\ket{\phi},k)} \sigma\), completing the proof.

\textbf{Case~5.} \(s=\tau x \gets f(\overline{u})\), \(f \notin \Theta^{-1}(\dom(\D_0))\) and \(f \notin \Fun(\St_0)\): if \(f \notin \SCC(f_{\mathrm{cur}})\), we define \(\ctx_1\) and \(\ctx_2\) as follows: 
\[
\ctx_1 = 
\begin{cases}
\ctx_0[\Qpool \mapsto \Qpool_0 \setminus \overline{p},\; 
       \rho_{\mathrm{loc}} \mapsto \rho_{\mathrm{loc},0}[x \mapsto \overline{p}], \\ 
       \qquad \Vactive \mapsto V_{\activelabel,0} \cup \overline{p}, 
       \Valloc \mapsto V_{\alloc,0} \cup \base(\overline{p})], &x \notin \dom(\rho_0); \\
      
\ctx_0[\Vactive \mapsto V_{\activelabel,0} \cup (\rho_{\mathrm{loc},0}(x) \setminus \rho_{\mathrm{loc},0}(\fp(f_\mathrm{curf})))], & x \in \dom(\rho_0).
\end{cases}
\]
We also let \(\ctx_2=\ctx_1\).  
Let \(\ctx_3 = \textsc{Compile\_Fun}(\K, \ctx_2, f, \{\rho_2(\mathrm{get\_Q}(\overline{u})), \rho_2(x)\})\).

If \(f \notin \Fun(\St_3)\), then \((\ctx, C_s) = \textsc{Compile\_Com}(\ctx_3, c)\).  
Otherwise, we distinguish two subcases based on the flag of \(x\):

\begin{itemize}
    \item If \(x.\flag = 2\), then
    \begin{align*}
    \ctx = \ctx_3\bigl[ &C_3 \mapsto C_3; C_s, \; V_{\activelabel,3} \mapsto V_{\activelabel,3} \cup Q_{3,f_{\mathrm{cur}}}[+] \bigr].
      \end{align*}
\item Otherwise,
    \[
    \ctx = \ctx_3\bigl[ C_3 \mapsto C_3; C_s \bigr].
    \]
\end{itemize}
In both cases, the compiled command \(C_s\) is given by
\[
C_s = \Theta(f)\bigl( \getC(\overline{u}),\; \rho_3(\getQ(\overline{u})),\; \rho_{3,\mathrm{loc}}(x),\; E_{\args,f} \bigr).
\]

Since \((\ctx_0, \ket{\phi}, k, \mathcal{D}_{\SCC(f_{\mathrm{cur}})}) \in \mathcal{V}_0\), Lemma~\ref{lem:state-valid} yields
\((\ctx_2, \ket{\phi}, k, \mathcal{D}_{\SCC(f_{\mathrm{cur}})}) \in \mathcal{V}_0\) and \(\ctx_2 \approx_{(\ket{\phi}, k)} \sigma_0\)
when \(f \in \SCC(f_{\mathrm{cur}})\); the case \(f \notin \SCC(f_{\mathrm{cur}})\) follows by the
same argument as that for \(\ctx_1\) in the proof of the lemma.
To apply Theorem~\ref{theo_fun}, we now verify the preconditions it requires. Assume that $\ctx_2$ satisfies the structural properties of validity.

First, by definition we clearly have
\(f \notin \Theta^{-1}(\dom(\D_2)) \cap \Fun(\St_2)\), $\rho_{2}(x) \subseteq V_{\activelabel,f_{\mathrm{cur}},2}$ 
and \(\ctx_{r,2} = \ctx_{r,3}\). 
Without loss of generality, we assume that all function arguments are local variables. A global argument can be replaced by a fresh local temporary that holds its value; the two are semantically equivalent. Hence, proving the local case suffices. Therefore, we have \(\rho_{2}(\getQ(\overline{u})) \cup \rho_{\mathrm{loc},2}(x) \subseteq \cod(\rho_{\mathrm{loc},2})\). 
Since the validity of \(\Gamma\) ensures \(\var(u) \cap \{x\} = \emptyset\),
and the validity of \(\ctx_2\) guarantees that \(\rho_{\mathrm{loc},2}\) is injective,
we obtain
\[ 
\rho_{\mathrm{loc},2}(\getQ(\overline{u})) \cap \rho_{\mathrm{loc},2}(x) = \emptyset .
\] 
As in the previous case, one shows \(f \in \Succ(f_{\mathrm{cur}})\);
and if \(f \in \SCC(f_{\mathrm{cur}})\), then \(\ctx_2.\Gar_{f_{\mathrm{cur}}} \cap V_{\activelabel,f_{\mathrm{cur}},2} = \emptyset\). 
Moreover, if \(f \in \SCC(f_{\mathrm{cur}})\), the construction of \(\ctx_2\) implies 
\(\var(\idx(q)) \neq \emptyset\) for every \(q \in V_{\activelabel,f_{\mathrm{cur}},2} \setminus (\rho_{\mathrm{loc},2}(\re(f_{\mathrm{cur}})) \cap \rho_{\mathrm{loc},2}(x))\).

% and for any $q \in \rho_{l,2}(x))$, $\var(\idx(q))=\emptyset \rightarrow q \in \rho_{l,2}(\re(\curf))$. 

Finally, we prove that assuming that $\ctx_3$ satisfies the structural properties of validity, then for every 
\(h \in \Fun(\St_3) \cap \SCC(f_{\mathrm{cur}})\),
\[
\ctx_{r,3}.\Gar_h \subseteq \ctx_3.\Gar_h.
\]

Recall that
\(\Gar_h = Q_{\succeq f_{\mathrm{cur}}^*}[+] \cup Q_{\succeq h}
        \subseteq Q_{\succeq f_{\mathrm{cur}}^*}[+] \cup Q_{\succeq f_{\mathrm{cur}}^*}
        = \Gar_{f_{\mathrm{cur}}^*}\),
which consists exactly of the quantum registers (together with their
future extensions) associated with variables \(x\) in
\(\Fun(\St_{\succ f_{\mathrm{cur}}^*})\) that satisfy \(x.\flag = 2\).
We can easily show that \(\Fun(\St_{\succ f_{\mathrm{cur}}^*}) \subseteq \SCC(f_{\mathrm{cur}}) \cap \Fun(\St)\) by Condition~1(2);  
therefore \(Q_{\succeq f_{\mathrm{cur}}^*}\) is determined solely by the flag‑2
variables of the functions inside \(\SCC(f_{\mathrm{cur}}) \cap \Fun(\St)\) and
their local register mappings.

The static parameters \(G_{\mathrm{fun}}\) and the call graph
\(G_{\mathrm{call}}\) are never modified during compilation.
Hence, to compare \(\ctx_3.\Gar_{f_{\mathrm{cur}}^*}\) with \(\ctx_{r,3}.\Gar_{f_{\mathrm{cur}}^*}\),
it suffices to check whether \(\SCC(f_{\mathrm{cur}}) \cap \Fun(\St)\) acquires
new functions during the step from \(\ctx_3\) to \(\ctx_{r,3}\),
and whether any existing function newly contributes a flag‑2 variable
in its local mapping.

Note that if \(f \in \SCC(f_{\mathrm{cur}})\), the flag definition,
exactly as argued above, implies \(x.\flag \neq 0\).
We distinguish two cases.

\begin{itemize}
\item \textbf{Case 1:} \(x.\flag = 1\).
      The mutual exclusion of static flag (Lemma~\ref{lem:static-wf})
      forbids any variable in this SCC from having flag \(2\);
      consequently no variable contributes to \(\Gar_{f_{\mathrm{cur}}^*}\).
      Hence \(\ctx_3.\Gar_{f_{\mathrm{cur}}^*} = \ctx_{r,3}.\Gar_{f_{\mathrm{cur}}^*} = \emptyset\),
      and for every \(h \in \SCC(f) \cap \Fun(\St)\),
      \(\ctx_3.\Gar_h = \ctx_{r,3}.\Gar_h = \emptyset\).

\item \textbf{Case 2:} \(x.\flag = 2\).
      By the definition of flags, \(f_{\mathrm{cur}}\) has no direct callee inside
      \(\SCC(g)\) other than \(f\).
      So \(\Succ(f_{\mathrm{cur}}) \setminus \{f\} \cap \SCC(f_{\mathrm{cur}})= \emptyset\). Together with
      \(f \in \Fun(\ctx_3.\St)\), we have \((\Succ^*(\Succ(f_{\mathrm{cur}}) \setminus \Fun(\ctx_3.\St)) \cap \SCC(f_{\mathrm{cur}}))= \emptyset\).
      Lemma~\ref{lem:completion-inv} 
      gives \(\Fun(\ctx_3.\St) = \Fun(\ctx_{r,3}.\St)\),
      i.e., no new function from \(\SCC(f_{\mathrm{cur}})\) appears on the stack
      in \(\ctx_{r,3}\).
      The same lemma guarantees that compiling the remaining statements of
\(f_{\mathrm{cur}}\) does not enlarge the domains of the local variable mappings of
any other suspended function: for every \(h \in \Fun(\St) \setminus \{f_{\mathrm{cur}}\}\),
\[
\dom(\ctx_{r,3}.\rho_{\mathrm{loc},h}) \subseteq \dom(\ctx_{3}.\rho_{\mathrm{loc},h}),
\]
and moreover, for any \(x\) in the common domain with
\(\var(\idx(\rho_{\mathrm{loc},h,3}(x))) \neq \emptyset\),
\[
\ctx_3.\rho_{\mathrm{loc},h}(x) = \ctx_{r,3}.\rho_{\mathrm{loc},h}(x).
\]
Recall that variables with flag \(2\) are never input or output parameters;
by Conditions~4(a) and~4(b) of the validity of \(\ctx_3\), every such variable
satisfies \(\var(\idx(\rho_{\mathrm{loc},h,3}(x))) \neq \emptyset\).  Hence, if they remain
in the local domain after compilation, their associated quantum registers are
unchanged. 
      Therefore, no new function is added and no new flag‑2 variable
      appears in the local mappings of the other functions;
      only the current function \(f_{\mathrm{cur}}\) needs to be examined.
      By flag mutual exclusion, inside \(G_{f_{\mathrm{cur}}}\) the unique variable
      with flag \(2\) is \(x\), which is already recorded in
      \(\dom(\rho_{\mathrm{loc},2})\) and $\var(\rho_{\mathrm{loc},2}(x))=\emptyset$. Similarly, by Lemmas~\ref{lem:fun-inv} and~\ref{lem:completion-inv}, if it is still in $\dom(\ctx_3.\rho_{\mathrm{loc}})$ and $\dom(\ctx_{r,3}.\rho_{\mathrm{loc}})$, we have $\ctx_3.\rho_{\mathrm{loc}}(x)=\ctx_2.\rho_{\mathrm{loc}}(x)$ and $\ctx_{r,3}.\rho_{\mathrm{loc}}(x)=\ctx_3.\rho_{\mathrm{loc}}(x)$. 
      Clearly \(\dom(\rho_3) \subseteq \mathsf{Var}(G_{f_{\mathrm{cur}}})\) and 
      \(\dom(\rho_{r,3}) \subseteq \mathsf{Var}(G_{f_{\mathrm{cur}}})\).
      Consequently, the collection of variables that constitute
      \(Q_{\succeq f_{\mathrm{cur}}^*}\) and \(Q_{\succeq h^*}\) remains unchanged,
      and we obtain \(\ctx_{r,3}.\Gar_h \subseteq \ctx_3.\Gar_h\)
      for every \(h \in \SCC(f) \cap \Fun(\St)\).
\end{itemize}

% In both cases the two inclusions (or a direct inspection of the
% construction) yield
% \[
% \ctx_{r,3}.\Gar_{f_{\mathrm{cur}}} \subseteq  \ctx_3.\Gar_{f_{\mathrm{cur}}}.
% \]

Now applying Theorem~\ref{theo_fun} we obtain
\((\ctx_3, \ket{\phi}, k, \D_{\SCC(f_{\mathrm{cur}})}) \in \mathcal{V}\) and
\(\ctx_3 \approx_{(\ket{\phi}, k)} \sigma_0\).

Note that the application of Theorem~\ref{theo_fun} here is recursive:
Theorem~\ref{theo_fun} relies on Theorem~\ref{theo:body}, which in turn
relies on the current lemma, which again invokes Theorem~\ref{theo_fun}.
To see that this recursion is well‑founded, define the measure
\[
  \mu(\ctx) \triangleq
  \bigl| \dom(\Xi) \setminus \bigl( \Theta^{-1}(\dom(\ctx.\D)) \cup \Fun(\ctx.\St) \bigr) \bigr|,
\]
which counts the number of source functions that have not yet been
compiled (i.e., are neither in the declaration domain nor on the stack).
According to the compilation algorithm, whenever a function \(f\) is
processed, it is pushed onto the stack \emph{before} its body is
compiled; hence \(f\) enters \(\Fun(\St)\) immediately.  Consequently,
the measure strictly decreases whenever a recursive invocation again
reaches the present case (i.e., when \(f \notin \Fun(\St_3)\) and
\(f \notin \Theta^{-1}(\dom(\D))\)).  This provides a well‑founded order
and guarantees termination.

If \(f \notin \Fun(\St_3)\), then
\((\ctx, C_s) = \textsc{Compile\_Com}(\K, \ctx_3, s)\),
and the desired conclusion follows directly from other case.

If \(f \in \Fun(\St_3)\), we proceed as follows.
Set \(\ctx' = \ctx_3\) and \(D_2 = \emptyset\).
Define \(D_1\) according to the subcase:
\begin{itemize}
\item Subcase~1: \(D_1 = \rho(x) \cup \Gar_{f_{\mathrm{cur}}}\);
\item Subcase~2: \(D_1 = \rho(x)\).
\end{itemize}
Note that \(\mvq(s) = x\) and \(\rho = \rho_3\).
Hence, if \(x.\flag \neq 2\) then \(D_1 = \rho(x)\);
otherwise \(D_1 = \rho(x) \cup \Gar_{f_{\mathrm{cur}}}\).
Moreover,
\[
\cod(\rho_{\mathrm{loc}}) \setminus \cod(\rho'_{\mathrm{loc}}) = \emptyset \subseteq D_2,
\]
and if \(x \notin \dom(\rho')\) then \(\rho(x)=\rho_{\mathrm{loc}}(x)\subseteq  \Qpool'\).  
For any \(z\),
\[
\qv(C_s^z) \setminus \cod(\rho')
\subseteq \qv(C_s^z) \setminus \rho'(\mvq(s) \cup \rvq(s))
\subseteq \anc(F_f^z)
\subseteq D_2 \cup \anc(F_{\succeq f_{\mathrm{cur}}^*}^z).
\]
 Moreover, \(\anc(C_s^k) \setminus \mathcal{F}_{\succeq}^k \subseteq \cod(\rho') \subseteq \Qpool'\cup \cod(\rho')\). 
The context update can be rewritten as
\[
\begin{aligned}
\ctx = &\ctx'[C \mapsto C'; C_s,\;
   \Vactive \mapsto \Vactive' \cup (D_1 \setminus (\rho'_{\mathrm{loc}}(\fp(f_{\mathrm{cur}})) \cup \cod(\rho_{\mathrm{glob}}))),\;
           \Valloc \mapsto \Valloc' \cup D_2,\;\\ 
 &\rho'_{\mathrm{loc}} \mapsto \rho'_{\mathrm{loc}}[x \mapsto \rho_{\mathrm{loc}}(x) \mid x \notin \dom(\rho')],\;
           \Qpool \mapsto \Qpool' \setminus \rho_{\mathrm{loc}}(x)].
\end{aligned}
\]

The rest of the proof is essentially the same as in the previous case,
except for the containment
\[
\qv(C_s^k) \setminus (\rho'(\rvq(s)) \cup \rho'(\mvq(s)))
\subseteq \ctx'.A_{f_{\mathrm{cur}}} \setminus (\ctx'.V_{\activelabel,f_{\mathrm{cur}}} \cup \cod(\rho'_{\mathrm{loc},f_{\mathrm{cur}}})), 
\]
whose verification is particularly simple in this case.
Indeed, the validity of \(\ctx'\) gives
\[
\anc(F_f^k) \subseteq \anc(F_f^{k+1})=\anc(C_f^k) \subseteq \ctx'.A_f.
\] By the definition of the function compilation algorithm,
\(\pos(f) > \pos(f_{\mathrm{cur}})\) inside \(\Fun(\St')\). 
Consequently, 
\[
(\ctx'.A_f \setminus \rho'_{\mathrm{loc},f}(\fp(f) \cup \re(f))) \subseteq \ctx'.A_{f_{\mathrm{cur}}}. 
\]   
By Lemma~\ref{lem:A_f-disjointness}, 
\[
(\ctx'.A_f \setminus \rho'_{\mathrm{loc},f}(\fp(f) \cup \re(f)))
\cap (V'_{\activelabel,[f_{\mathrm{cur}}^*, f)} \cup \cod(\rho'_{\mathrm{loc},[f_{\mathrm{cur}}^*, f)})) = \emptyset.\]
Moreover, it is easy to prove that \(V'_{\activelabel,f_{\mathrm{cur}}} \cup \cod(\rho'_{\mathrm{loc},f_{\mathrm{cur}}})\) is contained in
\((V'_{\activelabel,[f_{\mathrm{cur}}^*, f)} \cup \cod(\rho'_{\mathrm{loc},[f_{\mathrm{cur}}^*, f)}))\). 
Hence 
\[
(\ctx'.A_f \setminus \rho'_{\mathrm{loc},f}(\fp(f) \cup \re(f)))
\cap (V'_{\activelabel,f_{\mathrm{cur}}} \cup \cod(\rho'_{\mathrm{loc},f_{\mathrm{cur}}})) = \emptyset,
\]
and so
\[
\anc(F_f^k) \subseteq \ctx'.A_{f_{\mathrm{cur}}} \setminus (V'_{\activelabel,f_{\mathrm{cur}}} \cup \cod(\rho'_{\mathrm{loc},f_{\mathrm{cur}}})).  
\]

Now we can derive
\[
\begin{aligned}
&\qv(C_s^k) \setminus (\rho'(\rvq(s)) \cup \rho'(\mvq(s))) \\
= \ & \qv\bigl(F_f^k[\mathrm{in} \mapsto \rho'(\rvq(s)),\;
                 \mathrm{out} \mapsto \rho'(\mvq(s))]\bigr)
   \setminus (\rho'(\rvq(s)) \cup \rho'(\mvq(s))) \\
\subseteq \ & \anc(F_f^k) \subseteq \ctx'.A_{f_{\mathrm{cur}}} \setminus (\ctx'.V_{\activelabel,f_{\mathrm{cur}}} \cup \cod(\rho'_{\mathrm{loc},f_{\mathrm{cur}}})). 
\end{aligned}
\]
% The first equality holds because we already know
% \(\anc(P^k) \cap \cod(\rho_3) = \emptyset\), which implies
% \(\anc(P^k) \cap (\rho'(\rvq(s)) \cup \rho'(\mvq(s))) = \emptyset\).
% (Here \(P_f\) is defined exactly as in the previous case.) 

Finally, let \(g\) be the bijection from \(\anc(F_f^k)\) to \(\anc(C_s^k)\)
defined by \(g(q) = q\) for all \(q \in \anc(F_f^k)\).
The subsequent verification of the remaining conditions is completely
analogous to the previous case.

% (Besides, note that if $x \notin \dom(\rho_0)$, we have $\rho_2(x) \in \xi_0$, thus $\rho_2(x)  \cap  (ucv_{all,0} \cup \rho_g) =\emptyset$.  Since $(ucv_{all,3} \cup \rho_g) =(ucv_{\prec f, 3} \cup \rho_g)\subseteq (ucv_{\prec f, 2} \cup \rho_g) \subseteq (ucv_{\prec f, 0} \cup \rho_g) \subseteq (ucv_{all,0} \cup \rho_g)$, and $\rho_2(x)=\rho_3(x)$(since $\rho_2(x) \subseteq \rho_3(x)$ and $\base(\rho_2(x)) \neq \rho_2(x)$), we have $\rho_3(x) \cap (ucv_{all,3} \cup \rho_g) =\emptyset$.  Therefore, we have \(E \triangleq \qv(C^{k_{\curf}})  \setminus \rho'(\mvq(s) \cup \rvq(s)) \cup \rho(x \setminus \dom(\rho_0)) \subseteq ctx'.A_{f} \setminus (ctx'.ucv_{\all} \cup \cod(\rho_g)) \) and  $(\qv(C^{k_{\curf}})  \setminus \rho'(\mvq(s) \cup \rvq(s))) \cap \cod(\rho') =\emptyset$;
% Thus  \(ctx_3.C_{\curf}^k \ket{\phi} \models \ket{0}_{E}\).)  

\end{proof}

For simplicity, we shall use \(x\) to denote \(\mathrm{mvq}(s)\) in the following. 

\begin{theorem}
\label{state_valid_2}
Under Notations~\ref{note:state_1},~\ref{note:state_2} and Assumption~\ref{hpy:com}, Condition~2 holds.  
\end{theorem}
\begin{proof}

By Lemma~\ref{theo_com} we have \(\ctx.A_{f_{\mathrm{cur}}} = \ctx'.A_{f_{\mathrm{cur}}}\); moreover, \(\fp(f_{\mathrm{cur}}) \subseteq \dom(\rho'_{\mathrm{loc},f_{\mathrm{cur}}})\) and the construction of \(\rho_{\mathrm{loc},f_{\mathrm{cur}}}\) gives \(\rho_{\mathrm{loc},f_{\mathrm{cur}}}(\fp(f_{\mathrm{cur}})) = \rho'_{\mathrm{loc},f_{\mathrm{cur}}}(\fp(f_{\mathrm{cur}}))\). The validity of \(\ctx'\) supplies \(\ket{\phi}\models \ket{0}_{\ctx'.A_{f_{\mathrm{cur}}} \setminus \rho'_{\mathrm{loc},f_{\mathrm{cur}}}(\fp(f_{\mathrm{cur}}))}\), hence \(\ket{\phi}\models \ket{0}_{\ctx.A_{f_{\mathrm{cur}}} \setminus \rho_{\mathrm{loc},f_{\mathrm{cur}}}(\fp(f_{\mathrm{cur}}))}\). 

Theorem~\ref{theo_com} gives
\(cv = cv' \setminus \bigl(D_1 \setminus (\rho'_{\mathrm{loc},f_{\mathrm{cur}}}(\fp(f_{\mathrm{cur}})) \cup \rho'_{\mathrm{glob}})\bigr)\). For the other components,
\[
\begin{aligned}
V_{\activelabel,\succeq f_{\mathrm{cur}}^*} \cup \cod(\rho_{\mathrm{loc},\succeq f_{\mathrm{cur}}^*})
&= V'_{\activelabel,\succeq f_{\mathrm{cur}}^*} \cup \bigl(D_1 \setminus (\rho'_{\mathrm{loc},f_{\mathrm{cur}}}(\fp(f_{\mathrm{cur}})) \cup \rho'_{\mathrm{glob}})\bigr)
   \cup \cod(\rho'_{\mathrm{loc},\succeq f_{\mathrm{cur}}^*}) \cup \rho_{\mathrm{loc},f_{\mathrm{cur}}}(x).
\end{aligned}
\]

Since \(\rho_{f_{\mathrm{cur}}}(x) \subseteq D_1\), we have \(\rho_{\mathrm{loc},f_{\mathrm{cur}}}(x) \subseteq D_1\).
If \(x \in \dom(\rho'_{f_{\mathrm{cur}}})\), then
\(\rho_{\mathrm{loc},f_{\mathrm{cur}}}(x) = \rho'_{\mathrm{loc},f_{\mathrm{cur}}}(x) \subseteq \cod(\rho'_{\mathrm{loc},f_{\mathrm{cur}}})
 \subseteq \cod(\rho'_{\mathrm{loc},\succeq f_{\mathrm{cur}}^*})\);
otherwise, \(\rho_{\mathrm{loc},f_{\mathrm{cur}}}(x) \in \Qpool'\).
The validity of \(\ctx'\) gives \(\Qpool' \cap \cod(\rho'_{f_{\mathrm{cur}}}) = \emptyset\), hence
\(\rho_{\mathrm{loc},f_{\mathrm{cur}}}(x) \cap (\rho'_{\mathrm{loc},f_{\mathrm{cur}}}(\fp(f_{\mathrm{cur}})) \cup \rho'_{\mathrm{glob}}) = \emptyset\),
and therefore
\(\rho_{\mathrm{loc},f_{\mathrm{cur}}}(x) \subseteq D_1 \setminus (\rho'_{\mathrm{loc},f_{\mathrm{cur}}}(\fp(f_{\mathrm{cur}})) \cup \rho'_{\mathrm{glob}})\). Consequently,
\[
V_{\activelabel,\succeq f_{\mathrm{cur}}^*} \cup \cod(\rho_{\succeq f_{\mathrm{cur}}^*})
\subseteq V'_{\activelabel,\succeq f_{\mathrm{cur}}^*} \cup \cod(\rho'_{\succeq f_{\mathrm{cur}}^*})
    \cup \bigl(D_1 \setminus (\rho'_{\mathrm{loc},f_{\mathrm{cur}}}(\fp(f_{\mathrm{cur}})) \cup \rho'_{\mathrm{glob}})\bigr).
\]

The validity of \(\ctx'\) yields 
\(cv' \cap \bigl(V'_{\activelabel,\succeq f_{\mathrm{cur}}^*} \cup \cod(\rho'_{\mathrm{loc},\succeq f_{\mathrm{cur}}^*}) \cup \rho'_{\mathrm{glob}}\bigr) = \emptyset\).
Combining this with the update for \(cv\) and the fact that \(\rho_{\mathrm{glob}}\) is unchanged,
we obtain
\[
cv \cap \bigl(V_{\activelabel,\succeq f_{\mathrm{cur}}^*} \cup \cod(\rho_{\mathrm{loc},\succeq f_{\mathrm{cur}}^*}) \cup \rho_{\mathrm{glob}}\bigr) = \emptyset .
\] 

The next condition we first examine how \(\rho_{\mathrm{loc},\succeq f_{\mathrm{cur}}^*} \cup \rho_{\mathrm{glob}}\) and \(Q_{\succeq f_{\mathrm{cur}}^*}[+]\) change.
\begin{itemize}
    \item If \(x \in \dom(\rho'_{f_{\mathrm{cur}}})\), then by definition the sets \(Q_{\succeq f_{\mathrm{cur}}^*}[+]\), \(\rho_{\succeq f_{\mathrm{cur}}^*}\)
          and \(\rho_{\mathrm{glob}}\) all remain unchanged; hence the condition follows directly from the validity of \(\ctx'\).
    \item Otherwise, by Theorem~\ref{theo_com} we have \(\rho'_{\mathrm{loc},f_{\mathrm{cur}}}(x) \subseteq \Qpool'\),
          \(\rho_{\mathrm{loc},\succeq f_{\mathrm{cur}}^*} = \rho'_{\mathrm{loc},\succeq f_{\mathrm{cur}}^*} \cup \rho_{\mathrm{loc},f_{\mathrm{cur}}}(x)\),
          while \(\rho_{\mathrm{glob}}\) is still unchanged.
          Hence
          \(\rho_{\mathrm{loc},\succeq f_{\mathrm{cur}}^*} \cup \rho_{\mathrm{glob}}
           = \rho'_{\mathrm{loc},\succeq f_{\mathrm{cur}}^*} \cup \rho'_{\mathrm{glob}} \cup \rho_{\mathrm{loc},f_{\mathrm{cur}}}(x)
           \subseteq \rho'_{\mathrm{loc},\succeq f_{\mathrm{cur}}^*} \cup \rho'_{\mathrm{glob}} \cup \Qpool'\).
          The situation for \(Q_{\succeq f_{\mathrm{cur}}^*}[+]\) is similar:
          \(Q_{\succeq f_{\mathrm{cur}}^*}[+] \subseteq Q'_{\succeq f_{\mathrm{cur}}^*}[+] \cup \rho_{\mathrm{loc},f_{\mathrm{cur}}}(x)[+]
           \subseteq Q'_{\succeq f_{\mathrm{cur}}^*}[+] \cup \Qpool'[+]\).
          The remainder of the proof follows exactly the same reasoning as the corresponding part in the verification of Condition~2 for \(\ctx_1\) in Lemma~\ref{lem:state-valid-2}.
\end{itemize}

By Theorem~\ref{theo_com} we have \(\Qpool = \Qpool' \setminus \rho'_{\mathrm{loc},f_{\mathrm{cur}}}(x) \).
Clearly \(\Qpool \subseteq \Qpool'\) and therefore
\(\Qpool[+] = \Qpool'[+] \setminus \rho'_{\mathrm{loc},f_{\mathrm{cur}}}(x) \).
Note that \(\cod(\rho_{\mathrm{loc},\succeq f_{\mathrm{cur}}^*}) = \cod(\rho'_{\mathrm{loc},\succeq f_{\mathrm{cur}}^*}) \cup \rho'_{\mathrm{loc},f_{\mathrm{cur}}}(x) \),
hence \(\cod(\rho_{\mathrm{loc},\succeq f_{\mathrm{cur}}^*})[+] = \cod(\rho'_{\mathrm{loc},\succeq f_{\mathrm{cur}}^*})[+] \cup \rho'_{\mathrm{loc},f_{\mathrm{cur}}}(x)[+]\).
Since \(\rho'_{\mathrm{loc},f_{\mathrm{cur}}}(x) \in \Qpool' \cup \cod(\rho'_{\mathrm{loc},\succeq f_{\mathrm{cur}}^*})\), we obtain
\[
\Qpool[+] \cup \cod(\rho_{\mathrm{loc},\succeq f_{\mathrm{cur}}^*})[+] = \Qpool'[+] \cup \cod(\rho'_{\mathrm{loc},\succeq f_{\mathrm{cur}}^*})[+].
\]

Observe that
\(D_1 \setminus (\rho'_{\mathrm{loc},f_{\mathrm{cur}}}(\fp(f_{\mathrm{cur}})) \cup \cod(\rho'_{\mathrm{glob}})) \subseteq \rho'_{\mathrm{loc},f_{\mathrm{cur}}}(x) \cup Gar_{f_{\mathrm{cur}}}\).
Consequently,
\(D_1[+] \subseteq \rho'_{\mathrm{loc},f_{\mathrm{cur}}}(x)[+] \cup Q_{\succeq f_{\mathrm{cur}}^*}[+]\).
Using \(V_{\activelabel,\succeq f_{\mathrm{cur}}^*} = V'_{\activelabel,\succeq f_{\mathrm{cur}}^*} \cup \bigl(D_1 \setminus (\rho'_{\mathrm{loc},f_{\mathrm{cur}}}(\fp(f_{\mathrm{cur}})) \cup \cod(\rho'_{\mathrm{glob}}))\bigr)\),
we obtain
\[
V_{\activelabel,\succeq f_{\mathrm{cur}}^*}[+] \subseteq V'_{\activelabel,\succeq f_{\mathrm{cur}}^*}[+] \cup \rho'_{\mathrm{loc},f_{\mathrm{cur}}}(x)[+] \cup Q_{\succeq f_{\mathrm{cur}}^*}[+]
\subseteq V'_{\activelabel,\succeq f_{\mathrm{cur}}^*}[+] \cup \cod(\rho'_{\mathrm{loc},\succeq f_{\mathrm{cur}}^*})[+].
\]

Thus
\[
\Qpool[+] \cup V_{\activelabel,\succeq f_{\mathrm{cur}}^*}[+] \cup \cod(\rho_{\mathrm{loc},\succeq f_{\mathrm{cur}}^*})[+]
\subseteq \Qpool'[+] \cup V'_{\activelabel,\succeq f_{\mathrm{cur}}^*}[+] \cup \cod(\rho'_{\mathrm{loc},\succeq f_{\mathrm{cur}}^*})[+].
\]

The validity of \(\ctx'\) gives
\(\Qpool' \cap \bigl(\Qpool'[+] \cup V'_{\activelabel,\succeq f_{\mathrm{cur}}^*}[+] \cup \cod(\rho'_{\mathrm{loc},\succeq f_{\mathrm{cur}}^*})[+]\bigr) = \emptyset\);
together with \(\Qpool \subseteq \Qpool'\) we conclude
\[
\Qpool \cap \bigl(\Qpool[+] \cup V_{\activelabel,\succeq f_{\mathrm{cur}}^*}[+] \cup \cod(\rho_{\mathrm{loc},\succeq f_{\mathrm{cur}}^*})[+]\bigr) = \emptyset.
\] 
Since \(\Qpool \subseteq \Qpool'\), the condition that
\(\forall\, q \in \Qpool,\; \var(\idx(q)) = \emptyset \Longrightarrow \base[q] = q\)
follows directly from the validity of \(\ctx'\).

Combining the three items above, Condition~2 is established. 

\end{proof}

\begin{theorem}
\label{state_valid_3.b}
Under Notations~\ref{note:state_1},~\ref{note:state_2} and Assumption~\ref{hpy:com}, Condition~3(b) holds.  
\end{theorem}
\begin{proof}
Take any \(h \in \{f_{\mathrm{cur}}\} \cup \SCC(f_{\mathrm{cur}}) \cap \Fun(\St) \setminus \Theta^{-1}(\dom(\D))\).
Since \(h \neq f_{\mathrm{cur}}\), by construction the compilation variables that appear in this condition (the code fragment, \(\Vactive\), etc.) are not modified; hence the condition remains valid. We consider \(f_{\mathrm{cur}}\) below.  

  Recall that \(\ctx.C_{f_{\mathrm{cur}}} = \ctx'.C_{f_{\mathrm{cur}}}; C_s\).
By construction \(V_{\activelabel,f_{\mathrm{cur}}} = V'_{\activelabel,f_{\mathrm{cur}}} \cup (D_1 \setminus (\rho'_{\mathrm{loc},f_{\mathrm{cur}}}(\fp(f_{\mathrm{cur}})) \cup \rho'_{\mathrm{glob}}))\).
From Lemma~\ref{theo_com}, for any \(\ket{\psi}\) with \(\ket{\psi} \models \ket{0}_{E}\),
\[
C_s^k \ket{\psi} \models \ket{\psi}_{\sys \setminus (D_1)}. \tag{1}
\]

Now take any \(\ket{\varphi}\) with \(\ket{\varphi} \models \ket{0}_{\ctx.A_{f_{\mathrm{cur}}} \setminus \rho_{\mathrm{loc},f_{\mathrm{cur}}}(\fp(f_{\mathrm{cur}}))}\).
Since \(\qv(C_s^z) \setminus (\rho'_{f_{\mathrm{cur}}}(\mvq(s) \cup \rvq(s))) \subseteq \ctx'.A_{f_{\mathrm{cur}}} \setminus (V'_{\activelabel,f_{\mathrm{cur}}} \cup \cod(\rho'_{\mathrm{loc},f_{\mathrm{cur}}}))\),
Lemma~\ref{lem:curf} and the validity of \(\ctx'\) together with
\(\ctx'.C_{f_{\mathrm{cur}}}^k \ket{\varphi} \models \ket{0}_{\rho_{\mathrm{loc},f_{\mathrm{cur}}}(\mvq(s) \setminus \dom(\rho_{f_{\mathrm{cur}},0}))}\) yield 
\[
\ctx'.C_{f_{\mathrm{cur}}}^k \ket{\varphi} \models \ket{0}_{E}. \tag{2}
\]

Applying (1) with \(\ket{\psi} = \ctx'.C_{f_{\mathrm{cur}}}^k \ket{\varphi}\) and using (2) gives
\[
\ctx.C_{f_{\mathrm{cur}}}^k \ket{\varphi}
= C_s^k \bigl( \ctx'.C_{f_{\mathrm{cur}}}^k \ket{\varphi} \bigr)
\models \bigl( \ctx'.C_{f_{\mathrm{cur}}}^k \ket{\varphi} \bigr)_{\sys \setminus (D_1)},
\]
and therefore
\[
\ctx.C_{f_{\mathrm{cur}}}^k \ket{\varphi}
\models \bigl( \ctx'.C_{f_{\mathrm{cur}}}^k \ket{\varphi} \bigr)_{(\ctx'.A_{f_{\mathrm{cur}}} \setminus (V'_{\activelabel,f_{\mathrm{cur}}} \cup \rho'_{\mathrm{loc},f_{\mathrm{cur}}}(\fp(f_{\mathrm{cur}})))) \setminus D_1}. \tag{3}
\]

Moreover, Condition~3(b) for \(\ctx'\) gives
\[
\ctx'.C_{f_{\mathrm{cur}}}^k \ket{\varphi}
\models \ket{\varphi}_{\ctx'.A_{f_{\mathrm{cur}}} \setminus (V'_{\activelabel,f_{\mathrm{cur}}} \cup \rho'_{\mathrm{loc},f_{\mathrm{cur}}}(\fp(f_{\mathrm{cur}})))}. \tag{4}
\]

Lemma~\ref{theo_com} gives \(\ctx.A_{f_{\mathrm{cur}}} = \ctx'.A_{f_{\mathrm{cur}}}\).
Expanding \(V_{\activelabel,f_{\mathrm{cur}}}\) we obtain
\begin{align*}
&\ctx.A_{f_{\mathrm{cur}}} \setminus (V_{\activelabel,f_{\mathrm{cur}}} \cup \rho_{\mathrm{loc},f_{\mathrm{cur}}}(\fp(f_{\mathrm{cur}})))\\ 
= \ & \ctx.A_{f_{\mathrm{cur}}} \setminus \bigl( V'_{\activelabel,f_{\mathrm{cur}}} \cup (D_1 \setminus (\rho'_{\mathrm{loc},f_{\mathrm{cur}}}(\fp(f_{\mathrm{cur}})) \cup \rho'_{\mathrm{glob}})) \cup \rho_{\mathrm{loc},f_{\mathrm{cur}}}(\fp(f_{\mathrm{cur}})) \bigr) \\
= \ & \ctx.A_{f_{\mathrm{cur}}} \setminus \bigl( V'_{\activelabel,f_{\mathrm{cur}}} \cup (D_1 \setminus \rho'_{\mathrm{loc},f_{\mathrm{cur}}}(\fp(f_{\mathrm{cur}}))) \cup \rho'_{\mathrm{loc},f_{\mathrm{cur}}}(\fp(f_{\mathrm{cur}})) \bigr) \\
&\qquad \tag{\(\rho_{\mathrm{loc},f_{\mathrm{cur}}}(\fp(f_{\mathrm{cur}})) = \rho'_{\mathrm{loc},f_{\mathrm{cur}}}(\fp(f_{\mathrm{cur}}))\) and \(\ctx'.A_{f_{\mathrm{cur}}} \cap \rho'_{\mathrm{glob}}=\emptyset\)}\\
\subseteq \ & \bigl( \ctx'.A_{f_{\mathrm{cur}}} \setminus (V'_{\activelabel,f_{\mathrm{cur}}} \cup \rho'_{\mathrm{loc},f_{\mathrm{cur}}}(\fp(f_{\mathrm{cur}}))) \bigr) \setminus D_1 \\
\subseteq \ & \ctx'.A_{f_{\mathrm{cur}}} \setminus (V'_{\activelabel,f_{\mathrm{cur}}} \cup \rho'_{\mathrm{loc},f_{\mathrm{cur}}}(\fp(f_{\mathrm{cur}}))).
\end{align*}

Combining (3) and (4) with this inclusion yields
\[
\ctx.C_{f_{\mathrm{cur}}}^k \ket{\varphi}
\models \ket{\varphi}_{\ctx.A_{f_{\mathrm{cur}}} \setminus (V_{\activelabel,f_{\mathrm{cur}}} \cup \rho_{\mathrm{loc},f_{\mathrm{cur}}}(\fp(f_{\mathrm{cur}})))},
\]
which is precisely Condition~3(b) for \(\ctx\).
\end{proof}

\begin{theorem} 
\label{state_valid_3.c} 
Under Notations~\ref{note:state_1},~\ref{note:state_2} and Assumption~\ref{hpy:com}, Condition~3(c) holds. 
\end{theorem}
\begin{proof}
  For any \(h \in \SCC(f_{\mathrm{cur}}) \cap \Fun(\St) \setminus \Theta^{-1}(\dom(\D))\) the following conditions hold.
For 3(c)(1), 3(c)(4) and 3(c)(5), if \(h \neq f_{\mathrm{cur}}\) the properties follow directly from the validity of \(\ctx'\);
hence we only need to verify the case \(h = f_{\mathrm{cur}}\).

\begin{enumerate}
  \item[3(c)(1)] The validity of \(\ctx'\) implies
  \[
  \fp(f_{\mathrm{cur}}) \cup \re(f_{\mathrm{cur}})  \subseteq \dom(\rho'_{\mathrm{loc},f_{\mathrm{cur}}}).
  \]
  By definition
  \[
  \rho_{\mathrm{loc},f_{\mathrm{cur}}} = \rho'_{\mathrm{loc},f_{\mathrm{cur}}}[x \mapsto \rho_{\mathrm{loc},f_{\mathrm{cur}}}(x) \mid x \notin \dom(\rho'_{f_{\mathrm{cur}}})],
  \]
  hence \(\dom(\rho'_{\mathrm{loc},f_{\mathrm{cur}}}) \subseteq \dom(\rho_{\mathrm{loc},f_{\mathrm{cur}}})\), and therefore
  \[
  \fp(f_{\mathrm{cur}})\cup \re(f_{\mathrm{cur}})  \subseteq \dom(\rho_{\mathrm{loc},f_{\mathrm{cur}}}) .
  \]

  \item[3(c)(2)] The validity of \(\ctx'\) gives
  \[
  \ctx'.A_h \cap \rho'_{\mathrm{glob}} = \emptyset .
  \]
  We already proved \(\ctx.A_h = \ctx'.A_h\) by Theorem~\ref{theo_com}, and \(\rho_{\mathrm{glob}}\) is unchanged, thus the condition follows directly from the validity of \(\ctx'\).

  \item[3(c)(3)] We need to show
  \[
  \bigl( \ctx.A_h \setminus \rho_{\mathrm{loc},h}(\fp(h)) \setminus \rho_{\mathrm{loc},h}(\re(h)) \bigr)
  \;\cap\;
  \bigl( V_{\activelabel,[h^*,\,h)} \cup \cod(\rho_{\mathrm{loc},[h^*,\,h)}) \bigr)
  = \emptyset .
  \]

  From the validity of \(\ctx'\),
  \[
  \bigl( \ctx'.A_h \setminus \rho'_{\mathrm{loc},h}(\fp(h)) \setminus \rho'_{\mathrm{loc},h}(\re(h)) \bigr)
  \;\cap\;
  \bigl( V'_{\activelabel,[h^*,\,h)} \cup \cod(\rho'_{\mathrm{loc},[h^*,\,h)}) \bigr)
  = \emptyset .
  \]
  We already know \(\ctx.A_h = \ctx'.A_h\). Moreover, from 
\[
\rho_{\mathrm{loc},f_{\mathrm{cur}}} = \rho'_{\mathrm{loc},f_{\mathrm{cur}}}[x \mapsto \rho_{\mathrm{loc},f_{\mathrm{cur}}}(x) \mid x \notin \dom(\rho'_{f_{\mathrm{cur}}})]
\]
and the fact that \(\rho_{\mathrm{loc},f_{\mathrm{cur}}}(x)=\rho'_{\mathrm{loc},f_{\mathrm{cur}}}(x)\) for \(x \in \dom(\rho'_{\mathrm{loc},f_{\mathrm{cur}}})\), together with the validity of \(\ctx'\)---which guarantees \(\fp(f_{\mathrm{cur}}) \cup \re(f_{\mathrm{cur}}) \subseteq \dom(\rho'_{\mathrm{loc},f_{\mathrm{cur}}})\)---and since the local variable mappings remain unchanged for \(h \neq f_{\mathrm{cur}}\), we obtain
\[
\rho_{\mathrm{loc},h}(\fp(h)) = \rho'_{\mathrm{loc},h}(\fp(h)), \qquad
\rho_{\mathrm{loc},h}(\re(h)) = \rho'_{\mathrm{loc},h}(\re(h)).
\] 
  Consequently,
  \[
  \ctx.A_h \setminus \rho_{\mathrm{loc},h}(\fp(h)) \setminus \rho_{\mathrm{loc},h}(\re(h))
  =
  \ctx'.A_h \setminus \rho'_{\mathrm{loc},h}(\fp(h)) \setminus \rho'_{\mathrm{loc},h}(\re(h)) .
  \]
  Finally, by construction we have 
  \[
  V_{\activelabel,[h^*,\,h)} \cup \cod(\rho_{\mathrm{loc},[h^*,\,h)})
  =
  V'_{\activelabel,[h^*,\,h)} \cup \cod(\rho'_{\mathrm{loc},[h^*,\,h)}),
  \] 
  and the required conclusion follows.

  \item[3(c)(4)] From the previous analysis we have \(\rho_{\mathrm{loc},f_{\mathrm{cur}}}(\fp(f_{\mathrm{cur}})) = \rho'_{\mathrm{loc},f_{\mathrm{cur}}}(\fp(f_{\mathrm{cur}}))\) and the validity of \(\ctx'\) yields \(\rho'_{\mathrm{loc},f_{\mathrm{cur}}}(\fp(f_{\mathrm{cur}})) \cap V'_{\activelabel,f_{\mathrm{cur}}} = \emptyset\).  
Since  
\[
V_{\activelabel,f_{\mathrm{cur}}} = V'_{\activelabel,f_{\mathrm{cur}}} \cup \bigl( D_1 \setminus ( \rho'_{\mathrm{loc},f_{\mathrm{cur}}}(\fp(f_{\mathrm{cur}})) \cup \rho_{\mathrm{glob}} ) \bigr)
\]
and the added part is clearly disjoint from \(\rho'_{\mathrm{loc},f_{\mathrm{cur}}}(\fp(f_{\mathrm{cur}}))\), we obtain \(\rho_{\mathrm{loc},f_{\mathrm{cur}}}(\fp(f_{\mathrm{cur}})) \cap V_{\activelabel,f_{\mathrm{cur}}} = \emptyset\). 
\item[3(c)(5)]  We need to prove
\[
\rho_{\mathrm{loc},f_{\mathrm{cur}}}(\re(f_{\mathrm{cur}})) \subseteq V_{\activelabel,\mathrm{pred}(f_{\mathrm{cur}})}.
\] 
Both sets are invariant from \(\ctx'\) to \(\ctx\), so the inclusion is immediate from the validity of \(\ctx'\). 

  \item[3(c)(6)] We need to prove
  \[
  \base\bigl( \cod(\rho_{\mathrm{loc},f_{\mathrm{cur}}}) \setminus \rho_{\mathrm{loc},f_{\mathrm{cur}}}(\fp(f_{\mathrm{cur}})) \setminus \rho_{\mathrm{loc},f_{\mathrm{cur}}}(\re(f_{\mathrm{cur}})) \bigr)
  \subseteq V_{\alloc,f_{\mathrm{cur}}}.
  \]
  Observe that \(\cod(\rho_{\mathrm{loc},f_{\mathrm{cur}}}) \subseteq \cod(\rho'_{\mathrm{loc},f_{\mathrm{cur}}}) \cup \bigl( \cod(\rho_{\mathrm{loc},f_{\mathrm{cur}}}) \setminus \cod(\rho'_{\mathrm{loc},f_{\mathrm{cur}}}) \bigr)\). Using the validity of \(\ctx'\) we obtain
  \[
  \begin{aligned}
  &\base\bigl( \cod(\rho_{\mathrm{loc},f_{\mathrm{cur}}}) \setminus \rho_{\mathrm{loc},f_{\mathrm{cur}}}(\fp(f_{\mathrm{cur}})) \setminus \rho_{\mathrm{loc},f_{\mathrm{cur}}}(\re(f_{\mathrm{cur}})) \bigr) \\
  \subseteq \ & \base\!\Bigl( \bigl( \cod(\rho'_{\mathrm{loc},f_{\mathrm{cur}}}) \cup ( \cod(\rho_{\mathrm{loc},f_{\mathrm{cur}}}) \setminus \cod(\rho'_{\mathrm{loc},f_{\mathrm{cur}}}) ) \bigr)
                           \setminus \rho'_{\mathrm{loc},f_{\mathrm{cur}}}(\fp(f_{\mathrm{cur}})) \setminus \rho'_{\mathrm{loc},f_{\mathrm{cur}}}(\re(f_{\mathrm{cur}})) \Bigr) \\
  \subseteq \ & \base\bigl( \cod(\rho'_{\mathrm{loc},f_{\mathrm{cur}}}) \setminus \rho'_{\mathrm{loc},f_{\mathrm{cur}}}(\fp(f_{\mathrm{cur}})) \setminus \rho'_{\mathrm{loc},f_{\mathrm{cur}}}(\re(f_{\mathrm{cur}})) \bigr)
            \;\cup\; \base\bigl( \cod(\rho_{\mathrm{loc},f_{\mathrm{cur}}}) \setminus \cod(\rho'_{\mathrm{loc},f_{\mathrm{cur}}}) \bigr) \\
  \subseteq \ & V'_{\alloc,f_{\mathrm{cur}}} \cup D_2 \\
  = \ & V_{\alloc,f_{\mathrm{cur}}}.
  \end{aligned}
  \]
\end{enumerate}
\end{proof}

\begin{theorem}
\label{state_valid_4.a}
Under Notations~\ref{note:state_1},~\ref{note:state_2} and Assumption~\ref{hpy:com}, Condition~4(a) holds. 
\end{theorem}
\begin{proof}
  For any \(h \in \SCC(f_{\mathrm{cur}}) \cap \Fun(\St)\), the following conditions hold.
When \(h \neq f_{\mathrm{cur}}\) none of the compilation variables in these conditions is modified, so the properties follow directly from the validity of \(\ctx'\); we therefore only check the case \(h = f_{\mathrm{cur}}\).

\begin{enumerate}
  \item[4(a)(1)] We prove
  \[
  \cod(\rho_{\mathrm{loc},f_{\mathrm{cur}}}) \setminus \rho_{\mathrm{loc},f_{\mathrm{cur}}}(\fp(f_{\mathrm{cur}})) \setminus \rho_{\mathrm{loc},f_{\mathrm{cur}}}(\re(f_{\mathrm{cur}})) \subseteq V_{\activelabel,f_{\mathrm{cur}}}.
  \]
By definition, we have \(\cod(\rho_{\mathrm{loc},f_{\mathrm{cur}}}) \subseteq \cod(\rho'_{\mathrm{loc},f_{\mathrm{cur}}}) \cup \bigl( \rho_{\mathrm{loc},f_{\mathrm{cur}}}(x) \setminus ( \rho'_{\mathrm{loc},f_{\mathrm{cur}}}(\fp(f_{\mathrm{cur}})) \cup \rho_{\mathrm{glob}} ) \bigr)\).
  Together with the validity of \(\ctx'\),
  \begin{align*}
  &\cod(\rho_{\mathrm{loc},f_{\mathrm{cur}}}) \setminus \rho_{\mathrm{loc},f_{\mathrm{cur}}}(\fp(f_{\mathrm{cur}})) \setminus \rho_{\mathrm{loc},f_{\mathrm{cur}}}(\re(f_{\mathrm{cur}})) \\
  \subseteq \ & \Bigl( \cod(\rho'_{\mathrm{loc},f_{\mathrm{cur}}}) \cup \bigl( \rho_{\mathrm{loc},f_{\mathrm{cur}}}(x) \setminus ( \rho'_{\mathrm{loc},f_{\mathrm{cur}}}(\fp(f_{\mathrm{cur}})) \cup \rho_{\mathrm{glob}} ) \bigr) \Bigr)
            \setminus \rho'_{\mathrm{loc},f_{\mathrm{cur}}}(\fp(f_{\mathrm{cur}})) \setminus \rho'_{\mathrm{loc},f_{\mathrm{cur}}}(\re(f_{\mathrm{cur}})) \\
  \subseteq \ & \bigl( \cod(\rho'_{\mathrm{loc},f_{\mathrm{cur}}}) \setminus \rho'_{\mathrm{loc},f_{\mathrm{cur}}}(\fp(f_{\mathrm{cur}})) \setminus \rho'_{\mathrm{loc},f_{\mathrm{cur}}}(\re(f_{\mathrm{cur}})) \bigr)
            \cup \bigl( D_1 \setminus ( \rho'_{\mathrm{loc},f_{\mathrm{cur}}}(\fp(f_{\mathrm{cur}})) \cup \rho_{\mathrm{glob}} ) \bigr) \tag{since \(\rho_{\mathrm{loc},f_{\mathrm{cur}}}(x) \subseteq D_1\)} \\
  \subseteq  \ & V'_{\activelabel,f_{\mathrm{cur}}} \cup \bigl( D_1 \setminus ( \rho'_{\mathrm{loc},f_{\mathrm{cur}}}(\fp(f_{\mathrm{cur}})) \cup \rho_{\mathrm{glob}} ) \bigr) \tag{by the validity of \(\ctx'\)} \\
  = \ & V_{\activelabel,f_{\mathrm{cur}}}.
  \end{align*}
 \item[4(a)(2)] For \(h \in \Theta^{-1}(\dom(\D)) \setminus \{f_{\mathrm{cur}}\}\), \(\cod(\rho_{\mathrm{loc},h}) \cap \cod(\rho_{\mathrm{loc},[f_{\mathrm{cur}}^*,h]}) = \emptyset\). For the second component, \(\cod(\rho_{\mathrm{loc},[f_{\mathrm{cur}}^*,h]})\) is either unchanged or becomes  
\[
\cod(\rho'_{\mathrm{loc},[f_{\mathrm{cur}}^*,h]}) \cup \rho_{\mathrm{loc},f_{\mathrm{cur}}}(x). 
\] In the latter case, \(\rho_{\mathrm{loc},f_{\mathrm{cur}}}(x) \subseteq \Qpool'\). By the validity of \(\ctx'\), we have \(\Qpool' \cap \cod(\rho'_{\mathrm{loc},h}) = \emptyset\). It follows that \(\rho_{\mathrm{loc},f_{\mathrm{cur}}}(x) \cap \cod(\rho'_{\mathrm{loc},h}) = \emptyset\). Hence the intersection is empty in all cases. Therefore the condition follows directly from the validity of \(\ctx'\).

  \item[4(a)(2)]
 We must check
  \[
  \forall q \in V_{\activelabel,f_{\mathrm{cur}}} \cup \rho_{\mathrm{loc},f_{\mathrm{cur}}}(\re(f_{\mathrm{cur}})),\;
  \var(\idx(q)) = \emptyset \;\Longrightarrow\; \base[q] = q .
  \] 

  Observe that
  \[
  \begin{aligned}
  V_{\activelabel,f_{\mathrm{cur}}}
  &= V'_{\activelabel,f_{\mathrm{cur}}} \cup \bigl( D_1 \setminus ( \rho'_{\mathrm{loc},f_{\mathrm{cur}}}(\fp(f_{\mathrm{cur}})) \cup \cod(\rho_{\mathrm{glob}}) ) \bigr) \\
  &\subseteq V'_{\activelabel,f_{\mathrm{cur}}} \cup \bigl( \rho_{\mathrm{loc},f_{\mathrm{cur}}}(x) \setminus \rho'_{\mathrm{loc},f_{\mathrm{cur}}}(\fp(f_{\mathrm{cur}})) \bigr) \cup Gar_{f_{\mathrm{cur}}} .
  \end{aligned}
  \]

 Now take any \(q \in V_{\activelabel,f_{\mathrm{cur}}} \cup \rho_{\mathrm{loc},f_{\mathrm{cur}}}(\re(f_{\mathrm{cur}}))\).
By definition, every element of \(Gar_{f_{\mathrm{cur}}}\) carries a non‑empty variables in its indics,
so it suffices to consider \(q \notin Gar_{f_{\mathrm{cur}}}\).
In this case \(q\) falls into one of the two possibilities obtained from the above inclusion (which follows the same argument as the proof for this condition in Lemma~\ref{lem:state-valid-4.a}): 
  \begin{enumerate}
  \item \(q \in V'_{\activelabel,f_{\mathrm{cur}}} \cup \rho'_{\mathrm{loc},f_{\mathrm{cur}}}(\re(f_{\mathrm{cur}}))\): the property follows directly from the validity of \(\ctx'\).
  \item \(q \in \rho_{\mathrm{loc},f_{\mathrm{cur}}}(x) \setminus \rho'_{\mathrm{loc},f_{\mathrm{cur}}}(\fp(f_{\mathrm{cur}})) \cup \rho'_{\mathrm{loc},f_{\mathrm{cur}}}(\re(f_{\mathrm{cur}}))\).
        \begin{enumerate}
        \item If \(x \in \dom(\rho'_{f_{\mathrm{cur}}})\), then
              \(\rho_{\mathrm{loc},f_{\mathrm{cur}}}(x) = \rho'_{\mathrm{loc},f_{\mathrm{cur}}}(x)\).
              The validity of \(\ctx'\) gives
              \[
              \rho'_{\mathrm{loc},f_{\mathrm{cur}}}(x) \setminus \rho'_{\mathrm{loc},f_{\mathrm{cur}}}(\fp(f_{\mathrm{cur}})) \cup \rho'_{\mathrm{loc},f_{\mathrm{cur}}}(\re(f_{\mathrm{cur}}))
              \subseteq V'_{\activelabel,f_{\mathrm{cur}}} \cup \rho'_{\mathrm{loc},f_{\mathrm{cur}}}(\re(f_{\mathrm{cur}})) ,
              \]
              and the conclusion follows.
        \item If \(x \notin \dom(\rho'_{f_{\mathrm{cur}}})\), then
              \(\rho_{\mathrm{loc},f_{\mathrm{cur}}}(x) \in \Qpool'\).
              By the validity of \(\ctx'\), every \(r \in \Qpool'\) with \(\var(\idx(r)) = \emptyset\) satisfies \(\base[r] = r\);
              hence the property holds for \(q\).
        \end{enumerate}
  \end{enumerate} 

  \item[4(a)(3)] We prove \[
  \forall q \in V_{\activelabel,\succeq f_{\mathrm{cur}}^*} \cap \rho_{\mathrm{loc},f_{\mathrm{cur}}}(\re(f_{\mathrm{cur}})),\;
  \var(\idx(q)) = \emptyset \;\Longrightarrow\; q \in  \rho_{\mathrm{loc},h}(\re(h)).
  \]  
For \(h \neq f_{\mathrm{cur}}\), the mapping \(\rho_{\mathrm{loc},h}\) is unchanged. For \(h = f_{\mathrm{cur}}\), the previous analysis yields \(\rho_{\mathrm{loc},f_{\mathrm{cur}}}(\re(f_{\mathrm{cur}})) = \rho'_{\mathrm{loc},f_{\mathrm{cur}}}(\re(f_{\mathrm{cur}}))\); hence the implication is guaranteed by the validity of \(\ctx'\). 
\end{enumerate}
\end{proof}

\begin{theorem}
\label{state_valid_4.b}
Under Notations~\ref{note:state_1},~\ref{note:state_2} and Assumption~\ref{hpy:com}, Condition~4(b) holds. 
\end{theorem}
\begin{proof}
    Take any \(h \in \SCC(f_{\mathrm{cur}}) \cap \Fun(\St)\). 
    \begin{enumerate}
  \item[4(b)(1)] We first prove
  \[
  V_{\activelabel,h} \setminus Gar_h \subseteq \cod(\rho_{\mathrm{loc},h}) .
  \]
  By definition, for every \(h \in \SCC(f_{\mathrm{cur}}) \cap \Fun(\St)\) we have
  \(Gar'_h \subseteq Gar_h\) since \(\rho'_{\mathrm{loc},\succeq f_{\mathrm{cur}}^*} \subseteq \rho_{\mathrm{loc},\succeq f_{\mathrm{cur}}^*}\). 
  For \(h \neq f_{\mathrm{cur}}\), both \(V_{\activelabel,h}\) and \(\cod(\rho_{\mathrm{loc},h})\) are unchanged,
  so the statement remains true for non-current functions.
  We now consider only \(h = f_{\mathrm{cur}}\).

  Since \(D_1 \subseteq \rho_{f_{\mathrm{cur}}}(x) \cup Gar_{f_{\mathrm{cur}}}\),
  \[
  D_1 \setminus \bigl( \rho'_{\mathrm{loc},f_{\mathrm{cur}}}(\fp(f_{\mathrm{cur}})) \cup \rho'_{\mathrm{glob}} \bigr)
  \subseteq \bigl( \rho_{f_{\mathrm{cur}}}(x) \setminus ( \rho'_{\mathrm{loc},f_{\mathrm{cur}}}(\fp(f_{\mathrm{cur}})) \cup \rho'_{\mathrm{glob}} ) \bigr) \cup Gar_{f_{\mathrm{cur}}}
  \subseteq \rho_{\mathrm{loc},f_{\mathrm{cur}}}(x) \cup Gar_{f_{\mathrm{cur}}} .
  \]

  Using the definition and the validity of \(\ctx'\),
  \[
  \begin{aligned}
  V_{\activelabel,f_{\mathrm{cur}}} \setminus Gar_{f_{\mathrm{cur}}}
  &= \Bigl( V'_{\activelabel,f_{\mathrm{cur}}} \cup \bigl( D_1 \setminus ( \rho'_{\mathrm{loc},f_{\mathrm{cur}}}(\fp(f_{\mathrm{cur}})) \cup \rho'_{\mathrm{glob}} ) \bigr) \Bigr) \setminus Gar_{f_{\mathrm{cur}}} \\
  &\subseteq \bigl( V'_{\activelabel,f_{\mathrm{cur}}} \cup (\rho_{\mathrm{loc},f_{\mathrm{cur}}}(x) \cup Gar_{f_{\mathrm{cur}}}) \bigr) \setminus Gar_{f_{\mathrm{cur}}} \\
  &\subseteq (V'_{\activelabel,f_{\mathrm{cur}}} \setminus Gar_{f_{\mathrm{cur}}}) \cup \rho_{\mathrm{loc},f_{\mathrm{cur}}}(x) \\
  &\subseteq (V'_{\activelabel,f_{\mathrm{cur}}} \setminus Gar'_{f_{\mathrm{cur}}}) \cup \rho_{\mathrm{loc},f_{\mathrm{cur}}}(x)
        \qquad (\text{since } Gar'_{f_{\mathrm{cur}}} \subseteq Gar_{f_{\mathrm{cur}}}) \\
  &\subseteq \cod(\rho'_{\mathrm{loc},f_{\mathrm{cur}}}) \cup \rho_{\mathrm{loc},f_{\mathrm{cur}}}(x) \\
  &= \cod(\rho_{\mathrm{loc},f_{\mathrm{cur}}}).
  \end{aligned}
  \] 

  \item[4(b)(2)] We prove \(\forall\, h \neq f_{\mathrm{cur}},\; Q_{\succeq f_{\mathrm{cur}}^*}[+] \cap V_{\activelabel,h} = \emptyset.\) The proof is identical to the proof of the same condition in Lemma~\ref{lem:state-valid-4.b}. 

% For \(h \neq \curf\), by definition \(V_{\activelabel,h} = ucv'_h\).
% Moreover, from the definition of \(Q\) we obtain
% \[
% Q_{\succeq \curf^*}[+] \subseteq Q'_{\succeq \curf^*}[+] \cup \bigl( \rho_\curf(x)[+] \bigr) .
% \]
% The validity of \(ctx'\) already gives
% \(Q'_{\succeq \curf^*}[+] \cap ucv'_h = \emptyset\); hence it remains to show
% \[
% \rho_\curf(x)[+] \cap ucv'_h = \emptyset .
% \]

% If \(x \in \dom(\rho')\), then \(Q_{\succeq \curf^*}[+] = Q'_{\succeq \curf^*}[+]\), and the claim trivially holds.
% Otherwise, \(x \notin \dom(\rho')\) and therefore \(\rho_\curf(x) \subseteq \xi'\). 
% Consequently,
% \[
% \rho_\curf(x)[+] \subseteq \xi'[+] .
% \]
% The validity of \(ctx'\) ensures \(\xi'[+] \cap ucv'_h = \emptyset\), which implies the desired intersection is empty.
% Thus the conclusion follows. 

\item[4(b)(3)] Assume
  \[
  Q_{\succeq f_{\mathrm{cur}}^*}[+] \cap V_{\activelabel,f_{\mathrm{cur}}} \neq \emptyset .
  \]
  We must prove
  \[
  \exists\, y \in \dom(\rho_{\mathrm{loc},f_{\mathrm{cur}}}),\;
  y.\flag = 2 \;\wedge\; \rho_{\mathrm{loc},f_{\mathrm{cur}}}(y) \in V_{\activelabel,f_{\mathrm{cur}}}
  \;\wedge\; Gar_{f_{\mathrm{cur}}} \subseteq V_{\activelabel,f_{\mathrm{cur}}}.
  \]

  By construction,
  \[
  V_{\activelabel,f_{\mathrm{cur}}} = V'_{\activelabel,f_{\mathrm{cur}}} \cup \bigl( D_1 \setminus ( \rho'_{\mathrm{loc},f_{\mathrm{cur}}}(\fp(f_{\mathrm{cur}})) \cup \rho'_{\mathrm{glob}} ) \bigr) .
  \]
  Hence the non‑emptiness assumption implies that at least one of the following holds:
  \begin{enumerate}
    \item \(Q_{\succeq f_{\mathrm{cur}}^*}[+] \cap V'_{\activelabel,f_{\mathrm{cur}}} \neq \emptyset\);
    \item \(Q_{\succeq f_{\mathrm{cur}}^*}[+] \cap \bigl( D_1 \setminus ( \rho'_{\mathrm{loc},f_{\mathrm{cur}}}(\fp(f_{\mathrm{cur}})) \cup \rho_{\mathrm{glob}} ) \bigr) \neq \emptyset\).
  \end{enumerate}

  \noindent\textbf{Case~(a).}
  From the inclusion
  \[
  Q_{\succeq f_{\mathrm{cur}}^*}[+] \subseteq Q'_{\succeq f_{\mathrm{cur}}^*}[+] \cup \bigl( \rho_{\mathrm{loc},f_{\mathrm{cur}}}(x)[+] \bigr) ,
  \]
  we consider two subcases:
  \begin{itemize}
    \item If \(x \in \dom(\rho'_{f_{\mathrm{cur}}})\), then \(Q_{\succeq f_{\mathrm{cur}}^*}[+] = Q'_{\succeq f_{\mathrm{cur}}^*}[+]\).
          The hypothesis becomes \(Q'_{\succeq f_{\mathrm{cur}}^*}[+] \cap V'_{\activelabel,f_{\mathrm{cur}}} \neq \emptyset\).
    \item If \(x \notin \dom(\rho'_{f_{\mathrm{cur}}})\), then \(\rho_{\mathrm{loc},f_{\mathrm{cur}}}(x) \subseteq \Qpool'\);
          consequently \(\rho_{\mathrm{loc},f_{\mathrm{cur}}}(x)[+] \subseteq \Qpool'[+]\).
          The validity of \(\ctx'\) gives \(\Qpool'[+] \cap V'_{\activelabel,f_{\mathrm{cur}}} = \emptyset\),
          so we must have \(Q'_{\succeq f_{\mathrm{cur}}^*}[+] \cap V'_{\activelabel,f_{\mathrm{cur}}} \neq \emptyset\). 
  \end{itemize}
Since \(Q'_{\succeq f_{\mathrm{cur}}^*}[+] \cap V'_{\activelabel,f_{\mathrm{cur}}} \neq \emptyset\), together with the validity of \(\ctx'\), it follows that there exists \(y \in \dom(\rho'_{\mathrm{loc},f_{\mathrm{cur}}})\) such that
\[
y.\flag = 2,\quad \rho'_{\mathrm{loc},f_{\mathrm{cur}}}(y) \in V'_{\activelabel,f_{\mathrm{cur}}},\quad Gar'_{f_{\mathrm{cur}}} \subseteq V'_{\activelabel,f_{\mathrm{cur}}}.
\]
Moreover, since there is at most one variable with flag=\(2\), for any \(x \neq y\), we have \(x.\flag \neq 2\), which implies \(Gar'_{f_{\mathrm{cur}}} = Gar_{f_{\mathrm{cur}}}\). Together with \(\rho'_{\mathrm{loc},f_{\mathrm{cur}}}(y) = \rho_{\mathrm{loc},f_{\mathrm{cur}}}(y)\) and \(V'_{\activelabel,f_{\mathrm{cur}}} \subseteq V_{\activelabel,f_{\mathrm{cur}}}\), we obtain
\[
\exists\, y \in \dom(\rho_{\mathrm{loc},f_{\mathrm{cur}}}),\quad
y.\flag = 2 \;\wedge\; \rho_{\mathrm{loc},f_{\mathrm{cur}}}(y) \in V_{\activelabel,f_{\mathrm{cur}}} \;\wedge\; Gar_{f_{\mathrm{cur}}} \subseteq V_{\activelabel,f_{\mathrm{cur}}}.
\] 
  \noindent\textbf{Case~(b).}
  \[
  Q_{\succeq f_{\mathrm{cur}}^*}[+] \cap \bigl( D_1 \setminus ( \rho'_{\mathrm{loc},f_{\mathrm{cur}}}(\fp(f_{\mathrm{cur}})) \cup \rho_{\mathrm{glob}} ) \bigr) \neq \emptyset .
  \]
  We distinguish two situations based on \(x.\flag\).

  \begin{itemize}
    \item \textbf{Subcase \(x.\flag \neq 2 \vee (\Call(s)=\emptyset)\).}
          Theorem~\ref{theo_com} gives \(D_1 = \rho_{f_{\mathrm{cur}}}(x)\) and \(Q_{\succeq f_{\mathrm{cur}}^*}[+] = Q'_{\succeq f_{\mathrm{cur}}^*}[+]\).
          Moreover,
          \[
          D_1 \setminus ( \rho'_{\mathrm{loc},f_{\mathrm{cur}}}(\fp(f_{\mathrm{cur}})) \cup \rho_{\mathrm{glob}} )
          \subseteq \rho_{\mathrm{loc},f_{\mathrm{cur}}}(x) .
          \]
          We show that \(\rho_{\mathrm{loc},f_{\mathrm{cur}}}(x) \cap Q'_{\succeq f_{\mathrm{cur}}^*}[+] = \emptyset\);
          therefore this subcase cannot occur.

          \begin{itemize}
            \item If \(x \notin \dom(\rho'_{f_{\mathrm{cur}}})\), then \(\rho_{\mathrm{loc},f_{\mathrm{cur}}}(x) \in \Qpool'\).
                  Condition~2 in the validity of \(\ctx'\) yields \(\Qpool' \cap Q'_{\succeq f_{\mathrm{cur}}^*}[+] = \emptyset\).
            \item If \(x \in \dom(\rho'_{f_{\mathrm{cur}}})\), then Condition~2 of \(\ctx'\) similarly gives
                  \(\rho'_{\mathrm{loc},f_{\mathrm{cur}}}(x) \cap Q'_{\succeq f_{\mathrm{cur}}^*}[+] = \emptyset\).
          \end{itemize}
          In either case we obtain \(\rho_{\mathrm{loc},f_{\mathrm{cur}}}(x) \cap Q'_{\succeq f_{\mathrm{cur}}^*}[+] = \emptyset\),
          contradicting the assumption.  Hence this subcase is impossible.

    \item \textbf{Subcase \(x.\flag = 2 \wedge (\Call(s)\neq \emptyset)\).}  By the definition, we have 
          \[
          V_{\activelabel,f_{\mathrm{cur}}} = V'_{\activelabel,f_{\mathrm{cur}}} \cup D_1, 
          \]  
          and \(D_1 = \rho_{\mathrm{loc},f_{\mathrm{cur}}}(x) \cup Gar_{f_{\mathrm{cur}}}\). 
          Hence by the definition of \(D_1\) we already have
          \(y = x \in \dom(\rho_{\mathrm{loc},f_{\mathrm{cur}}})\) with
          \(x.\flag = 2\), \(\rho_{\mathrm{loc},f_{\mathrm{cur}}}(x) \in V_{\activelabel,f_{\mathrm{cur}}}\),
          and \(Gar_{f_{\mathrm{cur}}} \subseteq D_1 \subseteq V_{\activelabel,f_{\mathrm{cur}}}\).
          The required conclusion follows.
  \end{itemize}
\end{enumerate}
\end{proof}

\begin{theorem}
\label{state_valid_4.f}
Under Notations~\ref{note:state_1},~\ref{note:state_2} and Assumption~\ref{hpy:com}, Condition~4(f) holds. 
\end{theorem}
\begin{proof}
     Take any \(h \in \SCC(f_{\mathrm{cur}}) \cap \Fun(\St)\). Similarly, we only need to check for \(h = f_{\mathrm{cur}}\). 
From the validity of \(\ctx'\) we have
\(\bigl( \cod(\rho'_{\mathrm{loc},f_{\mathrm{cur}}}) \setminus (\rho'_{\mathrm{loc},f_{\mathrm{cur}}}(\fp(f_{\mathrm{cur}})) \cup \rho'_{\mathrm{loc},f_{\mathrm{cur}}}(\re(f_{\mathrm{cur}}))) \bigr) \subseteq V'_{\alloc,f_{\mathrm{cur}}}\).
By Theorem~\ref{theo_com}, we have for any \(z\),
\(\base(\qv(C_s^z) \setminus \cod(\rho'_{\mathrm{loc},f_{\mathrm{cur}}})) \subseteq D_2 \cup \base(\anc(F_{\succeq f_{\mathrm{cur}}^*}^z))\).
Combining these, for any \(z\),
\[
\begin{aligned}
\anc(C_s^z)
   &= \qv(C_s^z) \setminus \bigl( \rho_{\mathrm{loc},f_{\mathrm{cur}}}(\fp(f_{\mathrm{cur}})) \cup \rho_{\mathrm{loc},f_{\mathrm{cur}}}(\re(f_{\mathrm{cur}})) \bigr) \\
   &= \bigl( \qv(C_s^z) \setminus \cod(\rho'_{\mathrm{loc},f_{\mathrm{cur}}}) \bigr)
      \cup \bigl( \cod(\rho'_{\mathrm{loc},f_{\mathrm{cur}}}) \setminus (\rho'_{\mathrm{loc},f_{\mathrm{cur}}}(\fp(f_{\mathrm{cur}})) \cup \rho'_{\mathrm{loc},f_{\mathrm{cur}}}(\re(f_{\mathrm{cur}}))) \bigr) \\
   &\subseteq \bigl( D_2 \cup \base(\anc(F_{\succeq f_{\mathrm{cur}}^*}^z)) \bigr) \cup V'_{\alloc,f_{\mathrm{cur}}}.
\end{aligned}
\]

From the validity of \(\ctx'\) we also have
\[
\base\bigl( \anc(\ctx'.C_{f_{\mathrm{cur}}}^z) \bigr)
   \subseteq V'_{\alloc,\succeq f_{\mathrm{cur}}^*} \cup \base\bigl( \anc(F_{\succeq f_{\mathrm{cur}}^*}^z) \bigr).
\]

Thus, for any \(z\),
\[
\begin{aligned}
\base\bigl( \anc(\ctx.C_{f_{\mathrm{cur}}}^z) \bigr)
   &= \base\bigl( \anc(\ctx'.C_{f_{\mathrm{cur}}}^z) \cup \anc(C_s^z) \bigr) \\
   &= \base\bigl( \anc(\ctx'.C_{f_{\mathrm{cur}}}^z) \bigr) \cup \base\bigl( \anc(C_s^z) \bigr) \\
   &\subseteq \bigl( V'_{\alloc,\succeq f_{\mathrm{cur}}^*} \cup \base(\anc(F_{\succeq f_{\mathrm{cur}}^*}^z)) \bigr)
      \cup \bigl( D_2 \cup \base(\anc(F_{\succeq f_{\mathrm{cur}}^*}^z)) \cup V'_{\alloc,f_{\mathrm{cur}}} \bigr) \\
   &= V_{\alloc,\succeq f_{\mathrm{cur}}^*} \cup \base\bigl( \anc(F_{\succeq f_{\mathrm{cur}}^*}^z) \bigr),
\end{aligned}
\]
using the definition \(V_{\alloc,f_{\mathrm{cur}}} = V'_{\alloc,f_{\mathrm{cur}}} \cup D_2\) and the fact that \(V_{\alloc,h} = V'_{\alloc,h}\) for all \(h \neq f_{\mathrm{cur}}\), which yields \(V_{\alloc,\succeq f_{\mathrm{cur}}^*} = V'_{\alloc,\succeq f_{\mathrm{cur}}^*} \cup D_2 \cup V'_{\alloc,f_{\mathrm{cur}}}\).

\end{proof}

\begin{theorem}
\label{theo:com_valid}
Under Notations~\ref{note:state_1},~\ref{note:state_2} and Assumption~\ref{hpy:com}, we have
\[
(\ctx, (\ket{\phi}, k, \D_{\SCC(f_{\mathrm{cur}})})) \in \mathcal{V},
\]
where the validity condition holds except possibly for the correctness of the cleanup circuit in Condition~3(a) concerning \(f_{\mathrm{cur}}\). 
\end{theorem}

\begin{proof} 
We prove the properties by checking each condition one by one.

\textbf{Condition~1.} Since we have already established that Condition~1 holds for \(\ctx'\), and since the step from \(\ctx'\) to \(\ctx\) does not alter any of the compilation variables involved in this condition, Condition~1 remains valid for \(\ctx\).

\textbf{Condition~2.}  By Lemma~\ref{state_valid_2}. 

\textbf{Condition~3(a).} The verification of this condition (apart from the correctness of \(\kappa_{f_{\mathrm{cur}}}\)) follows exactly the same argument as the proof of Condition~3(a) for \(\ctx_1\) in Lemma~\ref{lem:state-valid-3.a}.

% If \((x \setminus z) \in \dom(\rho'_l)\), then \(\rho_l = \rho'_l\); hence \(\rho_l\) remains injective by the validity of \(ctx'\).
% If \((x \setminus z) \notin \dom(\rho'_l)\), observe that \(\rho'_l(x \setminus z) \in \xi'\).  By the validity of \(ctx'\) we have \(\xi' \cap \cod(\rho'_l) = \emptyset\), and by definition
% \[
% \rho_l = \rho'_l\bigl[(x \setminus z) \mapsto \rho_l(x \setminus z)\bigr],
% \]
% so \(\rho_l\) is again injective.
% Moreover, the graph \(G_f\) and the residual statements \(R_{\mathrm{stmt}}\) are unchanged; therefore the validity conditions they satisfy follow directly from the validity of \(ctx'\).

\textbf{Condition~3(b).}  
By Lemma~\ref{state_valid_3.b}. 

\textbf{Condition~3(c).}  
By Lemma~\ref{state_valid_3.c}. 

\textbf{Condition~4(a).}  
By Lemma~\ref{state_valid_4.a}.

\textbf{Condition~4(b).}  
By Lemma~\ref{state_valid_4.b}. 

\textbf{Condition~4(c).} For any \(h \neq f_{\mathrm{cur}}\), the code \(C_h\) remains unchanged and Lemma~\ref{theo_com} gives \(\ctx'.A_h \subseteq \ctx.A_h\).  By the validity of \(\ctx'\), we have \(\anc(C_h^k) \subseteq \ctx'.A_h \subseteq \ctx.A_h\), so the condition holds for such \(h\).

For \(f = f_{\mathrm{cur}}\), we have \(\ctx.C_{f_{\mathrm{cur}}} = \ctx'.C_{f_{\mathrm{cur}}}; C_s\).  Hence
\[
\anc(\ctx.C_{f_{\mathrm{cur}}}^k) = \anc(\ctx'.C_{f_{\mathrm{cur}}}^k) \cup \anc(C_s^k).
\]
The validity of \(\ctx'\) yields \(\anc(\ctx'.C_{f_{\mathrm{cur}}}^k) \subseteq \ctx'.A_{f_{\mathrm{cur}}}\).  
Moreover, the construction of \(C_s\) ensures \(\anc(C_s^k)  \subseteq \ctx'.A_{f_{\mathrm{cur}}}\).  
Since \(\ctx.A_{f_{\mathrm{cur}}} = \ctx'.A_{f_{\mathrm{cur}}}\) by Lemma~\ref{theo_com}, we obtain
\[
\anc(\ctx.C_{f_{\mathrm{cur}}}^k) \subseteq \ctx.A_{f_{\mathrm{cur}}},
\]
which is exactly Condition~4(c).  

\textbf{Condition~4(d).} For any \(h \neq f_{\mathrm{cur}}\), the code \(C_h\) is unchanged and \(\ctx.B = \ctx'.B\); hence the required property follows directly from the validity of \(\ctx'\).
For \(h = f_{\mathrm{cur}}\) we have \[
\anc(\ctx.C_{f_{\mathrm{cur}}}^k) = \anc(\ctx'.C_{f_{\mathrm{cur}}}^k) \cup \anc(C_s^k). \]
Since 
\[
\anc(C_s^k) \setminus \bigl( \anc(F_{\succeq f_{\mathrm{cur}}^*}^k) \cup \anc(F_{\succeq f_{\mathrm{cur}}^*}^k)[+] \bigr)
\subseteq \Qpool' \cup \cod(\rho'_{\mathrm{loc},f_{\mathrm{cur}}}),
\]
the same analysis as in the cleanup-algorithm case (Lemma~\ref{lemma:con-4.d}) yields
\[
\bigl( \anc(C_s^k) \setminus \bigl( \anc(F_{\succeq f_{\mathrm{cur}}^*}^k) \cup \anc(F_{\succeq f_{\mathrm{cur}}^*}^k)[+] \bigr) \bigr)
\cap \ctx'.B[+] = \emptyset .
\]
The validity of \(\ctx'\) already gives
\[
\bigl(\anc(\ctx'.C_{f_{\mathrm{cur}}}^k) \setminus
\bigl( \anc(F_{\succeq f_{\mathrm{cur}}^*}^k) \cup \anc(F_{\succeq f_{\mathrm{cur}}^*}^k)[+] \bigr)\bigr) \cap \ctx'.B[+] = \emptyset .
\]
Together with \(\ctx.B = \ctx'.B\), we obtain the required conclusion. 

\textbf{Condition~4(e).}  
By Lemma~\ref{theo_com}, we have \(\ctx.A_{f_{\mathrm{cur}}^*} = \ctx'.A_{f_{\mathrm{cur}}^*}\).  
By definition, \(E_{\args,h} = E'_{\args,h}\) and \(E_{\args,f_{\mathrm{cur}}^*} = E'_{\args,f_{\mathrm{cur}}^*}\).  
The inclusion
\[
\var(\idx(\ctx.A_{f_{\mathrm{cur}}^*})) \subseteq E_{\args,f_{\mathrm{cur}}^*} = E_{\args,h}
\]
follows directly from the corresponding property for \(\ctx'\), namely
\(\var(\idx(\ctx'.A_{f_{\mathrm{cur}}^*})) \subseteq E'_{\args,f_{\mathrm{cur}}^*} = E'_{\args,h}\).

\textbf{Condition~4(f).} By Lemma~\ref{state_valid_4.f}.  

\textbf{Condition~4(g).} \(\Succ(f) \setminus \Call(R_{\mathrm{stmt},f}) \cap \SCC(f) \subseteq \Fun(\St)\): Since \(\St=\St'\) and \(R_{\mathrm{stmt},f}=RS'_{\mathrm{stmt},f}\) by the definition, the condition follows directly from the validity of \(\ctx'\).  

Therefore, we obtain 
$(\ctx, (\ket{\phi}, k, \D_{\SCC(f_{\mathrm{cur}})})) \in \mathcal{V}$. 
\end{proof}

\subsection{Correctness Proof of Main Theorem~\ref{theo:correctness-main}}
\label{app:proof-main}
This subsection proves the semantic equivalence between the source main program
  \(s_{\main}\) and the generated target main program \(C_{\main}\).
  \rev{Let \(\Xi,\Gamma,\G\) be the static parameters of the source program,
  and let \(\D\) be the target declaration family generated by
  \[
  \D=\textsc{Compile\_Prog}(\Xi,\Gamma,\G).
  \]}
  Let \(\Sigma\) denote the state space of source states whose domains consist of
  the global variables, and let \(\mathcal{H}\) denote the Hilbert space of the
  target quantum system.

  \rev{
  \begin{lemma}[Lemma~\ref{lem:semantic-determinism}]
  \label{lem:semantic-determinism-app}
  The operational semantics of both RQIMP and \(\RQC^{++}\) are deterministic.

  For RQIMP, let \(s\) be an RQIMP statement and let
  \(\sigma,\sigma_1,\sigma_2\) be source states. If
  \[
  \langle s,\sigma\rangle \to_{\Xi}^{*}
  \langle\downarrow,\sigma_1\rangle
  \quad\text{and}\quad
  \langle s,\sigma\rangle \to_{\Xi}^{*}
  \langle\downarrow,\sigma_2\rangle,
  \]
  then \(\sigma_1=\sigma_2\).

  For \(\RQC^{++}\), let \(C\) be an \(\RQC^{++}\) program and let
  \(\ket{\psi},\ket{\phi_1},\ket{\phi_2}\) be target quantum states. If
  \[
  \langle C,\ket{\psi}\rangle \rightarrow_{\D}^{*}
  \langle\downarrow,\ket{\phi_1}\rangle
  \quad\text{and}\quad
  \langle C,\ket{\psi}\rangle \rightarrow_{\D}^{*}
  \langle\downarrow,\ket{\phi_2}\rangle,
  \]
  then \(\ket{\phi_1}=\ket{\phi_2}\).
  \end{lemma}

  \begin{proof}
  The result follows by induction on the operational derivations. All expression
  evaluations and primitive operations are deterministic; classical conditionals
  select a unique branch, quantum conditionals have a uniquely defined coherent
  evolution, and every procedure name has a unique declaration. Sequential
  composition preserves determinism.
  \end{proof}
  }

  \begin{lemma}[Finite-approximant termination reflection]
  \label{lem:finite-reflection}
   Let \(\sigma_0\in\Sigma\) and
  \[
  \ket{\psi_0}
  =
  \mathcal{E}(\sigma_0,\rho_{\mathrm{glob}})
  \otimes\ket{0}_{\mathrm{rem}}
  \in\mathcal{H}.
  \]
  For every finite \(k\), if
  \[
  \langle C_{\main}^{\,k},\ket{\psi_0}\rangle
  \rightarrow_{\D}^{*}
  \langle\downarrow,\ket{\psi}\rangle
  \]
  for some target state \(\ket{\psi}\), then there exists a source state
  \(\sigma\) such that
  \[
  \langle s_{\main}^{\,k},\sigma_0\rangle
  \to_{\Xi}^{*}
  \langle\downarrow,\sigma\rangle.
  \]
  \end{lemma}

  \begin{proof}[Proof sketch]
 We give only the main proof idea here. The induction structure and proof method
  are similar to those used in
  Subsections~\ref{app:proof-cleanup}--\ref{app:proof-com}; the difference is that we
  \rev{simultaneously maintain termination reflection for finite syntactic
  approximants alongside the existing validity invariant, thereby obtaining a
  strengthened induction invariant}. More specifically, we
  first establish termination reflection for the function-body compilation
  algorithm \textsc{Compile\_Body}. Assume that the \(k\)-th finite approximants
  of all functions in the SCC containing the current function already satisfy
  termination reflection. We prove that, for source and target states satisfying
  the preceding validity and reachable statement-boundary state-correspondence
  conditions, if the \((k+1)\)-st finite approximant of the target program
  corresponding to a source function body \(s\) terminates, then the
  \((k+1)\)-st finite approximant of \(s\) also terminates. This result is proved
  by structural induction on \(s\).

  \begin{itemize}
    \item If \(s\) is a basic statement, such as an expression assignment or an
    in-place update, then both the source semantics and the finite target circuit
    generated by compilation terminate, so the result follows directly.

    \item If \(s\) is a sequential composition \(s_1;s_2\) or a conditional
    statement, the result follows by inversion of the terminating target
    execution, together with the structural induction hypothesis, the existing
    forward-correctness results, the state-correspondence relation, and
    \rev{the determinism of the source and target operational semantics
    (Lemma~\ref{lem:semantic-determinism-app}), which identifies the intermediate
    states obtained by inversion with the corresponding states supplied by
    forward correctness}.

    \item If \(s\) \rev{is a function-call statement invoking} \(f\), we distinguish the same
    compilation cases as in
    Subsections~\ref{app:proof-cleanup}--\ref{app:proof-com}:
    \begin{itemize}
      \item If the compilation of \(f\) has already been completed, the result
      follows from the termination-reflection property for completed functions
      recorded in \rev{the strengthened induction invariant}.

      \item If the call to \(f\) is recursive, we use the simultaneous induction
      hypothesis at unfolding depth \(k\) for all functions in the current SCC.

      \item If \(f\) has not yet been compiled \rev{and the call is nonrecursive}, we recursively apply the
      corresponding termination-reflection property of the function-compilation
      algorithm. As in
      Subsections~\ref{app:proof-cleanup}--\ref{app:proof-com}, this recursive argument is well founded
      with respect to the number of source functions that have not yet been
      compiled. When the compilation of the SCC containing \(f\) is completed,
      we perform a simultaneous induction on the unfolding depth \(z\) for all
      functions in that SCC. When \(z=0\), the result is vacuous because the
      zeroth target approximant is \(\mathsf{abort}\). For the induction step,
      if the \(z\)-th target approximant of a function call terminates, inversion
      shows that the \((z-1)\)-st target approximant of its function body
      terminates. By the induction hypothesis, the corresponding
      \((z-1)\)-st source approximant of the function body also terminates, and
      hence the \(z\)-th source approximant of the function call terminates.
      The resulting property is then recorded as \rev{the termination-reflection
      component of the strengthened induction invariant for the completed SCC}.
    \end{itemize}
  \end{itemize}

  Finally, instantiating this property with the main function body, the initial
  compilation context \(\ctx_0\), and the corresponding encoding of the initial
  state proves the lemma. 
  \end{proof}

\begin{theorem}[Theorem~\ref{theo:correctness-main}]
\label{theo:main}
\rev{The compiled target program is equivalent to the source program:}
\(
C_{\mathtt{main}} \;\approx\; s_{\mathtt{main}} .
\)
\end{theorem}
\begin{proof}
Let $\sigma_0 \in \Sigma$ be any initial global source state over the global
variables \(\overline{z}\). Choose the quantum initial state
$\ket{\psi_0}\in \mathcal{H}$ such that on the global variables it satisfies
\[
\ket{\psi_0}_{\cod(\rho_{\mathrm{glob}})} = \mathcal{E}(\sigma_0, \rho_{\mathrm{glob}}),
\]
and on \(\Qpool\) and its higher-index extensions it is $\ket{0}$.

    We consider two cases. First, suppose that the source program terminates:
    \(
      \langle s_{\main},\sigma_0\rangle
      \to_{\Xi}^{*}
      \langle\downarrow,\sigma\rangle
    \)
    for some \(\sigma\). By finite-unfolding adequacy, there exists
    \(k \geq 1\) such that
    \(
      \langle s_{\main}^{\,k},\sigma_0\rangle
      \to_{\Xi}^{*}
      \langle\downarrow,\sigma\rangle.
    \)   We use the standard entry-point convention that \(\main\) is the unique top-level
entry function and is not contained in any directed cycle of the call graph.
Hence no recursive induction hypothesis is needed for its component; following
the convention in Subsection~\ref{app:nota-def}, we take the auxiliary specification
family to be empty.  

The compilation process begins by initializing the static parameters \(\K\) and the compilation context; the resulting context is denoted by \(\ctx_0\), where \(\ctx_0.C=I\) and \(\rho_{\mathrm{loc}} = \emptyset\). Similar to the proof of Lemma~\ref{lemma:step_1}, we can show that this input context is valid and semantically consistent with $\sigma_0$, i.e.,
\[
(\ctx_0, (\ket{\psi_0}, k, \emptyset)) \in \mathcal{V},
\qquad
\ctx_0 \approx_{(\ket{\psi_0}, k)} \sigma_0.
\]
Let \(\ctx = \textsc{Compile\_Body}(\K, \ctx_0, s_{\mathtt{main}})\). 
Applying the correctness theorem for function bodies (Theorem~\ref{theo:body}) yields
\[
(\ctx, (\ket{\psi_0}, k, \emptyset)) \in \mathcal{V},
\qquad
\ctx \approx_{(\ket{\psi_0}, k)} \sigma.
\]

Following the proof of Lemma~\ref{lemma:step_2}, we can also obtain the following properties for the compilation state of \(\ctx\); in what follows, we omit the \(\ctx.\) prefix when no ambiguity arises:
\[
\cod(\rho_{\mathrm{loc}}) = \rho_{\mathrm{loc}}(\fp(\mathtt{main})\cup \re(\mathtt{main})) \cup Q_{\main}.
\]

Since \(\mathtt{main}\) is not involved in any recursive component, it contains no variable with \(\flag=2\) by definition. Recall that \(Q_{\mathtt{main}}\) denotes the set of quantum variables corresponding to variables in \(\mathtt{main}\) with \(\flag=2\). Hence \(Q_{\mathtt{main}}=\emptyset\). Since $\main$ has no input or output parameters, we have \(\rho_{\mathrm{loc}} = \emptyset\). 

Moreover, the validity of \(\ctx\) also yields
\[
V_{\activelabel,\main} \subseteq Gar_{\mathtt{main}} \cup \cod(\rho_{\mathrm{loc}}) \subseteq \ctx.Q_{\succeq \main} \cup Q_{\succeq \main}[+] \cup \cod(\rho_{\mathrm{loc}}).
\]
Since \(\cod(\rho_{\mathrm{loc}}) = \emptyset\) and functions in $\SCC(\main)$ contain no variable with $\flag = 2$, indicating $Q_{\succeq \main}=\emptyset$, we have \(V_{\activelabel,\main} = \emptyset\). Taken together, these properties imply that the execution of the compiled code results in no residual uncleaned temporary variables. Expanding the definition of state equivalence $\ctx \approx_{(\ket{\psi_0}, k)} \sigma$ now gives
\[
\langle \ctx.C^{\,k}, \ket{\psi_0} \rangle \rightarrow_{\ctx.\D}^{*} 
\langle \downarrow, \mathcal{E}(\sigma, \rho_{\mathrm{glob}}) \otimes \ket{0}_{\mathrm{rem}}\rangle.
\]

By the definition of \textsc{Compile\_Prog}, we have $C_{\main}=\ctx.C$. Combining this with $\D=\ctx.\D \cup \{\Theta(\mathtt{main}) \Leftarrow \ctx.C\}$, we have
\[
\langle C_{\main}^{\,k}, \ket{\psi_0} \rangle \rightarrow_{\D}^{*} 
\langle \downarrow, \mathcal{E}(\sigma, \rho_{\mathrm{glob}}) \otimes \ket{0}_{\mathrm{rem}}\rangle.
\]
By finite-unfolding adequacy for the target language, it follows that
    \[
      \langle C_{\main},\ket{\psi_0}\rangle
      \rightarrow_{\D}^{*}
      \left\langle
        \downarrow,\,
        \mathcal{E}(\sigma,\rho_{\mathrm{glob}})
        \otimes\ket{0}_{\mathrm{rem}}
      \right\rangle.
    \]

    Second, suppose that
    \(\langle s_{\main},\sigma_0\rangle\) diverges. Assume, for contradiction,
    that \(\langle C_{\main},\ket{\psi_0}\rangle\) terminates. By
    finite-unfolding adequacy for the target language, there exists some finite
    \(k \geq 1\) such that \(C_{\main}^{\,k}\) terminates from
    \(\ket{\psi_0}\). By Lemma~\ref{lem:finite-reflection}, there exists a source
    state \(\sigma'\) such that
    \[
      \langle s_{\main}^{\,k},\sigma_0\rangle
      \to_{\Xi}^{*}
      \langle\downarrow,\sigma'\rangle.
    \]
    Finite-unfolding adequacy for the source language then implies that
    \(s_{\main}\) terminates from \(\sigma_0\), contradicting the assumption.
    Therefore \(C_{\main}\) diverges from \(\ket{\psi_0}\) as well.

    Thus, from every initial source state, the source and target programs either
    diverge together or terminate with corresponding results. Hence
    \(C_{\main} \approx s_{\main}\).

% Now let $\{\sigma_k\}_{k \geq 0}$ and $\{\ket{\psi_k}\}_{k \geq 0}$ be the
% approximation-result sequences of the source program $s_{\main}$ and the
% target program $C_{\main}$, respectively.
% Define the encoding map $f: \Sigma \to \mathcal{H}$ by
% \[
% f(\sigma) = \mathcal{E}(\sigma, \rho_{\mathrm{glob}}) \otimes \ket{0}_{\mathrm{rem}},
% \]
% where $\mathcal{E}(\sigma, \ctx.\rho_{\mathrm{glob}}) = \bigotimes_{x \in \dom(\rho_{\mathrm{glob}})} \ket{\sigma(x)}_{\rho_{\mathrm{glob}}(x)}$ is the standard computational basis encoding of the classical state on the global variable registers.

% By the approximation-depth correctness established above, we have that for every $k \geq 0$ the quantum computation sequence satisfies
% \[
% \ket{\psi_k} = f(\sigma_k).
% \]
% By Lemma~\ref{lem:f-properties-app} and Lemma~\ref{lem:convergence-app}, convergence (divergence) of the source approximation-result sequence is preserved for its quantum encoding sequence. Consequently,
% \[
% C_{\main} \approx s_{\main}.
% \]
\end{proof}

\begin{theorem}[Target Well-Formedness]
\label{theo:target-wf}
Assume that \(\Xi\) and \(\Gamma\) are obtained from type checking a well-typed RQIMP source program, and let \(\G\) be its dependency graph. Let
\[
    \D=\textsc{Compile\_Prog}(\Xi,\Gamma,\G).
\]
Then every declaration generated in \(\D\) satisfies the well-formedness conditions of \(\RQC^{++}\) stated in Appendix~\ref{app:tar}. 
\end{theorem}

\begin{proof}
The generated \(\RQC^{++}\) program must satisfy three semantic
well-formedness conditions. First, the external-coin condition requires that
the quantum guard of a quantum-controlled command is not modified by its
branches. Second, branch locality requires that classical updates inside a
quantum-controlled region remain local and do not affect the classical state
outside the region. Third, procedure locality requires that the classical state
inside a procedure body be locally scoped, so that procedure calls do not
modify non-local classical state.

These conditions are ensured jointly by the source typing rules and the
compilation process. For the external-coin condition and branch locality, the
RQIMP typing rules require that quantum guards are not modified by their
branches and that classical updates inside quantum-controlled regions are
local. Therefore, when quantum-controlled conditionals are translated into
\(\RQC^{++}\) constructs, the generated branches still satisfy the target
branch restrictions.

For procedure locality, the source function-call rules allow functions to
return only quantum variables, so procedure calls cannot affect the caller
through non-local classical return values. During function compilation, every
classical variable used in a function body is either passed as a local
classical parameter or confined by a local block in the generated target
declaration, e.g., a \texttt{begin ... end} block. Moreover, the source type
system requires all global variables to have quantum mode. Together, these
facts ensure that all classical variables accessed inside generated procedure
bodies are local to the procedure. Similarly, for quantum-controlled loops and
while-statements, the source typing rules and the high-level transformation
ensure that classical updates are confined to the generated procedure scope.
Hence, when these constructs are lifted into target procedure calls, they still
satisfy procedure locality.

Finally, the compiler maintains distinct quantum registers for target gates
through the register pool \(\Qpool\), the local mapping
\(\rho_{\mathrm{loc}}\), and the indexed-renaming discipline for recursive
calls. Hence the generated \(\RQC^{++}\) declaration family satisfies the
external-coin, branch-locality, procedure-locality, and distinct-register
requirements of \(\RQC^{++}\).
\end{proof}

\begin{theorem}[Coherent Oracle Correctness]
  \label{theo:coherent-oracle}
  The basis-state correctness in Theorem~\ref{theo:main} extends linearly to
  finite coherent superpositions of basis states encoding well-typed source stores.
  In particular, let \(\{\sigma_i\}_{i=1}^m\) be a finite family of well-typed
  source stores such that, for each \(i\),
  \[
    \langle s_{\main}, \sigma_i\rangle \to_{\Xi}^{*}
    \langle \downarrow, \sigma_i' \rangle .
  \]
  Then, for any amplitudes \(\alpha_i\),
  \[
  \sum_{i=1}^m \alpha_i\,
    \mathcal{E}(\sigma_i,\rho_{\mathrm{glob}})\otimes\ket{0}_{\mathrm{rem}}
  \]
  is mapped by the generated target program to
  \[
  \sum_{i=1}^m \alpha_i\,
    \mathcal{E}(\sigma_i',\rho_{\mathrm{glob}})\otimes\ket{0}_{\mathrm{rem}} .
  \]
\end{theorem} 

\begin{proof}
By Theorem~\ref{theo:main}, the generated target program agrees with the source
semantics on every computational-basis encoding and restores all auxiliary
registers to \(\ket{0}\). \rev{By Theorem~\ref{theo:target-wf}, the generated
\(\RQC^{++}\) program is well formed; therefore, its denotational
semantics is unitary by Appendix~\ref{app:tar}.} The action on arbitrary finite
superpositions follows by linearity.
\end{proof}

It is worth noting that the correctness proof above is carried out at the
\(\RQC^{++}\) target-language level and is therefore independent of any
particular low-level storage bound. As explained in Appendix~\ref{app:high-tr},
when the generated \(\RQC^{++}\) program is passed to the downstream QRM backend,
the backend determines the finite execution bound and provides a finite QRAM
layout sufficient for the indexed variables accessed in that bounded execution.
Thus, the proof should be read as establishing correctness for any adequate
finite interpretation of the indexed variables, namely any finite storage layout
that covers the indices actually accessed during the bounded execution;
computing such a layout is part of the QRM backend rather than ReOC.

\end{document}